\documentclass[a4paper,11pt]{book}

\usepackage{amsmath,amsfonts,amssymb,psfrag}
\usepackage{mathrsfs, dsfont}
\usepackage{multirow,array,multirow,stmaryrd}
\usepackage{comment}
\usepackage{sectsty}
\usepackage[format=hang,margin=1.5cm,font=small,labelfont=bf]{caption}
\usepackage{subfigure}
\usepackage{color}
\usepackage{ifpdf}
\ifpdf
  \usepackage[pdftex]{graphicx}
  \usepackage{epstopdf}
\else
  \usepackage[dvips]{graphicx}
\fi
\usepackage{enumerate}
\usepackage{epsfig}
\usepackage{array}
\usepackage{bbm}
\usepackage{color}
\usepackage{a4wide}
\usepackage{cite}
\usepackage{lscape}

 \usepackage{fancyhdr}
\usepackage{titlesec}

\usepackage[dvips,bookmarks,breaklinks]{hyperref}		
\usepackage[all,cmtip]{xy}

\titleformat{\chapter}[display]
{\normalfont\huge\bfseries}{\chaptertitlename\ \thechapter}{20pt}{\Huge}
\titlespacing*{\chapter} {00pt}{0pt}{15pt}

\newcommand{\beq}{\begin{equation}}
\newcommand{\eeq}{\end{equation}}
\newcommand{\be}{\begin{equation}} 
\newcommand{\ee}{\end{equation}}
\newcommand{\bea}{\begin{eqnarray}}
\newcommand{\eea}{\end{eqnarray}}   
\newcommand{\ben}{\begin{eqnarray*}}
\newcommand{\een}{\end{eqnarray*}}                  
\newcommand{\ba}{\begin{aligned}}
\newcommand{\ea}{\end{aligned}}
\newcommand{\bt}{\begin{tabular}}
\newcommand{\et}{\end{tabular}}
\newcommand{\bc}{\begin{center}}
\newcommand{\ec}{\end{center}}

\newcommand{\cO}{\mathcal{O}}
\newcommand{\cT}{\mathcal{T}}
\newcommand{\cE}{\mathcal{E}}

\newcommand{\cC}{\mathcal{C}}
\newcommand{\cD}{\mathcal{D}}
\newcommand{\cL}{\mathcal{L}}

\newcommand{\cK}{\mathcal{K}}
\newcommand{\cN}{\mathcal{N}}
\newcommand{\cW}{\mathcal{W}}
\newcommand{\cG}{\mathcal{G}}
\newcommand{\cA}{\mathcal{A}}

\newcommand{\cB}{\mathcal{B}}
\newcommand{\cF}{\mathcal{F}}

\newcommand{\cR}{\mathcal{R}}

\newcommand{\cV}{\mathcal{V}}
 
\newcommand{\KK}{\mathcal{K}}

\newcommand{\cM}{\mathcal M}

\newcommand{\I}{\text{Im}}

\DeclareMathOperator{\vol}{vol}

\newcommand{\dd}{d}

\newcommand{\bbZ}{\mathbb{Z}}
\newcommand{\bbR}{\mathbb{R}}
\newcommand{\bbC}{\mathbb{C}}
\newcommand{\bbP}{\mathbb{P}}

\newcommand{\bbL}{\mathbb{L}}

\newcommand{\nn}{\nonumber}

\newcommand{\cref}{{\bf [check ref]}}

\newcommand{\tr}{\mathrm{Tr}\:}
\newcommand{\id}{\mathbf{1}}

\newcommand{\WV}{\mathcal{W}}      

\newcommand{\norm}[1]{\lVert #1\rVert}

\newcommand{\simga}{\sigma}

\newcommand{\zI}{z^{\text{\tiny I}}}        
\newcommand{\zII}{z^{\text{\tiny II}}}

\newcommand{\Li}{{\rm Li}}
\psfrag{n1}{$\nu_1$}
\psfrag{n2}{$\nu_2$}
\psfrag{n1'}{$\nu_1'$}
\psfrag{n2'}{$\nu_2'$}
\psfrag{n9}{$\nu_9$}
\psfrag{n10'}{$\nu_{10}'$}
\psfrag{t1}{$\tilde{\nu}_1$}
\psfrag{t2}{$\tilde{\nu}_2$}
\psfrag{t9}{$\tilde{\nu}_9$}
\psfrag{t1'}{$\tilde{\nu}_1'$}
\psfrag{t2'}{$\tilde{\nu}_2'$}
\psfrag{t10'}{$\tilde{\nu}_{10}'$}

\entrymodifiers={+!!<0pt,\fontdimen22\textfont2>}		

\def\Z{\mathds{Z}}

\def\C{\mathds{C}}

\def\P{\mathds{P}}

\def\mcal{\mathcal}

\newcommand\T{\rule{0pt}{2.6ex}}			
\newcommand\B{\rule[-1.2ex]{0pt}{0pt}}		
\newcommand{\Div}{D}
\newcommand{\Jiv}{J}

\makeatletter
\renewcommand*\env@matrix[1][*\c@MaxMatrixCols c]{%
  \hskip -\arraycolsep
  \let\@ifnextchar\new@ifnextchar
  \array{#1}}
\makeatother

\newcommand{\footnoteremember}[2]{
  \footnote{#2}
  \newcounter{#1}
  \setcounter{#1}{\value{footnote}}
}
\newcommand{\footnoterecall}[1]{%
  \footnotemark[\value{#1}]
}

\begin{document}

\baselineskip=17pt

\begin{titlepage}
\begin{flushright}
\parbox[t]{1.8in}{
BONN-TH-2011-12}
\end{flushright}

\begin{center}

\vspace*{ 0.5cm}

{\LARGE \bf Holomorphic Couplings In\\ Non-Perturbative String Compactifications\footnoteremember{PhDref}{Based on the author's PhD thesis, defended on May 10, 2011.}
}

\vskip 1cm

\begin{center}
 {\Large\bf Denis Klevers\,\footnote{From September 1, 2011: Deptartment of Physics and Astronomy, 209 South 33rd Street, University of Pennsylvania, Philadelphia PA 19104, USA.}}
\end{center}
\vskip -0.2cm

\emph{ Bethe Center for Theoretical Physics\\
	Universit\"at Bonn \\
Nussallee 12\\ 
D-53115 Bonn, Germany}\\[0.7Em]
\texttt{klevers@th.physik.uni-bonn.de}
\\[0.15cm]
\end{center}

\vskip 0.2cm

\begin{center} {\bf ABSTRACT } \end{center}

In this review article\footnoterecall{PhDref} \hspace{-0.1cm}we present an analysis of several aspects of four-dimensional, 
non-perturbative $\mathcal{N}=1$ compactifications of string theory. Our study focuses on brane 
dynamics and their effective physics as encoded in the holomorphic
couplings of the low-energy $\mathcal{N}=1$ effective action, most prominently the 
superpotential $W$. This article is divided into three parts. In part one
we derive the effective action of a spacetime-filling D5-brane in generic Type IIB 
Calabi-Yau orientifold compactifications. In the second part we invoke tools from string 
dualities, namely from F-theory, heterotic/F-theory duality and mirror symmetry, for a 
more elaborate study of the dynamics of $(p,q)$ 7-branes and heterotic five-branes. 
In this context we perform exact computations of the complete perturbative effective 
superpotential, both due to branes and background fluxes. Finally, in the third part 
we present a novel geometric description of five-branes in Type IIB and heterotic 
M-theory Calabi-Yau compactifications via a non-Calabi-Yau threefold $\hat{Z}_3$, that
is canonically constructed from the original five-brane and the Calabi-Yau threefold $Z_3$
via a blow-up. We use the blow-up threefold $\hat{Z}_3$ to derive open-closed Picard-Fuchs 
differential equations, that govern the complete effective brane and flux superpotential. 
In addition, we present first evidence to interpret $\hat{Z}_3$ as a flux compactification 
geometrically dual to the original five-brane by defining an $SU(3)$-structure on $\hat{Z}_3$, 
that is generated dynamically by the five-brane backreaction. 

\hfill July, 2011
\end{titlepage}

\phantom{23}
\vspace{10cm} 
{\Large\bf Acknowledgment }

This work is based on my Ph.D. thesis. First of all I am deeply grateful to my supervisor Albrecht Klemm 
not only for sharing his insights into physics and mathematics and for countless discussions but in 
particular also for his constant support, guidance and encouragement. I would like to express my 
gratitude to Hans-Peter Nilles for supporting me from the early stages of my study up to my PhD 
and beyond. I am grateful to Thomas Grimm for his expert advise, support and constant encouragement. 
I am also thankful to Tae-Won Ha for many fruitful discussions and for the enjoyable collaboration 
during my Ph.D. period. I thank my office mates and friends Murad Alim, Babak Haghighat, Daniel Lopes, 
Maximilian Poretschkin, Marco Rauch, Marc Schiereck, Piotr Sulkowski, Thomas Wotschke and Jose Miguel Zapata Rolon
for providing a stimulating atmosphere and for many valuable discussions. I would like to thank Ralph Blumenhagen, Cumrun
Vafa, Mirjam Cvetic, Jim Halverson, Michael Hecht, Daniel Huybrechts, Hans Jockers, Sheldon Katz, Dieter L\"ust, 
Paul McGuirk, Adrian Mertens, Andrea Puhm and Eric Zaslow for interesting and inspiring discussions and correspondence.

I am indebted to the ``Deutsche Telekom Stiftung'' for supporting my Ph.D. by a full Ph.D. scholarship and to 
the ``Bonn-Cologne Graduate School of Physics and Astronomy'' for a partial scholarship.

\frontmatter

\addtolength\topmargin{-50pt}
\addtolength\textheight{105pt}

\tableofcontents

\addtolength\topmargin{50pt}
\addtolength\textheight{-105pt}

\mainmatter

\chapter{Introduction}
\label{ch:Intro}

It is ``The Unreasonable Effectiveness of Mathematics in the Natural Sciences'', to say it in 
the words of Wigner's famous article \cite{Wigner}, that explains the tremendous success of 
the current mathematical description of the laws of nature in today's theories. This title
subsumes the astonishing and countlessly verified observation that these laws 
can be consistently mapped to the clear and elegant structures of mathematics. This
in particular applies to theoretical physics, founded by the works of Galileo and Newton.
Contemporary 
theoretical physics nowadays provides an extremely accurate description of almost all 
processes in nature and has led to countless predictions that already have been verified 
or will hopefully be verified in the (near) future. This huge success of the synthesis of 
physics and mathematics has led to the rise of a new branch in science, denoted 
mathematical physics. It appreciates the strengths of both fields and exploits the 
synergies of their combined study. One especially interesting topic in mathematical physics is String 
Theory, which is at the final frontier of our understanding of fundamental physics, aiming at the unification 
of all matter and forces in a single theory - a theory of everything.
Some aspects of this enormously rich theory are the content of this work, which addresses 
naturally both physical and mathematical questions.

\section{String Theory and its Implications}

The foundation of contemporary physics consists of two fundamental theories, Quantum
Theory, governing the physics on subatomic scales, and General Relativity, that
determines the large scale structure of our universe. 

Quantum Theory is applied and verified in almost all areas of physics and is widely 
accepted without any doubt. Its most advanced formulation is relativistic quantum field 
theory being a combination of special relativity, classical field and gauge theory as well as
quantum mechanics. It provides the appropriate framework to analyze interactions at 
high energies and has led to the formulation of a theory of what is currently known as elementary 
particles and their interactions, the Standard Model (SM) of elementary particle physics. The SM describes 
all dynamics of the fundamental particles, the six leptons and six quarks, 
that are governed by strong, weak and electro-magnetic interactions, mathematically encoded by the 
SM gauge group $SU(3)\times SU(2)\times U_Y(1)$. The origin of particle masses is 
explained by the Higgs mechanism exploiting vacuum degeneracy and broken gauge symmetry 
which implies the existence of the Higgs particle. Except for the Higgs particle the 
SM has been tested and verified to high precession in modern experiments, with the Large 
Hadron Collider in Geneva currently at the frontier, which is also designed to discover the Higgs within the 
next years. However, the SM suffers from several experimental as well as conceptual drawbacks that
to our current knowledge seem to rule it out as a fundamental theory of nature. 
First of all it is an experimental and conceptual drawback that the SM does not explain the existence 
of dark matter, which is part of the standard model of cosmology $\Lambda$CDM \cite{Trodden:2004st} 
and e.g.~explains the rotational curves of galaxies. Since dark matter has 
not yet been directly observed it has to consist of very weakly interacting particles, that are not covered 
by the SM particle content. In addition it is conceptually unsatisfying that the SM contains a number of 
19 free parameters that are not predicted by the theory itself and have to be fixed by experiment. 
Furthermore the SM suffers from a hierarchy problem addressing the question of its naturalness, 
namely why the electroweak breaking scale $M_{\rm weak}=246 GeV$ is so 
much smaller than the cutoff-scale $\Lambda$ of the quantum field theory. This is a 
particularly crucial problem if one assumes that no new physics happens up to the Planck 
scale $M_{\rm Pl}$ where gravitational interactions, which are not covered in the SM, 
become relevant and invalidate the SM description and prediction. 
Then $\Lambda=M_{\rm Pl}$ and since for example the corrections to the Higgs 
mass at one loop are proportional to $\Lambda^2$, a fine-tuning would be 
necessary in order to reproduce the Higgs mass experimentally favored at about 
$120$ GeV. $\Lambda=M_{\rm Pl}$ also dramatically contradicts the measured value of the vacuum energy 
or the cosmological constant by a 30-orders-of-magnitude discrepancy \cite{Trodden:2004st}.

A compelling solution to some of these problems is supersymmetry at the TeV scale. 
By the introduction of superpartners to every SM-particle, dark matter candidates by an LSP Neutralino or 
the gravitino can be provided 
and the quadratic divergence in the Higgs mass is traded by a logarithmic one. In addition 
the cut-off $\Lambda$ is then at the TeV-scale making a fine-tuning obsolete. Furthermore, the supersymmetric vacuum
has generically vanishing vacuum energy so that a small cosmological constant could be realized by small
SUSY-breaking effects. For more conceptual 
reasons a supersymmetric extension of the SM, the simplest one being the 
Minimal Supersymmetric Standard Model (MSSM), is also favored since it supports the idea of a 
Grand Unified Theory (GUT) that unifies all SM-interactions into a bigger, more fundamental gauge 
theory\footnote{In order to avoid the singularity of $\alpha=\frac{e^2}{4\Pi}$ due to the Landau-pole of QED the 
idea to replace the SM gauge theory by an asymptotically free GUT gauge theory at a high scale 
$\sim 10^{15}GeV$ is also favorable.}. 
However, even the MSSM suffers from theoretical problems like a small hierarchy or $\mu$-problem of the 
Higgs mass parameter $\mu$ and an even bigger number of free parameters in its most general form\footnote{The cMSSM with mSUGRA provides a more minimal
extension of the SM with only 5 new parameters \cite{Martin:1997ns}.}. More conceptually, also 
the MSSM is a quantum field theory which does not include gravity and thus can not be, even if one
accepts the concept of renormalization in a microscopic theory, a fundamental
and complete theory. 

In this light it is an even more unfortunate fact, that canonically 
quantized Einstein gravity  is non-renormalizable\footnote{We note the asymptotic safety 
program which pursues the formulation of Einstein gravity as a valid quantum theory with 
a UV fixed point.} and thus fails to be a predictive quantum field theory by the requirement 
of introducing infinitely many free parameters to absorb all divergences. One way out is to 
view the MSSM as well as Einstein gravity as an effective theory that is valid up to a scale $\Lambda$. 
This cuts off all momenta above $\Lambda$ and thus trivially regularizes all quantum loops 
restoring the predictive power of the quantum field theory. This, however, is equivalent to the statement 
that the MSSM and Einstein gravity are not valid above the cut-off scale and that new physics 
applies that is not yet discovered.   
Another argument for this conclusion can be provided within classical General Relativity (GR). 
On the one hand also classical GR is at least as predictive as the SM. It has strong 
experimental verifications mainly in an astrophysical and cosmological context like 
its early checks by light deflection, time dilatation, gravitational lensing, gravitational 
red shifting, its more sophisticated implication of Hubble's expanding universe and the 
Big Bang Theory. On the other hand, however, also GR seems to be incomplete 
as a fundamental theory in the sense that it predicts its own breakdown due to spacetime 
singularities. Starting with a massive object of radius smaller than its Schwarzschild 
radius, a gravitational collaps occurs yielding a black hole as the final state of the 
evolution. This is a formal spacetime singularity that exceeds the description in 
terms of classical differential geometry. Also physically this poses a severe problem that is
known as the information loss paradox. Since the interior of the black hole is screened from 
an exterior observer by a horizon, all information about its initial state and the matter falling into 
the black hole is lost. Furthermore the black hole is known, following the semi-classical description of 
Bekenstein and Hawking, to be a thermodynamical object emitting 
thermal radiation, denoted Hawking radiation. Consequently the formation of a black hole 
violates unitarity as it describes the evolution of a pure state into a mixed state which 
yields a further tension between GR and the concepts of quantum mechanics. This implies the 
necessity to obtain a more fundamental and microscopic understanding of black holes. It requires, 
even from the point of view of pure gravity, the introduction 
of quantum gravity in which the evolution of an initial state, for example a star, into a black 
hole state is described in a unitary theory which allows to determine the statistical origin
of black hole thermodynamics. Together with the drawbacks 
of our description of particle physics this inevitably requires the invention of novel 
physical and mathematical ideas and concepts.
 
In summary we have collected known arguments for the observation that the SM 
as a quantum field theory and GR both as a classical geometrical theory and as a quantum theory can 
not be fundamental. They have to be replaced at a fundamental 
high scale $\Lambda$, which is presumably around the Planck scale $M_{\rm Pl}$, by a microscopic, 
fundamental theory of nature that UV-completes their incomplete physical description. 
The basic requirement of this theory is to contain, in a well-defined limit, both 
theories at low energies/large lengths compared to $\Lambda$. 

It is the ultimate goal of String Theory to overcome all these obstacles of nowadays 
theories in one unique mathematical formulation of nature that is valid on the most 
fundamental scales \cite{greensuperstring,lust1989lectures,polchinski1998string,becker2007string}. 
The basic idea of String Theory to 
simultaneously avoid the inconsistencies both due to infinities in quantum loops and 
due to a lack of a quantum gravity description is comparatively simple. String Theory 
just introduces a new, fundamental length scale $\sqrt{\alpha'}$ in a way that the 
concepts of Quantum Theory and general covariance are maintained. 
This is achieved by replacing the common point-particle description of elementary 
particles by a description via one-dimensional strings of characteristic size 
$\sqrt{\alpha'}$. This in a certain way minimal length scale smooths the interaction vertices 
in quantum field theory as well as the black hole and big bang singularities in spacetime 
in a general covariant way. Thus both infinities in quantum loops and 
in spacetime itself are naturally regularized. Stated differently String Theory 
introduces a new paradigm: Not the spacetime physics is at the core of any physical 
question but the physics on the two-dimensional world-sheet that the string sweeps out 
in spacetime. Indeed, as long as this microscopic description of String Theory 
makes sense, all emerging spacetime physics is consistent as well, even if the spacetime 
contains singularities. More conceptually String Theory provides a novel way to probe 
geometries, that fundamentally differs from the way point-particles probe geometries. 
This in particular alters the concepts of classical differential and algebraic 
geometry to what is usually referred to as quantum or stringy geometry. Consistency of 
microscopic quantum strings and their interactions then naturally leads to a unified 
description of gauge fields \textit{and} gravity. This happens in the sense, that whenever 
a massless gauge field is present, which is an element in the Fock space
of quantized open strings or dual closed heterotic strings, also a massless spin two 
particle or graviton has to be present as well, which is always a closed string 
state\footnote{Here we omit more sophisticated versions of quantum gravity via the 
description in terms of quantum gauge theories in the context of AdS/CFT duality.}. 
Moreover, consistency of the quantum string, which means vanishing Weyl anomaly of the 
underlying two-dimensional conformal field theory, implies the spacetime physics of 
gauge theory and general relativity. However, the minimal Maxwell and Einstein equations 
are reproduced only in the point-particle limit $\alpha'\rightarrow 0$ of the full string 
equations of motion. Higher $\alpha'$-corrections indicate hints how String Theory changes 
both conventional gauge theory and gravity that allows for a reconciliation of them at 
quantum level. Furthermore, consistency at one-loop, in more detail modular invariance, or
equivalently the absence of spacetime tachyons requires spacetime supersymmetry. 
A milestone of the quantum gravity description via String Theory for example is the 
explanation of the Bekenstein-Hawking entropy of specific supersymmetric black holes \cite{Strominger:1996sh}. 
This all is a first demonstration of the appealing feature of String Theory, namely its strong predictive 
power, which has almost all of the mentioned concepts of contemporary physics as a 
consequence of consistency, not an ambiguous choice.

However, it happens that String Theory is not quite as predictive as it seems at first 
glance due to the choice of a compactification geometry. The string equations of motion 
require a ten-dimensional spacetime, at least in the formulation of what is called 
critical string theory. In order to make contact with the real, four-dimensional world 
requires to ``hide'' the additional six dimensions in a certain way. One very prominent and 
old idea first proposed by Kaluza and Klein is the concept of 
compactification\footnote{An alternative to compactification is given by realizing our four-dimensional
spacetime as a slice in a higher dimensional space \cite{Randall:1999vf}, so called brane-world scenarios. 
This idea can be further used to solve the hierarchy problem \cite{Randall:1999ee} which is naturally realized in
string theory with D3-branes on the warped deformed conifold \cite{Giddings:2001yu}.}. Here the unwanted extra dimensions are hidden 
in a tiny, compact six-dimensional ``internal'' manifold of size scale $\ell_c$, which has to be chosen 
sufficiently small in order to be consistent with today's experiments. 
Indeed such a spacetime configuration can arise as a vacuum state of String 
Theory and the allowed compactification geometries maintaining four-dimensional supersymmetry, 
to address problems like the hierarchy problem, have been determined \cite{Candelas:1985en} to be Calabi-Yau 
manifolds \cite{greensuperstring,gross2003calabi}. In this course the heavy String and Kaluza-Klein modes 
of masses of order $1/\ell_s$, $1/\ell_c$ can be consistently integrated out yielding a four-dimensional 
Wilsonian effective action which is then viewed as the Lagrangian from that one builds 
up a four-dimensional effective quantum field theory. Furthermore, in this 
dimensional reduction process from ten to four dimensions the four-dimensional physics is 
encoded by key geometrical quantities of the internal Calabi-Yau manifold, which is denoted 
as geometrization of physics. The laws of classical geometry that mathematically encode
these quantities have however in general to be replaced by their stringy geometry counterparts, 
like in the context of mirror symmetry discussed below, and are often constraint from 
strong string theoretic consistency conditions like for example tadpoles. 

However, this complete geometrization of four-dimensional physics has one immediate and 
serious drawback. In general a topologically fixed internal compactification geometry 
can undergo shape and size deformations without cost of energy. This reflects in the effective 
four-dimensional theory by the existence of massless scalar fields, so called moduli 
\cite{greensuperstring,Candelas1991,huybrechts2005complex}, whose VEV determines the shape and 
size of the compactification 
geometry. Since no massless scalar field has yet been observed in nature and since the same 
fields also determine the four-dimensional couplings like the fine-structure constant, that would 
then be allowed to slowly vary - another contradiction with experiment, in particular cosmology - 
the moduli have to be fixed by a potential, generated for example by fluxes as noted first in \cite{Gukov:1999ya}
and thoroughly reviewed in \cite{Grana:2005jc,Douglas:2006es,Blumenhagen:2006ci}. It is thus 
one crucial task of every attempt for predictive string model building to give a mechanism that 
dynamically generates such a potential. Once this conceptual problem is solved, one can try to 
search a four-dimensional string vacuum reproducing all aspects of the MSSM, for example. However, although 
remarkably good models are known a single model meeting all criteria to be accepted as a valid theory 
of particle physics has not been found yet. Luckily there is, at least to our current
understanding of string dynamics, a vast landscape \cite{Susskind:2003kw,Douglas:2003um} of string 
vacua\footnote{A first counting of flux vacua led to the prominent finite result of the order of 
$10^{500}$ string vacua \cite{Bousso:2000xa}. The search for a realistic model out of these 
seems to be theoretical impossible without a better understanding of the  dynamics or without an 
intelligent searching strategy \cite{Denef:2006ad}.}. It is the goal of String Phenomenology to 
single out the most promising string vacua to reproduce the observable particles and interactions. 
Prominent corners in the landscape for performing explicit searches are the heterotic string, 
both on smooth Calabi-Yau manifolds as pioneered in \cite{Candelas:1985en,Witten:1985bz} and on orbifolds starting with \cite{Dixon:1985jw,Dixon:1986jc}, 
cf.~\cite{Nilles:2008gq} for a review, Type II setups with D-branes 
\cite{Lust:2004ks,Blumenhagen:2005mu,Douglas:2006es,Blumenhagen:2006ci} and most recently also 
F-theory compactifications \cite{Denef:2008wq,Heckman:2010bq,Weigand:2010wm}. It is furthermore expected 
that a more elaborate understanding of string dynamics, in particular in a non-perturbative 
formulation of string theory by M-theory or using string dualities, sheds further light on 
the phenomenologically appealing corners in the string landscape or even dynamically selects MSSM string
models, probably when non-perturbative effects are taken into account. Indeed there are prominent examples how 
non-perturbative string and field theory physics naturally reproduces aspects of 
four-dimensional physics like de Sitter vacua \cite{Kachru:2003aw} and confinement \cite{Seiberg:1994rs}. In particular 
the inclusion of non-perturbative physics drastically affects even qualitative features of 
physics\footnote{See also the reasoning of Dine, Seiberg that the string vacuum of our universe is
likely strongly coupled \cite{Dine:1985he}.}.
  
Conceptually, all these approaches of string phenomenology have in common that they first appreciate 
the reduction of the complexity of the full string dynamics to a comparably trivial subsector of 
light modes and then exploit the translation of the four-dimensional physics to the structure of the 
internal geometry. This reduction of the complicated string dynamics giving us back a four-dimensional 
effective theory is one principle we follow. 
Concretely there are two central aims of this work. One aim is the detailed study of a wide class 
of compactification geometries and the second aim is to analyze the dynamics of 
additional fundamental objects in string theory, the dynamics of branes.
Branes often have to be included in a string compactification for consistency and crucially 
influence string dynamics and phenomenology. The results of this analysis provide on the one hand a 
better conceptual understanding of the background geometries in string theory and on the other hand, 
however up to now only in toy models, exact potentials to make quantitative statements about 
moduli stabilization. 

Finally we note that issues addressed in the context of studying a particular physical problem
in string theory are often related to deep mathematical questions and structures. Both fields, mathematics 
and string theory, influence each other severely, and underlying mathematical structures often even make it
possible to perform specific physical calculations, as demonstrated on the basis of selected examples in this work. 
Conversely, physical motivation can be very successful in making mathematical predictions as in enumerative geometry 
where specific symplectic invariants, the Gromov-Witten invariants that are related to the counting of holomorphic 
curves, are calculated using mirror symmetry, cf.~\cite{Candelas:1990rm}. A further example for the interplay of mathematics and physics is the question 
of finding supersymmetric vacua in string theory. This translates, for example, to the mathematical problem solved by Yau of finding a
Ricci-flat metric, which is a solution to the closed string equations of motion, on a given manifold with vanishing 
first Chern class and fixed K\"ahler class. For supporting a supersymmetric open string sector this question corresponds to the task
of finding calibrated cycles of minimal area in their homology class \cite{Becker:1995kb}. In this context the key physical coupling 
functions of effective field theories are geometrized by key geometrical quantities. The $\mathcal{N}=1$ superpotential, 
for example, is related to period integrals in the closed string case or to chain integrals, more generally to the 
holomorphic Chern-Simons functional, in the open string case. Finally string theory sometimes even guides to 
interesting ways to extend the notion of classical geometry, for example. Directions include generalized geometry, special 
structure manifolds, like SU(3)-structure manifolds, and mirror symmetry, that relates 
classical complex geometry with a quantum deformation of symplectic geometry in the closed string case or classical 
deformation theory and quantum deformation theory in the open string case.  
  
\section{D-Branes}

The relevance of D-branes for a fundamental formulation of String Theory has been 
discovered during the second superstring revolution in 1995 starting with the 
seminal work of \cite{Polchinski1995}, see e.g.~\cite{Polchinski1996,johnson2006d} 
for a review. They were noted to be dynamical objects 
of String Theory carrying R--R-charge that are as fundamental as the string 
itself. However D-branes are intrinsically non-perturbative and thus not visible as a 
perturbative string state, which explains their relatively late discovery. 
This can be directly seen by their tension 
$T\sim\frac1{g_s}=\frac{1}{g_o^2}$ identifying them as open string 
solitons\footnote{There are also closed string solitons with tension $T\sim\frac{1}{g_s^2}$, 
the so-called NS5-brane, cf.~the discussion in \cite{Mohaupt2000}.}. 
BPS D-branes provide precisely those BPS-states that are necessary for a complete 
understanding of string dualities \cite{Witten:1995ex}. D-branes yield new 
particle states \cite{Witten:1995ex,Strominger:1995cz} in an effective theory and contribute as non-perturbative effects, 
so-called D-instantons, to correlation functions, see e.g.~\cite{Blumenhagen:2009qh} for a review. Prominent 
examples in string dualities are the rise of non-perturbatively enhanced gauge symmetry 
in the Type IIA string due to massless D-branes wrapping vanishing cycles in the geometry 
that matches the enhanced gauge symmetry of a dual heterotic string \cite{Witten:1995ex} or Type I/heterotic 
duality \cite{Witten:1995ex}, where the fundamental heterotic string at small coupling has to be viewed as the 
Type I D1-brane at strong coupling. The study of D-branes has further led to the famous 
AdS/CFT-correspondence \cite{Maldacena:1997re,Gubser:1998bc,Witten:1998qj} that is the first explicit realization of a holographic 
description of gravity in terms of a supersymmetric gauge theory, see e.g.~\cite{Aharony:1999ti,DHoker:2002aw} for reviews. From a very conceptual 
point of view the discovery of D-branes even gives rise to the question, what the 
fundamental formulation of string theory really is - a theory of fundamental strings 
and branes, of branes only, like in M-theory \cite{Townsend:1995kk,Witten:1995ex}, or a more abstract theory
e.g.~of matrices \cite{Dijkgraaf:1997vv,Berenstein:2002jq}.

In general D-branes have often to be included in a given string compactification
for consistency. Either they are required to cancel tadpoles due to orientifold
planes and fluxes or they are produced dynamically via a conifold transition 
turning flux into D-branes and vice versa \cite{Gopakumar:1998ki,Vafa:2000wi}. Most important for direct model 
building applications however is the case of spacetime-filling D-branes in a background 
yielding an $\mathcal{N}=1$ supersymmetric effective theory in four dimensions. Then 
an effective action description of the D-brane dynamics can be invoked to analyze 
their light modes. These are described by a supersymmetric gauge theory that is 
localized to the D-brane world-volume \cite{Witten:1995im} and further reduced to four 
dimensions. The most obvious case to consider are D-branes in flat space, so 
called brane world scenarios, where the real four-dimensional world is located on intersections 
of branes with the standard model gauge symmetry 
and particle content, see \cite{Kiritsis:2003mc,Uranga:2003pz,Lust:2004ks,Blumenhagen:2005mu,Blumenhagen:2006ci}
for a review. 
Generally these setups are defined by a Calabi-Yau background geometry $Z_3$ 
with orientifold planes in order to cancel unphysical tadpoles, so called Calabi-Yau 
orientifolds, in which the D-branes wrap specific calibrated submanifolds to preserve 
supersymmetry \cite{Becker:1995kb}. In this context the realization of the standard model 
gauge group or a sensible GUT-sector with the correct spectrum of charged matter is 
of most immediate relevance. Recently substantial progress has been made by the formulation
of realistic GUT models with enhanced gauge symmetry to exceptional gauge groups\footnote{Before
exceptional gauge groups have only been accessible in the heterotic string or in M-theory with 
ADE-singularities.} in the non-perturbative formulation of Type IIB via F-theory, see 
\cite{Heckman:2010bq,Weigand:2010wm} for reviews. These setups exploit the localization of the 
brane gauge dynamics to cycles inside the Calabi-Yau, which may allow to consistently decouple 
the complex physics of gravity in a compact geometry\footnote{Although appealing due to the 
obvious simplifications of decoupled gravity, global questions like the embedding of the local 
setup into a global string vacuum or the moduli problem can not be addressed.} and to work 
entirely in an (ultra)local model, to address more model independent and detailed phenomenological 
questions 
In addition model building with 
branes in Type IIB has on the one hand side led to a realization of hierarchies in warped 
compactifications with deep throats \cite{Giddings:2001yu} which provides an embedding of 
RS-scenarios \cite{Randall:1999ee} into String Theory. On the other hand scenarios for controlled 
supersymmetry breaking via anti-D3-branes to a metastable de Sitter vacuum with fixed moduli 
\cite{Kachru:2003aw} and string embeddings of cosmological models including inflation can 
be addressed, see e.g.~\cite{McAllister:2007bg} for a review. 

For all these phenomenological applications an adequate description of D-brane dynamics is essential.
However, in general the complete dynamics of quantum D-brane can only be indirectly described 
via string dualities or in specific parameter regions where for example a perturbative string theory 
description in terms of Dirichlet-branes or a supergravity description in terms of 
$p$-branes is valid. The thorough analysis of brane dynamics is the central aim of this work, where we focus mostly on 
the dynamics of spacetime-filling D5-branes, NS5-branes and seven-brane. For this analysis 
methods from basically three different fields are applied, that are methods from 
the study of Effective Actions, from String Dualities like F-theory, heterotic/F-theory 
and mirror symmetry and finally from purely geometrical consideration, namely Blow-Up Geometries 
and $SU(3)$-structure Manifolds. 
 
\section{Effective Actions}

A very direct way to analyze and to make quantitative statements about the  dynamics of a D-brane
in a given string background is the determination of the general D-brane effective action. Generically
a consistent and controllable string compactification has to be supersymmetric. The setups we initially
study are defined by a Calabi-Yau threefold $Z_3$ with supersymmetric D-branes and orientifold
planes or heterotic bundles. Since these backgrounds respect a minimal account of supersymmetry 
we are dealing with four-dimensional $\mathcal{N}=1$ string vacua. 

In general every effective four-dimensional $\mathcal{N}=1$ supergravity theory contains 
one supergravity multiplet, a number of vector and of chiral multiplets \cite{Wess}. 
The complete action is determined by three coupling functions\footnote{We neglect Fayet-Iliopoulos  
terms for our discussion.} of the chiral superfields $\Phi$, the K\"ahler potential $K(\Phi,\bar{\Phi})$, 
the gauge kinetic function $f(\Phi)$ and the superpotential $W(\Phi)$ \cite{Wess}. In this 
work the derivation of the full effective action, i.e.~of the three coupling 
functions $K$, $f$, and $W$, is performed for the case of a spacetime-filling D5-brane 
\cite{Grimm:2008dq}. This completes the list of $\mathcal{N}=1$ 
effective actions for compactifications with D-branes, see \cite{Grana:2003ek} for the D3-brane, 
\cite{Jockers:2004yj,Jockers:2005zy} for the D7-brane and finally \cite{Grimm:2011dx,Kerstan:2011dy} for the D6-brane action.

Besides the immediate application of the effective action for a detailed analysis of a specific 
D5-brane model of particle physics, one can address more model-independent questions. One natural 
such direction is to focus on an universal sector of the brane theory, that is given by a number 
of chiral multiplets corresponding to the position modes of the D5-brane. 
The scalars from this geometrical sector of the D5-brane effective action can remain as massless fields in the 
four-dimensional effective action and yield a serious moduli problem, that potentially spoils 
every D-brane model right at the beginning. Thus, it is of essential importance to ask the 
question whether these fields remain flat directions of the potential at all orders in the fields and in $\alpha'$. Out of 
the three coupling functions of the $\mathcal{N}=1$ effective theory, the perturbative brane 
superpotential depending on these moduli is of central importance for this question 
\cite{Witten:1997ep,Kachru:2000ih,Kachru:2000an}. 
Thus, together with the flux superpotential a complete understanding of the Type IIB superpotential 
allows to systematically study moduli stabilization in the Type IIB string with D5-branes, 
that are complex structure moduli and the dilaton in the closed string and the brane moduli 
in the open string sector\footnote{The K\"ahler moduli are fixed by non-perturbative effects, 
cf.~the pioneering paper \cite{Kachru:2003aw}.}. We refer to 
\cite{Grana:2005jc,Douglas:2006es,Blumenhagen:2006ci,Denef:2008wq} for a review of moduli stabilization
in flux compactifications.

Furthermore, the superpotential is in general, due to its holomorphy, a controllable 
coupling since it is protected by a non-renormalization theorem of $\mathcal{N}=1$ 
supersymmetry from perturbative and non-perturbative corrections, see e.g.~\cite{Terning}. Thus we can trust our 
expressions for it obtained for example by the dimensional reduction of the tree-level D5-brane effective action. 
Having realized this conceptual importance of the superpotential, we will in the 
remainder of the work calculate explicitly the flux and brane superpotential using techniques from string dualities.
This way, we obtain \textit{exact} string theoretic results for a coupling function in the four-dimensional 
effective theory. In the context of string dualities we will also view the superpotential as a test field for the more 
elaborate formulations of non-perturbative string and brane dynamics like F-theory, 
heterotic/F-theory duality and the blow-up geometry. 
 
\section{String Dualities}

In order to understand string theory more thoroughly beyond perturbation theory and including 
non-perturbative effects, string dualities can be applied, see \cite{Schwarz:1996bh,Aspinwall:1996mn,Sen:1998kr}
for a selection of review articles. 
Dualities in general rely on the observation that identical physics can come in two or more seemingly 
different disguises like for example in the 
mentioned AdS/CFT-correspondence relating gauge with a gravity theory. An important class 
of dualities are S-dualities that relate strong coupling physics in one theory to weak 
coupling physics in another or even the same theory. Examples are S-duality between the 
Type I and the heterotic $SO(32)$ string or of the Type IIB string with itself, which is exploited 
for the formulation of F-theory \cite{Vafa:1996xn}.

Besides S-duality F-theory also incorporates a second concept, namely 
geometrization. This term is in general used to express that a given theory is mapped to a complete 
geometric description, so that the underlying geometrical structures are exploited to study the dynamical
processes in the theory. In F-theory both the description of a varying coupling constant and the 
physics of non-perturbative seven-branes in the Type IIB theory are geometrized by an elliptic curve, 
a two-torus, fibered over the spacetime manifold \cite{Vafa:1996xn}. This provides on the one hand one of the 
few known duality invariant description of physics, in this case of S-duality in Type IIB, and on the other hand
allows to study non-perturbative
seven-brane dynamics. Furthermore,F-theory compactifications on elliptic Calabi-Yau manifolds 
\cite{Vafa:1996xn,Morrison:1996na,Morrison:1996pp} correspond to fully consistent supersymmetric Type IIB 
compactifications with automatically canceled tadpoles due to a consistent inclusion of orientifold planes 
\cite{Sen:1996vd,Sen:1997gv} and even allows for background R--R, NS--NS and brane fluxes\footnote{The 
only additional consistency condition is the M-theory tadpole \cite{Sethi:1996es}. Furthermore fluxes induce a warping, 
however, the internal metric is still conformally Calabi-Yau \cite{Becker:1996gj}.}. Thus F-theory provides
the ideal setup to study many aspects of Type IIB compactifications and to answer questions about non-perturbative brane dynamics and
about moduli stabilization, cf.~\cite{Denef:2008wq}. For the latter issue one in particular exploits the calculability 
of the F-theory superpotential in compactifications to four dimensions which unifies the Type IIB flux and brane superpotential
of the geometrized seven-branes\footnote{We note that the seven-brane superpotential is formally a five-brane superpotential induced by brane flux.}. 
This is a major strategy we follow in this work and in \cite{Grimm:2009ef}. 

A second example of geometrization is heterotic/M-theory duality \cite{Witten:1995ex}, which implies, using further 
string dualities, heterotic/Type II duality \cite{Witten:1995ex,Kachru:1995wm} and its close cousin, heterotic/F-theory duality \cite{Vafa:1996xn,Morrison:1996na,Morrison:1996pp}. 
Focusing on the latter we observe a geometrization of the data of a heterotic compactification in terms 
of the purely geometrical data defining an F-theory compactification. In particular heterotic gauge dynamics as well
as the dynamics of distinguished heterotic five-branes, so called horizontal five-branes, are completely geometrized 
in F-theory. This provides yet another route to analyze five-brane dynamics and to calculate brane superpotentials via
string dualities that is of essential use for this work \cite{Grimm:2009ef,Grimm:2009sy}.

One of the oldest string dualities is mirror symmetry, see e.g.~\cite{Hori:2003ic} for a review.   
Conceptually it can be understood as a consequence of probing the spacetime geometry
with one-dimensional strings instead of point-particles \cite{Vafa:1991uz}, which changes the notions of classical 
geometry to string geometry, a special form of quantum geometry. 
A toy version of mirror symmetry is T-duality\footnote{See \cite{Giveon:1994fu,Alvarez:1994dn} for
a detailed review of T-duality.} that is well-known for the closed bosonic string where it relates the compactification 
on a circle of radius $R$ to a different compactification on a circle of radius $1/R$ by exchanging windings and
momenta \cite{greensuperstring,lust1989lectures,polchinski1998string,becker2007string}. 
Mirror symmetry for Calabi-Yau threefold compactifications of Type II string theory was noted first from a CFT point of 
view \cite{Dixon:1987bg,Lerche:1989uy} and is indeed conjectured to be a generalization of T-duality \cite{Strominger:1996it}.
It is most thoroughly formulated for the topological versions of Type II string theory, the A- and B-model, 
on Calabi-Yau manifolds \cite{Witten:1991zz}, where it states the equivalence of the A-model on a given Calabi-Yau 
threefold $\tilde{Z}_3$ with the B-model on its mirror Calabi-Yau threefold $Z_3$. 
Geometrically it identifies the classical complex geometry of $Z_3$, i.e.~the complex structure moduli space, 
with the quantum corrected symplectic geometry\footnote{This is defined by a certain deformation 
quantization of the classical intersection theory on $\tilde{Z}_3$ due to world-sheet instantons \cite{Dine:1986zy,Dine:1987bq}, 
where $\hbar$ is identified with the string scale $\alpha'$.} of $\tilde{Z}_3$, i.e.~the K\"ahler 
structure on $\tilde{Z}_3$.
The equivalence of the full Type IIA theory on $\tilde{Z}_3$ with the Type IIB theory on $Z_3$ is only conjectured,
first by Kontsevich and then by \cite{Strominger:1996it,Becker:1995kb,Nekrasov:2004js},
since it requires the equivalence of all perturbative but most notably non-perturbative effects like BPS-particles 
and D-instantons. 
Physically mirror symmetry allows the calculation of the world-sheet instanton 
corrected\footnote{There are no perturbative nor non-perturbative $g_s$-corrections, since 
$g_s$ resides in a hypermultiplet \cite{Strominger:1995cz}.} holomorphic $\mathcal{N}=2$ prepotential 
in the Type IIA vector multiplet sector of the effective four-dimensional theory
by matching with the classical Type IIB prepotential \cite{Candelas:1990rm}. 

Mirror symmetry can be extended in two directions. One direction is the extension to higher-dimensional
Calabi-Yau manifolds like Calabi-Yau fourfolds \cite{Greene:1993vm,Mayr:1996sh,Klemm:1996ts} and the other direction is to 
open string mirror symmetry defined by the inclusion of D-branes and a corresponding open string sector. The latter was 
formulated first by Kontsevich as the homological mirror symmetry conjecture and was interpreted in more physical 
terms in \cite{Vafa:1998yp}. 
In physical setups both extensions yield effective theories with four supercharges.
Closed mirror symmetry for fourfolds naturally occurs in two-dimensional Type II, three-dimensional
M-theory and four-dimensional F-theory compactifications with an $\mathcal{N}=1$ effective theory, which is the case
of primary interest in this work \cite{Grimm:2009ef}. Open mirror symmetry as used in Type II theories
breaks $\mathcal{N}=2$ supersymmetry of closed string mirror symmetry to $\mathcal{N}=1$ by the inclusion of D-branes. 
It extends the purely geometrical closed mirror symmetry to D-branes identifying BPS D6-branes 
wrapping special Lagrangian three-manifolds in $\tilde{Z}_3$ in Type IIA with BPS D3-, D5-, D7- and D9-branes in
Type IIB\footnote{See \cite{Grimm:2011dx} for a detailed discussion of open mirror symmetry at the classical level.}. 
Both generalizations of mirror symmetry allow to calculate exact holomorphic couplings in the $\mathcal{N}=1$
effective theory. For mirror symmetry on fourfolds this is the effective flux superpotential, 
e.g.~of the F-theory compactification, which is classical in one theory and receives instanton corrections in the dual theory
\cite{Klemm:2007in}.  
In the open string case mirror symmetry also relates the corresponding brane superpotentials on both sides, 
which allows as in the closed string case to compute the disk instanton corrected superpotential on the 
IIA side by the classical superpotential in Type IIB \cite{Ooguri:1999bv}. Furthermore it can be physically motivated in F-theory, 
that under certain circumstances, i.e.~for specific flux choices, the fourfold flux superpotential agrees 
with the brane superpotential in a dual Type II or heterotic compactification \cite{Grimm:2009ef,Grimm:2009sy,Jockers:2009ti}. 
Detailed discussions and explicit calculations in favor of this observation are the core of the technical part 
of this work following \cite{Grimm:2009ef,Grimm:2009sy,Grimm:2010gk}.

\section{Blow-Up Geometries and SU(3)-Structure Manifolds}

As we have just discussed, string dualities often unexpectedly relate two different physical descriptions.
It is the essential goal of the last part of this work to collect hints for a novel duality
between a compactification with a five-brane supported on a curve $\Sigma$ and a dual blow-up 
threefold geometry carrying an $SU(3)$-structure, on which the five-brane has dissolved into flux and 
a canonical non-K\"ahler form \cite{Grimm:2010gk}. This duality should apply equally to Type IIB 
Calabi-Yau orientifold compactifications with D5-branes \cite{Grimm:2008dq,Grimm:2010gk} as well as 
to heterotic or M-theory Calabi-Yau threefold compactifications with M5-branes \cite{Grimm:2009ef,Grimm:2009sy}.

This dual description was first noted in \cite{Grimm:2008dq} in the context of the study of the 
D5-brane superpotential as a tool to analyze its in general complicated dependence on both complex structure moduli of $Z_3$ 
and on the deformations of $\Sigma$. It is argued that the natural setup for this study is provided by 
a different threefold $\hat{Z}_3$, that is obtained by blowing up along $\Sigma$ in $Z_3$, since 
the deformations of the curve $\Sigma$ in $Z_3$ are unified with the complex structure moduli of $Z_3$
as pure complex structure deformations of $\hat{Z}_3$\footnote{We thank D. Huybrechts for pointing us in that direction and for patient explanation
of mathematical details of this mapping.}.  
It is crucial for this unification that there are no branes present on $\hat{Z}_3$ hinting to
a complete geometrization of the D5-brane. Indeed it could be argued \cite{Grimm:2008dq} 
that the geometry of $\hat{Z}_3$ can be used to calculate the D5-brane and flux superpotential
of the original Type IIB compactification on $Z_3$.

It was later realized in \cite{Grimm:2009sy} that this blow-up procedure is also applicable to five-branes
on curves $\Sigma$ in the heterotic string, for which the brane superpotential is formally identical to that of a D5-brane 
and should also be calculable from $\hat{Z}_3$. In the context of heterotic/F-theory duality the description of the
five-brane dynamics via $\hat{Z}_3$ is further confirmed by the agreement with the usual geometrization of a horizontal 
five-brane as a blow-up in the dual F-theory fourfold $X_4$ 
\cite{Morrison:1996na, Bershadsky:1996nh,Bershadsky:1997zs,Berglund:1998ej,Rajesh:1998ik,Diaconescu:1999it}. 
This is shown and exploited for concrete calculations in \cite{Grimm:2009sy} which are cross-checked 
by the calculations of F-theory flux superpotentials in \cite{Grimm:2009ef} in combination with mirror symmetry 
for Calabi-Yau fourfolds and branes.

Finally the blow-up $\hat{Z}_3$ is used in \cite{Grimm:2010gk} for direct calculations of both the
five-brane and flux superpotential, that are again cross-checked via open mirror symmetry and shown to be
in agreement with the available results in the literature \cite{Aganagic:2000gs,Aganagic:2001nx,Lerche:2001cw,Lerche:2002ck,Lerche:2002yw,Walcher:2006rs,Baumgartl:2007an,Morrison:2007bm,Jockers:2008pe,Krefl:2008sj,Knapp:2008uw,Knapp:2008tv,Baumgartl:2008qp,Alim:2009rf,Jockers:2009mn,Walcher:2009uj,Alim:2009bx,Aganagic:2009jq,Li:2009dz,Jockers:2009ti,Baumgartl:2010ad,Alim:2010za,Fuji:2010uq,Shimizu:2010us}
that they partially extend. 
Furthermore, first arguments are provided for the blow-up threefold $\hat{Z}_3$ to define a consistent Type IIB or heterotic 
compactification by defining an $SU(3)$-structure on $\hat{Z}_3$. Most remarkably, a canonical 
non-K\"ahler form is defined that seems to naturally encode the five-brane in the original setup.
This together with the performed calculational checks should be viewed as a first step to 
consistently treat the backreaction of five-branes and to establish a physical duality between the 
pure closed string compactification on the blow-up threefold $\hat{Z}_3$ 
and the original string compactification with five-branes. Although complete evidence is still 
lacking it is tempting to view the blow-up geometry 
$\hat{Z}_3$ as a full dual description, that can probably be interpreted as a compact version of the 
geometric transition of \cite{Gopakumar:1998ki,Vafa:2000wi}.
In this regard the aim of this work is to provide a self-contained presentation of the idea, calculational
use and the current stage of the physical interpretation of the blow-up threefold $\hat{Z}_3$.

\section{Outline}

This work is split into three independent and self-contained parts on 
Effective Actions, on String Dualities and on Blow-Up Geometries and 
$SU(3)$-Structure Manifolds. 

We start in this work with the discussion of effective actions in part I. In order to prepare for 
our presentation we begin with a review of orientifold compactifications 
of Type II string theory in chapter \ref{ch:EffActCYOrieCompact}. We begin with a brief discussion 
of the classification of possible Type II orientifolds in section \ref{sec:classifyOrie}. Then we 
proceed with a review of the $\mathcal{N}=1$ effective action of generic Type 
IIB Calabi-Yau orientifold compactifications $Z_3/\mathcal{O}$ in section \ref{sec:EffActionOrie}, 
that are $O3/O7$- and $O5/O9$-orientifold compactifications, where the latter is of particular relevance for this work. 
The chapter is concluded by determining the $\mathcal{N}=1$ characteristic data.
In chapter \ref{ch:EffActD5} we begin the analysis of brane dynamics by the computation of the effective action 
of a spacetime-filling D5-brane in generic $O5/O9$-Calabi-Yau orientifolds following \cite{Grimm:2008dq}. In section \ref{sec:DbranesinCY3Orie} 
we present a general review of supersymmetric D-branes in Calabi-Yau manifolds focusing on their low-energy effective 
dynamics and on the geometric calibration conditions for BPS D-branes.
The calculation of the D5-brane effective action is performed in section \ref{sec:D5branes} by a purely bosonic
reduction. We put special emphasis on the universal sector of the D5-brane action that is given by the open-closed geometric 
moduli as well as on the determination of the scalar potential. The calculation of the scalar potential reveals 
the surprising relevance of four-dimensional non-dynamical three-forms in order to recover 
the complete F-term potential in a purely bosonic reduction. 
In section \ref{sec:N=1dataD5} we work out the corrections of the D5-brane to the $\mathcal{N}=1$ characteristic 
data of the pure $O5/O9$-orientifold compactification of chapter \ref{ch:EffActCYOrieCompact}. The D5-brane induces a brane 
superpotential, that is given by a chain integral, as well as in an additional gauging of a chiral field of the bulk by the 
$U(1)$-gauge boson of the D5-brane and the corresponding D-term. The discussion of the effective action is concluded 
in section \ref{sec:extensionInfinite} where the scalar potential is determined on the space of all geometric deformations
of the D5-brane, including also massive deformations. This will be further discuss in chapter \ref{ch:blowup}. 

Part II is dedicated to a more sophisticated treatment of brane dynamics via string dualities. We start 
our discussion in chapter \ref{ch:HetFThyFiveBranes} by a detailed presentation of heterotic/F-theory duality 
and its use to analyze non-perturbative seven-branes in F-theory respectively five-branes in the heterotic string.
Firstly, in section \ref{sec:HetString+Fivebranes} we review heterotic string compactifications in their most
complete formulation via heterotic M-theory, which allows to systematically calculate corrections to the perturbative heterotic string. Emphasis
is put on the influence of five-branes on the heterotic B-field and tadpole conditions.
In particular we note the interpretation of heterotic five-branes as a singular
bundle effect, a so-called small instanton, and its use, following \cite{Grimm:2009sy}, to obtain a chain-integral superpotential
for the five-brane when evaluating the holomorphic Chern-Simons functional on the small instanton. A brief review
of the spectral cover construction of heterotic vector bundles concludes the heterotic string 
section. Then in section \ref{sec:FTheoryCompactifications} we turn to F-theory. We review Vafa's basic idea of F-theory, 
readily turn to the construction of F-theory compactifications via elliptic Calabi-Yau manifolds and finally 
comment on the F-theory flux superpotential in Calabi-Yau fourfold compactifications. There we emphasize that this 
flux superpotential contains both the Type IIB
flux and seven-brane superpotential and can consequently be used, as we explicitly demonstrate in this work, 
to calculate open-closed superpotentials in a unified framework. Finally we present a discussion of heterotic/F-theory
duality, first of the fundamental eight-dimensional duality and then in lower dimensions. Of most conceptual
importance for our study of brane dynamics and the calculation of superpotentials is the duality map of heterotic five-branes, some of which are geometrized in F-theory by blow-ups in the F-theory fourfold.

The actual tools for many calculations performed in this work are provided in chapter \ref{ch:MirrorSymm+FiveBranes}.
Here the basic geometric techniques of mirror symmetry are introduced. We begin by reviewing toric geometry in section
\ref{sec:mirror_toric_branes} which is the main tool to generate concrete examples of Calabi-Yau manifolds, their mirrors
and also of branes. Then in section \ref{sec:CSModuliSpace+PFO} we present a basic account on Calabi-Yau threefold mirror
symmetry in order to prepare for its higher-dimensional analogue for Calabi-Yau fourfolds in section \ref{sec:FFMirrors}.
For both sections the study of the complex structure moduli space via Picard-Fuchs differential equations is essential. 
In particular the existence of these differential equations allows the study of global properties of the complex structure
moduli space as performed, following \cite{Grimm:2009ef}, for Calabi-Yau fourfolds yielding a 
novel behavior of the fourfold periods at the universal conifold. The chapter is concluded in section \ref{sec:EnumGeo} 
by reviewing the enumerative meaning of mirror symmetry for the A-model. In particular the interpretation of the flux 
and brane superpotentials as generating functions for world-sheet- and disk-instantons is emphasized. This, but also
the classical terms, provide cross-checks for our results obtained in the B-model calculation of the superpotentials.   

Finally in chapter \ref{ch:Calcs+Constructions} the methods from heterotic/F-theory duality, F-theory and mirror symmetry
are applied for calculations of effective superpotentials. In section \ref{sec:Superpots+MirrorSymmetry}
the F-theory flux superpotential is calculated for selected Calabi-Yau fourfolds that possess
a clear Type IIB interpretation and a heterotic dual, cf.~\cite{Grimm:2009ef}. Both the Type IIB
flux and seven-brane superpotentials are determined and a first interpretation in terms of the heterotic dual 
is given.
A more systematic study of the application of mirror symmetry to heterotic/F-theory duality
is presented in section \ref{sec:SuperpotsHetF}. The central strategy there is to calculate the heterotic flux 
and five-brane superpotential from the F-theory dual geometry. Two examples emphasizing different aspects of 
heterotic/F-theory duality with five-branes are discussed.  

In part III of this work, the novel prescription of the dynamics of five-branes on curves $\Sigma$ via 
blow-up threefolds $\hat{Z}_3$ is discussed and applied for calculations \cite{Grimm:2008dq}. Chapter \ref{ch:blowup} begins with a conceptual
discussion of five-brane dynamics in section \ref{sec:N=1gensection}. There the formal treatment of the 
backreaction of five-branes naturally leads to the consideration of the open manifold
$Z_3-\Sigma$, which is replaced by the blow-up $\hat{Z}_3-E$ in section \ref{sec:5braneblowupsanddefs}, where $E$
denotes the exceptional divisor of the blow-up of $\Sigma$. The unification of closed and open
deformations of $(Z_3,\Sigma)$ is discussed in detail and probed via the pullback $\hat{\Omega}$ of the Calabi-Yau three-form 
$\Omega$ from $Z_3$ to $\hat{Z}_3$ in section \ref{sec:hatomega}. Finally, the flux and five-brane superpotentials are lifted to $\hat{Z}_3$
in section \ref{sec:potentialhatZ3minusD} and shown to obey Picard-Fuchs equations for the complex structure moduli of $\hat{Z}_3$.
Then in chapter \ref{ch:CalcsBlowUp} the blow-up $\hat{Z}_3$ is exploited to perform calculations of the effective
superpotentials. These calculations are performed for two examples, branes in the quintic in section \ref{sec:ToricBraneBlowup}
and branes in an elliptically fibered threefold over $\P^2$ in section \ref{sec:ToricBraneBlowupII}. In both cases the existence of 
a toric GKZ-system, in general constructed in section \ref{sec:generaltoricstructure}, is used to derive open-closed 
Picard-Fuchs equations, that contain the flux and the five-brane superpotential as solutions. In 
section \ref{sec:heteroticF+blowup} the geometrization of five-branes via the blow-up $\hat{Z}_3$ is compared to 
the geometrization of five-branes in heterotic/F-theory duality. This comparison is applied for an explicit matching 
of the geometric structures and the five-brane superpotentials.

In the concluding chapter \ref{ch:su3structur} first steps to view the threefold $\hat{Z}_3$ as the 
background of a string compactification are provided. For this purpose an $SU(3)$-structure, as reviewed
in section \ref{sec:SU(3)rev}, is constructed on $\hat{Z}_3$, that can be interpreted to be generated
by the five-brane backreaction. Firstly the classical K\"ahler geometry on
$\hat{Z}_3$ is discussed in section \ref{sec:blow-up_as_Kaehler}. Then, a minimal non-K\"ahler deformation of 
the K\"ahler structure on $\hat{Z}_3$ is presented in section \ref{sec:non-Kaehlertwist}, which can be
physically viewed as a dissolution of the original five-brane on $\Sigma$ into conventional three-flux 
localized around the divisor $E$ and into additional geometrical flux.

Finally in part IV, we present in chapter \ref{ch:conclusion} our conclusions and an outlook on future work.
This work has three appendices. Appendix \ref{app:PartEffActions} contains background material on the calculation of the D5-brane effective action, appendix \ref{app:PartGeoBackground}
summarizes a basic account of the geometrical properties of elliptic Calabi-Yau manifolds, ruled threefolds and the blow-up 
threefold $\hat{Z}_3$ and finally appendix \ref{app:PartDetailsExamplesTables} provides further details
of the calculations performed in section \ref{sec:Superpots+MirrorSymmetry} and in chapter \ref{ch:CalcsBlowUp},
as well as two further fourfold examples.

\part{Effective Actions}

\chapter{Orientifold Compactifications}
\label{ch:EffActCYOrieCompact}

In this chapter we review compactifications of Type II string theory on Calabi-Yau threefold orientifolds  
$Z_3/\mathcal{O}$. Our main focus is on Type IIB string theory and the corresponding four-dimensional 
effective action, where we closely follow the detailed discussion in the original works
\cite{Grimm:2004uq,Grimm:2004ua,Grimm:2005fa}. This provides the necessary background for our 
computation of the D5-brane effective action in chapter \ref{ch:EffActD5}.

We start our presentation in section \ref{sec:classifyOrie} with a brief review of the classification 
of possible Type II orientifolds, first from the perspective of the two-dimensional string sigma-model, 
then from the spacetime point of view of Type II string theory, where we readily focus on the geometry 
of orientifold planes. Then we proceed with a review of the $\mathcal{N}=1$ effective action of generic 
Type IIB Calabi-Yau orientifold compactifications $Z_3/\mathcal{O}$ in section \ref{sec:EffActOrie}, 
that are $O3/O7$- and the $O5/O9$-orientifold compactifications where the latter are of particular 
relevance for this work. We first discuss the expected massless spectrum in four dimensions, then present 
the actual Kaluza-Klein reduction to obtain the effective action and conclude with the organization of 
the fields and couplings into the $\mathcal{N}=1$ characteristic data.

\section{Classification of Orientifolds}
\label{sec:classifyOrie}

In order to emphasize the concept of orientifolds we start in section \ref{sec:OrieSigmaModel} with a 
brief summary of the construction of an orientifold theory from the point of view of an 
$\mathcal{N}=(2,2)$ supersymmetric field theory in two-dimensions. Then in a second step we specialize 
to \text{sigma-models} with a Calabi-Yau target space $X$ and understand the orientifolded SCFT as part of a 
Type II string theory compactified on $X$ in section \ref{sec:OrieTypeII}. We conclude in section 
\ref{sec:geoOriePlanes} with an analysis of the geometry of orientifold planes as calibrated 
submanifolds in a Calabi-Yau threefold $Z_3$.

\subsection{Orientifolds in $\mathcal{N}=(2,2)$ Two-dimensional Field Theories}
\label{sec:OrieSigmaModel}

In general an orientifold theory of a given two-dimensional $\mathcal{N}=(2,2)$ field theory is defined by dividing out a 
symmetry group of the theory, the orientifold action. This is an involutive parity symmetry $P$ of the 
$\mathcal{N}=(2,2)$ theory where $P$ is a combination $P=\mathcal{T}\circ \bf{\Omega}$ of the parity 
action $\bf{\Omega}$ of the $\mathcal{N}=(2,2)$ superspace with a canonically induced action on the 
superfields and an internal action $\mathcal{T}$ on the superfields so that the $\mathcal{N}=(2,2)$
Lagrangian is invariant. There are two different parity operations on $\mathcal{N}=(2,2)$ superspace, 
denoted A- and B-parity $\Omega_A$, $\Omega_B$ \cite{Brunner:2003zm}, that both contain world-sheet parity 
$\Omega_p$ exchanging left- and right-movers but act differently on the $\mathcal{N}=(2,2)$-Gra{\ss}mann 
variables. Both parities break half but different supersymmetries and are interchanged by 
mirror symmetry\footnote{In the two-dimensional field theory mirror symmetry is just the exchange of 
supercharges $Q_-$ and $\bar{Q}_-$.}. In the case of a single chiral superfield $\Phi(x,\theta)$ with 
canonical kinetic term $\bar{\Phi}\Phi$ the internal action on the superfields is trivial, i.e.~$\Phi$ 
is transformed as a function on superspace
\begin{equation}
	P_A\,:\,\quad \Phi(x,\theta)\,\mapsto\,\overline{\Phi(\Omega_A(x,\theta))}\,,\qquad P_B\,:\,\quad \Phi(x,\theta)
	\,\mapsto\,\Phi(\Omega_B(x,\theta))\,.
\end{equation}
For non-linear \text{sigma-models} on a K\"ahler target manifold $X$, in particular for Calabi-Yau backgrounds, 
the two possible parities have to be combined with a non-trivial internal transformation on the chiral 
superfields \cite{Brunner:2003zm}. From invariance of the K\"ahler potential $K$ on $X$, which determines 
the kinetic term of the $\Phi$ as $K(\Phi,\bar\Phi)$, it follows that the internal symmetry acts as an 
(anti-)holomorphic and isometric diffeomorpism\footnote{For K\"ahler manifolds with K\"ahler form $J$ an 
(anti-)holomorphic and isometric diffeomorphism is characterized by $f^* J=J$ ($f^*J=-J$).} $f$ on the 
coordinates $z^i$ of $X$, which are the chiral superfields in the language of the \text{sigma-model}, 
for (A-) B-parities,
\begin{equation}
	P_A\,:\,\quad \Phi^i(x,\theta)\,\mapsto\,\overline{h^i(\Phi(\Omega_A(x,\theta)))}\,,\qquad 
	P_B\,:\,\quad \Phi(x,\theta)\,\mapsto\,f^i(\Phi(\Omega_B(x,\theta)))\,.
\label{eq:SigmaInv}
\end{equation}
Here, $f:z^i\mapsto f^i(z)$ and $h:z^i\mapsto h^i(z)$ are holomorphic functions with 
$K(f(z),\bar{f}(z))=K(z,\bar{z})$ and $K(\overline{h(z)},h(z))=K(z,\bar{z})$ up to K\"ahler transformations. 
In the case of an orientifold the map $f$ is involutive $f\equiv \sigma$ so that $P$ takes the form 
$P=\sigma\circ\bf\Omega$ with $\mathcal{P}^2=\id$. The geometrical fix-point locus of $\sigma$ is the 
orientifold plane. Exploiting further the consistency of the Sigma-model at quantum level, in particular 
ensuring anomaly freedom of the parity symmetry $P$, one concludes in general that \cite{Brunner:2003zm}
\begin{equation}
	f^*[B_2]=-[B_2]
\label{eq:Borie}
\end{equation}
for the cohomology class $[B_2]$ of the B-field.

\subsection{Orientifolds of Type II Superstring Theory}
\label{sec:OrieTypeII}

When we consider this orientifolded $\mathcal{N}=(2,2)$ as part of full string theory we are more 
interested in spacetime physics. In the following we consider compactifications of Type II string theory 
on $(\mathbb{R}^{1,3}\times Z_3)/\sigma$ with $\sigma$ acting trivially on Minkowski space 
$\mathbb{R}^{1,3}$ and as \eqref{eq:SigmaInv} on the Calabi-Yau threefold $Z_3$. Thus, the orientifold planes are 
space-time filling. In this context the quotient by the orientifold, denoted by $\mathcal{O}$ in the 
physical string, can be understood as a consequence of the inclusion of unoriented strings. Moreover, 
consistency of the full string theory, most importantly cancellation of tadpoles, and spacetime 
supersymmetry are more restrictive than preserving $\mathcal{N}=2$ world-sheet supersymmetry. 

Firstly, the requirement of $\mathcal{N}=1$ supersymmetry in four dimensions puts no additional restriction 
on a possible B-parity orientifolds. However A-parity preserves spacetime supersymmetry only if $\sigma$ 
acts on the holomorphic three-form $\Omega$ on $Z_3$ by complex conjugation modulo a phase. This is a 
consequence of the condition $J^3\cong \Omega\wedge\bar{\Omega}$. Thus, we arrive at the following 
transformation properties of the key geometrical objects on $Z_3$,
\begin{equation}
	P_A\,:\,\quad \sigma^*\Omega=e^{2i\theta}\bar{\Omega}\,,\quad \sigma^* J=-J\,,\quad 
	P_B\,:\,\quad \sigma^*\Omega=\pm\Omega\,,\quad \sigma^* J=J\,,
\label{eq:geoOrie}
\end{equation}
where we followed the convention of \cite{Grimm:2004uq,Grimm:2004ua,Grimm:2005fa}. The sign in the 
transformation $P_B$ will correspond to the two possible orientifolds with a plus sign for 
$\mathcal{O}5/\mathcal{O}9$-planes and a minus sign for $\mathcal{O}3/\mathcal{O}7$-planes, respectively.

Secondly, in Type II string theory orientifold planes are charged with respect to the R--R-fields with possible 
negative charge. The cancellation of tadpoles requires net charge zero which can be achieved without breaking 
supersymmetry by the inclusion of calibrated D-branes. The maintenance of supersymmetry can already be seen 
in the $\mathcal{N}=(2,2)$ theory since the linear combinations of the $\mathcal{N}=(2,2)$ world-sheet supercharges preserved by A- and 
B-parity are also preserved by A- and B-branes, respectively. Thus, for a stable compactification with 
spacetime supersymmetry of Type IIA respectively Type IIB string theory only A-parity respectively B-parity 
can be used. However, we ignore the effect of D-branes, which will be discussed in chapter 
\ref{ch:EffActD5}, for the moment and focus here exclusively on the closed strings. For the two parities in 
\eqref{eq:geoOrie} the orientifold action $\mathcal{O}$, lifted to the Type II theory, reads 
\begin{equation}
	IIA\,:\,\quad \mathcal{O}=\sigma\circ(-1)^{F_L}\Omega_p\,,\quad IIB\,:\,\quad (O3/O7)\,\, 
	\mathcal{O}=\sigma\circ(-1)^{F_L}\Omega_p\,,\,\,(O5/O9)\,\, \mathcal{O}=\sigma\circ\Omega_p
\label{eq:typeIIorie}
\end{equation}
which makes it more convenient to infer the 
action on all spacetime fields in the theory. Here $F_L$ is the spacetime fermion number of the left moving 
sector, see e.g.~\cite{Dabholkar1997a} for a review.

The action of the orientifold \eqref{eq:typeIIorie} on the massless bosonic fields in the Type II theory is 
determined by first noting that \cite{polchinski1998string,Dabholkar1997a,Angelantonj2002}
\begin{eqnarray}
	\Omega_p&:&\,\quad\quad +1\,:\,\,\, \phi\,,\,\, g\,,\,\,C_1\,,\,\, C_2\,,\qquad -1\,:\,\,\,C_0\,,\,\,B_2\,,\,\,
	C_3\,,\,\,C_4\,,\\
	(-1)^{F_L}&:&\,\quad\quad +1\,:\,\,\, \phi\,,\,\, g\,,\,\,B_2\,,\quad\qquad\quad -1\,:\,\,\,C_0\,,\,\,C_1\,,\,\,
	C_2\,,\,\,C_3\,,\,\,\,C_4\,,
\label{eq:parityEV}
\end{eqnarray}  
where we denoted the eigenvalue $\pm 1$ under corresponding transformation. Then, from this together 
with \eqref{eq:typeIIorie} we obtain the results of table \ref{tab:orieTrafo}, where one can further 
use \eqref{eq:Borie} and that $\sigma$ is isometric.
\begin{table}[ht!]
         \begin{center}
        \begin{tabular}{|c|c|}
                \hline
                \rule[-0.2cm]{0cm}{0.6cm}         & $\mathcal{O}$-image \\
                \hline 
                \rule[-0.2cm]{0cm}{0.7cm} O6      & $\sigma^*\phi\,,-\sigma^*B_2\,,\sigma^*g\,,-\sigma^*C_1\,,\sigma^*C_3$  \\
                \hline
                 \rule[-0.2cm]{0cm}{0.7cm} O3/O7  & $\sigma^*\phi\,,-\sigma^*B_2\,,\sigma^*g\,,\sigma^*C_0\,,-\sigma^*C_2\,,\sigma^*C_4$\\
                \hline 
                 \rule[-0.2cm]{0cm}{0.7cm}  O5/O9 &  $\sigma^*\phi\,,-\sigma^*B_2\,,\sigma^*g\,,-\sigma^*C_0\,,\sigma^*C_2\,,-\sigma^*C_4$ \\
                \hline
        \end{tabular}
        \caption{\label{tab:orieTrafo}Transformation of massless bulk fields.}
         \end{center}
\end{table}
Dividing out the orientifold symmetry projects on the $\mathcal{O}$-invariant states of the theory, 
$\mathcal{O}(\Phi)=\pm\sigma^*\Phi\stackrel{!}{=}\Phi$, where the sign is fixed in table 
\ref{tab:orieTrafo}. Thus, table \ref{tab:orieTrafo} determines also the parity with respect to $\sigma^*$ of the 
remaining massless fields by the requirement of invariance, as summarized below in table 
\ref{tab:invstates}. On the level of the four-dimensional effective action this is reflected in an 
$\mathcal{N}=1$ theory truncated from the $\mathcal{N}=2$ theory obtained from Type II string theory 
compactified on $\mathbb{R}^{1,3}\times Z_3$, as demonstrated below.

\subsection{The Geometry of Orientifold Planes}
\label{sec:geoOriePlanes}

As mentioned above the fixed-point set of the involution $\sigma$ determines the orientifold planes. In 
both Type II theories the geometry of orientifold planes is specified by the calibration of their volume 
and the vanishing of characteristic differential forms of $\mathbb{R}^{1,3}\times Z_3$. In type IIA the fixed point set 
of $\sigma$ is a real three-cycle $L$ in $Z_3$ times $\mathbb{R}^{1,3}$ and the corresponding orientifold planes are 
$O6$-planes. This follows in local coordinates $z^i$ from \eqref{eq:geoOrie} since 
$\Omega=g(z)dz^1\wedge dz^2\wedge dz^3$ which identifies the action of $\sigma$ as the real involution 
$\sigma(z^i)=\bar{z}^{\bar\imath}$. In Type IIB the involution $\sigma$ respects the complex structure on $Z_3$ and 
the fixed point set are holomorphic cycles $C_{Op}$ in $Z_3$ times $\mathbb{R}^{1,3}$. Again we check in local coordinates 
that the two possible phases for the transformation of $\Omega$ in \eqref{eq:geoOrie} allow either (minus sign) 
for fixed points $C_{O3}$ and fixed divisors $C_{O7}$ in $Z_3$, which are $O3/O7$-planes, or (plus-sign) for 
fixed curves $C_{O5}$ and the entire Calabi-Yau fixed, $C_{O9}=\mathbb{R}^{1,3}\times Z_3$, which are $O5/O9$-planes. 
Equivalently, the orientifold planes can be specified by the geometric conditions
\begin{eqnarray}
	O6&:&\quad(J+iB_2)\vert_{L}=0\,,\quad \text{Im}(e^{-i\theta}\Omega)\vert_{L}=0\,,\quad 
	\text{Re}(e^{-i\theta}\Omega)\vert_L=\frac{\sqrt{\det(g+B_2)}}{\sqrt{\det(g)}}\vol_L\nn\\
	Op&:&\quad \Omega\vert_{L}=0\,,\quad e^{J+iB_2}\vert_{C_{Op}}=\frac{\sqrt{\det(g+B_2)}}{\sqrt{\det(g)}}\vol_{C_{Op}}\,,\qquad p=3,5,7,9\,,
\label{eq:calibrationOrie}
\end{eqnarray}
where we included a background B-field \cite{Marino:1999af} and denoted the pullback to the fixed point loci
$L$, respectively $C_{Op}$, by $\vert_L$, respectively $\vert_{C_{Op}}$. The volume forms on the corresponding 
cycles are denoted $\text{vol}_L$, $\text{vol}_{C_{Op}}$. In general, cycles obeying \eqref{eq:calibrationOrie} are of 
minimal volume in their homology class. This directly identifies $L$ as a special Lagrangian submanifold 
and the cycles $C_{Op}$ as holomorphic submanifolds of $Z_3$. They are calibrated with respect to the calibration 
forms $\text{Re}(e^{-i\theta}\Omega)$, $(J+iB_2)^{\frac{p-3}{2}}$, respectively. This is very similar to the 
BPS-bound obeyed by massive particles in a theory with enhanced supersymmetry with a saturation of the BPS-bound 
for BPS-particles. Indeed, the calibration conditions \eqref{eq:calibrationOrie} are indeed derived as the 
condition for BPS stable D-branes in background B-fields with trivial worldvolume gauge flux \cite{Marino:1999af}.

We conclude with a qualitative argument to infer the existence of the various D-branes and O-planes in Type II theory. 
For the case of $O9$-planes the geometric involution $\sigma$ in the orientifold in \eqref{eq:typeIIorie} acts trivially on 
ten-dimensional spacetime and only world-sheet parity is divided out. Thus, orientifolds with 
$O9$-planes just add unoriented strings to Type IIB theory in ten dimensions. However, it is well-known that 
this theory is inconsistent due to tadpoles that have to be canceled by the inclusion of 32 D9-branes. This 
introduces an open string sector and the theory is identified with Type I string theory \cite{Dabholkar1997a,johnson2006d}. 
By compactifying this theory on tori $T^n$ and by application of T-duality consistent toroidal string theory 
compactifications are obtained. Moreover, T-duality interchanges Dirichlet and Neumann boundary conditions 
and furthermore world-sheet and spacetime parity \cite{Polchinski1996,Dabholkar1997a, johnson2006d}. 
This way D-branes of all possible dimension and O-planes sitting at the edge of a hypercube 
$T^n/(\mathds{Z}_2)^n$ are obtained.

\section{Effective Action of Type IIB Calabi-Yau Orientifolds}
\label{sec:EffActionOrie}

In this section we review the $\mathcal{N}=1$ effective action of Type IIB orientifold compactifications,
that are $O3/O7$- and $O5/O9$-orientifolds, where we closely follow \cite{Grimm:2004uq,Grimm:2005fa}. We 
demonstrate a purely bosonic reduction and infer the fermionic action and fields by spacetime supersymmetry. 
In section \ref{sec:specOrie} we summarize the massless four-dimensional spectrum as obtained from a 
Kaluza-Klein ansatz of the ten-dimensional Type IIB supergravity fields. Then in section \ref{sec:EffActOrie} 
we present the bosonic Lagrangian of the full effective action, that we recast into the standard $\mathcal{N}=1$ 
form in section \ref{sec:N=1dataOrie}. There we extract the $\mathcal{N}=1$ chiral coordinates and the three 
coupling functions, the K\"ahler potential, the gauge kinetic function and the superpotential, which is 
trivial due to the absence of fluxes. 

Since the main focus of this work is on Type IIB string compactifications we omit a complete discussion 
of the mirror dual Type IIA orientifolds and their effective actions and refer the interested reader to 
the literature \cite{Grimm:2004ua,Grimm:2005fa}. 

\subsection{The Spectrum in Four Dimensions}
\label{sec:specOrie}

We determine the massless fluctuations of Type IIB theory around the $\mathcal{N}=1$ supersymmetric 
background $(\mathbb{R}^{1,3}\times Z_3)/\mathcal{O}$. Since a ten-dimensional massive field remains massive in 
four-dimensions we only consider the massless fields in ten dimensions, which agrees with the field 
content of ten-dimensional Type IIB supergravity truncated by the orientifold $\mathcal{O}$. Dimensional 
reduction yields the field content of the four-dimensional effective theory that will be assembled to 
$\mathcal{N}=1$ superfields. These $\mathcal{N}=1$ fields are obtained by demanding invariance under the 
orientifold action yielding a consistent truncation from the $\mathcal{N}=2$ fields of Type IIB on 
$\mathbb{R}^{1,3}\times Z_3$. The requirement of invariance immediately fixes the parity under $\sigma^*$ of the 
ten-dimensional massless fields. From table \ref{tab:orieTrafo} and \eqref{eq:Borie} we infer the action 
as given in table \ref{tab:invstates}.
\begin{table}[ht!]
\begin{center}
 \begin{tabular}{|c|c|c|}
 \hline
 \rule[-0.2cm]{0cm}{0.6cm}& R--R& NS--NS\\ \hline
O3/O7\rule[-0.2cm]{0cm}{0.6cm} & $\sigma^*(C_p)=(-1)^{p/2}C_p$&$	\sigma^*\phi=\phi$,  $\sigma^*g=g$, \\  \cline{1-2}
O5/O9\rule[-0.2cm]{0cm}{0.6cm}& $\sigma^*(C_p)=(-1)^{(p+2)/2}C_p$& $\sigma^*B_2=-B_2$\\\hline
 \end{tabular}
\caption{\label{tab:invstates}The required $\sigma^*$-parity of the $\mathcal{O}$-invariant states.}
\end{center}
\end{table}
The fields $C_p$ denote the form-fields comprising the R--R-sector. In the democratic formulation of Type 
IIB supergravity these are $C_{0}$, $C_{2}$, $C_{4}$ $C_{6}$ and $C_{8}$ with field strengths \cite{Bergshoeff:2001pv} 
\begin{equation} \label{eq:10dRRfieldstrength}
        G^{(p)}=\begin{cases}
                                dC_{0}&         p=1,\\
                                dC_{p-1}-dB_2\wedge C_{p-3}& \text{else} \ .
                       \end{cases}
\end{equation}
In addition the duality constraints $G^{(n)}=(-)^{\left\lfloor \frac n2\right\rfloor}\ast_{10}G^{(10-n)}$ 
are imposed, which read \cite{Bergshoeff:2001pv}   
\begin{equation}\label{eq:selfduality}
        G^{(1)}=\ast_{10}G^{(9)}\,,\quad G^{(2)}=-\ast_{10} G^{(8)}\,,\quad  G^{(3)}=-\ast_{10}G^{(7)}\,,\quad 
        G^{(4)}=\ast_{10}G^{(6)}\,,\quad G^{(5)}=\ast_{10}G^{(5)}\ ,
\end{equation}
where $\ast_{10}$ denotes the ten-dimensional Hodge star and $\left\lfloor \frac n2\right\rfloor$ the 
integer part of $\frac{n}2$. These conditions are necessary in order to drop the redundant degrees of 
freedom in the democratic formulation of Type IIB supergravity \cite{Bergshoeff:2001pv}. 

Having identified the invariant fields in ten-dimensions we perform the Kaluza-Klein reduction. As a 
first step in the dimensional reduction process we specify the background values of the fields in the 
theory. Since the standard Kaluza-Klein ansatz for the metric field of a direct product form is not a 
solution to the supergravity equations of motion in the presence of source terms induced by branes or 
fluxes \cite{Becker:1996gj,Verlinde2000,Chan2000,Giddings:2001yu} a warped product with warp factor $A(z)$ 
has to be considered,
\begin{equation}
	ds_{10}^2=e^{2A(z)}\eta_{\mu\nu}^{\rm SF}dx^\mu dx^\nu+2e^{-2A(z)}g_{i\bar\jmath}(z)dz^id\bar{z}^{\bar \jmath}\,.
\label{eq:warpedmetric}
\end{equation}  
Here we choose the string frame (SF) and the Calabi-Yau metric $g_{i\bar\jmath}$ on $Z_3$. However, 
we neglect the effect of the warp factor in this work which might be argued for in the large 
radius limit of $Z_3$ where $A$ approaches unity \cite{Giddings:2001yu,Grana2001,Grana2002}. In addition 
we exclude any background fluxes for the moment and refer to section \ref{sec:scalarpotderivation} and 
to \cite{Grimm:2004uq,Grimm:2005fa} for a complete discussion. 

Next we make a reduction ansatz for the fluctuations of the ten-dimensional fields of Type IIB theory around their 
background vacuum expectation values (VEV). For this purpose we expand by harmonic forms $\mathcal{H}^{p}(Z_3,g)$ on $Z_3$ 
that are in the kernel of the Laplacian on $Z_3$ and thus correspond to massless modes in four-dimensions. Fortunately the 
number of harmonic forms is topological, i.e.~does not depend on the metric $g$, and is counted by the cohomology 
groups\footnote{More precisely, this isomorphism can be characterized by saying that for every cohomology class $[\alpha]$ there 
exists a unique representative $\alpha_h$ with $\Delta \alpha_h=0$ or equivalently with $\norm{\alpha_h}^2=\int\alpha_h\wedge\star\alpha_h$ 
minimal in its class.} $H^{p}(Z_3)\cong\mathcal{H}^{p}(Z_3,g)$. According to Hodge theory on compact K\"ahler manifolds these 
groups admit a decomposition into Dolbeault cohomology groups, $H^{p}(Z_3,\mathbb{C})=\sum_{i+j=p}H_{\bar \partial}^{i,j}(Z_3)$ 
for $d=\bar\partial+\partial$, which is independent of the K\"ahler structure. In order to accommodate for the transformation 
properties of table \ref{tab:invstates} we have to introduce the $\pm1$-eigenspaces $H^{(i,j)}_{\pm}$ of $\sigma^*$ 
in $H^{(i,j)}$,
\begin{equation}
	H^{(i,j)}(Z_3)=H^{(i,j)}_{+}(Z_3)\oplus H^{(i,j)}_{-}(Z_3)\,. 
\label{eq:eigenspaceH}
\end{equation}
This is possible since the action of $\sigma$ is closed on the Dolbeault cohomology as a 
holomorphic map and since $(\sigma^*)^2=\id$. For a Calabi-Yau threefold the only 
non-vanishing Hodge numbers are $h^{(0,0)}=h^{(3,3)}=h^{(3,0)}=h^{(0,3)}=1$, $h^{(1,1)}=h^{(2,2)}$ 
and $h^{(2,1)}=h^{(1,2)}$. In terms of these numbers the Euler characteristic reads 
$\chi(Z_3)=2(h^{(1,1)}-h^{(2,1)})$. We introduce a basis $(\omega_\alpha ,\omega_a)$ of 
$H_+^{(1,1)},\, H_-^{(1,1)}$ with dual basis $( \tilde \omega^\alpha,\tilde \omega^a)$ 
of $H_+^{(2,2)},\,H_-^{(2,2)}$ such that 
\begin{equation}
  \int_{Z_3} \omega_\alpha\wedge \tilde{\omega}^\beta=\delta^\beta_\alpha\ ,\qquad
  \int_{Z_3}\omega_a\wedge \tilde{\omega}^b=\delta^b_a\ ,
\end{equation}
where $\alpha,\beta = 1,\ldots,h^{(1,1)}_+$ and $a,b = 1,\ldots,h^{(1,1)}_-$. These 
vector spaces can still be dual since $\sigma$ commutes with the Hodge star $*$, which 
implies $h^{(1,1)}_{\pm}=h^{(2,2)}_{\pm}$. Moreover, we denote by $(\alpha_K,\beta^K)$, 
$(\alpha_{\tilde K},\beta^{\tilde K})$ a real symplectic basis of $H^3_+$ and $H^3_-$, 
respectively. This basis is chosen such that the intersection pairings take the form
\begin{equation} \label{eq:def-alpha_beta}
        \int_{Z_3} \alpha_K\wedge \beta^L=\delta^L_K\ ,\qquad\int_{Z_3} \alpha_{\tilde
          K}\wedge \beta^{\tilde L}=\delta^{\tilde L}_{\tilde K}\ ,
\end{equation}
and vanish otherwise.  
Note that the holomorphic three-form $\Omega$ is contained in $H^3_+(Z_3)$, hence, 
$K=0,\dots, h^{(2,1)}_+$ but $\tilde K=1,\dots,h^{(2,1)}_{-}$ for $O5/O9$ orientifolds. 
Conversely for $O3/O7$-orientifolds $\Omega$ is an element of $H^-(Z_3)$, thus 
$\tilde{K}=0,\ldots,h^{(2,1)}_-$, but $K=1,\ldots,h^{2,1}_+$. Thus, to indicate the 
difference of the dimensionality we introduce different labels of the cohomology basis of 
$H^{3}_{\pm}$ in the $O5/O9$ and $O3/O7$ case. Furthermore, we introduced a basis 
$\chi_\kappa$, $\chi_k$ of $H^{(2,1)}_\pm$, respectively. We summarize our conventions in 
table~\ref{tab:cohombasisOrie}. We note that $b^3_{\pm}$ is always even dimensional as in the 
Calabi-Yau case.
\begin{table}[ht!]
        \begin{center}
        \begin{tabular}{|c||c|c||c|c||c|c|}
                \hline
                Type &\multicolumn{2}{c||}{\rule[-0.2cm]{0cm}{0.6cm}  $(\pm1)$-Eigenspace}& \multicolumn{2}{c||}{Basis} & \multicolumn{2}{c|}{Dimension}\\
                \hline  
               $O3/O7$ &\rule[-0.2cm]{0cm}{0.7cm}  $H^3_+(Z_3)$ & $H^3_{-}(Z_3)$ & $\alpha_\kappa,\beta^\kappa$   & $\alpha_{\tilde K},\beta^{\tilde K}$ & $2h^{(2,1)}_+$ & $2h^{(2,1)}_-+2$ \\
                \hline  
                $O5/O9$ & \rule[-0.2cm]{0cm}{0.7cm}  $H^3_+(Z_3)$ & $H^3_{-}(Z_3)$ & $\alpha_K,\beta^K$   & $\alpha_{k},\beta^{k}$ & $2h^{(2,1)}_++2$ & $2h^{(2,1)}_-$ \\
                \hline
                \hline
                \multirow{3}*{Both} &\rule[-0.2cm]{0cm}{0.7cm}   $H^{(1,1)}_{+}(Z_3)$ & $H^{(1,1)}_{-}(Z_3)$ & $\omega_\alpha$ & $\omega_a$ & $h^{(1,1)}_+$ & $h^{(1,1)}_-$ \\
                 \cline{2-7}
                 &\rule[-0.2cm]{0cm}{0.7cm}   $H^{(2,1)}_{+}(Z_3)$ & $H^{(2,1)}_{-}(Z_3)$ & $\chi_\kappa$ & $\chi_k$ & $h^{(2,1)}_+$ & $h^{(2,1)}_-$ \\
                \cline{2-7} 
                 &\rule[-0.2cm]{0cm}{0.7cm}    $H^{(2,2)}_{+}(Z_3)$ &$H^{(2,2)}_{-}(Z_3)$ & $\tilde{\omega}^\alpha$ &   $\tilde{\omega}^a$& $h^{(2,2)}_+$ & $h^{(2,2)}_-$\\
                \hline
        \end{tabular}
        \caption{\label{tab:cohombasisOrie}Basis of the cohomology groups.}
       \end{center}
\end{table}

To determine the four-dimensional bulk spectrum we use this cohomology basis and expand 
the NS--NS as well as the R--R fields. We start with the NS--NS sector which is universal 
for both the $O3/O7$- and $O5/O9$-orientifold. According to table \ref{tab:orieTrafo} the 
ten-dimensional dilaton $\phi$ is expanded by harmonic, $\sigma$-invariant functions on $Z_3$. 
For a compact manifold $Z_3$ only the constant function is harmonic and we obtain one massless 
mode, the four-dimensional dilaton $\phi(x)$. The ten-dimensional B-field is odd under 
$\sigma$ for both orientifolds. Consequently its zero-modes $b_a(\underline{x})$ enter the 
expansion by forms $H^{(1,1)}_-(Z_3)$ as
\begin{equation} \label{eq:NSNSexp}
        B_2=b^a(\underline{x})\omega_a\ .
\end{equation} 
The reduction of the metric tensor $g$ is most efficiently organized by the number of spacetime indices 
$\mu$, $\nu$. For two spacetime indices $\mu$, $\nu$ we obtain the four-dimensional metric 
tensor, which is proportional to the flat Minkowski metric $\eta_{\mu\nu}$. Since a Calabi-Yau 
threefold does not admit continuous isometries, i.e.~there are no Killing vector fields, no 
four-dimensional massless vector fields are obtained from $g_{i\mu}$. Finally, variations of 
the internal metric yield four-dimensional scalars. These scalars take values in the Calabi-Yau 
orientifold moduli space\footnote{It was argued in \cite{Brunner:2003zm} that the Calabi-Yau 
orientifold moduli space forms a smooth submanifold of the Calabi-Yau moduli space.}. This moduli 
space splits into two parts related to deformations of the complex structure and of the K\"ahler 
structure, that are encoded by variations of the holomorphic three-form $\Omega$ and the K\"ahler 
class $[J]$, respectively, as we discuss next. 

The three-form $\Omega$ is expanded in accord with 
\eqref{eq:geoOrie} as
\begin{equation} \label{eq:periodexpansionOrie}
   O3/O7\,\,:\,\, \Omega = X^{\tilde{K}}(\underline{z}) \alpha_{\tilde{K}} - \cF_{\tilde K}(\underline{z}) \beta^{\tilde K}\,,\qquad
   O5/O9\,\,:\,\, \Omega = X^K(z) \alpha_K - \cF_{K}(z) \beta^K\,,
\end{equation} 
where the $\underline{z}$ denote the complex structure moduli of $Z_3/\sigma$. In the effective 
four-dimensional theory the $\underline{z}(x)$ are complex scalar fields taking values in the 
complex structure moduli space $\cM^{\rm cs}$. We note that due to \eqref{eq:geoOrie} there are 
only $h^{(2,1)}_\pm$ allowed complex structure deformations for $O5/O9$- respectively 
$O3/O7$-orientifolds. The coefficient functions $(\underline{X},\underline{\mathcal{F}})$ 
are the periods of $\Omega$. They can be expressed as period integrals over the symplectic 
homology basis $(A_{K},B^K)$ dual to $(\alpha_K,\beta^K)$ as 
$\int_{A_L} \alpha_K=\delta^L_K=-\int_{B^K} \beta^L$,
\begin{equation} \label{eq:periodsDef}
  X^K = \int_{A_K} \Omega \ , \qquad \cF_K = \int_{B^K} \Omega\ 
\end{equation}
for $O5/O9$- and analogously for $O3/O7$-orientifolds. 
Then the first order variation of $\Omega$ under a change of complex
structure reads 
\begin{equation} 
        \partial_{z^\kappa}\Omega=\chi_{\kappa}-K_{\kappa}\Omega\ ,
\label{eq:variationofomega}
\end{equation}
for $O5/O9$-orientifolds and with $\kappa\rightarrow k$ also for $O3/O7$-orientifolds where 
$K_\kappa$ is the first derivative of the K\"ahler potential on $\cM^{\rm cs}$. The 
periods may still be used to define special coordinates as $t^k = X^k / X^0$ for 
$O3/O7$-orientifolds respectively $t^\kappa = X^\kappa / X^0$ for $O5/O9$-orientifolds.
Similarly we expand the K\"ahler form as
\begin{equation}
	J=v^\alpha(\underline{x})\omega_\alpha
	\label{eq:expJ}
\end{equation}
where the $v^\alpha(x)$ are real fields taking values in the K\"ahler moduli space 
$\cM^{\rm K}$ of $Z_3$. At classical level $\cM^{\rm K}$ is a cone specified by positivity 
conditions on the $\underline{v}$. We emphasize that the K\"ahler form is not complexified 
by the B-field on $Z_3/\sigma$ due to the different parity under $\sigma^*$. 

The infinitesimal variations of the internal Calabi-Yau metric of mixed respectively 
of purely holomorphic type similarly describe variations of the K\"ahler and complex 
structure. They are related to the variations of the forms $J$, $\Omega$ as
\begin{eqnarray} \label{eq:var-g}
				& & \delta g_{i\bar\jmath}=-i(\omega_{\alpha})_{i\bar\jmath}\delta v^{\alpha}\ ,
				\qquad\qquad\qquad\,\,\; \; \alpha=1,\ldots,h^{(1,1)}_+\,,\\ 
   O3/O7 &:& \delta g_{ij} = \frac{i\mathcal{V}}{\int \Omega \wedge \bar{\Omega}}\
   \Omega_{j}^{\ \bar \imath \bar \jmath} \, (\bar
   \chi_{\bar k})^{\phantom{i}}_{\bar \imath \bar \jmath i}\,  \delta
   \bar z^{\bar k} \ ,\qquad k=1,\ldots,h^{(2,1)}_-\,,\nonumber\\
   O5/O9 &:& \delta g_{ij} = \frac{i\mathcal{V}}{\int \Omega \wedge \bar{\Omega}}\
   \Omega_{j}^{\ \bar \imath \bar \jmath} \, (\bar
   \chi_{\bar \kappa})^{\phantom{i}}_{\bar \imath \bar \jmath i}\,  \delta
   \bar z^{\bar \kappa} \ ,\qquad \kappa=1,\ldots,h^{(2,1)}_+\,,\nonumber
\end{eqnarray}  
where we introduced the string-frame volume $\mathcal{V} = \int_{Z_3}
d^6 y \sqrt{g}$ of $Z_3$.  

Let us now turn to the R--R sector. The various R--R-fields are expanded according 
to table \ref{tab:invstates}. For $O5/O9$-orientifolds the expansion into invariant fields reads
\begin{eqnarray}
        \label{eq:RRO59exp}
        C_6 &=& A_{(3)}^K\wedge\alpha_K+\tilde{A}^{(3)}_K\wedge\beta^K +
        \tilde{c}^{(2)}_\alpha \wedge \tilde\omega^\alpha+h\, m_6\ ,  \\
        C_4 &=& V^{k}\wedge\alpha_{k}+U_{k}\wedge\beta^{k}+\tilde{\rho}^a_{(2)} \wedge
        \omega_a+\rho_a\tilde\omega^a\ , \nn\\
        \nn C_2 &=& \mathcal{C}_{(2)}+c^\alpha\omega_\alpha\ ,
\end{eqnarray}
where $m_6 = \Omega \wedge \bar \Omega/\int_{Z_3} \Omega \wedge \bar \Omega$ 
is a top form on $Z_3$ normalized such that $\int_{Z_3} m_6 =1$.
Similarly for $O3/O7$-orientifolds we obtain
\begin{eqnarray}
        \label{eq:RRO37exp}
        C_8 &=& \mathcal{L}_{(2)}\wedge m_6\,,\\
        C_6 &=& A_{(3)}^{\tilde K}\wedge\alpha_{\tilde K}+\tilde{A}^{(3)}_{\tilde K}\wedge\beta^{\tilde K} +
        \tilde{c}^{(2)}_a \wedge \tilde\omega^a\ ,\nn   \\
        C_4 &=& V^{\kappa}\wedge\alpha_{\kappa}+U_{\kappa}\wedge\beta^{\kappa}+\tilde{\rho}^\alpha_{(2)} \wedge\omega_\alpha
        +\rho_\alpha\tilde\omega^\alpha\ ,\nn\\
        \nn C_2 &=& c^a\omega_a\ ,\qquad C_0=l\,.
\end{eqnarray}
Here the fields $(\underline{A}_{(3)}, \underline{\tilde{A}}^{(3)})$ are
three-forms, $( \underline{\tilde{c}}^{(2)}, \underline{\tilde{\rho}}_{(2)},
\mathcal{C}_{(2)},\mathcal{L}_{(2)})$ are two-forms, $(\underline{V},\underline{U})$ 
are vectors and $(h,\underline{\rho},\underline{c},l)$ are scalars in spacetime $\mathbb{R}^{1,3}$. 
We note that $h$ and $\mathcal{C}_2$ are projected out in $O3/O7$-orientifolds since 
$C_2$, $C_6$ are odd under $\sigma$, whereas the scalar $l$ and its dual 
$\mathcal{L}_{(2)}$ are projected out in $O5/O9$-orientifolds as $C_0$, $C_8$ are odd. 
The scalar $l$ can be arranged as the four-dimensional axio-dilaton
\begin{equation}
	\tau=l+ie^{-\phi}
\label{eq:axiodilaton}
\end{equation}
which is of convenience in the study of the non-perturbative $SL(2,\mathds{Z})$-symmetry 
in setups with $O3/O7$-planes and F-theory. This non-perturbative symmetry is inherited 
from the ten-dimensional non-perturbative $SL(2,\mathds{Z})$-invariance of Type IIB string 
theory. It is of particular importance for the understanding of F-theory discussed in chapter
\ref{ch:HetFThyFiveBranes} since it requires new objects, $(p,q)$-strings and $(p,q)$ 7-branes. 
For the effective action derived for $O3/O7$-orientifolds we expect invariance  
of the $\mathcal{N}=1$ data under this duality group\footnote{Since the effective action will 
be derived from classical supergravity we obtain an even bigger duality group $SL(2,\mathbb{R})$. 
This is broken to the modular group by D$(-1)$-instantons.} which is explicitly checked in 
\cite{Grimm:2004uq}. 

Let us comment on the expansions \eqref{eq:RRO59exp}, \eqref{eq:RRO37exp}. Note that due 
to the duality constraints \eqref{eq:selfduality} not all degrees of freedom of the 
R--R-forms are physical. On the level of the four-dimensional effective action one can 
eliminate half of the degrees of freedom in \eqref{eq:RRO59exp}, \eqref{eq:RRO37exp} by 
introducing Lagrange multiplier terms. For example the duality of the gauge potentials 
$\underline{V}$, $\underline{U}$ amounts to standard $\mathcal{N}=1$ electric-magnetic 
duality in four dimensions. However, in order to couple the bulk fields to the D-brane 
sector as demonstrated in chapter \ref{ch:EffActD5}, it turns out to be useful to work 
with the democratic formulation. Only at the end of our analysis we choose a 
set of physical degrees of freedom and eliminate the remaining fields using 
\eqref{eq:selfduality}. 

In addition, we note that the expansions \eqref{eq:RRO59exp}, \eqref{eq:RRO37exp} also contain 
massless three-form fields $(\underline{A}_{(3)}, \underline{\tilde{A}}^{(3)})$ that are 
clearly non-dynamical in four space-time dimensions. However, we show in section 
\ref{sec:scalarpotderivation} that the inclusion of these three-forms is crucial to determine 
the scalar potential of compactifications with background fluxes and D-branes in a purely 
bosonic reduction. In case these are omitted a fermionic reduction has to be applied to derive 
the induced brane and flux superpotential as done, for example, for D5- and D7-branes 
in \cite{Jockers:2005zy,Martucci2006}.

\subsection{The Orientifold Effective Action}
\label{sec:EffActOrie}

In this section we determine the effective action for Type IIB $O3/O7$- and $O5/O9$-orientifold 
compactifications.  It is obtained from the Type IIB supergravity theory in a bosonic 
Kaluza-Klein reduction on $Z_3/\sigma$. Here we mainly follow the steps and conventions
used in \cite{Grimm:2004uq,Grimm:2005fa}.
 
The string-frame Type IIB action is used in its democratic form \cite{Bergshoeff:2001pv}
\begin{eqnarray} \label{eq:bulkaction}
        S_{\text{IIB}}^{\text{SF}}=\int\,\tfrac12 e^{-2\phi}R *_{10}
        1-\tfrac14
        \,e^{-2\phi}\left(8d\phi\wedge\ast_{10}d\phi-H\wedge\ast_{10}
        H\right)+\tfrac18\,\sum_{p\ \text{odd}}G^{(p)}\cdot G^{(p)},
\end{eqnarray}
where $H = dB_2$ denotes the NS--NS field strength, 
$G^{(p)}\cdot G^{(p)}=\pm G^{(p)}\wedge \ast_{10} G^{(p)}$
\footnote{According to the conventions 
in \cite{Bergshoeff:2001pv} a sign has to be included for a non-vanishing kinetic
term of the R--R form fields upon application the duality constraint \eqref{eq:selfduality}. 
It is $+1$, for $p<5$, and $-1$ for $p>4$.} and the R--R field strengths have been introduced in 
\eqref{eq:10dRRfieldstrength}. The corresponding action in ten-dimensional Einstein-frame with 
canonically normalized Einstein-Hilbert term is obtained by the Weyl rescaling 
$g^{\rm SF}_{10}=e^{\frac{\phi}{2}}g^{\rm EF}_{10}$ 
\cite{polchinski1998string}. We can use the duality constraints \eqref{eq:selfduality} imposed 
on the level of the equations of motion and the conventions from \cite{Bergshoeff:2001pv} like 
$(*_{10})^2=(-1)^{k+1}$ on $k$-forms to obtain the common Type IIB action from \eqref{eq:bulkaction}. 

Clearly the effective action is expected to differ in the R--R-sector for the $O3/O7$- and the 
$O5/O9$-setup. The NS--NS-sector will be qualitatively identical for both orientifolds. In 
particular, the dilaton is invariant in both cases as well as the expansion of the B-field 
\eqref{eq:Borie} and the K\"ahler form $J$ \eqref{eq:expJ} are identical whereas the complex 
structure variations \eqref{eq:var-g} of the metric are mapped onto each other by the exchange 
$H^{(2,1)}_+\leftrightarrow H^{(2,1)}_-$. We first present the complete actions in both cases, 
then comment on their derivation and the differences.

We begin with the $O5/O9$-setup. We readily insert the expansions of the 
fields in the NS--NS and in the R--R sector in \eqref{eq:RRO59exp} into the bulk action 
\eqref{eq:bulkaction} to obtain the action in four-dimensional Einstein-frame as 
\begin{eqnarray} \label{eq:O59action} 
  S^{(4)}_{O5/O9}\!\!&=&\!\!\int -\tfrac{1}{2} R *\mathbf{1} 
  - G_{\kappa \bar \lambda} \; dz^{\kappa} \wedge *d\bar z^{\lambda}
  +\tfrac{1}{4}\text{Im}\; \cM_{k l}\; 
    F^{k}\wedge *F^{l}
     +\tfrac{1}{4}\text{Re}\; \cM_{k l}\;
     F^{k}\wedge F^{l}  \!\!
    \\ 
  \!\!&\!\!& \!\!- G_{\alpha \beta} \; dv^\alpha \wedge *dv^\beta- G_{ab}\; db^a \wedge * db^b -d D \wedge * dD
   \nn \\ 
  \!\!&\!\!&\!\!- \tfrac{1}{4\mathcal{V}}e^{2D}(dh+\tfrac{1}{2}(d\rho_a b^a -\rho_a db^a))
  \wedge *(dh+\tfrac{1}{2}(d\rho_a b^a - \rho_a db^a)) - e^{2D} \mathcal{V} G_{\alpha \beta}\; dc^\alpha \wedge * dc^\beta\!\!\nonumber \\
  \!\!&\!\!&\!\!- \tfrac{1}{16 \mathcal{V}}e^{2D}G^{ab}(d \rho_a - \KK_{ac\alpha} c^\alpha db^c)
  \wedge *(d \rho_b - \KK_{bd\beta} c^\beta db^d)\ . \nn
\end{eqnarray} 
Here we have introduced the four-dimensional dilaton
\begin{equation}
	e^D=e^\phi\mathcal{V}^{-1/2}\,.
\label{eq:dilaton4d}
\end{equation}
Similarly for $O3/O7$-orientifolds we use the same expansions of the fields in the 
NS--NS sector and the expansion \eqref{eq:RRO37exp} of the R--R-fields to obtain
\begin{eqnarray}  \label{eq:O37action}
S^{(4)}_{O3/O7} &=&\int -\tfrac{1}{2} R *\mathbf{1} 
  - G_{k \bar l} \; dz^{k} \wedge *d\bar z^{l}+\tfrac{1}{4}\text{Im}\; \cM_{\kappa \lambda}\; 
    F^{\kappa}\wedge *F^{\lambda}
     +\tfrac{1}{4}\text{Re}\; \cM_{\kappa \lambda}\;
     F^{\kappa}\wedge F^{\lambda}\ 
  \nonumber \\
  &&  -G_{\alpha \beta} \; dv^\alpha \wedge *dv^\beta - G_{ab}\; db^a \wedge * db^b - dD \wedge * dD
    -\tfrac{1}{4}e^{2 D}\mathcal{V}\, dl \wedge * dl
  \nonumber \\
  && -   e^{2D} \mathcal{V}
  G_{ab}\left(dc^a-l db^a \right) \wedge *\left(dc^b-l db^b \right)\nn \\
  &&-\tfrac{1}{16 \mathcal{V}}e^{2D} G^{\alpha \beta}  \Big(d\rho_\alpha-
  \KK_{\alpha a b} c^a db^b \Big)
  \wedge
  *\Big(d\rho_\beta - \KK_{\beta cd} c^c db^d 
\Big)\ .  
\end{eqnarray}
There are several remarks in order.

In both cases the four-dimensional two-forms in \eqref{eq:RRO59exp} respectively 
\eqref{eq:RRO37exp} are traded for their dual scalars to express the action only 
in terms of chiral multiplets instead of linear multiplets. In technical terms 
this is achieved by adding Lagrange multipliers \cite{Dall'Agata2001} 
\begin{eqnarray}
	S_{O5/O9}^{\rm Lag}&=&\frac14\left( dV^{k}\wedge dU_{k}+d\mathcal{C}_{(2)} \wedge dh+d\tilde{\rho}_{(2)}^a\wedge d\rho_a
	+d\tilde{c}^{(2)}_\alpha\wedge dc^\alpha\right)\,,\\
	S_{O3/O7}^{\rm Lag}&=&\frac14\left( dV^\kappa\wedge dU_\kappa+d\mathcal{L}_{(2)} \wedge dl+d\tilde{\rho}_{(2)}^\alpha\wedge d\rho_\alpha
	+d\tilde{c}^{(2)}_a\wedge dc^a\right)\,,
\label{eq:LagrangeMultis}
\end{eqnarray} 
which add an irrelevant total derivative to the action. Then the equations of 
motion of the four-dimensional tensor fields consistently reproduce the duality 
constraints \eqref{eq:selfduality} after dimensional reduction. We refer to 
appendix \ref{app:F-termscalarpot} for a more detailed presentation of the dualization 
between the associated linear and chiral multiplets in a different context.

To further discuss the actions \eqref{eq:O59action}, \eqref{eq:O37action} it is 
convenient to organize the various terms according to their origin in the reduction 
from the four-dimensional $\mathcal{N}=2$ theory obtained by the reduction of Type 
IIB on $\mathbb{R}^{1,3}\times Z_3$ to the $\mathcal{N}=1$ orientifold theory.

The first term is the standard Einstein-Hilbert action in four-dimensions that is 
obtained by a Weyl rescaling $\mathcal{V}^{-1}\eta^{\rm EF}e^{-2\phi}=\eta^{\rm SF}$ 
between the four-dimensional String-frame (SF) and Einstein-frame (EF) metric. 
Note that $C_4$ is even (odd) for $O3/O7$ ($O5/O9$) setups but $h^{(3,0)}_+=0$ 
($h^{(3,0)}_-=0$). Thus, the four-dimensional graviphoton $V^0$ which is obtained 
from the reduction of $C_4\supset V^0\wedge\alpha_0$ in the $\mathcal{N}=2$ theory 
is projected out in both orientifolds. The second to fourth term are the reduction 
of kinetic and instanton action of the $\mathcal{N}=2$ vector multiplets to 
$\mathcal{N}=1$ chiral and vector multiplets. However, the $\mathcal{N}=2$ vector 
multiplets do not simply split into an equal number of $\mathcal{N}=1$ chiral and 
vector multiplets. Of each $\mathcal{N}=2$ vector multiplet either the chiral 
multiplet or the vector multiplet are projected out. Since the vector fields in the 
$\mathcal{N}=2$ vector multiplet are obtained from the reduction of $C_4$ whereas the 
complex scalars are the zero-modes in the complex structure variation of the internal 
metric captured by $\Omega$, the parity of $C_4$ and $\Omega$ with respect to the orientifold 
$\mathcal{O}$ will determine the remaining fields in the $\mathcal{N}=1$ orientifold 
theory. For $O5/O9$-orientifolds $h^{(2,1)}_+$ chiral multiplets with complex scalars 
$z^{\kappa}$ remain from the $h^{(2,1)}=h^{(2,1)}_++h^{(2,1)}_-$ vector multiplets of the 
$\mathcal{N}=2$ theory while the corresponding $\mathcal{N}=1$ vectors are 
projected out. Conversely $h^{(2,1)}_-$ vector multiplets remain while the corresponding 
$\mathcal{N}=1$ chiral multiplets are projected out. For $O3/O7$-orientifolds the 
opposite $\mathcal{N}=1$ multiplets survive. The scalar metric $G$ and the 
gauge kinetic and instanton couplings, encoded in the complex matrix $\mathcal{M}$ 
\cite{Suzuki1996,Ceresole1996}, are intimately related to the geometry of the complex 
structure moduli space $\cM^{\text cs}$ of the orientifold $Z_3/\sigma$. The scalar 
metric $G$ is a restriction of the $\mathcal{N}=2$ metric on the complex structure 
moduli space of $Z_3$. It is obtained from the K\"ahler potential $K_{\rm cs}$ as 
\begin{equation}\label{eq:csmetricOrie} 
G_{ \kappa\bar{\lambda}} = 
\frac{\partial}{\partial z^{\kappa}}
\frac{\partial}{\partial\bar z^{\lambda}}
\  K_{\rm cs}\ , \qquad
 K_{\rm cs} = -\ln\Big[ - i \int_{Z_3} \Omega \wedge \bar \Omega\Big] 
= -\ln i\Big[ Z^{K} \bar \cF_{K}    - \bar Z^{K} {\mathcal{F}}_{K} \Big]
\ ,
\end{equation}
for $O5/O9$-orientifolds and with $(\kappa,\lambda)\rightarrow (k,l)$ and 
$K\rightarrow \tilde{K}$ for $O3/O7$-orientifolds where we used the expansions 
\eqref{eq:periodexpansionOrie}. We note that the periods $(X^K,\cF_K)$ of $Z_3$ have 
to be evaluated on the subvariety $z^k=0$ ($z^\kappa=0$) for $O5/O9$ ($O3/O7$) setups. 
The definition of the complex matrix $\cM$ and its relation with the periods is 
summarized in appendix \ref{app:complexMatrixM}.

Similarly the remaining terms are related to reduction of the $\mathcal{N}=2$ 
double-tensor \cite{Brandt2000,Theis2003,Dall'Agata2004} or its dual chiral multiplet 
containing the dilaton $\phi$ and of the quaternionic K\"ahler manifold of the 
hypermultiplets. The double-tensor multiplet is composed of the four-dimensional 
two-forms and scalars $(B_2,\mathcal{C}_2,\phi,C_0)$. For $O5/O9$-orientifolds $B_2$ 
and $C_0$ are projected out whereas $B_2$ and $\mathcal{C}_2$ do not survive in 
$O3/O7$-orientifolds. In the quaternionic sector the parity of the corresponding 
R--R-forms determines which one of the two $\mathcal{N}=1$ chiral multiplets contained in the 
$\mathcal{N}=2$ hypermultiplet remains. This and the complete orientifold spectrum are 
summarized in table \ref{tab:N=1SpecOrie}. 
\begin{table}[!ht] 
\begin{center}
\begin{tabular}{|c|c|c||c|c|c||c|c|} \hline
\rule[-0.3cm]{0cm}{0.8cm}\hspace{-0.3cm}$\mathcal{N}=2$ &\multicolumn{2}{c||}{Type IIB}& \hspace{-0.2cm}$\mathcal{N}=1$& \multicolumn{2}{c||}{$O3/O7$}&\multicolumn{2}{c|}{$O5/O9$}\\
 \rule[-0.3cm]{0cm}{0.8cm}  \hspace{-0.2cm}multiplet &\multicolumn{2}{c||}{} & \hspace{-0.2cm}multiplet &\multicolumn{2}{c||}{}  &\multicolumn{2}{c|}{} \\ \hline 
 \rule[-0.3cm]{0cm}{0.85cm}\hspace{-0.3cm}gravity & 1\!\! &\!\!\!\! $(g_{\mu\nu},V^0)$\!\! & 
 \hspace{-0.1cm}gravity &1\!\!&$g_{\mu \nu} $ &1\!\!& $g_{\mu \nu}$ \\ \hline
 \rule[-0.3cm]{0cm}{0.85cm} 
 \multirow{2}*{\hspace{-0.3cm}vector}& \multirow{2}*{$h^{(2,1)}$} \!\!&\!\!\!\! \multirow{2}*{$(V^K,z^K)$} \!\!&\hspace{-0.1cm}vector &   $ h_+^{(2,1)}\! $&  $V^{\lambda} $& $ h_-^{(2,1)}\! $ & $V^{k} $ \\ \cline{4-8}
 \rule[-0.3cm]{0cm}{0.85cm} & & &
 \hspace{-0.1cm}chiral &   $h_-^{(2,1)}\!$& $z^{k} $ &   $h_+^{(2,1)}\!$& $z^{\lambda} $\\ \hline
\rule[-0.3cm]{0cm}{0.85cm} 
 \multirow{2}*{\hspace{-0.3cm}hyper}&  \multirow{2}*{$h^{(1,1)}$}\!\! &\!\!\!\!  \multirow{2}*{$(v^i,b^i,c^i,\rho^i)$}\!\! & \hspace{-0.1cm}chiral/linear &  $h^{(1,1)}_+\!$& $t^\alpha=( v^\alpha, \rho_\alpha )$ & $h^{(1,1)}_-\!$& 
 $P^a=( b^a, \rho_a )$ \\ \cline{4-8}
  \rule[-0.3cm]{0cm}{0.85cm}
 & & & \hspace{-0.1cm}chiral &$ h^{(1,1)}_-\!$ &$G^a=( b^a, c^a)$ & $h^{(1,1)}_+\!$& $t^\alpha=( v^\alpha, c^\alpha)$
 \\ \hline
 \rule[-0.3cm]{0cm}{0.85cm} 
   \hspace{-0.2cm}double-&  \multirow{2}*{1}\!\! &\!\!\!\! \multirow{2}*{ $(B_2,\mathcal{C}_2,\phi,l)$}\!\! & \multirow{2}*{\hspace{-0.1cm}chiral/linear} & 1\!\! & $\tau=(\phi,l)$ && \\
   \cline{5-8}
\rule[-0.3cm]{0cm}{0.85cm} \hspace{-0.2cm}tensor& & &  & && 1\!\! & $S=(\phi,C_2)$ \\
\hline
\end{tabular}
\caption{\label{tab:N=1SpecOrie} The $\mathcal{N} =1$ spectrum of the Type IIB orientifold 
compactifications. The split of the $\mathcal{N}=2$ multiplet to $\mathcal{N}=1$
multiplets is indicated. The labels of the $\mathcal{N}=2$ hypermultiplets
correspond to a basis $\omega^i$ of $H^{(1,1)}(Z_3)$.}
\end{center}
\end{table} The coupling functions for the remaining terms in the actions 
\eqref{eq:O59action}, \eqref{eq:O37action} are encoded by the K\"ahler sector of 
$Z_3/\sigma$, which are the classical intersection numbers $\mathcal{K}_{ijk}$ of
three divisors in $Z_3$ or their duals in $H^{(1,1)}(Z_3)$
and the metric on the K\"ahler moduli space $\cM^{\rm K}$. The restriction imposed 
by the orientifold are universal for both setups since $\sigma^*J=J$ and  
$\sigma^* B_2=- B_2$ do hold for all B-parities, c.f.~\eqref{eq:Borie} 
and \eqref{eq:geoOrie}. The decomposition $H^{(1,1)}=H^{(1,1)}_+\oplus H^{(1,1)}_-$ 
results in a decomposition of the intersection numbers $\mathcal{K}_{ijk}$. In the 
basis of table \ref{tab:cohombasisOrie} only $\mathcal{K}_{\alpha\beta\gamma}$ and 
$\mathcal{K}_{\alpha bc}$ can be non-zero while
\begin{eqnarray}\label{eq:intersectsOrie}
  \mathcal{K}_{\alpha \beta c}= \mathcal{K}_{abc} =\mathcal{K}_{\alpha b}
  =\mathcal{K}_{a}=0\ ,
\end{eqnarray}
is imposed by the orientifold projection. Here we used the definition 
$\mathcal{K}_{\alpha b}=v^\beta\mathcal{K}_{\alpha\beta b}$, 
$\mathcal{K}_{a}=v^\alpha v^\beta \mathcal{K}_{\alpha\beta a}$. In general the K\"ahler metric 
on $Z_3$ is in general given by \cite{Strominger1985,Candelas1991}
\begin{equation}	
	G_{ij}=\frac{1}{4\mathcal{V}}\int_{Z_3}\omega_i\wedge\ast\omega_j
	=-\frac{\mathcal{K}_{ij}}{4\mathcal{V}}+
  \frac{\mathcal{K}_i \mathcal{K}_j}{16\mathcal{V}^2}\,,
\label{eq:KaehlermetricGen}
\end{equation}
where $\ast$ denotes the Hodge-star on $Z_3$ and $\omega_i$ a basis of $H^{(1,1)}(Z_3)$. 
It can be derived from the K\"ahler potential $K^{\rm K}$ given by
\begin{equation}
	K^{\rm K}=-\ln(8\mathcal{V})
	=-\ln\big(\frac{i}{6}\mathcal{K}_{ijk}(t+\bar{t})^i(t+\bar t)^j
(t+\bar t)^k\big)\,,
\label{eq:KaehlerPotGen}
\end{equation}
where the $t^i=v^i+ib^i$ denote the complexified coordinates on the K\"ahler moduli 
space of $Z_3$. The metric on the K\"ahler moduli space $\cM^{K}$ of the orientifold 
is then obtained as
\begin{eqnarray} \label{eq:KaehlerMetricOrie}
  G_{\alpha \beta}=
  -\frac{\mathcal{K}_{\alpha \beta}}{4\mathcal{V}}+
  \frac{\mathcal{K}_\alpha \mathcal{K}_\beta}{16\mathcal{V}^2}\ , \qquad
  G_{a b}=- \frac{\mathcal{K}_{a b}}{4\mathcal{V}}\ , \qquad
G_{\alpha b}\ =\ G_{a \beta}\ =\ 0\ ,
\end{eqnarray}
where we introduced the intersections
\begin{equation}\label{eq:intsabbrev}
  \KK_{\alpha\beta}=\KK_{\alpha\beta\gamma}\; v^\gamma\ , 
\quad \ 
\KK_{ab}=\KK_{ab\gamma}\; v^\gamma\ ,\quad
\KK_{\alpha}=\KK_{\alpha \beta\gamma}\; v^\beta v^\gamma\ ,
  \quad \KK=\KK_{\alpha \beta \gamma} \; v^\alpha v^\beta v^\gamma
\ .
\end{equation}
Its inverse is given by
\begin{equation}\label{eq:KaehlerMetricinvers}
  G^{\alpha \beta}
  =  -\frac{2}{3} \mathcal{K} \mathcal{K}^{\alpha \beta} + 2 v^\alpha v^\beta\ ,
\qquad
  G^{a b} =
   - \frac{2}{3} \mathcal{K} \mathcal{K}^{a b}\ ,
\end{equation}
where $\KK^{\alpha \beta}$ and $\KK^{a b}$ are the formal inverses of 
$\KK_{\alpha \beta}$ and $\KK_{a b}$, respectively. 

Having defined the couplings of the reduced quaternionic sector we note that the 
corresponding orientifold actions \eqref{eq:O59action}, \eqref{eq:O37action} for 
the $O5/O9$- and the $O3/O7$-setup differ since the massless spectrum from the 
R--R-sector differs, which in particular yields different couplings of the B-field 
zero modes $b^a$.

\subsection{The $\mathcal{N}=1$ Couplings and Coordinates}
\label{sec:N=1dataOrie}

In this section we rewrite the four-dimensional effective action of the orientifold 
theory into the standard $\mathcal{N}=1$ supergravity form. We first determine the complex 
coordinates $M^I$ forming the bosonic part of the $\mathcal{N}=1$ chiral multiplets. Their 
kinetic terms are expressed by a K\"ahler potential $K(M,\bar M)$, while their
F-term scalar potential is encoded by a  holomorphic superpotential $W(M)$. The 
couplings of the $\cN=1$ vector multiplets are encoded by holomorphic gauge-kinetic 
coupling functions $f(M)$. Possible gaugings of the scalars $M^I$ will induce a 
D-term potential. The general form of the bosonic $\cN=1$ action is given by 
\cite{Wess,Gates:1983nr}
\begin{equation}\label{eq:N=1action}
  S^{(4)} = -\int \tfrac{1}{2}R * 1 +
  K_{I \bar J} \cD M^I \wedge * \cD \bar M^{\bar J}  
  + \tfrac{1}{2}\text{Re}f_{\kappa \lambda}\ 
  F^{\kappa} \wedge * F^{\lambda}  
  + \tfrac{1}{2}\text{Im} f_{\kappa \lambda}\ 
  F^{\kappa} \wedge F^{\lambda} + V*1\ ,
\end{equation}
where the scalar potential reads
\begin{equation}\label{eq:N=1pot}
V=
e^K \big( K^{I\bar J} D_I W {D_{\bar J} \bar W}-3|W|^2 \big)
+\tfrac{1}{2}\, 
(\text{Re}\; f)^{-1\ \kappa\lambda} D_{\kappa} D_{\lambda}
\ .
\end{equation}
Here we set the four-dimensional gravitational coupling $\kappa_4=1$ for simplicity. 
We denote by $K_{I \bar J}$ and $K^{I \bar J}$ the K\"ahler metric and its inverse, 
where the former is locally given as $K_{I\bar J} = \partial_I \bar\partial_{\bar J} K(M,\bar M)$. 
The scalar potential is expressed in terms of the K\"ahler-covariant derivative 
$D_I W= \partial_I W + (\partial_I K) W$ which promotes $\partial_I K$ to a connection 
one-form on the scalar manifold.

Before we begin our discussion we note that in case that the action \eqref{eq:N=1action} 
is deduced from a higher-dimensional theory the formal expression for the K\"ahler potential 
can in general already be inferred from Weyl rescaling arguments, e.g.~from the factor $e^K$ 
in front of the $\cN=1$ F-term potential in \eqref{eq:N=1pot}. However, the explicit form as a 
function of the $\mathcal{N}=1$ chiral superfields depends on the setup under consideration 
and is in general not analytically computable.  
We also note that we will not obtain, for the case of orientifold compactifications of Type IIB, 
the most general $\mathcal{N}=1$ effective action since we 
did not include fluxes and thus do not obtain an F-term scalar potential, i.e.~we find $W=0$. 
We will include fluxes and a non-trivial superpotential in section \ref{sec:D5branes}.

\subsubsection{The K\"ahler potential and $\cN=1$ coordinates \label{sec:Kpot}}

As a first step to identify the $\mathcal{N}=1$ characteristic data we have to find 
the appropriate complex coordinates for which the scalar metrics are manifestly K\"ahler. 
First, we note that the complex structure moduli $z^\kappa$, respectively $z^k$, already 
define K\"ahler coordinates for both orientifolds since the scalar metric in the actions 
\eqref{eq:O59action}, \eqref{eq:O37action} is already K\"ahler with K\"ahler potential 
\eqref{eq:csmetricOrie}. The other K\"ahler coordinates are different for the two setups
and are discussed next.

For $O5/O9$-orientifolds the additional complex fields in the chiral multiplets read 
\cite{Grimm:2004uq,Grimm:2005fa}
\begin{equation} \label{eq:N=1coordsO59}
   t^\alpha = e^{-\phi} v^\alpha-ic^\alpha\ ,\qquad P_a = \Theta_{ab}\, b^b + i \rho_a\ ,\qquad
   S =  e^{-\phi} \cV + i \tilde{h} -\tfrac14 (\text{Re} \Theta)^{ab} P_a (P+\bar
   P)_b \ ,
\end{equation}
where $v^\alpha,b^a$, $c^\alpha,\rho_a$ and $\tilde{h}=h-\frac12\rho_ab^a$ are given in 
\eqref{eq:NSNSexp}, \eqref{eq:expJ} and \eqref{eq:RRO59exp}. The complex symmetric tensor 
$\Theta$ appearing in \eqref{eq:N=1coordsO59} is given by $\Theta_{ab}=\cK_{ab \alpha} t^\alpha$ 
and $(\text{Re} \Theta)^{ab}$ denotes the inverse of $\text{Re} \Theta_{ab}$. 
More conceptually, the coordinates \eqref{eq:N=1coordsO59} can be obtained
by probing the cycles in $Z_3/\sigma$ by D-branes and are then formally extracted from \cite{Haack2000,Haack:2001jz,Becker2002,Grimm:2004uq,Grimm:2005fa}
\begin{equation}
	\text{Im}(\varphi^{\text ev})-i\mathcal{A}=t^\alpha\omega_\alpha-P_b\tilde{\omega}^b-Sm_6\,.
\label{eq:N=1coordsO59compact}
\end{equation}
Here we introduced the polyforms of even 
forms $\varphi^{\rm ev}$, $\mathcal{A}$ in the NS--NS- respectively R--R-sector,
\begin{equation} \label{eq:evenpolyforms}
   \varphi^{\text ev} =e^{-\phi} e^{- B_2 + iJ}\ ,\qquad \mathcal{A} = e^{- B_2} \wedge \sum_{q=0,2,4,6,8} C_q \ , 
\end{equation} 
which have to be interpreted, as \eqref{eq:N=1coordsO59compact}, as formal polynomials 
of forms in the even cohomology $H^{\text{ev}}(Z_3)$ of different degrees.
Then the reduction ans\"atze \eqref{eq:NSNSexp}, \eqref{eq:expJ} and \eqref{eq:RRO59exp} 
have to be inserted to obtain the coordinates \eqref{eq:N=1coordsO59} as the coefficients  
of a basis expansion of \eqref{eq:N=1coordsO59compact}.  Roughly, the forms $\varphi^{\text ev}$ and $\mathcal{A}$
are the Dirac-Born-Infeld and Chern-Simons actions of D$p$-instantons, $p=1,3,5$, wrapping the even-dimensional cycles
and the coordinates \eqref{eq:N=1coordsO59} are the complexified volumes of the corresponding 
cycles in the Type IIB $O5/O9$-background.

For $O3/O7$-orientifolds besides the complex structure moduli $z^k$ the additional K\"ahler
coordinates read \cite{Grimm:2004uq,Grimm:2005fa},
\begin{equation} \label{eq:N=1coordsO37}
  \tau = l+ie^{-\phi} \ , \quad  G^a =c^a-\tau b^a\ ,\quad
  T_\alpha =  i( \rho_\alpha - \tfrac{1}{2} \mathcal{K}_{\alpha ab}c^a b^b) + \tfrac{1}{2} e^{-\phi} \mathcal{K}_\alpha
               + \frac{i}{2(\tau-\bar{\tau})}\mathcal{K}_{\alpha bc}G^b(G-\bar{G})^c\ , 
\end{equation}
where $\mathcal{K}_\alpha$ is introduced in \eqref{eq:intsabbrev}. This data is again 
summarized as
\begin{equation}
	\text{Re}(\varphi^{\text ev})-i\mathcal{A}=-i\tau-iG^a\omega_a-T_\alpha\tilde{\omega}^\alpha\,,
\label{eq:N=1coordsO37compact}
\end{equation} 
which is interpreted as the complexified volume of even dimensional cycles wrapped by 
D$p$-instantons, $p=-1,1,3$, in the Type IIB $O3/O7$-background.

The definition of the appropriate $\mathcal{N}=1$ coordinates prepares us for the determination
of the $\mathcal{N}=1$ coupling functions. As mentioned before, we do not obtain 
superpotential a $W$ due to the absence of fluxes and D-branes, $W=0$. We first 
determine the K\"ahler potential $K$. In general, the full $\cN=1$ K\"ahler potential 
is determined by integrating the kinetic terms of the complex scalars $M^I$.

For $O5/O9$-orientifolds it takes the form \cite{Grimm:2004uq,Grimm:2005fa}
\begin{equation} \label{eq:O59Kaehlerpot} 
  K \ =\ K_{\rm cs}(z,\bar z) + K^{\rm q}(S,t,P)\ , \qquad K_{\rm cs} = -\ln\Big[-i\int\Omega \wedge \bar \Omega \Big]
\end{equation}
where we introduce the K\"ahler potential
\begin{eqnarray} \label{eq:O59kaehlerpotMq}
 K^{\rm q}& =& - \ln\Big[\tfrac{1}{48}\mathcal{K}_{\alpha \beta \gamma}(t+\bar t)^\alpha 
        (t+\bar t)^\beta (t+\bar t)^\gamma  \Big]
      - \ln\Big[S + \bar S + \tfrac{1}{4} (P + \bar P)_a (\text{Re}\Theta^{-1})^{ab} 
        (P + \bar P)_b \Big]\nn\\
     &=& -2\ln\big[\sqrt{2} e^{-2\phi}\mathcal{V}\big]\ . 
\end{eqnarray}
We note that the scalar metric corresponding to $K$ is block-diagonal in the fields
$\underline{z}$ and $(S,\underline{P},\underline{t})$. Thus, we infer that the 
scalar manifold of the $O5/O9$-orientifold effective theory is locally a direct product 
$\mathcal{M}^{\rm cs}\times \mathcal{M}^{\rm q}$. The first factor is the special K\"ahler 
manifold with local coordinates $z^\kappa$. It is realized as a submanifold of the complex 
structure moduli space of $Z_3$, which itself is special K\"ahler. The second factor 
$\mathcal{M}^{\rm q}$ is a K\"ahler submanifold of dimension $h^{(1,1)}+1$ in the quaternionic 
manifold of the $\mathcal{N}=2$ hypermultiplets. Locally it is a fibration of the scalar 
manifold of $\underline{P}$, $S$ over the K\"ahler moduli space of the orientifold $Z_3/\sigma$. 

For $O3/O7$-orientifolds the K\"ahler potential reads \cite{Grimm:2004uq,Grimm:2005fa}
\begin{equation} \label{eq:O37Kaehlerpot} 
  K \ =\ K_{\rm cs}(z,\bar z) + K^{\rm q}(\tau,T,G)\ , \qquad 
  K_{\rm cs} = -\ln\Big[-i\int\Omega \wedge \bar \Omega \Big]
\end{equation}
where we introduce the K\"ahler potential
\begin{eqnarray} \label{eq:O37kaehlerpotMq}
 K^{\rm q}& =& - \ln\Big[-i(\tau-\bar \tau)\Big]
      - 2\ln\Big[\frac16 e^{-\frac32\phi}\mathcal{K}_{\alpha\beta\gamma}v^\alpha v^\beta v^\gamma\Big]
      = -2\ln\big[\sqrt{2} e^{-2\phi}\mathcal{V}\big]\ . 
\end{eqnarray}
We note that the scalar manifold of the $O3/O7$-orientifold effective theory is again 
locally a direct product $\mathcal{M}^{\rm cs}\times \mathcal{M}^{\rm q}$. However, the 
first factor is a different special K\"ahler submanifold with local coordinates $z^k$ 
of the complex structure moduli space of $Z_3$ than in the $O5/O9$ case. The second factor 
$\mathcal{M}^{\rm q}$ is a K\"ahler submanifold of the quaternionic manifold of the 
$\mathcal{N}=2$ hypermultiplets. Locally it is a fibration of the scalar manifold of 
$\underline{T}$, $\underline{G}$ over the dilaton moduli space. 
We note that the full K\"ahler potential $K$ is invariant, up to a K\"ahler transformation, 
under the action of $SL(2,\mathbb{R})$ on $\tau$ as required by $SL(2,\mathbb{R})$-duality of the 
ten-dimensional Type IIB theory.

In general a first check of the consistency of the K\"ahler coordinates \eqref{eq:N=1coordsO59},
\eqref{eq:N=1coordsO59} with the corresponding K\"ahler potentials \eqref{eq:O59Kaehlerpot}, 
\eqref{eq:O37Kaehlerpot} is provided by a comparison with the expected result from Weyl rescaling,
\begin{equation}
	K=-2\ln\big[\sqrt{2} e^{-2\phi}\mathcal{V}\int\Omega\wedge\bar{\Omega}\big]\,.
\label{eq:WeylRescalingKahler}
\end{equation}
This is in perfect agreement with the above results. However, we emphasize that the real 
difficulty in the determination of the $\mathcal{N}=1$ characteristic data is to express 
the general expression \eqref{eq:WeylRescalingKahler} as a function of appropriate 
coordinates so that all scalar kinetic terms, as well as some mixing terms, in the action are reproduced.

We conclude by rephrasing the $\mathcal{N}=1$ K\"ahler potential in a more formal and compact form for 
both orientifolds. We note that both K\"ahler potentials \eqref{eq:O59kaehlerpotMq}, \eqref{eq:O37kaehlerpotMq} 
can be written in a unified way as \cite{Grimm:2004uq,Grimm:2005fa}
\begin{equation} \label{eq:KaehlerpotMqUnified}
  K^{\rm q}\ =\ -2 \ln \Phi_B\ ,\qquad \Phi_B:=i\left\langle \varphi^{\rm ev},\bar{\varphi}^{\rm ev}\right\rangle
\end{equation}
up to a K\"ahler transformation where we introduced the pairing \cite{Hitchin2003}
\begin{equation}
	\big<\varphi,\psi \big>=\int_{Z_3}\sum_m(-1)^m\varphi_{2m}\wedge\psi_{6-2m}
\label{eq:pairingPolyform}
\end{equation}
on the space of polyforms. The only difference between the two orientifold setups 
then reduces to the choice of variables on which \eqref{eq:KaehlerpotMqUnified} 
depends. For the case of $O5/O9$-orientifolds $K^{\rm q}$ is a function of 
$\text{Im}(\varphi^{\rm ev})$ only, in particular the real part has to be understood as 
a function of $\text{Im}(\varphi^{\rm ev})$.
For $O3/O7$-orientifolds, however, \eqref{eq:KaehlerpotMqUnified} is a function of 
$\text{Re}(\varphi^{\rm ev})$ and the imaginary parts is determined by 
$\text{Re}(\varphi^{\rm ev})$.

\subsubsection{The Gauge Kinetic Function \label{sec:gaugeKin}}

We conclude with the determination of the gauge kinetic coupling $f_{\kappa \lambda}$ 
which is a holomorphic function of the chiral superfields. By comparison of the 
actions \eqref{eq:O59action}, \eqref{eq:O37action} with the general $\mathcal{N}=1$ 
action \eqref{eq:N=1action} we obtain \cite{Grimm:2004uq,Grimm:2005fa}
\begin{equation} \label{eq:gaugeKinOrie}
 O5/O9\ :\ \ f_{k l}
=-\tfrac{i}{2} \,
\bar{\cM}_{k l}\Big|_{z^{k}=0=\bar z^{l}}\ ,\qquad O3/O7\ :\ \ f_{\kappa \lambda}
=-\tfrac{i}{2} \,
\bar{\cM}_{\kappa \lambda}\Big|_{z^{\kappa}=0=\bar z^{\lambda}}
\end{equation}
where 
${\cM}$ is introduced in appendix \ref{app:complexMatrixM} and a submatrix of the 
$\mathcal{N}=2$ gauge kinetic coupling matrix. It has to be evaluated on the submanifold 
${z^{k}=\bar z^{l}}=0$ or ${z^{\kappa}=\bar z^{\lambda}}=0$, respectively, in the complex 
structure moduli space of $Z_3$. Its holomorphicity in the complex structure 
deformations $z^\kappa$, $z^k$ follows since $f$ can be rewritten for both orientifolds 
in terms of the holomorphic $\mathcal{N}=2$ prepotential $\mathcal{F}(z)$ evaluated 
on the submanifold $\mathcal{M}^{\rm cs}$ \cite{Grimm:2004uq,Grimm:2005fa},
\begin{eqnarray}\label{eq:gaugeKinprepot}
  O5/O9\ :\ \ f_{k l}(z^\kappa)=-\tfrac{i}{2} 
\mathcal{F}_{kl}\Big|_{z^{k}=0=\bar z^{l}}\ ,\qquad O3/O7\ :\ \ f_{\kappa \lambda} (z^k)
=-\tfrac{i}{2} 
\mathcal{F}_{\kappa \lambda}\Big|_{z^{\kappa}=0=\bar z^{\lambda}}\ .
\end{eqnarray} 

\chapter{The D5-Brane Effective Action}
\label{ch:EffActD5}

In this chapter we begin with the analysis of brane dynamics for the example of D5-branes in generic 
$O5/O9$-Calabi-Yau orientifolds. We present a detailed computation of the $\mathcal{N}=1$ effective 
action of a spacetime-filling D5-brane wrapping a curve $\Sigma$, mainly following the original work \cite{Grimm:2008dq}. 
First in section \ref{sec:DbranesinCY3Orie} we start with a general review of supersymmetric D-branes 
in Calabi-Yau manifolds, where we focus on their low-energy effective dynamics as encoded by the 
Dirac-Born-Infeld and Chern-Simons action and on the geometric calibration conditions on the internal 
cycles supporting a BPS D-brane. Then the actual calculation of the D5-brane effective action is 
performed in section \ref{sec:D5branes} by a purely bosonic reduction. This is divided into four parts, 
in which we first count the bosonic four-dimensional massless spectrum both of the closed and open string 
sector, then present some special identities on the space of open-closed geometric moduli before we 
actually derive the D5-brane effective action, where we put special emphasis on the determination of the 
scalar potential. The calculation of the scalar potential reveals the ad-hoc surprising relevance of 
keeping non-dynamical three-forms in the four-dimensional action in order to recover the complete 
F-term potential in a purely bosonic reduction. Next in section \ref{sec:N=1dataD5} we work out the 
$\mathcal{N}=1$ characteristic data of the D5-brane action, in particular in comparison with 
$O5/O9$-orientifold result of chapter \ref{ch:EffActCYOrieCompact}. We determine new $\mathcal{N}=1$ 
chiral coordinates and the K\"ahler potential that is derived explicitly at large volume of $Z_3$. 
Then the effective superpotential is read off, which is a projection of the familiar Type IIB 
flux superpotential and the D5-brane superpotential, that is given as a chain integral. 
The $\mathcal{N}=1$ data is completed by the determination of the D5-brane gauge kinetic function, 
the gaugings of chiral fields and the corresponding D-terms. 
Finally in section \ref{sec:extensionInfinite} we extend, following \cite{Grimm:2010gk}, the reduction 
of \cite{Grimm:2008dq} to the  full geometric deformation space of the D5-brane which is infinite 
dimensional and does include massive fields. This is proposed to be the natural domain of the 
D5-brane superpotential which encodes the general deformation theory of the curve $\Sigma$ in 
Calabi-Yau threefold $Z_3$. We concluded by a calculation of the corresponding F-term scalar potential 
as a functional of the infinite dimensional space of massive modes. Further details on some calculations
performed in the context of the derivation of the D5-brane action can be found in appendix \ref{app:PartEffActions}.

\section{BPS D-Branes in Calabi-Yau Manifolds}
\label{sec:DbranesinCY3Orie}

In general the existence of D-branes as dynamical objects in string theory
can be inferred from first principles, namely the spectrum of BPS-particles 
and string dualities. It is a basic fact found in textbooks, see 
e.g.~\cite{becker2007string}, that the fundamental string state with 
two oscillator modes, winding number $m$ and no momentum along a compact 
spacetime direction, here an $S^1$, has a mass proportional to $m$. 
The winding number $m$ is the quantized charge with respect to the NS--NS 
B-field and indeed this state is identified as a BPS-state in the theory 
\cite{Dabholkar1997a}. In combination with the non-perturbative 
$SL(2,\mathds{Z})$-symmetry of Type IIB theory, that acts on the 
B-field $B_2$ and the R--R-form $C_2$ as a doublet, this predicts a 
whole family of BPS-states with charge $(p,q)$ under $(B_2,C_2)$ for any pair 
of relative primes $p$, $q$ \cite{Witten:1995im}. In particular one expects 
fundamental states with charge $(0,1)$ that are charged under $C_2$. Since $C_2$
couples electrically to string-like objects the fundamental object corresponding 
to these states is called a D-string. Analogously, the strings associated to 
BPS-states with charge $(p,q)$ are denoted $(p,q)$-strings. Then by T-duality 
one expects BPS-states charged 
under all R--R-forms $C_p$ in the theory which correspond to quantized 
$p+1$-dimensional membranes. However, these states do not have a perturbative 
description in terms of quantized fundamental string excitations since all 
perturbative states do not couple to the R--R-forms but their field strengths. 
Consequently, these states are invisible in the spectrum of the perturbative 
string. However, they have convenient descriptions in various limits of string theory.

On the one hand, at low-energies membranes with non-zero electric R--R-charge are found  
in the effective supergravity theory as solitons, i.e.~stable finite energy 
solutions of the classical equations of motions with a conserved 
(topological) charge called $p$-branes, see e.g.~\cite{becker2007string} for a summary. 
Also the fundamental string occurs on an equal footing as a solitonic solution 
with non-zero NS--NS-charge. It can further be shown that these solutions indeed obey a BPS-bound.
On the other hand, membranes are visible at small coupling in the perturbative string as 
$9-p$ Dirichlet boundary conditions on the string endpoints, denoted by D$p$-branes.

Both descriptions of membranes have different regimes of validity, however. The 
requirement of the supergravity description is low curvature, which 
in more technical terms means, that the Schwarzschild radius $r_S$ of 
the $p$-brane solution is big compared to the string scale $\sqrt{\alpha'}$.
This requirement yields \cite{Mohaupt2000}
\begin{equation}
	n_p g_S>> 1
\label{eq:SUGRAregime}
\end{equation}
where $n_p$ is the integral number of fundamental $p$-brane charge of 
the solution. 
The requirement for the validity of the perturbative string is of course
small string coupling $g_S$ so that we obtain
\begin{equation}
	n_p g_S<< 1\,,
\label{eq:PerpSTregime}
\end{equation}
which is the opposite to \eqref{eq:SUGRAregime}.
Despite their different validity, however, it is believed that both the 
$p$-brane and the D$p$-brane describe one and the same fundamental object 
simply denoted as a D$p$-brane. This conjecture is well established by various 
checks invoking techniques from different corners of string theory starting with the 
CFT analysis in the groundbreaking work \cite{Polchinski1995}, see 
\cite{Blumenhagen:2005mu,Douglas:2006es,Blumenhagen:2006ci,Denef:2008wq} 
for reviews.  

For our purposes we treat spacetime-filling D-branes and their dynamics as a 
BPS-background state in Type IIB string theory with small 
fluctuations around it. Here we will 
mainly be restricted to low energies which means that we are working
with Type IIB supergravity and to scales larger than the thickness
of the D-brane where the description in terms of the D-brane effective 
action is valid. Then the BPS condition on the D-brane is translated
into BPS calibration conditions \cite{Becker:1995kb} on the embedding 
of the D-brane worldvolume $\mathcal{W}_{p+1}$ and the D-brane worldvolume fields 
into the Type IIB background under consideration. By additional 
consistency conditions of the setup imposed on the one hand by $\mathcal{N}=1$ 
supersymmetry in extended four-dimensional Minkowski space and on the
other hand by cancellations of tadpoles we will consider 
Calabi-Yau threefolds $Z_3$ with $O5/O9$-orientifolds.
 
Following this program we start by introducing the effective action of
a single D-brane. The spectrum of light fields on the D-brane can in 
general be thought of as a reduction of the $\mathcal{N}=1$ 
Super-Yang-Mills theory in ten dimensions to the worldvolume $\mathcal{W}_{p+1}$
\cite{Witten:1995im}. 
It consists of an $\mathcal{N}=1$ supersymmetric $U(1)$-gauge theory on the 
submanifold $\mathcal{W}_{p+1}$. In addition there are adjoint-valued 
scalars corresponding to fluctuations $\delta \iota$ of the embedding 
$\iota\,:\, \mathcal{W}_{p+1}\rightarrow M_{10}$ into ten-dimensional 
spacetime. The effective action of these fields consists of two parts, 
one part being the Dirac-Born-Infeld action 
\cite{Polchinski1996,polchinski1998string,johnson2006d}
\begin{equation}
	S^{\rm SF}_{\rm DBI}=-T_p\int_{\mathcal{W}_{p+1}}e^{-\phi}\sqrt{\det{\iota^*(g+B_2)-\ell F}}
\label{eq:DBIgeneral}
\end{equation}
in string-frame units, where $T_p$ denotes the D$p$-brane tension  
and $\ell=2\pi\alpha'$. The second part of the effective action is 
the Chern-Simons action encoding the coupling of the D-brane to 
the R--R-sector,
\begin{equation}
	S_{\rm CS}=-\mu_p\int_{\mathcal{W}_{p+1}}\sum_q C_q\wedge e^{\ell F- B_2}\,,
\label{eq:CSgeneral}
\end{equation}
where $\mu_p$ denotes the D-brane charge. The formal sum over all 
R--R-potentials indicates a coupling of the D-brane to forms of degree
lower than $C_{p+1}$ in a background of worldvolume flux/instantons 
\cite{Douglas:1995bn}. We note that the presence of the B-field $B_2$  in the 
combination $\mathcal{B}:=B_2-\ell F$ can be understood from gauge invariance 
in the bulk. The string sigma-model action for world-sheets $\mathcal{W}_2$ 
with boundary contains the terms
\begin{equation}
	S_{\sigma}\supset \frac1{\ell}\int_{\mathcal{W}_2} B_2
	+\int_{\partial \mathcal{W}_2} A\,,
\end{equation}   
where $A$ denotes the gauge field on the D-brane. Since this is not 
invariant under the gauge transformation $\delta B_2=d\Lambda$ alone, one 
has to demand $\delta A=\frac1{\ell}\Lambda$. Thus, $\mathcal{B}$ is the only
gauge invariant combination \cite{Witten:1995im}. 

Next we formulate the BPS conditions a D-brane in a Calabi-Yau manifold 
has to meet to define a supersymmetric background 
\cite{Becker:1995kb,Marino:1999af}. The most immediate condition is of course $\mu_p=T_p$
which is enforced by the SUSY algebra for every BPS state. For vanishing 
background fields the BPS-conditions further imply for a spacetime-filling D6-brane in Type IIA 
to wrap a special Lagrangian submanifold $L$, which can roughly be thought 
of as a real locus of real dimensions $d$ in a complex manifold of complex 
dimension $d$. In terms of the complex structure $I$ on $Z_3$ this is 
expressed by $I(TL)\subset NL$. A trivial example is a special 
Lagrangian $L$ in $\mathbb{C}$ which is just e.g.~the real axis, 
where $I$ is the usual multiplication by $i$. Equivalently $L$ is defined as the volume-minimizing 
representative within its homology class such that the volume form $\vol_L$ 
on $L$ is calibrated as \cite{Marino:1999af}
\begin{equation}
	\iota^*(J)=0\,,\qquad e^{-i\theta_{D_6}}\iota^*(\Omega)=\frac{\sqrt{\det(g
	+\mathcal{B})}}{\sqrt{\det(g)}}\vol_{L}\,.
\label{eq:DbraneIIACalibration}
\end{equation}
The constant angle $\theta_{D_6}$ is a priori free and determines the covariantly
constant spinor corresponding to the supercharge conserved by the BPS D-brane. More
precisely $\theta_{D_6}$ determines the linear combination of the two covariantly 
constant spinors of the Type II compactification on $Z_3$ \cite{Becker:1995kb} that 
is unbroken by the D-brane background and thus yields a supercharge in four 
dimensions generating $\mathcal{N}=1$ spacetime supersymmetry. All additional D-branes 
in the setup have to be calibrated with respect to the same calibration form 
$e^{-i\theta_{D_p}}\iota^*(\Omega)$ in order to preserve the same covariantly constant spinor
and thus preserve $\mathcal{N}=1$ spacetime supersymmetry. The identical argumentation
applies to orientifold planes, which are $O6$-planes in the Type IIA case, that have 
to be included for tadpole cancellation discussed at the end of this section. 
Then, the free angle $\theta_{D_6}$ has to equal the angle $\theta$ in the orientifold calibration 
\eqref{eq:calibrationOrie}.  

In Type IIB a spacetime-filling BPS D-brane is supported along holomorphic submanifolds in $Z_3$ \cite{Becker:1995kb}. 
This in particular yields a natural choice of complex structure on $\Sigma$ by aligning 
it with the ambient complex structure using the embedding $\iota$ which then defines 
a holomorphic map obeying ${\partial \bar{z}^{\bar \imath}(u^a)}/{\partial u}= 
{\partial z^{ i}(u^a)}/{\partial \bar u}=0$, where we introduced complex coordinates 
$\underline{u}$ on the worldvolume $\mathcal{W}_{p+1}$. In other words, D-branes in 
Type IIB test the even dimensional homology of $Z_3$, that are D3-branes on points , 
D5-branes on holomorphic curves $\Sigma$, D7-branes on holomorphic divisors $D$ or 
D9-branes on the entire Calabi-Yau threefold $Z_3$. A holomorphic submanifold is 
volume minimizing since its volume form is just proportional to (powers of) the 
pullback of $J$, a well-known mathematical fact for complex submanifolds of K\"ahler 
manifolds \cite{Griffiths:1978yf}. In the presence of background fields this is 
replaced by the more general calibration conditions \cite{Marino:1999af}
\begin{equation}
	\iota^*(\Omega)=0\,,\qquad e^{-i\theta_{D_p}}\iota^*e^{J+i\mathcal{B}}\vert_{D_p}=
	\frac{\sqrt{\det(g+\mathcal{B})}}{\sqrt{\det(g)}}\vol_{D_p}\,,\qquad p=3,5,7,9.
\label{eq:DbraneIIBCalibration}
\end{equation}
where $\vol_{D_p}$ denotes the volume form on the analytic cycle wrapped by the 
D$p$-brane. Again we note that in the presence of $Op$-orientifold planes, a 
necessary ingredient to cancel tadpoles, the angle $\theta_{D_p}$ has to equal the 
angle in the orientifold calibration \eqref{eq:calibrationOrie}, which is $\theta_{D_p}=0$.
Then the calibrations \eqref{eq:DbraneIIBCalibration} are even more restrictive.
Explicitly, we evaluate the real and imaginary part of the calibration conditions 
to obtain 
\begin{eqnarray} \label{eq:DbraneIIBCalibrationExpanded}
        D5&:&du^2\sqrt{-\det\left(\iota^{\ast}\left(g_{10}+\mathcal{B}\right)
        \right)}=\iota^{\ast} J\,,\qquad\qquad\qquad\qquad\,\,\,\, \,\,\,\,\,\,\,\,\,\,  \iota^*\mathcal{B} =0\ ,\\
        D7&:&du^4\sqrt{-\text{det}\left(\iota^{\ast}\left(g_{10}+\mathcal{B}\right)
        \right)}=\tfrac12\iota^{\ast} (J^2-\mathcal{B}^2)\,,\qquad \qquad\,\,\,\,\,\,  \iota^*(\mathcal{B}\wedge J) =0\ ,\nn\\
        D9&:&du^6\sqrt{-\text{det}\left(\iota^{\ast}\left(g_{10}+\mathcal{B}\right)
        \right)}=\tfrac16\iota^{\ast} (J^3-J\wedge\mathcal{B}^2)\,,\qquad   \iota^*(\mathcal{B}\wedge J^2-\mathcal{B}^3) =0\ .\nn
\end{eqnarray}
Once again, these formulas are given in the string frame and can be
translated to the Einstein frame by multiplying the first column of equation in 
\eqref{eq:DbraneIIBCalibrationExpanded} by $e^{\phi}$. 

Clearly, the calibration conditions \eqref{eq:DbraneIIACalibration}, 
\eqref{eq:DbraneIIBCalibration} only hold in the supersymmetric background 
configuration or vacuum. Consequently, in order to understand the \textit{dynamics} of D-branes we 
have to allow for small variations of the fields about this vacuum that explicitly 
violate the conditions \eqref{eq:DbraneIIACalibration}, \eqref{eq:DbraneIIBCalibration}. 
This, however, is part of the derivation of a low-energy effective action as performed 
in section \ref{sec:D5branes} for D5-branes.

We conclude this section by a brief discussion of tadpoles. D$p$-branes couple
electrically via the Chern-Simons-action \eqref{eq:CSgeneral} to $C_{p+1}$. This 
yields localized source terms in the equation of motion of $C_{p+1}$ of the form 
\begin{equation}
	dF_{8-p}=\sum_i\mu_p\delta_i^{(9-p)}+\text{induced charges}
\label{eq:C_pEOM}
\end{equation} 
for each D$p$-brane on the support of $\delta_i^{(9-p)}$, that is a
$(9-p)$-form\footnote{The analogue of delta-function 
as a $k$-form is properly called a delta-current, as introduced in chapter \ref{ch:blowup}.} 
on the normal space to the cycle wrapped by the D-brane. Here we also included possible charges 
induced from higher-dimensional D$q$-branes, $q>p$, 
due to worldvolume fluxes \cite{Douglas:1995bn}. Since the left-hand-side of \eqref{eq:C_pEOM} is 
exact, the sum on the right has to vanish in cohomology, which goes by the name 
of tadpole cancellation. This implies that we have to include negatively charged 
orientifolds planes in the same homology class as the D-branes to meet the 
requirement of tadpole cancellation in a supersymmetric fashion.

\section{Dynamics of D5-Branes in Calabi-Yau Orientifolds}
\label{sec:D5branes}

In this section we derive the effective action of a spacetime-filling D5-brane wrapped on a curve 
$\Sigma$. We turn on bulk and brane fluctuations around the vacuum defined in section 
\ref{sec:DbranesinCY3Orie} yielding fields in the four-dimensional action. We discuss this 
spectrum of fields in detail in section \ref{sec:specBrane}, where we put special emphasis on the 
universal sector of D5-brane fluctuations associated to geometric deformations of the curve $\Sigma$. 
As a first step to understand the structure of the combined geometrical open-closed deformation modes 
we derive some special identities between light open-closed fields in section \ref{sec:relations}. 
Then, in section \ref{sec:reductionD5}, we perform the actual reduction of the D5-brane action, where 
we restrict to bosonic fields only and infer the fermionic action from supersymmetry. We conclude the 
dimensional reduction in \ref{sec:scalarpotderivation} by a detailed derivation of the scalar potential, 
for which the couplings of non-dynamical four-dimensional three-forms are essential.

\subsection{The Four-dimensional Spectrum}
\label{sec:specBrane}

Here we discuss the four-dimensional spectrum emerging 
from compactification of Type IIB theory with D5-branes. We 
start our discussion by fixing the background geometry of our setup.
As before in section \ref{sec:EffActOrie} we consider the direct 
product of a compact Calabi-Yau orientifold $Z_3/\sigma$ and 
flat Minkowski space $\mathbb{R}^{1,3}$ for which the metric in the
string frame reads 
\begin{equation} \label{eq:metric}
        ds^2_{10}=\eta^{\text{SF}}_{\mu\nu}dx^\mu dx^\nu
        +2g_{i\bar \jmath}dz^id\bar{z}^{\bar \jmath}\,.
\end{equation}
Since we are interested in compactifications which allow the inclusion of
space-time filling D5-branes we have to add $O5$-planes to preserve $\cN=1$ 
supersymmetry in four dimensions. This fixes the orientifold projection to 
the $O5/O9$ case in \eqref{eq:typeIIorie} with the central properties
\begin{equation} \label{eq:O5actioReview}
        \mathcal{O} = \Omega_p \sigma^*\ ,\qquad \quad 
        \sigma^* J = J \ ,\qquad \qquad \sigma^* \Omega = \Omega\ , 
\end{equation}
for an isometric, holomorphic involution $\sigma$. The spectrum 
consists of two classes of fields. Firstly, there are zero modes
arising from the expansion of the ten-dimensional closed string fields 
into harmonics on $Z_3$. Secondly, one finds zero modes arising from the D5-brane
effective actions \eqref{eq:DBIgeneral} and \eqref{eq:CSgeneral}. 
In the following we discuss both sets of fields in turn, where we 
only briefly summarize the closed string content and refer to section 
\ref{sec:specOrie} for a more detailed discussion.  

\subsubsection{Closed String Spectrum}\label{sec:closedStringSpectrum}

We start by recalling the following transformation behavior of the bulk
fields under the orientifold from table \ref{tab:invstates} of section 
\ref{sec:specOrie}
\begin{equation} \label{eq:O5transfRev}
  \sigma^*g_{10}=g_{10}\ , \qquad \sigma^* B_2 = -B_2\ , \qquad \sigma^* \phi
  = \phi\ , \qquad \sigma^* C_p = (-1)^{(p+2)/2} C_p\ .
\end{equation}
This yields the following reduction ansatz for the massless four-dimensional
fields in terms of the cohomology groups of table \ref{tab:cohombasisOrie},
\begin{eqnarray} \label{eq:expJBPhi}
        &\delta g_{i\bar\jmath}=-i(\omega_{\alpha})_{i\bar\jmath}\delta v^{\alpha}\ 
        \,\,\,(\alpha=1,\ldots,h^{(1,1)}_+)\,,\qquad \delta g_{ij} = 
        \frac{i\mathcal{V}}{\int \Omega \wedge \bar{\Omega}}\ 
        \Omega_{j}^{\ \bar \imath \bar \jmath} \, (\bar
   			\chi_{\bar \kappa})^{\phantom{i}}_{\bar \imath \bar \jmath i}\,\delta
   			\bar z^{\bar \kappa} \ \,\,\, (\kappa=1,\ldots,h^{(2,1)}_+)\, ,&\nn\!\!\!\!\\
   			&J=v^\alpha(\underline{x})\omega_\alpha\ ,\qquad 
   			B_2=b^a(\underline{x})\omega_a\ , \qquad
        \phi=\phi(\underline{x})\ ,&\!\!\!\!
\end{eqnarray}
in the NS--NS-sector and accordingly in the R--R-sector
\begin{eqnarray}
        \label{eq:RRO59expRev}
         C_6 &=& A_{(3)}^K\wedge\alpha_K+\tilde{A}^{(3)}_K\wedge\beta^K +
        \tilde{c}^{(2)}_\alpha \wedge \tilde\omega^\alpha+h\, m_6\ ,  \\
        C_4 &=& V^{k}\wedge\alpha_{k}+U_{k}\wedge\beta^{k}+\tilde{\rho}^a_{(2)} \wedge
        \omega_a+\rho_a\tilde\omega^a\ , \nn\\
        \nn C_2 &=& \mathcal{C}_{(2)}+c^\alpha\omega_\alpha\ .
\end{eqnarray}
Here the fields $(\underline{A}_{(3)}, \underline{\tilde{A}}^{(3)})$ are 
three-forms, 
$( \underline{\tilde{c}}^{(2)}, \underline{\tilde{\rho}}_{(2)},\mathcal{C}_{(2)})$ 
are two-forms, $(\underline{V}, \underline{U})$ are vectors and 
$(h,\underline{\rho},\underline{c},\underline{v},\underline{z},\underline{b},\phi)$ 
are scalars in four-dimensional Minkowski space $\mathbb R^{1,3}$.

\subsubsection{Open String Spectrum}
\label{openspectrum}

Let us now include space-time filling D5-branes into our setup. In general,
they can be arranged in a complicated way as long as the consistency constraints 
for the compactification are met. We consider a stack of $N$ D5-branes on 
a curve $ \Sigma$ in $Z_3$. If $ \Sigma$ is in the fix-point set of 
the involution $\sigma$, the D5-branes lie on top of an orientifold 
five-plane and $ \Sigma$ is its own image with respect to $\sigma$. More generally $\Sigma$ 
can be mapped to a curve $ \Sigma'=\sigma( \Sigma)$ which is not pointwise 
identical to $\Sigma$.

In this work we will mostly focus on the simplest situation, for which $N=1$, 
$\Sigma\cap  \Sigma' =\emptyset$ and $ \Sigma,  \Sigma'$ are in different homology 
classes. Hence, we consider one D5-brane on $ \Sigma$ and its image brane on 
$\Sigma'$. For this situation the pair of the D5-brane and its image D5-brane is 
merely an auxiliary description of a single smooth D5-brane wrapping a cycle in 
the orientifold $Z_3/\mathcal{O}$. On $Z_3$ it is natural to define the curves
\begin{equation}
    \Sigma_+ =  \Sigma+ \Sigma'\ , \qquad  \Sigma_- =  \Sigma -  \Sigma'\ ,
        \label{eq:sigmaPM}
\end{equation}
where $ \Sigma_+$ is the union of $ \Sigma$ and $ \Sigma'$ while $ \Sigma_-$ 
contains the orientation reversed cycle $\Sigma'$. Clearly, one finds that 
$\sigma(\Sigma_\pm)=\pm \Sigma_\pm$.

We first discuss the degrees of freedom from the gauge theory on the D5-brane 
due to $U(1)$ Wilson lines $a_I(x)$. They arise from non-trivial one-cycles on the 
six-dimensional D5-brane world-volume. These enter the expansion of the $U(1)$ 
gauge boson $A(\xi)$ on the D5-brane as 
\begin{equation} \label{eq:gauge-expansion}
        A(\underline{x},\underline{u})=A_{\mu}(\underline{x})dx^{\mu}P_{-}(\underline{u})
        +a_I(\underline{x})A^I(\underline{u})+\bar{a}_{\bar I}(\underline{x})\bar{A}^{\bar I}(\underline{u})\ ,
\end{equation}
where $A_\mu(\underline{x})$ denotes the four-dimensional gauge-field.
Here we introduce real coordinates $\xi$ on the world-volume of the D5-brane where 
we distinguish coordinates $\xi=(\underline{x},\underline{u})$ for the Minkowski 
space and the two-cycle $\Sigma_+$, respectively. We denote complex coordinates 
for $\Sigma_+$ by $u$, $\bar{u}$ in the complex structure induced by the ambient 
space $Z_3$. The one-forms 
$A^I=A_{\bar u}^I d\bar{u},\,\bar{A}^{\bar I}=\bar{A}_u^{\bar I}du$ denote a 
basis of the Dolbeault cohomology $H_{-}^{0,1} (\Sigma_+ )$ and 
$H_{-}^{1,0} (\Sigma_+ )$, respectively, and $P_{-}$ is the step function equaling 
$1$ on $\Sigma$ and $-1$ on $\Sigma'$. Note that generally the $U(1)$ field 
strength $F=dA$ can admit a background flux $\langle F\rangle=f$. Since $F$ is 
negative under $\sigma$, it enjoys the expansion 
\begin{equation} \label{eq:D5flux}
  f=f^a\iota^{\ast}\omega_a=f^a(\iota^{\ast}\omega_a)_{u\bar u}du\wedge d\bar{u}\ ,
\end{equation}
where $\iota^{\ast}\omega_a$ denotes the pullback of the basis elements $\omega_a$ of
$H_-^{(1,1)}(Z_3)$ introduced in Table \ref{tab:cohombasisOrie}. As we already 
recalled in section \ref{sec:DbranesinCY3Orie}, $F$ naturally combines with the 
NS--NS B-field into the combination $\mathcal{B}=\iota^{\ast}(B_2)- \ell F$ with 
$\ell =2\pi \alpha'$. Thus it is convenient to consider variations of 
$\mathcal{B}$ around the background value
\eqref{eq:DbraneIIBCalibrationExpanded} in the calibrated vacuum. We denote
them by
\begin{equation}
	\mathcal{B}^a(\underline{x})=b^a(\underline{x})-\ell f^a\,, \qquad \mathcal{B}^{\Sigma}=\mathcal{B}^a\int_{\Sigma_-}\iota^* \omega_a\,,
\label{eq:calBperturbation}
\end{equation}
where we singled out the combination $\mathcal{B}^\Sigma$ of that four-dimensional field 
which corresponds to the class of the cycle $\Sigma_-$.

There is a universal sector in the worldvolume theory governing the dynamics of 
the D5-brane. It contains the geometric fluctuations of the embedding map 
$\iota:\Sigma_+ \hookrightarrow Z_3$. These fluctuations are described by sections
$\hat{\zeta}$ of the normal bundle to $\Sigma_+$. To first order massless fields 
$\zeta(\underline{x})$ correspond to holomorphic sections of the holomorphic normal 
bundle $H^{0}(\Sigma_+,N_{Z_3}\Sigma_+)$ of $\Sigma_+$ and their conjugates $\bar\zeta$. 
This gives rise to the expansion
\begin{equation} \label{eq:zetaExpansion}
        \hat\zeta= \zeta + \bar \zeta
        =\zeta^A(\underline{x}) \,s_A+\bar{\zeta}^{\bar A}(\underline{x})\, 
        \bar{s}_{\bar A}=\hat \zeta^{\cA} \hat s_{\cA}\ ,
\end{equation}
where we introduced a basis $\hat{s}_{\cA}$ of real sections $\hat\zeta$ on 
$H^{0}(\Sigma_+,N_{Z_3}\Sigma_+)\oplus\overline{H^{0}(\Sigma_+,N_{Z_3}\Sigma_+)}$ 
for convenience. We note that the group of holomorphic sections 
$H^{0}(\Sigma_+,N_{Z_3}\Sigma_+)$ is finite due to the strong requirement of 
holomorphicity. This is in contrast to the fact that a real vector bundle $E$ has in 
general infinitely many real sections in $\mathcal{C}^\infty(E)$. However, a real 
section of $N_{\mathbb{R}}\Sigma_+$ in general corresponds to a massive mode and is 
therefore neglected in our description of the effective action. However, as we will 
see in chapter \ref{ch:blowup}, the question for the appropriate truncation of 
$\mathcal{C}^\infty(N_{\mathbb{R}}\Sigma_+)$ is mathematically sophisticated. On 
the one hand side there is in general a brane superpotential for the modes $\zeta$ 
that may obstruct them at a higher order. On the other hand, when varying the 
complex structure on $Z_3$ a non-holomorphic section in 
$\mathcal{C}^\infty(N_{\mathbb{R}}\Sigma_+)$ may become massless at some point in the 
complex structure moduli space and contribute to the effective action. Due to this 
complications we restrict our considerations in this chapter, with the exception of
section \ref{sec:extensionInfinite}, to the finite group 
$H^{0}(\Sigma_+,N_{Z_3}\Sigma_+)$ to count first order massless fields and postpone 
the more difficult question of a global open-closed moduli space. The discussion of the 
complete open-closed moduli space and the brane superpotential will be of central 
importance in this work though, in particular in chapter \ref{ch:blowup}.

We conclude by summarizing the $\mathcal N=1$ field content in four dimensions 
emerging from the bulk and the brane sector in Table \ref{tab:fieldcontent}. The
precise organization of these fields into $\mathcal N=1$ complex coordinates is 
performed in section \ref{sec:N=1dataD5}. 

\begin{table}
\begin{center}
 \begin{tabular}{|c||c|c|c|c|c|}
 \hline
 \rule[-0.2cm]{0cm}{0.6cm}  multiplet & \multicolumn{2}{c|}{{closed}} & & 
 \multicolumn{2}{c|}{{open}}\\ \cline{1-3} \cline{5-6}
 \multirow{4}*{chiral } &\T\B $h^{(1,1)}_+$ & $
     \rule[-0.2cm]{0cm}{0.6cm} \quad  t^\alpha=(v^\alpha,c^\alpha)$ \quad \ &&
        \multirow{2}*{\quad $h^0_+(\Sigma_+,N\Sigma_+)$\quad} & \multirow{2}*{$\zeta^A$}\\ 
        \cline{2-3}
        & \T\B $h^{(1,1)}_- $ & 
         \rule[-0.2cm]{0cm}{0.6cm}  $P_a=(b^a,\rho_a)$ & & &\\ \cline{2-3} \cline{5-6}
        & \T\B $1$ &
         \rule[-0.2cm]{0cm}{0.6cm}  $S=(\phi,h)$ 
        &&\multirow{2}*{$h^{(1,0)}(\Sigma_+)$}&\multirow{2}*{$a_I$}\\
        \cline{2-3} 
        &\T\B  $h^{(2,1)}_+$ & $z^\kappa$ & & &   \\ \cline{2-3}
\cline{1-3} \cline{5-6}
vector & \T\B $h^{(2,1)}_-$& $V^{\tilde K}$ && $1$ & $A$\\
\hline

 \end{tabular}
\caption{\label{tab:fieldcontent}
The $\mathcal{N}=1$ spectrum of the D5-brane effective theory.}
\end{center}
\end{table}

\subsection{Special Relations on the $\mathcal{N}=1$ Moduli Space} 
\label{sec:relations}

In this section we discuss a subtlety in the decomposition 
\eqref{eq:zetaExpansion}. The notion of  $\zeta^A$ being a complex
scalar field depends on the background complex structure chosen on the
ambient Calabi-Yau $Z_3$, i.e.\ on the split \eqref{eq:zetaExpansion}, 
${N^\mathbb{R}}_{Z_3}\Sigma_+\otimes\mathbb{C}
=N_{Z_3} \Sigma_+\oplus \overline{N_{Z_3} \Sigma_+}$, 
into holomorphic and anti-holomorphic parts. To explore this dependence 
further it is natural to consider the contractions of the $s_A$ with 
the holomorphic $(3,0)$-form $\Omega$, the $(2,1)$-forms $\chi_\kappa$ 
introduced in \eqref{eq:expJBPhi} and their complex conjugates. In the 
background complex structure defined at $z_0$ we find, in the cohomology 
of $Z_3$ as well as in the cohomology of $\Sigma$, that
\begin{equation}
  s_A \lrcorner\Omega(z_0) = 0 \ , 
   \qquad \quad   s_A
  \lrcorner\bar \chi_\kappa(z_0) =0 \ ,
  \qquad \quad s_A \lrcorner \bar \Omega(z_0) = 0 \ .
\label{eq:backgroundcs}
\end{equation}
These contractions vanish on $Z_3$ since there are no non-trivial 
$(2,0)$-forms in $H^2(Z_3)$. Moreover, they also vanish on $\Sigma$ for a 
supersymmetrically embedded D5-brane. In addition, since every two-form 
pulled back to $\Sigma$ has to be proportional to the $(1,1)$-K\"ahler form $J$, 
only $s_A \lrcorner\chi_\kappa$ can be a non-trivial $(1,1)$-form on 
$\Sigma$. Note, however, that also $s_A \lrcorner\chi_\kappa$  is trivial in the 
cohomology of $Z_3$ due to the primitivity of $H^{(2,1)}(Z_3)$. 

However, in the four-dimensional effective theory we also have to allow 
for possible fluctuations around the supersymmetric background 
configuration, including those corresponding to complex structure 
deformations of $Z_3$. The holomorphic three-form $\Omega$ as well as 
the complex scalars $\zeta$ are then functions of the complex structure 
parameters $z^\kappa$. In a different complex structure on $Z_3$, the 
notion of holomorphic and anti-holomorphic coordinates on $Z_3$, 
encoded by the type of $\Omega(z)$, has not to be aligned with 
the splitting into complex scalars \eqref{eq:zetaExpansion} in general.
To measure this discrepancy, we consider the pullback $\iota^\ast(s_A\lrcorner\Omega(z))$  
on $\Sigma$. For $z=z_0+\delta z$ near a background complex structure 
$z_0$ we expand $\Omega(z)$ to linear order in $\delta z$ as
\begin{equation} 
        \iota^\ast(s_A\lrcorner\Omega(z))= (1-K_{\kappa}\delta z^ {\kappa})
         \iota^\ast(s_A\lrcorner\Omega(z_0))+
         \iota^\ast(s_A\lrcorner\chi_\kappa(z_0))\delta z^\kappa
          =\iota^\ast(s_A\lrcorner\chi_\kappa(z_0))\delta z^\kappa,
        \label{eq:expansion-omega-a}
\end{equation}
where we used \eqref{eq:variationofomega} and \eqref{eq:backgroundcs}.
In other words, the form $s_A\lrcorner\Omega$ is a $(2,0)$-form on $\Sigma$ in 
the complex structure $z$ but a $(1,1)$-form on $\Sigma$ in the complex 
structure $z_0$ to linear order in the complex structure variation $\delta z$. 
However, a similar argument shows that
\begin{equation}
(s_A\lrcorner\bar\Omega)(z)=(s_A\lrcorner\bar\chi) (z^\kappa)= 0\ ,
\label{eq:zeta-omega-bar-zero}
\end{equation}
even to linear order in $\delta z^\kappa$. These forms only appear at higher
order in the complex structure variations.

The above considerations allow us to describe the metric deformations of the
induced metric $\iota^\ast g$ on the two-cycle $\Sigma_+$. In general, both the
complex structure deformations of $Z_3$ and the fluctuations of the embedding
map $\iota$ contribute. Here, we discuss only those variations
$\delta(\iota^\ast g)$ originating from complex structure deformations and
postpone the analysis of all possible metric variations to section
\ref{sec:DBIreduction}. Analogously to \eqref{eq:expJBPhi} the complex structure
deformations on $\Sigma_+$ are encoded in the purely holomorphic metric variation
\begin{equation} \label{eq:internalmetricvariation}
 \iota^\ast(\delta g)_{uu}= \frac{2i v^{\Sigma}}{\int\Omega\wedge\bar{\Omega}}
 \iota^\ast( s_A\lrcorner\Omega)_{uu}(\iota^\ast g)^{u\bar u}
  \iota^\ast(\bar{s}_{\bar B}\lrcorner \bar{\chi}_{\bar{\kappa}})_{\bar{u}u}
  \mathcal{G}^{A\bar B}\ \delta \bar{z}^{\bar \kappa}\ .
\end{equation}
Here we have introduced the volume of the holomorphic two-cycle $\Sigma_+$ as
\begin{equation}
 v^{\Sigma}=\int_{\Sigma_+}d^2u\sqrt{g}=\int_{\Sigma_ +}\iota^\ast J
\end{equation}
and a natural hermitian metric $\mathcal{G}_{A\bar B}$ given by
\begin{equation} \label{eq:openmetric}
 \mathcal{G}_{A\bar B}=-\frac{i}{\mathcal V}\int_{\Sigma_+}s_A\lrcorner \bar{s}_{\bar B}\lrcorner (J)\iota^\ast J.
\end{equation}
We will show later on that $\mathcal{G}_{A\bar B}$ can be obtained by 
dimensional reduction, cf.~section \ref{sec:DBIreduction}. It is identified 
with the metric for the moduli $\underline{\zeta}$ on the open string moduli space and 
is independent of the coordinates $u,\bar u$ on $\Sigma$.

The metric variation \eqref{eq:internalmetricvariation} can be explained by 
application of some useful formulas for the open string moduli space. 
First, we use the fact that $H^{(1,1)}(\Sigma_+)$ is spanned by the pullback 
$\iota^\ast J$. This is exploited to rewrite the pullback of any closed 
$(1,1)$-form $\omega$ to $\Sigma_+$ in cohomology, cf.~\eqref{eq:pullbackformula}. 
In particular we obtain
\begin{equation}
        \iota^\ast(s_A\lrcorner \chi_{\kappa})
        =\frac{\iota^\ast J}{v^{\Sigma}}\int_{\Sigma_+}\iota^\ast(s_A\lrcorner \chi_{\kappa})\ ,
\end{equation}
which can be written after multiplication with 
$\mathcal V^{-1}\mathcal G^{A\bar B}g(s_C,\bar s_{\bar B})$ and by using 
\eqref{eq:openmetric} as 
\begin{equation} \label{eq:metricidentity}
 \int_{\Sigma_+} \iota^\ast(s_A \lrcorner \chi_{\kappa})
 =-\frac{v^{\Sigma}}{\mathcal V}\int_{\Sigma_+}g(s_A,\bar{s}_{\bar B})\mathcal{G}^{\bar{B}C} \iota^\ast(s_C\lrcorner\chi_{\kappa}).
\end{equation}
We evaluate this for every choice of $s_A$ and compare the coefficients on 
both sides to relate the metric on the normal bundle $N_{Z_3}\Sigma$ and the 
metric $\mathcal G^{A\bar B}$.  

Thus, the identity \eqref{eq:metricidentity} allows us to infer the metric variations 
\eqref{eq:internalmetricvariation} from the complex structure deformations 
on $Z_3$. First, we consider the pullback to $\Sigma_+$ of the metric variations 
$\delta g_{ij}$ in \eqref{eq:expJBPhi} of the ambient Calabi-Yau $Z_3$
\begin{equation} \label{eq:pullbackDeltag}
        \iota^\ast(\delta g)_{u u}=\frac{i\mathcal{V}}{\int \Omega \wedge \bar{\Omega}}\
   \Omega_{u}^{\ \bar \imath \bar \jmath} \, (\bar
   \chi_{\bar \kappa})^{\phantom{i}}_{\bar \imath \bar \jmath u}\,  \delta
   \bar z^{\bar \kappa}.
\end{equation}
Then we replace, motivated by \eqref{eq:metricidentity}, the inverse metric $g^{i\bar \jmath}$ 
occurring in the contraction of $\bar{\chi}_{\bar \kappa}$ and $\Omega$ by 
$s^i_As^{\bar \jmath}_{\bar B}\mathcal G^{A\bar B}$ to obtain our ansatz for the 
induced metric deformation on $\Sigma_+$ given in \eqref{eq:internalmetricvariation}.

However, there are some remarks in order. Since there are no $(2,0)$-forms 
on $\Sigma_+$ in the background complex structure $z_0$, the form 
$\iota^\ast(s_A\lrcorner\Omega)$ should vanish identically. Thus, in 
order to make sense of the metric variation \eqref{eq:internalmetricvariation} 
we have to evaluate it, following the logic of \eqref{eq:expansion-omega-a}, 
in the complex structure $z=z_0+\delta z$. 
Applying this to \eqref{eq:internalmetricvariation} we expand $\delta(\iota^\ast g)$ to linear
order in $\delta z$, i.e.~$\iota^\ast(\delta g)_{uu}(z)=
\iota^\ast(\delta g)_{uu}(z_0)+\iota^\ast(\delta g)_{u\bar{u}}(z_0)\cdot\delta z$, to obtain
\begin{equation} \label{eq:internalmetricvariation2}
 \iota^\ast(\delta g)_{u\bar{u}}(z_0)=
 \frac{2i v^{\Sigma}}{\int\Omega\wedge\bar{\Omega}} 
 \iota^\ast( s_A\lrcorner\chi_{\kappa})_{\bar{u}u}(\iota^\ast g)^{u\bar u}
  \iota^\ast(\bar{s}_{\bar B}\lrcorner \bar{\chi}_{\bar{\kappa}})_{\bar{u}u}
  \mathcal{G}^{A\bar B}\ \delta z^{\kappa}\delta \bar{z}^{\bar \kappa}\ .
\end{equation}
Here we emphasize the change in type from purely holomorphic indices $\delta
g_{uu}$ at $z$ to mixed type $\delta g_{u\bar u}$ at $z_0$. It is important to
note that there are neither any metric deformations linear in the complex structure
parameter $\delta z$ nor any of pure type.

These calculations stress that the analysis of the open string moduli space
crucially depends on the chosen background complex structure encoded by the moduli 
$z^\kappa$. It is hence natural that the complex structure parameters $z^\kappa$ 
of $Z_3$ and the open string moduli $\zeta^A$ should be treated on an equal 
footing to characterize the structure of the $\cN=1$ field space. This has let to
the the introduction of the blow-up proposal in \cite{Grimm:2008dq} that was further
exploited in \cite{Grimm:2009sy} and ultimately in \cite{Grimm:2010gk}. 
It will be used in chapter \ref{ch:blowup} to unify the open-closed fields and
to understand and derive the effective brane superpotential. In the next sections we derive 
the general four-dimensional effective D5-brane action and show that the brane superpotential 
is naturally encoded by the forms $s_A\lrcorner\Omega$ and $s_A\lrcorner\chi_\kappa$ 
that are sensible both to the complex structure on $Z_3$ and the open string 
deformations.

We conclude by noting that the analysis of $\delta(\iota^* g)$ and the 
induced variation \eqref{eq:internalmetricvariation2} can also be derived
in two alternative ways, one based on the consideration of all vector fields
in $\mathcal{C}^{\infty}(\Sigma,N_{Z_3}\Sigma)$ and a different by 
analyzing the variation of $\vol(\Sigma)$ under a change of complex structure. 
For the first way we introduce a basis $s_a$ of the infinite dimensional
Hilbert space of all sections in $\mathcal{C}^{\infty}(\Sigma,N_{Z_3}\Sigma)$. The 
natural metric $\mathcal{G}_{a\bar b}$ on these sections is formally identical 
to the integral in \eqref{eq:openmetric}. Then we introduce holomorphic vielbeins
$e_n$ of the normal bundle $N_{Z_3}\Sigma$ obeying the elementary relation
\begin{equation}
	g_{i\bar \jmath}e^{i}_n\bar{e}^{\bar\jmath}_{\bar m}=\eta_{n\bar m}\,,\qquad
	g^{i\bar \jmath}=\eta^{n\bar m}e^{i}_n\bar{e}^{\bar\jmath}_{\bar m}\,,
\label{eq:vielbeinDef}
\end{equation} 
where $g_{i\bar\jmath}$ denotes the hermitian metric on $N_{Z_3}\Sigma$ and $\eta_{n\bar m}$ the
flat hermitian metric. We readily expand the holomorphic vielbeins with respect to the basis of sections $s_a$
as $e_n^i=\sum c_n^as_a^i$. Upon integrating the first relation
in \eqref{eq:vielbeinDef} over $\Sigma_+$, we obtain
\begin{equation}
		\, \eta_{n\bar m}=\frac{1}{v_\Sigma}\int_{\Sigma_+}\eta_{n\bar m}\iota^*(J)
		=\frac1{v_\Sigma}\int_{\Sigma_+}
		g_{i\bar \jmath}\,e^{i}_n\bar{e}^{\bar\jmath}_{\bar m}\,\iota^*(J)
		=i\frac{\mathcal{V}}{v_\Sigma}\,\mathcal{G}_{a\bar b}\,c_n^a\bar{c}_{\bar m}^{\bar b}\,,
\label{eq:relEtaG}
\end{equation}
which is readily inserted into the second equation in \eqref{eq:vielbeinDef} to obtain
\begin{equation}
	g^{i\bar\jmath}=\frac{v_{\Sigma}}{\mathcal{V}}\mathcal{G}^{a\bar b}s^n_a s^{\bar m}_{\bar b}\,.
\label{eq:relgG}
\end{equation}
This is then again used in \eqref{eq:pullbackDeltag} to replace, to quadratic order in $s_a$, the 
metric $g^{i\bar \jmath}$. By expanding in the complex structures to second order, we 
reproduce \eqref{eq:internalmetricvariation2} as before, however, with the basis of sections $s_a$
of $\mathcal{C}^{\infty}(\Sigma,N_{Z_3}\Sigma)$ instead of $s_A$. Nonetheless,
the relation \eqref{eq:internalmetricvariation2} is still a good approximation for
the purpose of deriving the effective action since most of the modes associated to
the $s_a$ are massive, cf.~section \ref{sec:extensionInfinite}, and are consistently 
integrated out in the effective Lagrangian maintaining only the light fields, that are 
precisely counted by the finite number of sections $s_A$ in $H^0(\Sigma,N_{Z_3}\Sigma)$.  

For brevity, we describe the second method only in words. Since a variation $\delta(\iota^* g)$
violates the calibrations \eqref{eq:DbraneIIBCalibrationExpanded} and thus 
the volume minimization property of $\Sigma$, the volume of $\Sigma$ will increase by
$\delta\vol(\Sigma)$. Analyzing this variation to appropriate order in the fields 
it is possible to extract \eqref{eq:internalmetricvariation2} or to directly obtain the 
corresponding potential term in the dimensional reduction of Dirac-Born-Infeld action 
\eqref{eq:DBIpotential}. Indeed we follow a similar logic in section \ref{sec:extensionInfinite} 
to determine the potential on the infinite dimensional field space of massive
deformations of $\Sigma$.

\subsection{Reduction of the D5-Brane Action}
\label{sec:reductionD5}

Now we are prepared to derive the four-dimensional effective action of the
D5-brane in a Calabi-Yau orientifold with $O5/O9$-planes, where we mainly follow
\cite{Grimm:2008dq}. The action is obtained by reducing the 
bulk supergravity action $S_{\text{IIB}}$ as in section \ref{sec:EffActOrie}
and the effective D-brane actions using a Kaluza-Klein reduction ansatz. 
The string-frame Type IIB action is used in its democratic form 
\eqref{eq:bulkaction}. In addition, one includes the string-frame D5-brane 
action 
\begin{equation} \label{eq:D5DBI+CS}
        S^{\text{SF}}_{\text{D5}}=-\mu_5\int_{\mathcal{W}}d^6\xi
        e^{-\phi}\sqrt{-\text{det}\left(\iota^{\ast}\left(g_{10}+B_2\right)-\ell F\right)}
        +\mu_5\int_{\mathcal{W}}\sum_{q\ \text{even}} \iota^\ast(C_q) \wedge e^{\ell F-\iota^\ast(B_2)} \  ,
\end{equation}
which is the specialization of the general DBI-action \eqref{eq:DBIgeneral} 
and the general CS-action \eqref{eq:CSgeneral} to BPS D5-branes. For convenience we put $\mathcal{W}\equiv \mathcal{W}_6$.
The Kaluza-Klein reduction of the bulk action \eqref{eq:bulkaction} on the 
orientifold background has been carried out in \cite{Grimm:2004uq} and
and is reviewed in chapter \ref{ch:EffActCYOrieCompact}. 
Thus we mainly concentrate in the following on the reduction of the D5-brane action
\eqref{eq:D5DBI+CS} and later include the contributions entirely due
to bulk fields in the determination of the $\cN=1$ characteristic functions.

\subsubsection{Dirac-Born-Infeld Action \& Tadpole Cancellation \label{sec:DBIreduction}}

In the following we perform the Kaluza-Klein reduction of the
Dirac-Born-Infeld action given in \eqref{eq:D5DBI+CS}.
Firstly, we expand the determinant using  
\begin{equation} \label{eq:taylor}
        \sqrt{\text{det}\left(\mathfrak{A}+\mathfrak{B}\right)}=
         \sqrt{\text{det}\mathfrak{A}}\cdot\left[1+\tfrac{1}{2}\tr
        \mathfrak{A}^{-1}\mathfrak{B}+\tfrac{1}{8}\left(\left(\tr\mathfrak{A}^{-1}\mathfrak{B}\right)^2
         -2\tr\left(\mathfrak{A}^{-1}\mathfrak{B}\right)^2\right)+\ldots\right]\ .
\end{equation}
The matrix $\mathfrak{A}$ contains the fields in the calibrated background 
configuration of $Z_3/\sigma\times \mathbb{R}^{1,3}$ for which the conditions 
\eqref{eq:DbraneIIBCalibrationExpanded} are obeyed. Additionally, $\mathfrak{B}$ 
contains the fluctuation fields around this background and is small compared to $\mathfrak{A}$. These 
fluctuations are precisely the variations of the embedding $\iota$ of the 
two-cycle $\Sigma_+$ parametrized by the fields $\underline{\zeta}$ of 
\eqref{eq:zetaExpansion}, the Wilson lines $\underline{a}$ introduced in 
\eqref{eq:gauge-expansion} as well as the perturbations 
\eqref{eq:calBperturbation} about the calibrated NS--NS B-field 
defined in \eqref{eq:DbraneIIBCalibrationExpanded} and about the background complex 
structure. The fluctuations of all fields involve the variation $\delta\iota$ of 
the embedding $\iota$ since they have to be pulled back to worldvolume $\mathcal{W}$
of the D5-brane. These pullbacks are treated in the normal coordinate expansions 
\cite{AlvarezGaume:1981hn,Friedan:1980jm,Jockers:2004yj}, 
which is a standard technique in differential geometry 
to analyze families of diffeomorphisms and embeddings using the Riemann normal coordinates. 
In the case at hand the family of embeddings $\iota$ is parametrized by the sections 
$\underline{\zeta}$ of the normal bundle. Applying the normal coordinate expansion to the metric 
\eqref{eq:metric} and the NS--NS B-field \eqref{eq:expJBPhi} on the D5-brane 
world-volume as well as taking the metric variation $\delta (\iota^\ast g)_{u\bar u}$ 
of \eqref{eq:internalmetricvariation2} into account we obtain 
\begin{eqnarray}
        \iota^{\ast}g_{10}&=& \mathcal{V}^{-1} e^{2\phi}\eta_{\mu
          \nu}dx^{\mu}\cdot dx^{\nu} +(\iota^\ast g+\delta (\iota^\ast g))_{u\bar u} du\cdot d\bar{u}+
          g(\partial_{\mu}\zeta,\partial_{\nu}\bar{\zeta})dx^{\mu}\cdot dx^{\nu}\;,\nn\\
\iota^{\ast}B_2-\ell \cF &=&\cB^a\iota^{\ast}\omega_a-\ell F +
      \cB^a\, \iota^{\ast}\omega_a(\partial_{\mu}\zeta,\partial_{\nu}\zeta)dx^{\mu}\wedge dx^{\nu}\;,
\label{eq:pullbackvariation}
\end{eqnarray}
where $\ \cdot \ $ is the symmetric product and $\cV$, $g_{u\bar u}$ are the string 
frame volume and the induced hermitian metric on $\Sigma_+$. Note that the Minkowski 
metric $\eta$ is rescaled to the four-dimensional Einstein frame\footnote{Recall
that the four-dimensional metric in the Einstein frame $\eta$ is related to the 
string frame metric $\eta^{\rm SF}$ via $\eta = e^{-2\phi} \cV\, \eta^{\rm SF}$.}. 
In the expansions \eqref{eq:pullbackvariation} we readily identify the background 
fields and variations as
\begin{eqnarray}
        \mathfrak{A}&=&
                    \begin{pmatrix}
                                \mathcal{V}^{-1}e^{2\phi}\eta_{\mu\nu} & 0&0\\
                        0 & 0& g_{u\bar u}\\
                        0& g_{u\bar u}& 0
                     \end{pmatrix},\\
        \mathfrak{B}&=&\begin{pmatrix}
                (2g+\mathcal{B}^a\omega_a)(\partial_{\mu}\zeta,\partial_{\nu}\bar{\zeta})
                -\ell F_{\mu\nu}& -\ell\partial_{\mu} \bar{a}_{\bar J}\bar{A}_u^{\bar J} &
                -\ell\partial_{\mu}a_IA_{\bar u}^I\\
                -\ell\partial_{\nu} \bar{a}_{\bar J}\bar{A}_u^{\bar J}&0
                &(\delta g+\tfrac12\cB^a\omega_a)_{u\bar u}\\
                -\ell\partial_{\nu}a_IA_{\bar u}^I& (\delta g-\tfrac12\cB^a\omega_a)_{u\bar u}& 0
                  \end{pmatrix}\;,
\end{eqnarray}
where we omitted the pullback $\iota^\ast$ for notational convenience.
Only the terms
\begin{equation}
        \tfrac12\tr\mathfrak{A}^{-1}\mathfrak{B}-\tfrac14\tr\big((\mathfrak{A}^{-1}\mathfrak{B} )^2\big)
\end{equation}
of the Taylor expansion \eqref{eq:taylor} contribute to the effective action up to 
quadratic order in the fields. We insert the result into the first part of 
\eqref{eq:D5DBI+CS} and use the calibrations \eqref{eq:DbraneIIBCalibrationExpanded} 
to obtain the four-dimensional action 
\begin{equation} \label{eq:DBI}
         S^{\rm EF}_{\text{DBI}}
         =\text{-}\mu_5\int\Big[\tfrac{\ell^2e^{-\phi}}{4}  v^{\Sigma}F\wedge \ast
         F+\tfrac{\ell^2e^{\phi}}{\mathcal{V}}\mathcal{C}^{I\bar J}
         da_I\wedge \ast d\bar{a}_{\bar
         J}+\tfrac12e^{\phi}\mathcal{G}_{A\bar B}d\zeta^A\wedge\ast d\bar{\zeta}^{\bar
         B}+V_{\rm DBI}\ast 1\Big]
\end{equation}
in the four-dimensional Einstein frame.
The potential term in \eqref{eq:DBI} is of the form
\begin{equation}
  V_{\rm DBI} = \frac{e^{3\phi}}{2\mathcal{V}^2}\Big(v^{\Sigma}
  +\frac{2i\mathcal{G}^{A\bar B}}{\int\Omega\wedge\bar{\Omega}}\int_{\Sigma_+}s_A\lrcorner 
  \chi_{\kappa}\int_{\Sigma_+}\bar s_{\bar B}\lrcorner \bar \chi_{\bar\kappa}
  \delta z^{\kappa}\delta \bar{z}^{\bar \kappa}+\frac{(\mathcal{B}^{\Sigma})^2}{8v^{\Sigma}} \Big)\ .
  \label{eq:DBIpotential}
\end{equation}  
In the following we will successively discuss the separate terms appearing in the action 
$S_{\text{DBI}}$.

The first term in \eqref{eq:DBI} is the kinetic term for the $U(1)$ gauge boson
$A$ on the D5-brane. The gauge coupling is thus given by 
$1/g_{\rm D5}^2 = \tfrac 12\mu_5 \ell^2 e^{-\phi} v^{\Sigma}$, where $v^{\Sigma}$ 
is the volume of the two-cycle $\Sigma_ +$ evaluated via the calibration 
\eqref{eq:DbraneIIBCalibrationExpanded}. The second term is the kinetic term 
for the Wilson line moduli $a^I$. The corresponding metric takes the form 
\begin{equation}
   \mathcal{C}^{I\bar J}=\frac12\int_{\Sigma_ +}A^I\wedge\ast_2\bar{A}^{\bar J}
   =\frac i2\int_{\Sigma_ +}A^I\wedge\bar{A}^{\bar J}\,,
\end{equation}
where we have used $\ast_2 \bar{A}^{\bar J}=i\bar{A}^{\bar J}$ on the
basis of $(1,0)$-forms introduced in \eqref{eq:gauge-expansion}. The third term 
in \eqref{eq:DBI} contains the field space metric for the deformations $\zeta^A$ 
which is of the form
\begin{equation}  \label{eq:metrics}
        \displaystyle\mathcal{G}_{A\bar
 B}=-\frac{i}{2\mathcal{V}}\int_{\Sigma_+}s_A\lrcorner \bar s_{\bar
 B}\lrcorner\left(J\wedge
 J\right)=\frac{\mathcal{K}_\alpha}{2\mathcal{V}}\mathcal{L}_{A\bar
 B}^{\alpha}\ , \qquad \quad
        \displaystyle\mathcal{L}_{A\bar
 B}^\alpha=-i\int_{\Sigma_+}s_A\lrcorner\bar{s}_{\bar
 B}\lrcorner \tilde\omega^\alpha \ ,
\end{equation}
where we have used $\mathcal{K}_\alpha=\int \omega_\alpha\wedge J\wedge J$ and 
$J\wedge J=\mathcal{K}_\alpha\tilde{\omega}^\alpha$.

Finally, let us comment on the potential terms $V_{\rm DBI}$. In fact, the 
first of the three terms represents an NS--NS tadpole and takes the form of a
D-term. As mentioned in section \ref{sec:DbranesinCY3Orie} a consistent 
compactification with D-branes requires cancellation of R--R as well as  NS--NS 
tadpoles. Hence the two-cycle $\Sigma_+$ wrapped by the D5-brane has to lie 
in the same homology class as an $O5$-plane arising from the fix-points of $\sigma$. 
Consequently, we have to add the contribution of the orientifold plane
\begin{equation} \label{eq:orieaction}
        S_{\text{ori}}^{\text{SF}}=\mu_5\int_{\mathcal{W}_{\text{orie}}}d^6\xi
        e^{-\phi}
        \sqrt{-\text{det}\left(\tilde\varphi^{\ast}\left(g_{10}+B\right)\right)}\
        \quad  \rightarrow
        \quad S_{\text{ori}}^{\text{EF}}  = \mu_5 \int\frac{e^{3\phi}}{2\mathcal{V}^2}v^{\Sigma}\ast 1\ ,
\end{equation}
to the action \eqref{eq:DBI}. Here we applied the calibration condition 
\eqref{eq:calibrationOrie} for orientifold planes to obtain the 
two-cycle volume $v^{\Sigma}$. Having rescaled to the Einstein frame one compares 
$S_{\text{ori}}^{\text{SF}}$ with \eqref{eq:DBI} to observe a precise cancellation 
of the D-term potential of the D5-brane.
This cancellation in $V_{\rm DBI}$ again confirms, from a low-energy 
effective action point of view, that both the $O5$-plane and the D5-brane have to 
wrap volume minimizing cycles.

The last two terms in $V_{\rm DBI}$ describe deviations from the calibration 
conditions \eqref{eq:DbraneIIBCalibrationExpanded}. The first potential term 
accounts for the metric deformations \eqref{eq:internalmetricvariation2} induced 
by the change of the ambient complex structure and the second term describes 
the field fluctuation $\mathcal{B}^a$ of the NS--NS B-field defined in 
\eqref{eq:calBperturbation}. Later on, we will show that this term is actually 
a D-term which is consistent with the analysis of \cite{Douglas:2000ah}. Clearly, 
both terms vanish at the supersymmetric ground state defined by the calibration conditions 
\eqref{eq:DbraneIIBCalibrationExpanded}. Let us comment on the dimensional reduction 
yielding these two terms. The evaluation of $\text{Tr}\mathfrak{A}^{-1}\mathfrak{B}$ 
in the expansion \eqref{eq:taylor} of the DBI-action yields a term given by 
\begin{equation}
        \delta \cL_{\delta g}=\frac{ie^{3\phi}v^{\Sigma}\mathcal{G}^{A\bar B}}{\cV^2\int \Omega\wedge\bar{\Omega} }\delta z^\kappa\delta \bar{z}^{\bar \kappa}\int_{\Sigma_+}\iota^\ast( s_A\lrcorner\chi_{\kappa})_{\bar{u}u}(\iota^\ast g)^{u\bar u}(\iota^\ast g)^{u\bar u}
  \iota^\ast(\bar{s}_{\bar B}\lrcorner \bar{\chi}_{\bar{\kappa}})_{\bar{u}u}\ \iota^\ast J\ . 
  \label{eq:DBImetricVar}
\end{equation}
originating from the variation $\delta(\iota^* g)$. Indeed, this is the only 
contribution to the four-dimensional effective action originating from the 
metric variation \eqref{eq:internalmetricvariation2} that is relevant at our 
low order analysis in the fields. As discussed before, cf.~section 
\ref{sec:relations}, the $(1,1)$-form $\iota^\ast J$ is essentially the only 
non-trivial element in the cohomology $H^{(1,1)}(\Sigma_+)$. Thus, we rewrite 
the pullback of any closed $(1,1)$-form $\omega$ to $\Sigma_+$ as
\begin{equation}
        \iota^\ast \omega=\frac{\int_{\Sigma_+} \omega}{v^{\Sigma}}\iota^\ast J
\label{eq:pullbackformula}
\end{equation}
in cohomology, where we used again the calibration $\int_{\Sigma_+} \iota^\ast J=v^{\Sigma}$. 
In particular, we can apply this to the closed $(1,1)$-forms $s_A\lrcorner \chi_{\kappa}$ 
in \eqref{eq:DBImetricVar} to obtain the second term of $V_{\rm DBI}$ given in \eqref{eq:DBIpotential}. 
Considering the fluctuation $\mathcal{B}^a$, the only contribution arises from the term
$\text{Tr}(\mathfrak{A}^{-1}\mathfrak{B})^2$ in \eqref{eq:taylor}. Restricting to the 
relevant quadratic term in $\mathcal{B}^a$, this yields  
\begin{equation}
\delta \cL_{\cB}=  \frac{e^{3\phi}}{16 \cV^2 }\, \cB^a \cB^b
\int_{\Sigma_+}(\iota^\ast \omega_{a})_{u\bar u}\ (\iota^\ast\omega_{b})_{u\bar u}\ 
g^{u\bar u}g^{u\bar u}\iota^\ast J\ .
\end{equation}
Again we use \eqref{eq:pullbackformula} to expand 
$P_-\ \cB^a \iota^\ast \omega_a=\cB^{\Sigma}\, {\iota^\ast J}/{v^{\Sigma}}$
in the cohomology $H^{(1,1)}(\Sigma_+)$, where $P_-$ again denotes the sign function
introduced in \eqref{eq:gauge-expansion}, to obtain the geometrical dependence of 
the volume $v^{\Sigma}$ of the cycle as given in \eqref{eq:DBIpotential}.
Later on in section \ref{sec:N=1dataD5}, we show explicitly that the above results 
of the dimensional reduction are necessary to match the F- and D-term potential 
arising from a brane superpotential $W_{\rm brane}$ and a gauging of a shift symmetry by the $U(1)$ 
vector $A$ from the D5-brane gauge theory, respectively.

\subsubsection{Chern-Simons Action}

Let us now turn to the dimensional reduction of the Chern-Simons part of the D5-brane 
action. For this purpose we need the normal coordinate expansion of the pullback of all
R--R fields \eqref{eq:RRO59expRev} to the world-volume of the D5-brane. Here we only 
display the relevant terms for the reduction of the Chern-Simons action which read
\begin{eqnarray}
   (\iota^* C_p)_{i_1\ldots i_p} &=& \tfrac{1}{p!}C_{i_1 \ldots i_p} +\tfrac{1}{p!}\zeta^n\partial_nC_{i_1 \ldots i_p}-\tfrac{1}{(p-1)!}\nabla_{i_1}\zeta^nC_{ni_2 \ldots i_p}+\tfrac{1}{2p!}\zeta^n\partial_n(\zeta^m\partial_mC_{i_1 \ldots i_p}) \\
 &-&\tfrac{1}{(p-1)!}\nabla_{i_1}\zeta^n\zeta^m\partial_m C_{ni_2 \ldots i_p}+\tfrac{1}{2(p-2)!}\nabla_{i_1}\zeta^n\nabla_{i_2}\zeta^m C_{nmi_3 \ldots i_p}+\tfrac{p-2}{2p!}R_{ni_1 m}^{j}\zeta^n\zeta^m C_{ji_2 \ldots i_p} \ ,\nonumber
\end{eqnarray} 
where the indices $i_{n}$ label the coordinates  $\xi^{i_n}$ on the D5-brane world volume.
Inserting this expansion into the Chern-Simons part of \eqref{eq:D5DBI+CS}, one obtains up 
to second order 
\begin{eqnarray} \label{eq:CS}
        S_{\text{CS}} &=& \mu_5\int \Big[\tfrac{\ell^2}{4}  c^\Sigma F\wedge F 
        -\tfrac{\ell}{2} d(\tilde{\rho}_{(2)}^\Sigma-
        \mathcal{C}_{(2)}\mathcal{B}^\Sigma)\wedge A 
        +\tfrac{i}{4}\, \mathcal{L}_{A\bar B}^\alpha\,
        d\tilde{c}^{(2)}_\alpha\wedge (d\zeta^A\bar\zeta^{\bar B}
        -d\bar{\zeta}^{\bar{B}}\zeta^A)\nonumber \\
        &\phantom{=}&\phantom{\mu_5} 
         -\ell^2 \tfrac{i}{2} \mathcal{C}^{I\bar J}\, 
          d\mathcal{C}_{(2)}\wedge (da_I\bar{a}_{\bar J}-d\bar{a}_{\bar J}a_I )
          - \tfrac{i}{4} \cL_{a b A\bar B}\,
        d(\cB^a \tilde{\rho}_{(2)}^b)\wedge(d\zeta^A\bar{\zeta}^{\bar B}- d\bar{\zeta}^{\bar B}\zeta^A) \nonumber \\
        &\phantom{=}&\phantom{\mu_5}
        +\tfrac12(\mathcal{N}_{\mathcal AK}\, A_{(3)}^K
        +\mathcal{N}_{\mathcal A}^K\,  \tilde A ^{(3)}_K)\wedge d\hat \zeta^{\mathcal A}
        -\tfrac\ell2 \hat \zeta^{\mathcal A} (\mathcal{N}_{\mathcal Ak}\, dV^{k}
         + \mathcal{N}_{\mathcal A}^{k}\, dU_{k})\wedge F
        \Big]\ ,
        \label{eq:eff-cs-action}
\end{eqnarray}
where $\cB^\Sigma,\cB^a$ are introduced in \eqref{eq:calBperturbation} and we similarly 
define $\tilde{\rho}_{(2)}^\Sigma = \int_{\Sigma_-} C_4$ as well as 
$c^\Sigma=\int_{\Sigma_+}C_2$. In the $S_{\text{CS}}$ we further used the abbreviations
\begin{equation} \label{eq:n-tilde}
    \displaystyle \mathcal{N}_{\mathcal AK}=\int_{\Sigma_+} \hat s_{\mathcal A}
     \lrcorner\alpha_K\ , \qquad 
    \mathcal{N}_{\mathcal A}^K = \int_{\Sigma_+}\hat s_{\mathcal A}
    \lrcorner\beta^K\ ,\qquad \mathcal{N}_{\mathcal Ak}=\int_{\Sigma_-} \hat s_{\mathcal A}
    \lrcorner\alpha_{k}\ , \qquad \mathcal{N}_{\mathcal A}^{k} =
     \int_{\Sigma_-} \hat s_{\mathcal A} \lrcorner\beta^{k}\ ,   
\end{equation}
where the forms and their orientifold parity are given in table
\ref{tab:cohombasisOrie}. We emphasize that depending on the parity of the 
integrand under the involution $\sigma$ in general both combinations $\Sigma_+$ 
and $\Sigma_-$ occur in \eqref{eq:n-tilde}. However, terms involving $\Sigma_-$ 
can be translated to $\Sigma_+$ via the sign function $P_{-}$. In addition we 
evaluated the coupling 
\begin{equation}  \label{eq:symbols-cs-eff-action}
  \cL_{ab A\bar B}=-i \int_{\Sigma_+}s_A\lrcorner \bar s_{\bar
    B}\lrcorner(\omega_a \wedge \omega_b) = \cL^\alpha_{A \bar B} \cK_{\alpha
    a b}\ ,
\end{equation}
where $\cL^{\alpha}_{A \bar B}$ was introduced in \eqref{eq:metrics} and $\cK_{\alpha
ab} = \int_Y \omega_\alpha\wedge \omega_a \wedge \omega_b$ are the only non-vanishing 
triple intersection numbers involving the negative $(1,1)$-forms $\omega_a$ of Table 
\ref{tab:cohombasisOrie}. We note that both in the action \eqref{eq:CS} and in the 
definitions \eqref{eq:n-tilde} we used the expansion $\hat \zeta = \hat \zeta^\cA \hat s_\cA$ 
into a real basis $\hat s_\cA$ introduced in \eqref{eq:zetaExpansion}. Clearly, the 
expressions involving $ \hat \zeta^\cA$, $\hat s_\cA$ are readily rewritten into 
complex coordinates $\zeta^A,\bar \zeta^A$.  

Let us now discuss the  interpretation of the different terms appearing in the
action \eqref{eq:CS}. The first term in $S_{\text{CS}}$ corresponds to the theta-angle 
term of the gauge theory on the D5-brane. Thus it is identified with the imaginary
part of the gauge-kinetic function. The second term is a Green-Schwarz term
which indicates the gauging of the scalar fields dual to the two-forms 
$\underline{\tilde{\rho}}_{(2)}$ and $\cC_{(2)}$ by the D5-brane vector field $A$.
In fact, we show explicitly in section \ref{sec:N=1dataD5} that this term 
indeed induces a gauging of one chiral multiplet in the four-dimensional spectrum 
and that the corresponding D-term is precisely the one encountered in the reduction 
of the DBI action in section \ref{sec:DBIreduction}.

The interpretation of the remaining terms in \eqref{eq:CS} is of more
technical nature. The third, fourth and fifth terms are mixing terms which 
contribute in the kinetic terms of the scalars $\underline{c}$, $h$ and $\underline{\rho}$ 
dual to the two-forms $\tilde c^{(2)}_\alpha$, $\cC_{(2)}$ and $\underline{\tilde\rho}_{(2)}$. 
In section \ref{sec:N=1dataD5} they help us to identify the correct complex
coordinates which cast the kinetic term into the standard $\cN=1$ form.
The sixth term contains the four-dimensional three-forms $\underline{A}_{(3)}$ and 
$\underline{\tilde A}^{(3)}$. We show in the next section \ref{sec:scalarpotderivation} 
that these terms are crucial in the calculation of the complete scalar potential of the 
D5-brane effective action. Finally, the last term in $S_{\text{CS}}$ indicates a mixing of 
the field strength on the D5-brane with the $U(1)$ bulk vector fields $\underline{V},
\underline{U}$. The precise form of the redefined gauge-couplings is discussed in 
appendix \ref{app:kinmix}.

\subsection{The Scalar Potential}
\label{sec:scalarpotderivation}

In this section we compute the scalar potential of the four-dimensional
effective theory. We find two types of potentials according to their 
origin in the $\mathcal{N}=1$ language, i.e.~an F-term potential $V_{\rm F}$ and D-term 
potential $V_{\rm D}$. In the context of the D5-brane effective action the first 
one arises from the coupling of the three-forms $\underline{A}_{(3)}$, 
$\underline{\tilde{A}}^{(3)}$ in the Chern-Simons action and their kinetic 
terms from the bulk as well as from a completion term from the potential 
$V_{\rm DBI}$ in \eqref{eq:DBIpotential}. The second potential arises entirely 
from the bulk and the potential $V_{\rm DBI}$. We note that the potential due to 
background R--R and NS--NS fluxes $F_3=\langle dC_2 \rangle$ and 
$H_3=\langle dB_2 \rangle$ has already been studied in \cite{Grimm:2004uq}. 
Thus, we emphasize here the additional contributions in the presence 
of the space-time filling D5-branes. 

A first contribution to the scalar potential is induced by the couplings 
of the three-forms $\underline{A}_{(3)}$ and 
$\underline{\tilde A}^{(3)}$
in the Chern-Simons action \eqref{eq:CS}. Here it is crucial to keep 
these forms in the spectrum despite the fact that a 
massless three-form has no propagating degrees of freedom in  four dimensions. 
Moreover, if the degrees of freedom in these forms as well as the induced 
potential are treated quantum mechanically, as described
in \cite{Beasley:2002db}, one is able to account for the possible R--R 
three-form flux 
\begin{equation} \label{eq:R--Rflux}
     F_3 =m^K \alpha_K -  e_K \beta^K\ ,
\end{equation} 
where the flux quanta $(m^K,e_K)$ are interpreted as labeling the discrete
excited states of the system and $(\alpha_K,\beta^K)$ is the real symplectic 
basis introduced in \eqref{eq:def-alpha_beta}. This is in accord with the fact 
that the duality condition $G^{(3)} = (-1) *_{10} G^{(7)}$ given in 
\eqref{eq:selfduality} relates the three-form containing $F_3$ to the seven-form 
$F_7$ containing $(dA_{(3)}^K,d\tilde{A}^{(3)}_K)$.
 
Let us collect the terms involving the non-dynamical three-forms $\underline{A}_{(3)}$ 
and $\underline{\tilde{A}}^{(3)}$. The first contribution arises from the effective bulk 
supergravity action containing the kinetic term $ \tfrac14 \int dC_6\wedge *dC_6$ for the 
R--R-form field $C_6$.\footnote{The factor $\tfrac14$ arises due to the fact that we can 
eliminate $dC_2$ contributions in this analysis by the duality condition 
\eqref{eq:selfduality}.} Together with the contribution from the effective Chern-Simons action 
\eqref{eq:eff-cs-action} we obtain
\begin{equation}
      S_{A_{(3)}}=\int \big[ \tfrac14 e^{-4\phi} \cV^2  d\vec{A}_{(3)}\wedge *E\, d\vec{A}_{(3)} 
	+\tfrac12\mu_5\,\vec{\mathcal{N}}^T d\vec{A}_{(3)} \big]\ ,
\label{eq:cs-term-containing-a3}
\end{equation}
where the factor $e^{-4\phi} \cV^2$ arises due to the rescaling to the four-dimensional 
Einstein frame. Note that we have introduced a vector notation to keep the following 
equations more transparent. More precisely, we define the matrix $E$, the vector-valued 
forms $\vec{A}_{(3)}$ and the vector $\vec{\mathcal{N}}$, cf.\ \eqref{eq:n-tilde}, as
\begin{equation}
        E= \begin{pmatrix}
                \int \alpha_K\wedge *\alpha_L & \int \alpha_K\wedge *\beta^L\\
                \int\beta^K\wedge * \alpha_L & \int \beta^K\wedge * \beta^L
        \end{pmatrix}, \qquad \vec{A}_{(3)}=
        \begin{pmatrix}
                \underline{A}_{(3)}\\
                \underline{\tilde{A}}^{(3)}
        \end{pmatrix},\qquad 
        \vec{\mathcal{N}}=\hat \zeta^\cA 
              \begin{pmatrix}
                \mathcal{N}_{\cA K}\\
                \mathcal{N}^K_\cA
        \end{pmatrix}.
        \label{eq:e-matrix}
\end{equation}
Our aim is to integrate out the forms $dA_{(3)}^K$ and $d\tilde{A}^{(3)}_K$ similar to 
\cite{Beasley:2002db} by allowing for the discrete excited states labeled by the 
background fluxes $(e_K, m^K)$ in \eqref{eq:R--Rflux}. In fact, we can treat these as 
formally dualizing the three-forms $\underline{A}_{(3)}$ and $\underline{\tilde{A}}^{(3)}$ 
into constants $(e_K, m^K)$ \cite{Louis:2002ny}. We thus add to $S_{A_{(3)}}$ the Lagrange 
multiplier term $\tfrac12(e_K dA^K_{(3)}+ m^Kd\tilde{A}^{(3)}_K)$ such that 
\begin{equation}
  S_{A_{(3)}}\,\rightarrow\,S'_{A_{(3)}}=\int \big[\tfrac14e^{-4\phi} \cV^2  d\vec{A}_{(3)}\wedge *E\, d\vec{A}_{(3)} 
	+\tfrac12(\mu_5\,\vec{\mathcal{N}}+\vec{m})^T d\vec{A}_{(3)} \big]\ ,
\end{equation}
where we abbreviated $\vec{m}=(e_K, m^K)^T$. Formally replacing  $\vec{F}_{(4)}=d\vec{A}_{(3)}$ 
by its equations of motion, one finds the scalar potential  
\begin{equation}
       S'_{A_{(3)}}=\int V_{ A_{(3)}} \ast 1\,,\qquad V_{ A_{(3)}}=\frac{e^{4\phi}}{4\mathcal
        V^2}\left(\mu_5\vec{\mathcal{N}}+\vec{m}\right)^T E^{-1} \left(\mu_5\vec{\mathcal{N}}
	+\vec{m}\right).
        \label{eq:potential-from-a4}
\end{equation}
Here we used the results of appendix \ref{app:complexMatrixM} to determine the inverse 
matrix given by 
\begin{equation}
        E^{-1}=\begin{pmatrix}
                \int \beta^K\wedge * \beta^L &  -\int\beta^K\wedge * \alpha_L\\
                -\int \alpha_K\wedge *\beta^L & \int \alpha_K\wedge *\alpha_L
        \end{pmatrix}.
        \label{eq:matrix-d}
\end{equation}
It is convenient to rewrite the potential $V_{A_{(3)}}$ in \eqref{eq:potential-from-a4} in 
a more compact form 
\begin{equation}
        V_{ A_{(3)}}=\frac{e^{4\phi}}{4 \mathcal{V}^2}\int_{Z_3} \mathcal{G}\wedge *
        \mathcal{G},\qquad \quad  \mathcal{G}=F_{3}+\mu_5 \hat \zeta^\cA \left(\mathcal{N}^K_{\cA}\alpha_K
	- \mathcal{N}_{\cA K}\beta^K\right)\ ,
        \label{eq:v-written-in-g}
\end{equation}
where we identified $F_3$ as given in \eqref{eq:R--Rflux}. 
We note that the scalar potential contains the familiar contribution from the R--R fluxes 
$F_3$. In addition, there are terms linear and quadratic in the D5-brane deformations 
$\hat{\underline{\zeta}}$. 

In order to prepare for the derivation of $V_{ A_{(3)}}$ from this superpotential, it is 
necessary to introduce the holomorphic and anti-holomorphic variables $\zeta^A$ and 
$\bar \zeta^A$. Therefore we expand the three-form $\mathcal G$ in a complex basis 
$(\Omega,\chi_\kappa,\bar{\chi}_{\bar\kappa},\bar\Omega)$ of 
$H^3_+(Y)=H^{(3,0)}\oplus H^{(2,1)}_+\oplus H^{(1,2)}_+\oplus H^{(0,3)}$. Explicitly, we 
find the expansion
\begin{eqnarray}
    \cG =\big({\textstyle \int}\Omega\wedge\bar\Omega \big)^{-1}
        \big[I\, \bar{\Omega} +G^{\bar\kappa \kappa} \bar I_{\bar\kappa}\, \chi_\kappa 
        - G^{\bar\kappa \kappa}I_\kappa\, \bar{\chi}_{\bar\kappa}-\bar I \Omega \big]\ ,
        \label{eq:expansion-g-dolbeault}
\end{eqnarray}
where the coefficient  functions are given by 
\begin{equation}
        \label{eq:omega-wedge-g}
        I=\int_{Z_3}\Omega\wedge\mathcal G = \int_{Z_3} \Omega\wedge F_3- \mu_5 \int_{\Sigma_+} 
	\zeta\lrcorner\Omega\ , \qquad \
        I_\kappa=\int_{Z_3}\chi_\kappa\wedge\mathcal G = \int_{Z_3} \chi_\kappa\wedge F_3-\mu_5 
	\int_{\Sigma_+}\zeta\lrcorner\chi_\kappa\ .
\end{equation}
Here we have used the relations \eqref{eq:expansion-omega-a} and 
\eqref{eq:zeta-omega-bar-zero} as well as the familiar metric $G_{\kappa \bar\kappa }$ 
on the space of complex structure moduli space of $Z_3/\sigma$. Using its 
explicit form   
\begin{equation}
	 G_{\kappa\bar\kappa}=-\frac{\int\chi_\kappa\wedge\bar\chi_{\bar\kappa}}
	{\int\Omega\wedge\bar\Omega}\ ,
\end{equation}
the expansion \eqref{eq:expansion-g-dolbeault} together with \eqref{eq:omega-wedge-g} is 
readily checked. Finally, we insert \eqref{eq:expansion-g-dolbeault} into 
\eqref{eq:v-written-in-g} and use $\ast\Omega=-i\Omega,\ \ast\chi_\kappa=i\chi_\kappa$ 
to cast the complete potential $V_{\rm F}$ into the form
\begin{equation}  \label{eq:v-final}
      V_{\rm F}= \frac{i e^{4\phi}}{2\mathcal V^2 \int\Omega\wedge\bar\Omega}
      \Big[G^{\kappa\bar\kappa} I_\kappa \bar I_{\bar \kappa} + |I|^2
      +2\mu_5\ \mathcal{G}^{A\bar B}e^{-\phi}\int_{\Sigma_+}s_A\lrcorner \chi_{\kappa}
      \int_{\Sigma_+}\bar s_{\bar B}\lrcorner \bar \chi_{\bar\kappa}
      \delta z^{\kappa}\delta \bar{z}^{\bar \kappa}\Big]\ .
\end{equation}
Here we have added the potential term in \eqref{eq:DBIpotential} originating from the 
reduction of the Dirac-Born-Infeld action. Once we have computed the $\cN=1$ K\"ahler metric of the 
effective theory in appendix \ref{app:F-termscalarpot}, we will derive in section \ref{sec:suppot} 
the potential $V_{\text{F}}$ from a from a flux and brane superpotential depending on the complex 
structure and the D5-brane deformations. This will show that $V_{\rm F}$ is indeed an F-term potential
as indicated by the notation.

Let us now turn to the remaining potential terms arising from the DBI action \eqref{eq:DBI} 
and the NS--NS fluxes $H_3$. For simplicity, we only discuss electric NS--NS flux such 
that $H_3$ admits the expansion 
\begin{equation} \label{eq:NSNSflux}
   H_3 = -\tilde e_K \beta^K\ .
\end{equation}
It was shown in \cite{Grimm:2004uq} that the electric fluxes $\tilde e_K$ result in a 
gauging of the scalar $h$ dual to $\cC_{(2)}$ in \eqref{eq:RRO59expRev}. The effect of magnetic 
fluxes $\tilde m^K \alpha_K$ is more involved since they directly gauge the two-from $\cC_{(2)}$ 
as demonstrated in \cite{Grimm:2004uq}. In order to be able to work with the scalar $h$, we 
do not allow for the additional complication and set $\tilde m^K=0$. Together with the last 
term in \eqref{eq:DBI} we find the potential
\begin{equation}
   V_{\rm D} =\mu_5 \frac{e^{3\phi}}{\mathcal{V}^2}
   \frac{(\mathcal{B}^{\Sigma})^2}{16v^{\Sigma}} + \frac{e^{2\phi}}{4 \cV^2}
   \int_Y H_3 \wedge * H_3 \ ,\label{eq:Dtermpot}
\end{equation}
which turns out to be a D-term potential due to the gauging of two chiral 
multiplets\footnote{Recall that the first potential term in the DBI action \eqref{eq:DBIpotential} 
is canceled by the contribution \eqref{eq:orieaction} of the $O5$-planes.} as demonstrated in 
section \ref{sec:gaugeKin+rest}.

\section{D5-Brane $\mathcal{N}=1$ Effective Couplings and Coordinates}
\label{sec:N=1dataD5}  

In this section we bring the four-dimensional effective action of the D5-brane 
and bulk fields into the standard $\mathcal{N}=1$ supergravity form of \eqref{eq:N=1action}. 
First in section \ref{sec:N=1coords+KpotD5} we determine the $\mathcal{N}=1$ chiral coordinates
and the large volume expression for the K\"ahler potential. Then in section \ref{sec:suppot}
we read off the effective superpotential by matching the F-term scalar potential in 
\ref{sec:scalarpotderivation}, cf.~also appendix \ref{app:F-termscalarpot}. 
We conclude in section \ref{sec:gaugeKin+rest} with a determination of the D5-brane gauge 
kinetic coupling function, the gaugings of chiral multiplets and the D-term potential, that
is again in perfect agreement with the results directly obtained in the reduction.
The kinetic mixing between bulk and brane gauge fields can be found in appendix \ref{app:kinmix}.
We note that the $\mathcal{N}=1$ characteristic data of the D5-brane agrees with the results 
of chapter \ref{ch:EffActCYOrieCompact} if the D5-brane fields are frozen out. 

\subsection{The K\"ahler Potential and $\cN=1$ Coordinates}
\label{sec:N=1coords+KpotD5}

We first define the $\cN=1$ complex coordinates $M^I$ which are the bosonic 
components of the chiral multiplets. They define those complex coordinates for which
the scalar metric is K\"ahler. We note that the $M^I$ consist of the D5-brane 
deformations $\zeta^A$ and Wilson lines $a_I$ introduced in section \ref{sec:specBrane}.
In addition there are the complex structure deformations $z^\kappa$ as well as
the complex fields 
\begin{eqnarray} \label{eq:N=1coordsD5}
   t^\alpha &=& e^{-\phi} v^\alpha - i c^\alpha +\tfrac12\mu_5\, \cL^\alpha_{A\bar B} \zeta^A
   \bar \zeta^{\bar B}\ ,\nonumber\\
   P_a &=& \Theta_{ab}\, \mathcal{B}^b + i \rho_a\ ,\\
   S &=&  e^{-\phi} \cV + i \tilde{h} -\tfrac14 (\text{Re} \Theta)^{ab} P_a (P+\bar
   P)_b +\mu_5\,\ell^2\,\mathcal{C}^{I\bar J}a_{I}\bar a_{\bar J}\ ,\nonumber
\end{eqnarray}
where $\underline{v},\underline{b}$, $\underline{c},\underline{\rho}$ as well as 
$\underline{\mathcal B}$ are given in \eqref{eq:expJBPhi}, \eqref{eq:RRO59expRev} 
and \eqref{eq:calBperturbation} as well as $\tilde h=h-\tfrac12\rho_a\mathcal{B}^a$. 
The complex symmetric tensor appearing in \eqref{eq:N=1coordsD5} is given by
$\Theta_{ab}=\cK_{ab \alpha} t^\alpha$ and $(\text{Re} \Theta)^{ab}$ denotes the
inverse of $\text{Re} \Theta_{ab}$. The function $\cL^\alpha_{A \bar B}$
is defined in \eqref{eq:metrics}. Note that we recover the $\cN=1$ coordinates 
\eqref{eq:N=1coordsO59} of the bulk $O5/O9$ setup discussed in section 
\ref{sec:N=1dataOrie} if we set $\underline{\zeta}=\underline{a} = 0$. The 
completion \eqref{eq:N=1coordsD5} by the open string fields is inferred from the couplings in the D5-brane 
action \eqref{eq:DBI} and \eqref{eq:CS}.

The full $\cN=1$ K\"ahler potential is determined by integrating the kinetic
terms of the complex scalars 
$M^I=(S,\underline{t},\underline{P},\underline{z},\underline{\zeta},\underline{a})$. 
It takes the form 
\begin{equation} \label{eq:kaehler-pot}
    K =-\ln\big[ -i\int\Omega\wedge\bar\Omega \Big]+K_q\ ,\qquad  
   K_q=-2\ln\big[\sqrt{2}e^{-2\phi}\mathcal V \Big]\ ,
\end{equation}
where $K_q$ has to be evaluated in terms of the coordinates
\eqref{eq:N=1coordsD5}. In contrast to general compactifications with 
$O3/O7$-orientifold planes, cf.~section \ref{sec:N=1dataOrie}, this can be done 
explicitly for $O5$-orientifolds yielding 
\begin{equation} \label{eq:Kaehlerpart}
        K_q
       = -\ln \big[ \tfrac{1}{48} \cK_{\alpha \beta \gamma} \Xi^\alpha\,
      \Xi^\beta\, \Xi^\gamma \big] 
      - \ln \big[S +\bar S + \tfrac{1}{4} (\text{Re} \Theta)^{ab} (P +\bar P)_a (P
     +\bar P)_b - 2\mu_5\,\ell^2\,\mathcal{C}^{I\bar J}a_{I}\bar a_{\bar J}\big]\ ,
\end{equation}
where we write
\begin{equation}
  \Xi^\alpha = t^\alpha +\bar t^\alpha -
        \mu_5\,\cL^{\alpha}_{A\bar B} \zeta^A \bar \zeta^{\bar B} \ .
                \label{eq:xi-def}
\end{equation}
Note that the expression \eqref{eq:kaehler-pot} for $K$ can already be
inferred from general Weyl rescaling arguments, e.g.~from the factor $e^K$ in
front of the $\cN=1$ potential \eqref{eq:N=1pot}. However, the explicit form
\eqref{eq:Kaehlerpart} displaying the field dependence of $K$ has to be
derived by taking derivatives of $K$ and comparing the result with the bulk
and D5-brane effective action.
Let us also note that the expression \eqref{eq:Kaehlerpart} reduces to the 
results found in \cite{Kors:2003wf,Lust:2004cx} in the orbifold limit.

\subsection{The Superpotential}
\label{sec:suppot}

Having defined the $\mathcal{N}=1$ chiral coordinates as well as 
the K\"ahler potential we are prepared to deduce the effective superpotential 
$W$. 
Using the general supergravity formula \eqref{eq:N=1pot} for the scalar 
potential expressed in terms of $W$ we are able, as presented below, to deduce the 
superpotential $W$ entirely by comparison to the scalar potential 
$V_{\rm F}$ in \eqref{eq:v-final} as derived from dimensional reduction. This 
indeed identifies $V_{\rm F}$ as an F-term potential of the $\mathcal{N}=1$ effective 
theory as indicated by the notation.

The superpotential $W$ yielding $V_{\rm F}$ consists of two parts, a truncation 
of the familiar flux superpotential for the closed string 
moduli \cite{Gukov:1999ya} and a contribution encoding the dependence on the 
open string moduli of the wrapped D5-brane,
\begin{equation} \label{eq:effsuperpot} 
 W=\int_{Z_3} F_3\wedge \Omega+\mu_5\int_{\Sigma_+}\zeta\lrcorner\Omega\ ,
\end{equation}
where we introduced the R--R-flux $F_3$.
Now, it is a straightforward but lengthy calculation summarized in appendix 
\ref{app:F-termscalarpot} to obtain the F-term contribution of the scalar potential 
\eqref{eq:N=1pot}. The detailed calculations yield the positive definite F-term 
potential 
\begin{equation}
 V=\frac{ie^{4\phi}}{2\mathcal V^2 \int\Omega\wedge\bar\Omega}\left[ \left|
     W\right|^2 +D_{z^\kappa}WD_{\bar z^{\bar\kappa}}\bar W
     G^{\kappa\bar\kappa} 
     +\mu_5\ \mathcal{G}^{A\bar B}e^{-\phi}\int_{\Sigma_+}s_A\lrcorner 
     \Omega\int_{\Sigma_+}\bar s_{\bar B}\lrcorner \bar \Omega \right]\,.
\label{eq:F-termpotential}
\end{equation} 
Here the covariant derivatives with respect to the complex structure 
coordinates $z^\kappa$ and the open string moduli $\zeta^A$ read
\begin{equation}
       D_{z^\kappa}W = \int F_3\wedge\chi_\kappa +\mu_5\int\zeta\lrcorner\chi_\kappa\ ,
	\qquad
        D_{\zeta^A}W = \mu_5\int s_A\lrcorner\Omega+\hat{K}_{\zeta^A}W\ .\label{d_zW}
\end{equation}
Furthermore, we have to use the first order expansion of $s_A\lrcorner \Omega$ 
discussed in \eqref{eq:expansion-omega-a} to obtain a form of type $(1,1)$ that
can be integrated over $\Sigma_+$ yielding a potentially non-vanishing result,
\begin{equation}
        \int_{\Sigma_+}s_A\lrcorner \Omega
	=\int_{\Sigma_+}s_A\lrcorner\chi_{\kappa}\delta z^\kappa.
\end{equation}
Inserting this into \eqref{eq:F-termpotential}, the F-term potential perfectly 
matches the scalar potential $V_{\rm F}$ of \eqref{eq:v-final} obtained by 
dimensional reduction of the D5-brane as well as the bulk supergravity action.

The superpotential \eqref{eq:effsuperpot} is the perturbative superpotential
of the Type IIB compactification. However, in the form \eqref{eq:effsuperpot}
it is just the leading term in the expansion of the chain integral\footnote{In this section we set $\mu_5=1$.}
~\cite{Witten:1997ep,Kachru:2000an,Aganagic:2000gs,Aganagic:2001nx}   
\begin{equation}
        W_{\rm brane}=\int_\Gamma\Omega\ ,
        \label{eq:chainIIB}
\end{equation}
where $\Gamma$ is a three-chain with boundary given 
as $\partial\Gamma=\Sigma-\Sigma_0$, where $\Sigma_0$ 
is a fixed reference curve in the same homology class as $\Sigma$. $W_{\rm brane}$ 
depends on the closed string complex structure moduli through the holomorphic three-form $\Omega$ 
and on the open string fields through the deformation parameters 
of the curve $\Sigma$. Using the general power series expansion of a functional about a 
reference function, we recover our result for the superpotential
\eqref{eq:effsuperpot} to linear order.\footnote{The general
Taylor expansion is given by 
$F[g]=\sum_{k=0}^{\infty}\int\,dx_1\cdots dx_k\frac{1}{k!}\left.\frac{\delta^k
    F[g]}{\delta g(x_1)\cdots \delta g(x_k)}\right|_{g=\tilde g}\delta
g(x_1)\cdots \delta g(x_k)$. 
For $W$ as a functional of the embedding $\iota$ and $\delta \iota\equiv\zeta$
as well as $\tilde g=\iota$ we to first order derive the 
second term of \eqref{eq:effsuperpot}).} It is one central aim of this work
to study and exactly calculate this brane superpotential in various setups and 
invoking different physical and mathematical techniques.

We conclude with a discussion of the derivation and the special structure 
of the F-term potential. We first note that the potential \eqref{eq:F-termpotential} 
is positive definite unlike the generic F-term potential of supergravity. 
This is due to the no-scale structure \cite{Cremmer:1983bf,Ellis:1983sf,Barbieri:1985wq} 
of the underlying $\mathcal{N}=1$ data. Indeed, the superpotential 
\eqref{eq:effsuperpot} only depends on $\underline{z}$ and $\underline{\zeta}$ and is independent of the 
chiral fields $S$, $\underline{P}$, $\underline{a}$ and $\underline{t}$. Consequently,
the $\mathcal{N}=1$ covariant derivative $D_{M^I}W$ of the superpotential simplifies 
to $K_{M^I}W$ when applied with respect to the fields 
$M^I=(S,\underline{P},\underline{a},\underline{t})$. The K\"ahler potential 
\eqref{eq:kaehler-pot} for these fields has the schematic form
\begin{equation} \label{eq:noscalekaehlerpot}
 K=-m\ \text{ln}(t+\bar
t+f(\zeta,\bar\zeta))-n\ \text{ln}(S+\bar{S}+g(P+\bar P,t+\bar t)+h(a,\bar a))\ 
\end{equation}
with $m=3$ and $n=1$, where we concentrate on the one-modulus case for each chiral 
multiplet in order to clarify our exposition. The generalization to an arbitrary 
number of moduli is straightforward, cf.~appendix \ref{app:F-termscalarpot}, 
where also the functions $f$, $g$ and $h$ can be found. Then the contributions of the 
fields $M^I=(S, P, a,t, \zeta)$ to 
the scalar potential $V$ are found to take the characteristic form 
given by
\begin{equation}\label{eq:noscalescalarpot}
 K^{I\bar J}D_{M^I}W D_{\bar{M}^{\bar J}}\bar{W}
 =|\partial_\zeta W|^2K^{\zeta\bar\zeta}+(n+m)|W|^2 
\end{equation}
as familiar from the basic no-scale type models of supergravity.\footnote{This no-scale 
structure will be  clarified further, extending 
the example of \cite{Grimm:2004uq}, in appendix \ref{app:F-termscalarpot} using the dual 
description of $S+\bar S$ in terms of a linear multiplet $L$.} Consequently, this turns 
the negative term $-3|W|^2$ in \eqref{eq:N=1pot} into the positive definite term 
$|W|^2$ of \eqref{eq:F-termpotential} for the case $n=1$ and $m=3$. A similar 
structure for the underlying $\mathcal{N}=1$ data has been found for D3- and 
D7-branes as shown in 
\cite{DeWolfe:2002nn,Camara:2003ku,Grana:2003ek,Camara:2004jj,Jockers:2004yj,Jockers:2005zy}.\footnote{See 
\cite{Correia:2007sv} for a similar discussion in heterotic 
M-theory.} In particular, this form for the scalar potential $V$ on the complex structure and 
D-brane deformation space implies that a generic vacuum is de 
Sitter, i.e.~has a positive cosmological constant, while in a supersymmetric 
vacuum both $V$ and $W$ vanish. However, the potential depends on the K\"ahler 
moduli only through an overall factor of the volume and thus drives the 
internal space to decompactify.

\subsection{The Gauge-Kinetic Function, Gaugings and D-term Potential}
\label{sec:gaugeKin+rest}

In the following we discuss the terms of the four-dimensional effective
action arising due to the $U(1)$ vector multiplets in the spectrum. Firstly, 
there are the kinetic terms of the D5-brane vector $A$ and the 
vectors $\underline{V}$ arising from the expansion
\eqref{eq:RRO59expRev} of the R--R form $C_4$.
The gauge-kinetic function is determined from the actions \eqref{eq:DBI} and 
\eqref{eq:CS} and reads
\begin{equation} \label{eq:gauge-couplingD5bulk}
    f_{\Sigma \Sigma}(t^\Sigma) = \tfrac12\mu_5 \ell^2\, t^\Sigma \ , \qquad  
   \quad f_{kl}(z^\kappa) = - \tfrac{i}{2} \bar \cM_{kl} 
    = - \tfrac{i}{2} \cF_{kl}\big|_{z^k=0=\bar{z}^{\bar k}}\ ,
\end{equation}
where the complex matrix $\mathcal{M}$ is defined in appendix \ref{app:complexMatrixM}. 
Here $ f_{\Sigma \Sigma}$ is the gauge-coupling function for the D5-brane
vector $A$ and $f_{kl}$ is the  gauge-coupling function for 
the bulk vectors $\underline{V}$ discussed in \eqref{eq:gaugeKinOrie}. As reviewed in 
section \ref{sec:N=1dataOrie}, we note that the latter can be expressed via
$ \cF_{kl} = \partial_{z^{k}} \partial_{z^{l}} \cF$ 
as the second derivative of the $\cN=2$ prepotential $\cF$ with respect to the 
$\cN=2$ coordinates $z^{k}$ which have then to be set to zero in the orientifold 
set-up. This ensures that the gauge-coupling function is 
holomorphic in the coordinates $z^\kappa$ which would not be the case for the 
full $\cN=2$ matrix $ \bar \cM_{K L} $ given in \eqref{eq:matrix-a-b-m}.

There are some remarks in order. Firstly, we note that the gauge-kinetic function 
encoding the mixing between the D5-brane vector and the bulk vectors is discussed 
in appendix \ref{app:kinmix}. Secondly, we observe that the quadratic dependence of $f_{\Sigma \Sigma}$ 
on the open string moduli $\underline{\zeta}$ through the coordinate $\underline{t}$ 
in \eqref{eq:N=1coordsD5} is not visible on the level of the effective action. 
These corrections as well as further mixing with the open string moduli are due 
to one-loop corrections of the sigma model and thus not covered by our bulk 
supergravity approximation nor the Dirac-Born-Infeld or Chern-Simons actions of 
the D5-brane.

Let us now turn to the terms in the scalar potential induced by the gauging of
global shift symmetries and compare to the potential $V_{\rm D}$ in \eqref{eq:Dtermpot}. 
There are two sources for such gaugings. The first gauging arises due to the source 
term proportional to $d(\tilde{\rho}^\Sigma -\mathcal{C}_{(2)}\mathcal{B}^\Sigma)\wedge A$ 
in \eqref{eq:CS}. It enforces a gauging of the scalars dual to the two-forms 
$\tilde{\rho}^\Sigma$ and $\mathcal{C}_{(2)}$. In fact, eliminating $d\tilde{\rho}^\Sigma$ 
and $d\mathcal{C}_{(2)}$ by their equations of motion, the kinetic terms of the dual 
scalars $\rho_a$ and $h$ contain the covariant derivatives
\begin{equation} \label{eq:gaugedfields}
        \cD \rho_a=d\rho_a+\mu_5\ell\delta_a^\Sigma A\ ,\qquad \quad \cD
        h=dh+\mu_5\ell\mathcal{B}^\Sigma A\ ,
\end{equation}
where $A$ is the $U(1)$ vector on the D5-brane. Rearranging this into $\mathcal{N}=1$
coordinates we observe that the signs in the covariant derivative of 
$h$ and $\underline{\rho}$ 
arrange\footnote{The plus sign in the covariant derivative of $h$ arises due to the minus 
sign in the duality conditions \eqref{eq:selfduality}.} to ensure that the complex scalar $S$ defined in \eqref{eq:N=1coordsD5} remains neutral 
under $A$. However, the gaugings \eqref{eq:gaugedfields} imply a charge for the chiral 
field $P_\Sigma$. It is gauged by the D5-brane vector $A$. Its covariant derivative is 
given by
\begin{equation} \label{eq:gaugingA}
        \cD P_\Sigma =dP_\Sigma +i\mu_5\ell A\ .
\end{equation}

The second gauging arises in the presence of electric NS--NS
three-form flux $\tilde e_K$ introduced in \eqref{eq:NSNSflux}. 
It was shown in \cite{Grimm:2004uq}, that the scalar $h$ is 
gauged by the bulk $U(1)$ vectors $\underline{V}$ arising in the 
expansion \eqref{eq:RRO59expRev}
of $C_4$. This forces us to introduce the covariant derivative 
\begin{equation} \label{eq:gaugingS}
  \cD S = d S - i\tilde e_{\tilde K} V^{\tilde K}\ .
\end{equation}
The introduction of magnetic NS--NS three-form flux is more involved and leads
to a gauged linear multiplet $(\phi ,\, \cC_{(2)})$ as described in \cite{Grimm:2004uq}. 

Having determined the covariant derivatives \eqref{eq:gaugingA} and \eqref{eq:gaugingS} 
it is straightforward to evaluate the D-term potential. Recall the general formula
for the D-term \cite{Wess}
\begin{equation} \label{eq:D-term_gen}
K_{I\bar J} \bar X^{\bar J}_k = i \partial_I D_k \ ,
\end{equation}
where $X^{I}$ is the Killing vector of the $U(1)$ transformations
defined as $\delta M^I = \Lambda^k_0 X_k^J \partial_J M^I$. For the 
gaugings \eqref{eq:gaugingA} and \eqref{eq:gaugingS} we find the Killing vectors 
$X^{P_\Sigma}=i\mu_5\ell$ and $X^S_{\tilde K} = - i\tilde e_{\tilde K}$
which are both constant. Integrating \eqref{eq:D-term_gen} one evaluates the D-terms using
$K_{P_\Sigma}=\tilde{K}_{P_\Sigma}$ and $K_S$ given in \eqref{eq:firstderivatives_2} respectively 
above \eqref{eq:firstderivatives} in appendix \ref{app:F-termscalarpot} as 
\begin{equation}
  D = - \tfrac14 \mu_5 \ell e^{\phi} \cB^\Sigma \cV^{-1}  \ ,\qquad \quad 
  D_{\tilde K} = \tfrac{1}{2} \, \tilde e_{\tilde K}\,e^{\phi}\cV^{-1}\ .
\end{equation}  
Inserting these D-terms into the $\cN=1$ scalar potential \eqref{eq:N=1pot} and using the
gauge-kinetic functions \eqref{eq:gauge-couplingD5bulk}, we precisely recover the D-term potential
$V_{\rm D}$ in \eqref{eq:Dtermpot} found by dimensional reduction.

\section{Extension to Infinite Degrees of Freedom}
\label{sec:extensionInfinite} 

In this concluding section we slightly extend, following \cite{Grimm:2010gk}, our discussion of the 
D5-brane effective action to the full geometric deformation space of the D5-brane on $\Sigma$, that 
is given by the Hilbert space of all sections in $\mathcal{C}^{\infty}(\Sigma,N_{Z_3}\Sigma)$. It 
contains both the light fields $\zeta^A$ in $H^0(\Sigma,N_{Z_3}\Sigma)$ discussed before but also an
infinite tower of massive modes\footnote{This is in contrast to the usual lore of dimensional reduction, 
however, we refer to section \ref{branedeformationsI} for a mathematical explanation of the fact, that the
natural domain of the brane superpotential $W_{\rm brane}$ is $\mathcal{C}^{\infty}(\Sigma,N_{Z_3}\Sigma)$.}. 
It is the content of the following discussion to show
that the massive modes already receive a mass at leading order and consequently decouple
consistently from the effective action. We note a similar and mathematically more self-contained
discussion in section \ref{sec:N=1gensection}.

In words we exploit for our effective action analysis the mathematical result of 
\cite{simons1968minimal,mclean1998deformations} that the volume of a holomorphic curve $\Sigma$, 
$\text{Vol}(\Sigma)$, for any deformation of $\Sigma$ by an infinitesimal 
displacement $\epsilon s$, for $s\in\mathcal{C}^{\infty}(\Sigma,N_{Z_3}\Sigma)$, 
increases quadratically  as
\begin{equation} \label{eq:volvarD5}
 	\left.\frac{d^2}{d \epsilon^2}\text{Vol}(\Sigma_{\epsilon s})\right\vert_{\epsilon=0}
 	=\frac12\int_{\Sigma}\norm{\bar{\partial} s}^2\, \vol_{\Sigma}\,,
\end{equation}
where $\Sigma_{\epsilon s}$ denotes the deformed curve. Here $\norm{\bar{\partial} s}^2$ 
denotes a contraction of all indices, i.e.~the tangential directions and the directions normal to $\Sigma$,
via the metric.
We prove the statement that this second variation of the volume \eqref{eq:volvarD5} is a 
part of the F-term potential of the D5-brane effective action. We first obtain 
this potential by dimensional reduction of the the Dirac-Born-Infeld action 
of the D5-brane. Then we use the D5-brane superpotential $W_{\rm brane}$ in \eqref{eq:chainIIB} and a 
generalization of the K\"ahler metric of \eqref{eq:kaehler-pot} \cite{Grimm:2008dq} to 
the infinite dimensional space $\mathcal{C}^\infty(\Sigma,N_{Z_3}\Sigma)$ to deduce the 
same potential as an F-term potential when gravity is decoupled.

We start again from the Dirac-Born-Infeld action \eqref{eq:D5DBI+CS} of a single D5-brane 
in $Z_3$. We perform the dimensional reduction for the 
background of a D5-brane wrapping a holomorphic curve $\Sigma$. 
In a background with vanishing B-field and gauge flux $F$ the action \eqref{eq:D5DBI+CS} 
is just the volume of the wrapped curve. Thus, the variation of $S_{\text DBI}^{\text SF}$ 
under a deformation along $s$ is captured, up to second order in the variation parameter 
$\epsilon$, by \eqref{eq:volvarD5} and reads  
\begin{equation} \label{eq:VDBI}
 	V_{\text{DBI}}\supset \frac{\mu_5e^{3\phi}}{\mathcal{V}^2}\left.\frac{d^2}{d \epsilon^2}\text{Vol}(\Sigma_{\epsilon s})\right\vert_{\epsilon=0}=\frac{\mu_5e^{3\phi}}{2\mathcal{V}^2}\int_{\Sigma}\norm{\bar{\partial} s}^2\, \vol_{\Sigma}\,.
\end{equation} 
Here we used the formula \eqref{eq:volvarD5} and further a Weyl-rescaling to the 
four-dimensional Einstein-frame to obtain the right factors of the dilaton $\phi$ and 
the compactification volume $\mathcal{V}$.

In the following we deduce this potential from the $\mathcal{N}=1$ formulation of the 
D5-brane effective action. Indeed, the term \eqref{eq:VDBI} is an F-term potential of 
the form
\begin{equation}
	V_{\text F}=e^K K^{a\bar b}\partial_{u_a}W_{\rm brane}\partial_{\bar{u}_{\bar b}}\bar{W}_{\rm brane}
\label{eq:Ftermpot}
\end{equation}
for the fields $u^a(x)$ associated to the expansion $s=u^as_a$ in a basis of 
$\mathcal{C}^{\infty}(\Sigma,N_{Z_3}\Sigma)$. In order to evaluate $V_F$ we need the 
K\"ahler metric for the modes $u^a$ as well as a more tractable form of the brane 
superpotential $W_{\rm brane}$.  The K\"ahler metric for the $u^a$ as deformations in 
the infinite dimensional space $\mathcal{C}^{\infty}(\Sigma,N_{Z_3}\Sigma)$ is a 
straightforward generalization of the K\"ahler metric of \eqref{eq:kaehler-pot} originally 
considered for the modes counted by $H^0(\Sigma, N_{Z_3}\Sigma)$. It reads
\begin{equation}
	K_{a\bar b}=\frac{-i\mu_5e^\phi}{4\mathcal{V}}\int_{\Sigma}s_a\lrcorner \bar{s}_{\bar b}\lrcorner (J\wedge J)
	=\frac{i\mu_5e^\phi}{\int\Omega\wedge \bar\Omega}\int_{\Sigma}(\Omega_a)_{ij}(\bar{\Omega}_{\bar b})^{ij} \iota^*(J)\,,
\label{eq:Kahlermetric}
\end{equation}
where we used the abbreviation $\Omega_a=s_a\lrcorner \Omega$. For details of this 
equality we refer to appendix \ref{app:potcalc}. Firstly, we Taylor expand $W_{\rm brane}$ 
to second order in the brane deformations $u^a$ around the holomorphic curve $\Sigma=\Sigma_h$
\begin{equation} \label{eq:Wsecondorder}
	W_{\rm brane}=\int_{\Gamma_h}\Omega+\frac12u^au^b\int_{\Sigma}s_a\lrcorner ds_b\lrcorner \Omega+\mathcal{O}(u^3)
\end{equation}
where $s\lrcorner$ denotes the interior product with $s$ and $s_a$ denotes a section of 
$N_{Z_3}\Sigma$ that is not required to be holomorphic. $\Gamma_h$ is a chain ending on 
the holomorphic curve $\Sigma_h$, $\partial \Gamma_h=\Sigma_h-\Sigma_{0}$. Secondly, 
introducing the abbreviation $\Omega_a=s_a\lrcorner \Omega$ the variation of 
\eqref{eq:Wsecondorder} with respect to $u_a$ reads
\begin{equation}
 	\partial_{u_a}W_{\rm brane}=-\mu_5\int_{\Sigma}\bar\partial s\lrcorner \Omega_a\,.
\end{equation}  
In addition we rescaled the superpotential $W_{\rm brane}\mapsto \mu_5 W_{\rm brane}$ 
to restore physical units as before \cite{Grimm:2008dq}. In order to evaluate the contraction 
\eqref{eq:Ftermpot} we have to exploit that the $\Omega_a$ form a basis of sections of a 
specific bundle on $\Sigma$. Indeed, the isomorphism of 
$KZ_3\vert_{\Sigma}=T^* \Sigma\otimes N^*\Sigma$ which is a consequence of the normal 
bundle sequence of $\Sigma$ tells us that the $\Omega_a$ form a basis of sections of 
$\Omega^{(1,0)}(\Sigma,N^*\Sigma)$ with the property that $s_a\lrcorner \Omega_a=0$. We 
can use this basis to represent any other section. In particular, the contraction 
$\bar\partial s\lrcorner J$ is a section of $\Omega^{(0,1)}(\Sigma,\overline{N^*\Sigma})$ 
that we can expand in the basis $\bar{\Omega}_{\bar a}$ as
\begin{equation} \label{eq:basisexpansion1}
	\bar\partial s\lrcorner J=\frac{-i\mu_5e^\phi}{\int\Omega\wedge \bar\Omega}\,\Omega_a\,K^{a\bar b}
	\int_{\Sigma}\bar{\partial} s\lrcorner \bar\Omega_{\bar b}\,.
\end{equation} 
Again we refer to appendix \ref{app:potcalc} for the details of this calculation.
Finally, we calculate the F-term potential \eqref{eq:Ftermpot} as
\begin{eqnarray} \label{eq:Ftermpotfinal}
	V_{\rm F}=e^K\mu_5^2\int_{\Sigma}\bar\partial s\lrcorner \big(\Omega_a K^{a\bar b}
	\int_{\Sigma}\partial \bar{s}\lrcorner \bar{\Omega}_b\big)
	= \frac{\mu_5e^{3\phi}}{2\mathcal{V}^2}\int_{\Sigma}||\bar\partial s||^2 \vol_{\Sigma}\,.
\end{eqnarray}
Here we used in the second equality the identity \eqref{eq:basisexpansion1} as well as 
$e^K=\frac{ie^{4\phi}}{2\mathcal{V}^2\int\Omega\wedge\bar\Omega}$, cf.~appendix 
\ref{app:potcalc}. The norm $||\cdot||^2$ denotes as before the contraction of all indices 
using the metric. 

This F-term potential is in perfect agreement with contribution \eqref{eq:VDBI} to the scalar 
potential $V_{\rm DBI}$ that we obtain from the reduction of the DBI-action \eqref{eq:D5DBI+CS} 
using the variation \eqref{eq:volvarD5} of the calibrated volume.

\part{String Dualities}

\chapter[Heterotic/F-Theory Duality and Five-Brane Dynamics]{Heterotic/F-Theory Duality and \newline Five-Brane Dynamics}
\label{ch:HetFThyFiveBranes}

In this chapter we introduce heterotic/F-theory duality and its application to 
analyze non-perturbative heterotic string compactifications. In general, the 
strength of F-theory is the geometrization of physics, for the analysis of both 
non-perturbative Type IIB compactifications with seven-branes and of dual 
heterotic string compactifications with five-branes. We will mostly be interested
in four-dimensional $\mathcal{N}=1$ compactification, for which this geometrization
provides strong geometrical tools to study the coupling functions of the 
four-dimensional effective actions. We will put particular emphasis on the geometrization 
of branes in F-theory, both for the Type IIB theory and for the dual heterotic string,
which we will exploit explicitly in chapter \ref{ch:Calcs+Constructions} for concrete calculations
of the corresponding effective superpotentials.

In section \ref{sec:HetString+Fivebranes} we start with a discussion of the heterotic string,
its strongly coupled formulation as heterotic M-theory and its compactifications.
We first review heterotic M-theory compactifications on a Calabi-Yau threefold 
$Z_3\times S^1/\mathds{Z}_2$ to four-dimensions in the presence of spacetime-filling five-branes wrapped 
on a curve $\Sigma$. Then we briefly comment on the massless spectrum including the geometric moduli
and discuss in some detail the small instanton transition of a smooth bundle into a heterotic five-brane.
Then we present the full heterotic superpotential and briefly review the essential steps of
the spectral cover construction of heterotic vector bundles $E$ on elliptic Calabi-Yau threefolds $Z_3$.
Next in section \ref{sec:FTheoryCompactifications} we review the basics of F-theory. We start
with a brief discussion of Vafa's original motivation for F-theory before we proceed to a more elaborate
construction of F-theory vacua from elliptic Calabi-Yau manifolds. There we readily focus on 
four-dimensional compactifications on Calabi-Yau fourfolds and the induced effective Gukov-Vafa-Witten
flux superpotential. Finally we study heterotic/F-theory duality in section \ref{sec:HetFDuality}.
After reviewing the underlying eight-dimensional duality between F-theory on an elliptic $K3$ and
the heterotic string on $T^2$, we use the adiabatic argument to construct pairs of dual 
lower-dimensional $\mathcal{N}=1$ theories, in particular in four dimensions. We emphasize the duality 
map for the heterotic bundle $E$ and the heterotic five-branes, that are mapped, as we discuss in 
detail, either to three-branes or to blow-ups in F-theory.

\section{Heterotic String Compactifications with Five-Branes} 
\label{sec:HetString+Fivebranes}

In this section we review the construction of heterotic M-theory compactifications to four-dimensions.
As detailed in section \ref{sec:hetM} an $\cN=1$ compactification is determined, in the absence of 
fluxes, by a choice of a Calabi-Yau threefold $Z_3$, a holomorphic stable vector bundle
$E$ and a number of five-branes wrapped on curves $\Sigma$ in $Z_3$ that are located at points in the 
M-theory interval $S^1/\mathds{Z}_2$. We note that heterotic M-theory can be used to systematically 
calculate corrections to these backgrounds. Furthermore, we emphasize the importance of a cancellation 
of tadpoles that restricts the second Chern class of $E$ and the choice of five-brane curves $\Sigma$.
Then in section \ref{sec:hettransition} we present a brief account on the determination of the 
massless spectrum, including the geometrical and bundle moduli. We emphasize the relation of the
bundle moduli and five-brane moduli via a small instanton transition of the smooth heterotic bundle 
$E$ to a singular configuration, that is more appropriately described as a heterotic five-brane. 
Next in section \ref{sec:het_superpot} we discuss the perturbative heterotic superpotential,
that is a sum of the heterotic flux superpotential, the holomorphic Chern-Simons functional of
the bundle $E$ and a superpotential for heterotic five-branes, that takes the form of a chain integral. 
The existence of the latter is inferred from the effect of the small instanton transition on
the holomorphic Chern-Simons functional. We conclude in section \ref{sec:spectralcover} with a 
presentation of the essential steps in the spectral cover construction of heterotic vector bundles 
on elliptically fibered Calabi-Yau manifolds. There we summarize also some results for $E_8$-bundles, 
that have to be constructed by different means.

\subsection{Heterotic M-Theory Compactification}
\label{sec:hetM}

The formulation of the ten-dimensional $E_8\times E_8$ heterotic string theory at 
strong coupling $g_S$ is given by eleven-dimensional supergravity, which is an 
effective description of the low-energy interactions of the light modes of M-theory, 
on $\mathbb{R}^{1,9}\times S^1/\mathds{Z}_2$ \cite{Horava:1995qa}. The heterotic 
string coupling is related to the radius $R$ of $S^1$ in the same way the string 
coupling of Type IIA is related to $R$,
\begin{equation}
	g_S=R^{3/2}\,,
\label{eq:g_SRadiusRelation}
\end{equation}
which is deduced in \cite{Witten:1995ex} by the equivalence of the corresponding 
effective actions. We note that at small $R$ compared to the eleven-dimensional 
Planck length $\kappa^{2/9}$, the 
eleven-dimensional supergravity description breaks down due to large curvature
effects and the perturbative heterotic string description applies. 
However, in order to treat non-perturbative solitonic string effects in the 
heterotic theory, like NS5-branes, it is appropriate to consider the M-theory 
formulation which is denoted heterotic M-theory.

We begin by giving a definition of M-theory on the orbifold $S^1/\mathds{Z}_2\times 
\mathbb{R}^{1,9}$. Heterotic M-theory in eleven dimensions is known by its effective 
action. Its Lagrangian is systematically constructed in \cite{Horava:1996ma} as an 
expansion in the gravitational coupling $\kappa^{2/3}$ from eleven-dimensional 
supergravity in the bulk of $S^1/\mathds{Z}_2$ coupled to two ten-dimensional 
Super-Yang-Mills theories on the boundaries of $S^1/\mathds{Z}_2$. Here $\kappa$ and the 
gauge coupling $\lambda$ are related as 
\begin{equation}
	\lambda^2=2\pi(4\pi \kappa^2)^{2/3}\,.
\label{eq:kappaLambda}
\end{equation}
In this expansion the kinetic terms of the gauge bosons, proportional to 
$\frac{1}{\lambda^2}$, appear at order $\kappa^{2/3}$ relative to the gravitational term, 
that scales as $\frac{1}{\kappa^2}$. Furthermore, a consistent and gauge invariant effective 
theory requires the inclusion of Green-Schwarz terms \cite{Horava:1996ma} even in the minimal
Lagrangian. This is in contrast to the weakly coupled heterotic string where the Green-Schwarz
term is not present in the minimal supergravity and Super-Yang-Mills theory and only appears 
when quantum loops are taken into account, where it guarantees cancellation of gauge and 
gravitational anomalies. Thus, only to zeroth order in $\kappa^{2/3}$, i.e.~without the boundary 
theory, heterotic M-theory exists as a consistent classical theory and at any higher order 
in $\kappa^{2/3}$ quantum effects, that are one-loop effects like anomalies at first order, have to be 
included to guarantee consistency, which is consequently gauge invariance at first order \cite{Horava:1996ma}.
The use of $\kappa^{2/3}$ as an expansion parameter is further exploited in \cite{Witten:1996mz},
where a strong coupling expansion is applied to calculate corrections to the
perturbative heterotic string backgrounds with $\mathcal{N}=1$ spacetime supersymmetry found 
in \cite{Candelas:1985en}. These are defined by a Calabi-Yau threefold $Z_3$ and a 
stable holomorphic vector bundle $E$ and appear as lowest order backgrounds 
in the M-theory context as reviewed below\footnote{We note that one already obtains 
corrections within the heterotic string at weak coupling for non-standard embeddings 
\cite{Witten:1985bz,Witten:1986kg,greensuperstring}, see e.g.~\cite{Ovrut:1999xu}
for a review. See \cite{Nilles:1998uy} for a detailed discussion of phenomenological 
applications.}. 
The correction terms to these backgrounds are organized in a double expansion 
in dimensionless parameters \cite{Lukas:1997fg,Lukas:1998hk}
\begin{equation}
	\epsilon_s=\Big(\frac{\kappa}{4\pi}\Big)^{2/3}\frac{2\pi R}{\mathcal{V}^{2/3}}\,,
	\qquad \epsilon_R=\frac{\mathcal{V}^{1/6}}{\pi R}\,,
\label{eq:doublExp}
\end{equation}
which are roughly the compactification scale $\mathcal{V}^{1/6}$ and the interval size $R$.
This expansion is valid for all types of embeddings and in particular in the presence 
of M5-branes. Besides its conceptual relevance the crucial point of the M-theory description
and the strong coupling expansion is the supersymmetric treatment of 
a non-trivial Bianchi identity of the field strength $G_4$ of the three-form $C_3$ in 
eleven-dimensional supergravity. This is in general non-trivial, in particular 
in the presence of five-branes \cite{Witten:1996mz} that appear as singular sources of the Bianchi
identity. In addition, the heterotic M-theory formalism allows to treat the backreaction of
five-branes on the geometry and the background fields, in particular $G_4$, cf.~\cite{Lukas:1998hk} 
for an analysis to first order in $\epsilon_s$.

Let us review the construction of heterotic compactifications to four dimensions with
$\mathcal{N}=1$ supersymmetry in the M-theory context. This means that we consider the
background geometry $S^1/\mathds{Z}_2\times Z_3\times \mathbb{R}^{1,3}$. In order to
obtain a supersymmetric background the supersymmetry variations of all fermionic fields 
in the theory have to vanish identically in the background. To lowest order in $\epsilon_s$ 
this yields a Killing spinor equation from the eleven-dimensional gravitino which is solved 
by vanishing background flux $G_4=0$ and by $Z_3$ being a Calabi-Yau threefold 
\cite{Candelas:1985en}. Although a first order effect in $\epsilon_s$, one usually also 
includes the two $E_8$ gauge theories on the boundary in the construction of the 
heterotic background\footnote{In the weakly coupled heterotic string the $E_8\times E_8$
gauge symmetry is inferred from the requirement of modular invariance implying anomaly cancellation 
in the low-energy effective theory \cite{Candelas:1985en}.}. 
The supersymmetry variation of the ten-dimensional gauginos 
yields the condition on the VEV of the gauge-fields on $Z_3$, which can be 
non-vanishing without breaking four-dimensional Lorentz invariance. The 
gauge-fields have to define a vector bundle $E=E_1\oplus E_2$ with 
structure group being a subgroup of $E_8\times E_8$ such that the curvature 
$\mathcal F$ obeys
\begin{equation}
	\mathcal F_{ij}=\mathcal F_{\bar \imath\bar\jmath}=0\,,
	\qquad g^{i\bar \jmath}\mathcal F_{i\bar \jmath}=0\,.
\label{eq:DonUhlYau}
\end{equation} 
The first two conditions turn $E$ into a holomorphic vector bundle and the 
second condition is the Donaldson-Uhlenbeck-Yau equation. The latter condition 
requires the class of the field strength $\mathcal F$ to be primitive, 
i.e.~$g^{i\bar \jmath}\mathcal \mathcal F_{i\bar \jmath}
\sim \mathcal F\wedge *J\sim \mathcal{F}\wedge J^2=0$, which is a 
restrictive condition rendering a direct construction of a bundle $E$ obeying it 
technically challenging. A construction is possible only under particular circumstances, 
one of which is when $Z_3$ is elliptically fibered as will be discussed below in section 
\ref{sec:spectralcover}. However, the existence of $E$ obeying \eqref{eq:DonUhlYau} was 
proven by Donaldson for the rank two case in \cite{donaldson1985anti} and by Uhlenbeck 
and Yau in \cite{uhlenbeck1986existence} for arbitrary rank in the context of finding 
solutions to the Hermitian Yang-Mills equation,
\begin{equation}
	J^2\wedge \mathcal F=\frac{2\pi}{\mathcal{V}}\mu(E)\vol_{Z_3}\id\,,\qquad 
	\mu(E)=\frac{1}{\text{rk}(E)}\int_{Z_3}c_1(E)\wedge J^2\,,
\label{eq:HYM}
\end{equation}
where $\text{rk}(E)$ and $\mu(E)$ denote the rank and the slope of $E$, respectively.
Then the second condition in \eqref{eq:DonUhlYau}, $\mathcal F\wedge J^2=0$, implies the special 
case of zero slope $\mu(E)=0$. In general, \eqref{eq:HYM} has a unique solution 
if $E$ is slope-(semi-)stable with respect to $J$, i.e.~for any subbundle $V\subset E$ 
the condition
\begin{equation}
	\mu(V)\leq \mu(E)
\label{eq:semistable}
\end{equation} 
holds, or if $E$ is a direct sum of (semi-)stable bundles\footnote{We note that a stable
bundle requires $H^0(Z_3,E)\cong H^3(Z_3,E)=0$ to guarantee the absence of maps $\mathcal{O}
\rightarrow E$ \cite{Andreas:1998zf}.}. 
  
This setup is further constrained by the cancellation of tadpoles in heterotic 
M-theory. This condition is derived from the Bianchi identity of the field 
strength $G_4$ that takes the form \cite{Horava:1995qa,Witten:1996mz,Lukas:1998hk}
\begin{equation}
	dG_4=4\sqrt{2}\pi\Big(\frac{\kappa}{4\pi}\Big)^{\frac23}\left[J^{(0)}\delta(x^{11})
	+J^{(N+1)}\delta(x^{11}-\pi R)+\tfrac12\sum_{n=1}^N \delta_{\Sigma_i}(\delta(x^{11}-x_n)
	+\delta(x^{11}+x_n))\right]
\label{eq:HetMTadpole}
\end{equation}
where $\delta(x^{11}-x_n)$, $\delta_{\Sigma_i}$ denote delta-functions on $S^1/\mathds{Z}_2$, 
with coordinate $x^{11}$, respectively in $Z_3$, that are supported on curves $\Sigma_i$ wrapped by five-branes.
We note that a supersymmetric five-brane preserving four-dimensional Minkowski invariance 
localizes to a point $x_n$ in $S^1/\mathds{Z}_2$ and to a holomorphic curve in $Z_3$ \cite{Becker:1995kb}. 
The sources $J^{(0)}$, $J^{(N+1)}$ contribute boundary terms from the two ten-dimensional gauge
theories on the boundary,
\begin{equation}
	J^{(0)}=-\frac1{16\pi^2}\left.\Big(\text{tr}\, \mathcal F^2-\frac12\text{tr}\, R^2\Big)\right\vert_{x^{11}=0}\,,\qquad 
	J^{(N+1)}=-\frac1{16\pi^2}\left.\Big(\text{tr}\, \mathcal F^2-\frac12\text{tr}\, R^2\Big)\right\vert_{x^{11}=\pi R}\,,
\label{eq:trF2-trR2}
\end{equation}
where $R$ denotes the eleven-dimensional curvature. The symbol ``tr'' denotes the trace in the 
vector representation of $\text{O}(1,9)$ for $R$ and the $\frac1{30}$ of the trace in the adjoint 
of $E_8$ for the gauge field $\mathcal{F}$, respectively.
Whereas \eqref{eq:HetMTadpole} determines the actual form $G_4$ pointwise it can be evaluated
in the cohomology of $Z_3$ yielding the tadpole condition,
\begin{equation}
	\lambda(E_1)+\lambda(E_2)+\sum_i\left[\Sigma_i \right]=c_2(Z_3)\ ,
\label{eq:anomaly}
\end{equation} 
where $\lambda(E)$ is the fundamental characteristic class of the vector bundle
$E$, which, for example, is $c_2(E)$ for $SU(N)$ bundles and $c_2(E)/30$ for
$E_8$ bundles\footnote{Here we used the Chern characters as $\text{ch}_2(Z_3)=-c_2(Z_3)$, 
$\text{ch}_2(E)=-c_2(E)$ for a compactification on Calabi-Yau threefolds $Z_3$ assuming $c_1(E)=0$. 
In addition we ignore the numerical coefficients in \eqref{eq:HetMTadpole}.}. 
This tadpole requires the inclusion of gauge background bundles $E$ over $Z_3$ with 
structure group contained in the ten-dimensional heterotic gauge group \cite{Candelas:1985en}. 
Note that in many heterotic compactifications the inclusion of five-branes is not a choice, 
but rather a requirement for tadpole cancellation as demonstrated explicitly in the case
of elliptic threefolds $Z_3$ in \cite{Friedman:1997yq}.
The condition \eqref{eq:anomaly} then dictates consistent choices of the cohomology classes 
$\left[\Sigma_i \right]$ of the curve $\Sigma_i$ in the presence of a non-trivial 
vector bundle $E$ to match the curvature of the threefold $Z_3$ as 
measured by the second Chern class $c_2(Z_3)$.  In particular, it implies that 
$\Sigma_i$ corresponds to an effective class in $H_2(Z_3,\mathds{Z})$ \cite{Donagi:1999gc}.
We note that the sources in \eqref{eq:HetMTadpole} are of order $\epsilon_s$. Thus the 
solution $G_4=0$ above \eqref{eq:DonUhlYau} is consistent at zeroth order.
However, since the inclusion of five-branes and their backreaction is the central point in this work,
an appropriate treatment of the non-trivial Bianchi identity with localized sources will be 
essential as demonstrated in chapter \ref{ch:blowup}.

There is one further condition on the first Chern-class of $E$, sometimes denoted
as the K-theory constraint. It can be derived from the requirement of anomaly 
cancellation in the sigma-model world-sheet theory \cite{Witten:1985mj,Freed:1986zx} 
and reads
\begin{equation}
	c_1(E)=0\mod 2\,\,\quad\Leftrightarrow\,\,\quad c_1(E)\in H^{2}(Z_3,2\mathds{Z}).
\label{eq:KTheoryCond}
\end{equation}
Geometrically this conditions guarantees that the bundle $E$ admits spinors, since 
\eqref{eq:KTheoryCond} is equivalent to a trivial second Stiefel-Whitney class 
$w_2\in H^2(Z_3,\mathds{Z}_2)$ for holomorphic bundles.

\subsection{Charged Matter, Moduli and Small Instanton Transitions}
\label{sec:hettransition}

The low-energy effective theory obtained by compactification of the heterotic string
or heterotic M-theory is very roughly an $\mathcal{N}=1$ gauge theory with 
additional chiral multiplets \cite{Candelas:1985en}.
In the perturbative string all four-dimensional fields arise from the ten-dimensional
gauge theory and the moduli of the background geometry. The four-dimensional gauge 
symmetry $G$ arises from the ten-dimensional gauge symmetry as the commutant of the 
background bundle $E$ with structure group $H$ within $E_8\times E_8$. In the group-theoretical 
decomposition of the adjoint of one $E_8$ into representations of $G\times H$, 
\begin{equation}
	\mathbf{248}\,\,\rightarrow\,\,(\mathbf{adj}(G),\mathbf{1})\oplus 
	(\mathbf{1},\mathbf{adj}(H))\bigoplus_{i} (\mathbf{R}_i,\mathbf{Q}_i)\,,
\label{eq:grouptheoDecomp}
\end{equation}
this corresponds to the fields in the adjoint of $G$, $\mathbf{adj}(G)$.
Then charged chiral multiplets occur in the representations $\mathbf{R}_i$ under the 
four-dimensional gauge symmetry $G$. Their chirality $n(\mathbf{R}_i)$ in the four-dimensional 
theory is calculated by an index theorem for the Dirac-operator of the gauginos on $Z_3$ 
that take values in $\mathbf{Q}_i$ \cite{greensuperstring,Andreas:1998zf,Ovrut:1999xu},
\begin{equation}
	n(\mathbf{R}_i)=\chi(E_i)=\sum_n (-1)^nh^n(Z_3,E_i)=\int_{Z_3}\text{Td}(Z_3)\text{ch}(E_i)=\frac12\int_{Z_3}c_3(E_i)\,.
\label{eq:chirality}
\end{equation}
Here we assumed $c_1(E_i)=0$, which will be the case in the examples constructed later 
on, and introduced the bundle $E_i$ associated to the representation $\mathbf{Q}_i$.
To the index only $h^2(Z_3,E_i^*)=h^1(Z_3,E_i)$ by Serre duality contribute since 
$h^3(Z_3,E_i)=h^0(Z_3,E_i)=0$ for stable bundles. Indeed, the zero modes of the 
ten-dimensional gauginos are given by the cohomology groups $H^1(Z_3,E_i)$, 
$H^1(Z_3,E_i^*)$ \cite{greensuperstring}. We note that the charged fields are interpreted as matter 
fields in more phenomenological applications.
In addition there are neutral chiral multiplets. They correspond to the moduli 
of the gauge bundle $E$, that are related to the adjoint $\mathbf{adj}(H)$ in 
\eqref{eq:grouptheoDecomp} and are counted by $H^1(Z_3,\text{End}(E))$, and to the moduli of $Z_3$,
counted by $H^{(2,1)}(Z_3)$ and $H^{(1,1)}(Z_3)$. In phenomenological applications it is
the issue of moduli stabilization to fix these fields in order to avoid 
neutral massless scalars in four dimension.

In the non-perturbative heterotic string, there are additional fields in the low-energy effective
theory due to spacetime-filling five-branes. Assuming a single five-brane wrapping a holomorphic
curve $\Sigma_i$ of genus $g$ in $Z_3$ one obtains an additional gauge group $U(1)^g$, a
universal sector of a chiral multiplet with bosonic fields $(x^{11},a)$ and a number
of chiral multiplets for the deformations of $\Sigma_i$ inside $Z_3$. Here the gauge theory 
as well as the scalar $a$ arise from the reduction of the self-dual two-form by the $2g$ 
one-forms respectively the harmonic volume form on the curve $\Sigma_i$ wrapped
by the five-brane \cite{Witten:1997sc,Klemm:1996bj}, whereas $x^{11}$ denotes the position of the five-brane
in the interval $S^1/\mathds{Z}_2$. In the following we neglect additional massless states from 
intersections of five-branes as well as the gauge enhancement to $U(N)^g$ in the case of multiple
M5-branes, see e.g.~\cite{Lukas:1998hk,Ovrut:1999xu} for a detailed discussion.

In summary, the analysis of the moduli space of the heterotic string on $Z_3$ requires the
study of three a priori  very different pieces.  Firstly, we have the geometric
moduli spaces of the threefold $Z_3$ consisting of the complex structure as
well as the K\"ahler moduli space.  Secondly, there are the moduli of the
bundle $E$ which parametrize different gauge-field backgrounds on
$Z_3$. Finally, if five-branes are wrapped on non-rigid curves $\Sigma_i$, the
deformations of $\Sigma_i$ within $Z_3$ of the various five-branes have to be taken
into account.  Thus, the global moduli space is in general very complicated since it
is a fibration of the three different individual moduli spaces.  This problem, however, 
becomes more tractable if one focuses Calabi-Yau threefolds $Z_3$ admitting additional 
structure, like e.g.~an elliptic fibration. 
In this case it was shown in \cite{Friedman:1997yq} that there exist elegant constructions 
of stable holomorphic vector bundles $E$, as we review in section \ref{sec:spectralcover}. 
Moreover, the moduli space of five-branes on elliptically fibered $Z_3$ has been discussed 
in great detail in \cite{Donagi:1999jp}. Qualitatively it is generically true, however, that 
the heterotic moduli space on a fixed Calabi-Yau threefold $Z_3$ admits several different 
branches corresponding to the number and type of five-branes present as well as to the 
topology of the bundle $E$. There are distinguished points in this moduli space 
corresponding to enhanced gauge symmetry \cite{Witten:1995gx,Ganor:1996mu} of the heterotic 
string that allow for a clear physical interpretation and that we now discuss in more detail. 
It will turn out that at these points an interesting transition happens where a
five-brane completely dissolves into a finite size instanton of the bundle $E$
and vice versa. Thus, this transition connects different branches of the heterotic 
moduli space with a different number of five-branes and in particular with different topological
type of the vector bundle $E$ as the second Chern-class $c_2(E)$ will jump in this process.
Consequently this can be viewed as a heterotic extension of the familiar topology changes
in Type II string compactifications that connect Calabi-Yau threefolds $Z_3$ of different 
Euler characteristic and  Hodge numbers \cite{Greene:1995hu}, see 
e.g.~\cite{Greene:1996cy,Hori:2003ic} for a review.

Let us discuss, following \cite{Grimm:2009sy}, this small instanton transition by starting 
with a Calabi-Yau threefold $Z_3$ with $c_2(Z_3)\neq 0$ and no five-branes.
Thus, the anomaly condition \eqref{eq:anomaly} forces us to turn on a
background bundle $E$ with non-trivial second Chern class $c_2(E)$ in order to
cancel $c_2(Z_3)$. Then the bundle is topologically non-trivial and carries
bundle instantons characterized by the topological second Chern number
\cite{Witten:1985bz}
\begin{equation}
 	\left[c_2\right]=-\int_{Z_3}J\wedge \mathcal F\wedge \mathcal F\,,
\end{equation}
where $J$ denotes the K\"ahler form on $Z_3$ and $\mathcal F$ the field strength 
of the background bundle. The heterotic gauge group $G$ in
four dimensions is generically broken and given by the commutant of the
holonomy group of $E$ in $E_8\times E_8$. Varying the moduli of 
$E$ one can ask whether it is possible to restore parts or all of the broken gauge 
symmetry by flattening out the bundle as much as possible~\cite{Aspinwall:1996mn}. 
To show how this can be achieved, one first decomposes $c_2(E)$ into its 
irreducible components each of which being dual to an irreducible curve $[\Sigma_i]$ 
in $Z_3$ modulo chains. Since the invariant $\left[c_2\right]$ has to be kept fixed, 
the best we can do is to consecutively split off the components of $c_2(E)$ and to 
localize the curvature form of $E$ on the corresponding curves $\Sigma_i$, i.e.~the 
curvature becomes singular with support in $\Sigma_i$. This should be contrasted
with the generic situation, where the curvature form is smooth on $Z_3$ and 
meets the condition $[\Sigma_i] \subset c_2(E)$ only in cohomology, i.e.~up to exact 
forms. In this localization limit the holonomy of $E$ around each individual
curve $\Sigma_i$ becomes trivial\footnote{The curvature is flat away from $\Sigma_i$
and only monodromy effects persist.} and the gauge group $G$ enhances accordingly.
Having reached this so-called small instanton configuration at the boundary of
the moduli space of the bundle, the dynamics of (this part of) the gauge bundle
can be effectively described by a five-brane on $\Sigma_i$ \cite{Witten:1995gx}.
In particular since $E$ is holomorphic, the corresponding curves $\Sigma_i$ are 
holomorphic in $Z_3$ and the wrapped five-brane is supersymmetric. 

Small instanton configurations thus allow for supersymmetric transitions between
branches of the moduli space with different numbers of five-branes and topologies of $E$, that
consequently map bundle moduli to five-brane moduli and vice versa
\cite{Buchbinder:2002ji}. Note that this transition is completely consistent with
\eqref{eq:anomaly} since we have just shifted irreducible components between the two 
summands $c_2(E)$ and $\left[\Sigma_i \right]$. Thus, we are in the following allowed to 
think about the small instanton configuration of $E$ as a five-brane. In
particular, doing this transition for all components of $c_2(E)$ the full
perturbative heterotic gauge group $E_8\times E_8$ can be restored.  Turning
this argument around, a heterotic string with full perturbative $E_8\times E_8$ gauge
symmetry on a threefold $Z_3$ with non-trivial $c_2(Z_3)$ has to contain
five-branes to cancel the anomaly according to \eqref{eq:anomaly}.  We mention that
the small instanton transition will be of particular importance once we are working 
in the framework of heterotic/F-theory duality, cf.~chapter \ref{ch:Calcs+Constructions}.
In particular in our concrete examples of section \ref{sec:SuperpotsHetF} we will precisely 
encounter a situation with full perturbative $E_8\times E_8$ gauge symmetry, where the 
complete heterotic anomaly is canceled by five-branes, which will guide us to the interpretation 
of the F-theory flux superpotential in terms of a five-brane superpotential for a particular class 
of heterotic five-branes.

\subsection{The Heterotic Superpotential}
\label{sec:het_superpot}

In the following we discuss the perturbative superpotential for the heterotic string. 
It consists of three parts, the heterotic flux-superpotential induced by a possible 
flux $H_3$ of the B-field, the holomorphic Chern-Simons superpotential for the bundle 
$E$ and the superpotential induced by five-branes, which will be of particular importance
for this work. The three superpotentials can formally be unified as follows. We reduce the 
identity \eqref{eq:HetMTadpole} on the interval $S^1/\mathds{Z}_2$ to obtain the Bianchi 
identity for $H_3$ on $Z_3\times R^{1,3}$ as\footnote{In addition we set the numerical 
constants as well as $\kappa$ to unity in \eqref{eq:HetMTadpole}.} 
\begin{equation} \label{eq:dH3}
   dH_3 = \tr (R^2) - \tr (\cF^2) + \sum_i \delta_{\Sigma_i}\ .
\end{equation} 
Then the superpotential can be expressed in terms of $H_3$ and reads, motivated by 
\cite{Gukov:1999gr, Behrndt:2000zh},
\begin{equation}
\label{eq:hetallpots}
 	W_{\rm het}=\int_{Z_3}\Omega\wedge H_3= W_{\rm flux} + W_{\rm CS} + W_{\rm brane}\ ,
\end{equation}
where the different terms are associated to the various contributions in $H_3$
in \eqref{eq:dH3}\footnote{According to the heterotic Bianchi identity, a term involving
the Chern-Simons form $\omega^{L}_3(R)$ should be present. However, to our knowledge such
a term is not discussed in the literature.}. Since a more rigorous treatment to obtain the 
superpotentials on the right in \eqref{eq:hetallpots} will be demonstrated 
in section \ref{sec:N=1gensection} using the notion of currents, we will just 
list them in the following.

The superpotential $W_{\rm flux}$ is due to background fluxes $H_3$.
In general a non-trivial background flux $H_3$ has to be in $H^3(Z_3,\mathds{Z})$ due 
to the flux quantization condition. It is expanded as $H_3=N^K\alpha_K-M_i\beta^i$ 
in the integral basis $\alpha_K$, $\beta^K$ of $H^3(Z_3,\mathds{Z})$ with integer 
flux numbers $N^i$, $M_i$. Then the flux superpotential \cite{Gukov:1999ya} takes the form
\begin{equation}
\label{eq:fluxpot}
 	W_{\rm{flux}}=\int_{Z_3}\Omega \wedge H_3  = M_KX^K-N^K\mathcal{F}_K\,,
\end{equation}
where we used the period expansion $\Omega=X^K\alpha_K-\mathcal{F}_K\beta^K$, 
cf.~\eqref{eq:periodexpansionOrie}. The superpotential $W_{\rm flux}$ is in general
a complicated function on the complex structure moduli of $Z_3$. Its complete moduli 
dependence is encoded in the periods $(\underline{X},\underline{\mathcal{F}})$. 
In order to analyze their dependence one benefits from the techniques of algebraic 
geometry that are applicable for a wide range of examples, see~\cite{Polchinski:1995sm,Taylor:1999ii,Mayr:2000hh,Curio:2000sc} 
and \cite{Douglas:2006es,Blumenhagen:2006ci,Denef:2008wq} for reviews.  This is due to the fact that
$(\underline{X},\underline{\mathcal{F}})$ obey differential equations, the so-called 
Picard-Fuchs equations, 
that can be determined and solved explicitly and thus allow to fix the complete moduli 
dependence of $W_{\rm{flux}}$ once the flux numbers are given. We will review these
techniques in section \ref{sec:CSModuliSpace+PFO}.
To end our discussion of the flux superpotential, let us stress that strictly speaking we have excluded
background fluxes in section \ref{sec:hetM}. Indeed, there is a 
back-reaction of $H_3$ which renders $Z_3$ to be non-K\"ahler \cite{Strominger:1986uh}. Since 
our main focus will be on the five-brane superpotential, we will not be concerned with this 
back-reaction for the moment. However, we will come back to this point in chapter \ref{ch:su3structur} 
and present a more rigorously treatment of the flux $H_3$ in the context of a compactification on a 
manifold with SU(3)-structure.

The second term in \eqref{eq:hetallpots} denotes the heterotic holomorphic Chern-Simons 
functional \cite{Witten:1985bz}\footnote{The same functional was derived in \cite{Witten:1992fb} as
the superpotential for B-branes wrapping the entire Calabi-Yau threefold $Z_3$.}
\begin{equation} \label{eq:hCS}
	W_{\rm CS}=\int_{Z_3}\Omega\wedge(A\wedge\bar\partial A+\frac{2}{3}A\wedge A\wedge A)
\end{equation}
by inserting the Chern-Simons form $\omega_3^{\rm YM}$ for $H_3$ in \eqref{eq:hetallpots}. The functional
$W_{\rm CS}$ depends on both the complex structure moduli of $Z_3$ and the bundle moduli of $E$ through 
the connection $A$. As can be readily shown its extremal points are the holomorphic vector bundles $E$ 
over $Z_3$ of the same topological type, cf.~\cite{Thomas:2001ve}. 

Finally the third term comprises the five-brane superpotential. The brane superpotential has the 
following properties. It depends holomorphically on the complex structure moduli 
of $Z_3$, as well as on the (obstructed) deformations corresponding to holomorphic sections of 
$N_{Z_3}\Sigma$.\footnote{More precisely, denoting by $u$ an open string deformation we expect a superpotential 
$W_{\rm brane}=u_a^{n+1}$ if the deformation along the direction $s^a$ is obstructed at order $n$ 
\cite{Kachru:2000ih,Kachru:2000an}.} Furthermore, the F-term supersymmetry conditions of $W_{\rm brane}$, 
i.e.~its critical points, correspond to holomorphic curves. This determines $W_{\rm brane}$ uniquely up to a constant 
as done in \cite{Witten:1997ep} for M-theory on a Calabi-Yau threefold $Z_3$ 
with a spacetime-filling M5-brane wrapped on a curve $\Sigma$,
\begin{equation}
\label{eq:chainMtheory}
 	W_{\rm brane}=\int_{\Gamma(\underline{u})}\Omega(\underline{z})\,.
\end{equation}
Here $\Gamma(\underline{u})$ denotes a three-chain bounded by the deformed curve $\Sigma_{\underline{u}}$ 
and a reference curve $\Sigma_{0}$ that is in the same homology class, i.e.~$\partial \Gamma(\underline{u})=\Sigma_{\underline{u}}-\Sigma_0$. 
It depends on both the deformation $\underline{u}$ of the five-brane on $\Sigma$ as well as the complex 
structure moduli $\underline{z}$ of $Z_3$ due to the holomorphic 
three-form $\Omega$. We note that the superpotential \eqref{eq:chainMtheory} is formally identical to the result obtained
by dimensional reduction of the D5-brane action \cite{Grimm:2008dq}, as demonstrated in section \ref{sec:suppot}, and can 
directly be obtained from the Type IIB result via dualities. 
A third and more rigorous way to obtain and treat the chain integral \eqref{eq:chainMtheory} will be presented 
in section \ref{sec:fivebranesuperpotential} using the language of currents. 

We conclude by investigating the consequences of the small instanton 
transition of section \ref{sec:hettransition} for the heterotic 
superpotentials \eqref{eq:hetallpots}. The small instanton transition implies 
a transition between bundle and five-brane moduli~\cite{Buchbinder:2002ji}. 
However, also away from the supersymmetric configuration of a holomorphic 
vector bundle and a five-brane on a holomorphic curve, the transition applies.
Then the corresponding obstructed deformation fields of the non-holomorphic
vector bundle and curve should be identified.
Since both types of deformations are generally obstructed by a superpotential, 
cf.~section \ref{sec:fivebranes}, also the superpotentials for bundle and 
five-brane have to be connected by the transition. 
To see how the two superpotentials \eqref{eq:hCS} and \eqref{eq:chainMtheory} are
mapped onto each other in the transition, let us assume a single instanton
solution $\mathcal{F}$ with $\mathcal{F}\wedge \mathcal{F}$ dual to an
irreducible curve $\Sigma$. Displaying the explicit moduli dependence of the
configuration $\mathcal{F}$ \cite{Tong:2005un},  in the small instanton limit
$\mathcal{F}\wedge \mathcal{F}$ reduces to the delta function $\delta_{\Sigma}$
of four real scalar parameters. They describe the position moduli of the instanton 
normal to the curve in the class $\left[\Sigma\right]$ on which it is localized. 
Inserting the gauge configuration $\mathcal{F}$ into $W_{\text{CS}}$, the holomorphic 
Chern-Simons functional is effectively dimensionally reduced to the curve $\Sigma$, 
see \cite{Aganagic:2000gs} for a similar argumentation in the B-model. In the
vicinity of $\Sigma$ we may write the holomorphic three-form as $\Omega=d\omega$
which we insert into \eqref{eq:hCS} in the background
$\mathcal{F}\wedge\mathcal{F}$ to obtain, after a partial integration,
\begin{equation}
 	W_{\rm{CS}}=\int_{\Sigma}\omega\,.
\end{equation}
Adding a constant given by the integral of
$\omega$ over the reference curve $\Sigma_{0}$ this precisely matches the chain
integral \eqref{eq:chainMtheory}.  Applying the above discussion, we can think about
the M5-brane deformations in $W_{\rm{M5}}$ as the bundle deformations describing the
position of the instanton configuration $\mathcal{F}$.

We will verify this matching explicitly from the perspective of the
F-theory dual setup later on. There we will on the one hand identify some of
the fourfold complex structure moduli with the heterotic bundle moduli, on the
other hand, however, show that part of the F-theory flux superpotential
depending on the same complex structure moduli really calculates the
superpotential of a five-brane on a curve. This way, employing
heterotic/F-theory duality, we show in the case of an example the equivalence
of the small instanton/five-brane picture.

\subsection{Spectral Cover Construction}
\label{sec:spectralcover}

In this section we present a basic account on the construction of vector bundles
on elliptic Calabi-Yau manifolds. Although the constructions we discuss are valid
in any complex dimension, we directly focus to the most relevant case of Calabi-Yau
threefolds. Instead of delving into the mathematical details of the construction of
\cite{donagi1997principal,Friedman:1997ih,Friedman:1997yq}, we will just focus on the general idea of the
spectral cover approach and on giving the formulas of the Chern classes of $E$.
These are essential for the calculations performed in section \ref{sec:SuperpotsHetF}.
We will however restrict to the case of $SU(N)$ bundles and $E_8$-bundles, where
the construction of the latter using either parabolics or by embedding into $dP_9$ 
\cite{Friedman:1997yq} will not be discussed for brevity\footnote{The method of embedding
into del Pezzo surfaces allows the construction of bundles with exceptional structure group 
$E_{6,7,8}$, whereas the construction via parabolics applies to ADE-bundles \cite{Friedman:1997yq}.}, 
see e.g.~\cite{Andreas:1998zf} for a review.

The basic strategy of the spectral cover is to obtain stable holomorphic
bundles $E$ on an elliptic threefold $Z_3$ roughly speaking by fibering the
stable bundles on the fiber torus so that they globally fit into a stable
bundle on the threefold $Z_3$ \cite{donagi1997principal,Friedman:1997yq}. This way, the topological
data of the bundle $E$ can be determined in terms of the cohomology of the
two-dimensional base $B_2$ and the section of the elliptic fibration. 
More precisely one first defines a stable bundle on each 
elliptic fiber of $Z_3$, which is, if we focus to the case of an $SU(N)$-bundle, specified 
by $N$ line bundles on $T^2$, or, in the dual picture, $N$ points $Q_i$ on the 
dual torus \cite{Friedman:1997yq} obeying $\sum_i Q_i=0$\footnote{Strictly speaking, $0$ should 
be replaced by $p$ which is the marked point on $T^2$.}. Fibering these $N$ points over $B_2$ 
specifies a divisor in $Z_3$ that is an $N$-fold ramified cover of the base $B_2$, called the spectral cover
divisor or spectral cover for short. Concretely, for an elliptic Calabi-Yau threefold $Z_3\rightarrow B_2$ with base manifold
$B_2$ the Weierstrass form\footnote{This
projectivization of the affine Weierstrass equation corresponds to $\P^2(1,2,3)[6]$.} reads
\begin{equation} \label{eq:def-p0}
  p_0 =  y^2 + x^3 + f x z^4 + g z^6
\end{equation}
in an ambient space $\P(O_{B_2}\oplus \mathcal{L}^2\oplus\mathcal{L}^3)$ for $\mathcal{L}=K_{B_2}$\footnote{We 
refer to appendix \ref{app:chernEllipticCY} for details on elliptic Calabi-Yau manifolds.}. The data of the $N$ 
points on each elliptic fiber is specified by the zeros of the section \cite{Friedman:1997yq,Berglund:1998ej}
\begin{equation} \label{eq:def-p+}
 \mathcal{C}\,:\quad p_+ = b_0 z^N + b_2 x z^{N-2} + b_3 y z^{N-3} + \ldots + \left\{ \begin{array}{l} b_N x^{N/2}\\ b_N y x^{(N-3)/2}\end{array} \right. \ ,
\end{equation}
where one distinguishes the cases $N$ even and $N$ odd. Here the coefficients in $p_+$ are chosen such that it
defines a section of $\mathcal{O}(\sigma)^N\otimes\mathcal{M}$, where $\sigma:B_2\rightarrow Z_3$ denotes the section of the elliptic
fibration and $\mathcal{M}$ an arbitrary line bundle on $B_2$ with first Chern class $c_1(\mathcal{M})=\eta$. 
Here and in the following we furthermore use the notation $\sigma=c_1(\mathcal{O(\sigma)})$ which is the Poincar\'e dual
to the section $\sigma$.
The coefficients $b_i$ are sections of a line bundle on $B_2$. In general, they depend on moduli fields $u_i$
which encode the deformations of the spectral cover and, hence, the bundle $E$.

Having defined the spectral cover divisor $\mathcal{C}$ the vector bundle $E$ is obtained by a sequence of formal steps, 
that we only mention and refer to \cite{Friedman:1997yq,Andreas:1998zf} for details and a review. One constructs the 
so-called Poincare line bundle $\mathcal{P}_B$ over the fiber product $Z_3\times_{B_2}Z_3$, in which the subspace 
$Y=\mathcal{C}\times_{B_2} Z_3$ is canonically embedded and on which $\mathcal{P}_B$ is trivially defined by restriction. 
Roughly speaking the first factor $Z_3$ contains the moduli space of $E$ whereas the second factor $Z_3$ is the space over which $E$ 
will be constructed. Concretely, $E$ is given by
\begin{equation}
	E=(\pi_2)_*(\pi^*_1\mathcal{N}\otimes \mathcal{P}_B)
\label{eq:spectralcoverConstr}
\end{equation} 
where $\mathcal{N}$ is a for the moment arbitrary line bundle on $\mathcal{C}$ and $\pi_1:\,Y\rightarrow \mathcal{C}$
as well as $\pi_2:\, Y\rightarrow Z_3$ denote the two canonical projections. We note that the push-forward $\pi_2$ maps 
a line bundle on $Y$ at a generic point on $Z_3$
to a rank $N$ vector bundle $E$, that degenerates on the ramification divisor of the spectral cover $\pi:\,\mathcal{C}\rightarrow B_2$.
The requirement of vanishing first Chern class $c_1(E)$ then fixes the line bundle $\mathcal{N}$ as
\begin{equation}
	c_1(\mathcal{N})=-\frac12\big(c_1(\mathcal{C})-\pi^*(c_1(B_2))\big)+\gamma=\frac12(N\sigma+\eta+c_1(B_2))+\gamma\,,
\label{eq:etaChernclass}
\end{equation}
where $\gamma$ is a class in $H^{(1,1)}(\mathcal{C},\Z)$ obeying $\pi_*(\gamma)=0$. Thus $\mathcal{N}$ is not completely
fixed and the freedom in its definition is encoded by the class\footnote{In the approach via parabolics, the class $\gamma$
is set to zero by construction \cite{Friedman:1997yq,Andreas:1998zf}.} $\gamma$.
The most important achievement of this construction for our purposes is that it allows the determination of the Chern classes of 
$E$. One readily calculates the second Chern class $c_2(E)$ using the Hirzebruch-Grothendieck-Riemann-Roch theorem to obtain 
\cite{Friedman:1997yq}
\begin{equation}
	c_2(E)=\eta\sigma-\frac{1}{24}c_1(\mathcal{L})^2(N^3-N)-\frac{N}8\eta(\eta-N c_1(\mathcal{L})-\frac12\pi_*(\gamma^2)\,,
\label{eq:SU(N)secondChernclass}
\end{equation}
where we recall from appendix \ref{app:chernEllipticCY} that $c_1(B_2)=c_1(\mathcal{L})$. We note that nothing prevents us from 
setting $N=1$, which as a bundle with structure group $SU(1)$ seems not to make sense. However, 
as we will discuss and exploit in section \ref{sec:SuperpotsHetF} a bundle with $SU(1)$ structure 
group has to be interpreted as a horizontal five-brane in the heterotic theory \cite{Friedman:1997yq,Berglund:1998ej}.

Another case of interest is an $E_8$-bundle. We omit its construction and just state its second 
Chern-class that reads
\begin{equation}
	\lambda(E)=\frac{c_2(E)}{60}=\eta_i\sigma-15\eta^2+135\eta c_1(B_2)-310c_1(B_2)^2\,,
	\label{eq:lambda-e8}
\end{equation} 
where the notation of the $SU(N)$ case still applies.
Thus, we conceptually summarize the second Chern class $c_2(E)$ of the two bundles $E$ under 
consideration schematically as
\begin{equation}
	\lambda(E)=\eta\sigma+\pi^*(\omega)\,,
\label{eq:c2E}
\end{equation}  
where $\eta$ and $\omega$ are classes in $H^2(B_2,\mathds{Z})$ respectively $\sigma\cdot H^2(B_2,\Z)$. 
The meaning of the class $\eta$ will be physically clarified in heterotic/F-theory duality, where it 
can be constructed entirely from the fibration data of the base $B_3\rightarrow B_2$ of the dual F-theory,
see section \ref{sec:het-Fdual}.

\section{F-Theory Compactifications} 
\label{sec:FTheoryCompactifications}

Here we present a basic account on F-theory. We start in section \ref{sec:FtheoryConstruction}
by reviewing the original idea underlying F-theory which is the geometrization of the axio-dilaton
and the $SL(2,\Z)$-invariance of Type IIB via an auxiliary two-torus, that is allowed
to vary over spacetime defining an elliptic fibration. The dynamics of seven-branes is then encoded 
in the degeneration loci and type of this elliptic fibration. In section \ref{sec:ellFourfolds}
we proceed to construct F-theory compactifications to lower dimension by specifying an elliptic 
Calabi-Yau manifold. We readily focus on four-dimensional compactifications on Calabi-Yau fourfolds
and comment on consistency conditions from tadpole cancellation. Furthermore we discuss flux 
quantization on fourfolds and the form of allowed four-flux $G_4$ in F-theory. In addition we present 
a brief identification of the geometric F-theory moduli with the Type IIB fields and moduli. 
Finally in section \ref{sec:F-flux_sup} we discuss in some detail the structure of the F-theory
flux superpotential as encoded, for a fixed four-flux $G_4$, by the fourfold periods and 
comment on its splitting into flux and seven-brane superpotential of the underlying Type IIB theory.

\subsection{Basics of F-theory Constructions}
\label{sec:FtheoryConstruction}

F-theory provides a geometrization of $\mathcal{N}=1$ Type IIB backgrounds with backreacted
seven-branes and a holomorphically varying axio-dilaton \cite{Vafa:1996xn}
\begin{equation}
	\tau=C_0+\frac{i}{g_s}=C_0+i e^{-\phi}
\label{eq:tau}
\end{equation}
Due to the genuine description of non-perturbative effects like $(p,q)$-strings and string junctions
rich gauge dynamics can be obtained allowing even for exceptional groups $E_k$ for $k>6$ 
\cite{Gaberdiel:1997ud,Gaberdiel:1998mv}, that have been accessible before only in the heterotic theory.   

Starting with the simplest setup of a backreacted D7-brane in flat space it is the basic question of F-theory 
to find a compact geometry, which is a solution to the equations of motion of Type IIB string theory, 
in particular the Einstein equations, and that extends the local solution of a single D7-brane. 
This compact solution is found in \cite{Greene:1989ya} in the context of stringy cosmic strings and applied to D7-branes in 
the seminal work \cite{Vafa:1996xn}. The D7-brane is a magnetic source of the axio-dilaton $\tau$ 
through the R--R-form $C_0$ and a source of gravity. Consequently a solution to the Type IIB
effective action is determined by a solution of $\tau$ and the metric $g$, which for symmetry
reasons is of the form of a direct product $\mathds{R}^{1,7}\times B_1$ where the space $B_1$ transverse to
the D7-brane worldvolume is determined in the following. The solution of $\tau$ is given in complex
coordinates on $B_1$ by
\begin{equation}
	\tau(z)\sim\frac{1}{2\pi i}\log(z)
\label{eq:D7monodromy}
\end{equation} 
which is consistent with the monodromy $C_0\mapsto C_0+1$ due to the $C_0$-charge of the D7-brane.
The holomorphicity of $\tau$ is a consequence of a BPS-condition implying that half of the supersymmetries
are preserved \cite{Vafa:1996xn}. The metric $g$ is found to describe a conical space centered at $z=0$ 
with a deficit angle of $\pi/6$ \cite{Greene:1989ya, Vafa:1996xn}. It is crucial to note that the 
$SL(2,\mathds{Z})$-symmetry of Type IIB acting on $\tau$ renders the energy of the solution \eqref{eq:D7monodromy} 
finite\footnote{The domain of integration is reduced from the complex plane to $\mathcal{F}$, the fundamental 
domain of the torus.}. The $SL(2,\mathds{Z})$-action in particular implies, that $\tau(z)$ is not a well-defined
function on $B_1$, but a section of an $SL(2,\mathds{Z})$-bundle over $B_1$. More invariantly, $\tau$ can
be described by the modular parameter of an elliptic curve $\mathcal{E}$, i.e.~a two-torus, that is fibered holomorphically 
over $B_1$. To obtain a space $B_1$ of finite volume one considers a multicenter solution of 24 D7-branes 
yielding a deficit angle of $4\pi$, i.e.~$B_1$ curls itself up to form the compact space of 
$S^2\cong \mathds{P}^1$. Physically this is consistent since the metric is well-defined around $1/z$ and 
the net D7-charge on the $S^2$ is zero since a loop encircling all 24 D7-branes is contractible to a point
yielding a trivial monodromy for $C_0$. Furthermore, this already indicates that we are no more 
allowed to think of 24 D7-branes since the total monodromy and charge are zero. The resolution of this paradox 
is again the $SL(2,\mathds{Z})$-invariance of the solution that allows for more general seven-branes, denoted 
$(p,q)$ 7-branes\footnote{A $(p,q)$7-brane is an object on which a $(p,q)$-string can end, cf.~sections 
\ref{sec:specOrie} and \ref{sec:DbranesinCY3Orie}.}, having a different monodromy and charge. In particular a 
seven-brane is only specified up to its conjugacy class under $SL(2,\mathds{Z})$. We note that $\tau(z)$ varies 
holomorphically over $\mathds{P}^1$ where $g_s$ is not necessarily small and diverges, $\tau \rightarrow i\infty$, 
at the location of a seven-brane. 
This and the presence of non-perturbative seven-branes indicates that this eight-dimensional Type IIB vacuum 
is non-perturbative. It is denoted as an F-theory compactification to eight dimensions.

Geometrically the F-theory setup is described by a fibration of an elliptic curve over $B_1=\mathds{P}^1$
where the generic elliptic fiber $\mathcal{E}$ degenerates at the 24 loci of the seven-branes. However, the total space
of the fibration remains smooth. A smooth complex surface which is an elliptic fibration over $\mathds{P}^1$
is a two-dimensional Calabi-Yau manifold, which is $K3$. This can be seen as follows. First we note that 
every two-torus $T^2$ is algebraic. Indeed by means of the Weierstrass $\mathfrak{p}$-function, every point $u$ on the 
lattice quotient $T^2=\mathds{C}/L$ is mapped bijectively to the projective plane curve $\mathcal{E}=\{y^2=4x^3-g_2x-g_3\}$ via 
$u\mapsto(\mathfrak{p}(u),\mathfrak{p}'(u))\equiv (x,y)$, due to the algebraic differential equation obeyed 
by $\mathfrak{p}$ \cite{freitag2006funktionentheorie}. In a fibration, the curve $\mathcal{E}_z$ depends on the point $z$ 
in $\mathds{P}^1$. Then the discriminant $\Delta=g_2^3-27g_3^2$ of $\mathcal{E}_z$ has to vanish to first order at 24 points, 
which fixes the degree of $g_2$, $g_3$ as polynomials in the local coordinate $z$ on $\mathds{P}^1$ to be eight
and twelve so that the total space of the fibration, denoted as $X_2$, obeys the Calabi-Yau condition.  
Thus, we see that an F-theory vacuum in eight dimensions is in one-to-one correspondence with an elliptic $K3$-surface. 
Analogously, lower-dimensional F-theory vacua are obtained using the adiabatic argument \cite{Vafa:1995gm} by 
compactifying on $n$-dimensional elliptically fibered Calabi-Yau manifolds $X_n$, where the base $B_1$ is 
replaced by a complex $n-1$-dimensional Fano variety $B_{n-1}$ 
\cite{Vafa:1996xn,Morrison:1996na,Morrison:1996pp}. This explains the relevance of elliptically fibered Calabi-Yau 
manifolds as discussed next in section \ref{sec:ellFourfolds}. In all these cases, the relation of the F-theory setup
specified by the elliptic Calabi-Yau manifold $X_n$ to the Type IIB physics is made precise in \cite{Sen:1996vd,Sen:1997gv} 
by identifying the Type IIB manifold as the double cover of the base $B_{n-1}$ branched over the divisor wrapped by the $O7$-plane.  

We conclude this general discussion by noting the M-theory description of F-theory, see \cite{Denef:2008wq} 
for a detailed derivation. Using the adiabatic argument \cite{Vafa:1995gm} for M-theory on an elliptically 
fibered Calabi-Yau $n$-fold $X_n$ and for the application of fiberwise T-duality, F-theory is identified 
with M-theory on $X_n$ in the limit $\vol(T^2)\rightarrow 0$, where $T^2$ denotes the class of the generic 
elliptic fiber. This M-theory description in particular yields an alternative and independent explanation 
why F-theory vacua have $\mathcal{N}=1$ spacetime supersymmetry.

\subsection{Elliptic Calabi-Yau Manifolds and Seven-Branes in F-theory}
\label{sec:ellFourfolds}

As we have seen in an F-theory compactification on a Calabi-Yau $n$-fold $X_n$ to $(11-2n,1)$-dimensional 
Minkowski space the axio-dilaton $\tau$ is described as the complex structure modulus of an elliptic curve 
fibered over the Type IIB target manifold $B_{n-1}$ that is a K\"ahler manifold with positive curvature,
\begin{equation} \label{FonX4}
   \text{F-theory on}\ X_n \ =\ \text{Type IIB on}\ B_{n-1}\ .
\end{equation} 
The Calabi-Yau $X_n$ geometrizes non-perturbative seven-branes by non-trivial monodromies of $\tau$ around 
degeneration loci of the elliptic curve, which includes D7-branes and O7-branes as special cases. 
Here we systematically discuss compactifications of F-theory with a focus on four dimensional vacua and on 
the consistency condition imposed by tadpoles inherited from the M-theory description of F-theory \cite{Denef:2008wq}.

Let us study the F-theory geometry of an elliptically fibered Calabi-Yau $n$-fold 
$X_n \rightarrow B_{n-1}$ with a section.  This section can be used to express  
$X_n$ as an analytic hypersurface in the projective bundle 
$\mathcal{W}=\mathds{P}(\mathcal{O}_{B_{n-1}}\oplus \mathcal{L}^2\oplus \mathcal{L}^3)$ 
with coordinates $(z,x,y)$, which is a fiber-bundle over $B_{n-1}$ with generic fiber 
$\mathds{P}^2(1,2,3)$. The hypersurface constraint can be brought to the Weierstrass form
\begin{equation}
\label{eq:Weierstrass}
 	y^2 = x ^3 + g_2(\underline{u}) x z^4 + g_3(\underline{u}) z^6\ .
\end{equation}
Here $\mathcal{L}=K^{-1}_{B_{n-1}}$ for $X_n$ being Calabi-Yau and 
$g_2(\underline{u})$, $g_3(\underline{u})$ are sections of  $\mathcal{L}^4$ and $\mathcal{L}^6$ 
for \eqref{eq:Weierstrass} to be a well-defined constraint equation. Locally on the base $B_{n-1}$
they are functions in local coordinates $\underline{u}$ on $B_{n-1}$. We refer to appendix \ref{app:chernEllipticCY}
for details.

F-theory defined on $X_n$ automatically takes care of a consistent inclusion of spacetime-filling seven-branes,
as we have seen in section \ref{sec:FtheoryConstruction}. These are supported 
on the in general reducible divisor $\mathbf{\Delta}$ in the base $B_{n-1}$ determined by the degeneration 
loci of \eqref{eq:Weierstrass} given by the discriminant
\begin{equation}
 	\mathbf{\Delta}=\{\Delta:=27g_2^2+4g_3^3=0\}\,.
\end{equation}
The degeneration type of the fibration specified by the order of vanishing of $g_2$, $g_3$ and $\Delta$ 
along the irreducible components $\mathbf{\Delta}_i$ of the discriminant have an ADE--type classification 
that physically specifies the four-dimensional gauge group $G$. It can be determined explicitly using 
generalizations of the Tate formalism \cite{Bershadsky:1996nh}. 
For the order of vanishing of $\Delta$ at most one the Calabi-Yau manifold $X_n$ is smooth corresponding
to a single seven-brane on $\Delta$. This is the $I_1$ locus of the elliptic fibration
\cite{Bershadsky:1996nh}. However, this is not the generic and interesting situation in F-theory
in general and in this work. Indeed in all our examples we will consider a singular $X_n$ with a 
rich gauge symmetry generated. 

The weak string coupling limit of F-theory is given by 
$\text{Im}\, \tau \rightarrow \infty$ and yields a consistent orientifold setup with D7-branes and $O7$-planes on a 
Calabi-Yau manifold \cite{Sen:1997gv,Sen:1996vd}. In general, as the axio-dilaton of Type IIB string theory $\tau$ 
corresponds to the complex structure of the elliptic fiber, it can be specified by the value of the 
classical $SL(2,\mathds{Z})$-modular invariant $j$-function which is expressed through the 
functions $g_2$ and $g_3$ in \eqref{eq:Weierstrass} as
\begin{equation} \label{eq:def-j}
  j(\tau) = \frac{4(24\,g_2)^3}{\Delta}\ , \qquad \Delta = 27\, g_2^2 + 4 g_3^3\ .
\end{equation}
The function $j(\tau)$ admits a large $\I\, \tau$ expansion 
$j(\tau) = e^{-2\pi i \tau}+744+\cO(e^{2\pi i \tau})$ from which we can directly read off the 
monodromy \eqref{eq:D7monodromy} of $\tau$ around a D7-brane, for example. 

There are further building blocks necessary to specify a consistent F-theory setup. This is due to the 
fact that a four-dimensional compactification generically has a tadpole of the form 
\cite{Vafa:1995fj,Becker:1996gj,Sethi:1996es}
\begin{equation}
\frac{\chi(X_4)}{24}=n_3+\frac{1}{2}\int_{X_4}G_4\wedge G_4\,,
\label{eq:3tadpole}
\end{equation}
which can be deduced by considering the dual M-theory formulation. There a tadpole of 
the the three-form $C_3$ is induced due to the Green-Schwarz term $\int C_3 \wedge X_8$, the coupling  
to the M2-brane and the Chern-Simons term $\int C_3 \wedge (dC_3)^2$.
In the case that the Euler characteristic $\chi(X_4)$ of $X_4$ is non-zero a given number $n_3$ of 
spacetime-filling three-branes on points in $B_3$ and a specific amount of quantized four-form 
flux $G_4$ have to be added in order to fulfill \eqref{eq:3tadpole}. 
In addition, non-trivial fluxes on the seven-brane worldvolume contribute as \cite{Bershadsky:1997zs}
\begin{equation}
\frac{\chi(X_4)}{24}=n_3+\frac{1}{2}\int_{X_4}G_4\wedge G_4+\sum_i\int_{\mathbf{\Delta}_i}c_2(E_i)\,,
\label{eq:3tadpolemodified}
\end{equation}
where $E_i$ denotes the corresponding gauge bundle\footnote{It has been argued in \cite{Denef:2008wq} 
that also the brane fluxes should be describable by transcendental flux $G_4$.} localized on the 
discriminant component $\mathbf{\Delta}_i$. The cancellation of tadpoles in F-theory compactifications 
on Calabi-Yau fourfolds thus is restrictive for global model building since the total amount of allowed 
flux is bounded by the Euler characteristic of $X_4$ that is, in known explicit constructions, 
at most $1820448$ \cite{Klemm:1996ts}. 

The choice of $G_4$ is further constrained
by the F-theory consistency conditions on allowed background fluxes. The first constraint comes from 
the quantization condition for $G_4$, which depends on the second Chern class of $X_4$ as 
\cite{Witten:1996md}
\begin{equation} 
G_4 + \frac{c_2(X_4)}{2} \in H^4(X_4,\mathds{Z})\ 
\label{eq:fluxcondition1}
\end{equation}  
and has been deduced from anomaly freedom of the M2-brane theory.
More restrictive is the condition that $G_4$ has to be primitive,
i.e.~orthogonal to the K\"ahler form of $X_4$. In the F-theory limit of vanishing elliptic 
fiber this yields the constraints 
\begin{equation} 
\int_{X_4} G_4 \wedge J_i\wedge J_j=0\ . 
\label{eq:fluxcondition2} 
\end{equation}
for every generator $J_i$, $i=1,\ldots,h^{(1,1)}(X_4)$ of the K\"ahler cone. 
There is one caveat in order when working in the K\"ahler sector of $X_4$ and when
evaluating topological constraints like \eqref{eq:3tadpole}, \eqref{eq:fluxcondition1} and 
\eqref{eq:fluxcondition2}. In the case that $X_4$ is singular, which happens for enhanced 
gauge symmetry, it is not possible to 
directly work with the singular space since the topological quantities such as the Euler 
characteristic and intersection numbers are not well-defined. 
Thus the above constraints can be naively evaluated only in the case of a smooth $X_4$, which
corresponds to the physically simplest situation with single D7-branes\footnote{Our examples in 
chapter \ref{ch:Calcs+Constructions} are significantly more complicated and admit seven-branes 
with rather large gauge groups. Technically this is a consequence of working with fourfolds $X_4$ 
with few complex structure moduli, which typically have in the order of thousand elements 
of $H^{(1,1)}(X_4)$ of which many correspond to blow-ups of singular elliptic fibres 
signaling the presence of enhanced gauge groups.}. 
To remedy the problems when working with
singular $X_4$ we systematically blow up the singularities\footnote{In the cases considered in this 
work this is done using the methods of toric geometry \cite{Bershadsky:1996nh,Candelas:1996su,Candelas:1997eh}.} 
to obtain a smooth geometry \cite{Bershadsky:1996nh}
for which the constraints \eqref{eq:3tadpole}, \eqref{eq:fluxcondition1} and \eqref{eq:fluxcondition2} 
are valid. The resulting smooth geometry still 
contains the information about the gauge-groups on the seven-branes and allows to analyze the 
compactification in detail. 

Let us comment on the effect of three-branes and fluxes on the F-theory gauge group.
For a generic setup with three-branes and fluxes, the four-dimensional gauge symmetry as 
determined by the seven-brane content is not affected. However, if the three-brane happens to 
collide with a seven-brane, it can dissolve, by a similar transition as discussed in section 
\ref{sec:hettransition}, into a finite-size instanton on the seven-brane worldvolume that 
breaks the four-dimensional gauge group $G$. During this transition the number $n_3$ of 
three-branes jumps and a bundle $E_i$ over the seven-brane worldvolume is generated describing 
the gauge instanton \cite{Denef:2008wq}. 
However, we will not encounter this since we restrict our discussion to the case 
that the gauge bundle on the seven-branes is trivial and no three-branes sit on top of 
their worldvolumes.

We conclude by a discussion of the complex structure moduli space of elliptic fourfolds $X_4$, that
is of central importance in this work, and the interpretation in terms of the Type IIB moduli.
Consider a smooth elliptic Calabi-Yau fourfold $X_4$ with a number of 
$h^{(3,1)}(X_4)$ complex structure moduli, that may be obtained from a singular 
fourfold by multiple blow-ups\footnote{This affects only the number of K\"ahler 
moduli, which we will not discuss in the following.}. In order to compare to the Type IIB weak 
coupling picture the complex structure moduli can be split into three classes \cite{Denef:2008wq}:
\begin{enumerate}
	\item[(1)]One complex modulus that physically corresponds to the complex axio-dilaton $\tau$ and 
	that geometrically parametrizes the complex structure of the elliptic fiber.
\item[(2)] A number of $\sum_i h^{(2,0)}(\mathbf{\Delta}_i)$ complex structure moduli corresponding 
to the deformations of the seven-branes wrapped on the discriminant loci in $B_3$.
\item[(3)] $h^{(2,1)}(Z_3)$ complex structure moduli corresponding to the deformations of the 
basis and its double covering Calabi-Yau threefold $Z_3$ obtained in the orientifold limit.
\end{enumerate}
For a more detailed analysis of this and the organization in terms of the low-energy effective action of
F-theory we refer to \cite{Grimm:2010ks}.

\subsection{The Flux Superpotential}
\label{sec:F-flux_sup}

Next we discuss the perturbative F-theory superpotential, which is given by a flux superpotential. 
We emphasize only the physical and qualitative aspects of the superpotential and refer to 
chapter \ref{ch:MirrorSymm+FiveBranes} for a presentation of the expected mathematical structure 
and of the tools to calculate it efficiently.

The F-theory flux superpotential is generated upon switching on four-form flux $G_4$ in the M-theory 
perspective of F-theory \cite{Denef:2008wq,Haack:2001jz}.  
In an M-theory compactification on $X_4$ one encounters the famous Gukov-Vafa-Witten superpotential 
\cite{Gukov:1999ya} 
\begin{equation} \label{eq:GVW-super}
  W_{G_4}(\underline{z}) = \int_{X_4}  G_4 \wedge \Omega_4=
N^a\,  \Pi^b(\underline{z})\, \eta_{ab}, \qquad a,b=1,\ldots b^4(X_4)\ , 
\end{equation}
which directly applies to the F-theory setup, once the condition \eqref{eq:fluxcondition2} is met.
Here $\Omega_4$ is the holomorphic $(4,0)$ form on $X_4$ that depends on the complex structure
moduli $\underline{z}$ of $X_4$ counted by $h^{(3,1)}(X_4)$. For even second Chern class $c_2(X_4)$, 
cf.~\eqref{eq:fluxcondition1}, we expanded the flux $G_4=N^a \hat \gamma_a$ into a cohomology basis 
$\hat \gamma_a$ of the horizontal cohomology $H^{4}_H(X_4,\mathds{Z})$, which is a subgroup of 
$H^{4}(X_4,\mathds{Z})$ taking \eqref{eq:fluxcondition2} 
into account. We define a dual basis $\gamma^a$ of the integral homology group $H_{4}^H(X_4,\Z)$ in order
to define the flux quantum numbers $N^a=\int_{\gamma_a} G_4 $ that are integral. Then the whole complex structure 
dependence of $W_{G_4}$ is encoded by the complex structure dependence of the fourfold periods 
\begin{equation}
 \Omega_4(\underline{z})=\Pi^a(\underline{z}) \hat \gamma_a\,,\qquad 
 \Pi^a(\underline{z})=\int_{\gamma^a} \Omega_4(\underline{z})\,,
\label{eq:fourformExp}
\end{equation}
as in the Calabi-Yau threefold case. We introduce the topological metric 
\cite{Greene:1993vm,Mayr:1996sh,Klemm:1996ts} 
\begin{equation}
   \eta_{ab}=\int_{X_4} \hat \gamma_a \wedge
\hat \gamma_b\ ,\qquad \quad \int_{\gamma_a} \hat \gamma_b=\delta^a_{b}\ ,
\end{equation}
that parametrizes the intersection of four-cycles in $X_4$. We note that in contrast to 
$H^3(Z_3,\mathds{Z})$ of Calabi-Yau threefolds the fourth cohomology group of $X_4$ does 
not carry a symplectic structure which necessitates the introduction of $\eta_{ab}$. This 
technically complicates mirror symmetry on fourfolds compared to the threefold case, as 
we will review in section \ref{sec:mirror_toric_branes}.

One goal of this work is to explicitly compute \eqref{eq:GVW-super}
for specific elliptically fibered Calabi-Yau fourfolds, see section \ref{sec:Superpots+MirrorSymmetry}. 
Exploiting the geometrization of both bulk and brane dynamics in the complex geometry of $X_4$ 
we match the results of this calculation with the superpotentials in the language of  
Type IIB theory, which are the flux superpotential and the seven-brane superpotential.
Thus, let us present a brief review of the Type IIB superpotential in setups with seven-branes and 
$O7$-planes. The superpotential is induced by three-from fluxes $F_3$ 
and $H_3$ of the R--R and NS--NS sectors, as well as by two-form fluxes $F_2$ 
for the field strength of the $U(1)$ gauge-potential $A$ on the internal part 
of the seven-brane worldvolume. Thus $F_2$ defines an element of $H^2(D,\mathds{Z})$, where 
$D$ denotes the complex surface wrapped by the seven-brane.
The respective superpotentials are then given by \cite{Gukov:1999ya,Giddings:2001yu,Thomas:2001ve,Denef:2008wq}
\begin{equation} \label{eq:oc_super}
  W_{\rm flux}(\underline{z}) = \int_{Z_3} (F_3 - \tau H_3)\wedge \Omega_3  \ , \qquad \quad 
  W_{\rm brane}(z,\zeta) = \int_{\Gamma_5} F_2 \wedge \Omega_3\ ,
\end{equation}
where $\tau$ is the axio-dilaton and $\Omega$ is the holomorphic three-form on the Calabi-Yau 
threefold $Z_3$. These are well-defined e.g.~for the weak coupling limit with D7-branes in $O3/O7$-orientifolds 
since both $\Omega$ and the fluxes $F_3$, $H_3$ have negative parity with respect to the orientifold 
involution $\sigma$ and thus the integrals in \eqref{eq:oc_super} can be non-zero, see section \ref{sec:OrieTypeII}. 
We emphasize that $W_{\rm flux}$ only depends on the complex structure deformations of $Z_3$ due to the appearance 
of $\Omega_3$, while $W_{\rm brane}$ will also depend on the deformation 
moduli $\zeta$ of the seven-brane. To see the latter, one notes that $\Gamma_5$ is a five-chain which 
ends on the divisor $D$, i.e.~one has $ D\subset \partial\Gamma_5$, and thus carries the information about the 
embedding of the seven-brane into $Z_3$. We note that the seven-brane superpotential is related to a 
localized five-brane charge on the worldvolume of the seven-brane\footnote{It is important to emphasize
that the five-brane charge is only locally non-trivial, i.e.~the class of $F_2$ is trivial in $Z_3$. This 
is necessary in a setup with $O7$-planes since the five-brane charge in the quotient geometry $Z_3/\sigma$ is 
projected out \cite{Denef:2008wq}.} and thus has the form of the five-brane superpotential encountered
in \eqref{eq:chainIIB} in Type IIB and in \eqref{eq:chainMtheory} in heterotic string theory upon taking the
Poincare dual of $F_2$ in $\Gamma_5$. Thus, we can interpret the seven-brane superpotential as a special
case of the five-brane superpotential, where a description of the five-brane curve $\Sigma$ in terms of
seven-brane flux $[F_2]=\Sigma$ is applicable \cite{Grimm:2009ef}, see also 
\cite{Alim:2009bx,Alim:2009rf,Alim:2010za,Jockers:2009ti,Jockers:2009mn,Baumgartl:2010ad,Fuji:2010uq,Shimizu:2010us} for a 
similar use of the seven-brane superpotential. 

It is the great advantage of the F-theory formulation, that the calculation
of the superpotential \eqref{eq:oc_super} can be performed in a fully consistent string vacuum 
including the backreaction of seven-branes and orientifold planes. Indeed, one expects, as we will demonstrate
explicitly in chapter \ref{ch:Calcs+Constructions}, a match of the F-theory superpotential with the superpotentials
\eqref{eq:oc_super} for specific flux choices $G_4$ and seven-branes,
\begin{equation} \label{eq:SuperpotLimit}
  \int_{X_4}  G_4 \wedge \Omega \quad \rightarrow \quad \int_{Z_3} (F_3-\tau H_3) \wedge \Omega_3 
                  + \sum_{m} \int_{\Gamma^{m}_5} F^m_2 \wedge \Omega_3\ .
\end{equation}
where $m$ labels all seven-branes on divisors $\mathbf{\Delta}_m$ carrying two-form fluxes $F_2^m$.

Let us conclude by emphasizing once more that it is crucial in the context of this naive matching to 
consider fluxes in $H_H^{4}(X_4,\mathds{Z})$. We note that already by a pure counting 
of the flux quanta in $H^{4}(X_4,\bbZ)$, as well as in $F_3,H_3 \in H^3(Z_3,\mathds{Z})$ and 
in $F_2^m \in H^{2}(\mathbf{\Delta}_m,\mathds{Z})$ one will generically encounter 
a mismatch. This can be traced back to the fact that not all fluxes $G_4$ are actually 
allowed in an F-theory compactification, since in the duality between M-theory on $X_4$ and F-theory on $X_4$ 
the K\"ahler class of the elliptic fiber is sent to infinity turning one of the dimensions of the elliptic fiber 
into a space-time dimension \cite{Denef:2008wq,Grimm:2010gk}.

\section{Heterotic/F-Theory Duality} 
\label{sec:HetFDuality}

Let us now come to a more systematic discussion of heterotic/F-theory duality.
Since the fundamental duality that underlies heterotic/F-theory duality in any dimensions 
is the eight-dimensional equivalence of the heterotic string compactified on $T^2$ 
and F-theory on elliptic $K3$ we begin with a summary of this duality in section \ref{sec:hetF8d}. 
In this most simple setup, the duality can be readily checked by comparison of the moduli
space on both sides of the duality.
Then we apply the adiabatic argument in section \ref{sec:het-Fdual} to obtain heterotic/F-theory
dual setups in lower dimensions, with particular emphasis on four-dimensional setups.
We present a detailed and explicit discussion of the unified description of both the heterotic 
Calabi-Yau threefold $Z_3$ and the bundle $E$ in terms of the complex geometry of the F-theory 
fourfold $X_4$. In this context we put special emphasis on the split of the F-theory
fourfold constraint in the stable degeneration limit, that gives back the heterotic threefold 
$Z_3$ as well as the spectral cover data of $E$. We conclude in section \ref{sec:F_blowup}
by considering the duality map between F-theory and heterotic moduli and of heterotic five-branes 
on curves $\Sigma$. We review that a horizontal five-brane maps to a blow-up in $X_4$
along the five-brane curve $\Sigma$, where we explicitly construct the blow-up both locally and 
globally as a complete intersection. We end our discussion with the details of the duality map for 
the moduli in both theories. It is important to note that the F-theory complex structure moduli also 
have to account for the deformation modes of the curve $\Sigma$ for consistency if a horizontal 
five-brane on $\Sigma$ is included.

\subsection{Heterotic/F-theory in Eight Dimensions}
\label{sec:hetF8d}

The fundamental duality underlying heterotic/F-theory duality in any dimension is the eight-dimensional
duality of the heterotic string on $T^2$ and F-theory on an elliptic $K3$ \cite{Vafa:1996xn}. It
can be checked by matching the moduli and the gauge symmetry on both sides of the duality 
\cite{Vafa:1996xn,Morrison:1996pp}. In addition heterotic/F-theory duality agrees with the duality of M-theory
on $K3$ and the heterotic string on $T^3$ \cite{Witten:1995ex}, when $K3$ is taken to be elliptic
and one further compactifies the eight-dimensional heterotic/F-theory setup on an additional $S^1$.

On the heterotic side the torus $T^2$ is defined by the Weierstrass equation \eqref{eq:Weierstrass} in 
$\mathds{P}^2(1,2,3)$. A gauge background on $T^2$ obeying \eqref{eq:DonUhlYau} is defined by a
flat connection $F_{u\bar{u}}$ for complex coordinates $u$ on $T^2$, since 
$g_{u\bar u}=\frac{v}{2\tau_2}$ for\footnote{If we construct $T^2$ 
as the lattice quotient by $L=\mathds{Z}\oplus\tau\mathds{Z}$ with coordinates $(x_1,x_2)$ on $\mathds{C}$, 
we define $z=x_2+\tau x_1$. The connection with the algebraic representation \eqref{eq:Weierstrass} is established
via $\tau=\frac{\int_{\gamma_2}\Omega_1}{\int_{\gamma_1}\Omega_1}$, where $\gamma_i$ denote the A- and B-cycle and
$\Omega_1$ the holomorphic $(1,0)$-form on $T^2$. Locally it is $\Omega_1=dz$.} $\tau_2=\text{Im}(\tau)$ and $v$ its volume.
This is solved by switching on $16$ Wilson lines $A^a_i=C_i^a$, $i=1,2$ and $a=1,\ldots, 16$, that are defined up 
to periodicity on $T^2$ \cite{Buchbinder:2002ji}. This is an $U(1)^{16}$ gauge bundle $E$ on $T^2$ 
with bundle moduli space\footnote{We note that in the construction of $SU(n)$-bundles on elliptic Calabi-Yau manifolds $Z_n$, 
this moduli space occurs as the fiber of the moduli space of bundles on $Z_n$, cf.~the spectral cover 
construction in section \ref{sec:spectralcover}.} parametrized by the real parameters $C_i^a$, up to permutations by $S_{15}$,
\begin{equation}
	\mathcal{M}(E)= T^{30}/S_{15}\times T^2\cong \mathds{P}^{15}\times T^2\,.
\label{eq:modulihetU(1)^16}
\end{equation}
For this gauge background, the eight-dimensional gauge symmetry contributed from the ten-dimensional gauge bosons of the
heterotic string is $U(1)^{16}$.
We note that more general bundles $E$ on $T^2$ with different structure groups have to be described by other means like an embedding 
of $T^2$ into $dP_9$ for $E_8$-bundles \cite{Friedman:1997yq}.

From the CFT-point of view the heterotic string on $T^2$ can be analyzed directly and the global moduli space
of the heterotic string on $T^2$ is identified as the moduli space of (complexified) metrics on the Narain-lattice $\Gamma^{18,2}$ of 
signature $(18,2)$
\cite{Narain:1985jj,Narain:1986am},
\begin{equation}
		\mathcal{M}_{\text{het}}=O(18,2,\mathds{Z})\backslash O(18,2)/(O(18)\times O(2))\times \mathds{R}_+
\label{eq:modulihetT^2}
\end{equation}
that takes into account the complex structure $\tau$, the K\"ahler structure $\rho$ of $T^2$ and the heterotic dilaton,
that corresponds to $\mathds{R}_+$. 
Here the homogeneous space counts the number of metrics on the lattice $\Gamma^{18,2}$, whereas the discrete group $SO(18,2,\mathds{Z})$ 
is the physical T-duality group. Taking the reduction of the ten-dimensional metric and the B-field into account, the heterotic
gauge symmetry in eight dimensions has a maximal rank of $20$, which at a generic point in $\mathcal{M}_{\text{het}}$ is 
$U(1)^{20}$. It enhances to non-abelian gauge groups for special symmetric points in the moduli space $\mathcal{M}_{\text{het}}$ 
yielding up to $G=E_8\times E_8\times G'$ for $G'=SU(2)\times U(1)^3,\,SU(3)\times U(1)^3,\,SU(2)^4,\,SU(3)^2$ gauge symmetry\footnote{
Only the first two gauge groups $G'$ appear in \cite{Kachru:1995wm}.} 
in eight dimensions \cite{Kachru:1995wm,becker2007string}. Geometrically this means that we have a trivial gauge background on 
$T^2$, which we are allowed to have since the heterotic tadpole \eqref{eq:anomaly} is trivial on $T^2$, and the moduli $\tau$ 
and $\rho$ of $T^2$ take special symmetric values.

On the F-theory side, the geometry of the elliptic $K3$-surface is obtained by fibering the heterotic two-torus $T^2$ over the base 
$B_1=\mathds{P}^1$ as in section \ref{sec:FtheoryConstruction}. Then, a first step to check the duality is to match the moduli and 
gauge symmetry of the heterotic string from the F-theory perspective.  It is already very promising that the complex structure
moduli space of a generic, elliptic $K3$-surface is given by the complex $18$-dimensional manifold
\begin{equation}
	\mathcal{M}_{\text cs}=O(18,2,\mathds{Z})\backslash O(18,2)/(O(18)\times O(2))\,, 
\label{eq:moduliFOnK3}
\end{equation}
which is the moduli space of an algebraic $K3$-surface with Picard-number $\rho=2$ for the elliptic fiber and the base.
This manifold is identical with the first factor in \eqref{eq:modulihetT^2}.
In addition we have a factor $\mathds{R}_+$ for the real K\"ahler volume of the base $\mathds{P}^1$ that is identified with
the heterotic dilaton. Since the volume of the elliptic fiber is physically irrelevant and formally sent to zero/infinity, 
it does not contribute a K\"ahler class. In addition, it is important to note that in F-theory K\"ahler moduli are in general 
not complexified which is consistent with the $\mathcal{N}=1$ coordinates \eqref{eq:N=1coordsO37} for $O7$-orientifolds
where the $v^\alpha$, measuring volumes of curves, are not complexified, but the quadratic combination $\mathcal{K}_\alpha$. 

The heterotic gauge symmetry is matched by the singularity type of the elliptic fibration of $K3$. For a smooth $K3$ there 
is an $U(1)^{20}$ gauge theory originating from $18$ elements in $H^{(1,1)}(K3)$, since both the class
of the elliptic fiber and the class of the base $\mathds{P}^1$ do not contribute vector fields \cite{Grimm:2010ks}\footnote{In the 
notation of \cite{Grimm:2010ks} the 18 elements on $H^{(1,1)}(K3)$ are blow-up modes. In the lift from M- to F-theory the class of 
the elliptic fiber lifts to the eighth dimension in Minkowski space, $\mathds{R}^{1,6}\rightarrow \mathds{R}^{1,7}$, and the
K\"ahler class of the base $\mathds{P}^1$ yields a K\"ahler coordinate $T_\alpha$ of the Type IIB theory.}, and from $H^{(2,0)}(K3)\oplus H^{(0,2)}(K3)$.   
In the case of enhanced gauge symmetry $G$ the type of the degeneration of the elliptic fibration of $K3$ in F-theory has to match
the heterotic gauge group $G$. For the example of $G=E_8\times E_8\times U(1)^4$ the corresponding singular F-theory geometry is given in 
the affine patch $z=1$ by \cite{Morrison:1996pp}
\begin{equation}
	y^2=x^3+\alpha xs^4+(s^5+\beta s^6+s^7)\,,\qquad \alpha,\,\beta\in \mathds{C}\,,
\label{eq:K3E8sing}
\end{equation}
which is obtained by specializing the complex structure deformations in $g_2$, $g_3$ in \eqref{eq:Weierstrass} accordingly. We note 
that \eqref{eq:K3E8sing} has an $E_8$ singularity at $s=0,\infty$ on the $\mathds{P}^1$-base. Indeed for $s\rightarrow 0$ we obtain the 
constraint $y^2=x^3+s^5$ of an $E_8$ singularity, where we ignored the irrelevant deformations $s^6$, $s^7$, $xs^4$. Similarly the $E_8$
singularity at infinity is visible in coordinates $\tilde{s}=1/s$. For the case of the singular $K3$ \eqref{eq:K3E8sing} the Picard lattice
is at least 18-dimensional, of which 16 are the vanishing cycles of the $E_8\times E_8$-singularity and of which two are the class of the
base and the elliptic fiber, that is sent to zero/infinity. Further enhancement to $E_8\times E_8\times G'$ for 
$G'=SU(2)\times U(1)^3,\,SU(3)\times U(1)^3,\,SU(2)^4,\,SU(3)^2$ are possible\footnote{However it is not clear to the author how to realize
the enhancement of more than two $U(1)$ factors of $E_8\times E_8\times U(1)^4$ since only 18 of the generators of $H^{(1,1)}(K3)$ are allowed to
correspond to the Cartan generators of non-abelian gauge groups \cite{Grimm:2010ks}.}.

\subsection{Heterotic/F-theory Duality in Lower Dimensions}
\label{sec:het-Fdual}

After establishing heterotic/F-theory duality in eight dimensions, lower dimensional
versions of the duality can be constructed using the adiabatic argument \cite{Vafa:1995gm}.
The general idea is to consider a family of dual eight-dimensional theories parameterized 
by a manifold $B_{n-2}$. Upon a slow variation of the parameters, the pairs of dual theories
at every point $p\in B_{n-2}$ should glue together to form globally consistent duality
between the heterotic string and F-theory in lower dimensions \cite{Vafa:1996xn}. Geometrically
we obtain pairs of smooth dual geometries $X_{n}$ and $Z_{n-1}$ on the F-theory, respectively, 
heterotic side with generic fiber of an elliptic $K3$, respectively, an elliptic curve. For 
technical reasons $B_{n-2}$ is chosen to be a complex K\"ahler manifold 
and in order to preserve $\mathcal{N}=1$ supersymmetry, $X_n$ and also $Z_{n-1}$ obey
the Calabi-Yau condition. We summarize the fibration structure of the heterotic/F-theory dual 
geometries schematically as
\begin{eqnarray}
			\text{Heterotic}\,\,\quad Z_{n-1}\,\,\left\{\,\,\begin{matrix}T^2\\ \\ \downarrow\\B_{n-2}\end{matrix}\right.& \begin{matrix}\leftarrow\joinrel\relbar\joinrel\relbar\joinrel\rightarrow\\ \\ \phantom{\downarrow}\\ 				\phantom{B_{n-2}}\end{matrix}\,\,\,\,\,\, \left.\begin{matrix}\phantom{a}\vspace{-1.1cm}\\ \left.\begin{matrix}T^2\\ \downarrow\\ \P^1\end{matrix}\,\,\right\}K_3\\
			\hspace{-1cm}\downarrow\\
			\hspace{-1cm}B_{n-2}
			\end{matrix}\,\,\right\}\,\, X_n \quad\,\,\text{F-theory}
			\label{eq:HetFDualityScheme}
\end{eqnarray}
This fibration structure implies that the base of $X_n$ with respect to the elliptic fibration 
on the F-theory side is a holomorphic $\P^1$-fibration over $B_{n-2}$ forming a manifold $\tilde{B}_{n-1}$.
In other words $\tilde{B}_{n-1}$ is the total space of the
projective bundle $\mathds{P}(\mathcal{O}_{B_{n-2}}\oplus L)$ where the line bundle
$L=\mathcal{O}_{B_{n-2}}(-\Gamma)$ is associated to an effective divisor $\Gamma$ in $B_{n-2}$.
There are two distinguished classes in $H^2(\tilde{B}_{n-1},\mathds{Z})$, namely the  class of the 
hyperplane of the $\mathds{P}^1$-fiber denoted by $r=c_1(\mathcal{O}(1))$ and of the line bundle $L$ 
with $c_1(L)=t$.
We note that, in contrast to the $K3$-fibration of $X_n$, the $\P^1$-fibration of 
$\tilde{B}_{n-1}$ in \eqref{eq:HetFDualityScheme} is not fixed by the geometrical data of the heterotic 
Calabi-Yau $Z_{n-1}$. As we see below, it is fixed by the topology of the heterotic vector bundle $E$.

In order to make connection to the heterotic gauge bundle $E$ we define two holomorphic sections 
denoted $C_0$, $C_\infty$ of the fibration $p:\tilde{B}_{n-1}\rightarrow B_{n-2}$ corresponding to 
the first and second coordinate of $\mathds{P}(\mathcal{O}_{B_{n-2}}\oplus L)$ set to zero,
\begin{equation}
 	C_\infty=C_0-p^*\Gamma\,.
 	\label{eq:cinfty}
\end{equation}
In terms of these divisors the perturbative gauge group $G=G_1\times G_2$, 
where we denote the group factors from the first $E_8$ as $G_1$ and from the second 
$E_8$ as $G_2$, is realized by seven-branes over $C_0$ and $C_\infty$ with singularity type $G_1$
and $G_2$, respectively \cite{Morrison:1996na,Morrison:1996pp,Bershadsky:1997zs}. 
These sections are simply the higher dimensional analog of the points $u=0,\infty$ 
on $\P^1$, cf.~\eqref{eq:K3E8sing}, where the singularities of $K3$ dual to the heterotic gauge symmetry 
are located in the eight-dimensional duality of section \ref{sec:hetF8d}.
Conversely, singularities that do not descend from the eight-dimensional theory have to 
correspond to new physics in the lower-dimensional theory. In particular, components of 
the discriminant on which $\Delta$ vanishes of order greater than one that project 
onto codimension two subvarieties $\Sigma_i$ in $B_{n-2}$ correspond to heterotic 
five-branes on the same subvarieties in $Z_{n}$ 
\cite{Morrison:1996na,Bershadsky:1996nh,Bershadsky:1997zs}. Consequently, the corresponding 
seven-branes induce a gauge symmetry that maps to a non-perturbative effect due to the five-branes 
on the heterotic side.

Applying the duality map for the case $n=3$ with base $B_1=\P^2$ we obtain the duality
of the heterotic string on $Z_2=K3$ and F-theory on an elliptic $K3$-fibered Calabi-Yau 
threefold $X_3$ \cite{Morrison:1996na,Morrison:1996pp}.
The base $\tilde{B}_2$ on the F-theory side is given by a Hirzebruch surface\footnote{Here
we do assume that there are now heterotic five-branes present. A five-brane at a point in $B_1$ 
corresponds to a blow-up of the same point in $\mathds{F}_k$ yielding new K\"ahler moduli and tensor 
multiplets in six dimensions \cite{Vafa:1996xn,Morrison:1996na,Morrison:1996pp}.} 
$\mathds{F}_k=\P(\mathcal{O}_{\P^1}\oplus\mathcal{O}_{\P^1}(k))$, which is a $\P^1$-bundle over $\P^1$. Then the 
F-theory Calabi-Yau threefold $X_3$ is constructed as a Weierstrass form of the elliptic 
fibration over $\mathds{F}_k$. The complex structure moduli of $X_3$ encode both the complex structure 
moduli of $K3$ and the bundle moduli of $E$ on the heterotic side \cite{Morrison:1996pp}.
Introducing the K\"ahler classes $k_b$, $k_f$ of the base and the fiber of $\mathds{F}_k$ one has the 
matching
\begin{equation}
	\frac{1}{e^{2\phi}}=\frac{k_b}{k_f}
\label{eq:DilatonMap6dHetF}
\end{equation} 
of the heterotic dilaton $\phi$ forming a six-dimensional tensor multiplet and of $k_bk_f$ that is related to
the overall volume $k_b(k_f-nk_b)$ of $\mathds{F}_k$ yielding a universal hypermultiplet in six dimensions 
\cite{Vafa:1996xn,Andreas:1998zf}.  
Finally, the integer $k$ specifying the base $\mathds{F}_k$ is determined from the topology of the heterotic
vector bundle $E$. We note that in a heterotic compactification on $K3$ the tadpole condition 
\eqref{eq:anomaly} evaluated on the single four-cycle $K3$ itself implies the condition
\begin{equation}
	n_1+n_2+n_5=24\,,
\label{eq:K3Hetanomaly}
\end{equation}
where the right hand side denotes the Euler characteristic $24=\int_{K3}c_2(K3)$. The
integers $n_1$, $n_2$ denote the instanton numbers in the heterotic gauge-bundle $E$ over $K3$,
where we split $E$ into $E=E_1\oplus E_2$, and the integer $n_5$ takes into account possible five-branes.
From the analysis of phase transitions in the heterotic string as well as in the K\"ahler moduli space
of $\mathds{F}_k$ it is possible to infer the relation $n_{1/2}=12\pm k$ for $n_5=0$ \cite{Morrison:1996na}.
Thus, once the topology of $E$ is fixed (as well as the number of five-branes is put to zero) the 
complete F-theory geometry is determined and vice versa. For a more refined analysis one applies the spectral 
cover construction of $E$ reviewed in section \ref{sec:spectralcover}.

We conclude by noting that a more detailed matching of the F-theory and heterotic fields 
can be inferred by working out the number and type of multiplets in the $\mathcal{N}=2$ theory 
in six dimensions on both sides of the duality. This analysis can be performed e.g.~by 
compactifying on an additional $T^2$ for which F-theory becomes dual to Type IIA string theory 
on $X_3$ and heterotic/F-theory duality becomes the duality of the heterotic string on $K3\times T^2$ 
and Type IIA on $X_3$, cf.~\cite{Louis:1996ya,Aspinwall:1996mn} for a review.

Next we construct a pair of dual heterotic/F-theory setups yielding a four-dimensional theory \cite{Bershadsky:1997zs}.
Then heterotic/F-theory duality states the equivalence of the heterotic string on an elliptic Calabi-Yau threefold 
$Z_3$ and F-theory on an elliptic K3-fibered Calabi-Yau fourfold $X_4$. According to \eqref{eq:HetFDualityScheme} 
the three-dimensional base $\tilde{B}_3$ of the elliptic fibration of $X_4$ is ruled over the 
base $B_2$ of the heterotic threefold $Z_3$. As we will make more precise below the complex structure moduli
of $X_4$ map to the complex structure moduli of $Z_3$ as well as the bundle moduli of $E$. The K\"ahler sector
of $X_4$ in general contains the class of the elliptic fiber, several K\"ahler blow-ups in the elliptic fibration 
of $X_4$, corresponding to non-abelian gauge symmetry, and the classes of $H^{(1,1)}(\tilde{B}_3)$, which
are $h^{(1,1)}(B_2)$ K\"ahler classes of $B_2$, one additional class\footnote{
This provides a universal tensor multiplet containing the dilaton \cite{Morrison:1996na,Morrison:1996pp,Bershadsky:1996nh,Bershadsky:1997zs},
see \eqref{eq:DilatonMap4dHetF}.}  of the generic $\P^1$-fiber in $\tilde{B}_3$
as well as several blow-ups in $\tilde{B}_3$ corresponding to heterotic five-branes. 
We note that these qualitatively different K\"ahler classes have 
also qualitatively different meaning for the heterotic theory. On the one hand,
singularities within the elliptic fibration of the $K3$-fiber correspond to the heterotic perturbative 
gauge symmetry with $h^{(1,1)}(K3)-2$ Cartan elements, which is consistent with the statement below \eqref{eq:cinfty}. 
On the other hand, gauge symmetry that is not already visible in eight dimensions, i.e.~in the $K3$-fiber, has to 
correspond to non-perturbative effects on the heterotic side.
Analogously to \eqref{eq:DilatonMap6dHetF}, the heterotic dilaton $\phi$ is obtained as
\begin{equation}
	e^{-2\phi}=\frac{\vol(B_2)}{k_f}\,,
\label{eq:DilatonMap4dHetF}
\end{equation}
where $k_f$ denotes the K\"ahler class of the $\P^1$-fiber in $\tilde{B}_3$ and $\vol(B_2)$ the volume of $B_2$, see \cite{Haack2000}
for a similar analysis in heterotic/M-theory duality. 

It turns out that again the fibration data of $\tilde{B}_3$ is crucial for the 
construction of the stable vector bundle $E$ on $Z_3$ in the dual heterotic theory.
To analyze this issue in a more refined way it is necessary to use the methods developed in 
\cite{Friedman:1997yq}, in particular the spectral cover.
For example, consider the heterotic string with an $E_8\times E_8$-bundle on $Z_3$. 
Besides the required singularities of the elliptic fibration of $X_4$ at the divisors $C_0$ and $C_\infty$ 
matching the perturbative heterotic gauge group $G$ only the base $B_2$ of the $K3$-fibration is fixed by duality. 
The threefold $\tilde{B}_3=\P(\mathcal{O}_{B_2}\oplus L)$ can be freely specified by choosing the 
$\mathds{P}^1$-fibration over $B_2$ that is fixed by a line bundle $L$ on $B_2$, whose Chern-class we denote by $c_1(L)=t$, cf.~the discussion above \eqref{eq:cinfty}.  
Then, the heterotic bundle $E=E_1\oplus E_2$ is given in terms of the cohomology of $\tilde{B}_2$ as \cite{Friedman:1997yq}
\begin{equation}
	\eta(E_1)=6c_1+t\,,\quad \eta(E_2)=6c_1-t\,,
\label{eq:etaE8}
\end{equation}
which uniquely determines the $\eta$-classes, cf.~section \ref{sec:spectralcover}, of the 
two bundles by the choice of $\mathds{P}^1$-fibration. In particular, we note that the heterotic 
anomaly \eqref{eq:anomaly} is trivially fulfilled without the inclusion of any horizontal five-branes, i.e.~five-branes
wrapping curves in $B_2$. However, a number of five-branes wrapping the elliptic fiber have to be included as deduced in \cite{Friedman:1997yq}.
We note that the duality map \eqref{eq:etaE8} remains valid also for $SU(n)$-bundles where $X_4$ is singular as concluded in 
\cite{Andreas:1997ce}. Also in this case five-branes on the elliptic fiber are required by the heterotic anomaly cancellation 
\eqref{eq:anomaly}.

Again we refer to the literature for a more detailed
analysis of the four-dimensional $\mathcal{N}=1$ spectrum and the translation to the heterotic side, see 
\cite{Curio:1997rn,Andreas:1997pd,Mohri:1997uk,Andreas:1998zf} and \cite{Grimm:2010ks} for a detailed derivation
of the F-theory effective action in a dimensional reduction on $X_4$. Upon a further compactification on $T^2$ 
heterotic/F-theory duality becomes heterotic/Type II duality in two dimensions which can alternatively be used for 
an discussion of the massless spectrum and effective action \cite{Haack:2000di,Haack2000}.

Finally, since we are in this work mainly interested in the complex structure moduli of $X_4$ 
and their map to the heterotic moduli, let us conclude with a more detailed discussion of this aspect of
heterotic/F-theory duality. As mentioned above, in the absence of five-branes the complex structure moduli of
$X_4$ split into complex structure moduli as well as bundle moduli of the heterotic 
compactification.
This splitting can be made directly visible by investigation of the Weierstrass equation defining the 
F-theory Calabi-Yau geometry. In eight dimensions, for simplicity, an $E_8\times E_8$-bundle over $T^2$
on the heterotic side yields the Weierstrass equation for the elliptic $K3$ on the F-theory side as 
\cite{Morrison:1996pp}
\begin{equation}
	y^2=x^3+\sum_{n=-4}^{4}f_{4-n}s^{4-n}t^{4+n}xz^4+\sum_{n=-6}^6g_{6-n}s_1^{6-n}s_2^{6+n}z^6
\label{eq:K3E8deformed}
\end{equation}
which contains two deformed $E_8$ singularities at $s_1=0$, $s_2=0$ in projective coordinates on the $\P^1$-base of $K3$, 
cf.~\eqref{eq:K3E8sing}. Based on the observation that the heterotic geometry $T^2$ is encoded
only by the terms 
\begin{equation}
	p_0:=-y^2+x^3+f_0x\tilde{z}^4+g_0\tilde{z}^6\,,
\label{eq:HetGeoInX_n}
\end{equation}
upon identifying the projective coordinate $\tilde{z}$ of $T^2$ as $\tilde{z}=\frac{z}{s_1s_2}$, it was concluded in 
\cite{Morrison:1996pp} that the information about the heterotic bundle $E$ is encoded in the terms in
\eqref{eq:K3E8deformed} with different powers in $s_1$, $s_2$. This reflects the fact that precisely these terms contain
the information about the singularities of the elliptic fibration, cf.~\eqref{eq:K3E8sing} for the case of an $E_8\times E_8$
singularity. This observation was made rigorous and adapted to the case of toric hypersurfaces in \cite{Berglund:1998ej}, 
which allows a systematic application to compactifications to lower dimensions .

More precisely in the duality between the heterotic string on $Z_3$ with bundle $E=E_1\oplus E_2$ and 
F-theory on $X_4$ the constraint $P$ of $X_4$ is shown to split as \cite{Berglund:1998ej}
\begin{equation} \label{eq:BMform}
  P = p_0 + p_+ +  p_- =0 \ ,
\end{equation} 
where $p_0  = 0$ specifies the threefold $Z_3$ as in \eqref{eq:HetGeoInX_n}, $p_+:=\sum_{i>0} p_{+,i}v^i = 0$  
the bundle $E_1$, and $p_-:= \sum_{i<0} p_{-,i}v^i=0$ the bundle $E_2$. This is also referred to as the stable degeneration
limit of $X_4$, cf.~\cite{Friedman:1997yq}. Here the coordinate $v$ is the coordinate on 
the $\P^1$-base in the $K3$-fibers, that is related to the projective coordinates $(s_1:s_2)$ on $\P^1$ as $v=s/t$. We note 
that the description of the heterotic bundle $E$ via \eqref{eq:BMform} agrees with the spectral cover construction of 
\cite{Friedman:1997yq} if applicable, so that $p_+$ is for example given by the spectral cover divisor \eqref{eq:def-p+} 
in the case of an $SU(N)$-bundle.
However, the split \eqref{eq:BMform} applies even for bundles not describable in the spectral cover approach like bundles
with exceptional structure group. Another advantage of \eqref{eq:BMform} is the fact that the map of the complex structure
moduli of $X_4$ to the heterotic bundle moduli can be directly studied. 

In fact, this will allow us in our concrete examples of section \ref{sec:Superpots+MirrorSymmetry} to show 
that some complex structure moduli of $X_4$, that are also the moduli of seven-branes are precisely mapped 
to the coefficients of the spectral cover $p_+$ of an $SU(1)$-bundle or heterotic five-brane. Consequently, switching on 
appropriate four-form fluxes $G_4$ in F-theory generates a superpotential for these fields which we will determine explicitly
and which in the dual heterotic theory corresponds to the five-brane superpotential \eqref{eq:chainMtheory}.

\subsection{Five-Branes in Heterotic/F-Theory Duality: Blowing Up in F-Theory}
\label{sec:F_blowup}

In this section we will discuss the F-theory dual of five-branes, in particular 
horizontal five-branes, in very much detail following \cite{Grimm:2009sy}. 
This is of particular importance on the one hand side 
for conceptual reasons since it turns out that the F-theory dual to the 
heterotic string with maximal perturbative gauge symmetry $G=E_8\times E_8$ has to be analyzed 
more thoroughly and then naturally yields a description of the F-theory dual of horizontal five-branes. 
On the other hand a complete understanding of the heterotic string requires the inclusion of 
five-brane dynamics, in particular in the light of the small instanton transition reviewed in section 
\ref{sec:hettransition}. Furthermore, as we will see, it will be precisely the
horizontal five-branes introduced in this section that are geometrized by blow-ups into 
exceptional divisors $E$ in $\tilde{B}_3$ for which our analysis and calculation of the 
superpotential will be performed in sections \ref{sec:SuperpotsHetF} and \ref{sec:heteroticF+blowup}.

In general there are two qualitatively different types of five-branes on curves $\Sigma$ in an
elliptically fibered Calabi-Yau threefold. This is reflected in the 
decomposition of the class $\left[\Sigma\right]$ of the five-brane curve as
\begin{equation}
 	\Sigma=n_f F+\Sigma_B\ ,
\end{equation}
where $\Sigma_B$ denotes a curve in the base $B_2$ of the elliptic fibration, $F=\pi^*([p])$
denotes the elliptic fiber over a point $p$ in $B_2$, and $n_f$ is a positive integer.  
This is a split into five-branes vertical to the projection $\pi:\,Z_3\rightarrow B_2$, 
where the integer $n_f$ counts the number of five-branes wrapping the elliptic fiber,
and into horizontal five-branes on $\Sigma_B$ in the base $B_2$.  Both cases lead to 
different effects in the F-theory dual theory. Vertical five-branes correspond to 
spacetime filling three-branes at a point in the base $\tilde{B}_3$ of the F-theory fourfold $X_4$
\cite{Andreas:1997ce,Friedman:1997yq}. Conversely, horizontal five-branes on
the curve $\Sigma_B$ map completely to the geometry of the F-theory side. They map to
seven-branes supported on a component $\Delta_i$ of the discriminant\footnote{As we see in the following
$\Delta$ vanishes of order greater than one on $\Delta_i$ \cite{Morrison:1996na,Bershadsky:1996nh,Bershadsky:1997zs}.} 
in the fourfold base $\tilde{B}_3$ which projects under $p:\tilde{B}_3\rightarrow B_2$ onto the 
curve $\Sigma_B$ in $B_2$ \cite{Morrison:1996na, Bershadsky:1996nh,Bershadsky:1997zs} and that has to
be blown-up in $\tilde{B}_3$ into a divisor $E$ \cite{Berglund:1998ej,Rajesh:1998ik,Diaconescu:1999it}. 
Consequently, they correspond to seven-branes that induce a non-perturbative gauge symmetry.

This duality map for branes is in particular consistent with the small instanton 
transition of a brane, i.e.~a coherent sheaf, into a smooth bundle on both sides of 
the duality. Indeed, the three-branes undergoing a three-brane/instanton 
transition in F-theory are precisely the dual of the transition of a vertical five-brane 
into a finite size instanton breaking the gauge group on the F-theory and the heterotic 
side accordingly. However, this transition is not of relevance for our considerations 
since our analysis does not depend on the specific four-dimensional gauge symmetry.  
 
Let us now perform a detailed analysis of the F-theory geometry dual to a horizontal five-brane. 
For simplicity we consider the enhanced symmetry point with $G=E_8\times E_8$ due to small instantons/five-branes such 
that the heterotic bundle is trivial. For this setup the following considerations are
the most comprehensible. For general vector bundles $E$ on the heterotic side, an analysis 
of the local F-theory geometry near the five-brane curve $\Sigma_B$ is possible \cite{Diaconescu:1999it}
applying the method of stable degeneration
\cite{Aspinwall:1997ye,Friedman:1997yq}. However, since the essential point in
this analysis is the locally trivial heterotic gauge bundle, the results of
\cite{Diaconescu:1999it} carry immediately over to our situation with a globally trivial gauge bundle. 

As follows in general, using the adjunction formula, the canonical bundle of the
ruled base\footnote{If we are working entirely with fourfolds $X_4$ we drop the 
superscript $\tilde{B}_3$ and denote the base of the elliptic fibration by $B_3$.} 
$B_3=\P(\mathcal{O}_{B_2}\oplus L)$ can be determined from the general formula for
the total Chern class of a projective $\P^1$-bundle as $c_1(B_3)=c_1(B_2)+2r+t$, 
cf.~appendix \ref{app:chernEllipticCY}.
Then we use \eqref{eq:cinfty} to obtain
\begin{equation}
	K_{B_3}=-2C_0+p^*(K_{B_2}-\Gamma)
	=-C_0-C_\infty+p^*(K_{B_2})\,.
\label{eq:ccB3}
\end{equation}
From this we deduce the classes $F$, $G$ and $\mathbf{\Delta}$ of the divisors
defined by\footnote{Here we follow the common convention to denote 
$(g_2,g_3)$ by $(f,g)$ in the standard Weierstrass form \ref{eq:Weierstrass}.} $f$, $g$ and 
$\Delta$ as sections of $K_{B_3}^{-4}$, $K_{B_3}^{-6}$ and $K_{B_3}^{-12}$, respectively.
To match the heterotic gauge symmetry $G=E_8\times E_8$, there have to be $II^*$
fibers over the divisors $C_0$, $C_\infty$ in $B_3$. Since $II^*$ fibers
require that $f$, $g$ and $\Delta$ vanish to order $4$, $5$ and $10$ over $C_0$
and $C_\infty$ \cite{Bershadsky:1996nh}, their divisor classes split accordingly with
remaining parts 
\begin{eqnarray}
	F'&=&F-4(C_0+C_\infty)=-4p^*(K_{B_2})\,,\nn\\
	G'&=&G-5(C_0+C_\infty)=C_0+C_\infty-6p^*(K_{B_2})\,,\\
	\mathbf{\Delta}'&=&\mathbf{\Delta}-10(C_0+C_\infty)=2C_0+2C_\infty-12p^*(K_{B_2})\,.\nn
\label{eq:rest}
\end{eqnarray}
This generic splitting implies that the component $\Delta'$ can locally be
described as a quadratic constraint in a local normal coordinate $s$ to $C_0$ or
$C_\infty$, respectively. Thus, $\Delta'$ can be understood locally as a double
cover over $C_0$ respectively $C_\infty$ branching over each irreducible curve
$\Sigma_i$ of $\Delta'\cdot C_0$ and $\Delta'\cdot C_\infty$. In fact, near
one irreducible curve $\Sigma_i$ intersecting say $C_0$ the splitting \eqref{eq:rest} implies
that the sections $f,g$ take the form 
\begin{equation}
\label{eq:brane-g5}
	f=s^4f'\,,\quad g=s^5(g_5+s g_6)\equiv s^5 g'
\end{equation}
with $f'$ denoting a section of $KB_3^{-4}$ and $g_5$, $g_6$ sections of 
$KB_3^{-6}\otimes L$, $KB_3^{-6}$, respectively. 
The discriminant then takes the form $\Delta=k^{10}\Delta'$ where $\Delta'$ is calculated from $f'$ and $g'$.
Thus, the intersection curve is given by $g_5=0$ and the degree of the discriminant 
$\Delta$ rises by two over $\Sigma_i$ with
$f'$ and $g'$ vanishing of order zero and one. Precisely the singular curves 
$\Sigma_i$ in $X_4$ that appear in $g$ as above are the locations of the small instantons/horizontal
five-branes in $Z_3$ \cite{Rajesh:1998ik,Diaconescu:1999it} on the heterotic
side. In the fourfold $X_4$ the collision of an $II^*$ and an $I_1$ singularity
over $\Sigma_i$ induces a singularity of $X_4$ exceeding Kodaira's
classification of singularities.  Thus, it requires a blow-up
$\pi:\hat{B}_3\rightarrow B_3$ in the three-dimensional base of the curves
$\Sigma_i$ into divisors $E_i$. This blow-up can be performed without
violating the Calabi-Yau condition since the shift in the canonical class of
the base, $K_{\tilde{B}_3}=\pi^*K_{B_3}+E_i$, can be absorbed into a
redefinition of the line bundle $\mathcal{L}'=\pi^*\mathcal{L}-E_i$ entering
\eqref{eq:Weierstrass} such that
$K_{X_4}=p^*(K_{B_3}+\mathcal{L})=p^*(K_{\hat{B}_3}+\mathcal{L}')=0$. 

To describe this blow-up explicitly let us restrict to the local neighborhood 
of one irreducible curve $\Sigma_i$ of the intersection of $\mathbf{\Delta}$ and $C_0$. 
We note  that the curve $\Sigma_i$ in $B_2$ 
is given by the two constraints
\begin{equation}
	h'_1:=v s=0\ ,\quad h'_2:= g_5=0\ ,
\label{eq:curveB2}
\end{equation}
for $s$ and $g_5$ being sections of the normal bundle $N_{B_3}C_0$ and of 
$KB_3^{-6}\otimes L$, respectively. Then if $X_4$ is given as a hypersurface 
$P=0$ we obtain the blow-up as the complete intersection \cite{Griffiths:1978yf}
\begin{equation} \label{eq:4fold_blowup}
  P = 0 \ , \qquad Q = l_1 h_2' - l_2 h_1' = 0 \ ,
\end{equation} 
where we have introduced coordinates $(l_1,l_2)$ parameterizing the 
$\mathds{P}^1$-fiber of the exceptional divisor. For a detailed discussion of the blow-up 
construction we refer to section \ref{sec:geometricblowups}.

However, at least in a local description, we can introduce a local normal 
coordinate $t$ to $\Sigma_i$ in $B_2$ such that $g_5=tg_5'$ for a section 
$g_5'$ which is non-vanishing at $t=0$. Then by choosing a local coordinate 
$\ell$ of the $\mathds{P}^1$-fiber of the exceptional divisor we can solve 
the blow-up relation $Q$ of \eqref{eq:4fold_blowup} to obtain $s=\ell t$. 
This coordinate transformation can be inserted into the constraint $P=0$ 
of $X_4$ to obtain the blown-up fourfold $\hat{X}_4$ as a hypersurface\footnote{By abuse 
of notation since we directly construct the blown-up fourfold $\hat{X}_4$ 
as the F-theory dual of a heterotic setup with five-branes, cf.~section \ref{sec:SuperpotsHetF},
we denote $\hat{X}_4$ simply by $X_4$.}. 
The polynomials $f',g'$ of this hypersurface are given by
\begin{equation}
  \label{eq:f'g'}
  f' =\ell^4 f \ , \qquad g' =\ell^5 (g_5 + \ell t\, g_6 + \ldots)
\end{equation}
In particular, calculating the discriminant $\Delta'$ of $\hat{X}_4$ 
it can be demonstrated that the $I_1$ singularity no longer hits the 
$II^*$ singularity over $C_0$ \cite{Diaconescu:1999it}. This way we 
have one description of $\hat{X}_4$ as the complete intersection 
\eqref{eq:4fold_blowup} and another as a hypersurface. Both are 
of importance for the explicit examples discussed in sections 
\ref{sec:Example1}, \ref{sec:Example2} and in particular section \ref{sec:non-CYblowup}. 

To draw our conclusions of the blow-up in F-theory, we summarize what we just discussed.
The F-theory counterpart of a heterotic string with full perturbative gauge
group is given by a fourfold with $II^*$ fibers over the sections $C_0$,
$C_\infty$ in $B_3$. The component $\Delta'$ of the discriminant enhances the
degree of $\Delta$ on each intersection curve $\Sigma_i$ such that a blow-up in
$B_3$ becomes necessary. On the other hand, each blow-up corresponds to a small
instanton in the heterotic bundle \cite{Morrison:1996na,Aspinwall:1996mn}, that 
we previously identified in section \ref{sec:hettransition} as a horizontal 
five-brane on the curve $\Sigma_i$ in the heterotic threefold $Z_3$ or as an 
$SU(1)$-bundle in section \ref{sec:spectralcover}. Indeed, this agrees also
with the observation mentioned above that a vertical
component of the discriminant with degree greater than one corresponds to a
horizontal five-brane \cite{Bershadsky:1997zs} as the degree of $\Delta'$ on
$C_0$ and $C_\infty$ is two, as we saw explicitly.

We finish this discussion by a briefly looking at the moduli map between F-theory
and its heterotic dual, where we focus on the fate of the five-brane moduli in
the just mentioned blow-up process. We note that only now where we have discussed
all degrees of freedom on both sides of the duality, in particular five-branes,
we can expect a complete and consistent moduli map. Indeed if five-branes were 
excluded a possible mismatch can occur as pointed out in \cite{Rajesh:1998ik}.
The first step in establishing the moduli analysis is to relate the dimensions 
of the various moduli spaces in both theories with each other. Then, as is argued
in \cite{Rajesh:1998ik}, the relation of the fourfold Hodge numbers $h^{(3,1)}(X_4)$ 
and $h^{(1,1)}(X_4)$ counting complex structure and
K\"ahler deformations, respectively, to $h^{(2,1)}(Z_3)$, $h^{(1,1)}(Z_3)$ and
the bundle moduli and characteristic data has to be modified in the presence of
five-branes compared to the results in \cite{Andreas:1997ce,Curio:1998bva} obtained
before without five-branes. The extra contributions are due to deformation moduli of the curve
$\Sigma_i$ supporting the five-brane, counted by
$h^{0}(\Sigma_i,N_{Z_3}\Sigma_i)$, as well as the $n_5^{\text{hor}}$ blow-ups in $B_3$
that increase $h^{(1,1)}(B_3)$. Consequently, one obtains \cite{Rajesh:1998ik}
\begin{eqnarray}
	h^{(3,1)}(X_4)&=&h^{(2,1)}(Z_3)+I(E_1)+I(E_2)+h^{(2,1)}(X_4)+1+\sum_i h^{0}(\Sigma_i,N_{Z_3}\Sigma_i)\,,\nn\\
	h^{(2,1)}(X_4)&=& n_o\nn \\
		h^{(1,1)}(X_4)&=&1+h^{(1,1)}(Z_3)+\rm{rk}(G)+n_5^{\text{hor}}\,,
\label{eq:modulimap}
\end{eqnarray}
where now $X_4$ is the fourfold including the blow-ups in the base $B_3$.
Here the sum index $i$ runs over all irreducible curves $\Sigma_i$, rk$(G)$ denotes the rank of the four-dimensional 
non-abelian gauge group and $n_o$ is the number of chiral multiplets odd with respect to $\tau$, which denotes the 
involution on $Z_3$ promoted from $T^2$ mapping $y\mapsto -y$. The index $I(E_{1,2})$ counts a topological invariant of the
bundle moduli and is given by \cite{Friedman:1997yq,Andreas:1999zv}
\begin{equation}
	I(E_i) = \operatorname{rk}(E_i)+\int_{
	B_2}\left(4(\eta_i\sigma-\lambda_i)+\eta_i
	c_1(B_2)\right)\ .
	\label{eq:index}
\end{equation}
In general it counts the difference $n_e-n_o$ of even and off chiral multiplets with respect to $\tau$. 
It reads, by application of an index theorem,
\begin{equation}
	I=-\sum_0^3(-1)^i h^i(Z_3,\text{ad}(E))_e=n_e-n_o\,,
\end{equation}
where we note that the usual index $\sum_0^3(-1)^i h^i(Z_3,\text{ad}(E))$ vanishes by Serre duality 
\cite{Friedman:1997yq,Andreas:1998zf}.

The map for $h^{(3,1)}(X_4)$ reflects the fact that the
four-dimensional gauge symmetry $G$ is on the heterotic side determined by the
gauge bundle $E$ whereas on the F-theory side $G$ is due to the seven-brane
content defined by the discriminant $\Delta$ that is sensitive to a change of
complex structure. For an explicit demonstration of this map exploiting the
techniques of \cite{Berglund:1998ej} we refer to sections \ref{sec:Superpots+MirrorSymmetry} 
and \ref{sec:Example2} and our works \cite{Grimm:2009ef,Grimm:2009sy}.

Let us now discuss how \eqref{eq:modulimap} changes during the blow-up procedure. 
To actually perform the blow-up along the curve $\Sigma_i$ it is necessary to 
first degenerate the constraint of $X_4$ such that $X_4$ develops the singularity 
over $\Sigma_i$ described above. This requires a tuning of the coefficients 
entering the fourfold constraint thus restricting the complex structure of 
$X_4$ accordingly which means $h^{(3,1)}(X_4)$ is lowered. Then, we perform the 
actual blow-up by introducing the new K\"ahler class associated to the exceptional 
divisor $E_i$. Thus, we end up with a new fourfold with 
decreased $h^{(3,1)}$ and with $h^{(1,1)}(\hat{B_3})$ increased by one. This is also clear 
from the general argument \cite{Bershadsky:1997zs} that, enforcing a given gauge 
group $G$ in four dimensions in F-theory, the allowed complex structure moduli of
$X_4$, which is then singular according to $G$, have 
to respect the form of $\Delta$ dictated by the singularity type $G$. Since the 
blow-up which is dual to the heterotic small instanton/five-brane transition enhances 
the gauge symmetry $G$, the form of the discriminant becomes more restrictive, thus 
fixing more complex structures. Conversely, the blow-down can be understood as 
allowing for new complex structure moduli to be switched on that decrease the singularity 
type of the elliptic fibration.

Similarly, we can understand \eqref{eq:modulimap} from the heterotic side. For every 
small instanton transition between a smooth bundle $E$ and a five-brane, the bundle looses parts of its 
moduli since the small instanton is on the boundary of the bundle moduli space. 
Consequently, the index $I$ reduces accordingly. In the same process, the five-brane 
in general gains position moduli counted by $h^{0}(\Sigma_i,N_{Z_3}\Sigma_i)$, 
that have to be added to \eqref{eq:modulimap}.

We close the discussion of moduli by making a more refined and illustrative statement 
about the heterotic meaning of the K\"ahler modulus of the exceptional divisor $E_i$. 
To do so, we have to consider heterotic M-theory on $Z_3\times S^1/\mathds{Z}_2$, 
see section \ref{sec:hetM}. In this picture the instanton/five-brane transition can be 
understood \cite{Douglas:1996xp} as a spacetime-filling five-brane wrapping $\Sigma_i$ 
and moving onto on of the end-of-the-world branes of $S^1/\mathds{Z}_2$ where one perturbative 
$E_8$ gauge group is located. There, it dissolves into a finite size instanton of the 
heterotic bundle $E$. With this in mind the distance of the five-brane $x^{11}$ on the interval 
$S^1/\mathds{Z}_2$ away from the end-of-world brane complexified by the axion $a$ of 
the self-dual two-form, cf.~section \ref{sec:hettransition}, precisely maps to the  
K\"ahler modulus of the divisor $E_i$ resolving $\Sigma_i$ in $B_3$.

\chapter{Mirror Symmetry and Five-Branes}
\label{ch:MirrorSymm+FiveBranes}

This chapter is devoted to the study of mirror symmetry for Calabi-Yau manifolds, 
in particular for Calabi-Yau three- and fourfolds, and to mirror symmetry with D-branes. 
Since essentially all calculations of $\mathcal{N}=1$ coupling functions performed in this
work involve the use of mirror symmetry and the geometrical tools related to its study, this
chapter provides the technical core of this work.

In its weak formulation mirror symmetry states the equivalence of the complex structure 
moduli space of $X$ and the (instanton corrected) K\"ahler moduli space of its mirror
$\tilde{X}$. As was pointed out in~\cite{Witten:1991zz} this equivalence can be
formulated in physical terms by considering topological field theories called
the A- and B-model with target spaces $(\tilde{X},X)$, respectively. These theories are 
consistent cohomological truncations of some particular $\mathcal{N}=(2,2)$ superconformal field
theories. Their physical observables are the vertical subspace of the de Rham
groups\footnote{We keep our discussion as general as 
possible. For concreteness set $n=3,4$ for Calabi-Yau three- and fourfolds.} $H^{(p,p)}(\tilde X)$, 
$0\leq p\leq n$, and the horizontal subspace of $H^p(X,\bigwedge^q TX)$, $p+q=n$, respectively.  
In particular their marginal deformations coincide with the cohomology groups $H^{(1,1)}(\tilde X)$ 
for the A-model and $H^1(X,TX)$ for the B-model that are, in geometrical terms,
precisely the infinitesimal directions on the K\"ahler and complex structure
moduli space of $\tilde{X}$ and $X$, respectively. Therefore, the physical
statement of mirror symmetry is the equivalence of the A-model constructed from
$\tilde X$ and the B-model constructed from $X$. We note that for our purposes there 
is a favored point in the moduli space of marginal deformations to perform the matching between
the A- and B-model. This is the large radius point in the K\"ahler moduli space of the A-model, 
which is identified with a point of maximal unipotent monodromy in the complex structure moduli 
space of the B-model.

Our final goal is the calculation of the periods of $X$ and special holomorphic quantities 
$F^0$ and $F^0(\gamma)$ for Calabi-Yau three- and fourfolds, respectively. They are identified 
with the holomorphic superpotential of $\mathcal N=1$ effective actions, the flux superpotential 
\eqref{eq:oc_super} of a Type IIB compactification on $X$ in the threefold case and
the Gukov-Vafa-Witten superpotential \eqref{eq:GVW-super} arising in M- and F-theory compactifications. 
At the same time $F^0$ and $F^0(\gamma)$ are generating functions of the genus zero Gromov-Witten 
invariants of the mirror three- respectively fourfold. In the case of F-theory on elliptic fourfolds 
the superpotential \eqref{eq:GVW-super} is further identified with the flux and seven-brane 
superpotential in the limit \eqref{eq:SuperpotLimit} to Type IIB for particular $G_4$-flux choices. 
Thus some of the functions $F^0(\gamma)$ can be further interpreted as generating functions of 
disk instantons in a dual type IIA theory. However, although the A-model perspective is essential for 
a complete understanding of mirror symmetry, we will be rather brief and use, at large volume, the 
enumerative interpretation of the A-model side rather as a technical tool and for cross-checks of 
our B-model calculations\footnote{We refer to 
\cite{Marino:2004eq,Neitzke:2004ni,Klemm:2005tw,Vonk:2005yv} for a detailed discussion of the 
topological string on Calabi-Yau threefolds and \cite{Greene:1993vm,Mayr:1996sh,Klemm:1996ts} 
for an extension to higher dimensions.}.

Following mainly the guideline of \cite{Grimm:2009ef} and \cite{Grimm:2009sy} we begin in 
section \ref{sec:mirror_toric_branes} with the introduction of toric geometry which provides the 
central geometrical tools in order to construct a broad class of examples of Calabi-Yau 
manifolds realized as toric hypersurfaces. There we also present the toric 
realization of mirror symmetry both in the closed string case
as well as in the open string or brane case. In section \ref{sec:CSModuliSpace+PFO} we
provide a basic account on mirror symmetry for Calabi-Yau threefolds. There we focus almost
exclusively on the B-model side by reviewing the complex geometry of $Z_3$ as encoded
mathematically by the variations of Hodge structures. This yields the
concepts of special geometry and the Picard-Fuchs equations from which the 
Calabi-Yau threefold periods can be obtained. In this context we put special emphasis on the
use of the toric GKZ-system and the structure of the periods at large volume as fixed by
the A-model classical intersection data.
Then we proceed to the higher dimensional case of mirror symmetry for Calabi-Yau fourfolds 
in section \ref{sec:FFMirrors}. This involves a more systematic review of the B-model observables
and the Frobenius algebra underlying both the A- and B-model operator algebra. Finally we
use this structure to establish the mirror map for Calabi-Yau fourfolds. There we  
present, following \cite{Grimm:2009ef}, novel results from analytic continuation and global 
monodromy analysis of the periods of the Calabi-Yau fourfold that allow us to fix the integral 
basis of $H^4_H(X_4,\Z)$ and the classical terms in the periods at large volume/large complex 
structure. We conclude with section \ref{sec:EnumGeo} where we present a 
basic account of the A-model perspective on the prepotentials $F^0$, the disk amplitude and the 
generating functions $F^0(\gamma)$. In this context we emphasize the enumerative interpretation 
of the three- and fourfold flux superpotential and of the brane superpotential.

\section{Toric Calabi-Yau Hypersurfaces and Toric Branes}
\label{sec:mirror_toric_branes}

In this section we briefly introduce a basic account on toric geometry that is an inevitable tool for 
the rest of this work. We review the construction of mirror pairs $(\tilde{X},X)$ 
of toric Calabi-Yau hypersurfaces in section \ref{sec:CY_hyper}, where we introduce basic notions
of toric geometry. We  review the general formulas to obtain the Hodge numbers of 
$(\tilde{X},X)$. In section \ref{sec:toricBranes} we extend our discussion to toric branes, mainly following \cite{Hori:2000kt,Aganagic:2000gs,Aganagic:2001nx}. We discuss the relevance
of the symplectic quotient construction to conveniently visualize toric branes in toric Calabi-Yau 
threefolds and present the classical open mirror map to obtain a mirror pair of toric branes.

\subsection{Toric Mirror Symmetry}
\label{sec:CY_hyper}

A powerful tool to construct Calabi-Yau manifolds $X$ and their mirrors $\tilde X$ for an 
arbitrary complex dimension $n$ is by realizing them as hypersurfaces in toric ambient spaces. 
These hypersurfaces are specified by reflexive polyhedra \cite{Batyrev:1994hm} that encode all information
necessary to define the hypersurface. Since the cases of most interest for us are Calabi-Yau three-
and fourfolds we will restrict to these cases, but note, that all formulas given below apply for a
general complex dimension\footnote{We give the caveat that mirror symmetry in even dimension
relates the Type IIA/B with itself while it interchanges Type IIA and IIB with each other in odd
dimensions.} $n$. Threefold mirror pairs are in the following denoted as $(\tilde{Z}_3,Z_3)$.

We start our considerations for a Calabi-Yau threefold\footnote{For 
convenience we denote quantities in Type IIA always with `$\sim$' in order to omit them for 
their mirror quantities on the Type IIB side where we will mainly work.} $\tilde Z_3$ in Type IIA which is mirror dual 
to Type IIB on the mirror Calabi-Yau threefold $Z_3$. We realize the compact Calabi-Yau $\tilde Z_3$ 
as a hypersurface in a toric ambient variety $\P_{\tilde{\Delta}}$. The toric ambient space is constructed 
from a reflexive polyhedron $\Delta_4^{\tilde Z}$ that is encoded by a set of $k$ vectors $\ell^{(i)}$ forming 
a basis of relations among the $m=k+4$ vertices $\tilde{v}_j$ in $\Delta_4^{\tilde Z}$. Then the construction of the
Calabi-Yau constraint for $\tilde{Z}_3$ involves the pair of reflexive polyhedra $\Delta_4^{\tilde Z}$ 
and $\Delta_4^{Z}$ in lattices $N$, $M$, that are dual, i.e.~$\Delta_4^{\tilde Z}=(\Delta_4^Z)^\ast$. 
In general, the dual polyhedron $\Delta^\ast$ of a given polyhedron $\Delta$ in a lattice $M$ is defined 
as the set of points $p$ in the real span $N_\mathbb{R}=N\otimes \mathbb{R}$ of the dual lattice $N$ of $M$ 
such that
\begin{equation}
 	\Delta^\ast=\{p\in N_\mathbb{R}|\langle q,p\rangle\geq -1\text{ for all } q\in\Delta \}.
\end{equation}
$\P_{\tilde\Delta}$ can be represented as the quotient $(\mathbb{C}^{m}-\text{SR})//G$ defined by 
dividing out the isometry or gauge group $G=(U(1))^k$ and by imposing vanishing D-terms or moment 
maps\footnote{The symplectic quotient description is equivalent to the homogeneous coordinate
description $\tilde{x}_j\mapsto \lambda_i^{i\ell^{(i)}_j}\tilde{x}_j$ for $\lambda_i\in\C$ of the toric 
variety $\P_{\tilde\Delta}$.}, 
\begin{equation} \label{eq:ToricVariety}
 	\tilde{x}_j\mapsto e^{i\ell^{(i)}_j\phi_i}\tilde{x}_j\,,\qquad \sum_{j=1}^{k+4} \ell^{(i)}_j |\tilde x_j|^2= r^i\,.
\end{equation}
Here, the $r^i$ denote the real volumes of distinguished, effective curves in $\P_{\tilde\Delta}$. They form 
the basis of the Mori cone of curves $C_i$ each of which is associated to one charge vector $\ell^{(i)}$. 
The basis of charge vectors, thus the generators of the Mori cone, are determined by the triangulation of 
$\Delta_4^{\tilde Z}$ that correspond to the different phases of $\P_{\tilde\Delta}$ \cite{Witten:1993yc}. 
This then fixes the form of the D-term constraints in \eqref{eq:ToricVariety} from which the Stanley-Reisner 
ideal SR can be read off \cite{Denef:2008wq}. 
Geometrically the toric variety can be viewed as follows. Upon solving the second condition in \eqref{eq:ToricVariety} 
and using coordinates $p_j=|\tilde x_j|^2$ and angles $\theta_j$ the toric variety $\P_{\tilde\Delta}$ 
can be visualized as a $T^{m-k}$-fibration over a real $(m-k)$-dimensional base $\mathbb{B}_{m-k}$ \cite{Leung:1997tw, Aganagic:2000gs}. 
The degeneration loci of the $T^{m-k}$-fibration where one or more $S^1$ shrink are on the boundary of 
$\mathbb{B}_{m-k}$ which is determined by $p_j=0$ or intersections thereof since $p_j\geq 0$.

The action of $G$ on the coordinates is generated infinitesimally 
by $k$ vector fields
\begin{equation}
 	V^{(i)}=\sum_j\ell^{i}_j\tilde{x}_j\frac{\partial}{\partial \tilde{x}_j}\,.
\end{equation}
Using these vector fields it is straight forward to construct forms on $\P_{\tilde\Delta}$ from invariant forms 
of $\C^{k+4}$. Of particular importance is the holomorphic top-form $\Delta_{\P_{\tilde\Delta}}$ on $\P_{\tilde\Delta}$ 
that is constructed from the holomorphic top-form $\Delta_{\C}=d\tilde{x}_1\wedge\ldots\wedge d\tilde{x}_{k+4}$ 
on $\C^{k+4}$ as
\begin{equation} \label{eq:toricMeasure}
 	\Delta_{\P_{\tilde\Delta}}=V^{(1)}\lrcorner \ldots\lrcorner V^{(k)}\lrcorner (\Delta_{\C})\,,
\end{equation}
where $\lrcorner$ denotes the interior product defined by contracting a form with a vector. For the example 
of projective space $\P^n$ this yield the unique holomorphic section of 
$\Omega_{\P^n}^n(n+1):=\Omega_{\P^n}^n\otimes\mathcal{O}(n+1)$ given by
\begin{equation}\label{eq:MeasurePn}
 	\Delta_{\P^n}=\sum_{i=1,n+1}(-1)^{i-1}x_i dx_1\wedge\ldots \widehat{dx_i}\ldots \wedge dx_n\,,
\end{equation}
where the $x_i$ denote the homogeneous coordinates on $\P^n$ and $\widehat{dx_i}$ indicates, that $dx_i$ is 
omitted. 

The Calabi-Yau threefold $\tilde Z_3$ is then given as the hypersurface $\{\tilde P=0\}$ in 
$\P_{\tilde{\Delta}}$ where $\tilde P$ is given as the polynomial \cite{Batyrev:1994hm} 
\begin{equation} \label{eq:Z3typeIIA}
 	\tilde P(\underline{\tilde{x}},\underline{\tilde{a}})=\sum_{q\in \Delta^Z_4\cap M}\tilde a_q\prod_{i}\tilde x_i^{\langle \tilde v_i,q\rangle+1}
\end{equation}
in the $m$ projective coordinates $\underline{\tilde{x}}$ of $\P_{\tilde{\Delta}}$ associated to each vertex $\tilde v_j$. 
This formula provides a direct way to count the number of complex structure parameters $\underline{\tilde{a}}$ 
(up to automorphisms of $\P_{\tilde{\Delta}}$) by counting the integral points $q\in\Delta_4^Z$.
Furthermore, $\tilde Z_3$ is Calabi-Yau since \eqref{eq:Z3typeIIA} contains the monomial 
$\tilde x_1\ldots\tilde x_m$ corresponding to the origin in $\Delta_4^Z$ so that $\tilde{P}$ is a 
section of the anti-canonical bundle $K_{P_{\tilde{\Delta}}}^\ast=\mathcal{O}(\sum_i D_i)$, 
where $D_i=\{\tilde x_i=0\}$ denotes a toric divisor.

For the case of hypersurfaces in toric varieties, (closed string) mirror symmetry \cite{Batyrev:1994hm} 
is realized in a very elegant way. The mirror threefold $Z_3$ on the Type IIB side is obtained 
by simply interchanging the roles of $\Delta_4^{\tilde Z}$ and $\Delta_4^Z$ so that \eqref{eq:Z3typeIIA} 
describes $Z_3$ as the hypersurface in the toric variety $\P_{\Delta}$ associated to the polyhedron $\Delta_4^Z$, 
\begin{equation} \label{eq:Z3typeIIB}
 	 P(\underline{x},\underline{a})=\sum_{p\in \Delta^{\tilde Z}_4\cap N}a_p\prod_{i}x_i^{\langle v_i,p\rangle+1}.
\end{equation}
Here, we again associated the projective coordinates $x_i$ to each vertex $v_i$ of $\Delta_4^Z$.  
In general there can be a discrete orbifold symmetry group $\Gamma$ such that $Z_3$ is an orbifolded 
hypersurface. 
Indeed, the necessary requirements for mirror symmetry, $h^{(1,1)}(Z_3)=h^{(2,1)}(\tilde Z_3)$ and 
$h^{(2,1)}(Z_3)=h^{(1,1)}(\tilde Z_3)$, are fulfilled for this construction. This is obvious from 
Batyrev's formula for the Hodge numbers \cite{Batyrev:1994hm} of a given $n$-fold $X$ in a toric 
ambient space specified by an $n+1$-dimensional polyhedron $\Delta_{n+1}^{X}$ and its dual $\Delta_{n+1}^{\tilde{X}}$
\begin{eqnarray} 
  h^{(n-1,1)}(X_n) &=& 
h^{(1,1)}(\tilde X)\label{eq:Hodgenumbers1}\\ &=&
\ l(\Delta_{n+1}^{\tilde{X}}) - (n+2) -\sum_{\dim \tilde\theta =n} l'(\tilde\theta)+
              \sum_{\text{codim} \tilde\theta_i =2} l'(\tilde \theta_i) l'(\theta_i)\ ,\nn\\
  h^{(1,1)}(X_n) &=& h^{(n-1,1)}(\tilde X_n)  \label{eq:Hodgenumbers2}\\ &=&  
\ l(\Delta_{n+1}^X) - (n+2) -\sum_{\dim \theta =n} l'(\theta)+
              \sum_{\text{codim}  \theta_i =2} l'(\theta_i) l'(\tilde \theta_i)\ .\nn
\end{eqnarray}
In this expression $\theta$ ($\tilde \theta$)
denote faces of $\Delta^X_{n+1}$ ($\Delta^{\tilde X}_{n+1}$), while the sum is over 
pairs $(\theta_i,\tilde \theta_i)$ of dual faces. The $l(\theta)$ and $l'(\theta)$ count 
the total number of integral points of a face $\theta$ and the number inside the face $\theta$, 
respectively. Finally, $l(\Delta)$ is the total number of integral points in the polyhedron $\Delta$. 
Using these formulas one notes that polyhedra with a small 
number of points will correspond to Calabi-Yau manifolds with few K\"ahler moduli, i.e.~small 
$h^{(1,1)}$, and many complex structure moduli $h^{(n-1,1)}$. Since $h^{(1,1)}$ and
$h^{(n-1,1)}$ are exchanged by mirror symmetry Calabi-Yau manifolds with small $h^{(n-1,1)}$ are 
obtained as mirror manifolds of hypersurfaces specified by a small number of lattice points in the polyhedron. 

For the case of Calabi-Yau fourfolds $(\tilde{X}_4,X_4)$ the complete list of model dependent Hodge numbers is
$h^{(1,1)}(X_4)$, $h^{(3,1)}(X_4)$, $h^{(2,1)}(X_4)$ and $h^{(2,2)}(X_4)$, However only three of these are independent 
due to the Hirzebruch-Riemann-Roch index theorem implying \cite{Klemm:1996ts} 
\begin{equation} \label{eq:HodgeRel}
 	h^{(2,2)}(X_4)=2(22 + 2h^{(1,1)}(X_4) + 2h^{(3,1)}(X_4) - h^{(2,1)}(X_4))\,. 
\end{equation}
Therefore, only $h^{(2,1)}(X_4)$ has to be calculated in addition to fix the basic topological data of 
$(\tilde{X}_4,X_4)$. Analogously to \eqref{eq:Hodgenumbers1} it is readily given by the mirror symmetric expression
\begin{equation} \label{eq:Hodgenumbers3}
	h^{(2,1)}(X_4)= h^{(2,1)}(\tilde{X}_4)=\sum_{\text{codim} \tilde\theta_i =3}l'(\tilde \theta_i) l'(\theta_i)\,.
\end{equation}
This finally enables us to calculate the Euler number of fourfolds by
\begin{equation}\label{eq:FFEulerNumb}
 	\chi(X_4)=\chi(\tilde{X}_4)=6(8+h^{(3,1)}+h^{(1,1)}-h^{(2,1)})\,.
\end{equation}

\subsection{Toric Branes}
\label{sec:toricBranes}

Now we are prepared to add an open string sector to the Calabi-Yau threefold mirror pair $(\tilde{Z}_3,Z_3)$.
We will consider supersymmetric branes and their deformations. In local toric Calabi-Yau threefolds mirror 
symmetry with calibrated branes, so-called Harvey-Lawson branes, has been studied intensively in 
\cite{Aganagic:2000gs, Aganagic:2001nx}. In this case the toric variety $\P_{\tilde{\Delta}}$ itself defines 
a non-compact Calabi-Yau threefold. To fix our notation to include this case we will work in the following 
with a toric space of dimension $m-k$ denoted by $\P_{\tilde{\Delta}}$, where $m-k=3$ in the 
non-compact case, and $m-k=4$ in the compact case. 

In the type IIA theory supersymmetric branes wrap special Lagrangian cycles $L$. In the toric ambient space 
$\P_{\tilde\Delta}$ one describes such a three-cycle $L$ by $r$ additional, so-called brane charge vectors 
$\hat \ell^{(a)}$ restricting the $|\tilde{x}_j|^2$ and their angles $\theta_i$ as \cite{Aganagic:2000gs}
\begin{equation} \label{eq:ABranes}
 	\sum_{j=1}^{k+4} \hat\ell^{(a)}_j |\tilde x_j|^2=c^a\ ,\qquad \theta_i=\sum_{a=1}^r\hat\ell^{(a)}_i\phi_a\ ,
\end{equation}
for angular parameters $\phi_a$. To fulfill the `special' condition of $L$, which is 
equivalent to $\sum_i\theta_i=0$, one demands $\sum_j\hat\ell_j^{(a)}=0$. 
Wrapping a brane on such a cycle we obtain a so-called toric brane, in this case a toric D6-brane,
a specific brane type well studied in the context of toric mirror symmetry in non-compact Calabi-Yau manifolds 
\cite{Aganagic:2000gs, Aganagic:2001nx}. The great advantage of the description of toric branes in terms 
of toric data similar to the charge vectors $\hat{\ell}^{(i)}$ is the technically straightforward construction
of the mirror branes, as presented below.

However, before discussing the mirror side we comment on the non-compact case. 
For non-compact Calabi-Yau threefolds 
$\tilde Z_3=\mathbb{C}^{k+3}//G$ the Calabi-Yau condition 
$\sum_j\ell^{(i)}_j=0$ has to hold. 
Then A-branes introduced by \eqref{eq:ABranes} are graphically represented 
as real codimension $r$ subspaces of the toric base $\mathbb{B}_{3}$. 
The case which was considered for the non-compact examples in 
\cite{Aganagic:2000gs, Aganagic:2001nx} is $r=2$ where the non-compact three-cycle $L$ 
is represented by a straight line ending on a point when projected onto the base $\mathbb{B}_3$.
The generic fiber is a $T^2$ so that the topology of $L$ is just $\mathbb{R}\times S^1\times S^1$. 
However, upon tuning the moduli $c^a$ it is most convenient to move the $L_a$ to the boundary 
of $\mathbb{B}_3$ where two $\{p_j=0\}$-planes intersect. Then one of the two moduli is frozen, and 
one $S^1$ pinches such that the topology becomes $\mathbb{C}\times S^1$. 
These D6-branes are mirror dual to non-compact D5-branes which intersect a Riemann surface at a point. 
We note, that this is the local geometry we encounter in section \ref{sec:Superpots+MirrorSymmetry}.

Let us now turn to the mirror Type IIB description \cite{Aganagic:2000gs,Hori:2000kt}. To include also
non-compact Calabi-Yau threefolds and D-branes into the setup it is convenient to use another representation
of the mirror Calabi-Yau $Z_3$, 
\begin{equation} \label{eq:HVmirror}
 	P=\sum_{j=0}^m a_jy_j\ ,\qquad \prod_{j=0}^m (a_jy_j)^{\ell^{(i)}_j}=z^i\ ,\qquad i=1,\ldots, k\ ,
\end{equation}
Here the $z^i$ denote the complex structure moduli of $Z_3$ that are related to the complex numbers 
$\underline{a}$ by the second relation in \eqref{eq:HVmirror}. Classical mirror symmetry is established by the
relation $z^i=e^{2\pi it^i}$ to the (complexified) K\"ahler moduli $t^i$ of $\tilde Z_3$.
We note that we introduced a further coordinate $y_0$ in \eqref{eq:HVmirror} for which we also have 
to include a zeroth component of the charge vectors as $\ell^{(i)}_0=-\sum^m_{j=1} \ell^{(i)}_j$.  
The relation to the constraint \eqref{eq:Z3typeIIB} is established by identifying the $y_i$ with monomials 
$m_i(\underline{x})$ in the coordinates $\underline{x}$, $y_i\mapsto m_i(\underline{x})$, also known as the etal\'e map, 
that are associated to the integral points $\tilde{v}_i$ in $\Delta_4^{\tilde Z}$ by the mirror construction of Batyrev 
\cite{Hosono:1993qy,Batyrev:1994hm}. Then the two representations \eqref{eq:Z3typeIIB} and \eqref{eq:HVmirror} 
for $Z_3$ agree for the case of compact Calabi-Yau mirror pairs $(\tilde{Z}_3,Z_3)$.
In the non-compact case, $Z_3$ is similarly given as
\begin{equation}\label{eq:localMirror}
 	xz=P(\underline{x})\,,
\end{equation}
where we emphasize the use affine coordinates $x,z$ of $\mathds{C}$ and $x_i\in\mathbb{C}^*$, respectively.

Finally, toric holomorphic submanifolds $\Sigma$ in $\P_{\Delta}$, that support 
calibrated B-branes mirror dual to the A-branes in \eqref{eq:ABranes}, are specified by the constraints 
\begin{equation} \label{eq:BBrane}
 	\prod_{j=0}^m (a_jy_j)^{\hat\ell^{(a)}_j}=u^a,\quad a=1,\ldots, r\ . 
\end{equation}
The classical open mirror map takes the form $u^a=\epsilon^ae^{-c^a}$, where the phases 
$\epsilon^a$ are dual to the Wilson line background of the flat U(1)-connection on 
the special Lagrangian $L$ and complexify the moduli $c^a$ in \eqref{eq:ABranes} to the open moduli $u^a$ 
\cite{Strominger:1996it}. This can also be re-expressed in terms of the coordinates 
$\underline{x}$ using again the etal\'e map $y_i\mapsto m_i(x_j)$. Intersected with the Calabi-Yau 
constraint $P=0$ the constraint \eqref{eq:BBrane} describes families of submanifolds of codimension $r$ in $Z_3$.
For the configuration $r=2$, $\Sigma$ is a curve in $Z_3$ that we call a toric curve and 
which is precisely the geometry we are interested in for the study of five-branes.
Other cases can be considered as well leading to mirrors given by D7-branes 
on divisors ($r=1$) or D3-branes on points ($r=3$).

\section{Mirror Symmetry for Calabi-Yau Threefolds}
\label{sec:CSModuliSpace+PFO}

In this section we discuss, in the light of mirror symmetry, the geometric structure underlying 
the complex structure moduli space of a Calabi-Yau threefold $Z_3$. Here we restrict almost 
entirely to the B-model side which is the main focus in this work in general. We refer for example 
to~\cite{Marino:2004eq,Neitzke:2004ni,Klemm:2005tw,Vonk:2005yv} for a more detailed discussion 
including the A-model and symplectic geometry side. We begin in section \ref{sec:csModuli} by 
reviewing the description of deformations of complex structures as found in standard textbooks like 
\cite{kodaira2005complex,huybrechts2005complex} or in the original works 
\cite{kodaira1958deformations,kodaira1958deformationsII}. Then in section \ref{sec:HodgeStructure}
we discuss how this analysis is captured by variations of Hodge structures applied 
to the Calabi-Yau case. This yields differential relations among the periods of a
Calabi-Yau threefold, that lead to special geometry, reviewed in section \ref{sec:specialGeometry}, 
and finally to Picard-Fuchs equations that govern the complex structure dependence 
of the periods. We present the basic techniques in order to obtain the Picard-Fuchs equations both 
in the generic hypersurface case as well as in the toric case in section \ref{sec:PFO+3foldMirrorSymmetry}. 
We conclude with the structure of the solutions to the Picard-Fuchs system at the point of maximal 
unipotent monodromy \cite{morrison1998picard} and the identification of those linear combinations of
solutions that correspond to the periods of $\Omega$ with respect to an integral symplectic
basis of $H^3(Z_3,\Z)$. As we demonstrate this is the only occasion where the comparison with the
A-model is needed.

\subsection{Deformations of Complex Structures} \label{sec:csModuli}

We consider a complex manifold $Z_3$ as a real manifold equipped with a background 
complex structure $I:TZ_3\rightarrow TZ_3$ with $I^2=-\id$ for which the Nijenhuis-tensor 
$N$ vanishes ensuring integrability\footnote{In the following we denote the (anti-)holomorphic 
tangent bundle just by ($\overline{TZ_3}$) $TZ_3$.} $[T^{(0,1)}Z_3,T^{(0,1)}Z_3]\subset T^{(0,1)}Z_3$. 
The complex structure determines the Dolbeault operator $\bar{\partial}$ and vice versa. 
Finite deformations of $I$ are described by elements $A(\underline{t})$ in $\Omega^{(0,1)}(Z_3,TZ_3)$ 
that are $(0,1)$-forms taking values in $TZ_3$ and depend on parameters $\underline{t}$ denoting 
coordinates on a parameter manifold\footnote{We choose $M$ so that the Kodaira-Spencer map 
$T_0M\rightarrow H^1(Z_3,TZ_3)$ is bijective at a point $0\in M$.} $M$. In fact, if we perturb 
$\bar{\partial}$ by $A$ and demand $(\bar{\partial}+A)^2=0$, the deformation $A$ has to obey 
the Maurer-Cartan equation
\begin{equation}\label{eq:MaurerCeq}
 	\bar\partial A+\frac12[A,A]=0\,.
\end{equation}
If we write $A$ as a formal power series in $t$, $A=A_1(\underline{t})+A_2(\underline{t})+\ldots$ 
we obtain 
\begin{equation} \label{eq:MaurerCorder}
	\bar{\partial} A_1=0\,,\qquad \bar{\partial} A_n+\frac12\sum_{i=1}^{n-1}[A_i,A_{n-i}]=0\,,\, n>1\,,
\end{equation}
where $A_n(\underline{t})$ denotes a homogeneous polynomial in $t$ of degree $n$.
Taking coordinate transformations into account that trivially change the complex structure $I$, 
we learn that first order or infinitesimal deformations of $I$ for which $t\sim 0$ are in one-to-one 
correspondence with classes $[v]=[A_1]$ in the cohomology group $H^1(Z_3,TZ_3)$, called the 
Kodaira-Spencer class of $A_1$. However, deformation classes $A_1$ lift to finite deformations 
only if we can recursively solve \eqref{eq:MaurerCorder} for the $A_n$ at every finite order in $n$. 
We immediately identify the necessary condition for $[A_1]=[v]$ being integrable with the integrability 
condition
\begin{equation}\label{eq:obstructionclass}
\bar \partial A_2=-\frac12[A_1,A_1],
\end{equation}
which means that $[A_1,A_1]$ has to be $\bar\partial$-exact in order to find a solution. Thus, one 
associates to every class $v$ in $H^1(Z_3,TZ_3)$ the class $[v,v]$ in $H^2(Z_3,TZ_3)$ on the right 
hand side of \eqref{eq:obstructionclass} called the obstruction class. It necessarily has to vanish 
in order to define a finite $A(\underline{t})$ obeying \eqref{eq:MaurerCorder} with $[A_1]=[v]$. 
This is in particular the case if $H^2(Z_3,TZ_3)=0$. However, this is not a necessary condition for 
the existence of $A(\underline{t})$ since the obstruction classes for integrating an infinitesimal 
$A_1$ can be zero even for $H^2(Z_3,TZ_3)\neq 0$. Indeed, this is the generic case for Calabi-Yau 
manifolds, where $h^2(Z_3,TZ_3)=h^{(1,1)}(Z_3)$. It is the content of the classical theorem by Tian 
and Todorov that for every Calabi-Yau manifold all commutators $\sum_i[A_i,A_{n-i}]$ in the recursive 
equations \eqref{eq:MaurerCorder} are exact so that a finite $A(\underline{t})$ exists for every 
infinitesimal deformation $[A_1]=[v]$. Thus, complex structure deformations of a Calabi-Yau manifold 
are generically unobstructed such that there is a global moduli space of complex structures of complex 
dimension\footnote{Here we use the isomorphism $H^1(Z_3,TZ_3)= H^{(2,1)}(Z_3)$ by contraction with the 
$(3,0)$-form $\Omega$.} $h^{(2,1)}$. In physics, this is reflected by the fact that, as long as 
background fluxes are absent, there is no scalar potential in the effective theory of a Calabi-Yau 
compactification for the fields $t(x)$ associated to complex structure deformations.

\subsection{Hodge Structures for Picard-Fuchs Equations} \label{sec:HodgeStructure}

For Calabi-Yau manifolds the infinitesimal study of the complex structure moduli 
space can be carried out by the study of the variation of the Hodge structure on its 
cohomology groups\footnote{See e.g.~\cite{griffiths1970summary,Green:1993qv,Voisin2002} 
for a review and a discussion of the local Torelli theorem underlying this statement.}.
This analysis is particularly convenient since there is an unique non-vanishing holomorphic 
three-form $\Omega$ on a Calabi-Yau manifold. On the one hand, this enables us to map the 
infinitesimal deformations in $H^1(Z_3,TZ_3)$ simply to forms in $H^{(2,1)}(Z_3)$ which 
are identified with distinguished periods $X^i$. On the other hand, the change of the Hodge 
type of $\Omega$ under a complex structure deformation allows us to study the variation of 
the Hodge structure explicitly. We note that we will later apply the variations of Hodge structures 
in section \ref{sec:hatomega} to also study the complex structure deformations of projective K\"ahler 
threefolds with $h^{(3,0)}=1$, that are a blow-up of a Calabi-Yau threefold along a curve $\Sigma$.

If we consider $H^3(Z_3)$ over every point of the complex structure moduli space $\mathcal M^{\rm{cs}}$, we
obtain a holomorphic vector bundle over $\mathcal M^{cs}$ which we denote by $\mathcal H^3(Z_3)$. It is endowed with
a locally constant frame given by $H^3(Z_3,\Z)$. We define a decreasing filtration on $H^3(Z_3)$, the Hodge filtration, 
which equips $H^3(Z_3)$ with a (pure) Hodge structure 
\begin{eqnarray}
        \nn F^3H^3(Z_3) &=& H^{(3,0)}(Z_3)\ ,\\
        \nn F^2H^3(Z_3) &=& H^{(3,0)}(Z_3)\oplus H^{(2,1)}(Z_3)\ ,\\
        \nn F^1H^3(Z_3) &=& H^{(3,0)}(Z_3)\oplus H^{(2,1)}(Z_3)\oplus H^{(1,2)}(Z_3)\ ,\\
        F^0H^3(Z_3) &=& H^{(3,0)}(Z_3)\oplus H^{(2,1)}(Z_3)\oplus H^{(1,2)}(Z_3)\oplus H^{(0,3)}(Z_3)=H^3(Z_3)\ ,
        \label{eq:closed-hodge-filtration}
\end{eqnarray}
where we recover the familiar decomposition of the de Rham group $H^3(Z_3)$ into $(p,q)$-forms for K\"ahler manifolds.
This filtration is decreasing since $F^mH^3(Z_3)$ is contained in $F^{m-1}H^3(Z_3)$ for all $m$.
It is convenient to study the filtration $F^mH^3(Z_3)$ instead of the individual groups $H^{(p,q)}(Z_3)$ because the $F^mH^3(Z_3)$ 
form a holomorphic subbundle $\mathcal F^m$ of $\mathcal H^3(Z_3)$ \cite{Griffiths:1968,Griffiths:1968II,Griffiths:1969fr,Griffiths:1969frII}, 
but $H^{(p,q)}(Z_3)$ do not. The bundle $\mathcal H^3(Z_3)$ has a flat connection $\nabla$ which is called the 
Gau{\ss}-Manin connection \cite{griffiths1970summary}. It has the so-called Griffiths transversality property 
\cite{Griffiths:1968,Griffiths:1968II}
\begin{equation}
        \nabla\mathcal F^m\subset \mathcal F^{m-1}\otimes \Omega^1_{\mathcal M^{\rm{cs}}}\ .
        \label{eq:closed-griffiths-transversality}
\end{equation}

This together with $h^{(3,0)}=1$ is one of the main ingredients for the formulation of the $\mathcal N=2$ special geometry for 
Calabi-Yau manifolds, as we will see below. Starting in $F^3H^3(Z_3)=H^{(3,0)}{Z_3}$ generated by $\Omega$, we can study the 
variation of the complex structure by analyzing how $\Omega$ changes under complex structure deformations $\delta z$ in the diagram
\begin{equation}
	H^{(3,0)}(Z_3)=\mathcal{F}^3\stackrel{\nabla}{\rightarrow} \mathcal{F}^2
	\stackrel{\nabla}{\rightarrow}\mathcal{F}^1\stackrel{\nabla}{\rightarrow} \mathcal{F}^0=H^3(Z_3)\,.
\label{eq:VariationsOfHodgeStructure}
\end{equation}
This diagram encodes the variations of Hodge 
structures \cite{Voisin2002} with respect to complex structure deformations $\delta z$. The form $\Omega$ and its derivatives 
$\nabla^k\Omega$, $k\leq 3$, span the spaces $F^3H^{3-k}(Z_3)$ and thus the complete space $H^3(Z_3)=F^0H^3(Z_3)$.
Consequently a fourth order derivative of any element of $H^3(Z_3)$ is expressible as a linear combination of 
$\nabla^{k}\Omega$ for $k\leq 3$. These linear relations between the derivatives of $\Omega$ up to fourth order yield 
the Picard-Fuchs equations which are for this very reason of maximal order four, see e.g.~\cite{morrison1998picard} for 
a mathematically thorough presentation of one-modulus Calabi-Yau hypersurfaces.

\subsection{Calabi-Yau Threefold Special Geometry} \label{sec:specialGeometry}

Next we exploit the special geometrical properties of $\mathcal{M}_{\text cs}$ which reflect the 
structure of the variations of Hodge structures in terms of differential relations among the periods
of the Calabi-Yau threefold $Z_3$. In particular it is special to the threefold case \cite{Candelas1991}
that there exists a holomorphic prepotential $F^0$ governing the structure of all periods. This is denoted as 
special geometry or special K\"ahler geometry. From the physics point of view this is a consequence 
of $\mathcal{N}=2$ supersymmetry in the four dimensional effective theory, since in general the 
vector multiplet moduli space of any $\mathcal{N}=2$ theory is encoded by a prepotential 
\cite{deWit:1984pk,Cremmer:1984hj,deWit:1984px}.

We start by exploiting Griffiths transversality \eqref{eq:closed-griffiths-transversality} to obtain
$\frac{\partial}{\partial z_i}\Omega=c_i(z)\Omega+\chi_i$ for a function $c_i(z)$ of the complex
structure moduli and a basis $\chi_i$ of $H^{(2,1)}(Z_3)$. This implies 
\begin{equation}
	0=\int_{Z_3}\Omega\wedge \frac{\partial}{\partial z_i}\Omega\,.
\label{eq:GriffithsTransversalityRelation}
\end{equation}
Recalling the period expansion $\Omega=X^K\alpha_K-\mathcal{F}_K\beta^K$ we note that the
basis $(\alpha_K,\beta^K)$ of $H^3(Z_3,\Z)$ can be chosen such that the $X^K$ form projective coordinates
on $\mathcal{M}_{\rm cs}$, where we use that $\Omega$ is only defined up to multiplication with $e^f$ for a 
function $f$ on the complex structure moduli space\footnote{More precisely $\Omega$ is a section of the 
vacuum line bundle $\mathcal{L}$ over $\mathcal{M}_{\rm cs}$, see \cite{Hori:2003ic}.}. Mathematically 
the existence of these coordinates reflects the isomorphism $H^1(Z_3,TZ_3)\cong H^{(2,1)}(Z_3)$
and the local Torelli theorem. Then since the dimension of $\mathcal{M}_{\rm cs}$ is $h^{(2,1)}$ the periods 
$\underline{\mathcal{F}}$ have to be functions of the $\underline{X}$ and can be written, again by using Griffiths 
transversality \eqref{eq:GriffithsTransversalityRelation}, as derivatives of a single function
\begin{equation}
	\mathcal{F}_L=\frac12\frac{\partial}{\partial X^L}(X^K\mathcal{F}_K)\equiv \frac{\partial}{\partial X^L}
	\mathcal{F}\,.
\label{eq:prepotProjective}
\end{equation}
This is the projective prepotential $\mathcal{F}$ which is a homogeneous (local) function of degree two in the $\underline{X}$
and consequently a section of the square of the vacuum line bundle $\mathcal{L}^2$ over $\mathcal{M}_{\rm cs}$\footnote{
The integrability condition $\partial_K\mathcal{F}_L-\partial_L\mathcal{F}_K=0$ follows from the special structure of 
the Riemann tensor that is characteristic for special K\"ahler geometry, see e.g.~\cite{Klemm:2005tw}.}. We divide 
by one coordinate, say $X^0$, to define affine coordinates $t^a=X^a/X^0$, $a\neq 0$, and introduce the 
prepotential
\begin{equation}
	F^0(\underline{t})=(X^0)^{-2}\mathcal{F}(\underline{X})\,.
\label{eq:prepotAffine}
\end{equation}
It will be the special coordinates $\underline{t}$ that are relevant in the context of mirror symmetry 
as we will discuss below and more generally in section \ref{sec:bmodel}.
According to \eqref{eq:closed-griffiths-transversality} the first non-trivial pairing are the Yukawa couplings
\cite{greensuperstring,polchinski1998string}, that take in the normalization $\Omega\rightarrow \frac{1}{X^0}\Omega$ 
the form
\begin{equation}
	C_{ijk}(\underline{t})=\int_{Z_3}\Omega\wedge \frac{\partial^3}{\partial t^i\partial t^j\partial t^k}
	= \frac{\partial^3}{\partial t^i\partial t^j\partial t^k}F^0(\underline{t})
\label{eq:YukawaCouolingCY3}
\end{equation}
for $i,j,k=1,\ldots h^{(2,1)}$, that form sections of the bundle $Sym^3T\mathcal{M}_{\rm cs}\otimes \mathcal{L}^2$.

Next we study the hermitian geometry of $\mathcal{M}_{\rm cs}$ and $\mathcal{L}$, emphasizing aspects of special geometry.
Firstly, we note that there is a natural hermitian metric on the vacuum line bundle $\mathcal{L}$ spanned by $\Omega$
\cite{Hori:2003ic},
\begin{equation}
	h=i\int\Omega\wedge\bar\Omega\,.
\label{eq:HermitMetricL}
\end{equation}
Here we used a standard construction in mathematics to obtain the hermitian metric of a line bundle $\mathcal{L}$
once sections $s_i$ generating $\mathcal{L}$ everywhere are specified, cf.~\cite[p.~166]{huybrechts2005complex}.
As we have noted before in \eqref{eq:csmetricOrie} the metric $h$ also defines the K\"ahler potential on 
the complex structure moduli space $\mathcal{M}_{\rm cs}$ itself, that reads
\begin{equation}
	K_{\rm cs}=-\log(i\int\Omega\wedge \bar \Omega)=-\log i(\bar{X}^K\mathcal{F}_K-X^I\bar{\mathcal{F}}_I)
	=-\log i\norm{X^0}^2((t-\bar t)^i(F+\bar F)_i+2\bar{F}-2F)\,,
\label{eq:KaehlerpotCS}
\end{equation}
where we used the period expansion of $\Omega$.
Similarly $h$ is used to construct the hermitian/Chern connection and the Chern curvature form $R$ on the 
holomorphic line bundle $\mathcal{L}$, cf.~\cite[p.~177,p.~186]{huybrechts2005complex}, that takes the form
\begin{equation}
		A_i=\partial_i\log(h)=-\partial_i K_{\rm cs}=\frac{i}{h}(\bar{X}^K\partial_i\mathcal{F}_K-\bar{\mathcal{F}}_{\bar i})\,,
		\quad R_{i\bar \jmath}=(\bar{\partial}\partial \log(h))_{i\bar{\jmath}}=\partial_i\bar{\partial}_{\bar{\jmath}}K_{\rm cs}\,.
\label{eq:Connection+CurvatureL}
\end{equation}
The existence of the bundle $\mathcal{L}$, whose curvature form is simultaneously the K\"ahler form on its base 
manifold is one central property of $\mathcal{N}=2$ special geometry\footnote{A very simple example is projective 
space $\P^n$ with the Fubini-Study metric that is simultaneously the curvature tensor on $\mathcal{O}_{\P^n}(1)$.}.
With these definitions at hand we can determine the constant $c_i(z)$ in $\frac{\partial}{\partial z_i}\Omega$ as
\begin{equation}
	\frac{\partial}{\partial X^i}\Omega=\chi_i-\frac{\partial}{\partial X^i}K_{\rm cs}\Omega\,,\qquad c_i=-\partial_i K_{\rm cs}\,.
\label{eq:CSDerivativeOmega}
\end{equation} 

We conclude by presenting the consequences of the existence of the prepotential $F^0$ for the structure of
the period vector $\Pi:=(\underline{X},\underline{\mathcal{F}})$. Upon normalizing $\Omega$ by $X_0$ the
period vector can be expressed as
\begin{equation}
	\Pi=X^0\begin{pmatrix}
						1\\
						t^i\\
						\partial_i F^0\\
						2F^0-t^i\partial_i F^0
				\end{pmatrix}\,.
\label{eq:periodVectorCY3}
\end{equation}
We note that the period vector is only defined up to symplectic transformations in the fundamental representation of 
$Sp(b^3,\Z)$ that leave the pairing \eqref{eq:HermitMetricL} invariant. Geometrically this corresponds to the freedom in
the choice of symplectic basis of $H^3(Z_3,\Z)$. Thus $\Pi$ forms a section of a symplectic vector bundle over 
$\mathcal{M}_{\rm cs}$ of rank $b^3$. Since this bundle is flat the remaining gauge freedom will lead to a monodromy 
action $\Pi\mapsto M\Pi$, $M\in Sp(b^3,\Z)$ around singularities in $\mathcal{M}_{\rm cs}$, which are the regular singular 
points of the Picard-Fuchs equations \cite{griffiths1970summary} introduced next.

\subsection{Picard-Fuchs Equations, Toric GKZ-Systems and the Mirror Map} \label{sec:PFO+3foldMirrorSymmetry}

In the context of toric hypersurface the structure implied by the variations of 
Hodge structures \eqref{eq:VariationsOfHodgeStructure} can be explicitly studied. 
In fact in the case of an algebraic representation $P(\underline{x};\underline{a})=0$ 
of a family of Calabi-Yau hypersurfaces residue integral expressions for the 
three-form $\Omega$ and its periods are known. These can be used to find Picard-Fuchs 
equations governing their complex structure dependence. We start by presenting a 
very brief review of the general techniques. 

The explicit analysis of the holomorphic three-form $\Omega$ is based on the 
algebraic representation of $\Omega$ and its periods $\Pi^k$ by the residue 
integral expressions \cite{Griffiths:1969fr,Griffiths:1969frII,griffiths1970summary,Griffiths:1978yf}
\begin{equation} \label{eq:residueZ3}
 	\Omega(\underline{z})=\int_{S^1_P}\frac{\Delta_{\mathbb{P}_{\Delta}}}{P(\underline{x},\underline{z})}\,,
 	\quad \Pi^k(\underline{z})=\int_{\Gamma^k\times {S^1_P}}\frac{\Delta_{\mathbb{P}_{\Delta}}}{P(\underline{x},\underline{z})}\,,
\end{equation}
where $Z_3$ is given as the zero locus of a polynomial constraint $P(\underline{x},\underline{z})$ 
in coordinates $\underline{x}$ in a toric ambient space $\mathbb{P}_{\Delta}$ and the holomorphic 
top-form $\Delta_{\mathbb{P}_{\Delta}}$ was introduced in \eqref{eq:toricMeasure}. 
Here $\Gamma^k$, $k=1,\ldots, b_3$ with $b_3=2h^{(2,1)}+2$, denote a basis of  $H_3(Z_3,\mathds{Z})$
and $S^1_P$ denotes a small $S^1$ surrounding the zero locus of $P(\underline{x},\underline{z})$ 
in the normal direction. The $\underline{z}$ are the complex structure moduli of $Z_3$. 
Then one performs the so-called Griffiths-Dwork reduction method\footnote{See 
e.q.~\cite{morrison1998picard,cox1999mirror} for a review and \cite{Libgober:1993hq} for an 
extension and algorithm for complete intersections.}, also referred to as reduction of pole
order \cite{Griffiths:1969fr}, to obtain differential operators $\mathcal{D}_a$ obeying
\begin{equation}
\label{eq:uptoexact}
 	\mathcal{D}_a(\underline{z})\Omega(\underline{z})=d\alpha_a
\end{equation}
for a two-form $\alpha_a$. Integration over $\Gamma^k$ 
yields homogeneous linear differential equations for the periods $\Pi^k(\underline{z})$, the 
Picard-Fuchs equations, that directly reflect the diagram \eqref{eq:VariationsOfHodgeStructure}.

In the toric context the determination of the Picard-Fuchs equations\footnote{There is one caveat in order since
the toric means, in contrast to the Griffiths-Dwork reduction method, not necessarily yield differential equations 
of minimal order.} simplifies significantly due to the toric symmetries of $\P_{\Delta}$ and the requirement of 
invariance of the residue integrals \eqref{eq:residueZ3}.
For this analysis it is necessary to use instead of the coordinates $\underline{z}$ on $\mathcal{M}_{\rm cs}$
first the redundant parameterization of the complex structure variables of $Z_3$ by the coefficients 
$\underline{a}$ on which the toric symmetries of $\P_{\Delta}$ act canonically.  The 
infinitesimal version of theses symmetries acting on the $\underline{a}$ give rise to 
very simple differential operators ${\cal L}_l(\underline{a})$ and  ${\cal Z}_k(\underline{a})$  
called the Gelfand-Kapranov-Zelevinski or short GKZ differential system of the toric 
complete intersection \cite{gelfand1989hypergeometric,gelfand1990generalized}. The operators 
either annihilate $\Omega(\underline{a})$ or, if the toric symmetries are broken in a specific 
way, annihilate $\Omega(\underline{a})$ up to an exact form as in \eqref{eq:uptoexact}. The latter 
operators, the ${\cal Z}_k(\underline{a})$, express the fact that $\Omega(\underline{a})$ depends 
only on specific combinations of the $\underline{a}$, which are the genuine complex structure 
deformations $\underline{z}$ defined in \eqref{eq:HVmirror}. From the operators ${\cal L}_k(\underline{a})$ 
it is possible in simple situations to obtain a complete system of Picard-Fuchs operators 
${\cal D}_a(\underline{z})$ \cite{Hosono:1993qy}. 

Let us give the explicit representation of the operators forming the GKZ-system.
Both the operators ${\cal L}_l(\underline{a})$ and ${\cal Z}_k(\underline{a})$ are completely 
determined by the toric data encoded in the points $\tilde{v}_j$ and relations $\ell^{(a)}$ 
of $\Delta_4^{\tilde{Z}}$.\footnote{We note that the following discussion applies to any Calabi-Yau 
manifold of complex dimensions $n$ by replacing $\Delta_4^{\tilde{Z}}\rightarrow \Delta_{n+1}^{\tilde{Z}}$.}
The first set of operators is given by
\begin{equation} \label{eq:pfo}
 	\mathcal{L}_{i}=\prod_{\ell^{(i)}_j>0}\left(\frac{\partial}{\partial a_j}\right)^{\ell^{(i)}_j}
 	-\prod_{\ell^{(i)}_j<0}\left(\frac{\partial}{\partial a_j}\right)^{-\ell^{(i)}_j}\,,\qquad i=1,\ldots, k\,.
\end{equation}
These operators annihilate $\Omega(\underline{a})$ in \eqref{eq:residueZ3} and its periods 
$\Pi^k(\underline{a})$ identically as can be checked as a simple consequence of the second 
relation in \eqref{eq:HVmirror}. In other words, the differential operators ${\cal L}_i$ 
express the trivial algebraic relations between the monomials entering $P(\underline{x};\underline{a})$. 
The second differential system of operators encoding the automorphisms of $\P_{\Delta}$ is 
given by
\begin{equation} \label{eq:Zs}
 	\mathcal{Z}_{j}=\sum_n(\bar{v}_n)^j\vartheta_n-\beta_j\,,\qquad j=0,\ldots,4\,.
\end{equation}
Here $\beta=(-1,0,0,0,0)$ is the so-called exponent of the GKZ-system and 
$\vartheta_n=a_n\frac{\partial}{\partial a_n}$ denote logarithmic derivatives. We have 
embedded the points $\tilde{v}_n$ into a hypersurface at distance $1$ away from the origin by 
defining $\bar{v}_n=(1,\tilde{v}_n)$ so that all zeroth components are $(\bar{v}_n)^0=1$. 
The operators $\mathcal{Z}_j$ simply represent the torus symmetries of $\P_{\Delta}$ on the 
periods $\Pi^k(\underline{a})$ that are functions of the parameters $\underline{a}$. In detail, 
the operator $\mathcal{Z}_0$ expresses the effect of a rescaling of the constraint 
$P\mapsto \lambda P$ by the homogeneity property $\Pi^k(\lambda \underline{a})=\lambda \Pi^k(\underline{a})$. 
It is often convenient to absorb this scaling freedom by performing the redefinition 
$\Omega(\underline{a})\mapsto a_0\Omega(\underline{a})$ and 
$\Pi^k(\underline{a})\mapsto a_0\Pi^k(\underline{a})$ so that the periods become invariant 
functions under the overall rescaling $\underline{a}\mapsto \lambda \underline{a}$. Accordingly, 
we obtain a shift $\vartheta_0\mapsto \vartheta_0-1$ in all the above operators, such that the 
$\mathcal{Z}_0=\sum_{n}\vartheta_n$ in particular. The operators $\mathcal{Z}_j$, $j\neq 0$, 
express rescalings of the coordinates like e.~g.~$x_m\mapsto \lambda x_m$, $x_n\mapsto \lambda^{-1}x_n$ 
that can be compensated by rescalings of the monomial coefficients $\underline{a}$. In particular 
these coordinate rescalings leave the measure $\Delta_{\P_{\Delta}}$ on $\P_{\Delta}$ invariant. 
Examples for the operators $\mathcal{Z}_i$ and the corresponding action on the coordinates $x_j$ 
are given in section \ref{sec:Superpots+MirrorSymmetry} and chapter \ref{ch:CalcsBlowUp}.

The coordinates $\underline{z}$ introduced in \eqref{eq:HVmirror} are solutions to the 
$\mathcal{Z}_j$-system and generally read
\begin{equation} \label{eq:algCoords}
 	z^i=(-)^{\ell^{(i)}_0}\prod_{j=0}^{k+4} a_j^{\ell^{(i)}_j}\,,\quad i=1,\ldots k.
\end{equation}
They indeed turn out to be appropriate coordinates at the point of maximally 
unipotent monodromy in the complex structure moduli space of $Z_3$, which is centered at $\underline{z}=0$. 
Indeed the Picard-Fuchs operators $\mathcal{D}_a$ extracted from the GKZ-system in the coordinates 
$\underline{z}$ is solved explicitly using the Frobenius method \cite{Hosono:1993qy,Hosono:1994ax}, cf.~section 
\ref{sec:matching} for more details. The obtained solutions have the well-known log-grading in 
terms of powers of $(\log(z^i))^k$, $k=0,\ldots, 3$ with a unique holomorphic power series solution $X^0$.
This is expected by the known shift symmetry of the NS-NS B-field, $b^i\mapsto b^i+1$, in the Type IIA theory, 
which implies on the B-model side a maximal logarithmic degeneration of periods near the point $z_i=0$ 
\cite{Candelas:1990rm,Hosono:1993qy}. 

In order to relate these solutions with the geometrical periods $\Pi$ in \eqref{eq:periodVectorCY3} of $\Omega$
it is convenient to exploit mirror symmetry. It relates the complex geometry of $Z_3$ and $H^3(Z_3)$ to the 
symplectic geometry of $\tilde{Z}_3$ and $H^{\rm ev}(\tilde Z_3)=\bigoplus_{i=0 }^3H^{(i,i)}(\tilde Z_3)$.
In particular an integral basis of $H^{3}(Z_3,\Z)$ is determined by the integral basis of 
$H^{\rm ev}(\tilde Z_3,\Z)$ upon demanding a matching of the A- and B-model operator algebra. This fixes the
leading structure of the periods of $\Omega$ in the expansion by an integral basis of $H^{3}(Z_3,\Z)$ that 
is mirror dual to $H^{\rm ev}(\tilde Z_3,\Z)$, see e.g.~\eqref{eq:periodStrucutreLV} for the general matching 
at large volume. It is special for the threefold case that there is a canonical basis of $H^{\rm ev}(\tilde Z_3,\Z)$ 
given by products of the K\"ahler cone generators $J_i$ on $\tilde{Z}_3$, 
$(1,J_i,a_k^{i_1i_2}J_{i_1}J_{i_2},\frac{1}{\mathcal{V}}J^3)$. The constant coefficients $a_k^{i_1i_2}$ 
are determined by demanding duality between $H^{(1,1)}(\tilde{Z}_3)$ and $H^{(2,2)}(\tilde Z_3)$, which just involves
the calculation of the classical intersections on $\tilde{Z}_3$. For the marginal deformations the mirror 
map is then established at large volume as
\begin{equation}
	t_A^i\equiv t^i=\frac{X^i}{X^0}=\frac{1}{2\pi i}(\log(z^i)+p(\underline{z}))\,,
\label{eq:MirrorMapThreefold}
\end{equation}
which identifies, in geometrical terms, the K\"ahler moduli $t^i_A$ on $Z_3$ with the affine complex structure 
moduli $t^i$ of $Z_3$. On the level of the cohomology basis or in other words on the level of the A- and B-model 
operators this corresponds to $J_i\leftrightarrow \beta^{(1)}_i=\theta_i\Omega=\chi_iz^i-\theta^iK_{\rm cs}$, 
cf.~the general mirror map \eqref{eq:Atobeta_map}. We note that this maps 
$H^{(p,p)}(\tilde Z_3)$ to $F^{3-p}H^3(Z_3)$ in order to appreciate the holomorphicity of the vector bundles 
$\mathcal{F}^m$. In order to extract the more intuitive matching of 
$H^{(3-p,p)}(Z_3)=F^{p}H^{3-p}(Z_3)\cap \overline{F^{p}H^3(Z_3)}$ we have to perform an 
anholomorphic projection map. 

Thus, we have seen that the knowledge of the classical terms fixes the coefficients of the 
leading order in $\log(z^i)$. The subleading terms are restricted further by demanding integral monodromy in 
$Sp(b^3,\Z)$ under the Peccei-Quinn symmetry $t_A^i\mapsto t_A^i+1$. This allows to unambiguously identify the 
solutions to the Picard-Fuchs system with the periods with respect to the integral basis of $H_3(Z_3,\mathds{Z})$.
We refer to section \ref{sec:FFMirrors} for a more systematic derivation of the structure of the periods and 
conclude by summarizing the periods at large complex structure. The prepotential is given by \cite{Klemm:2005tw} 
\begin{equation} \label{eq:pre_largeV}
  F^0 = -\tfrac{1}{3!} \cK_{ijk}\, t^i t^j t^k - \tfrac{1}{2!}\cK_{ij}\, t^i t^j + \cK_{i} t^i + \tfrac12 \cK_0  
  + \sum_{\beta} n_\beta^0\,\text{Li}_3(q^\beta)
\end{equation}
where $q^\beta = e^{2\pi id_j t^j}$ for a class $\beta=d_i\beta$ with entries  $d_i\in\Z_{\ge 0}$, cf.~section
\ref{sec:EnumGeo}. The last term contains
the quantum corrections by string world-sheet instantons to the A-model and will further be discussed in section 
\ref{sec:EnumGeo} on enumerative geometry. Here 
\begin{align} \label{eq:classic_terms_threefold}
  &\cK_{ijk} = \int_{\tilde Z_3} J_i \wedge J_j \wedge J_k\ ,\qquad 
  &&\cK_{ij}=\frac{1}{2}\int_{\tilde Z_3} \imath_*(c_1(J_j)) \wedge J_i \ , \\ 
  &\cK_j = \frac{1}{2^2 3!}\int_{\tilde Z_3} c_2(T_{\tilde Z_3})\wedge J_j\ , \qquad 
  &&\cK_0= \frac{\zeta(3)}{(2 \pi i)^3}\int_{\tilde Z_3} c_3(T_{\tilde Z_3}) \ , \nn
\end{align}
are determined by the classical intersections of the mirror Calabi-Yau threefold $\tilde Z_3$ of $Z_3$.
Note that $c_1(J_j)$ denotes the first Chern class of the divisor line bundle associated to $J_j$ and 
$\imath_*=P_{\tilde Z_3}i_*P^{-1}_{J_j}$ is the Gysin homomorphism where $P_{\tilde Z_3}$ 
($P_{\tilde J_j}$) is the Poincar\'e-duality map on $\tilde Z_3$ ($J_j$) and $i_*$ is the push-forward 
on the homology. Thus, $\imath_*(c_1(J_j))$ is a four-form.
We insert the prepotential $F^0$ into the general expression of the period vector \eqref{eq:periodVectorCY3} 
to obtain all integral periods of $Z_3$ as
\begin{equation}
	\Pi=X^0\begin{pmatrix}
				1 \\
				t^i\\
				-\tfrac{1}{2} \cK_{ijk}\, t^j t^k - \cK_{ij}\, t^j + \cK_{i} + \sum_{\beta} n_\beta^0d_i\,\text{Li}_2(q^\beta)\\
				\tfrac{1}{3!} \cK_{ijk}\, t^i t^j t^k  + \cK_{i} t^i + \cK_0 + \sum_{\beta} n_\beta^0\,(2\text{Li}_3(q^\beta)-d_it^i\text{Li}_2(q^\beta))
				\end{pmatrix}\,.
\label{eq:periodVectorCY3LV}
\end{equation}
Ultimately the quantum corrected Yukawa couplings $C_{ijk}(t)$ are evaluated from \eqref{eq:YukawaCouolingCY3} as
\begin{equation}
	C_{ijk}(t)=\partial_{i}\partial_j\partial_kF^0(t)=-\mathcal{K}_{ijk}+\sum n_\beta^0d_id_jd_k \frac{q^\beta}{1-q^\beta}\,,
\label{eq:YukawaCouolingCY3Full}
\end{equation}
where we used the elementary identity $Li_0(x)=\frac{x}{1-x}$. We immediately observe that the subleading terms do 
not affect the quantum corrections to the A-model intersections. However, the subleading terms are essential in the
determination of the $\mathcal{N}=2$ K\"ahler potential $K_{\rm cs}$ in \eqref{eq:KaehlerpotCS} for the vector multiplet 
moduli and for the flux and brane superpotentials, as we will discuss in sections \ref{sec:specialGeometry} and 
\ref{sec:CY3superpotsEnumGeo}.

We conclude by noting that it is the aim of chapter \ref{ch:blowup} to extend the use of the Picard-Fuchs equations, 
in particular the GKZ-system, the coordinates at large volume as well as the concept of periods to the open string sector 
of a five-brane. This is achieved by formulating a GKZ-system for a threefold $\hat{Z}_3$ obtained as a blow-up of the curve $\Sigma$ 
wrapped by the brane and by analyzing its solutions.

\section{Mirror Symmetry for Calabi-Yau Fourfolds} 
\label{sec:FFMirrors}

In this section we describe mirror symmetry for Calabi-Yau fourfolds $(\tilde{X},X)$.
The key quantities we analyze are holomorphic functions $F^0(\gamma)$, that  
are the fourfold analogue of the holomorphic prepotential $F^0$ \eqref{eq:prepotAffine} 
on Calabi-Yau threefolds in the sense, that the (covariant) double derivatives of $F^0(\gamma)$ 
with respect to the moduli $t^a$ yield the three-point correlators $C^{(1,1,2)}_{ab\gamma}$ of 
the underlying topological field theory.  These three-point correlators together with the two-point 
correlation functions are the fundamental couplings of the topological A- respectively B-model that
encode all other correlators in the theory. Whereas the classical two-point coupling is exact, the three-point couplings
encode, as in the Yukawa couplings \eqref{eq:YukawaCouolingCY3Full} in the threefold case, 
to leading order the classical intersections of two divisors with an element of $H_4(\tilde{X})$ 
on the A-model side, that receive quantum corrections due to world-sheet instantons.
As in the threefold case it is the power of mirror symmetry for Calabi-Yau fourfolds to calculate 
these couplings exactly by relating the classical geometric calculations in the B-model on $X_4$ 
to the quantum cohomology ring of A-model on $\tilde{X}_4$.

We begin in section \ref{sec:bmodel} by introducing the B-model operator ring, that is formed
by the operators representing elements of $H^p(X,\bigwedge^q TX)$, together with their two-point 
and three-point  correlators, that are given by period integrals. In the large radius limit the 
A-model is defined by another ring, namely the vertical part of the classical cohomology ring 
$H^{(*,*)}(\tilde X)$. To prepare the analysis of mirror symmetry we note in section \ref{sec:frobenius} 
that both rings with the two- and three-point correlators just mentioned carry a natural
Frobenius structure. Mirror symmetry then identifies the quantum cohomology ring of 
the A-model away from the large volume limit with the B-model ring. For the precise 
identification one has to specify matching points in the moduli spaces. The natural candidate 
for our present purposes is the large radius point/large complex structure point in the A- 
respectively B-model. We describe in section \ref{sec:matching} the details of this matching,
namely the use of the mirror map, properties of the Picard-Fuchs system near the point of maximal
unipotent monodromy and the classical intersection ring of $\tilde{X}$. We then exploit the 
precise matching to obtain enumerative predictions for the A-model and to construct an 
integral basis of the horizontal cohomology of $H^{(n,n)}(X)$.
For the latter it is a particularly important step to determine the 
classical terms in the leading logarithms of the fourfold periods. We conclude our discussion 
in section \ref{sec:analyticCont} by fixing these classical terms by means of analytic 
continuation to other points on the fourfold complex structure moduli space, namely
the universal conifold, and a monodromy analysis.
Our presentation follows mainly \cite{Greene:1993vm,Mayr:1996sh,Klemm:1996ts} and \cite{Grimm:2009ef}.

Since our discussion can at several points be generalized to arbitrary complex
dimensional Calabi-Yau $n$-folds \cite{Greene:1993vm} we denote a mirror pair of
Calabi-Yau $n$-folds by $(\tilde{X},X)$ and identify with the fourfold case by
putting $n=4$ and writing $(\tilde X_4,X_4)$ instead.

\subsection{States and Correlation Function of the B-model}
\label{sec:bmodel}

In the B-model one considers a family of $n$-folds $X_z$ fibered over the complex
structure moduli space ${\cal M}_{\rm cs}$, $X_z\rightarrow \mathcal{M}_{\rm cs}$. The
states\footnote{We use in the following the same symbol for states and
operators.} in the B-model are elements $B^{(j)}_k$ of the cohomology groups
$H^j(X_z,\bigwedge^j T)$. Their cubic forms are defined as 
\begin{equation} 
C(B^{(i)}_a,B^{(j)}_b,B^{(k)}_c)=\int_X \Omega_n(B^{(i)}_a\wedge B^{(j)}_b\wedge B^{(k)}_c) \wedge \Omega \ ,
\label{eq:threepoint}    
\end{equation}   
and vanish unless $i+j+k=n$.  Here $\Omega_n$ is the unique holomorphic
$(n,0)$-form and $\Omega_n(B^{(i)}_a\wedge B^{(j)}_b\wedge B^{(j)}_c)$ is the
contraction of the $n$ upper indices of $B^{(i)}_a\wedge B^{(j)}_b\wedge
B^{(j)}_c\in H^n(X_z,\bigwedge^n TX_z)$ with $\Omega_n$, which produces an
anti-holomorphic $(0,n)$-form on $X_z$. Note that this is the standard isomorphism
$H^i(X_z,\bigwedge^j TX_z)\cong H_H^{(n-j,i)}(X_z)$ obtained by contraction with $\Omega_n$. 
It is important to emphasize that not all elements of $H^{n}(X_z)$ can be reached as variations of $\Omega_n$ with
respect to the complex structure, 
in contrast to the Calabi-Yau threefold case, which makes it necessary to introduce the 
primary horizontal subspace $H^{n}_H(X_{z})$ of $H^{n}(X_{z})$. Its complement
is the vertical subspace\footnote{This is completely 
analogous to the two-dimensional case of $K3$, where $H^{(1,1)}_V$ spans the Picard group.} 
$H^{n}_V(X_{z})$, that is generated by powers of the K\"ahler cone generators 
in $H^{(1,1)}(X_{z})$ \cite{Greene:1993vm}. We denote the image of $B^{(i)}_k$ in 
$H^{(n-i,i)}_H(X_z)$ by $b^{(i)}_k=\phantom{}^\Omega(B^{(i)}_k)$ and the inverse\footnote{The 
inversion is just the contraction 
$(b^{(i)}_k)^{\Omega}=\tfrac{1}{||\Omega||^2}\bar{\Omega}^{a_1\ldots a_{n-i}b_1\ldots b_i}(b^{(i)}_k)_{a_1\ldots a_{n-i}\bar{b}_1\ldots\bar{b}_i}$ 
such that $(\Omega)^{\Omega}=1$. Formally it is the multiplication with the inverse 
$\mcal{L}^{-1}$ of the K\"ahler line bundle $\mcal{L}=\langle\Omega\rangle$, see 
e.g.~\cite{Hori:2003ic}.} by $B^{(i)}_k=(b^{(i)}_k)^\Omega$.  We define the 
hermitian metric
\begin{equation} 
G(B^{(i)}_c,\bar B^{(i)}_d)=\int_X b^{(i)}_c \wedge \bar b^{(i)}_d  \ .    
\label{eq:twopoint}
\end{equation} 
The image $b^{(i)}_a$ of the states $B^{(i)}_a$  lies in the
horizontal subspace $H^{(n-j,i)}_{H}(X_z)$ and the elements $b^{(i)}_a$ form a
basis of this space. For $B^{(1)}_c\in H^1(X_z, TX_z)$ the image spans all of
$H^{(n-1,1)}(X_z)$ and \eqref{eq:twopoint} is the Weil-Petersson metric on $\mathcal{M}_{\rm cs}$.

The integrals \eqref{eq:threepoint} and  \eqref{eq:twopoint} are calculable by 
period integrals of the holomorphic $n$-form $\Omega_n$. It is in general hard to 
integrate the periods directly. They encode however, as in the Calabi-Yau threefold case 
in section \ref{sec:HodgeStructure}, the variations
of Hodge structures of the family $X_z\rightarrow \cal{M}_{\rm cs}$, which leads to
the Picard-Fuchs differential equations on $\mathcal{M}_{\rm cs}$. 
The periods are therefore determined as solutions of the latter up to linear 
combination. The precise identification as addressed in section \ref{sec:matching}
is an important problem, in particular for physical applications like the determination
of the flux superpotential \eqref{eq:GVW-super} on the fourfold $X$.

For a given base point $z=z_0$ in the complex structure moduli space $\mathcal{M}_{\rm cs}$
with fiber $X_{z_0}$, one fixes a graded topological basis
$\hat{\gamma}^{(p)}_{a}$ of the primary horizontal subspace
$H^{n}_H(X_{z_0},\mathds{Z})$. Here $a=1,\ldots,h_H^{(n-p,p)}$ labels the basis
$\hat{\gamma}^{(p)}_a$ for fixed $p=0,\ldots,n$ of each graded piece
$H^{(n-p,p)}(X_z)$.  In addition, these forms can be chosen to satisfy\footnote{The generalization from fourfolds to $n$-folds is the obvious
one.}
\begin{eqnarray} \label{eq:def-eta}
  \int_{X_4} \hat{\gamma}^{(i)}_{a} \wedge  \hat{\gamma}^{(4-i)}_{b} &=& \eta^{(i)}_{a b}\ ,\\
   \int_{X_4} \hat{\gamma}^{(i)}_{a} \wedge \hat{\gamma}^{(j)}_{b}\ \  &=& 0  \quad\text{for}\quad i+j> 4\ .
     \nn
\end{eqnarray}
Later on in section \ref{sec:matching} we will identify this grading by $p$
with the natural grading on the observables of the A-model which are given by the
vertical subspaces $H^{(p,p)}_V(\tilde X)$ of the mirror cohomology. 

Note that the $\hat\gamma^{(p)}_a$ basis serves as a local frame of the vector
bundle $\mathcal H^{n}_H(X)\rightarrow\mathcal{M}_{\rm cs}$ over the moduli space whose
fiber at the point $z\in \mathcal{M}_{\rm cs}$ is $H^n_H(X_z)$. Thus, the bundle 
$\mathcal{H}^n_H(X)$ is flat and admits a flat connection $\nabla$ denoted the Gauss-Manin
connection. However, the individual cohomology groups $H^{(n,n-p)}(X_z)$ form no holomorphic 
vector bundles over ${\cal M}_{\rm cs}$ since holomorphic and anti-holomorphic coordinates 
are mixed under a change of complex structure $z$. Analogously to the threefold case, 
cf.~section \ref{sec:HodgeStructure}, only the horizontal parts of 
$F^k=\oplus_{p=0}^k H^{(n-p,p)}(X_z)$ form holomorphic vector bundles for which we introduce 
frames $\beta^{(k)}_a$ specified by the basis expansion  
\begin{equation} \label{eq:FiltBasis}
     \beta^{(k)}_{a} = 
       \hat{\gamma}^{(k)}_{a} + 
       \sum_{p>k}\Pi^{(p,k)\ c}_{a} \hat{\gamma}_{c}^{(p)}\ .
\end{equation}
In special coordinates $t^a$ at the point of maximal unipotent monodromy, cf. \eqref{eq:flatcoords}, this 
can be written as $\{\beta^{(0)}=\Omega_n,\beta^{(1)}_{a}=\partial_a\Omega_n,\ldots \}$. 
We note that for this basis choice we fixed the overall 
normalization of each $\beta^{(k)}_a$ such that the coefficient of $\hat{\gamma}^{(k)}_a$ 
is unity. This is needed to obtain the right inhomogeneous flat coordinates on 
$\mathcal{M}_{\rm cs}$ and to make contact with enumerative predictions for the 
A-model, see section \ref{sec:matching}. For grade $k=0$ it corresponds to the 
familiar gauge choice in the vacuum line bundle $\mathcal{L}$ generated by $\Omega_n$, 
cf.~section \ref{sec:specialGeometry} for the threefold case and \eqref{eq:flatcoords} for
fourfolds. We also introduce a basis of integral homology cycles 
$\gamma^{(p)}_{a}$ dual to $\hat \gamma_{a}^{(p)}$ obeying
\begin{equation} \label{eq:dualbasis}
 	\int_{\gamma^{(i)}_{a}}\hat{\gamma}^{(j)}_{b} 
     = \delta^{ij} \delta_{a b}\ .
\end{equation}
Then, by construction of the basis \eqref{eq:FiltBasis}, the period matrix $P$ defined by 
period integrals takes an upper triangular form in this basis, 
\begin{equation} 
 	 P=\int_{\gamma^{(p)}_{a}} \beta^{(k)}_{b}=
	\begin{cases}
         \Pi^{(p,k)\, a}_{b}\ ,& p>k\ ,\\ 
 	 \delta_{a b}\ ,& p=k\ ,\\
	0\ ,& p<k \ ,
	\end{cases}
\label{eq:ffperiods} 
\end{equation}
where $(p,k)$ is the bi-grade of the non-trivial periods $\Pi^{(p,k)\, a}_{b}$.
Since $\hat\gamma^{(p)}_{a}$ are topological and thus are locally constant
on $\mathcal{M}_{\rm cs}$, the moduli dependence of \eqref{eq:FiltBasis} is captured by the
moduli dependence of the period matrix $P\equiv P(z)$\footnote{By means of
$\nabla_a\beta^{k}_b=\nabla_a\Pi_b^{(k+1,k)\, c}\beta^{(k+1)}_c$ \cite{Mayr:1996sh} the 
periods of $\Omega_n$ fix the periods $P$.}. For Calabi-Yau fourfolds we 
summarize the basis choices and the periods in Table \ref{tab:basis_summary}.
 \begin{table}[t]
\begin{center}
	\begin{tabular}{|c|c|c|c|c|c|}
	\hline
Dimension & $1$ & $h^{(3,1)}(X_4)$ & $h^{(2,2)}(X_4)$ &$h^{(3,1)}(X_4)$& $1$ \\ \hline	
 Basis	of& $\hat{\gamma}^{(0)}_{a_0}$ & $\hat{\gamma}^{(1)}_a$ & $\hat{\gamma}^{(2)}_\beta$& $\hat{\gamma}^{(3)}_a$& $\hat{\gamma}^{(4)}_{a_0}$ \\
 $H^4_H(X_4,\mathds{Z})$& $\beta^{(0)}_{a_0}$ &  $\beta^{(1)}_a$ &  $\beta^{(2)}_\beta$&$\beta^{(3)}_a$&$\beta^{(4)}_{a_0}$ \\ \hline
\end{tabular}
\caption{Topological $\hat{\gamma}$ and moduli-dependent $\beta$ basis 
of $H^4_H(X_4,\mathds{Z})$}
\label{tab:basis_summary}
\end{center}
 \end{table}
The $(n,0)$-form $\Omega_n$ can
always be expanded over the topological basis 
$\hat{\gamma}^{(p)}_{a}$ of $H_H^n(X_z,\mathds{Z})$ as
\begin{equation}
 	\Omega_n=\Pi^{(p,0)\, a}_{1} \ \hat{\gamma}_{a}^{(p)} \equiv \Pi^{(p)\, a} \ \hat{\gamma}_{a}^{(p)}\, ,
\label{eq:omegaexpansion} 
\end{equation}
where we introduced a simplified notation $\Pi^{(p)\, a}$ for the periods 
$\Pi^{(p,0)\, a}_1$ of the holomorphic $n$-form already given in 
\eqref{eq:FiltBasis} for $k=0$. For an arbitrarily normalized $n$-form $\Omega_n$, the 
periods $(X^{0},X^{a})=(\Pi^{(0)\, 1},\Pi^{(1)\, a})$ with
$a=1,\ldots,h^{(n,1)}(X_{z_0})$ at the fixed reference point $z_0$ form for all
Calabi-Yau $n$-folds homogeneous projective coordinates of the complex moduli
space $\mcal{M}_{\rm cs}$. The choice of inhomogeneous coordinates set by
\begin{equation} \label{eq:flatcoords}
	t^{a}=\frac{X^{a}}{X^0}=\frac{\int_{\gamma^{(1)}_{a}}\Omega}{\int_{\gamma^{(0)}_{1}}\Omega}\,, 
\end{equation} 
which agrees with the basis choice in \eqref{eq:FiltBasis}, corresponds to a
gauge in the K\"ahler line bundle $\langle\Omega_n\rangle$. At the point of
maximal unipotent monodromy $X^0$ and $X^a$ are distinguished by their 
monodromy, as $X^0$ is holomorphic and single-valued and $X^a\sim \text{log}(z^a)$ 
has monodromy $X^a\mapsto X^a+1$. Below the $t^a$ in \eqref{eq:flatcoords} are 
identified with the complexified K\"ahler parameters of the mirror $\tilde X$. 

The $t^a$ defined in \eqref{eq:flatcoords} are flat coordinates for the Gauss-Manin 
connection $\nabla_a$, i.e.~the latter becomes  just the ordinary differential
$\frac{\partial}{\partial t_a}$. This can be seen from the gauge choice
reflecting in the basis \eqref{eq:FiltBasis} combined with the Griffiths 
transversality constraint $\nabla_a(\mathcal{F}^k)\subset \mathcal{F}^{k+1}/\mathcal{F}^{k}$ which together 
imply \cite{Greene:1993vm} $\nabla_a t^b=\delta_a^b$, i.e~$\nabla_a=\partial_a$.  
In these coordinates the three-point coupling becomes 
\begin{equation} 
C^{(1,k,n-k-1)}_{abc}=C((\beta^{(1)}_a)^{\Omega},(\beta^{(k)}_b)^{\Omega},(\beta^{(n-k-1)}_c)^{\Omega})=\int_X \beta_{c}^{(n-k-1)}\wedge \partial_{a} 
\beta_b^{(k)} \,  .
\label{eq:3point} 
\end{equation} 
Here we use \eqref{eq:FiltBasis}, \eqref{eq:ffperiods} and the fact that 
$(\beta^{(1)}_a)^\Omega \wedge (\beta_b^{(k)})^\Omega$ can be replaced  
by $(\nabla_a \beta_b^{(k)})^\Omega$ \cite{Greene:1993vm} under the integral 
\eqref{eq:threepoint} to obtain \eqref{eq:3point}. 
This triple coupling is a particularly important example of \eqref{eq:threepoint}. 
Furthermore, one can show from 
the properties of the Frobenius algebra that all other triple couplings in 
\eqref{eq:threepoint} can be expressed in terms of \eqref{eq:3point}, see 
section~\ref{sec:frobenius}.   

The holomorphic topological two-point couplings of \eqref{eq:def-eta} in this 
basis trivially read 
\begin{equation} 
\eta^{(k)}_{ab}=\int_X     \beta_b^{(n-k)} \wedge   \beta_a^{(k)}   
\label{eq:toptwopoint} 
\end{equation} 
since only the lowest $\hat{\gamma}^{(p)}$ for $p=k$ in the upper-triangular
basis transformation \eqref{eq:FiltBasis} contributes to the integral due to the
second property of \eqref{eq:def-eta}.  In particular $\eta^{(k)}_{ab}$ is moduli
independent. From the above it is easy to see the basis expansion at grade
$k+1$,
\begin{equation} 
\label{eq:alphader}
\partial_a \beta_b^{(k)} =C^{(1,k,n-k-1)}_{abc}\eta^{cd}_{(n-k-1)}
\beta_d^{(k+1)}\ ,
\end{equation}     
where $\eta^{cd}_{(p)}$ is the inverse of $\eta_{cd}^{(p)}$.  

Let us finish this section with some comments on general properties of the
periods and their implications on the $\mathcal{N}=1$ effective action.  The
period integrals \eqref{eq:ffperiods} obey differential and algebraic relations,
which are different from the special geometry relations of Calabi-Yau threefold
periods in section \ref{sec:specialGeometry}. The differential relations have 
however exactly the same origin namely the Griffiths
transversality constraints $\int_X \Omega_n \wedge \partial_{i_1}\ldots
\partial_{i_k} \Omega_n=0$ for $k<n$. However since $a \wedge b=(-1)^n b \wedge a
$ for $a,b$ real $n$-forms one has additional algebraic relations from
$\int_X \Omega_n\wedge \Omega_n=0$ between the periods $\Pi^{(p)\,a}$ for $n$ even like 
$n=4$, which are absent for $n$ odd, in particular the threefold case. $\mathcal{N}=2$
compactifications of Type II on Calabi-Yau threefolds to four dimensions inherit
the special geometry structure induced by the above differential relations in the vector moduli
space. In contrast for a generic $\mathcal{N}=1$ supergravity theory in 4d there is no special
structure beyond K\"ahler geometry.

\subsection{The Frobenius Algebras}
\label{sec:frobenius}

As was mentioned above the B-model operators form a Frobenius algebra. Since
also the A-model classical cohomology operators as well as its quantum
cohomology operators form such an algebra it is worthwhile to describe the
general structure before discussing the precise matching via mirror symmetry 
in the next section.

A Frobenius algebra has the following structures. It is a graded vector space
${\cal A}=\oplus_{i=1}^n {\cal A}^{(i)}$ with ${\cal A}^{(0)}=\mathbb{C}$ 
equipped with a non-degenerate symmetric bilinear form  $\eta$ and a cubic form 
\begin{equation}
  C^{(i,j,k)}:{\cal A}^{(i)}\otimes {\cal A}^{(j)}\otimes {\cal A}^{(k)}\rightarrow \mathbb{C}\ ,
\end{equation}
$i,j,k\ge 0$ and the following properties:
\begin{itemize}
\item[(F1)] Degree: $C^{(i,j,k)}=0$ unless $i+j+k=n$ 
\item[(F2)] Unit: $C^{(0,i,j)}_{1bc}=\eta^{(i)}_{bc}$
\item[(F3)] Nondegeneracy: $C^{(1,i,j)}$ is non-degenerate in the second slot 
\item[(F4)] Symmetry: $C^{(i,j,k)}_{abc}=C^{(\sigma(i,j,k))}_{\sigma(abc)} $ 
						under any permutation of the indices.
\item[(F5)] Associativity:
  $C^{(i,j,n-i-j)}_{abp}\eta^{pq}_{(n-i-j)}C^{((i+j),k,(n-i-j-k))}_{qef}=C^{(i,k,n-i-k)}_{aeq}\eta^{qp}_{(n-i-k)} C^{(i+k,j,n-i-j-k)}_{pbf}$
where the sum over common indices is over a basis of the corresponding spaces.
\end{itemize}
The product 
\begin{equation} \label{eq:FrobeniusOPE}
{\cal O}^{(i)}_a \cdot {\cal O}^{(j)}_b = C^{(i,j,i+j)}_{abq} \eta^{qp}_{(i+j)} {\cal O}^{(i+j)}_p \ 
\end{equation}     
defines now the Frobenius algebra for a basis of elements ${\cal O}_k^{(i)}$ of
${\cal A}^{(i)}$ which is easily seen using (F4) to be commutative. Note that the 
associativity implies that $n$-point correlators can be factorized in various ways 
in the three-point functions. Also the three-point correlators are not independent 
and by associativity, non-degeneracy and symmetry it can be shown\footnote{See 
\cite{Klemm:1996ts} for an explicit inductive proof.} that all three-correlators can 
be expressed in terms of the $C^{(1,r,n-r-1)}$ correlators defined in \eqref{eq:3point}.

It is easy to see that the $B^{(i)}$ operators of the B-model with the correlators 
defined by \eqref{eq:threepoint} or equivalently \eqref{eq:3point} and \eqref{eq:toptwopoint} 
fulfill the axioms of a Frobenius algebra.      

Let us consider  the A-model operators. As already mentioned
such an operator corresponds to an element in the vertical subspace 
$H^{(*,*)}_{V}(\tilde X)$. It is generated by the K\"ahler cone generators
$J_i$, $i=1,\ldots, h^{(1,1)}(\tilde X)$  and is naturally graded, 
\begin{equation}
A^{(p)}_\alpha =a_\alpha^{i_1,\ldots, i_p} J_{i_1}\wedge \ldots \wedge J_{i_p}\,\in H^{(p,p)}(\tilde X)\,.  
\label{eq:amodeloperators}
\end{equation} 
For the classical A-model the correlation functions are simply  
the intersections
\begin{equation} 
C_{abc}^{0\, (i,j,k)}=C(A^{(i)}_a A^{(j)}_b A^{(k)}_c)= \int_{\tilde X}  A^{
  (i)}_a\wedge  A^{(j)}_b \wedge A^{(k)}_c\ .
\label{eq:classicalamodeltriplecouplings}  
\end{equation} 
They vanish unless $i+j+k=n$. The topological metric is similarly 
defined by 
\begin{equation} 
\eta^{(k)}_{ab} = \int_{\tilde X}  A^{(k)}_a\wedge  A^{(n-k)}_b \  
\label{eq:amodeltopologicametric}  
\end{equation} 
and together with  \eqref{eq:classicalamodeltriplecouplings} this 
defines a Frobenius algebra. Clearly the $A^{(p)}_k$ are not freely 
generated by the $J_i$. The products $J_{i_1}\wedge \ldots \wedge J_{i_p}$  
are set to zero if their pairings \eqref{eq:amodeltopologicametric} with 
all other cohomology elements vanish. This is easily calculated using 
toric techniques and reflects geometrical properties of $\tilde X$ like 
for instance fibration structures. Likewise we calculate the quartic
intersections, i.e.~the intersection numbers, on a fourfold $X_4$ as
\begin{equation} \label{eq:FourfoldclassicalIntersections}
  \mathcal{K}^0_{abcd}=\int_{\tilde X_4}A_{a}^{(1)}\wedge A_{b}^{(1)}\wedge
A_{c}^{(1)}\wedge  A_{d}^{(1)}
=C^{0\, (1,1,2)}_{ab\alpha}\eta^{\alpha\beta}_{(2)}C^{0\,(1,1,2)}_{cd\beta}
\end{equation}
using the algebra relation \eqref{eq:FrobeniusOPE}. Thus, we see that the 
four-point couplings are bilinears in the triple intersections.

The classical intersections are extended using mirror symmetry to the quantum
cohomological intersections\footnote{We denote both the operators of the
classical algebra and of quantum cohomology algebra by
$A^{(p)}_k$.}
\begin{equation} 
C(A^{(i)}_a A^{(j)}_b A^{(k)}_c)= C_{abc}^{0\, (i,j,k)}+
{\rm instanton\ corrections}\ , 
\label{eq:quantumamodeltriplecouplings}  
\end{equation} 
where the instanton corrections due to holomorphic curves with meeting
conditions on the homology cycles dual to the $A^{(p)}_k$. Again
the correlators \eqref{eq:quantumamodeltriplecouplings} vanish unless $i+j+k=n$. 
Note that  the $C(A^{(i)}_a A^{(j)}_b A^{(k)}_c)$ depend via the instantons 
on the complexified  K\"ahler parameters of $\tilde X$, while 
$\eta^{(k)}_{ab} = \int_{\tilde X}A^{(k)}_a\wedge  A^{(n-k)}_b$ is still 
purely topological. There are no instanton corrections present
because in the moduli space for the two-pointed sphere not all zero modes are 
saturated due to the one remaining conformal Killing field on the sphere. 

\subsection{Matching of the A-model and B-model Frobenius Algebra} 
\label{sec:matching}

We now describe the matching of the B-model Frobenius algebra with the $A$ model
quantum Frobenius algebra. At the large radius point of the K\"ahler structure, 
the correlation functions of the classical $A$-model can be calculated using
toric intersection theory. We will match this information with the leading
logarithmic behavior of the periods at the corresponding point in the 
complex structure moduli space, the point of large complex structure, 
which is characterized by its maximal logarithmic degeneration, which 
leads to a maximal unipotent monodromy.

Let us now discuss the Picard-Fuchs differential operators associated to 
the mirror Calabi-Yau $X$ at the large complex structure point. In general
the Picard-Fuchs system can be obtained also in the fourfold case by applying
the method of reduction of pole order to the residue integral for $\Omega_4$, that reads
\begin{equation} 
\Omega_4=\int_{\epsilon_1} \ldots  \int_{\epsilon_s} \prod_{k=1}^s \frac{a_0^{(k)}}{P_k}\Delta \,, 
\label{eq:residuumform} 
\end{equation}  
for Calabi-Yau fourfold hypersurfaces and complete intersections 
$\{P_1= \ldots =P_s=0\}$ with  $dP_1\wedge\ldots \wedge dP_s\neq 0$ 
in a toric variety $\mathds{P}_\Delta$ of dimensions $s+4$.
Here $\epsilon_i$ are paths in $\mathds{P}_\Delta$, which encircle $P_i=0$ and $\Delta$ 
is a measure invariant under the torus actions in $\mathds{P}_\Delta$. The parameter $a_0^{(k)}$ denotes 
a distinguished coefficient in the defining constraint $P_k$, cf.~section \ref{sec:PFO+3foldMirrorSymmetry}
for the analogous Calabi-Yau threefold expression.
Since we are dealing throughout our paper with mirror pairs $(\tilde{X},X)$ that 
are given torically, the derivation of the Picard-Fuchs operators simplifies 
significantly, since they are completely encoded by the toric data.
To the Mori cone generators $\ell^{(a)}$ on the A-model side one associates  
the canonical GKZ-system of differential operators $\mathcal{L}_a$ on the B-model side
given in \eqref{eq:pfo}, where the derivative is taken with respect to the coefficients 
$a_i$ of monomials in the constraint $f=0$ defining $X$. By the methods described in 
\cite{Hosono:1993qy} we obtain linear Picard-Fuchs operators 
$\mathcal{D}_a({\underline \theta},{\underline z})$, which are written in terms 
of the logarithmic derivatives $\theta_a=z^a\frac{\partial}{\partial z^a}$ with respect 
to the algebraic complex variables $z_a$ defined by \eqref{eq:algCoords} that vanish at 
the large complex structure point. 

We extract the leading term in $\theta_a$ of the 
differential operators as the formal limit ${\cal D}_i^{\rm lim}({\underline \theta})
=\lim_{\underline{z}\rightarrow 0} {\cal D}_i({\underline \theta},{\underline z})$, $i=1,\ldots, r$ 
and consider the algebraic ring  
\begin{equation} 
{\cal R}=\mathbb{C}[{\underline \theta}]/({\cal J}=\{ {\cal D}_1^{\rm lim},\ldots,{\cal
  D}_r^{\rm lim}\})\ .
\label{eq:ring}
\end{equation}  
This ring is graded by the degree $k$ in $\theta$ and we denote the ring at grade $k$
by $\mathcal{R}^{(k)}$ which is generated a number $h^H_{(n-k,k)}(X)=h^V_{(k,k)}(\tilde X)$ 
of elements for $k=0,\ldots,n$. We note that for $k=0,1,n-1,n$ there is no 
difference between the vertical (horizontal) homology and the full 
homology groups. Let us explain in more detail how this 
ring connects  the $A$- and the B-model structure  at large radius:

\begin{itemize} 
\item[(A)] 
The construction of the ring $\mathcal{R}$ is up to a normalization equivalent to the 
calculation of the intersection numbers of the $A$-model. In particular 
the $n$-point intersections appear as coefficients of the up to a 
normalization unique ring element of maximal degree, ${\cal R}^{(n)}=\sum_{i_1\leq\ldots \leq
 i_n} C^{0}_{i_1, \ldots, i_n } 
\theta_{i_1}\ldots \theta_{i_n}$ and similar the ${\cal R}^{(k)}$ 
are generated by ${\cal R}^{(k)}_\alpha=  \sum_{i_1\leq\ldots \leq
 i_k }a_{\alpha}^{i_1, \ldots,i_{k}}\theta_{i_1}\ldots \theta_{i_k}$, $\alpha=1,\ldots,h^V_{(k,k)}(\tilde X)$, 
where the coefficients obey 
$a_\alpha^{i_1, \ldots,i_{k}} \kappa_{i_1, \ldots,i_{k},j_1,\ldots j_{n-k}}=C^{0}_{\alpha,j_1,\ldots j_{n-k}}$,
cf.~\cite{Mayr:1996sh}.    
\item[(B)] 
The ring $\mathcal{R}^{(k)}$ is in one-to-one correspondence to the solutions 
to the Picard-Fuchs equations at large radius. As discussed before the solutions 
are characterized by their monodromies around this point, 
i.e.~they are graded by their leading logarithmic structure. 
To a given ring element ${\cal R}^{(k)}_a=\sum_{|{\underline \alpha}|=k}
\frac{1}{(2\pi i)^k} m_{a\,\underline \alpha} \theta_1^{\alpha_1}\ldots
\theta^{\alpha_h}_h$ in ${\cal R}^{(k)}$ we associate a solution of the form 
\begin{equation}
\label{eq:generalformofsolution} 
   \tilde{\Pi}^{(k)}_a=X_0({\underline z})\Big[\mathbb{L}^{(k)}_a  + 
                 {\cal O}(\log(\underline{z})^{|\alpha|-1})\Big]\ ,
\end{equation} 
with leading logarithmic piece of order $k$, 
\begin{equation} \label{eq:LeadingLog}
\mathbb{L}^{(k) }_a= \sum_{|{\underline \alpha}|=k} \frac{1}{(2\pi i)^k} 
\tilde m_{a\,\underline \alpha} \log^{\alpha_1}(z^1)\ldots \log^{\alpha_h}(z^h)\,, 
\end{equation}
by assigning $\tilde m^{a}_{\underline\alpha} (\prod_i \alpha_i !) = m^{a}_{\underline \alpha}$. 
In particular we map the unique element $1$ of $\mathcal{R}^{(0)}$ to the unique 
holomorphic solution $X_0({\underline z})=1+{\cal O}(\underline{z})$. The above map follows from the 
fact that all ${\cal D}_i^{\rm lim}$ in the ideal $\mathcal{J}$ annihilate the leading
logarithmic term $\mathbb{L}^{(k) }_a$ for $\Pi^{(k)}_a$ to be a solution. But this is just
the same conditions as for the element $\mathcal{R}^{(k)}$ to be normal to $\mathcal{J}$ in \eqref{eq:ring}.
\end{itemize} 
The two facts (A) and (B) imply mirror symmetry at the level of the classical 
couplings. Furthermore, mirror symmetry can be proven for toric hypersurfaces by matching the 
toric intersection calculation with the toric derivation of the 
Picard-Fuchs operators as it was argued in the threefold case in \cite{Hosono:1994ax} and in
\cite{Greene:1993vm,Mayr:1996sh,Klemm:1996ts} in the fourfold case.  
The identification   
\begin{equation}  
\theta_i \leftrightarrow J_i
\label{eq:matchingmap}  
\end{equation} 
provides a map between ${\cal R}^{(k)}_a$ and the classical A-model operators
$A^{(k)}_a$ defined in \eqref{eq:amodeloperators}. This provides also the matching 
of the A- and B-model Frobenius algebra at the large radius limit by identifying 
the periods of $\Omega_n$ and the solutions of the Picard-Fuchs system as follows. 

To a given element ${\cal R}^{(k)}_a$ we can associate an A-model operator  
$A^{(k)}_a$ by the replacement \eqref{eq:matchingmap}. Similarly 
we can construct the dual B-model operators $\beta^{(k)}_a$ in $F^{k}$ by applying the 
elements of the ring $\mcal{R}^{(k)}$ as differentials in the coordinates $\underline{z}$ 
to the holomorphic form $\Omega_n$. We obtain the mirror map
\begin{equation}  \label{eq:Atobeta_map}
\xymatrix{
A^{(k)}_a\ar@{|->}[r]&\beta^{(k)}_a|_{\underline{z}=0}={\mcal R}^{(k)}_a \Omega_n|_{\underline{z}=0} \ .}
\end{equation}
This determines the Frobenius algebra of the B-model completely as we show now. However, 
to relate the two- and three-point functions to the period integrals of the $\beta^{(k)}_a$ along the 
lines of section \ref{sec:bmodel} we have to specify the topological basis $\gamma^{(k)}_a$,
which simultaneously fixes the structure of the periods $\Pi^{(k,l)\,b}_a$, 
in terms of the A-model operator $A^{(k)}_a$ as well. 

First we select a basis of 
solutions denoted $\Pi^{(k)\,a}(\underline{z})$ of the Picard-Fuchs system with leading logarithm 
$\mathbb{L}^{(k)\,a}$ that is dual to the $A^{(k)}_a$ at large radius as
\begin{equation} \label{eq:periodStrucutreLV}
   \Pi^{(k)\,a}(\underline{z})=\mathbb{L}^{(k)\, b }(\underline{z})+\mathcal{O}((\log(\underline{z}))^{k-1})\,,
   \qquad{\mcal R}^{(k)}_a\,  \mathbb{L}^{(k)\, b }=\delta^{b}_a \ ,
\end{equation}
in the limit $\underline{z}\rightarrow 0$ \cite{Mayr:1996sh}. Then, the basis $\gamma^{(k)}_a$ is fixed
by identifying the periods in the expansion \eqref{eq:omegaexpansion} of 
$\Omega_n=\Pi^{(k)\,a}(\underline{z})\hat{\gamma}^{(k)}_a$ with the solutions as 
$\Pi^{(k,0)\ a}_1\equiv \Pi^{(k)\,a}(\underline{z})$ 
that associates the $\mathbb{L}^{(k)\, a}$ to the $\hat \gamma^{(k)}_a$. With these definitions 
the requirements \eqref{eq:ffperiods} on the upper triangular form of the basis $\beta^{(k)}_a$ are 
trivially fulfilled since $\beta^{(q)}_b={\mcal R}^{(q)}_b \Omega_n= 
{\mcal R}^{(q)}_b (\Pi^{(q)\,a}\hat{\gamma}^{(q)}_a+\ldots)\rightarrow\hat{\gamma}^{q}_b+\ldots$ 
where the dots indicate forms $\hat{\gamma}^{(k)}_a$ at grade $k>q$ with higher logarithms.
For concrete calculations it is convenient to present the solution to \eqref{eq:periodStrucutreLV} 
\cite{Mayr:1996sh},
\begin{equation}
	\mathbb{L}^{(k)\,b}=\sum_{\underline{a}}\mathbb{L}^{(k)\, b}_{a_1\ldots a_k}l_{a_1}\cdots l_{a_k}\,,\qquad\mathbb{L}^{(k)\, b}_{a_1\ldots a_k}=\mathcal{K}_{a_1\ldots a_kb_{k+1}\ldots b_{n}}\hat{a}_{b}^{b_{k+1}\ldots b_n}
\label{eq:PoincareDualOperator}
\end{equation}
where we used the abbreviation $\log(z^{a_i})=l_{a_i}$ and introduced the operator  that is Poincare 
dual to $A^{(k)}_c$, $\hat{A}^{(n-k)}_{c}=\hat{a}_{c}^{b_{1}\ldots b_{n-k}}J_{b_1}\cdots J_{b_{n-k}}$ 
obeying $\int_{\tilde X}A^{(k)}_a\wedge\hat{A}^{(n-k)}_{c}=\delta_{ac}$.
In other words, the solution $\mathbb{L}^{(k)\,b}$ is the Poincare dual to $A^{(k)}_b$ under the map
$\log(z^{a_i})\mapsto J_{a_1}$. For Calabi-Yau fourfolds we express 
\eqref{eq:PoincareDualOperator} on the grade two operators $k=2$, using the definition of the 
operator algebra relation \eqref{eq:FrobeniusOPE} in terms of the three-point function 
$C^{(1,1,2)}_{ab\alpha}$, as
\begin{equation}
	\mathbb{L}^{(2)\, \alpha}_{a_1a_2}=\mathcal{K}_{a_1a_2a_3a_4}\hat{a}_{\alpha}^{a_3a_4}
	=\int_{\tilde{X}_4}J_{a_1}J_{a_2}J_{a_3}J_{a_4}=C^{0\,(1,1,2)}_{a_1a_2\beta}\eta^{\beta\gamma}_{(2)}
	\int_{\tilde{X}_4}A_\gamma^{(2)}\wedge \hat{A}^{(2)}_{\alpha}=C^{0\,(1,1,2)}_{a_1a_2\beta}\eta^{\beta\alpha}_{(2)}\,.
\label{eq:fourfoldPoincareDual}
\end{equation}
Thus, the coefficients of $\mathds{L}^{(2)\,\alpha}$ are given by the classical intersections \eqref{eq:classicalamodeltriplecouplings} 
and the topological metric \eqref{eq:amodeltopologicametric} of the A-model, which is essential for the matching
of intersections of $\tilde{X}_4$ at large complex structure of $X_4$. We will also make extensive use of this relation
in our explicit examples in chapter \ref{ch:Calcs+Constructions}. 

After this technical preparations we present explicitly that the matching \eqref{eq:Atobeta_map} of the A- and B-model operators 
as implies of the complete Frobenius algebra. This is checked easily by matching by the fundamental coupling $C^{(1,1,n-2)}_{abc}$ on
both sides, which determines, as mentioned above, all other correlators. 
We restrict our attention to the fourfold case for which we evaluate \eqref{eq:3point} as
\begin{equation}  
C^{(1,1,2)}_{ a b \gamma}=\partial_{t^a} \Pi_{b}^{(2,1) \, \delta }
\eta^{(2)}_{\delta \gamma}=\partial_{t^a} \partial_{t^b} \Pi^{(2)\, \delta}
\eta^{(2)}_{\delta \gamma}=:\partial_{t^a} \partial_{t^b} F^{g=0}(\gamma)  \ ,
\label{eq:simple3point} 
\end{equation}   
where $a,b = 1,\ldots,h^{(3,1)}$ and $\gamma=1,\ldots, h^H_{(2,2)}$. 
Here we  used the upper triangular form \eqref{eq:FiltBasis} of
$\beta^{(k)}_a$ and the intersection properties \eqref{eq:def-eta} of the $\hat
\gamma_i$ for the first equality. Then we replaced 
$\partial_{t_a}\beta^{(0)}=\beta^{(1)}_a$ for general dimension $n$ as follows from 
\eqref{eq:alphader} and used the property (F2) in flat coordinates.
If we now let $\underline{z}\rightarrow 0$ and use the flat coordinates \eqref{eq:flatcoords}, 
which are given by \eqref{eq:generalformofsolution} as $t^i\sim
\log(z^i)+\text{hol.}\rightarrow\log(z^i)$, we see using \eqref{eq:fourfoldPoincareDual} 
that in this limit the classical intersection $C^{0\,  (1,1,2)}_{ a b \gamma}$ of 
\eqref{eq:classicalamodeltriplecouplings} are reproduced. Once the matching is 
established in  this large radius limit we can define the full quantum cohomological 
Frobenius structure by \eqref{eq:simple3point} in the coordinates \eqref{eq:flatcoords}. 
The latter can then be viewed as the classical topological intersections deformed by 
instanton corrections.

For the Calabi-Yau fourfold case at hand the intersections are obtained from the second derivative 
of the holomorphic quantities $F^0(\gamma)$ introduced in \eqref{eq:simple3point} 
for each basis element $\beta^{(2)}_\gamma$, $\gamma=1,\ldots h_H^{(2,2)}(X_4)$.
These are the analogues of the holomorphic prepotential $F^0$ familiar from the threefold case
and they are obtained as the special case $k=1$ from 
the general generating functionals\footnote{We note that the terms $b^{0}_{a\gamma }$, 
$a^{0}_{\gamma}$ are irrelevant for the quantum cohomology, but important for the 
large radius limit of the superpotential \eqref{eq:GVW-super}.} \eqref{eq:g=0multicovering} in section 
\ref{sec:EnumGeo}. The relation \eqref{eq:simple3point} tells us that we obtain the $F^0(\gamma)$ simply 
from the Picard-Fuchs equation as double-logarithmic solutions, that we will identify below.

However as  mentioned above the identification using the ring structure fixes the 
solutions of the Picard-Fuchs system so far only up to normalizations.
The normalization of the unique holomorphic solution is determined by the fact that 
the leading term in $X_0(\underline{z})=1+{\cal O}(\underline{z})$  has to be one. Also the dual period can be 
uniquely normalized by the classical $n$-point intersections.
The single logarithmic solutions are normalized to reproduce the effect of a shift 
of the NS-background B-field on $t^i= \int_{\cal C}(B+i J)$, where ${\cal C}$ 
is a generator of $H_2(\tilde X_4,\mathds{Z})$ and $J$ is the K\"ahler form. The shift 
is then $t^i\rightarrow t^i+1$ which corresponds to the monodromy around $z_i=0$ and implies 
according to \eqref{eq:flatcoords} that $\tilde{m}_{a\,\underline{\alpha}}=1$ for $|\alpha|=1$ in 
\eqref{eq:generalformofsolution}.
All other $t$-dependent quantities are restricted further by the monodromy of the period vector 
$\Pi$ of the holomorphic $(4,0)$-form, $\Pi=(\Pi^{(0)},\Pi^{(1) *},\Pi^{(2) *},\Pi^{(3) *}, \Pi^{(4)})^T$ 
around  $\underline{z}=0$.  Let $\Sigma$ be the matrix representing the intersection 
$K=\int_{X_4}\Omega_4\wedge \bar \Omega_4=\Pi\Sigma \Pi^\dagger$. Using \eqref{eq:omegaexpansion}, \eqref{eq:def-eta} 
it is easy to see that the  anti-diagonal of $\Sigma$ is given by the blocks 
$(1,(\eta^{(1)})^T, \eta^{(2)},\eta^{(1)},1)$. The monodromies act by $\Pi\rightarrow M_i\Pi$, 
where $M_i$ is a  $(h^H_4\times h^H_4)$ matrix.  The monodromy invariance of 
$K$ and $\hat \gamma_a^{(p)}\in H^4_H(X_4,\mathds{Z})$ implies 
\begin{equation} \label{eq:monodromyconstraint}
M_i^T \Sigma M_i=\Sigma
\end{equation}
with $M_i$ an integer matrix. However the monodromy at $\underline{z}\rightarrow 0$ is not enough to
fix the integral basis completely and the monodromy at other points in the moduli space and analytic 
continuation have to be applied to fix undetermined constants in the solutions to the Picard-Fuchs system.
This analysis is performed in the next section.

We conclude with some remarks about the enumerative geometry of the prepotentials $F^0(\gamma)$
of \eqref{eq:simple3point}. As in the threefold case there is an enumerative geometry or counting
interpretation of mirror symmetry in higher dimensions for the A-model
\cite{Klemm:2007in}. The results and formulas necessary for our analysis are
summarized in section \ref{sec:EnumGeo} to which we refer at several points.  As
will be discussed there, the label $\gamma$ is associated to a flux choice that is necessary to 
reduce to a counting of curves with the prepotential $F(\gamma)$ as a generating function,
cf.~\eqref{eq:g=0multicovering}.  The prepotentials furthermore have a
$\Li_2$-structure and it is possible to calculate the genus zero BPS
invariants $n^0_\beta(\gamma)$ of section \ref{sec:EnumGeo} for a suitable basis of
$H^{(2,2)}_{V}(\tilde X_4)$ and class $\beta$ in $H_2(\tilde X,\mathds{Z})$. This is in particular 
part of the calculations performed in chapter \ref{ch:Calcs+Constructions}.

\subsection{Fourfold Conifold: Monodromies and Classical Terms} \label{sec:analyticCont} 

In general the analysis of the global properties of the complex structure moduli space of $X_4$ 
is tedious since it requires analytic continuation of the periods which is numerically challenging.
In general one expects due to the flatness of the Hodge bundle only monodromy effects that
occur at the singular points in the moduli space. Upon demanding integrality of the corresponding 
monodromy matrix $M_i$ around these points the integral basis is fixed. In this section we explicitly present
such an analytic continuation to the universal conifold of the one modulus example of the sextic Calabi-Yau fourfold.
  
Before delving into these technicalities of this analysis we note that useful information about some of the irrational 
constants that appear for example in the leading logarithmic solution follow already from the Frobenius method 
\cite{Hosono:1994ax,libgober1998chern}.  
By this method the leading logarithmic solution can be obtained by applying the operator 
$D^{(4)}=\frac{1}{4! (2 \pi i)^4} \mathcal{K}_{i_1 i_2 i_3 i_4}\partial_{\rho_{i_1}}
\partial_{\rho_{i_2}}\partial_{\rho_{i_3}} \partial_{\rho_{i_4}}$ on the 
fundamental solution
\begin{equation} 
\omega_0({\underline{z}},{\underline{\rho}}) 
= \sum_{{\underline{n}} } c({\underline{n}},{\underline{\rho}}) z^{{\underline{n}}+{\underline{\rho}}}
\end{equation} 
with 
\begin{equation} 
 c({\underline{n}},{\underline{\rho}})=
\frac{\Gamma(-\sum_\alpha l_0^{(\alpha)} ( n_\alpha+\rho_\alpha)+1)}{\prod_i \Gamma(\sum_{\alpha} l_i^{(\alpha)}( n_\alpha+\rho_\alpha)+1)}\  
\end{equation}
and setting ${\underline{\rho}}=0$.    
The general leading logarithmic solution, i.e.~with all possible admixtures 
of lower logarithmic solutions, for $X^0:=\omega_0({\underline{z}})|_{{\underline \rho}=0}$ reads
\begin{equation}  \label{eq:Pi4}
\Pi^{(4)}=X^0(\tfrac{1}{4!}\mathcal{K}_{ijkl} t^{i}t^{j}t^{k}t^{l} +\tfrac{1}{3!} a_{ijk} t^{i}t^{j}t^{k} 
+\tfrac{1}{2!} a_{ij} t^{i}t^{j}+a_{i} t^{i}+a_0)\,,
\end{equation}
 where as in the threefold case $\mathcal{K}_{ijkl}$ is the 
classical top intersection.
It was observed in~\cite{Hosono:1994ax} for the threefold case that the Frobenius 
method reproduces some of the topological constants \eqref{eq:pre_largeV} in the leading and subleading logarithmic terms
of $F^0$. In particular we deduce the relations
$\int_{\tilde Z_3} c_2\wedge J_i=\frac{3}{\pi^2} \mathcal{K}_{i j k}\partial_{\rho_{j}} \partial_{\rho_{k}} c({\underline{0}},{\underline{\rho}})_ {{\underline \rho}=0}$ 
and $\int_{\tilde Z_3} c_3=\frac{1}{3!\zeta(3)} \mathcal{K}_{i j k}\partial_{\rho_{i}}\partial_{\rho_{j}} \partial_{\rho_{k}} c({\underline{0}},
{\underline{\rho}})_ {{\underline \rho}=0}$, where we denote by $c_i$ the $i$-th Chern class of 
the Calabi-Yau threefold $Z_3$, cf.~\eqref{eq:classic_terms_threefold}. If we generalize these to fourfolds, we obtain
\begin{equation}
 	\int_{\tilde X_4} \frac{3}{4} c_2^2+c_4
=\frac{1}{4!\zeta(4)} \mathcal{K}_{i j k l}\partial_{\rho_{i}}\partial_{\rho_{j}} \partial_{\rho_{k}} \partial_{\rho_{l}} c({\underline{0}},
{\underline{\rho}})_ {{\underline \rho}=0}\,.
\end{equation}
These constants are expected to appear as coefficients of the subleading logarithms in \eqref{eq:Pi4}. 

Similar as in the threefold case one can also use the induced K-theory charge 
formula~\cite{Minasian:1997mm,Cheung:1997az} in combination with central charge formula
\begin{equation} 
\vec Q\cdot \vec \Pi=-\int_{\tilde X_4} e^{-J} {\rm ch}(A)\sqrt{{\rm td}(\tilde X_4)}=Z(A)\ 
\end{equation} 
for $A$ denoting the bundle on the brane wrapping $\tilde X_4$
and mirror symmetry to obtain information about the subleading logarithmic terms in the periods.

Let us apply, however, a more direct argument and use properties of the simplest
Calabi-Yau fourfold, the sextic in $\mathds{P}^5$. The mirror sextic 
has the Picard-Fuchs equation, see e.g. \cite{Klemm:2007in},
\begin{equation}
\theta^5 - 6 z \prod_{k=1}^5(6 \theta+k)\ . 
\end{equation}    
We can easily construct solutions at $z=0$ using the Frobenius 
method, but let us first give a different basis of logarithmic 
solutions namely
\begin{equation}
\hat \Pi_k=\frac{1}{(2\pi i)^k}\sum_{l=0}^k\left(k\atop l\right) \text{log}(z)^l s_{k-l}(z)\ ,
\end{equation}
where
\begin{eqnarray}
   X^0 &=& s_0=1+720z+748440 z^2+\ldots \ ,\quad
   s_1 = 6246 z+ 7199442z^2+\ldots\ , \nn \\  
   s_2 &=& 20160z+327001536 z^2+\ldots,\quad
   s_3 =-60480 z-111585600 z^2+\ldots \nn \\ 
   s_4 &=&-2734663680z^2
-57797926824960 z^3+\ldots\ .
\end{eqnarray}
The point is that under the mirror 
map one obtains $\hat \Pi_k=t^k+{\cal O}(q)$, so that these solutions correspond
to the leading volume term of branes of real dimension $2k$. 
The ``conifold'' locus of the sextic is at $\Delta=1-6^6 z=0$. Near that point the 
Picard-Fuchs equation has the indicials $\left(0,1,2,3,\frac{3}{2}\right)$. It 
is easy to construct solutions and we choose a basis in which 
the solution  to indicial $k\in \mathds{Z}$ has the next power $z^4$.
The only unique solution posses a branch cut and reads 
\begin{equation}
\nu=\Delta^\frac{3}{2}+\frac{17}{18} \Delta^\frac{5}{2}
+\frac{551}{648} \Delta^\frac{7}{2}+\ldots\ .
\end{equation}

The situation at the universal conifold is crucial for mirror 
symmetry in various dimensions $n$. At this point the non-trivial 
monodromy affects a cycle of topology $\mathbb{T}^n$ that 
corresponds to the solution $X^0$, i.e. the zero-dimensional brane 
in the $A$-model, which is uniquely defined at $z=0$, and a cycle of topology 
$\mathbb{S}^n$ that corresponds to the solution $\Pi^{(n)}$, i.e.\ the top
dimensional brane in the $A$-model. The topological intersection 
between these cycles is $1$ and their classes in homology are 
the fiber and the base of the SYZ-fibration respectively \cite{Strominger:1996it}. 
In odd dimensional Calabi-Yau manifolds, like threefolds $Z_3$, the conifold monodromy
acts on the vector $\Pi_{\text{red}}=(\Pi^{(3)},X_0)^T$ by a matrix as
\begin{equation}
	M_{2\times 2} 
=\begin{pmatrix}[cc] 1&0\\ 1&1 \end{pmatrix}.
\label{eq:monodromyConi3fold}
\end{equation}
This corresponds to the Lefschetz formula  with vanishing $\Pi^{(3)}$, 
i.e. the quantum volume of $\tilde Z_3$ vanishes. 

In four dimensions we have a monodromy of order two and the only 
way to  obtain an integral idempotent monodromy compatible with the 
intersection \eqref{eq:monodromyconstraint} is 
\begin{equation}
	M_{2\times 2}=\begin{pmatrix}[cc] 0&1\\ 1&0 \end{pmatrix}\,.
\label{eq:monodromyConi}
\end{equation} 
It is noticeable that the zero- and  the highest dimensional brane get 
exchanged by the conifold monodromy in even dimensions.  
This implies the identification $X^0=\eta-c \nu$ and\footnote{The sign is chosen so 
that the $t^4$ term in $\Pi^{(4)}$ has a positive sign.}  
$\Pi^{(4)}=\eta+c \nu$. Here $\eta$ is a combination of solutions at 
$\Delta=0$  without branch cut. We can determine the latter by analytic 
continuation of $X^0$ to the conifold. While the precise combination is 
easily obtained, the only constant that matters below is $c$, which turns 
out to be $c=\frac{1}{\sqrt{3}\pi^2}$. 

Now we can determine the combination 
which corresponds to the correct integral choice of the geometric period 
$\Pi^{(4)}$ as
\begin{equation}
 \Pi^{(4)}=2 c \nu +X^0\ 
\label{eq:canonical}
\end{equation}
from the uniquely defined periods $(X^0,\nu)$ at $z=0$ and 
$\Delta=0$. The analytic continuation of $\nu$ to $z=0$  fits nicely with our 
expectation from above and fixes most of the numerical coefficients in \eqref{eq:Pi4} 
universally. We obtain
\begin{equation}
  a_0=\frac{\zeta(4)}{2^4 (2\pi i)^6} \int_{\tilde X_4} 5 c_2^2\ , \qquad 
\end{equation}
and
\begin{equation}
   a_{i} =- \frac{\zeta(3)}{(2\pi i)^3} \int_{\tilde X_4} c_3 \wedge J_i \, ,\quad a_{ij}=\frac{\zeta(2)}{2(2\pi i)^2} \int_{\tilde X_4} c_2 \wedge J_i \wedge J_j\, , 
   \quad a_{ijk} = \tilde c \int_{\tilde X_4} \imath_*(c_1(J_i)) \wedge J_j \wedge J_k\ ,
\end{equation}
where as before $c_i = c_i (T_{\tilde X_4})$ and $c_1(J_i)$ denotes the first Chern class of the 
divisor associated to $J_i$ which is mapped to a four-form via the Gysin homomorphism $\imath_*$ 
of the embedding map of this divisor into $\tilde X_4$. This is the generalization of 
\eqref{eq:classic_terms_threefold} to the case of Calabi-Yau fourfolds. To be precise, the 
coefficients $a_{ijk}$ are not fixed by the sextic example, because it turns out to be zero 
in this case, and for the canonical choice of $\Pi^{(4)}$ \eqref{eq:canonical}. This does not mean 
that it is absent in general. Rather it implies that it is physically irrelevant for the sextic 
because the divisibility of the correctly normalized solution, which is cubic in the logarithms, 
allows an integral symplectic choice of the periods in which this term can be set to zero. This 
might not be in general the case and  other hypersurface in weighted projective space indicate 
that $\tilde c=1$. It is similarly possible to use the orbifold monodromy to fix the exact 
integral choice of the other periods. The principal form of the terms should again follow 
from the Frobenius method.

\section{Basics of Enumerative Geometry} \label{sec:EnumGeo}

In this section we describe the relevant
enumerative quantities for the A-model, which are calculated in this work in the B-model using
mirror symmetry. We start with a discussion of the genus zero and genus one closed string
Gromov-Witten invariants in section \ref{sec:csGW} before we review the open string
Gromov-Witten invariants in section \ref{sec:osGW}. 
Then we apply the definitions
and structures of sections \ref{sec:csGW} and \ref{sec:osGW} to the Calabi-Yau
threefold superpotentials $W_{\rm flux}$ and $W_{\rm brane}$ in section \ref{sec:CY3superpotsEnumGeo}.
Similarly we proceed with the flux superpotential on fourfolds in section \ref{sec:CY4superpotsEnumGeo}.
It will be this enumerative meaning of the B-model calculations for the A-model that will provide 
important cross-checks for the computations of the superpotentials in chapter \ref{ch:Calcs+Constructions}. 

\subsection{Closed Gromov-Witten Invariants} \label{sec:csGW}

First we review the theory of closed Gromov-Witten invariants, i.e.\ the theory
of holomorphic maps
\begin{equation}
	\phi:\Sigma_g \rightarrow \tilde X
\end{equation}
from an oriented closed curve $\Sigma_g$ into a  Calabi-Yau manifold $\tilde X$, where
we do not consider marked points for simplicity.
It can be defined mathematically rigorously in general and explicitly calculated
using localization techniques if $\tilde X$ is represented for example by a
hypersurface in a toric variety. Here $g$ is the genus of the domain curve $\Sigma_g$ and
we denote by $\beta\in H_2(\tilde X,\mathds{Z})$ the homology class of the image
curve. The multi-degrees of the latter are measured with respect to an ample polarization 
$L$ of $\tilde X$, i.e.~$\beta=\sum_{i=1}^{h^{(1,1)}}
d_i \beta_i$ for ${\rm deg} (\beta)=\int_{\beta} c_1(L)=\sum_{i=1}^{h^{(1,1)}} d_i t_i$ with
$d_i\in\mathds{Z}_+$.  In string theory and in the context of the mirror
symmetry the volume of the curve $\beta_i$ is complexified by an integral over
the antisymmetric two-form field $B$.  Thus, one defines the complexified closed
K\"ahler moduli $t^i=\int_{\beta_i} (B+ic_1(L))$.   

For smooth $\tilde X$ the virtual (complex) dimension of the moduli space of 
these maps $\phi$ is computed by an index theorem and reads  
\begin{equation}
{\rm vir} \ {\rm dim} \ {\cal M}_g(\tilde X,\beta)= \int_\beta c_1(\tilde X)+ ({\rm dim}
\tilde X-3)(1-g)\ .
\label{eq:virtdim} 
\end{equation}            
In particular for Calabi-Yau fourfolds one obtains ${\rm vir} \ {\rm dim} \
{\cal M}_g(\tilde X_4,\beta)=1-g$. Thus in order to define genus zero
Gromov-Witten invariants one requires an incidence relation of the image curve with
$k=({\rm dim}(\tilde X)-3)$ surfaces to reduce the dimension to zero
in order to arrive at a well-defined counting problem. For fourfolds one thus
needs one incidence surface and we denote the dual cycle of the surface by
$\gamma\in H^{(2,2)}(\tilde X_4)$.
Note that for $\dim \tilde X\ge 4$ and $g\ge 2$ the
dimension of the moduli space is negative and
no holomorphic maps exist. The Calabi-Yau threefolds are critical in the sense
that the dimension of the moduli space for all genera is zero. Thus, in general
invariants associated to the maps are non-zero for all values of $g$.

We define a generating function for each genus $g$ Gromov-Witten invariant as follows:
\begin{equation}
F^g(\gamma_1)=\sum_{\beta\in H_2(\tilde X,\mathds{Z})}
r^g_\beta(\gamma_1,\dots,\gamma_k)  q^\beta \ .
\end{equation}
They
are labeled by $g$, $\beta$ and for $\dim \tilde X\ge 4$ also by cycles
$\gamma_i$ dual to the incidence surfaces. 
Here $q^\beta$ is a shorthand notation for $q^\beta=\prod_{i=1}^{h^{(1,1)}} e^{2\pi it_i
d_i}$. We note  that this is not just a formal power series\footnote{This is 
important for the interpretation of such terms in the effective action. In fact,
analyticity allows to define such terms beyond the large radius limit in terms
of period integrals on the mirror geometry.}, but rather has  finite region of
convergence for large volumes of the curves $\beta_i$, i.e.\ for ${\rm
Im}(t^i)\gg 0$. This puts a bound on the growth of the Gromov-Witten invariants
$r^g_\beta(\gamma_1)$. 
The contributions of the
maps is divided by their automorphism groups and the associated Gromov-Witten
invariants $r^g_\beta(\gamma_1,\ldots, \gamma_k)$ are in general rational.

Although the discussion of \eqref{eq:virtdim} indicates that the Gromov-Witten
theory on higher dimensional Calabi-Yau manifolds  
is less rich than in the threefold case, one has a remarkable integrality
structure associated to the 
invariants. In particular at genus zero one can define integer  
invariants $n^g_\beta(\gamma_1,\ldots \gamma_k)\in \mathds{Z}$ 
for arbitrary ${\rm  dim}(\tilde X)=k+3$  dimensional manifolds as
\begin{equation} 
F^0(\gamma_1,\ldots, \gamma_k)=\tfrac{1}{2}C^{0\, (1,1,n-2)}_{ab\gamma_1\cdots\gamma_k} 
t^a t^b + b^{0}_{a \gamma_1\cdots\gamma_k }t^a + a^{0}_{\gamma_1\cdots\gamma_k} +  \sum_{\beta>0} n^g_\beta(\gamma_1,\ldots,\gamma_k) {\rm Li}_{3-k} (q^\beta)\ ,
\label{eq:g=0multicovering}
\end{equation}  
where ${\rm Li}_{p}(q)=\sum_{d=1}^\infty \tfrac{q^d}{d^{p}}$ and $C^{0\,
  (1,1,n-2)}_{ab\gamma_1\cdots\gamma_k}$ are the classical triple intersections.   
For threefolds an analogous formula, see \eqref{eq:closedBPS} below, was found 
in \cite{Candelas:1990rm} and
the multicovering was explained in~\cite{Aspinwall:1991ce}. Note that
$b^{0}_{a \gamma_1\cdots\gamma_k }$, $a^{0}_{\gamma_1\cdots\gamma_k}$ are
irrelevant for the quantum cohomology, as the latter is defined by the 
second derivative of $F^0(\gamma_1,\ldots, \gamma_k)$, cf.~\eqref{eq:simple3point}.

Genus one Gromov-Witten invariants exist on Calabi-Yau manifold of all 
dimensions with the need of incidence conditions as discussed above. For fourfolds 
the authors of \cite{Klemm:2007in} define the following integrality 
condition    
\begin{equation}
F^1 =
\sum_{\beta>0} n^1_\beta \tfrac{\sigma(d)}{d} q^{d\beta} 
 +\tfrac{1}{24} \sum_{\beta>0} n^0_\beta(c_2(\tilde X_4)) \log(1-q^\beta) 
-\tfrac{1}{24} \sum_{\beta_1,\beta_2} m_{\beta_1,\beta_2} \log(1-q^{{\beta_1+\beta_2}})\, .  
\end{equation} 
Here the $m_{\beta_1,\beta_2}$ are so called meeting invariants, which
are likewise integers as the $n^g_\beta(\cdot)$
and the function $\sigma$ is defined by $\sigma(d)=\sum_{i|d}i$.

For Calabi-Yau threefolds there is also the general counting formula for BPS-states that is 
obtained by evaluating a one-loop Schwinger computation for M2-branes in the M-theory lift
of Type IIA \cite{Gopakumar:1998jq}
\begin{equation}
F(\lambda,q)=\sum_{g=0}^\infty \lambda^{2g-2}F^g = \sum_{k=1}^\infty \sum_{\beta>0,g\ge
  0} n^g_\beta\tfrac{1}{k} \left(2 \sin \tfrac{k
  \lambda}{2}\right)^{2g-2} q^{k\beta}\ .
\label{eq:closedBPS}
\end{equation}   
Here $g$ not only corresponds to genus of the curve wrapped by the M2-branes but
also to the left spin of the M2-brane in the five-dimensional theory in a fixed 
basis of the representation of the little group for massive particles, $SO(4)\cong SU(2)\times SU(2)$.

\subsection{Open Gromov-Witten Invariants} \label{sec:osGW} 

Let us come now to the open Gromov-Witten invariants on Calabi-Yau 
threefolds $\tilde Z_3$. They arise in the open topological A-model on $\tilde Z_3$. 
We consider a Calabi-Yau threefold $\tilde Z_3$ together with a special Lagrangian
submanifold $L$ and consider a map from an oriented open Riemann surface,
i.e.~a Riemann surface with boundary,
\begin{equation} 
\psi:\Sigma_{g,h}\rightarrow (\tilde Z_3,L) 
\end{equation}      
into the Calabi-Yau threefold $\tilde Z_3$. Here the Riemann surface is mapped with
a given winding number into $L$ such that the $h$ boundary circles $B_i$ of 
$\Sigma_{g,h}$ are mapped on non-trivial elements $\vec \alpha =(\alpha_1,\ldots ,\alpha_h)\in
H^1(L,\mathds{Z})^{\oplus h}$. As in the closed case we do not 
consider marked points. For threefolds the moduli space  
${\cal  M}_{g,h}(\tilde Z_3,L,\beta,\vec \alpha, \mu)$ with the Maslov index $\mu$
 has virtual dimension zero \cite{liu2002moduli}. If
$H^1(L,\mathds{Z})$ is non-trivial, the special Lagrangian has
classical geometric deformation moduli. The open string moduli $\underline{u}$ are 
 complexifications of the geometric moduli by the Wilson-Loop integrals 
of the flat $U(1)$ gauge connection on the brane.      

The open BPS-state counting formula analogous to \eqref{eq:closedBPS}
is derived in \cite{Ooguri:1999bv} by counting degeneracies of open M2-branes 
ending on an M5-brane wrapping L or of D4-branes wrapping L in the Type IIA picture.
It reads
\begin{eqnarray} 
F(\underline{t},\underline{u})&=&\sum_{g,h}\lambda^{2g-2+h}\,F^{g}_h(\underline{t},\underline{u})
=\sum_{g=0,h=1}^\infty\sum_{\alpha_n^i=0}^\infty \lambda^{2g-2+h}\, F^{g,\alpha^i_n}_{n_1,\ldots,n_h} (\underline{t}) 
\prod_{n=1}^h{\rm tr} \bigotimes_{i=1}^{b^1(L)}  U_i^{\alpha_n^i}\nn\\
&=&i \sum_{n=1}^{\infty}\sum_{\cal R}\sum_{\beta>0,r\in\mathds{Z}/2}
\tfrac{n^g_{\beta,{\cal R}}}{2 n \sin\left(\tfrac{n\lambda}{2}\right)}
q^{nr}_\lambda q^{n\beta}\, {\rm Tr}_{\cal R} \bigotimes_{i=1}^{b^1(L)}U_i^n\ .  
\end{eqnarray}
Here $\alpha^i = (\alpha^i_1,\ldots,\alpha^i_h)\in H^1(L,\mathds{Z})^{\oplus h}$ and 
$q_\lambda = e^{2\pi i\lambda}$. The instanton numbers counting
BPS-particles of the M2-brane ending on the M5-branes in the representation $\mathcal{R}$ and 
of spin $g$ are given by the integers $n^g_{\beta,{\cal R}}$. The class $\beta$ corresponds to the 
bulk charge as in the closed string formula. The holonomies of the gauge field on the D4-brane 
along non-trivial one-cycles in $H_1(L)$ are given by $\text{tr}\, U_i$, $i=1,\ldots, b_1(L)$, where the
trace is taken in the fundamental representation. 
The integers $\alpha^i_n$ are the winding numbers of the $n$-th boundary along the $i$-th element of $H^1(L,\Z)$. The open string moduli $u$ enter the definition of the matrices $U_i$. 
In particular the disk amplitude, which gives rise to the 
superpotential, is given by
\begin{equation} 
W_{\rm brane}=F^0_{h=1}=\sum_{\beta, m} n^m_\beta{\rm Li}_2(q^\beta Q^m)
\label{eq:wopgromovwitten}
\end{equation}     
with $Q^m=e^{2\pi i m_iu^i}$. Comparison with \eqref{eq:g=0multicovering} suggest 
that the counting problem of specific disks amplitudes can
be mapped to the counting of rational curves in fourfolds since the integrality
structure is the same given by the $\Li_2$-double-covering.

\subsection{Closed and Open Superpotentials on Calabi-Yau Threefolds} \label{sec:CY3superpotsEnumGeo}

Let us first discuss the flux superpotential $W_{\rm flux}$ as it occurs in Type IIB and heterotic 
compactifications, cf.~sections \ref{sec:suppot} respectively \ref{sec:het_superpot}, and its meaning
for mirror symmetry. As noted before it takes in both setups the same form and can be 
evaluated in terms of the periods $(X^K,\mathcal{F}_K)$ of the holomorphic three-form $\Omega$ as
\begin{equation} \label{eq:flux_threefold}
  W_{\rm flux} = \hat N_K X^K(\underline{z}) - \hat M^K \mathcal{F}_K(\underline{z})\ ,
  \qquad  X^K = \int_{A_K} \Omega \ , \qquad \mathcal{F}_K = \int_{B^K} \Omega\ \,,
\end{equation}
where $(\hat M_K,\hat N^K) = (M_K -\tau \tilde M_K,N^K - \tau \tilde N^K)$ are complex numbers 
in an $\mathcal{N}=2$ Type IIB theory formed from the flux quantum numbers $(M_K,N^K)$ of $F_3$ and 
$(\tilde M_K,\tilde N^K)$ of $H_3$. Whereas in the $O3/O7$-orientifold setup both $F_3$ and $H_3$ fluxes
contribute to $W_{\rm flux}$, in the $O5/O9$-orientifold setup of section \ref{sec:suppot} there 
are only fluxes $F_3$ due to the orientifold projection $\mathcal{O}$. Similarly in the heterotic setup of 
\ref{sec:het_superpot} only NS--NS fluxes are present. As before we have introduced the symplectic basis 
$(A_K,B^K)$ of three-cycles in $H_3(Z_3,\Z)$. 

As we have reviewed in section \ref{sec:CSModuliSpace+PFO} 
the complete complex structure moduli dependence of $W_{\rm flux}$ is determined, once the flux is specified, 
from the periods obeying the Picard-Fuchs differential system. The use of the Picard-Fuchs system makes it 
even possible to evaluate the flux superpotential deep inside the complex structure moduli space, where 
conventional supergravity breaks down due to strong curvature effects for example from singularities in $Z_3$ like
the conifold. Thus, $W_{\rm flux}$ in general inherits the characteristic properties of special geometry, in particular the 
existence of the prepotential. In other words, even in $\mathcal{N}=1$ effective actions, the flux superpotential 
enjoys remnants of an underlying $\mathcal{N}=2$ structure.

In the context of mirror symmetry and the enumerative interpretation of the A-model 
a few further general observations can be made. Although the concrete form of the flux 
superpotential $W_{\rm flux}$, will highly depend on the point at which it is evaluated 
on the complex structure moduli space, we can make statements about its structure at 
particular points in th moduli space. One particularly distinguished point is the large 
complex structure point which by mirror symmetry corresponds to a large volume 
compactification of Type IIA string theory. As we have seen in section \ref{sec:CSModuliSpace+PFO}
the solutions at this point have a characteristic grading by powers of $\log(\underline{z})$.
Mirror symmetry then maps the logarithmic terms to classical large-volume contributions on the Type 
IIA side while the regular terms in the periods $(\underline{X},\underline{\mathcal{F}})$ encode the closed 
string world-sheet instantons corrections\footnote{Recall that the mirror map takes the form $z^i=e^{2\pi it^i} + \ldots$, 
where $t^i = X^i/X^0$ on the IIB side which is identified with the world-sheet volume complexified with the NS-NS B-field on the 
Type IIA side.}. Inserting the form of the prepotential \eqref{eq:pre_largeV} into the flux 
superpotential \eqref{eq:flux_threefold} one finds the characteristic structure
\begin{eqnarray} \label{eq:flux_ori}
   W_{\rm flux} \!\!&\!\!=&\!\!\! X^0\big[\hat N_0 + \hat{M}^0\mathcal{K}_0-\hat{M}^i\mathcal{K}_i+(\hat N_i-\hat{M}^j\mathcal{K}_{ij}+\hat{M}^0\mathcal{K}_i) t^i 
   -  \tfrac12\hat M^i \cK_{ijk}\, t^j t^k+\tfrac1{3!}\hat{M}^0\mathcal{K}_{ijk}\,t^i t^j t^k \nn\\ 
   \!\!&\!\!+&\!\!\!(\hat{M}^i-\hat{M}^0 t^i)\sum_{\beta} d_i n_\beta^0\,\text{Li}_2(q^\beta)
   +2\hat{M}^0\sum_{\beta} n_\beta^0\,\text{Li}_3(q^\beta)\big]\ .
\end{eqnarray}
This expression directly shows that in addition to a cubic polynomial of classical terms, also  instanton 
correction terms proportional to $\text{Li}_k(z) = \sum_{n=1}^\infty \frac{z^n}{n^k}$ for $k=2,3$ 
are induced by non-vanishing fluxes $\hat{M}^i$, $\hat{M}^0$. Thus, we can directly read off the Gromov-Witten
invariants $n_\beta^0$ of \eqref{eq:closedBPS} from the flux superpotential $W_{\rm flux}$ for particular flux choices. 
Conversely, given a few invariants $n_\beta^g$, we can determine the flux numbers for 
any given superpotential $W_{\rm flux}$ or compare to superpotentials obtained from different setups,
like e.g.~Calabi-Yau fourfolds, or different string theories using string dualities. Indeed, this
will be our strategy to explicitly relate Calabi-Yau fourfold superpotentials to dual Type IIB or heterotic superpotentials 
in sections \ref{sec:Superpots+MirrorSymmetry} and \ref{sec:SuperpotsHetF}.  

Let us now turn to the superpotential for the open string sector that is given, for a curve $\Sigma_{\underline{u}}$, 
by the chain integral
\begin{equation}
	W_{\text brane}=\int_{\Gamma(\underline{u})}\Omega(\underline{z})\,,\qquad \partial\Gamma(\underline{u})\supset \Sigma_{\underline{u}}\,,
\end{equation}
which we encountered in the Type IIB context in section \ref{sec:suppot} as a D5-brane superpotential, 
in section \ref{sec:het_superpot} in the heterotic context as a small instanton/five-brane superpotential
and in F-theory setups in section \ref{sec:F-flux_sup} as a seven-brane superpotential. 

Ideally one would like to explicitly compute the functional dependence of $W_{\text brane}$ on the
brane deformations $\underline{u}$ of the curve $\Sigma_{\underline{u}}$ and the complex structure 
moduli $\underline{z}$ e.g.~by evaluating, as in the closed string case, a set of \textit{open-closed} 
Picard-Fuchs equations. Indeed, one way to achieve this is to lift the setup to an F-theory compactification
on a Calabi-Yau fourfold, as we will demonstrate in section \ref{sec:Superpots+MirrorSymmetry}. Another, more 
direct and mathematically canonical procedure is discussed in part III of this work \cite{Grimm:2008dq,Grimm:2010gk}. 
There a constructive method is used to directly compute the superpotential $W_{\rm brane}$ for five-branes on 
a curve $\Sigma_{\underline{u}}$ on the complex geometry side  by mapping the deformation problem of the curve  
$\Sigma_{\underline{u}}$ in $Z_3$ to the deformation problem of complex structures on a non-Calabi-Yau threefold $\hat{Z}_3$, 
that is canonically obtained by blowing up $Z_3$ along the curve 
$\Sigma_{\underline{u}}$. 

Before performing any calculations we infer some crucial properties of $W_{\text brane}$ by applying 
mirror symmetry at the large complex structure/large volume point. Recall that under mirror symmetry, 
a Type IIB compactification with D5- or D7-branes is mapped to a Type IIA compactification with D6-branes
wrapping special Lagrangian cycles $L$ in the mirror Calabi-Yau space $\tilde Z_3$, cf.~section \ref{sec:mirror_toric_branes}.
Thus, in order for mirror symmetry with branes to hold \cite{Vafa:1998yp} the superpotentials have to 
agree on both sides.
However, on the A-model side the moduli of $L$ are counted by elements in $H^1(L,\Z)$ and are generically
unobstructed \cite{mclean1998deformations}. In contrast, the deformations of the curve $\Sigma_{\underline{u}}$ 
are in general obstructed, which is a basic fact in classical geometry \cite{kodaira1962theorem}, and 
reflected in physics by the non-trivial superpotential $W_{\text brane}$ \cite{Kachru:2000an,Kachru:2000ih}.
Consequently, the superpotential in the A-model has to be induced entirely by quantum 
corrections\footnote{In mathematical terms this equivalence can be formulated as a matching of classical 
obstruction theory of $\Sigma$ on the B-model side with quantum obstructions $L$ in the A-model.}, which 
are string world-sheet discs ending on $L$. 
As was reviewed in detail in section \ref{sec:osGW} this superpotential induced by the open string world-sheets 
reads 
\begin{equation} 
   W_{\rm brane}= C_{i} t^i \hat t + C_{ij} t^i t^j +C \hat t^2+  \sum_{\beta,\, m} n_{\beta}^m\, {\rm Li}_2(q^\beta Q^m)\ ,\qquad Q^m = e^{2\pi im_iu^i}\ .
\label{eq:wgromovwitten}
\end{equation}     
with constants $C,C_i,C_{ij}$ and the Gromov-Witten invariants $n^m_{\beta}$ determined by the brane 
geometry and $\tilde{Z}_3$, as well as the flux $F_2$ in the seven-brane context.

\subsection{Flux Superpotentials on Calabi-Yau Fourfolds} \label{sec:CY4superpotsEnumGeo}

In this section we discuss the F-theory flux superpotential and recall how 
mirror symmetry for Calabi-Yau fourfolds \cite{Greene:1993vm,Mayr:1996sh,Klemm:1996ts,Grimm:2009ef} 
allows to relate it to the enumerative geometry of the A-model. In this route we demonstrate 
some general features of the flux superpotential. 

Recall, that the F-theory superpotential is induced by four-form flux $G_4$ and 
given by \cite{Gukov:1999ya} 
\begin{equation} \label{eq:fluxpotFourfoldExpanded}
W_{G_4}(\underline{z})=\sum_kN^{(k)\,a}\,  \Pi^{(k)\,b}(\underline{z})\, \eta^{(k)}_{ab}=\mathbf{N}\Sigma\Pi\ ,
\end{equation} 
where $\underline{z}$ collectively denote the $h^{(3,1)}(X_4)$ complex structure deformations 
of $X_4$, $\Pi^{(k)}_a$ the periods and the integers $N^{(k)\, a}$ the flux quanta. Both
periods and flux quanta are summarized in vectors $\mathbf{N}$, $\Pi$, where $\Sigma$ denotes a
$h^4_H\times h^4_H$-matrix containing the topological metric\footnote{Recall that in contrast to 
$H^3(Z_3,\Z)$ of Calabi-Yau threefolds the fourth cohomology group of $X_4$ does not carry a symplectic 
structure which necessitates the introduction of $\eta^{(k)}_{ab}$.} $\eta^{(k)}_{ab}$ in \eqref{eq:def-eta}.
Here we further used the expansions into an integral basis $\hat{\gamma}^{(k)}_a$ of $H^4_H(X_4,\Z)$  as
\begin{equation}
    \Omega_4=\sum_k\Pi^{(k)\,a}\hat{\gamma}^{(k)_a}\,,\quad  \Pi^{(k)\, a} = \int_{\gamma^{(k)}_{a}} \Omega_4\,,
    \qquad G_4= \sum_k N^{(k)\, a}\ \hat{\gamma}^{(k)}_{a}\ , \quad  N^{(k)\, a} = \int_{\gamma^{(k)}_{a}} G_4 \,.
\end{equation} 
We refer to section \ref{sec:FFMirrors} and in particular \eqref{eq:FiltBasis} for 
more details on the notation. 

In F-theory setups the flux $G_4$ is restricted by the two conditions \eqref{eq:fluxcondition1}
and \eqref{eq:fluxcondition2}. The latter condition implies that $G_4$ is an element in the primary horizontal 
subgroup 
\begin{equation}
H^{(2,2)}(X_4)=H^{(2,2)}_V(X_4)\oplus  H^{(2,2)}_H(X_4)\ . 
\end{equation} 
A corollary of this statement is that the Chern classes are in
the vertical subspace, so that half integral flux quantum numbers are not
allowed if condition \eqref{eq:fluxcondition2} is met. In general, it is an important 
open problem to have a description of four-flux $G_4$ on a generic Calabi-Yau 
fourfold. Formally, this can be solved by mirror symmetry established via the map \eqref{eq:Atobeta_map}.                     
This implies that one can think of the integral basis $\hat \gamma_a$ in terms of their corresponding 
differential operators ${\cal R}_{a}^{(k)}$ acting on $\Omega_4$. 
In particular, 
this formalism allows us to express the flux $G_4$ in an integral basis in the form
\begin{equation} 
  G_4=\sum_{k=0}^4 \sum_{p_k} N^{p_k (k)} \left. {\cal R}_{p_k}^{(k)}\Omega_4\right|_{\underline{z}=0}\ .
\end{equation} 
The representation of the integral basis as differential operators 
will be particularly useful in the identification of the 
heterotic and F-theory superpotential, cf.~chapter \ref{ch:Calcs+Constructions} and section \ref{sec:non-CYblowup}.

Once the flux $G_4$ is fixed, also in the fourfold case the flux superpotential 
\eqref{eq:fluxpotFourfoldExpanded} is completely determined in terms of the periods $\Pi$.
As we have discussed in section \ref{sec:matching} the periods $\Pi$ can in principle be 
determined from the Picard-Fuchs differential system which allows for an analytic continuation 
of $W_{G_4}$ deep into the complex structure moduli space of $X_4$. However, it is the most 
important task on the B-model side to find the fourfold periods which are evaluated with 
respect to an integral basis of $H^H_4(X_4,\mathds{Z})$. Moreover, an intrinsic definition 
of the integral basis $\gamma^{(k)}_{a}$ seems to be technically impossible, due to the absence
of a symplectic basis as in the threefold case, and mirror symmetry and analytic continuation, 
like the monodromy analysis at the  conifold in section \ref{sec:analyticCont}, have to be 
used in order to construct an integral basis. 

In the context of mirror symmetry it is meaningful to comment on the structure of \eqref{eq:fluxpotFourfoldExpanded}
at distinguished points in the complex structure moduli space. Again the large complex
structure/large volume point is of particular importance since an interpretation as
classical volumes and quantum instantons on the A-model side is possible.

For a toric hypersurface $X_4$ the point of maximal unipotent monodromy is the origin in the Mori cone
coordinate system $\underline{z}$ introduced in section \ref{sec:PFO+3foldMirrorSymmetry} as in the threefold case. 
Geometrically at the point $\underline{z}=0$  several cycles $\gamma^{(k)}_a$ hierarchically vanish which
is encoded in the grading of the solutions to the Picard-Fuchs system by powers in $(\log(z^i))^k$, $k=0,4$, see
\eqref{eq:LeadingLog}. According to the map \eqref{eq:Atobeta_map} and the condition \eqref{eq:periodStrucutreLV} there is 
one analytic solution $X^0(z)=\int_{\gamma_0} \Omega_4$ corresponding to the 
fundamental period, $h^{(3,1)}(X_4)$ logarithmic periods $X^a(z)=\int_{\gamma_a} \Omega_4 \sim X^0(z) \log(z_{a})$,
$h^{(2,2)}_H(X_4)$ double logarithmic solutions, $h^{(3,1)}(X_4)$ triple logarithms and one quartic logarithms.
Noting that $t^a\sim \log(z^a)$ at this point we can use these flat coordinates to write the leading logarithmic
structure of the period vector as
\begin{equation} 
\Pi^T=\Big(\int_{\gamma^{(0)}} \Omega_4,\ldots, \int_{\gamma^{(4)}_{b_H^4}} \Omega_4 \Big) 
\sim X^0\big(1,\, t^a,\,\tfrac{1}{2} C^{0\,\delta}_{ab} t^a t^b,\,
\tfrac{1}{3!} C^{0\,a}_{bcd} t^b t^c t^d ,\, \tfrac{1}{4!} C^{0}_{abcd} t^a t^b t^c t^d \big) \ .    
\label{eq:4f-period} 
\end{equation}   
Here we have introduced the constant coefficients $C_{ab}^{0\,\delta}:=\eta^{(2)\ \delta\gamma}
C^{0\, (1,1,2)}_{ab\gamma}$, $C_{abc}^{0\,0}=\eta^{(1)\ ed} \mathcal{K}^{0}_{abcd}$ that are related to 
the classical three-point function $C^{0\,,(1,1,2)}_{ab\gamma}$ and the intersection numbers 
$\mathcal{K}^0_{abcd}$, cf.~\eqref{eq:fourfoldPoincareDual}. These couplings can be calculated in 
the classical cohomology ring of $\tilde{X}_4$ in the basis \eqref{eq:amodeloperators} via the 
integrals \eqref{eq:classicalamodeltriplecouplings} and \eqref{eq:FourfoldclassicalIntersections}. 
In particular, the grading $(\{k\})=(0,1,2,3,4)$ in powers 
of $t^a$ corresponds to a grading of $\gamma_a\in H_4(X_4)$ which matches the grading of the dual 
cohomology group $H_H^4(X_4,\mathds{Z} )$ in the fixed complex structure given by the point $\underline{z}$.    
 We note that the periods \eqref{eq:4f-period} contain instanton
corrections that we suppressed for convenience, that are however crucial for the A-model. 

For applications of fourfolds for example to F-theory the instanton corrections in particular help to identify the 
physical meaning of the periods, like e.g.~the interpretation in terms of the flux or brane superpotential of
the underlying Type IIB theory in the limit \eqref{eq:SuperpotLimit}. Indeed the comparison of
the enumerative data of the double logarithmic periods $F^0(\gamma)$ with the Ooguri-Vafa double-covering
\eqref{eq:wgromovwitten} will allow us in section \ref{sec:Superpots+MirrorSymmetry} to identify periods
corresponding to $W_{\text brane}$ for specific flux choices $G_4$.

\chapter[Constructions and Calculations in String Dualities]{Constructions and Calculations in \newline String Dualities}
\label{ch:Calcs+Constructions}

In this chapter we present explicit calculations of the effective superpotentials
in F-theory and in heterotic/F-theory dual setups, where we mainly follow 
\cite{Grimm:2009ef} and \cite{Grimm:2009sy}. First in section \ref{sec:Superpots+MirrorSymmetry}
we explicitly calculate the F-theory flux superpotential for four-dimensional F-theory compactifications
on elliptic Calabi-Yau fourfolds and extract the Type IIB flux and seven-brane superpotential
along the lines of the discussion in section \ref{sec:F-flux_sup}. Before delving into the
details of the calculations we first present a general strategy to obtain elliptic Calabi-Yau fourfolds
$X_4$ with a little number of complex structure moduli. This is necessary in order to
work with a technically controllable complex structure moduli space of $X_4$, which we will mainly study 
using Picard-Fuchs equations and cross-checks from mirror symmetry. We put particular emphasis on the 
toric realization of examples with few moduli and the toric means to analyze possible fibration structures.
Then we start with the construction of a concrete Calabi-Yau fourfold, that we realize as a toric Calabi-Yau
hypersurface. We focus on one main example and refer to appendix \ref{app:furtherexamples} for two further examples.
For this example we perform a detailed analysis of the seven-brane dynamics as encoded by the discriminant
of the elliptic fibration, comment on the Calabi-Yau threefold geometry of its heterotic dual and finally 
calculate, using fourfold mirror symmetry, the Type IIB flux and seven-brane superpotential. This is 
technically achieved by specifying appropriate four-flux $G_4$, that singles out linear combination of 
fourfold periods, that are then identified with the Type IIB Calabi-Yau threefold periods upon matching 
the classical terms and the instanton corrections on the fourfold with the Calabi-Yau threefold results. 
The Type IIB seven-brane superpotential is similarly identified by a matching of the brane disk instanton 
invariants with the fourfold instanton invariants.    

Then in section \ref{sec:SuperpotsHetF} we compute the heterotic superpotential for a Calabi-Yau threefold
compactification on $Z_3$ from its F-theory dual setup. Before dealing with concrete examples we comment 
on the general matching strategy of the F-theory flux superpotential with the heterotic superpotential. 
We emphasize the duality map for heterotic five-branes and their superpotential to the F-theory side. 
Then we study in detail two explicit examples
of heterotic string compactifications. Here we essentially take the same threefold geometries that we 
used in section \ref{sec:ffConstruction} in order to obtain Calabi-Yau fourfolds $X_4$ with a small number 
of complex structure moduli. This surprising coincidence goes back to the rich fibration structure of $X_4$ 
and its mirror $\tilde{X}_4$, which is, on the one hand, a consequence of the requirement of a little number 
of moduli but, on the other hand, also natural for heterotic/F-theory dual setups, see the schematic
diagram \eqref{eq:dualityDiagramm} and \cite{Berglund:1998ej}.

The first example of a heterotic threefold we consider, denoted by $\tilde{Z}_3$, has a small number of 
K\"ahler moduli, which allows to explicitly calculate intersection numbers and topological indices as 
necessary to e.g.~concretely construct heterotic bundles $E$ via the methods of section \ref{sec:spectralcover}. 
We equip $\tilde{Z}_3$ with an $E_8\times E_8$ bundle and explicitly construct the F-theory dual
geometry $X_4$, once in the absence and once in the presence of horizontal heterotic five-branes.
Indeed we are able to check the duality map \eqref{eq:modulimap} between F-theory and heterotic moduli 
explicitly for both cases. The corresponding heterotic superpotentials can however not be computed
due to a huge amount of complex structure moduli of $\tilde{Z}_3$ and $X_4$. The converse
situation applies in the second example, which is basically a heterotic compactification on the mirror
threefold $Z_3$ of $\tilde{Z}_3$. For this example we are not able to explicitly demonstrate the matching 
of moduli, but directly study the complex geometry of $Z_3$ and its F-theory dual fourfold $X_4$. 
This allows both to extract the heterotic bundle and five-brane moduli and to calculate the heterotic
flux and five-brane superpotential explicitly. Finally, we specify the corresponding flux $G_4$ that, 
together with the knowledge of the periods from section \ref{sec:Superpots+MirrorSymmetry}, determines the 
dual heterotic superpotential completely.

Before we start let us for future reference summarize the fibration structures of the 
Calabi-Yau three- and fourfolds we consider in the following, both in sections 
\ref{sec:Superpots+MirrorSymmetry} and \ref{sec:SuperpotsHetF}:
\begin{eqnarray}
\xymatrix{Z_3 \ar@<4pt>@{<->}@/^0.7cm/[rrrrr]^{MS} \ar@{->>}[d]^{\pi_{Z}}  & \ar@_{(->}[l]\,\mathcal{E}\,\,   \ar@<2pt>[r]^{Het/F} & \ar@<2pt>[l]\,\, K3\,  \ar@^{(->}[r]  & X_4 \ar@{->>}[d]^{\pi_{X}}\ar@{<->}[r]^{MS}&\tilde{X}_4\ar@{->>}[d]^{\pi_{\tilde{X}}}&\ar@_{(->}[l]\,\tilde{Z}_3\\
         B_2^{Z}& &  &B_2^{Z}&\P^1& }.
         \label{eq:dualityDiagramm}
\end{eqnarray}
Here we used the abbreviations $MS$ and $Het/F$ for the action of mirror symmetry respectively 
heterotic/F-theory duality. In words, starting from a heterotic string compactification on an elliptic 
threefold $\pi_Z:\,Z_3\rightarrow B_2^{Z}$ with generic elliptic fiber $\mathcal{E}$ we obtain the F-theory 
dual elliptic $K3$-fibered fourfold $\pi_{X}:\,X_4\rightarrow B_{2}^{Z}$ as the fourfold mirror to the Calabi-Yau
threefold fibered fourfold $\pi_{\tilde{X}}:\,\tilde{X}_4\rightarrow \P^1$ with generic fiber $\tilde{Z}_3$,
which in turn is the mirror of $Z_3$.

\section{F-theory, Mirror Symmetry and Superpotentials}
\label{sec:Superpots+MirrorSymmetry}

In this section we explicitly perform the computation of the F-theory flux superpotential 
\eqref{eq:fluxpotFourfoldExpanded}. The class of Calabi-Yau fourfolds $X_4$ 
that we consider here have to have, for technical reasons, a low number of complex structure 
moduli. We outline in section \ref{sec:Kreuzermethods} a strategy, cf.~\cite{Mayr:1996sh,Klemm:1996ts}, 
to construct such examples of fourfolds $X_4$ as in \eqref{eq:dualityDiagramm} as the mirror dual
to a fourfold $\tilde X_4$ with a small number of K\"ahler moduli, that itself is realized as 
a Calabi-Yau threefold fibration $\tilde{Z}_3$ over a $\mathds{P}^1$-base. In addition we 
discuss toric means to identify interesting fibration structures like an elliptic or 
$K3$-fibration, which is of particular importance both for constructions of F-theory and 
heterotic/F-theory dual geometries. 

Then in section \ref{sec:ffConstruction} we fix a concrete 
Calabi-Yau fourfold $\tilde{X}_4$ by specifying the threefold fiber $\tilde{Z}_3$, that is given 
for the example at hand as an 
elliptic fibration over $\P^2$. This guarantees a small number of only four complex structure
moduli in the mirror $X_4$. We emphasize that $\tilde{Z}_3$ can be viewed as a compactification of the 
local geometry $K_{\P^2}\rightarrow \P^2$ which was studied in \cite{Aganagic:2001nx} in the 
context of mirror symmetry with D5-branes on the local mirror geometry given by a Riemann surface 
$\Sigma$. We exploit this fact in our analysis of the seven-brane content 
of the F-theory compactification on $X_4$, where the local brane geometry of $\Sigma$ can
be made visible as an additional deformation modulus of the discriminant of the elliptic 
fibration of $X_4$. Finally in section \ref{sec:ApplicationsMirrorSymmetryToF} we determine the
solutions of the Picard-Fuchs system for $X_4$ and obtain the linear combination of solutions 
$F^0(\gamma)$, which depends on this distinguished deformation modulus and which we thus identify as 
the Type IIB seven-brane superpotential. In addition, we check this assertion further by mirror 
symmetry, namely a comparison of the fourfold instanton invariants of $F^0(\gamma)$ with the disk 
instantons\footnote{It is already the required matching 
of the integral structure of closed fourfold invariants in $W_{G_4}$ and the open disk instanton invariants 
in $W_{\rm brane}$, which has a $Li_2$-structure, that necessarily 
identifies the appropriate linear combination of fourfold periods with a specific prepotential 
$F^0(\gamma)$, cf.~section \ref{sec:EnumGeo}.} in the limit of the local brane geometry considered 
in \cite{Aganagic:2001nx}. Analogously 
we determine the Type IIB flux superpotential by a matching of the classical terms and the world-sheet 
instanton corrections from the fourfold periods. This explicitly demonstrates the split of the 
F-theory flux superpotential into the Type IIB flux and brane superpotential as required in 
\eqref{eq:SuperpotLimit} and thus confirms the unified description of Type IIB open-closed deformations 
and obstructions in F-theory. We conclude with an independent check via the heterotic dual on $Z_3$, 
compare to the diagram in \eqref{eq:dualityDiagramm}.

\subsection{Constructing Elliptic Fourfolds: Strategy}
\label{sec:Kreuzermethods}

In the following we discuss the construction of elliptically fibered 
Calabi-Yau fourfolds $X_4$ for which we are able to compute the F-theory flux 
superpotential \eqref{eq:fluxpotFourfoldExpanded} explicitly. As outlined before our strategy is to 
find examples of elliptic fourfolds $X_4$ that are constructed as toric hypersurfaces and 
that admit a small number of complex structure moduli, such that the Picard-Fuchs equations 
and the fourfold periods can be explicitly determined. 
One main aim of the following discussion is to understand the various arrows and fibrations 
in the the diagram \eqref{eq:dualityDiagramm}.

The Calabi-Yau fourfolds $X_4$ studied in this paper are be obtained, according to 
\eqref{eq:dualityDiagramm}, as mirror duals to a Calabi-Yau threefold fibration $\tilde{X}_4$ over $\P^1$. 
Denoting by $\tilde Z_3$ the generic Calabi-Yau threefold fiber we summarize this structure 
in a schematic diagram of the form
\begin{equation} \label{eq:tildeYoverP1}
	\xymatrix{\text{fiber}  \ar[r]  & \text{total space} \ar[d] \\
         & \text{base}  }\,,   
  \qquad\qquad \xymatrix{ \tilde Z_3  \ar[r]  & \tilde X_4 \ar[d] \\
         &\P^1 } \,.
\end{equation}
It can now be shown using toric means, see the discussion below or the original literature 
\cite{Avram:1996pj,Berglund:1998ej}, that the mirror Calabi-Yau fourfold $X_4$ is elliptically 
fibered given that the mirror Calabi-Yau threefold $Z_3$ is elliptically fibered. Then,
upon taking a threefold $Z_3$ with a little number of complex structure moduli, its mirror
$\tilde Z_3$ has a small number of K\"ahler moduli, which is inherited by $\tilde X_4$.
In addition, it is also guaranteed that the mirror fourfold $X_4$ has a small number of complex
structure moduli, as required in order to be able to concretely work on the complex 
structure moduli space of $X_4$. Thus the examples we consider have the following structure,
\begin{equation}
   \xymatrix{ \mathcal{E}  \ar[r]  & Z_3 \ar[d] \\
         & B_2^Z  }\,,   
  \qquad\qquad \xymatrix{ \mathcal{E}  \ar[r]  & X_4 \ar[d] \\
         & B_3^X } \,,
         \label{eq:Z3X4}
\end{equation} 
where $\cE$ denotes again the generic elliptic fiber and $B_2^{Z}$ and $B_3^X$ are the two-
and three-dimensional base manifolds of the elliptic fibrations, respectively.

In the concrete examples we construct in this section we in addition consider 
Calabi-Yau threefolds $\tilde Z_3$ which themselves admit an elliptic fibration and
are obtained as compactifications of local Calabi-Yau threefolds 
$K_{B_2^{\tilde Z}}\rightarrow B_{2}^{\tilde Z}$. For these geometries the requirement 
of a small number $h^{(1,1)}(\tilde{Z}_3)$ of K\"ahler moduli can readily be realized
by the choice of a trivial base $B_2^{\tilde{Z}}$ like $\P^2$ or $\mathds{F}_n$, as we choose 
in the main example of section \ref{sec:ffConstruction} and in appendix 
\ref{app:PartDetailsExamplesTables}. We also restrict to elliptic fibrations with generic 
fiber $\cE = \P^2(1,2,3)[6]$, which has only one section $\sigma$ \cite{Klemm:1996ts} 
and which is both the generic elliptic fiber\footnote{We note that the polyhedron of $\bbP^2(1,2,3)$
is self-dual.} of $Z_3$ and $X_4$ in \eqref{eq:Z3X4}. 
The general recipe to describe such an elliptic fibration by appropriate toric data is described in
detail in \cite{Klemm:1996ts}. We apply this to construct the Calabi-Yau fourfold $\tilde{X}_4$ 
which by construction has a small number of K\"ahler moduli. Then, we invoke toric mirror
symmetry, as reviewed in section \ref{sec:mirror_toric_branes}, to readily obtain the desired
elliptic fourfold $X_4$.

Before we conclude we recall a general theorem of \cite{Avram:1996pj} to analyze the fibration 
structure of a given Calabi-Yau manifold. This is of particular use if we are interested in quickly 
checking a desired fibration structure of a given mirror pair of Calabi-Yau manifolds, like 
e.g.~the pair of Calabi-Yau fourfolds $(\tilde X_{4},X_{4})$ just constructed, and in particular 
in order to understand the relations in \eqref{eq:dualityDiagramm} more thoroughly in concrete 
examples. The main statement of this theorem is the observation that it suffices to analyze the 
reflexive polyhedra polyhedra $(\Delta^{\tilde X}_{5},\Delta^X_{5})$ of either $\tilde{X}_4$ or its mirror 
$X_4$ without explicitly computing the intersection numbers \cite{Avram:1996pj}. 
In fact, this is essential if the considered Calabi-Yau manifolds have a huge number of K\"ahler 
moduli, as in our examples of the Calabi-Yau fourfolds $X_4$, which makes a determination of the intersection 
numbers technically challenging. 

The basic theorem can now be formulated as follows, where we restrict to the Calabi-Yau fourfold 
case for simplicity.
Suppose $(\tilde X_{4},X_{4})$ are given as hypersurfaces in the toric varieties 
constructed from the reflexive pair 
$(\Delta_5^X,\Delta^{\tilde X}_5)$ in the pair of dual lattices $(M,N)$.
Then, \cite{Avram:1996pj} gives two equivalent conditions for the existence of a 
Calabi-Yau fibration structure of the given fourfold $X_{4}$ once in terms of $\Delta^X_5$ and 
another time in terms of its dual $\Delta^{\tilde X}_5$. Assume there exists a $(n-k)$-dimensional 
lattice hyperplane in $N$ through the origin such that $\Delta_k^{F}:=H\cap\Delta^{X}_5$ is a 
$k$-dimensional reflexive polyhedron. Then this is equivalent with the existence of a projection 
$P$ to a $k$-dimensional sublattice of $M$ such that $P\Delta^{\tilde X}_5$ is a $k$-dimensional 
reflexive polyhedron $\Delta^{\tilde F}_k$ which is the dual of $\Delta^{F}_k$. If these conditions 
are satisfied, then the Calabi-Yau manifold $X_{4}$ which is obtained as a hypersurface of 
$\Delta^X_5$ has a Calabi-Yau fibration whose $(k-1)$-dimensional fiber $F_{k-1}$ is encoded by 
$\Delta^F_k$. The crucial point of these two equivalent criteria is that we can turn things around 
and analyze $X_{4}$ by not looking at hyperplanes $H$ in the complicated polyhedron $\Delta^X_5$, 
but at projections $P$ in $\Delta^{\tilde X}_5$ which is simple by construction. In both cases the 
base of the fibration can be found by considering the quotient polyhedron $\Delta^X_5/\Delta^F_k$ 
\cite{Candelas:1997eh}. Here this quotient polyhedron is obtained by first determining the quotient 
lattice in $M\supset\Delta_5^X$ by dividing out the lattice generated by the integral points of 
$\Delta^F_k$. Then the integral points of $\Delta^X_5/\Delta^F_k$ are the equivalence classes of 
integral points in $\Delta^X_5$ in this quotient lattice. Schematically the analysis of the 
fibration structure can be summarized as
\begin{equation} \label{eq:fibrations}
\rule[-2.0cm]{0cm}{4.2cm}\begin{array}{|c|l|c|l|c|}
   \cline{1-1}\cline{3-3} \cline{5-5}  \rule[-3mm]{0mm}{5mm}
   	 \rule[-3mm]{0mm}{8mm} \text{Fibration structure} & &(\Delta_5^{\tilde X},\tilde X_4)  & \leftrightarrow&   (\Delta_5^{ X}, X_4) \\ \cline{1-1}\cline{3-3} \cline{5-5}
	\tilde{X}_4\text{ admits} & &\text{Injection} & \leftrightarrow&\text{Projection} \rule[-3mm]{0mm}{8mm}  \\
   	\text{CY}_{m-1}-\text{fiber } \tilde{f}_{m-1}& & \Delta^{\tilde{f}}_{m}=\tilde{H}\cap \Delta_5^{\tilde{X}}&  & \Delta^f_{m}=P\Delta_5^X\\
	\cline{1-1}\cline{3-3} \cline{5-5}
	X_4\text{ admits} & &\text{Projection} &\leftrightarrow &\text{Injection} \rule[-3mm]{0mm}{8mm}\\
   	\text{CY}_{k-1}-\text{fiber } F_{k-1}&  & \Delta^{\tilde F}_{k}=P\Delta_5^{\tilde{X}}&  &\Delta^{F}_{k}=H\cap \Delta_5^{X}\\
\cline{1-1}\cline{3-3} \cline{5-5}
\end{array}
\end{equation}
where the arrow `$\leftrightarrow$' indicates the action of mirror symmetry interchanging 
projection and injection. Clearly, this analysis can be also used to determine Calabi-Yau 
fibers $\tilde f_{m-1}$ of the mirror $\tilde X_4$. In general, it is not the case that 
mirror symmetry preserves fibration structures. However, in the constructions which we will 
analyze in the section \ref{sec:ffConstruction}, we will find that both $X_4$ and $\tilde X_4$ 
admit an intriguingly rich fibration structure 

We conclude by mentioning two appealing byproducts of the geometric structure of the discussed 
Calabi-Yau manifolds. One interesting consequence of the fibration structure
indicated in \eqref{eq:dualityDiagramm} is the simultaneous use also for heterotic/F-theory 
duality, where the Calabi-Yau fourfold $X_4$ defines the F-theory setup of a dual heterotic
compactification on $Z_3$ \cite{Berglund:1998ej}, see chapter \ref{ch:HetFThyFiveBranes} 
for more details on heterotic/F-theory duality. Indeed it can be shown explicitly using the 
simple rules \eqref{eq:fibrations} that $X_4$ also admits a $K3$-fibration with a heterotic 
dual on the elliptic threefold $Z_3$. We will exploit this briefly in section \ref{sec:ffConstruction}
and come to more systematic computations in heterotic/F-theory duality in section 
\ref{sec:SuperpotsHetF}.

Secondly, for $X_4$ obtained from an elliptic Calabi-Yau threefold $\tilde Z_3$ according to
\eqref{eq:dualityDiagramm} it is possible to take the local limit of $(\tilde{Z}_3,Z_3)$,
which is then also promoted to the fourfold mirror pair\footnote{We note that in our explicit
example in this section, also $X_4$ is a $Z_3$-fibration over $\P^1$, that makes the lifting
of the local limit to the fourfolds even more obvious.} $(\tilde{X}_4,X_4)$.
We probe the local geometry of $Z_3$, being a conic over a Riemann surface $\Sigma$ as reviewed
in section \ref{sec:mirror_toric_branes}, by placing a heterotic five-brane on the non-compact fiber
that hits $\Sigma$ in one point. Then the F-theory compactification on $X_4$ in this local limit 
reproduces, in an appropriate further limit, precisely the setup of the
local brane geometry of \cite{Aganagic:2001nx}, where the same geometry and the brane superpotential
were studied in the B-model. Thus, by invoking the results of \cite{Aganagic:2001nx} from
local mirror symmetry, this serves as a cross-check for our computation of the brane superpotential
in section \ref{sec:ApplicationsMirrorSymmetryToF} and \ref{sec:Example2}. In particular the compact fourfold $X_4$ 
will provide a canonical extension of the local
results to the compact Calabi-Yau mirror pair $(\tilde{Z}_3,Z_3)$. Alternatively,
we can directly take the limit of $(\tilde{Z}_3,Z_3)$ to the non-compact geometry 
in a Type II description where similar results can be obtained, cf.~also 
\cite{Alim:2009rf,Alim:2009bx}.

\subsection{Constructing Elliptic Fourfolds: Concrete Examples} \label{sec:ffConstruction}

In the following we exemplify in detail our construction for a main example 
of an elliptic fourfold $X_4$. It is specified by the choice of the base manifold 
$B_2^{\tilde{Z}}$ of the elliptic fibration of $\tilde{Z}_3$ that we here take to be
the most simple one, namely $B_2^{\tilde{Z}}=\P^2$. We refer to appendix \ref{app:furtherexamples} 
for further examples.

We start with the discussion of the non-compact Calabi-Yau threefold 
$\mathcal{O}_{\P^2}(-3)$ and its mirror $\Sigma$ and discuss in some detail mirror
pairs of toric branes. Then we compactify the non-compact fiber to an elliptic 
curve $\mathcal{E}$ to obtain the compact threefold $\tilde{Z}_3$ and its mirror $Z_3$,
that is elliptically fibered as well. We put strong emphasis on the
Weierstrass model of $Z_3$ and check that, in the limit of large
elliptic fiber, we recover the local B-model geometry, a conic over $\Sigma$.
Finally we obtain $\tilde{X}_4$ by fibering $\tilde{Z}_3$ over $\P^1$
and determine the elliptic mirror fourfold $X_4$ as in diagram \eqref{eq:dualityDiagramm}.
We study the geometry of $X_4$ in great detail and identify dynamical seven-branes
by analyzing the discriminant of the elliptic fibration of $X_4$. These can be understood 
as a lift of the brane in the non-compact geometry $\Sigma$.
This allows us to single out those complex structure moduli of $X_4$ that map to
seven-brane moduli in Type IIB. Then, we briefly exploit the existence of a heterotic dual
compactification on $Z_3$ to determine, following section \ref{sec:het-Fdual}, the 
splitting of the Weierstrass constraint of $X_4$ into the heterotic threefold 
constraint $p_0$ and the bundle data $p_\pm$. This allows an explicit identification
of the complex structure moduli of $X_4$ with complex structure moduli and brane 
respectively bundle moduli in the heterotic string. A more thorough discussion of the heterotic dual 
and the dual heterotic five-brane is given in section \ref{sec:SuperpotsHetF}.

\subsubsection{A toric Calabi-Yau threefold with D-branes}

In the following we discuss the local Calabi-Yau threefold over 
$B_2^{\tilde{Z}}=\mathds{P}^2$, i.e.~${\cal O}(-3)\rightarrow \mathds{P}^2$,
in the presence of toric branes. 
Then, as a next step, we consider the elliptically fibered Calabi-Yau threefold in 
the weighted projective space $\mathds{P}^4(1,1,1,6,9)$ that contains the non-compact 
geometry in the limit of large elliptic fiber.  

\begin{figure}[!ht]
\begin{center} 
\includegraphics[height=6cm]{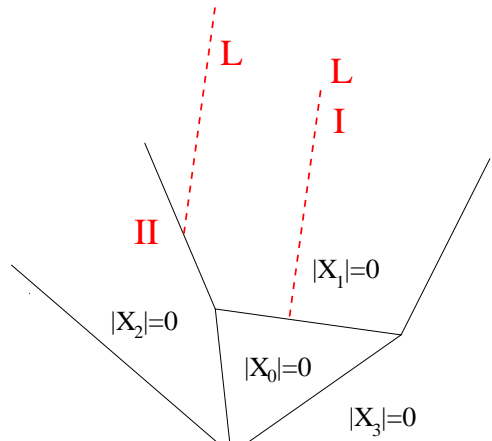}
\vspace*{-.5cm} 
\caption{\small  \label{two_branes_phases} Toric base $\mathbb{B}_3$ and Harvey-Lawson Lagrangians for
  non-compact $\mathds{P}^2$  }
\end{center} 
\end{figure}

In \cite{Aganagic:2001nx} the non-compact ${\cal O}(-3)\rightarrow \mathds{P}^2$ 
Calabi-Yau threefold with non-compact Harvey-Lawson branes was considered. 
The local Calabi-Yau is defined as the toric variety $\P_{\tilde{\Delta}}$ characterized 
by the polyhedron
\begin{equation}\label{eq:localp2}
	\begin{pmatrix}[c|ccc|c|c]
	    	&   &  \tilde{\Delta}_3    &   	&  \ell^{(1)}& \\ \hline
		v_1 & 0 & 0 &  1 	   &-3 &  X_0\\
		v^b_1 & 1 & 1 &  1 	   & 1 & X_1\\
		v^b_2 &-1 & 0 &  1 	   & 1 & X_2\\
		v^b_3 & 0 &-1 &  1 	   & 1 & X_3\\
	\end{pmatrix},
\end{equation}
where the superscript $^b$ denotes the two-dimensional basis $\mathds{P}^2$ and the $X_i$ 
denote homogeneous coordinates. The D-term constraint \eqref{eq:ToricVariety} for this 
geometry reads
\begin{equation}
-3|X_0|^2+|X_1|^2+|X_2|^2+|X_3|^2=r\, 	
\end{equation}
where $r$ denotes the K\"ahler modulus of $\P^2$ and $\P_{\tilde{\Delta}}$ can be viewed 
as a $(S^1)^3$-fibration over a three-dimensional base $\mathbb{B}_3$. The degeneration 
loci of the fiber, $|X_i|=0$, are shown in figure \ref{two_branes_phases}. The brane is 
defined torically by the brane charge vectors  
\begin{equation} \label{eq:def-ellhat}
    \hat\ell^{(1)}=(1,0,-1,0) \ , \qquad  \hat\ell^{(2)}=(1,0,0,-1)\ . 
\end{equation}
This leads to the two constraints
\begin{equation} \label{ABrane}
 	|X_0|^2-|X_2|^2=c^1\,,\qquad |X_0|^2-|X_3|^2=c^2\,,
\end{equation}
where the $c^a$ denote the open string moduli.
The brane geometry is $\mathbb{C}\times
S^1$ and can be described by a one dimensional half line in the three real 
dimensional toric base geometry $\mathbb{B}_3$ ending on a line where two of the three
$\mathbb{C}^*$-fibers degenerate. The A-brane has two inequivalent brane phases I and II as 
indicated in Figure \ref{two_branes_phases}.\footnote{Note that our phase II is precisely 
phase III of \cite{Aganagic:2001nx}. The phase II of \cite{Aganagic:2001nx} has been omitted 
since it is equivalent to phase I by symmetry of $\mathds{P}^2$.} 

Mirror symmetry for this geometry was analyzed in \cite{Aganagic:2001nx} where the disk 
instantons of the A-model were calculated exploiting the fact that the mirror geometry of 
$\mathcal{O}(-3)\rightarrow \mathds{P}^2$ effectively reduces to the Riemann surface 
$\Sigma$ defined by $P(x_1,x_2)=0$ in \eqref{eq:localMirror}. The D6-brane is mapped under 
mirror symmetry to a D5-brane which intersects $\Sigma$ in a point. It will be this 
D5-brane picture which can be reformulated as a seven-brane with flux and embedded into 
an F-theory compactification below.

\subsubsection{The compact elliptic Calabi-Yau threefold}

This local Calabi-Yau threefold can easily be embedded into a compact Calabi-Yau threefold $\tilde{Z}_3$. 
The compactification can be understood as a replacement of the non-compact $\mathds{C}$-fiber
in $\mathcal{O}(-3)\rightarrow \P^2$, that is dual to the divisor associated to $v_1$ in $\tilde{\Delta}_3$, 
by an elliptic fiber. Here we choose the generic fiber to be the elliptic curve in $\mathds{P}^2(1,2,3)$ 
which we fiber over the $\mathds{P}^2$-base the same way as the non-compact $\mathbb{C}$-fiber before. 
Thus, the polyhedron of this compact threefold $\tilde Z_{3}$, its charge vectors, the homogeneous 
coordinates $\tilde{x}_i$ as well as the corresponding monomials for the mirror geometry $Z_3$, 
cf.~\eqref{eq:mirror3foldellp2}, are given by\footnote{Besides the chosen $(2,3)$, which leads 
to an elliptic fibration with one section, the values $(1,2)$ and $(1,1)$ are also admissible in the 
sense that these choices lead to reflexive polyhedra. The corresponding elliptic fibration has two and 
three sections, respectively \cite{Klemm:1996ts}.}
\begin{equation}\label{eq:3foldellp2}
	\begin{pmatrix}[c|cccc|cc|c|c]
	    	&   &  \Delta_4^{\tilde Z} &   &    &  \ell^{(1)} & \ell^{(2)} &  & \\ \hline
		v_0 & 0 & 0 & 0 & 0 	  &  0  &  -6 & \tilde x_0 & zxyu_1u_2u_3\\
		v_1 & 0 & 0 & 2 & 3 	  &  -3   &1  & \tilde x_1 & z^6 u_1^6 u_2^6 u_3^6\\
		v^b_1 & 1 & 1 & 2 & 3 	&  1   & 0  & \tilde x_2 & z^6 u_3^{18}\\
		v^b_2 &-1 & 0 & 2 & 3 	&  1   & 0  & \tilde x_3 & z^6 u_1^{18}\\
		v^b_3 & 0 &-1 & 2 & 3 	&  1   & 0  & \tilde x_4 & z^6 u_2^{18}\\
	        v_2 & 0 & 0 &-1 & 0 	&  0   & 2  & \tilde x_5 & x^3\\
		v_3 & 0 & 0 & 0 &-1 	&  0   & 3  & \tilde x_6 & y^2 
	\end{pmatrix}.
\end{equation}
Here the points $v_1,v_2,v_3$ carry the information of the elliptic fiber where we added the 
inner point $v_1$ in order to recover $\mathds{P}^2(1,2,3)$, in particular its 
homogeneous coordinate $\tilde{x}_1$ with weight one under the new $\mathbb{C}^*$-action 
$\ell^{(2)}$. Furthermore, applying the insights of \eqref{eq:fibrations}, the elliptic 
fibration structure of $\tilde{Z}_3$ is obvious from the fact, that the polyhedron of 
$\mathds{P}^2(1,2,3)$ occurs in the hyperplane $H=\{(0,0,a,b)\}$, but also as a projection 
$P$ on the (3-4)-plane is found that indicates an elliptic fibration of the mirror $Z_3$, 
too.

The polyhedron \eqref{eq:3foldellp2} describes the degree $18$ hypersurface in the
weighted projective space $\mathds{P}^4(1,1,1,6,9)$ considered in \cite{Candelas:1994hw} 
that is blown up along the singular curve $\tilde{x}_2=\tilde{x}_3=\tilde{x}_4=0$ with 
exceptional divisor $v_1$. Its Euler number is $\chi=-540$ whereas $h^{(1,1)}=2, \ h^{(2,1)}=272$. 
Denoting the toric divisors $\tilde x_i =0$ by $D_i$, the
two K\"ahler classes $J_1 = D_2$ and $J_2 = 3 D_2 + D_1$ correspond to the
Mori vectors $\ell^{(1)}$ and $\ell^{(2)}$ in \eqref{eq:3foldellp2}. They represent a curve 
in the hyperplane class of the $\mathds{P}^2$ base and a curve in the elliptic
fiber, respectively. The triple intersections of the dual divisors and the intersections with
the second Chern class are respectively computed to be\footnote{In performing these 
toric computations we have used the Maple package {\tt Schubert}.}
\begin{eqnarray} \label{eq:intersectionsY}
  {\mcal C}_0 &=& 9 J_2^3+3 J_2^2 J_1 + J_2 J_1^2\ , \\
  {\mcal C}_2 &=& 102\, J_2+36\, J_1\ .\nn
\end{eqnarray}
In this notation the coefficients 
of the top intersection ring $\mcal C_0$ are the cubic
intersection numbers $J_i\cap J_j\cap J_k$, while the coefficients
of $\mcal C_2$ are $[c_2(T_{\tilde Z_3})]\cap J_i$. 

Mirror symmetry for this example has been studied in \cite{Hosono:1993qy,Candelas:1994hw}. 
In order to construct the mirror pair $(Z_3,\tilde Z_3)$ as well as their constraints 
\eqref{eq:Z3typeIIA}, \eqref{eq:Z3typeIIB} we need the dual polyhedron 
\begin{equation}\label{eq:mirror3foldellp2}
	\begin{pmatrix}[c|cccc|c]
	      &    &  \Delta_4^{Z} &   & &   \\ \hline
	v_1   &  0 &   0 &  1 &  1 & z\\
	v_1^b &-12 &   6 &  1 &  1 & u_1\\
	v_2^b &  6 & -12 &  1 &  1 & u_2\\
 	v_3^b &  6 &   6 &  1 &  1 & u_3\\
	v_2   &  0 &   0 & -2 &  1 & x\\
 	v_3   &  0 &   0 &  1 & -1 & y
	\end{pmatrix},
\end{equation}
where again the basis was indicated by a superscript $^b$. 
Again we added the inner point $v_1$ to recover the polyhedron of $\mathds{P}^2(1,2,3)$ as 
the injection with $H=\{0,0,a,b\}$, thus confirming the elliptic fibration of the mirror $Z_3$.
Here we distinguish between the two-dimensional basis $B_2^Z=\mathds{P}^2$ and the elliptic 
fiber by denoting the homogeneous coordinates of $\mathds{P}^2(1,2,3)$ by $(z,x,y)$ and of 
$B_2^Z$ by $(u_1,u_2,u_3)$. The elliptic fibration structure reflects in particular in the 
constraint of $Z_3$ which takes a Weierstrass form\footnote{In order to prepare for a
heterotic/F-theory duality analysis, we renamed the constraint $P$ in \eqref{eq:Z3typeIIB} to $p_0$.}
\begin{equation} \label{eq:Weierst3fold}
 	p_0:=a_6 y^2 +a_5 x^3+ a_0 z x y u_1 u_2 u_3+z^6 (a_3 u_1^{18}+a_4 u_2^{18}+a_1 u_1^6 u_2^6 u_3^6+ a_2 u_3^{18})=0\,.
\end{equation}
The generic elliptic fiber can be seen by setting the coordinates $\underline{u}$ of the basis $B_2^Z$ 
to some reference point, such that $p_0$ takes the form of a degree six hypersurface in $\mathds{P}^2(1,2,3)$. 
The base itself is obtained as the section $z=0$ of the elliptic fibration over $B_2^Z$.

The complex structure dependence of $Z_3$ is evident from the dependence of $p_0$ on the 
parameters $\underline{a}$ which are coordinates on $\mathds{P}^6$. However, they redundantly 
parameterize the complex structure of $Z_3$ due to the symmetries of $\mathds{P}^4(1,1,1,6,9)$. 
Indeed there is a $(\mathbb{C}^\ast)^6/(\mathbb{C}^\ast)^2$ rescaling symmetry of the coordinates 
that enables us to eliminate four of the $\underline{a}$ recovering the two complex structure parameters 
that match $h^{(1,1)}(\tilde Z_3)=h^{(2,1)}(Z_3)=2$. The appropriate coordinates $\underline{z}$ obeying 
$\underline{z}=0$ at the large complex structure/large volume point are completely determined by the phase 
of the A-model, i.e.~the choice of charge vectors $\ell^{(i)}$ of $\Delta_4^{\tilde{Z}}$. They are 
given in general by \eqref{eq:algCoords} which we readily apply for the situation at hand to 
obtain\footnote{Compared to the general definition \eqref{eq:algCoords} we changed a superscript to a 
subscript for convenience.}
\begin{equation} \label{eq:zLargeRadius}
 	z_1=\frac{a_2a_3a_4}{a_1^3}\ ,\qquad z_2=\frac{a_1a_5^2a_6^3}{a_0}\ .
\end{equation}
Thus, we can use the $(\mathbb{C}^\ast)^4$ action and the overall scaling 
to set $a_i=1,\, i=2,\ldots,6$ for five parameters to obtain
\begin{equation} \label{eq:MirrorEtale}
 	p_0  =  y^2+x^3+z x y m_1(\underline{u}) +z^6 m_6(\underline{u}) \ ,
\end{equation}
where we have abbreviated 
\begin{equation} \label{eq:defm_i}
 	m_1(\underline{u}) = z_2^{-1/6} z_1^{-1/18} u_1 u_2 u_3\ ,\qquad \quad m_6(\underline{u}) =u_1^{18}+u_2^{18}+u_3^{18}+z_1^{-1/3}u_1^6 u_2^6 u_3^6\ .
\end{equation}

Alternatively, this result can be obtained more directly by the mirror 
construction  \eqref{eq:HVmirror}. In this case one needs the 
following assignment of coordinates $y_i$ to points of $\Delta_4^{\tilde Z}$ and monomials
\begin{equation}  \label{eq:etalep2}
 	\begin{pmatrix}[c|c|c] 	 	
	y_0 &v_0  &  a_0\, z x y u_1 u_2 u_3 \\
	y_1 &v_1  &  a_1\, z^6 u_1^6 u_2^6 u_3^6 \\
	y_2 &v^b_1&  a_2\, z^6 u_3^{18} \\ 
	y_3 &v^b_2&  a_3\, z^6 u_1^{18} \\ 
	y_4 &v^b_3&  a_4\, z^6 u_2^{18} \\
	y_5 &v_2  &  a_5\, x^3 \\
	y_6 &v_3  &  a_6\,  y^2 
 	\end{pmatrix}\ . 
\end{equation}
This defines the etal\'{e}-map that solves the constraints of \eqref{eq:HVmirror} automatically 
when \eqref{eq:zLargeRadius} holds. By setting $a_0=z_2^{-1/6}z_1^{-1/18}$, $a_1=z_1^{-1/3}$
and $a_i=1,\, i=2,\ldots,6$, we solve \eqref{eq:zLargeRadius} and $P=\sum_jy_j$ immediately 
reproduces $p_0$ in \eqref{eq:MirrorEtale}.

Next we show that \eqref{eq:MirrorEtale} indeed gives back the local geometry which is a 
conic over a genus one Riemann surface $\Sigma$ \cite{Aganagic:2001nx}.
The local limit in the A-model geometry is given as a double scaling limit in which
the elliptic fiber decompactifies. This corresponds to $z_2\rightarrow 0$ in the B-model geometry. 
Indeed we parameterize $z_2$ by $\varepsilon\equiv z_2$ such that the local limit 
is given by $\varepsilon\rightarrow 0$. At the end we should obtain an affine equation, thus, 
using the two $\mathbb C^*$-action we set the coordinates $z$ and $u_3$ to one.
By redefining the coordinates $x$ and $y$ as follows \cite{Kachru:1995fv}
\begin{equation} \label{eq:epslimit}
	y\rightarrow \varepsilon^{-1/2}y+k_1^{1/2}\quad ,\quad x\rightarrow \varepsilon^{-1/3}x+k_2^{2/3},
\end{equation}
the hypersurface equation $p_0=0$ becomes
\begin{equation}
	p_0=\frac{1}{\varepsilon} \tilde p_0+k_1^2+k_2^2+m_6=0
\end{equation}
where we set $z=1$ and $u_3=1$. Now we split this equation as 
\begin{equation}
	\tilde p_0=\varepsilon\quad  ,\quad k_1^2+k_2^2+m_6 =-1.
\end{equation}
If we now take the $\varepsilon\rightarrow 0$ limit we obtain, after appropriately redefining the $k_i$, 
the equation for the local geometry of the form
\begin{equation}
	uv=H(x,y)=x+1-\phi \frac{x^3}{y}+y.
\end{equation}
We observe that the Riemann surface defined by $H(x,y)=0$ is isomorphic to the surface $m_6=0$ up to 
isogeny, i.e.~the homology lattice differs only by integral multiples.

As discussed in section \ref{sec:het-Fdual} considering heterotic string theory on the 
elliptically fibered Calabi-Yau threefold $Z_3$ is expected to be dual to F-theory 
on $X_4$ if the fourfold admits a $K3$ fibration. This is automatic in the construction
in \eqref{eq:tildeYoverP1} by fibering the mirror $\tilde{Z}_3$ over $\P^1$ \cite{Berglund:1998ej}. 
We have shown that $Z_3$ is indeed an elliptic fibration, and will confirm in the next 
section that $X_4$ is a $K3$ fibration. However, it is crucial to point out that there will 
be a large heterotic non-perturbative gauge group from the blown-up singularities of the elliptic 
fibration of $Z_3$. Indeed by calculation of the discriminant of \eqref{eq:MirrorEtale} 
one notes that the elliptic fibration not only degenerates 
over the curves $m_6=0$ and $432 m_6 + m_1^6=0$ in the base of $Z_3$, but also over 
many curves described by the additional coordinates corresponding to the inner points in $\Delta_4^{Z}$. 
Let us point out that we will similarly find a large gauge group in the F-theory compactification on 
$X_4$. However, the identification of the moduli of the heterotic gauge bundle $E$ with the complex 
structure moduli of $X_4$ can still be performed by focusing on the heterotic perturbative 
gauge symmetry. This is technically achieved by extracting the spectral cover constraint $p_+$ of \eqref{eq:def-p+}
in the splitting \eqref{eq:BMform} of the constraint $P$ of the fourfold $X_4$, as demonstrated in
section \ref{sec:Example2}.

Before continuing with the construction of the Calabi-Yau fourfold, let us close 
with another comment on the use of the vectors $\hat \ell^{(1)}$ and $\hat \ell^{(2)}$ 
given in \eqref{eq:def-ellhat}. On the compact threefold they translate to 
\begin{equation}
   \hat \ell^{(1)} = (0,1,0,-1,0,0,0)\ ,\qquad \hat \ell^{(2)} = (0,1,0,0,-1,0,0)\ , 
\end{equation}
due to the new origin in the polyhedron \eqref{eq:3foldellp2}.
In fact, applying \eqref{eq:BBrane} and using \eqref{eq:etalep2}, they define the divisors 
\begin{equation}\label{eq:divisors}
 	   z_1^{-1/3}  u_1^6 u_2^6 u_3^6 =\hat{z}_1  u_1^{18}\,,\qquad 
 	   z_1^{-1/3}  u_1^6 u_2^6 u_3^6 =\hat{z}_2  u_2^{18}\,,
\end{equation}
in the compact $Z_3$. Here we introduced the moduli $\hat{z}_a$ corresponding to the 
charge vector $\hat{\ell}^{(a)}$. 

Note that in our F-theory compactification of the next section 
we will find seven-branes, that are localized on components of the discriminant of the elliptic 
fibration, which possess additional moduli. These additional fields correspond precisely to either 
$\hat z_1$ or $\hat z_2$ and parametrize deformations of the seven-brane constraint by the
terms in \eqref{eq:divisors}. Hence, $\hat z_i$ can be interpreted as deformations of the seven-brane 
divisors in $X_4$, or as spectral cover moduli in the heterotic dual theory as we will see in section 
\ref{sec:Example2}. Upon turning on a brane-flux
$F_2$ on these divisors, the moduli $\underline{\hat{z}}$ can get obstructed by the brane superpotential 
\eqref{eq:oc_super}. Upon lifting the brane flux $F_2$ to a $G_4$-flux on $X_4$, see e.g.~\cite{Denef:2008wq},
this is mapped to the flux superpotential \eqref{eq:fluxpotFourfoldExpanded} on $X_4$ matching
the brane superpotential in the limit \eqref{eq:SuperpotLimit}.

\subsubsection{The elliptically fibered Calabi-Yau fourfold} 

Having discussed the threefold geometry $(Z_3,\tilde{Z}_3)$, we are now in the position to construct and analyze 
the elliptically fibered Calabi-Yau fourfold $X_4$ which is used as an F-theory compactification.

We start by constructing the mirror $\tilde X_4$ first. According to \eqref{eq:tildeYoverP1} it is 
obtained by fibering the Calabi-Yau threefold $\tilde Z_3$ over a $\P^1$. The fibration data can be specified
in such a way that one of the D-brane vectors $\hat \ell^{(i)}$ of the local model \eqref{eq:localp2} 
appears as a new charge vector of the polyhedron defining $\tilde{X}_4$. As we demonstrate later on, this 
new charge vector dictates the location of the moving seven-brane, while a second additional vector not 
used in the construction of the fourfold controls the volume of the $\mathds{P}^1$-basis of the dual 
fourfold $\tilde{X}_4$ in \eqref{eq:tildeYoverP1}. The flux superpotential and the corresponding 
four-flux are determined in section \ref{sec:ApplicationsMirrorSymmetryToF}. In the following 
we exemplify our constructions in detail and list all toric and geometrical data necessary to reproduce our results. 

The Calabi-Yau fourfolds $(X_4,\tilde X_4)$ are realized as hypersurfaces in 
a toric ambient space described by a dual pair of reflexive 
polyhedra $(\Delta_5^X,\Delta_5^{\tilde X})$. The 
reflexive polyhedron $\Delta_5^{\tilde X}$ describes a fibration of the toric 
variety constructed from $\Delta_4^{\tilde Z}$ over $\P^1$ and is specified as 
\begin{equation}
	\Delta_5^{\tilde X} =
	\left(
	\begin{array}{rccc|r}
		\multicolumn{4}{c|}{\Delta_4^{\tilde Z}} & 0\\ \hline
		\text{-}1 & 0 & 2 & 3 & \text{-}1\\
		0 & 0 & 2 & 3 & \text{-}1\\
		0 & 0 & 2 & 3 & 1
	\end{array}
	\right).
	\label{eq:delta5}
\end{equation}
By construction, one finds $\Delta_4^{\tilde Z}$ by intersecting the hyperplane 
$\tilde H = (p_1,p_2,p_3,p_4,0)$ with $\Delta_5^{\tilde X}$. Following 
\eqref{eq:fibrations} this indeed identifies $\tilde X_4$ as a $\tilde Z_3$-fibration, 
and by performing the quotient $\Delta_5^{\tilde X}/\Delta_4^{\tilde Z}$ the base is 
readily shown to be the toric variety $((-1),(1))$, i.e.~a rational curve $\P^1$. The 
additional points which do not lie on $\tilde H$ determine the fibration structure of the
$\tilde{Z}_3$-fibration. Firstly, they are chosen such that the mirror $X_4$ is 
elliptically fibered\footnote{The fact that $\tilde X_4$ is also elliptically 
fibered in the example at hand is not crucial in the construction, cf.~\eqref{eq:dualityDiagramm}. 
In particular, the construction also applies e.g.~for the quintic hypersurface fibered over 
$\P^1$, since the mirror quintic admits an elliptic fibration with generic elliptic fiber 
being a torus in $\bbP^2$.} which means, that using the projection to the third and fourth 
coordinate one finds the polyhedron of a torus in $\P^2(1,2,3)$ just as in the threefold 
case in \eqref{eq:3foldellp2}.  Secondly, they can be arranged such that one 
charge vector of the Calabi-Yau fourfold is of the form $(\hat\ell^{(1)},-,-,-)$, i.e.~contains
the brane charge vector $\hat{\ell}^{(1)}$. As we see below this will imply a lift of the toric
brane of \eqref{eq:divisors} to F-theory on the mirror fourfold $X_4$.

Before proceeding with the concrete example let us note an alternative perspective on the construction 
of $\Delta_5^{\tilde{X}}$. In fact, $\Delta_5^{\tilde{X}}$ can be understood more thoroughly from the 
perspective of the GKZ-system obtained using the blow-up procedure \cite{Grimm:2008dq,Grimm:2010gk} 
presented in chapter \ref{ch:blowup}. In this context the connection of the fourfold geometry 
$\Delta_5^{\tilde X}$ with the brane charge vectors can be understood as a consequence of 
heterotic/F-theory duality \cite{Grimm:2009sy}.
We note that adding this vector to form a higher-dimensional non-reflexive polyhedron was first proposed in 
\cite{Mayr:2001xk,Lerche:2001cw,Alim:2009rf} in the context of the B-model and then extended to the compact
case in \cite{Jockers:2009ti,Alim:2010za}, where a connection with heterotic/F-theory duality was exploited.

We begin by choosing the open string vector $\hat\ell^{(1)}$ to construct\footnote{We could have used also $\hat\ell^{(2)}$, reproducing the same local D5-brane limit. } the $\P^1$-fibration in \eqref{eq:delta5}. 
The Calabi-Yau fourfold $\tilde X_4$ is then realized as a hypersurface in the toric space described by the polyhedron $\Delta_5^{\tilde X}$.
Its topological numbers are computed to be 
\begin{equation}
       h^{(3,1)}=2796\ ,\quad  h^{(1,1)}=4\ ,\quad h^{(2,1)}=0\ ,\quad
  h^{(2,2)} = 11244\,,\quad \chi = 16848\, .
\end{equation}
Here we first used \eqref{eq:Hodgenumbers1}, \eqref{eq:Hodgenumbers2} as well as \eqref{eq:Hodgenumbers3} 
and next applied \eqref{eq:HodgeRel}, \eqref{eq:FFEulerNumb}.

Next, we note that $\Delta_5^{\tilde X}$ has three triangulations, which correspond to non-singular Calabi-Yau 
phases which are connected by flop transitions. In the following we consider two of these phases in detail. 
These phases match, as we will show explicitly, the two brane phases in figure \ref{two_branes_phases} in the local Calabi-Yau threefold geometry.

To summarize the topological data of the 
Calabi-Yau fourfold for the two phases of interest, we  
specify the generators of the Mori cone $\ell^{(i)}_{I}$ and $\ell^{(i)}_{II}$ for
$i=1,\ldots4$,
\small
\begin{equation} 
	\begin{pmatrix}[c|ccccc|cccc|cccc]
	    	& && \Delta_5^{\tilde X} &&  &           \ell_I^{(1)} & \ell_I^{(2)} &
                \ell_I^{(3)} & \ell_I^{(4)} & \ell_{II}^{(1)} &
                \ell_{II}^{(2)} & \ell_{II}^{(3)} & \ell_{II}^{(4)} \\ \hline
		v_0   & 0 & 0 & 0 & 0 & 0 	    & 0  &  -6   & 0  &0    &  0    & -6&  0  & 0    \\
		v^b_1 & 0 & 0 & 2 & 3 & 0 	    &-2  & 1   &-1  & -1    & -3    & 0&  1  &  -2    \\
		v^b_2 & 1 & 1 & 2 & 3 & 0 	    & 1  &  0   & 0  & 0    &  1    & 0&  0  &  0     \\
		v^b_3 &-1 & 0 & 2 & 3 & 0 	    & 0  &  0   & 1  & -1   &  1    & 1& -1  &  0    \\
		v^b_4 & 0 &-1 & 2 & 3 & 0 	    & 1  &  0   & 0  & 0    &  1    & 0&  0  &  0 \\
		v_1   & 0 & 0 &-1 & 0 & 0 	    & 0  &  2   & 0  & 0    &  0    & 2&  0  &  0 \\
		v_2   & 0 & 0 & 0 &-1 & 0 	    & 0  &  3   & 0  & 0    &  0    & 3&  0  &  0 \\
   \hat v_1   &-1 & 0 & 2 & 3 &-1	  & 1  &  0   &-1  & 1    &  0    & -1&  1 &  0  \\
	 \hat v_2   & 0 & 0 & 2 & 3 &-1 	&-1  &  0   & 1  & 0    &  0    & 1& -1  &  1 \\
	 \hat v_3   & 0 & 0 & 2 & 3 & 1 	& 0  &  0   & 0  & 1    &  0    & 0&  0  &  1  
	\end{pmatrix}.
	\label{eq:vl-for-p2}
\end{equation}
\normalsize
The charge vectors are best identified in the phase II. Here $\ell^{(1)}_{II}$ and $\ell^{(2)}_{II}$
are the extensions of the threefold charge vectors $\ell^{(1)},\, \ell^{(2)}$ in \eqref{eq:3foldellp2} 
to the fourfold. The brane vector $\hat\ell^{(1)}$ is visible in phases II as a subvector of $\ell^{(3)}_{II}$. 
The remaining vector $\ell^{(4)}_{II}$ arises since we had to complete the polyhedron 
such that it becomes reflexive implying that $\tilde X_4$ is a Calabi-Yau manifold. It is the class
of the $\P^1$ in the base of the Calabi-Yau threefold fibration of $\tilde{X}_4$.
Phase I is related to phase II by a flop transition of the curve associated to 
$\ell^{(3)}_I$. Hence, in phase I the brane vector is identified with $-\ell^{(3)}_I$.
Furthermore, in the flop transition we identify 
\begin{equation} \label{eq:ell_flop}
  \ell^{(3)}_{II}=-\ell^{(3)}_{I}\ ,\qquad \ell^{(1)}_{II}=\ell^{(1)}_I + \ell^{(3)}_I\ ,\qquad 
   \ell^{(2)}_{II}=\ell^{(2)}_I + \ell^{(3)}_I\ ,\qquad  \ell^{(4)}_{II}=\ell^{(4)}_I + \ell^{(3)}_I\ .
\end{equation}
Note that the charge vectors $\ell^{(i)}_I$ and $\ell^{(i)}_{II}$ are chosen to be generators of the Mori cone 
of $\tilde X_4$, that is dual to K\"ahler cone. The generators of the latter for phase I are then given by 
\begin{equation} \label{eq:KCPhaseI}
      J_1=D_2\ ,\quad   J_2=D_1+2D_2+D_3+2D_9\ ,\quad J_3=D_3+D_9, \quad J_4=D_9\ ,
\end{equation}
where $D_i:=\{x_i=0\}$ are the nine toric divisors associated to the points $\Delta_5^{\tilde X}$
which differ from the origin. In phase II we find analogously
\begin{equation} \label{eq:KCPhaseII}
	J_1=D_2,\quad J_2=D_1 + 2 D_2 + D_3 + 2 D_9 \ ,\quad  J_3=D_1 + 3 D_2 + 2 D_9,\quad J_4=D_9 \ .
\end{equation}
The generators $J_i$ provide a distinguished integral basis of $H^{(1,1)}(\tilde X_4)$ since in the 
expansion of the K\"ahler form $J$ in terms of the $J_i$ all coefficients are positive and 
parameterize physical volumes of cycles in $\tilde X_4$. The 
$J_i$ are also canonically used as a basis in which one determines the topological data of 
$\tilde X_4$. The complete set of topological data of $\tilde X_4$ including the intersection ring as well 
as the non-trivial Chern classes are summarized in appendix \ref{app:FurtherP2}.

The polyhedron $\Delta_5^{\tilde X}$ has only few K\"ahler classes which makes it possible to 
identify part of the fibration structures directly from the intersection numbers.
However, an analogous analysis is not possible for the mirror manifold $X_4$ since the dual polyhedron $\Delta_5^{X}$ 
has $2796$ K\"ahler classes. Therefore, we apply the methods reviewed in section \ref{sec:Kreuzermethods} 
for analyzing both $\tilde X_4$ and $X_4$. As already mentioned above, $\Delta_5^{\tilde X}$ intersected with the 
two hyperplanes
\begin{equation}
	H_1 = (0,0,p_3,p_4,0)\ ,\qquad H_2=(p_1,p_2,p_3,p_4,0)
\end{equation}
yields two reflexive polyhedra corresponding to a generic elliptic fiber and the generic three-dimensional 
Calabi-Yau fiber $\tilde Z_3$. The fibration structure of the mirror $X_4$ is studied by 
identifying appropriate projections to $\Delta^k_{\tilde{F}}\subset\Delta_5^{\tilde X}$. Three relevant projections 
$P_i$ are 
\begin{equation}
	P_1(\underline{p})=(p_3,p_4)\ , \qquad P_2(\underline{p})=(p_1,p_2,p_3,p_4)\ , \qquad P_3(\underline{p})=(p_3,p_4,p_5)\ ,
\end{equation}
where $\underline{p}=( p_1,\dots,p_5)$ denote the columns in the polyhedron $\Delta_5^{\tilde X}$.
Invoking the theorem of section \ref{sec:Kreuzermethods}, we see from $P_1$ 
that $X_4$ is also elliptically fibered and 
since the polyhedron of $\P^2({1,2,3})$ is self-dual, the fibration is of $\P^2({1,2,3})$-type.
In addition, it is clear from $P_2$ that $X_4$ is Calabi-Yau threefold fibered.
The fiber threefold is $Z_3$, the mirror to $\tilde Z_3$. The fact, that the threefold 
fibers of $X_4$ and $\tilde X_4$ are mirror manifolds is special to this example since the subpolyhedra
obtained by $H_2$ and $P_2$ are identical. Finally, we note that $X_4$ is $K3$ fibered as inferred 
from the projection $P_3$. This ensures the existence of a heterotic dual theory by 
fiberwise applying the duality of F-theory on $K3$ to heterotic strings on $T^2$, as reviewed in 
section \ref{sec:HetFDuality}.  

The hypersurface constraint for $X_4$ depends on four complex structure moduli $\underline{z}$. This dependence 
is already captured by only introducing $12$ out of the many coordinates needed to specify a non-singular 
$X_4$. This subset of points in $\Delta_5^X$ is given by
\small
\begin{equation} \label{eq:Delta5P2}
         \Delta_5^X \supset	\left(
\begin{array}{c|rrrrr|c}
v_1   &   0 &  0 & 1 & 1 & 0 & z \\
v_2   & -12 & 6 & 1 & 1 & 0 & u_1\\
v_3   &  6  &-12 & 1 & 1 & 0 & u_2\\
v_4   &  6  &  6 & 1 & 1 & 0 & u_3\\
v_5   &  0  &  0 &-2 & 1 & 0 & x\\
v_6   &  0  &  0 & 1 &-1 & 0 & y\\
v^b_1 & -12 &  6 & 1 & 1 &-6 & x_1\\
v^b_2 & -12 &  6 & 1 & 1 & 6 & x_2\\
v^b_3 &  6  &-12 & 1 & 1 &-6 & x_3\\
v^b_4 &  6  &  6 & 1 & 1 &-6 & x_4\\
v^b_5 &  0  & -6 & 1 & 1 & 6 & x_5\\
v^b_6 &  0  &  6 & 1 & 1 & 6 & x_6
\end{array}
\right)
\end{equation}
\normalsize
where we have omitted the origin. Note that we have listed in \eqref{eq:Delta5P2} 
the vertices of $\Delta_5^X$ and added the inner points $v_1$ 
and $v_2$ to list all points necessary to identify the polyhedron 
$\Delta_4^Z$ with vertices \eqref{eq:mirror3foldellp2} in the hyperplane orthogonal to $(0,0,0,0,1)$.
Thus we directly observe the Calabi-Yau fibration with generic fiber $Z_3$. The base of 
this fibration is given by the points labeled by a superscript $^b$.
Note that $v_1$ is also needed to display the elliptic fibration.
The base of the elliptic fibration is obtained by performing the quotient 
$\Delta_3^{\text{base}}=\Delta_5^X/(P_1\Delta_5^{\tilde X})^*$ which 
amounts to simply dropping the third and fourth 
entry in the points of $\Delta_5^X$.

In addition, one can also see the elliptic fibration directly on the 
defining polynomial $P$ of $X_4$ 
which can be written in a Weierstrass form. Indeed if we apply \eqref{eq:Z3typeIIB} 
for the points displayed in \eqref{eq:Delta5P2} of $\Delta_5^X$ and all points 
$p$ of $\Delta_5^{\tilde X}$ that are not on codimension one faces we obtain a 
hypersurface of the form\footnote{The polynomial $P$ can 
be easily brought to the standard Weierstrass form by completing the square and 
the cube, i.e.\ $\tilde y=y+\frac12\tilde m_1 xz$ and $\tilde x=x-\frac{1}{12}\tilde m_1^2z^2$.}
\begin{equation} \label{eq:preWeier}
	P=a_6 y^2+ a_5 x^3+\tilde m_1 (\underline{x},\underline{u}) x y z+\tilde m_6 (\underline{x},\underline{u}) z^6=0\, .
\end{equation}
Here $\underline{x},\underline{u}$ are the homogeneous coordinates on the base of the elliptic fibration, while 
$x$, $y$, and $z$ are the homogeneous coordinates of the $\mathds{P}^2(1,2,3)$-fiber.
The polynomials $\tilde m_1$ and $\tilde m_6$ are given by
\begin{eqnarray}  \label{eq:def-tildem1}
 	\tilde m_1(\underline{x},\underline{u})&=&a_{0} u_1 u_2 u_3 x_1 x_2 x_3 x_4 x_5 x_6\,,\\ 
 	\tilde m_6 (\underline{x},\underline{u})&=&
	 u_1^{18}\, (a_7 x_1^{24} x_2^{12} x_3^6 x_4^6 + a_3 x_1^{18} x_2^{18} x_5^6 x_6^6 )+a_4 u_2^{18}\, x_3^{18} x_5^{12}
	 + a_2 u_3^{18}\, x_4^{18} x_6^{12} \nn\\
	&&+ u_1^6 u_2^6 u_3^6\, ( a_1 x_1^6 x_2^6 x_3^6 x_4^6 x_5^6 x_6^6+ a_9 x_2^{12} x_5^{12} x_6^{12} + a_8  x_1^{12} x_3^{12} x_4^{12})\ , \label{eq:def-tildem6}
\end{eqnarray}
where the coefficients $\underline{a}$ encode the complex structure deformations of $X_4$. 
However, since $h^{(3,1)}(X_4)=h^{(1,1)}(\tilde X_4)=4$ there are only four complex structure 
parameters rendering six of the $\underline{a}$ redundant. It is also straightforward to recover 
from $\tilde m_1$, $\tilde m_6$ of the fourfold $X_4$ the corresponding threefold data $m_1$, $m_6$ in 
\eqref{eq:MirrorEtale} and \eqref{eq:defm_i}, upon fixing the coordinates of the $\P^1$-base of the 
$Z_3$-fibration as $\underline{x}=1$.

For the different phases we identify the complex structure moduli in the hypersurface constraint $P$
by using the charge vectors $\ell^{(i)}$ in \eqref{eq:vl-for-p2} and by applying the general formula 
\eqref{eq:algCoords}. For phase I one finds 
\begin{equation} \label{eq:zI}
  \zI_1=\frac{a_2a_4a_7}{a_1^2a_8}\,,\quad \zI_2=\frac{a_1 a_5^2 a_6^3}{ a_{0}^6} \,,\quad \zI_3=\frac{a_3 a_8}{a_1 a_7}\,,\quad 
  \zI_4=\frac{a_7 a_9}{a_1a_3}\ ,
\end{equation}
while for the phase II one finds in accord with \eqref{eq:ell_flop} that
\begin{equation}\label{eq:zIzII}
  \zII_1 = \zI_1 \zI_3\ , \quad 
  \zII_2 = \zI_2 \zI_3 \ , \quad 
  \zII_3 = (\zI_3)^{-1} \ ,\quad 
  \zII_4 = \zI_4 \zI_3 \ .
\end{equation}
In order to prepare for a comparison with the constraint $p_0$ in \eqref{eq:MirrorEtale} of the threefold $Z_3$ we chose the gauge $a_i=1,\, i=2,\ldots,6$ and $a_8=1$, such that
\begin{equation} \label{eq:id_ai}
   a_0^6 = \frac{1}{(\zII_1)^{{1}/{3}} \zII_2 \zII_3 }\ ,\quad
 a_1 = \frac{1}{(\zII_1)^{1/3}}\ ,\quad
   a_7 =  \zII_3 (\zII_1)^{1/3}\ ,\quad
  a_9 = \frac{\zII_4}{(\zII_1)^{2/3}} 
\ .
\end{equation}
It is straightforward to find the similar expression for phase I by inserting \eqref{eq:zIzII} into 
this expression for $a_0,a_1$ and $a_7,a_9$.

Having determined the defining equation $P$ for the Calabi-Yau fourfold we readily
evaluate the discriminant $\Delta(X_4)$ of the elliptic fibration. Using \eqref{eq:def-j} for 
a Calabi-Yau fourfold with constraint in the Weierstrass form \eqref{eq:Weierstrass} we find 
\begin{equation} \label{eq:disc_X4}
  \Delta(X_4) = - \tilde m_6 (432 \tilde m_6 + \tilde m_1^6)\ .
\end{equation} 
We conclude that there are seven-branes on the divisors $\tilde m_6= 0$ and $432 \tilde m_6 + \tilde m_1^6=0$ 
in the base $B_3^X$. The key observation is that in addition to a moduli independent part $\tilde m_6^{0}$ the 
full $\tilde m_6$ is shifted as 
\begin{equation} \label{eq:m6shift}
	\tilde m_6 = \tilde m_6^0+a_1 (u_1 u_2 u_3 x_1 x_2 x_3 x_4 x_5 x_6)^6+
	  a_7 u_1^{18} x_1^{24} x_2^{12} x_3^6 x_4^6
	 + a_9 u_1^6 u_2^6 u_3^6 x_2^{12} x_5^{12} x_6^{12}\ .
\end{equation}
The moduli dependent part is best interpreted in the phase II with $a_1,a_7$ and $a_9$ given 
in \eqref{eq:id_ai}. In fact, when setting the fourth modulus to $\zII_4=0$, one notes
that the deformation of the seven-brane locus $\tilde m_6 = 0$ is precisely parameterized 
by $\zII_3$. By setting $x_i=1$ one fixes a point in the base of $X_4$ viewed as fibration 
with generic fiber $Z_3$. One is then in the position to compare the shift 
in \eqref{eq:m6shift} with the first constraint in \eqref{eq:divisors}
finding agreement if one identifies $\hat z_1 = \zII_3 (\zII_1)^{1/3}$. 

In the next section we will exploit this further by showing that the open string BPS numbers 
of the local model with D5-branes of \eqref{eq:localp2}
are recovered in the $\zII_3$-direction. The 
shift of the naive open modulus $\hat z_1$ by the closed complex structure modulus $\zII_1$ 
fits then nicely with a similar redefinition made for the local models in \cite{Aganagic:2001nx}. 
This leaves us with the interpretation that indeed $\zII_3$ deforms the seven-brane locus 
and matches an open string modulus in the local picture. As we will 
show in the next section, a $\zII_3$-dependent superpotential is induced upon 
switching on fluxes on the seven-brane or equivalently by specifying four-flux $G_4$. The 
superpotential can be computed explicitly and is matched with the results for a D5-brane in 
the local Calabi-Yau.

A second interpretation of the shifts in the discriminant \eqref{eq:m6shift} by the monomials proportional to 
$\zII_3,\zII_4$ is given by the heterotic dual theory on $Z_3$ and the encoding of the spectral cover 
data of the heterotic bundle $E$ by the fourfold constraint of $X_4$ as discussed in section \ref{sec:het-Fdual}. To see this, 
we bring $P$ into the form \eqref{eq:BMform} by an appropriate coordinate redefinition. Setting 
$v=x_1^6 x_3^6 x_4^6 x_2^{-6}$, $\tilde u_1 = u_1 x_1 x_2$, $\tilde u_2=u_2 x_3$, 
$\tilde u_3=u_3 x_4$, and picking the local patch $x_5=x_6=1$ one rewrites \eqref{eq:preWeier} 
as 
\begin{equation}
   P = p_0 + v p_+ + v^{-1} p_-\ ,
\end{equation}
where $p_0(y,x,z,\tilde u_1,\tilde u_2,\tilde u_3)=0$ is the threefold constraint \eqref{eq:Weierst3fold} 
of $Z_3$, and 
\begin{equation}
   p_+ = (a_7 \tilde u_1^{18} + a_8 \tilde u_1^{6} \tilde u_2^{6} \tilde u_3^{6}) z^6 \, , 
   \qquad p_- = a_9 \tilde u_1^{6} \tilde u_2^{6} \tilde u_3^{6} z^6 \, ,
\end{equation}
describe the two heterotic bundles $E_1\oplus E_2$ in $E_8\times E_8$.
Hence, in the local mirror limit in which $p_- \rightarrow 0$ \cite{Berglund:1998ej}, it is natural to interpret 
the modulus $\zII_3$ as a bundle modulus of an $E_1=\text{SU}(1)$ in the heterotic dual 
theory.
One might be surprised that an SU$(1)$-bundle carries any bundle moduli due to the trivial structure group. 
Indeed the adequate physical interpretation of this configuration is in terms of heterotic five-branes, 
cf.~section \ref{sec:spectralcover} and \cite{Berglund:1998ej}, as discussed in detail in section \ref{sec:Example2} 
and in \cite{Grimm:2009sy}.

Finally, as a side remark, let us note again that \eqref{eq:disc_X4} with \eqref{eq:def-tildem1} and \eqref{eq:def-tildem6} 
is not the full answer for the discriminant since we have set many of the blow-up coordinates 
in $X_4$ to unity. However, one can use the toric methods of \cite{Bershadsky:1996nh,Candelas:1996su,Candelas:1997eh}
to determine the full minimal gauge group in the absence of fluxes to be 
\begin{equation}
   G_{X_4}\ =\ E_8^{25}\ \times\ F_4^{69} \ \times \ G_2^{184}\ \times \ SU(2)^{276}\ .
\end{equation}
Groups of such large rank are typical for elliptically fibered Calabi-Yau geometries 
with many K\"ahler moduli corresponding to blow-ups of singular fibers \cite{Candelas:1997eh}.

\newpage

\subsection{Mirror Symmetry Applications to F-Theory} \label{sec:ApplicationsMirrorSymmetryToF}

In this section we calculate the flux and brane superpotentials \eqref{eq:oc_super}
of the Type IIB theory from the perspective of F-theory on the fourfold $X_4$ with
fluxes $G_4$ in the limit \eqref{eq:SuperpotLimit}. We perform this analysis by applying the methods of section \ref{sec:FFMirrors}
to the example discussed in section \ref{sec:ffConstruction}. We use the following strategy. 

First we identify the periods of the threefold fiber $Z_3$ of $X_4$ among the fourfold periods. 
This implies a matching of all instanton numbers as well as the classical intersections of the mirror 
$\tilde Z_3$. Then we explicitly identify fourfold periods that reproduce 
the physics of branes on the local geometry of $\tilde Z_3$ discussed in section 
\ref{sec:ffConstruction}, namely all disk instantons calculated in \cite{Aganagic:2001nx}.
This is equivalent to the statement that we calculate the flux superpotential 
\eqref{eq:flux_ori} and the seven-brane superpotential \eqref{eq:wgromovwitten} for a 
specific brane flux from the fourfold perspective of F-theory. For the mirror fourfold $\tilde{X}_4$ the closed 
BPS-states are encoded in $F^0(\gamma)$ so that this matching implies, in mathematical terms,
a map of the integral structure of $F^0(\gamma)$ to the integral structure in \eqref{eq:wgromovwitten}
encoded by the Ooguri-Vafa invariants. Thus, we explicitly show that 
there is an element $\hat{\gamma}\in H^{(2,2)}_H(Z_3)$ such that the complete enumerative geometry 
of the threefold mirror pair $(\tilde Z_3,Z_3)$ with and without branes is reproduced. 
We already note that the results presented below are of further importance since they provide a concrete check of 
heterotic/F-theory duality along the lines of section \ref{sec:het-Fdual} as 
the Calabi-Yau threefold $Z_3$ can also be promoted to the background geometry a heterotic string
compactification. The details of this analysis are found in section \ref{sec:SuperpotsHetF}.

Let us now perform the concrete calculations. The Calabi-Yau fourfold $X_4$ introduced in \eqref{eq:Delta5P2} 
has four complex moduli. We find that the moduli dependence of its periods is determined by a complete set 
of six Picard-Fuchs operators which are linear differential operators ${\cal D}_\alpha$, $\alpha=1,\ldots, 6$ of order 
$(3,2,2,2,3,2)$. These are obtained from the GKZ-system \eqref{eq:pfo} associated to the charge vectors $\ell_I^{(1)},\ell_I^{(2)},\ell_I^{(3)},\ell_I^{(4)},\ell_I^{(1)}+\ell_I^{(3)},\ell_I^{(3)}+\ell_I^{(4)}$, 
by the methods described in \cite{Hosono:1993qy}. We use logarithmic derivatives $\theta_a=z_a \frac{d}{d z_a}$
in the canonical complex variables \eqref{eq:algCoords} and present only the leading piece of the differential equations
${\cal D}_\alpha^{\rm lim}=\lim_{z_a\rightarrow 0} {\cal D}_\alpha(\theta_a,z_a)$, $a=1,\ldots, 4$. For the case at hand we obtain
\begin{equation} 
\begin{array}{rlrl}  
{\cal D}_1^{\rm lim}&=\theta_1^2(\theta_3 -\theta_1 - \theta_4), \ \ \qquad &\qquad
{\cal D}_2^{\rm lim}&=  \theta_2(\theta_2- 2 \theta_1- \theta_3 - \theta_4),\\[0.7Em]
{\cal D}_3^{\rm lim}&= (\theta_1 -\theta_3)(\theta_3-\theta_4), \ \ \qquad &\qquad
{\cal D}_4^{\rm lim}&=\theta_4 (\theta_1 - \theta_3 + \theta_4 ),\\[0.7Em] 
{\cal D}_5^{\rm lim}&=   \theta_1^2 (\theta_4 - \theta_3),\ \ \qquad &\qquad
{\cal D}_6^{\rm lim}&=  \theta_4 (\theta_1 - \theta_3)\ .
\end{array}
\end{equation} 
For the complete Picard-Fuchs system as well as the cohomology basis we extract from it
by calculating the ring $\mathcal{R}$ in \eqref{eq:ring} we refer to appendix \ref{app:FurtherP2}.  

The calculation of $\mathcal{R}$ is readily performed yielding $(1,4,6,4,1)$ 
generators of the ring ${\cal R}$ at grade $(0,1,2,3,4)$, which are  
\begin{equation} 
\begin{array}{l|c}  
\cR^{(0)} & 1 \\[2mm]
\cline{1-1}
\rule[.5cm]{0cm}{.2cm}\cR^{(1)}_a & \theta_1, \ \  \theta_2, \ \ \theta_3, \ \ \theta_4, \\[2 mm] 
\cline{1-1}
\rule[.5cm]{0cm}{.2cm} \cR^{(2)}_\alpha & \theta_1^2, \ \   (\theta_1+\theta_3) \theta_4, \ \ (\theta_1+\theta_3)\theta_3,\ \ (\theta_1+2 \theta_2) \theta_2, \ \
(\theta_2+\theta_4 ) \theta_2,\ \ (\theta_2+\theta_3) \theta_2\\[2 mm]
\cline{1-1}
\rule[.5cm]{0cm}{.2cm} \cR^{(3)}_a &  \left(\theta_3+\theta_4\right) \left(\theta_1^2+\theta_1 \theta_3+\theta_3^2\right),\ \ \theta_2 \left(\theta_3^2+3 \theta_2 \theta_3+5 \theta_2^2+\theta_1 \left(\theta_2+\theta_3\right)\right),\\ 
& \theta_2 \left(\theta_1 \left(\theta_2+\theta_4\right)+\theta_4 \left(\theta_3+3 \theta_2\right)+\theta_2 \left(\theta_3+6 \theta_2\right)\right),\ \ 
\theta_2 \left(\theta_1^2+2 \theta_1 \theta_2+4 \theta_2^2\right)\\[2 mm]
\cline{1-1}
\rule[.5cm]{0cm}{.2cm} \cR^{(4)} & \theta_4(\theta _1^2 \theta _2+3 \theta _1 \theta _2^2+9 \theta _2^3+\theta _1 \theta _2 \theta _3+3 \theta _2^2 \theta _3+\theta _2 \theta _3^2)\\
&+ \theta _2 \left(46 \theta _2^3+15 \theta _2^2 \theta _3+4 \theta _2 \theta _3^2+\theta _3^3+\theta _1^2 \left(2 \theta _2+\theta _3\right)+\theta _1 \left(11 \theta _2^2+4 \theta _2 \theta _3+\theta _3^2\right)\right)\ .
\end{array}
\label{eq:ringffp2}
\end{equation} 
These elements can be associated to solutions of the Picard-Fuchs equations and to a
choice of basis elements of the Chow ring as explained in section \ref{sec:matching}. 
At grade $k=2$ the leading solutions $\bbL^{(k)\, \alpha}$ of the full Picard-Fuchs system 
\eqref{PFOFFP2T1}, that obey the normalization $\mathcal{R}^{(k)}_\alpha \mathbb{L}^{(k)\, \beta}
=\delta^{\beta}_\alpha$ of \eqref{eq:periodStrucutreLV}, are then given by
\begin{eqnarray}
 &\bbL^{(2)\, 1}= l_1^2\,,\ \ \bbL^{(2)\, 2}=\frac{1}{2} l_4 \left(l_1+l_3\right)\,,\ \ \bbL^{(2)\, 3}=\frac{1}{2} l_3 \left(l_1+l_3\right)\,,&\nn\\
&\bbL^{(2)\, 4}=\frac{1}{7}  l_2\left(3 l_1-2 \left(l_3+l_4-l_2\right)\right) \,,\ \ \bbL^{(2)\, 5}=\frac{1}{7} l_2 \left(-2 l_1+l_2+6 l_4-l_3\right)\,,&\nn\\
&\bbL^{(2)\, 6}=\frac{1}{7}  l_2 \left(-2 l_1+l_2+6 l_3-l_4\right)\,,& \label{eq:Sbasis}
\end{eqnarray}
where we used the abbreviation $\text{log}(z_k)\equiv l_k$ and omitted the prefactor 
$X_0$. In comparison to the complete solutions $\Pi^{(2)\, \alpha}$ of the Picard-Fuchs 
equations we focused on the leading terms only and omitted terms of order $\mathcal{O}(\underline{l})$. 
The full solutions with the leading logarithms \eqref{eq:Sbasis} directly occur in the periods expansion of 
the holomorphic four-form $\Omega_4$.
Since we are calculating the holomorphic potentials $F^0(\gamma)$ 
of \eqref{eq:simple3point} and the corresponding BPS-invariants we have to change the basis 
of solutions such that to any operator $\mathcal{R}^{(2)}_\alpha$ in \eqref{eq:ringffp2} we 
associate a solution with leading logarithm determined by the classical intersection 
$C^{0\ (1,1,2)}_{ab\alpha}$ of the A-model defined in \eqref{eq:classicalamodeltriplecouplings} as
\begin{equation} \label{eq:intersecsLL}
 	\mathbb{L}^{(2)}_\alpha=\tfrac12X_0 C^0_{\alpha ab}l_al_b\,.
\end{equation}
As we have already proven in \eqref{eq:fourfoldPoincareDual} these solutions are related to the leading 
logarithms $\bbL^{(2)\,\alpha}$ as $\bbL^{(2)}_\alpha=\bbL^{(2)\, \beta}\eta^{(2)}_{\alpha\beta}$. 
Thus, we readily obtain the leading terms $\bbL^{(2)}_\alpha$ from the classical intersection data \eqref{eq:ringffp2} 
in $\mathcal{R}^{(4)}$.

As in the discussion below \eqref{eq:Atobeta_map} the choice of periods $\Pi^{(2)\, \alpha}$ 
with leading terms $\bbL^{(2)\, \alpha}$ corresponds to a particular choice of a basis 
$\hat \gamma^{(2)}_\alpha $ of $H^{(2,2)}_V(\tilde X_4)$. In fact, by construction one identifies
the flux basis at grade two as 
\begin{equation}
  \hat \gamma^{(2)}_\alpha =  \mathcal{R}^{(k)}_\alpha \Omega_4|_{z=0}\ .
\end{equation}
However, this choice of basis for $H^{(2,2)}_V(\tilde X_4)$ is not necessarily a basis of 
integral cohomology. An integral basis can, however, be determined by an appropriate 
basis change. We next identify some basis elements of the integral cohomology of $\tilde{X}_4$.
First we note that the K\"ahler generator $J_4$ is identified as the 
class of the Calabi-Yau threefold fiber $\tilde Z_3$, where we refer to appendix \ref{app:FurtherP2} 
for more details on this identification from the intersection numbers. Moreover, we identify the 
fourfold K\"ahler generators $J_i$ with the threefold generators 
$J_k(\tilde Z_3)$ as
\begin{equation}  \label{eq:exchangeJs}
   J_1+J_3\ \leftrightarrow\ J_1(\tilde Z_3)\ , \qquad J_2\ \leftrightarrow\ J_2 (\tilde Z_3)\ ,
\end{equation} 
by comparing the coefficient of $J_4$ in the intersection form $\cC_0(\tilde X_4)$, given 
in \eqref{IntX_TextEx} in appendix \ref{app:FurtherP2}, with the threefold intersections $\mathcal{C}_{0}(\tilde Z_3)$ 
in \eqref{eq:intersectionsY}. A subset of the basis elements of the fourfold integral basis are 
now naturally induced from the threefold integral basis. In order do so we identify 
the threefold periods $\partial_i  F_{\tilde Z_3}$, with derivatives in the directions 
$J_1(\tilde Z_3)$ and $J_2(\tilde Z_3)$, with an appropriate linear combination of the 
fourfold periods $\Pi^{(2)\, \alpha}$ \cite{Mayr:1996sh}. In other words we determine a 
new basis $\hat \gamma_i^{(2)}$ such that one fourfold prepotential equals $\partial_i  F_{\tilde Z_3}$,
\begin{equation} \label{eq:matchPrepot}
 	\partial_i F_{\tilde Z_3} = F^0(\hat{\gamma}_i^{(2)})|_{z_4=0} \equiv \Pi^{(2)}_i|_{z_4=0}\ . 
\end{equation}
In this matching both the classical part of the periods as well as the threefold 
BPS-invariants $n_{d_1,d_2}$ and the fourfold BPS-invariants $n_{d_1,d_2,d_1,0}(\gamma)$ have to 
match in the limit of large $\P^1$-base. 

The match \eqref{eq:matchPrepot} is most easily performed by first comparing the classical parts 
of the periods. In fact, using the classical intersections of $\tilde Z_3$ in \eqref{eq:intersectionsY}
one deduces that the leading parts of the threefold periods are
\begin{equation} \label{eq:threefoldperiods}
   \bbL_1(Z_3) =\tfrac{1}{2} X_0\tilde l_2 \big(2 \tilde l_1+3 \tilde l_2\big)\,\quad 
   \bbL_2(Z_3) =\tfrac{1}{2} X_0\big(\tilde l_1+3 \tilde l_2\big)^2\, ,
\end{equation}
where $\tilde l_i = \log \tilde z_i$ correspond to the two threefold directions 
$J_k(\tilde Z_3)$ in \eqref{eq:exchangeJs}. Using \eqref{eq:exchangeJs} and \eqref{eq:matchPrepot} 
one then finds the appropriately normalized leading fourfold periods as
\begin{equation} \label{eq:LeadPer}
   \bbL^{(2)}_2=\tfrac{1}{2} X_0l_2 \left(2 l_1+3 l_2+2 l_3\right)\,\quad \bbL^{(2)}_5=\tfrac{1}{2} X_0\left(l_1+3 l_2+l_3\right)^2\,.
\end{equation}
A direct computation also shows that the threefold BPS invariants $d_i n_{d_1,d_2}$ and 
fourfold BPS invariants $n_{d_1,d_2,d_1,0}(\hat{\gamma}_i)$ match in the large $\P^1$-base 
limit, such that \eqref{eq:matchPrepot} is established on the classical as well as quantum level.
This match then fixes corresponding integral basis elements of $H^{(2,2)}_V(\tilde{X}_4)$ as follows. 

First we determine those two ring elements $\tilde{\cR}^{(2)}_\alpha$, $\alpha=2,5$, such 
that we obtain \eqref{eq:LeadPer} from the new three-point couplings using \eqref{eq:intersecsLL} . 
Then we complete them into a new basis of ring elements $\tilde{\cR}^{(2)}_\alpha$ by choosing
\begin{eqnarray} \label{eq:newring}
 &\tilde{\cR}^{(2)}_1= \theta_1^2\,,\ \ \tilde{\cR}^{(2)}_2=\frac{1}{2}  \theta_4 \left( \theta_1+ \theta_3\right)\,,
 \ \ \tilde{\cR}^{(2)}_3=\frac{1}{2}  \theta_3 \left( \theta_1+ \theta_3\right)\,,&\nn\\
&\tilde{\cR}^{(2)}_4=\frac{1}{7}   \theta_2\left(3  \theta_1-2 \left( \theta_3+ \theta_4- \theta_2\right)\right) \,,
\ \ \tilde{\cR}^{(2)}_5=\frac{1}{7}  \theta_2 \left(-2  \theta_1+ \theta_2+6  \theta_4- \theta_3\right)\,,&\nn\\
&\tilde{\cR}^{(2)}_6=\frac{1}{7}   \theta_2 \left(-2  \theta_1+ \theta_2+6  \theta_3- \theta_4\right)\, .&
\end{eqnarray}
Consequently, these operators fix the two integral basis elements 
\begin{equation}
  \hat \gamma^{(2)}_2 = \tilde{\cR}^{(2)}_2 \Omega_4 |_{z=0}\ ,\qquad  \hat \gamma^{(2)}_5 = \tilde{\cR}^{(2)}_5 \Omega_4 |_{z=0}\ .
\end{equation}
which reproduce the corresponding part of the flux superpotential \eqref{eq:flux_ori} on $Z_3$ 
for $\hat{N}_i=\hat{M}^0=0$ when turning on four-form flux $G_4$ on $X_4$ in these directions,
\begin{eqnarray} \label{eq:FP2flux}
 	W_{\rm flux}\equiv M^1F^{0}(\hat{\gamma}_2^{(2)})+M^2F^{0}(\hat{\gamma}_5^{(2)})=\int_{X_4}\Omega_4\wedge G_4
	=M^1\Pi^{(2)}_2+ M^2\Pi^{(2)}_5\,
\end{eqnarray}
for the flux 
\begin{equation}
 	G_4=M^1\hat{\gamma}^{(2)}_2+M^2\hat{\gamma}^{(2)}_5\,.
\end{equation}
For the choices $M^{i}=\delta^{ij}$ we extract the threefold invariants $d_j n_{d_1,d_2}$ from this superpotential, 
i.e.~from the prepotentials $F^{0}(\hat{\gamma}_2^{(2)})$ and $F^{0}(\hat{\gamma}_5^{(2)})$. 
We note that the above grade two basis elements \eqref{eq:newring} map under $\theta_i \leftrightarrow l_i$ 
exactly to the leading solutions of the Picard-Fuchs system \eqref{eq:Sbasis}. Using the same identification 
we find $\bbL^{(2)\, 2}=X_0(l_1+l_3 ) l_4$ and $\bbL^{(2)\, 5}=X_0(l_2+l_4 ) l_2$ as the leading behavior 
of corresponding periods $\int_{\gamma^{2}_\alpha}\Omega=\Pi^{(2)\,\alpha}$ of the holomorphic four-form. 
This agrees with the naive expectation from the large base limit that a partial factorization of the periods 
occurs as $t_4\cdot t^{\tilde Z_3}_i$ where $t^{\tilde Z_3}_i$, $i=1,2$, denoted the two classes in $\tilde Z_3$ \cite{Mayr:1996sh}.

It is one crucial point of our whole analysis that we can extend this matching of threefold invariants 
even to disk invariants counting curves with boundaries on Lagrangian cycles $L$ in $\tilde Z_3$, cf.~section 
\ref{sec:osGW}. Having explained the F-theory origin of this fact via an 
analysis of the discriminant in \eqref{eq:m6shift} we now explicitly find the flux choice 
in $H^{(2,2)}_H(X_4)$ for which the flux superpotential \eqref{eq:fluxpotFourfoldExpanded} on $X_4$ reproduces 
the brane superpotential \eqref{eq:wgromovwitten}.

By construction the fourfold $\tilde X_4$ inherits 
information of the fiber $\tilde Z_3$ and in particular the local limit geometry $\mathcal{O}(-3)\rightarrow \mathds{P}_2$
for which the disk invariants have been computed in \cite{Aganagic:2001nx}. As noted earlier the brane 
data is translated to the F-theory picture of the fourfold $\tilde X_4$ by the Mori cone generator 
$\ell^{(3)}$ and its dual divisor $J_3$. Therefore, we expect to reproduce all classical terms as 
well as to extract the disk instantons of \cite{Aganagic:2001nx} from the Gromov-Witten invariants 
$n_{d,0,d+k,0}(\hat{\gamma}_3)$ of a fourfold period, that we construct via \eqref{eq:intersecsLL} from operators 
of \eqref{eq:Sbasis} of the form $\cR^{(2)}_{\gamma}=\theta_3(\theta_1+\theta_3)+\ldots$. However, the 
geometry is more complicated and the ring element $\cR^{(2)}_{\gamma}$ with this property is not unique. 
It takes the form
\begin{equation} 
\cR^{(2)}_{\gamma}=-\cR^{(2)}_1+\tfrac{1}{3}\cR^{(2)}_2+ \cR^{(2)}_3
=-\theta_1^2+\tfrac{1}{2}\theta_3(\theta_1+\theta_3)+\tfrac{1}{6}\theta_4(\theta_1 +\theta_3)
\end{equation} 
that is the most convenient choice by setting the arbitrary coefficients of $\cR^{(2)}_\alpha$, 
$\alpha=4,5,6$, to zero. We note that only the coefficient in front of $\cR^{(2)}_3$ was fixed to 
unity by the requirement of reproducing the disk instanton invariants. The two further coefficients 
$(-1,\tfrac13)$ were fixed by the requirement of reproducing the Gromov-Witten invariants $n_{d}$ of 
local $\mathds{P}^2$ \cite{Haghighat:2008gw} by the fourfold invariants $n_{d,0,d,0}\equiv n_d$, 
i.e.~for $m=0$, as explained below. The relation between $\cR^{(2)}_{\gamma} $ and the corresponding 
solution is established as $\hat \gamma={\cal R}_{\gamma} \Omega_4|_{z=0}$ and $\Pi^{(2)\gamma}=\int_{\gamma} \Omega_4$, 
i.e.~$\cR^{(2)}_\gamma\Pi^{(2)\gamma}=1$, so that    
\begin{equation}
\mathbb{L}^{(2)\, \gamma}=- X_0 l_1^2\ ,\quad
\mathbb{L}^{(2)}_{\gamma}=\tfrac{1}{6} X_0l_2 \left(8 l_1+9 l_2+2 l_3\right)\,.
\end{equation}
In other words, this implies that we have explicitly calculated the seven-brane superpotential \eqref{eq:wgromovwitten} 
from the fourfold superpotential \eqref{eq:fluxpotFourfoldExpanded} by turning on the flux $G_4=\hat{\gamma}$,
\begin{equation} \label{eq:FP2branepot}
 	W_{\rm brane}\equiv F^{0}(\hat{\gamma})=\int\Omega_4\wedge\hat{\gamma}=\Pi^{(2)}_\gamma\,.
\end{equation}

By listing the numbers $n_{d_1, 0,d_3, 0}(\hat{\gamma})$ extracted from $F^{0}(\hat{\gamma})$ we obtain the results of
table \ref{tab:FP2diskInst}. 
\begin{table}[htbp]
\centering
$
 \begin{array}{|c|rrrrrrr|}
\hline
\rule[-0.2cm]{0cm}{0.6cm}  \text{d}_1&d_3=0&d_3=1&d_3=2&d_3=3&d_3=4&d_3=5&d_3=6\\
\hline
0&  0& 1& 0& 0& 0& 0& 0\\ 
1&  1& n_1& -1& -1& -1& -1& -1\\ 
2&  -1& -2& 2 n_2& 5& 7& 9& 12\\ 
3&   1& 4& 12& 3 n_3& -40& -61& -93\\ 
4&   -2& -10& -32& -104& 4 n_4& 399& 648\\ 
5&   5& 28& 102& 326& 1085& 5 n_5& -4524\\ 
6& -13& -84& -344& -1160& -3708& -12660& 6 n_6\\
\hline
\end{array}
$
\caption{BPS invariants $n_{d_1, 0, d_3, 0}(\hat{\gamma})$ for the disks. With the 
identification $d_3-d_1=m$ (winding) and $d_1=d$ ($\mathds{P}^2$ degree) this agrees with Tab. 5 
in~\cite{Aganagic:2001nx}.\label{tab:FP2diskInst}}
\end{table}
The BPS invariants of the holomorphic disks depend only  on the relative
homology class of the latter. In the table $d_3-d_1=m$ labels the winding number 
of the disks and $d_1=d$ the degree with respect to canonical class 
of $\mathds{P}^2$, i.e. if the open string disk superpotential is 
defined in terms of the closed string parameter $q=e^{2\pi it}$ and the open string
string parameter $Q=e^{2\pi i\hat t}$ for the outer brane as 
\begin{equation} 
W_{\rm brane}=a_{tt} t^2 + a_{t \hat t} t \hat t  + a_{\hat t \hat t}\hat t^2 + a_t t + a_{\hat t} \hat t
+ a_0  + \sum_{d=1}^\infty \sum_{m=-d}^\infty n_{d,m} {\rm Li}_2(q^d Q^m),
\end{equation} 
then $n_{d_1,0,d_3,0}=n_{d_1,d_3-d_1}$. Note that the numbers $n_{d_1,0}$ are not
calculated in the framework of \cite{Aganagic:2001nx}. However it is natural and calculable  
in the topological vertex formalism \cite{Aganagic:2003db} that they should be identified 
with  $d n_d$, where $n_d$ is the closed string  genus zero BPS invariant, 
defined via the prepotential as $F^0=\sum_{d=1}^\infty n_d {\rm  Li}_3(q^d)$. 
The factor of $d$ stems from the fact that we identify $\frac{d}{dt} F^0\subset W_{\rm brane} $.
This interpretation as $n_{d,0,d,0}=d n_d$ could be consistently imposed and yields  
two further conditions as mentioned above.

To obtain the open BPS invariants of phase III of \cite{Aganagic:2001nx}, we use the phase II 
in \eqref{eq:vl-for-p2}. In this phase the fiber class is not realized as a generator of 
the K\"ahler cone. However we readily recover the classes of $\tilde Z_3$ as
\begin{equation} \label{eq:3foldMatchT2}
 	J_1\ \leftrightarrow\ J_1(\tilde Z_3)\,\qquad J_2+J_3\ \leftrightarrow\ J_2(\tilde Z_3)\,
\end{equation}
by comparison of the Mori cone \eqref{eq:vl-for-p2} with the Mori cone \eqref{eq:3foldellp2} 
of $\tilde Z_3$. Then we fix a basis $\mathcal{R}^{(2)}_\alpha$ of the ring at grade two as
\begin{eqnarray}
	\theta _1^2,\,\,\,\,2 \theta _2 \left(\theta _1+3 \theta _3\right),\,\,\,\,\theta _3 \left(\theta _1+3 \theta _3\right),\,\,\,\,\theta _1 \theta _4,\,\,\,\,\theta _2^2,\,\,\,\,\left(\theta _2+\theta _3\right) \left(2 \theta _3+\theta _4\right)\,,
\end{eqnarray}
from which we obtain a basis of dual solutions $\mathbb L^{(k)\,\alpha}$ to the Picard-Fuchs 
system \eqref{PFOFFP2T2}
\begin{eqnarray} \label{eq:LeadPerFFP2T2}
\nn	&\mathbb L^{(2)\, 1}=l_1^2,\,\,\,\,\mathbb L^{(2)\, 2}=\frac{1}{140} \left(l_1 \left(16 l_2+9 l_3\right)+3 \left(l_2 \left(6 l_3-5 l_4\right)-l_3 \left(l_3+5 l_4\right)\right)\right),&\\
	&\mathbb L^{(2)\, 3}=\frac{1}{70} \left(l_1 \left(9 l_2+16 l_3\right)-3 \left(l_3 \left(-6 l_3+5 l_4\right)+l_2 \left(l_3+5 l_4\right)\right)\right),\,\,\,\,\mathbb L^{(2)\, 4}=l_1 l_4,&\\
\nn	&\mathbb L^{(2)\, 5}=l_2^2,\,\,\,\,\mathbb L^{(2)\, 6}=\frac{1}{14} \left(l_2+l_3\right) \left(-3 l_1+l_3+5 l_4\right),&
\end{eqnarray}
Next we construct two solutions with leading logarithms matching the two threefold periods in 
\eqref{eq:threefoldperiods} for which we are able to match the threefold invariants $d_in_{d_1,d_2}$ 
in the large base limit as well. The leading logarithms of the corresponding fourfold periods read
\begin{eqnarray}
\nn	\mathbb L_4^{(2)}&=&\tfrac{1}{2} X_0\left(l_1+3 \left(l_2+l_3\right)\right){}^2,\\
	\mathbb L_6^{(2)}&=& \tfrac{1}{2} X_0\left(l_2+l_3\right) \left(2 l_1+3 \left(l_2+l_3\right)\right).
\end{eqnarray}
which is in perfect agreement with \eqref{eq:threefoldperiods} under the identification \eqref{eq:3foldMatchT2}.
We fix the corresponding operators $\tilde{\cR}^{(2)}_4$, $\tilde{\cR}^{(2)}_6$ by matching the 
above two leading logarithms with the classical intersections $C^0_{\alpha ab}$ via \eqref{eq:intersecsLL}. 
We complete them to a basis of $\tilde{\cR}^{(2)}$ as 
\begin{eqnarray}
\nn&\tilde{\cR}_1^{(2)}=\theta _1^2,\,\,\,\,\tilde{\cR}_2^{(2)}=\frac{1}{140} \left(\theta _1 \left(16 \theta _2+9 \theta _3\right)+3 \left(\theta _2 \left(6 \theta _3-5 \theta _4\right)-\theta _3 \left(\theta _3+5 \theta _4\right)\right)\right),&\\
\nn	&\tilde{\cR}_3^{(2)}=\frac{1}{70} \left(\theta _1 \left(9 \theta _2+16 \theta _3\right)-3 \left(\theta _3 \left(-6 \theta _3+5 \theta _4\right)+\theta _2 \left(\theta _3+5 \theta _4\right)\right)\right),\,\,\,\,\tilde{\cR}_4^{(2)}=\theta _1 \theta _4,&\\
\nn	&\tilde{\cR}_5^{(2)}=\theta _2^2,\,\,\,\,\tilde{\cR}_6^{(2)}=\frac{1}{14} \left(\theta _2+\theta _3\right) \left(-3 \theta _1+\theta _3+5 \theta _4\right),&
\end{eqnarray}
where again this basis relates to the leading periods \eqref{eq:LeadPerFFP2T2} by $\theta_i\leftrightarrow l_i$.
The corresponding integral basis elements of $H^{(2,2)}_H(X_4)$ read 
\begin{equation}
 	\hat{\gamma}^{(2)}_4=\tilde{\cR}^{(2)}_4\Omega_4|_{z=0}\ ,\qquad  \hat{\gamma}^{(2)}_6=\tilde{\cR}^{(2)}_6\Omega_4|_{z=0}\,.
\end{equation}

Furthermore, we determine the ring element $\mathcal{R}^{(2)}_\gamma$ that matches the open 
superpotential by turning on four-form flux in the direction 
$\hat{\gamma}=\mathcal{R}^{(2)}_\gamma\Omega_4|_{z=0}$. Again we fix
\begin{equation} \label{eq:braneflux2}
	\mathcal R^{(2)}_\gamma=a_1\mathcal R^{(2)}_2-\tfrac{1}{10}(1+6a_2)\mathcal R^{(2)}_3+\mathcal R^{(2)}_{4}+a_3\mathcal R^{(2)}_5+a_2\mathcal R^{(2)}_6
\end{equation}
by extracting the disk invariants from the solution associated to it via \eqref{eq:intersecsLL} 
which reads
\begin{eqnarray}
 	\nn\mathbb{L}^{(2)\,\gamma}&=&c(a_1) X_0l_2 \left(l_1+3 l_3\right)\ ,\\ \nn\mathbb{L}^{(2)}_\gamma&=&\tfrac{1}{6} \left(l_2+l_3\right) \left(2 l_1+3 \left(l_2+l_3\right)\right)-\tfrac{1}{10} \left(l_1+3 \left(l_2+l_3\right)\right) \left(3 l_1+29 l_2+29 l_3+10 l_4\right)\,.
\end{eqnarray}
Here we explicitly displayed the dependence on the three free parameters $a_i$ for 
$\mathbb{L}^{(2)\,\gamma}$ by $c(a_1)=\tfrac{7}{9+140 a_1}$ and evaluated $\mathbb{L}^{(2)}_\gamma$ 
for the convenient choice $\underline{a}=0$. The associated disk instantons are listed in table \ref{tab:phase3table}.

\begin{table}[htbp]
	\centering
	$
\begin{array}{|c|rrrrrrr|}
	\hline
	\rule[-0.2cm]{0cm}{0.6cm} \text{d}&k=0&k=1&k=2&k=3&k=4&k=5&k=6\\
	\hline
 0&0 & n_1 & 2n_2 & 3n_3 & 4n_4 & 5n_5 & 6n_6\\
 1&-1 & 2 & -5 & 32 & -286 & 3038 & -35870 \\
 2&0 & 1 & -4 & 21 & -180 & 1885 & -21952 \\
 3&0 & 1 & -3 & 18 & -153 & 1560 & -17910 \\
 4&0 & 1 & -4 & 20 & -160 & 1595 & -17976 \\
 5&0 & 1 & -5 & 26 & -196 & 1875 & -20644 \\
 6&0 & 1 & -7 & 36 & -260 & 2403 & -25812\\
 \hline
\end{array}
$
\caption{BPS invariants $n_{k, 0, i, 0}(\gamma)$ for the disks of the second triangulation. }
	\label{tab:phase3table}
\end{table}

\section{Heterotic/F-Theory Duality: Moduli and Superpotentials}
\label{sec:SuperpotsHetF}

In this section we present calculations in heterotic/F-theory duality for two concrete 
four-dimensional compactifications. These calculations serve as a direct check of heterotic/F-theory
duality on the level of matching the moduli as well as the holomorphic superpotentials on both sides.

Before delving into the details of the calculations we summarize in section \ref{sec:HetFModuli}, 
from a conceptual point of view, the matchings we perform in order to finally compute the 
heterotic superpotential, in particular the heterotic flux and five-brane superpotential, from its F-theory 
dual flux superpotential. Then in in section \ref{sec:Example1} we consider an F-theory Calabi-Yau 
fourfold geometry, that will have few K\"ahler moduli and many complex structure moduli. It is constructed directly
using the techniques of section \ref{sec:het-Fdual} from its heterotic dual, which is an elliptic Calabi-Yau threefold with base $\P^2$
and an $E_8\times E_8$ bundle with no perturbative heterotic gauge symmetry in four dimensions. 
In this case we can use toric geometry to compute explicitly in the K\"ahler sector, which allows us to 
evaluate both sides of the moduli map \eqref{eq:modulimap}. From this we can read off the number of deformations of the 
five-brane curve, if present, and check the anomaly formula \eqref{eq:anomaly}, from which we determine the number of 
vertical five-branes as well as the class of possible horizontal five-branes. A direct calculation of the superpotential
is however not possible due to a large dimension of the complex structure moduli space of $X_4$.
The second Calabi-Yau fourfold example, introduced in section \ref{sec:Example2}, will admit few complex 
structure moduli and many K\"ahler moduli. It will basically be the mirror geometry of section \ref{sec:Example1}. 
This allows us to identify the bundle moduli and five-brane moduli under duality by studying the Weierstrass constraint.
The analysis of the K\"ahler sector is technically challenging and omitted. We exploit the calculations of section 
\ref{sec:ApplicationsMirrorSymmetryToF} to identify the heterotic superpotential for a horizontal five-brane as the 
F-theory flux superpotential for a specific four-flux, that we determine.

\subsection{Heterotic Superpotentials from Compact Calabi-Yau Fourfolds} 
\label{sec:HetFModuli}

In this section we are mainly interested in the map of the complex structure moduli of $X_4$ to 
the moduli of the heterotic compactification. In fact, not only the moduli on both sides
map according to \eqref{eq:modulimap} but also their obstructions expressed by the superpotentials 
\eqref{eq:hetallpots} on the heterotic side and \eqref{eq:GVW-super} on the F-theory side. 
In particular we perform calculations in the complex structure moduli space of $X_4$ in 
section \ref{sec:Example2} to derive the heterotic superpotentials 
explicitly for a specific four-form flux $G_4$, that we determine.

Let us prepare for the following explicit calculations by a more formal study of the map of the 
dual superpotentials. Let us recall, that the heterotic superpotential 
\eqref{eq:hetallpots}, is given by 
\begin{equation} 
W_{\rm het}(\underline{t}^c,\underline{t}^g,\underline{t}^o) = W_{\rm flux}(\underline{t}^c) + 
W_{\rm CS}(\underline{t}^c,\underline{t}^g) + W_{\rm brane}(\underline{t}^c,\underline{t}^o)\,,
\end{equation} 
where $\underline{t}^c,\underline{t}^g$ and $\underline{t}^o$ denote the complex structure, 
bundle and five-brane moduli respectively\footnote{Here we denote the moduli by $\underline{t}$ 
in order to indicate a choice of appropriate flat $\mathcal{N}=1$ coordinates, which have to be 
chosen for $\mathcal{N}=1$ mirror symmetry and for the matching of the heterotic with the F-theory side.}.
As we have discussed in section \ref{sec:hettransition} the last two terms are not inequivalent, 
since by tuning the moduli $\underline{t}^g$ or $\underline{t}^o$ one can condense or 
evaporate five-branes and explore different branches of the heterotic moduli space. 
Clearly also the moduli spaces parametrized by $\underline{t}^c$ and $\underline{t}^g$ 
do not factorize globally in complex structure and bundle moduli since the notion of a
holomorphic gauge bundle on $Z_3$ depends on the complex structure of $Z_3$.
Similarly, $\underline{t}^c$ and $\underline{t}^o$ do not factorize as well as the
notion of a holomorphic curve in $Z_3$ does depend on the complex structure of
$Z_3$. 

The key point of heterotic/F-theory duality is then that we can map as in \eqref{eq:modulimap} 
the heterotic moduli $(\underline{t}^c,\underline{t}^g,\underline{t}^o)$ 
to the complex structure moduli $\underline{t}$ of $X_4$, which 
are encoded in the fourfold period integrals. This matching involves at least an 
explicit counting on both sides, as performed in section \ref{sec:Example1}, and 
involves calculations in the K\"ahler moduli space of $Z_3$.
Then, to make precise also the extension of the equivalence to the obstructions as  
\begin{equation}  \label{eq:supermatch}
W_{\rm het}(\underline{t}^c,\underline{t}^g,\underline{t}^o)=W_{G_4}(\underline{t})\ ,
\end{equation}
we need to establish a dictionary between the topological data 
on the heterotic side, which consist of the  heterotic flux quanta, 
the topological classes of gauge bundles and the class of the curves 
$\Sigma$ wrapped by five-branes, and the F-theory flux quanta.

The identification \eqref{eq:supermatch} implies then as a minimal check, that the periods of 
$\Omega$ and complex structure moduli of $Z_3$ arise as a subset of the periods and complex structure 
moduli of $\Omega_4$ in specific directions, cf.~also section \ref{sec:ApplicationsMirrorSymmetryToF} 
and \cite{Mayr:1996sh,Grimm:2009ef}. In order to study the duality map \eqref{eq:supermatch} 
further, one can include additional parts of the heterotic superpotential like the five-brane moduli 
and superpotential. One strategy to study this explicitly in concrete examples, as in section \ref{sec:Example2},
is by restricting the heterotic gauge bundle $E$ to be of trivial 
$SU(1)\times SU(1)$ type, which automatically implies the inclusion of heterotic five-branes 
to satisfy the anomaly cancellation condition \eqref{eq:anomaly}.
Then, it remains to single out the four-flux in F-theory that is dual to the curve $\Sigma$ supporting the 
heterotic five-brane.

In general, it is the challenging part of \eqref{eq:supermatch} to determine the integral F-theory flux that
is dual to a particular heterotic effect.
One immediate strategy to determine $G_4$ is to calculate either the classical terms 
or some of the instanton corrections in $W_{\rm het}$ and then to match with the fourfold periods. Then,
the $G_4$-flux is automatically integral and we can use the full periods of the fourfold and the map \eqref{eq:supermatch} 
to determine the instanton corrections and vice versa. 
Formally, the flux $G_4$ can always be expressed via differential operators acting on $\Omega_4$ as discussed in
section \eqref{sec:matching}. In particular, for the five-brane superpotential $W_{\rm brane}(t^o,t^c)$, which is the case
of most interest for us, one finds that the dual flux $G^{\rm M5}_4$ can be expressed as
\begin{equation} 
G_4^{\rm M5}=\sum_{p} N^{p (2)} \left. {\cal R}_{p}^{(2)}\Omega_4\right|_{\underline z=0}\,.
\end{equation}  
In addition, we note that for fluxes $G_4$ generated by operators $\cR^{(2)}$ the superpotential yields an
integral structure of the fourfold symplectic invariants at large
volume of the mirror $\tilde{X}_4$ of $X_4$ in terms of the double-covering formula \eqref{eq:g=0multicovering}
for $k=1$ where the co-dimension two cycle $\gamma\equiv\gamma(G_4)$ is specified
by the flux \cite{Grimm:2009ef}. 
Thus, by the matching \eqref{eq:supermatch} a similar integrality structure is predicted for 
the heterotic superpotentials in geometric phases of their parameter spaces.  For superpotentials from  five-branes wrapped
on a curve $\Sigma$  this matches naturally the disk multi-covering formula \eqref{eq:wopgromovwitten}
of \cite{Ooguri:1999bv}, that encodes the disk instantons ending on special Lagrangians $L$ mirror dual to $\Sigma$. 
It would be interesting to explore a generalization of this integral structure to 
the vector bundle sector of the holomorphic Chern-Simons functional of the heterotic theory.

Finally, we conclude by a geometric way to identify the flux $G^{\rm M5}_4$ which corresponds to the
five-brane superpotential given by the chain integral $\int_{\Gamma} \Omega$. The three-chain ${\Gamma}$ can be
mapped to a three-chain ${\Gamma}$ in the F-theory base $B_3$ whose boundary two-cycles $\Sigma$ lie in
the worldvolume of a seven-brane over which  the cycles of the F-theory
elliptic fiber degenerate.  By fibering the one-cycle of the elliptic fiber which
vanishes at the seven-brane locus over $\Gamma$, one obtains a transcendental
cycle in $H_4(X_4,\mathds{Z})$. Its dual form lies then in the horizontal part
$H^4_H(X_4,\mathds{Z})$ and therefore yields the sought-after flux $G_4$, see
\cite{Denef:2008wq} for a review on such constructions.  For a recent
very explicit construction of these cycles in F-theory compactifications on
elliptic K3 surfaces and Calabi-Yau threefolds see~\cite{Braun:2008ua,Braun:2009bh}.

\subsection{Example 1: Horizontal Five-Branes in $\P^4(1,1,1,6,9)[18]$}
\label{sec:Example1}

We begin the discussion of a first example of heterotic/F-theory dual four-dimensional 
theories. First we construct the setup on the heterotic side and then infer the F-theory
fourfold by duality.

Following section \ref{sec:hetM} the heterotic theory is specified by an elliptic Calabi-Yau threefold with a stable 
holomorphic vector bundle $E=E_1\oplus E_2$ obeying the heterotic anomaly constraint \eqref{eq:anomaly}.
We choose the threefold $\tilde{Z}_3$ as the elliptic fibration over the base $B_2=\P^2$ with 
generic torus fiber $\P_{1,2,3}[6]$, which is precisely the example \eqref{eq:3foldellp2} considered
in section \ref{sec:ffConstruction}\footnote{In order to ensure consistency of our notation we thus denote
the heterotic threefold by $\tilde{Z}_3$.}. However, since we are studying $\tilde{Z}_3$ in the context of 
heterotic/F-theory duality where we explicitly analyze the duality relation \eqref{eq:modulimap} 
of the moduli on both sides, in particular including heterotic horizontal five-branes, we directly study the 
K\"ahler geometry of $\tilde{Z}_3$. 

We recall that $\tilde{Z}_3$ is given as the hypersurface $p_0=0$ 
in the toric ambient space given in \eqref{eq:3foldellp2} with the class of the hypersurface $\tilde{Z}_3$ given by 
\begin{equation} \label{eq:classZ_3}
	[\tilde{Z}_3]=\sum D_i=6B+18H\,.
\end{equation} 
Here we denoted the two independent toric divisor classes $D_i$ by $H=D_2$ and $B=D_1$, that are the pullback of the hyperplane 
class of the $\mathds{P}^2$ base respectively the class of the base itself. They are identified with
the K\"ahler cone generators $J_1$, $J_2$ in section \ref{sec:ffConstruction} as $J_1\equiv H$ and $J_2\equiv 3H+B$ 
so that the intersections of $H$ and $B$ can be readily read off from \eqref{eq:intersectionsY}.
From the toric data the basic topological numbers of $Z_3$ are obtained as
\begin{equation}
	\chi(\tilde{Z}_3)=540\ ,\quad h^{(1,1)}(\tilde{Z}_3)=2\ ,\quad h^{(2,1)}(\tilde{Z}_3)=272\ .
	\label{eq:top-data-tfp2}
\end{equation}
We note that this little amount of K\"ahler classes allows for direct calculations in the K\"ahler 
sector, whereas a direct calculation in the complex structure sector seems technically impossible.
The second Chern-class of $\tilde{Z}_3$ is in general given in terms of the Chern classes $c_1(B_2)$, $c_2(B_2)$ 
and the section $\sigma:B_2\rightarrow \tilde{Z}_3$ of the elliptic fibration by the formula
$c_2(\tilde{Z}_3)=12 c_1(B_2)\sigma+11 c_1(B_2)^2+c_2(B_2)$.
To see this we refer to section \ref{sec:spectralcover} and recall that the toric variety \eqref{eq:3foldellp2} is in the case
at hand is the projective bundle $\P_{\tilde{\Delta}}=\P(\mathcal{O}\oplus K_{B_2}^{-2}\oplus K_{B_2}^{-3})$.
We further identify $\sigma=B$, $B_2=\P^2$ and thus obtain, using $K_{\P^2}=\mathcal{O}_{\P^2}(-3H)$ and $c_2(\P^2)=3H^2$, 
\begin{equation}
 	c_2(\tilde{Z}_3)=36H\cdot B+102 H^2\,.
\end{equation}

Next, in order to satisfy the heterotic anomaly formula \eqref{eq:anomaly}, we 
have to construct the heterotic vector bundle $E_1 \oplus E_2$ and 
to compute the characteristic classes $\lambda(E_i)$.  Since $\tilde{Z}_3$ is elliptically fibered the classes 
$\lambda(E_i)$ can be constructed using the basic methods of \cite{Friedman:1997yq} that were briefly 
reviewed in section \ref{sec:spectralcover}. We restrict $E_1\oplus E_2$ to be an $E_8\times E_8$ bundle over $Z_3$ 
and choose the classes $\eta_1,\eta_2\in H^{2}(B_2,\Z)$ in \eqref{eq:c2E}  as 
$\eta_1 = \eta_2 = 6c_1(B_2)$. Then, we use formula \eqref{eq:lambda-e8} for the second Chern class of $E_8$-bundles 
to obtain 
\begin{equation}
	\lambda(E_1)=\lambda(E_2)=18 H\cdot B-360 H^2\,.
\label{eq:secondChernclassP^2}
\end{equation}
The anomaly condition \eqref{eq:anomaly} then leads to conditions on the coefficients of the independent 
classes in $H^4(\tilde{Z}_3)$. For the class $H\cdot B$ contributed by the base via 
$\sigma\cdot H^2(\mathds{P}^2,\Z)$ the anomaly is trivially satisfied by the $\lambda(E_i)$ in \eqref{eq:secondChernclassP^2}. 
This implies that no horizontal five-branes are present. For the class of the fiber $F$ the anomaly 
forces the inclusion of vertical five-branes in the class 
\begin{equation}
	[\Sigma]=c_2(B_2)+91c_1(B_2)^2=822 H^2 \equiv n_f F\,.
\label{eq:verticalfivebranes}
\end{equation}
Since $F$ is dual to the base $B_2$ the number $n_f$ of vertical branes is determined by integrating 
$\Sigma$ over $\mathds{P}^2$ as 
\begin{equation}
	n_f=\int_{\mathds{P}^2}\Sigma= \Sigma \cdot B=822.
	\label{eq:nf}
\end{equation}
To conclude the heterotic side we compute the index $I(E_i)$ since it appears
in the identification of moduli \eqref{eq:modulimap} and thus is crucial for
the analysis of heterotic/F-theory duality. We readily use the formula
\eqref{eq:index} to obtain that $I(E_1)=I(E_2)=8+4\cdot 360+18\cdot3=1502$.   

Next we include horizontal five-branes to the setup by shifting the classes $\eta_i$ 
appropriately \cite{Rajesh:1998ik}. We achieve this by putting $\eta_2=6c_1(B)-H$. 
The class of the five-brane $\Sigma$ can then be determined analogous to the above 
discussion by evaluating \eqref{eq:lambda-e8} and imposing the anomaly \eqref{eq:anomaly}. 
It takes the form
\begin{equation}
 	[\Sigma]=91 c_1(B_2)^2 + c_2(B_2) - 45 c_1(B_2)\cdot H + 15 H^2 + H\cdot B=702 H^2+H\cdot B\,,
\end{equation}
which means that we have to include five-branes in the base on a curve $\Sigma$ in 
the class $H$ of the hyperplane of $\mathds{P}^2$. Additionally the number of five-branes 
on the fiber $F$ is altered to $n_f=702$. Accordingly, the shifting of $\eta_2$ changes 
the second index to $I_2=1019$, whereas $I_1=1502$ remains unchanged.

Let us now turn to the dual F-theory description. We first construct the
fourfold $X_4$ dual to the heterotic setup with no five-branes. In this case
the base $B_3$ of the elliptic fibration of the fourfold $X_4$ is $B_3=\P^1\times \P^2$.
This can be seen from the relation \eqref{eq:etaE8} of the classes $\eta_i$ and
the fibration structure of $B_3=\P(\mathcal{O}_{B_2}\oplus L)$ for $E_8$-bundles. Since both 
classes $\eta_i$ equal
$6c_1(B_2)$ we have $t=0$ and thus the bundle $L=\mathcal{O}_{\mathds{P}^2} $ is
trivial as well as the projective bundle
$B_3=\mathds{P}(\mathcal{O}_{\mathds{P}^2}\oplus\mathcal{O}_{\mathds{P}^2})$.
Then the fourfold $X_4$ is constructed as the elliptic fibration over $B_3$
with generic fiber given by $\mathds{P}_{1,2,3}[6]$.  Again $X_4$ is described
as a hypersurface in a five-dimensional toric ambient space $\P_{\Delta}^5$ as described
by the toric data in \eqref{eq:fourfold1} if one drops the point
$(3,2,-1,0,1)$ and sets the divisor $D$ to zero. The class of $X_4$ is then
given by 
\begin{equation}
	\left[X_4\right]= \sum_i D_i = 6 B + 18 H + 12 K\ , 
\label{eq:X4}
\end{equation}
where the independent divisors are the base $B_3$ denoted $B$, the pullback
of the hyperplane $H$ in $\mathds{P}^2$ and of the hyperplane $K$ in
$\mathds{P}^1$.  Then, the basic topological data reads  
\begin{equation}
	\chi(X_4)=19728\quad,\quad h^{(1,1)}(X_4)=3\quad ,\quad h^{(3,1)}(X_4)=3277\quad ,\quad h^{(2,1)}(X_4)=0.
	\label{eq:top-data-ffp2}
\end{equation}

Now we have everything at hand to discuss heterotic/F-theory duality along the
lines of section \ref{sec:F_blowup}, in particular the map of moduli
\eqref{eq:modulimap}. As discussed there, the complex structure moduli of the
F-theory fourfold are expected to contain the complex structure moduli of $\tilde{Z}_3$
on the heterotic side as well as the bundle and brane moduli of possible
horizontal five-branes. Indeed we obtain a complete matching by adding up all
contributions in \eqref{eq:modulimap},
\begin{equation}
	h^{(3,1)}(X_4)=3277=272+1502+1502+1\,,
\end{equation}
where it is crucial that no horizontal five-branes with possible brane moduli
are are present. 

Finally, to obtain the F-theory dual of the heterotic theory with horizontal
five-branes, we apply the recipe discussed in section \ref{sec:F_blowup}.
We have to perform the described geometric transition of first tuning the
complex structure moduli of the fourfold $X_4$ such that it becomes singular
over the five-brane curve $\Sigma$ which we then blow up into a divisor $E$. This way we
obtain a new smooth Calabi-Yau fourfold denoted by $\hat X_4$.  The toric data
of this fourfold are given by
\begin{equation}
	\Delta_5^{\hat X}=\left(
	\begin{array}{ccccc|c|c}
		-1 &  0 &  0  & 0  & 0    & 3D+3B+9H+6K & D_1\\
		0  & -1 &  0  & 0  & 0    & 2D+2B+6H+4K & D_2\\
		3  &  2 &  0  & 0  & 0    & B & D_3\\
		3  &  2 &  1  & 1  & 0    & H & D_4\\
		3  &  2 &  -1 & 0  & 0    & H-E & D_5\\
		3  &  2 &  0  & -1 & 0    & H & D_5\\
		3  &  2 &  0  & 0  & 1    & K & D_7\\
		3  &  2 &  0  & 0  & -1    & K+E & D_8\\ 
		3  &  2 &  -1  & 0  & 1   & E & D_9\\
	\end{array}
	\right),
	\label{eq:fourfold1}
\end{equation}
which agrees with \eqref{eq:vl-for-p2} up to a similarity transformation. Here we included 
the last point $(3,2,-1,0,1)$ and a corresponding divisor $D_9=E$ to perform the blow-up 
along the curve $\Sigma$. 

To demonstrate that \eqref{eq:fourfold1} indeed describes a blow-up of $X_4$ can be seen as follows. 
Since the curve $\Sigma$ in the heterotic theory is in the 
class $H$ we have to blow-up over the hyperplane class of $B_2=\mathds{P}^2$ in $B_3$. 
First we project the polyhedron $\Delta_5^{\hat{X}}$ to the base $B_3$ which is done just by
omitting the first and second column in \eqref{eq:fourfold1}. Then the last
point maps to the point $(-1,0,1)$ that subdivides the two-dimensional cone
spanned by $(-1,0,0)$ and $(0,0,1)$ in the polyhedron of $B_3$. This two-dimensional cone,
however, corresponds precisely to the curve $\Sigma=H$. Thus, upon
adding the point $(-1,0,1)$ the curve $\Sigma$ in $B_2$ is
removed from $B_3$ and replaced by the divisor $E$ corresponding to the new
point. Thus, we see that the toric data \eqref{eq:fourfold1} contain the
blow-up of $\Sigma$ in the base $B_3$ in the last three columns.

The fourfold $\hat{X}_4$ is then realized as a generic constraint $P = 0$ in the class
\begin{equation}
  [\hat{X}_4] = 6 B + 18 H + 12 K + 6 E\ .
\end{equation}  
Note that this fourfold has now three different triangulations which correspond to 
the various five-brane phases on the dual heterotic side.
The topological data for the new fourfold $\hat X_4$ is given by
\begin{equation}
	\chi(\hat X_4)=16848\,,\quad\quad h^{(1,1)}(\hat X_4)=4\,,\quad\quad h^{(3,1)}(\hat X_4)=2796\,,\quad
	\quad h^{(2,1)}(\hat X_4)=0\ ,
	\label{eq:xhat-top-data}
\end{equation}
where the number of complex structure moduli has reduced in the transition as
expected.

If we now analyze the map \eqref{eq:modulimap} of moduli in heterotic/F-theory
duality we observe that we have to put $h^0(\Sigma,N_{Z_3}\Sigma)=2$
in order to obtain a matching.  This implies, from the point of view of
heterotic/F-theory duality, that the horizontal five-brane wrapped on $\Sigma$ has
to have two deformation moduli. Indeed, this precisely matches the fact that
the hyperplane class of $\P^2$ has two deformations since a general hyperplane
is given by the linear constraint $a_1x_1+a_2x_2+a_3x_3=0$ in the three
homogeneous coordinates $\underline{x}$ of $\mathds{P}^2$. Upon the overall scaling it
thus has two moduli parameterized a $\mathds{P}^2$ with homogeneous
coordinates $\underline{a}$. 

This way we have found an explicit construction of an
F-theory fourfold $\hat{X}_4$ with complex structure moduli encoding the dynamics of
a horizontal heterotic five-brane.
Unfortunately, we are not able to check this matching further for example by calculation of
a brane superpotential from F-theory as done in section \ref{sec:ApplicationsMirrorSymmetryToF} 
for different examples. However, this is not a conceptual problem but merely a technical difficulty 
due to the large number of complex structure moduli of the fourfold $\hat{X}_4$ which makes the 
determination of the solutions to the Picard-Fuchs equations very hard. It would be interesting, 
however, to extract the F-theory superpotential for a subsector of the moduli including the two brane
deformations.\footnote{If one considers exactly the mirror of $\hat X_4$, as we
will in fact do in section \ref{sec:Example2}, it might be possible to embed this
reduced deformation problem into the complicated deformation problem of $\hat
X_4$ constructed in this section.} In the next section \ref{sec:Example2} we take a different 
route and consider an example with only a few complex structure moduli that is
constructed by using mirror symmetry. However in these cases the calculation of
the indices $I(E_i)$ becomes very hard due to a big number of K\"ahler moduli.

We conclude by noting that we will provide further evidence for the identification
of the complex structure moduli space of $\hat{X}_4$ with the brane moduli space in 
section \ref{sec:non-CYblowup}. Indeed, we
will show there that one can also construct $\hat X_4$ as a complete intersection starting with a heterotic 
non-Calabi-Yau threefold obtained by blowing up along the curve $\Sigma$ in $\tilde{Z}_3$. This exploits already on
the heterotic side
the map of deformations of $\Sigma$ to complex structure moduli of the blow-up threefold under the blow-up procedure discussed in chapter \ref{ch:blowup}.

\subsection{Example 2: Five-Brane Superpotential in Heterotic/F-Theory}
\label{sec:Example2}

Let us now discuss a second example for which the F-theory flux superpotential 
can be computed explicitly since the F-theory fourfold 
admits only few complex structure moduli.
Clearly, using mirror symmetry such fourfolds can be obtained 
as mirror manifolds of examples with few K\"ahler moduli.

To start with, let us consider heterotic string theory 
on the \textit{mirror} of the Calabi-Yau threefold $\tilde{Z}_3$ which we 
studied in the last section \ref{sec:Example1}. This mirror is 
the heterotic manifold $Z_3$ that we studied already in section \ref{sec:ffConstruction} 
in the context of mirror symmetry on fourfolds. 
As we noted there using the methods of \cite{Avram:1996pj} $Z_3$ is also 
elliptically fibered, such that it has an F-theory dual description and that bundles
can at least in principle be constructed explicitly using the spectral cover construction
of section \ref{sec:spectralcover}. The polyhedron of $Z_3$ is the dual polyhedron to 
\eqref{eq:3foldellp2} presented in \eqref{eq:mirror3foldellp2} and we recall its Weierstrass form,
\begin{equation}
	p_0=x^3+y^2+x y \tilde{z} a_0 u_1 u_2 u_3+\tilde{z}^6( a_1 u_1^{18}+ a_2 u_2^{18}+ a_3 u_3^{18}+ a_4 u_1^6 u_2^6 u_3^6).
	\label{eq:exp2-het-threefold}
\end{equation}
The coordinates $\underline{u}$ are the homogeneous coordinates of the twofold base $B_2$ and
$(x,y,\tilde{z})$ denote the homogeneous coordinate of $\P^2(1,2,3)$. 
Note that one finds that the elliptic fibration of $Z_3$ is highly degenerate over $B_2$. 
The threefold is nevertheless non-singular 
since the singularities are blown up by many divisors in the toric ambient space of $Z_3$. 
However, in writing \eqref{eq:exp2-het-threefold} many of the coordinates parameterizing 
these additional divisors have been set to one\footnote{The blow-down of these 
divisors induces a large non-perturbative gauge group in the heterotic string.}.
Turning to the perturbative gauge bundle $E_1 \oplus E_2$ we restrict in 
the following to the simplest bundle $SU(1)\times SU(1)$ which thus preserves the 
full perturbative $E_8 \times E_8$ gauge symmetry in four dimensions. 

To nevertheless satisfy the anomaly condition \eqref{eq:anomaly} one has to include 
five-branes. In particular, we consider a five-brane in  
$Z_3$ given by the equations
\begin{equation}
	h_1 := b_1u_1^{18}+b_2u_1^6u_2^6u_3^6=0\ , \qquad h_2 := \tilde{z} =0\ .
	\label{eqn:exp2-brane-heterotic}
\end{equation}
The curve $\Sigma$ wrapped by the five-brane is thus in the base $B_2$ of $Z_3$ and horizontal.
Unfortunately, it is hard to check the heterotic anomaly \eqref{eq:anomaly} explicitly as in the 
example of section \ref{sec:Example1} since there are too many K\"ahler classes 
in $Z_3$. 

However, one can proceed to construct the associated 
Calabi-Yau fourfold $X_4$ which encodes a consistent completion of the setup by duality.
The associated fourfold $X_4$ cannot be constructed as it was done in section~\ref{sec:Example1}. 
However, one can follow the strategy of \cite{Berglund:1998ej} summarized in diagram \eqref{eq:dualityDiagramm} to 
construct heterotic/F-theory dual geometries by employing mirror symmetry to first obtain the mirror fourfold 
$\tilde{X}_4$ of $X_4$ as the Calabi-Yau threefold fibration with generic fiber $\tilde{Z}_3$
being the mirror of the heterotic threefold $Z_3$.
This then naturally leads us to identify $X_4$ as the mirror to the fourfold \eqref{eq:fourfold1}
in section~\ref{sec:Example1}. This fourfold is also the main example discussed in 
detail in section \ref{sec:Superpots+MirrorSymmetry} and in \cite{Grimm:2009ef}.

In the following we check that this is indeed the correct identification by using the 
formalism of \cite{Berglund:1998ej,Diaconescu:1999it}.
The Weierstrass form of $X_4$ can be computed using the polyhedron \eqref{eq:Delta5P2} that is dual to 
\eqref{eq:fourfold1} yielding
\begin{equation} \label{eq:weierstrass-x4}
	P=y^2+ x^3+m_1 (\underline{u},\underline{w},\underline{k}) x y z+	m_6 (\underline{u},\underline{w},\underline{k}) z^6=0\,,
\end{equation}
where we abbreviated
\begin{eqnarray}  \label{eq:def-m1-m6}
	m_1(\underline{w},\underline{u},\underline{k})&=&a_{0} u_1 u_2 u_3 w_1 w_2 w_3 w_4 w_5 w_6 k_1 k_2\,,\\ 
\nonumber 	m_6 (\underline{w},\underline{u},\underline{k})&=&\phantom{+} a_1 (k_1 k_2)^6 u_1^{18} w_1^{18} w_2^{18} w_5^6 w_6^6+a_2 (k_1 k_2)^6 u_2^{18} w_3^{18} w_5^{12}\\
\nonumber 	&\phantom=&+a_3 (k_1 k_2)^6 u_3^{18} w_4^{18} w_6^{12}+ a_4 (k_1 k_2)^6 (u_1 u_2 u_3 w_1 w_2 w_3 w_4 w_5 w_6)^6\\
\nonumber	&\phantom=&+ b_1 k_2^{12} u_1^{18} w_1^{24} w_2^{12} w_3^6 w_4^6+b_2 k_2^{12} (u_1u_2u_3)^6 (w_1 w_3 w_4)^{12}\\
\nonumber	&\phantom =&+c_1 k_1^{12} (u_1 u_2 u_3)^6 (w_2 w_5 w_6)^{12}.
\end{eqnarray}
The coordinates $\underline{u}$ are the coordinates of the twofold base $B_2$ as before
and $\underline{w},k_1,k_2$ are the additional coordinates of the threefold base $B_3$\footnote{We note that
we slightly changed our labeling of the coordinates compared to \eqref{eq:preWeier}, \eqref{eq:def-tildem1}.}. 
Again, note that we have set many coordinates to one to display $P$. The chosen
coordinates correspond to divisors which include the vertices of $\Delta_5^X$
and hence determine the polyhedron completely.  In particular, one finds that
$(k_1,k_2)$ are the homogeneous coordinates of the $\P^1$-fiber over $B_2$.  The coefficients
$\underline{a},\underline{b},c_1$ encode the complex structure
deformations of $X_4$.  However, since $h^{(3,1)}(\hat X_4)=4$, there are
only four complex structure parameters rendering six of them redundant.  

As
the first check that $X_4$ is indeed the correct dual Calabi-Yau fourfold, we follow
the discussion of section \ref{sec:het-Fdual} and use the
stable degeneration limit \cite{Friedman:1997yq} and write $P$ in a local
patch with appropriate coordinate redefinition as  \cite{Berglund:1998ej}
\begin{equation}
	P=p_0+p_++p_-\, ,
\end{equation}
where
\begin{eqnarray} \label{eq:def-p0p+p-}
	 p_0&=& x^3+y^2+x y \tilde{z} a_0 u_1 u_2 u_3+\tilde{z}^6\left( a_1 u_1^{18}+ a_2 u_2^{18}+ a_{3} u_3^{18} + a_4 u_1^6 u_2^6 u_3^6\right)\ ,\\
	\nn p_+&=& v\tilde{z}^6 \left(b_1 u_1^{18}+b_2 u_1^6 u_2^6 u_3^6\right)\ ,\\
	\nn p_-&=& v^{-1}\tilde{z}^6 c_1 u_1^6 u_2^6 u_3^6\ .
\end{eqnarray}
The coordinate $v$ is the affine coordinate of the fiber $\P^1$.  In the stable
degeneration limit $\{p_0=0\}$ describes the Calabi-Yau threefold of the
heterotic string. In this case $p_0$ coincides with the constraint
\eqref{eq:Weierst3fold} of the threefold geometry $Z_3$ as discussed in section \ref{sec:ffConstruction}  
which directly identifies it as the dual heterotic Calabi-Yau threefold to $X_4$.  This shows in particular
that the geometric moduli of the heterotic compactification on $Z_3$ are correctly embedded in $X_4$.  
The
polynomials $p_\pm$ encode the perturbative bundles, and the explicit form
\eqref{eq:def-p0p+p-} shows that we are dealing with a trivial $SU(1)\times SU(1)$ bundle, which is
also directly checked by analyzing the polyhedron of $\hat X_4$
using the methods of \cite{Bershadsky:1996nh,Candelas:1996su,Candelas:1997eh}.  Indeed, one shows explicitly that over each
divisor $k_i=0$ in $B_3$ a full $E_8$ gauge group is realized. Since the full
$E_8 \times E_8$ gauge symmetry is preserved we are precisely in the
situation of section \ref{sec:F_blowup}, where we recalled from
\cite{Diaconescu:1999it} that a smooth $X_4$ has to contain a blow-up
corresponding to a horizontal heterotic five-brane.  We now check that this allows us
to identify the heterotic five-brane and its moduli explicitly in the duality to F-theory.

Let us begin by making contact to the formalism of section \ref{sec:F_blowup}.  First, to make the perturbative
$E_8\times E_8$ gauge group visible in the Weierstrass equation
\eqref{eq:weierstrass-x4}, we have to include new coordinates $(\tilde
k_1,\tilde k_2)$ replacing $(k_1,k_2)$. This can be again understood by
analyzing the toric data using the methods of
\cite{Candelas:1996su,Candelas:1997eh,Bershadsky:1996nh}.  We denote by $(3,2,\vec{\mu})$ the
toric coordinates of the divisor corresponding to $\tilde k_1$ in the
Weierstrass model. Then the resolved $E_8$ singularity corresponds to the
points\footnote{Note that we have chosen the vertices in the $\P_{1,2,3}[6]$ to
be $(-1,0),(0,-1),(3,2)$ to match the discussion in
refs.~\cite{Bershadsky:1996nh,Candelas:1996su,Candelas:1997eh}. However, if one explicitly
analyses the polyhedron of $X_4$ one finds that one has to apply a
$Gl(2,\Z)$ transformation to find a perfect match. This is due to the fact that
$X_4$, in comparison to its mirror $\tilde{X}_4$, actually contains the
dual torus as elliptic fiber.}
\begin{eqnarray}
	&(3,2,n \vec{\mu}),\ n=1,...,6\quad ,\quad (2,1,n \vec{\mu}),\ n=1,...,4\ ,&\\
	\nn&(1,1,n \vec{\mu}),\ n=1,2,3\quad ,\quad (1,0,n \vec{\mu}),\ n=1,2\quad,\quad (0,0,\vec{\mu})&
\end{eqnarray}
While $(3,2,6 \vec{\mu})$ corresponding to $k_1$ is a vertex of the
polyhedron, $(3,2,\vec{\mu})$ corresponding to $\tilde k_1$ is an inner point.
Using the inner point for $\tilde k_1$, the Weierstrass form $P$ changes
slightly, while the polynomials $p_0,p_+$ and $p_-$ can still be identified in
the stable degeneration limit.  

Next, to determine $g_5$ in \eqref{eq:brane-g5}, we
compute $g$ of the Weierstrass form in a local patch where $\tilde k_2=1$
\begin{equation}
	g=\tilde k_1^5 \left(b_1 u_1^{18}+b_2 u_1^6 u_2^6 u_3^6+ \tilde k_1
\left(a_1 u_1^{18}+a_2 u_2^{18}+\dots\right)\right)\ .
\end{equation}
The dots contain only terms of order zero or higher in $\tilde k_1$.
Comparing this with \eqref{eq:f'g'}, we note that the Calabi-Yau fourfold $X_4$ can be understood
as a blow-up\footnote{This blow-up can
be equivalently described as a complete intersection as we will discuss in 
sections \ref{sec:5braneblowupsanddefs}. A simple example of such a construction will be 
presented in section \ref{sec:non-CYblowup}.} along the curve $\tilde k_1=g_5=0$ in the base of $X_4$, where $g_5$ is given by
\begin{equation}
	g_5=b_1 u_1^{18}+b_2 u_1^6 u_2^6 u_3^6.
	\label{eqn:exp2-g5}
\end{equation}
This identifies $\{g_5=0\}$ with the curve of the five-brane $\Sigma$ in the base $B_2$
of $Z_3$, which is in perfect accord with \eqref{eqn:exp2-brane-heterotic}. Thus, one concludes
that $X_4$ is indeed a correct fourfold associated to $Z_3$ with the given horizontal
five-brane.  As we can see from \eqref{eqn:exp2-g5}, the five-brane has one
modulus.  If we compare $g_5$ with $p_+$, we see that $p_+=v\tilde{z}^6g_5$.  This
nicely fits with the bundle description of a five-brane as a small instanton.  
In fact, in our configuration, $p_+$ and $p_-$ should describe $SU(1)$-bundle since we have the full unbroken perturbative
$E_8\times E_8$-bundle as described above.  The $SU(1)$-bundles do not have any
moduli, such that the moduli space corresponds to just one
point~\cite{Friedman:1997yq}.  In the explicit discussion of the Weierstrass
form in our setting, $p_+$ has one modulus which corresponds to the modulus of
the five-brane.  

Finally, we consider the computation of the F-theory flux superpotential that has been carried out
already in \ref{sec:ApplicationsMirrorSymmetryToF} but in a different context, see also \cite{Grimm:2009ef}. 
Here, we do not need to recall all the details. As was discussed there, the different triangulations
of $\tilde{X}_4$ correspond to different five-brane phases. It was proposed that the
four-form flux, for the two possible five-brane phases, is given by the
basis element
\begin{equation}  \label{eq:expl_flux}
  \hat \gamma_1^{(2)} = (-\theta_1^2+\tfrac{1}{2}\theta_3(\theta_1+\theta_3)+\tfrac{1}{6}\theta_4(\theta_1 +\theta_3))
\Omega_4 |_{\underline z=0} \ , 
\end{equation}
and the element \eqref{eq:braneflux2} for the other phase, 
where the $\theta_i = z_i \frac{d}{dz_i}$ are the logarithmic derivatives as introduced in 
\eqref{eq:Atobeta_map}.
The moduli $z_1,z_2$ could be identified as the deformations of the complex
structure of the heterotic threefold $Z_3$, while $z_3$ corresponded to the
deformation of the heterotic five-brane.\footnote{The deformation $z_4$
describes the change in $p_-$.} Indeed, a non-trivial check of this identification was
provided in section \ref{sec:ApplicationsMirrorSymmetryToF} and in \cite{Grimm:2009ef}, 
where it was shown that the F-theory flux superpotential \eqref{eq:fluxpotFourfoldExpanded} 
in the directions \eqref{eq:expl_flux} matches with the superpotential for a five-brane 
configuration in a local Calabi-Yau threefold $K_{\P^2}\rightarrow \P^2$ obtained by 
decompactifying $Z_3$, where the non-compact five-brane is described
as a point on a Riemann surface in the base $B_2$ of $Z_3$ as discussed in section \ref{sec:ffConstruction}. 
Using heterotic F-theory duality as in section \ref{sec:F_blowup} then implies that the
flux \eqref{eq:expl_flux} actually describes a compact heterotic five-brane setup.

\part{Blow Up Geometries and SU(3)-Structure Manifolds}

\chapter{Five-Branes and Blow-Up Geometries }
\label{ch:blowup}

In this chapter we provide, following \cite{Grimm:2008dq,Grimm:2009sy,Grimm:2010gk}, a novel
geometric description of the dynamics of five-branes via a dual non-Calabi-Yau threefold $\hat{Z}_3$.
Starting with a spacetime-filling five-brane on a curve $\Sigma$ in a Calabi-Yau threefold $Z_3$,
as it may occur in Type IIB and heterotic M-theory compactifications, the basic motivation of 
\cite{Grimm:2008dq} was to find a natural description, where both the bulk and brane fields
associated to deformations of $Z_3$ respectively $\Sigma$ are treated equally in an unified
framework. To obtain such a description a very canonical procedure was proposed in \cite{Grimm:2008dq}, 
where the pair $(Z_3,\Sigma)$ is replaced by a new threefold $\hat{Z}_3$ with a distinguished divisor
$E$, that is obtained by blowing up along $\Sigma$ in $Z_3$. It was further argued, that both the 
geometric deformations of $(Z_3,\Sigma)$ are canonically unified as complex structure deformations of 
$\hat{Z}_3$ and it was suggested to use Picard-Fuchs equations for the complex structure deformations
of $\hat{Z}_3$ to a study the open-closed deformations of $(Z_3,\Sigma)$. Ultimately, both 
the flux and five-brane superpotential should be given as solutions to these Picard-Fuchs equations.

Indeed the details of this proposal and concrete examples of five-branes in Calabi-Yau threefolds $Z_3$,
their open-closed Picard-Fuchs equations and the superpotentials have been worked out in \cite{Grimm:2010gk}.
Furthermore it has been argued to view the geometry of the blow-up $\hat{Z}_3$ not only as a tool
for calculations but as being dynamically generated by the five-brane backreaction. Evidence for its
use to define a flux compactification of the Type IIB and heterotic string that is dual to the original
five-brane setup was provided by the definition of an $SU(3)$-structure on $\hat{Z}_3$.
It is the purpose of this chapter to introduce the physical and mathematical ideas that lead to the blow-up
threefold $\hat{Z}_3$ as the natural geometry to analyze five-brane dynamics. Simultaneously this
serves as a preparation both for the calculations of the Picard-Fuchs system on $\hat{Z}_3$ as well as 
for the physical interpretation of the blow-up. The details of the calculations of the
open-closed superpotential for two concrete examples and the definition of the $SU(3)$-structure are presented
in chapter \ref{ch:CalcsBlowUp} respectively in chapter \ref{ch:su3structur}.

We begin our discussion of five-brane dynamics in $\mathcal{N}=1$ Calabi-Yau compactifications
in section \ref{sec:N=1gensection}. We emphasize that the presence of five-brane sources for the 
flux is more thoroughly treated using the notion of currents and naturally leads to the consideration
of the open manifold $Z_3-\Sigma$. In addition we discuss geometric deformations of the five-brane curve 
$\Sigma$, where we distinguish between deformations leading, even at first order, to massive or 
light fields in the four dimensional effective action. We comment on the use
of the five-brane superpotential for determining higher order obstructions of the light fields.
Next in section \ref{sec:5braneblowupsanddefs} we are naturally led to the blow-up $\hat{Z}_3$,
that we construct explicitly both locally and globally as a complete intersection. We give a detailed
explanation for the unification of the bulk and brane geometrical deformations of $Z_3$ respectively 
$\Sigma$ \textit{and} their obstructions as pure complex structure deformations on $\hat{Z}_3$ and a
specific flux element. Then in section \ref{sec:hatomega} we analyze the open-closed deformation space
by studying the variation of the pullback $\hat{\Omega}$ of the Calabi-Yau three-form 
$\Omega$ of $Z_3$ to $\hat{Z}_3$ under a change of complex structure on $\hat{Z}_3$. There we also
discuss the general structure of the open-closed Picard-Fuchs equations as obtained from a residue
integral for $\hat{\Omega}$. Finally in section \ref{sec:potentialhatZ3minusD} we present the lift
of the flux and five-brane superpotential of $(Z_3,\Sigma)$ to the blow-up $\hat{Z}_3$ and argue that
both are obtained as solutions to the same open-closed Picard-Fuchs equations on $\hat{Z}_3$.

\section{Five-Brane $\mathcal{N}=1$ Effective Dynamics} 
\label{sec:N=1gensection}

In this section we discuss basic aspects of five-brane dynamics. Our 
point of view will be geometrical and appropriate to formulate the blow-up 
proposal in section \ref{sec:5braneblowupsanddefs}. We begin our discussion
in section \ref{sec:N=1branes} with a brief summary of heterotic and Type IIB 
string compactifications with five-branes focusing on global consistency 
conditions and the use of currents to describe the localized brane sources.
We are naturally led to work on the open manifold $Z_3-\Sigma$, where all
fields in the theory are well-behaved, even in the presence of the singular brane
source. The geometrical deformations of the five-brane around the supersymmetric 
configuration specified by a holomorphic curve $\Sigma$ are discussed in section 
\ref{sec:fivebranes}, where we present a physically motivated discussion of
both light and massive fields and their behavior under complex structure deformations
of $Z_3$. A purely mathematical analysis of analytic families of holomorphic curves
provides the necessary background to appreciate the use of the brane superpotential
to determine higher order obstructions, as discussed in section 
\ref{sec:fivebranesuperpotential}.

\subsection{Five-Branes, Currents and Open Manifolds} 
\label{sec:N=1branes}

In the following we consider four-dimensional $\mathcal{N}=1$ compactifications 
of Type IIB string theory and the heterotic string, where we are particularly
interested in including spacetime-filling five-branes.
The Type IIB setups are orientifold compactifications with D5-branes and $O5$-planes on 
a three-dimensional Calabi-Yau manifold $Z_3$ that is modded out by the orientifold 
involution, cf.~chapter \ref{ch:EffActCYOrieCompact}.
In the heterotic string we consider NS5-branes and vector bundles on 
a smooth Calabi-Yau threefold $Z_3$ as reviewed in section \ref{sec:HetString+Fivebranes}. 
In both compactifications global consistency conditions restrict the choice of 
valid configurations and link the discrete data counting 
branes and fluxes via tadpole cancellation conditions.

We briefly recall how tadpole cancellation arises both in the Type IIB string and 
the heterotic string. In the presence of magnetic sources like smooth heterotic bundles 
or localized branes, the Bianchi identity of the R--R-form $F_3$ and the NS--NS B-field take the form
\begin{eqnarray} \label{eq:dF_3}
    d F_3 &=& \delta_{\Sigma} + \sum_i \delta_{\Sigma_i}-2\sum_\alpha\delta_{\tilde \Sigma_\alpha}\,,\\
     d H_3 &=& \delta_{\Sigma} + \text{tr} \cR \wedge \cR - \tfrac{1}{30} \text{Tr}\mathcal{F}\wedge \mathcal{F}
     \label{eq:dH_3het}
\end{eqnarray}
Note that the $O5$-planes carry $-2$ times the charge of a D5-brane. 
In both theories these equations imply, on the level of cohomology, global tadpole cancellation conditions.
Read as equations for actual forms they imply in both theories that the 
wavefunction of the five-brane is sharply peaked\footnote{The delta-form $\delta_\Sigma$ is the wavefunction of 
a brane in eigenstate of the position space operator.} at the curve $\Sigma$ which is reflected by the delta-function 
$\delta_\Sigma$ in the Bianchi identity. In contrast to the global 
tadpole condition that fixes only the cohomology classes, the Bianchi identity fixes, up to gauge transformations, 
the actual forms $F_3$ and $H_3$ pointwise and implies that globally defined forms $C_2$ and $B_2$ with $F_3$ and
$H_3$ do not exist\footnote{For a discussion of the global structure of $C_2$, $B_2$ in terms of  $\check{\text{C}}$ech 
de Rham complexes we refer to \cite{Freed:1998tg}.}. However, in a local patch we can solve the Bianchi identity for 
the field strength and the potential explicitly as we present in the following.
Let us note here that one crucial point that leads to the blow-up proposal below is the appropriate 
mathematical treatment of the actual forms $H_3$ or $F_3$ that become singular near the brane.

In the vicinity of a single brane source the Bianchi identity reads
\begin{equation}
	d \sigma_3=\delta_{\Sigma}\,,
\label{eq:general}
\end{equation}
where $\sigma_3$ is identified with the singular part in the field strength $H_3$, $F_3$ in both theories. 
For the setups we will consider the other 
localized sources do not interfere with the following local analysis and 
can be treated similarly. Furthermore, we will ignore the smooth part 
of the heterotic bundle $c_2(E)$. 
Then equation \eqref{eq:general} is best treated in the theory of currents, see e.g.~\cite{Griffiths:1978yf}. 
In this context $\sigma_3$ can be understood as the Poincar\'e dual of a chain $\Gamma$ in the following way. 
First we associate a functional $T_\Gamma$ to every three-chain $\Gamma$ with boundary $\partial \Gamma=\Sigma$ by two defining properties. For any smooth three- and two-form $\eta_3$, $\varphi_2$ we have 
\begin{equation} \label{eq:currentChain}
	T_{\Gamma}(\eta_3)=\int_{\Gamma}\eta_3\,,\qquad \quad dT_{\Gamma}(\varphi_2)=\int_{\Gamma}d\varphi_2=\int_\Sigma \varphi_2=T_{\Sigma}(\varphi_2)\,.
\end{equation}
Such a map from smooth forms to complex numbers is usually denoted as a current and 
is a generalization of distributions to forms. In this language \eqref{eq:currentChain} 
is usually written as $dT_\Gamma=T_{\Sigma}$. This is precisely the dual of the expression 
\eqref{eq:general} on the level of currents. Indeed we can use $\sigma_3$ to define a current $T_{\sigma_3}$ that also enjoys $dT_{\sigma_3}=T_{\Sigma}$ as follows
\begin{eqnarray}
	T_{\sigma_3}(\eta_3) &=&\int_{Z_3}\sigma_3\wedge \eta_3\,,\\
	dT_{\sigma_3}(\varphi_2) &=& \int_{Z_3}\sigma_3\wedge d\varphi_2=\int_{Z_3}\delta_{\Sigma}\wedge\varphi_2=\int_{\Sigma}\varphi_2=T_{\Sigma}(\varphi_2)\,.\nn 
\end{eqnarray}
Thus we identify $\sigma_3$ and $\delta_\Sigma$ as the Poincare dual of $\Gamma$ and $\Sigma$ 
respectively. However, both $\sigma_3$ and $\delta_{\Sigma}$ are not forms in the usual sense. 
$\delta_{\Sigma}$ fails to be a form similar to the fact that the delta-function fails to be a function. 
$\sigma_3$ is not a form on $Z_3$. Though, it is a smooth form on the open space $Z_3-\Sigma$. 
This can be seen directly in a local analysis in the fiber of the normal bundle $N_{Z_3}\Sigma$, 
that is isomorphic to $\mathds{C}^2$. Let us summarize the essential results.

On $N_{Z_3}\Sigma\vert_p\cong \mathds{R}^4$ the form $\sigma_3$ is the unique rotationally invariant form on $\mathds{R}^4-\{0\}$ that is orthogonal to $dr$ and integrates to $1$ over a three-sphere $S^3_r$ of any radius $r$. In hyperspherical coordinates we obtain
\begin{equation}
\sigma_3=\frac{1}{2\pi^2}\vol_{S^3}\,,\quad \int_{S^3_r} \vol_{S^3}=1\,. 
\end{equation}
Thus, $\sigma_3$ is ill-defined at $r=0$ where the three-sphere $S^3_r$ degenerates. Consequently, we can deal with $\sigma_3$ rigorously by working on the open manifold $\mathds{R}^4-\{0\}$ where $\sigma_3$ is a smooth form and by taking boundary contributions into account in the following way. 

Whenever we have a bulk integral over $Z_3$ we replace it by an integral over the open manifold $Z_3-\Sigma$ as \cite{Freed:1998tg}
\begin{equation}
	\int_{Z_3}\mathcal{L}:=\lim_{\epsilon\rightarrow 0}\int_{Z_3-\mathcal{U}_\epsilon^{(4)}(\Sigma)}\mathcal{L}\,.
\label{eq:Z3-C}
\end{equation}
where we substract a tubular neighborhood $\mathcal{U}_{\epsilon}^{(4)}(\Sigma)$ of radius $\epsilon$ over $\Sigma$.
All integrands are regular when evaluated on this open manifold, even those including the 
singular form $\sigma_3$ in $H_3$, $F_3$. One has to consider two cases, either $\mathcal{L}$ is 
well-behaved in $\Sigma$ and thus the limit $\epsilon\rightarrow 0$ in \eqref{eq:Z3-C} 
just gives back the integral over $Z_3$. In the other case $\mathcal{L}$ contains the 
form $\sigma_3$ and a boundary term is produced by partial integration as follows
\begin{equation}
	\lim_{\epsilon\rightarrow 0}\int_{Z_3-\mathcal{U}_\epsilon^{(4)}(\Sigma)}\sigma_3\wedge d\varphi_2=\lim_{\epsilon\rightarrow 0}\int_{S^3_\epsilon(\Sigma)}\sigma_3\wedge \varphi_2=\int_{\Sigma}\varphi_2\,,
\label{eq:boundaryTerms}
\end{equation} 
where we used in the second equality that $\sigma_3$ is locally exact, $\sigma_3=\frac{1}{2\pi^2}\vol_{S^3}$, and  integrates $\sigma_3$ to $1$ in each $S^3_\epsilon$-fiber of the sphere bundle $S^3_\epsilon(\Sigma)=\partial \mathcal{U}_\epsilon^{(4)}(\Sigma) $ over $\Sigma$.

We conclude by discussing the global structure of this construction. Since the normal bundle $N_{Z_3}\Sigma$ is in general non-trivial, we have to take into account the effects of a non-trivial connection. As worked out in \cite{Freed:1998tg} the adequate globalization of $\sigma_3$ is related to the Thom-class $e_3/2$ of the normal bundle, see e.g.~\cite{Bott:1982sy} for a reference. The Thom class is the unique closed form $de_3=0$, that is gauge invariant under the SO$(4)$ structure on $N_{Z_3}\Sigma$ and that integrates to $1$ over any fiber $S^3_r$.  The basic idea is to smooth out the localized source of the five-brane \eqref{eq:general} using a smooth bump form $d\rho$ normalized to integral $1$ with $\rho(r)=-1$ around $r\sim 0$ and $\rho(r)=0$ for $r> 2\epsilon$ such that the support $supp(d\rho)\subset ]\epsilon,2\epsilon[$. Then
\begin{equation}
d\sigma_3=d\rho\wedge e_3/2\,
\end{equation}
approaches $\delta_{\Sigma}$ when taking the limit $\epsilon\rightarrow 0$. Thus, we identify the contribution of the five-brane as $\sigma_3=-d\rho\wedge e_2^{(0)}/2$ with $e_3=de_2^{(0)}$ locally, where a possible term $\rho\, e_3$ has been discarded since $e_3$ is not well-defined at the position $r=0$ of the brane. We note further that $e_2^{(0)}$ is not a global form since it is not gauge invariant under the SO$(4)$ action on the normal bundle. Then, we obtain the global expressions for the field strength $F_3$ and $H_3$ as
\begin{equation} \label{eq:globalForms}
	F_3=\left<F_3\right>+dC_2-d\rho\wedge e_2^{(0)}/2\,,\quad H_3=\left<H_3\right>+dB_2-d\rho\wedge e_2^{(0)}/2+\omega_3\,,
\end{equation}
where $\omega_3=\omega_3^{\rm{L}}-\omega_3^{\rm{G}}$ denotes the Chern-Simons form for $\text{tr} R^2-\frac{1}{30}\text{Tr} F^2$ and  $\left<F_3\right>$, $\left<H_3\right>$ are background fluxes in $H^3(Z_3,\mathds{Z})$. With these formulas at hand we immediately check that the reasoning of \eqref{eq:Z3-C} and the localization \eqref{eq:boundaryTerms} to the boundary of the open manifold $Z_3-\Sigma$ applies globally. Furthermore, the expansion \eqref{eq:globalForms} formally unifies the superpotentials as we will discuss in detail below in section \ref{sec:fivebranesuperpotential}.

Finally, we note that \eqref{eq:globalForms} implies that $C_2$ respectively $B_2$ have an anomalous transformation under the SO$(4)$ gauge transformations of $N_{Z_3}\Sigma$. This is necessary to compensate the anomalous transformation $\delta e_2^{(0)}$ so that $F_3$ respectively $H_3$ are gauge invariant. This anomalous transformation plays a crucial role for anomaly cancellation in the presence of five-branes \cite{Freed:1998tg}.

\subsection{Deformations and Supersymmetry Conditions}
\label{sec:fivebranes}

In this section we discuss the massive and light fields associated to geometric deformations 
of a five-brane. In many situations there is a superpotential for the light fields that 
obstructs deforming the brane at higher order. As a preparation of the following discussion, 
we refer the reader to the unobstructed case of the complex structure moduli of Calabi-Yau 
threefolds $Z_3$ as reviewed in section \ref{sec:CSModuliSpace+PFO}. This serves as an 
introduction to the necessary concepts that similarly apply in the more complicated case of 
brane deformations and superpotentials presented here.

\subsubsection{Brane Deformations I: Infinitesimal Deformations of Holomorphic Curves}
\label{branedeformationsI}

In the following we consider a five-brane wrapped on a curve $\Sigma$ in a 
given Calabi-Yau background $Z_3$. The five-brane preserves $\mathcal{N}=1$ 
supersymmetry if $\Sigma$ is a holomorphic curve, cf.~section \ref{sec:DbranesinCY3Orie}
for the Type II analysis. A holomorphic curve can be specified as curve of minimal 
volume in its homology class. In the language of calibrations this condition reads
\begin{equation} \label{eq:vol=J}
   \vol_{\Sigma }= J|_{\Sigma}\ 
\end{equation}
using the calibration by the K\"ahler form $J$ on $Z_3$. In the effective
four-dimensional theory the volume of the wrapped curve contributes terms to the 
scalar potential. However, the leading term for holomorphic curves is canceled, in 
a consistent compactification obeying \eqref{eq:dF_3} or \eqref{eq:dH_3het}, by 
contributions from the supersymmetric $O5$-planes in 
Type IIB, as demonstrated explicitly for the example of a D5-brane in 
section \ref{sec:D5branes}, or by bundle and curvature 
contributions in the heterotic string. 
Thus, this part of the vacuum energy cancels which is a necessary condition for 
supersymmetry. Indeed, this is easily seen, for example, in orientifold setups. As we 
have seen in chapter \ref{ch:EffActCYOrieCompact} the orientifold compactification 
preserve $\mathcal{N}=1$ supersymmetry in the effective theory if the geometric part of 
the orientifold projection is a holomorphic and 
isometric involution $\sigma$ acting on $Z_3$. Hence, the $O5$-planes, being the 
fix-point set of $\sigma$, wrap holomorphic curves inside $Z_3$ and are also 
calibrated with respect to $J$. Thus, they contribute the same potential in the 
vacuum but with opposite sign as argued in section \ref{sec:D5branes} \cite{Grimm:2008dq}.

\begin{figure}[htb]
\begin{center}
\includegraphics[width=.5\textwidth]{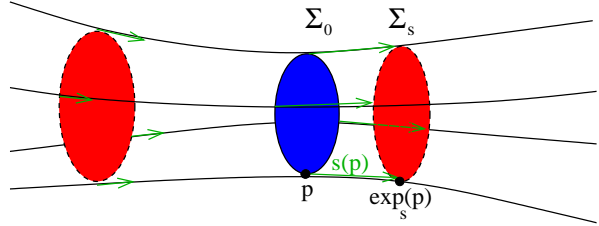}
\begin{quote}
\caption{Deformations of $\Sigma_0$ along geodesics in $Z_3$. The tangent vector $s(p)$
along the geodesic passing $p$ is a normal vector of $\Sigma_0$ at $p$.\vspace{-1.2cm}} 
\label{geodesicdef}
\end{quote}
\end{center}
\end{figure}

Let us now consider a general fluctuation of the supersymmetric $\Sigma \equiv \Sigma_0$ 
to a nearby curve $\Sigma_{s}$. From the above one expects the generation of a positive 
potential when deforming $\Sigma$ non-holomorphically. A deformation is described by a 
complex section $s$ of the normal bundle $N_{Z_3} \Sigma \equiv N_{Z_3}^{1,0} \Sigma$.
The split of the complexified normal bundle has been performed in a background complex 
structure of $Z_3$. Clearly, the space of such sections is infinite dimensional as is the 
space of all $\Sigma_{s}$. To make the identification between $\Sigma_{s}$ and $s$ more 
explicit, one recalls that in a sufficiently small neighborhood of $\Sigma_0$ the 
exponential map $exp_s$ is a diffeomorphism of $\Sigma_0$ onto $\Sigma_s$. Roughly 
speaking, one has to consider geodesics through each point $p$ on $\Sigma_0$ with 
tangent $s(p)$ and move this point along the geodesic for a distance of $||s||$ to 
obtain the nearby curve $\Sigma_s$ as depicted in figure \ref{geodesicdef}.
That the holomorphic curve $\Sigma_0$ is of minimal volume can now be seen 
infinitesimally. For any normal deformation $\Sigma_{\epsilon s}$ to $\Sigma_0$, 
i.e.~a deformation with infinitesimal displacement $\epsilon s$,
the volume increases quadratically \cite{simons1968minimal,mclean1998deformations} as
\begin{equation} \label{eq:volvar}
 	\left.\frac{d^2}{d \epsilon^2}\text{Vol}(\Sigma_{\epsilon s})\right\vert_{\epsilon=0}
 	=\frac12\int_{\Sigma}\norm{\bar{\partial} s}^2\, \vol_{\Sigma}\,,
\end{equation}
where $\norm{\bar{\partial} s}^2$ denotes the contraction of indices, both on the curve as 
well as on its normal bundle $N_{Z_3}\Sigma$ via the metric. In the effective action, 
\eqref{eq:volvar} is the leading F-term potential when deforming the D5-brane curve 
non-holomorphically as we derived explicitly in section \ref{sec:extensionInfinite}. 
A quadratic term of the form \eqref{eq:volvar} in the scalar potential implies 
that the four-dimensional fields corresponding to these non-holomorphic deformations $s$ 
acquire masses given by the value of the integral \eqref{eq:volvar}\footnote{The first 
variation of $\vol(\Sigma_u)$ vanishes by the First Cousin Principle 
\cite{mclean1998deformations}.}. When integrating out the massive deformations with
$\bar \partial s \neq 0$ the remaining sections are elements of 
\begin{equation}
   H^{0}(\Sigma,N_{Z_3} \Sigma) \equiv H^{0}_{\bar \partial}(\Sigma,N^{1,0}_{Z_3} \Sigma) \ \subset \  C^\infty(N_{Z_3}\Sigma) \ .
\end{equation}
Reversely only holomorphic sections $s \in H^{0}(\Sigma,N_{Z_3} \Sigma)$ deforming $\Sigma$ 
into a nearby curve $\Sigma_{s}$ can lead to massless or light fields in the effective theory. 
It is crucial to note that even for an $s \in H^{0}(\Sigma,N_{Z_3} \Sigma)$ the deformation 
might be obstructed at higher order and hence not yield a massless deformation. The higher 
order mass terms for these deformations can be studied by computing the superpotential as we 
will discuss throughout the next sections.

Before delving into the discussion of these holomorphic deformations, let us conclude with 
a discussion of the effect of complex structure deformations on the F-term potential 
\eqref{eq:volvar} and the hierarchy of masses of fields associated to brane deformations. 
Deformations generated by non-holomorphic vector fields $s$ do not obey the 
classical equations of motion\footnote{This can be seen readily by varying \eqref{eq:volvar}
with respect to $s$.}. The main complication is that there are infinitely many such 
off-shell deformations and it would be very hard to compute their full scalar potential. 
In contrast to $\mathcal{C}^{\infty}(N_{Z_3}\Sigma)$, the space $H^0(\Sigma,N_{Z_3}\Sigma)$
is finite dimensional. However, there is a distinguished finite dimensional subset of 
$\mathcal{C}^{\infty}(N_{Z_3}\Sigma)$ that should not be integrated out in the effective action. 
This is related to the fact that the dimension of $H^0(\Sigma,N_{Z_3}\Sigma)$ is not 
a topological quantity and will generically jump when varying the complex structure of $Z_3$.
For example, this can lift some of the holomorphic deformations $s \in H^0(\Sigma,N_{Z_3}\Sigma)$ 
since the notion of a holomorphic section is changed. 
Indeed, by deforming $\bar{\partial}$ by $A$ in $H^1(Z_3,TZ_3)$ we obtain
\begin{equation}\label{eq:csinducedmass}
	\left.\frac{d^2}{d \epsilon^2} \text{Vol}(\Sigma_{\epsilon s})\right\vert_{\epsilon=0}
	=\frac12\int_{\Sigma}\norm{As}^2 \vol_{\Sigma}=
	\frac12 
	|t|^2\int_{\Sigma}\norm{A_1s}^2 \vol_{\Sigma}
	+\mathcal{O}(t^4)\, ,
\end{equation}
where $A_1$ is the first order complex structure deformation of $Z_3$ as introduced above.
Here we used that the complex structure on $\Sigma$ is induced from $Z_3$ and $s$ is 
in $H^0(N_{Z_3}\Sigma)$ in the unperturbed complex structure on $Z_3$, $\bar\partial s=0$. 
This result is clear from the point of view of the new complex structure 
$\bar{\partial}'=\bar{\partial}+A$, since $\bar\partial's=A s \neq 0$. Thus $s$ is a section 
in $\mathcal{C}^{\infty}(N_{Z_3}\Sigma)$ in the new complex structure unless $s$ is in the 
kernel of $A$. Similarly, the corresponding field acquires a mass given by 
the integral \eqref{eq:csinducedmass}. However, the main difference 
to a generic massive mode in $\mathcal{C}^{\infty}(N_{Z_3}\Sigma)$ 
with mass at the compactification scale, cf.~eq. \eqref{eq:volvar} and \eqref{eq:Ftermpotfinal}, 
is the proportionality to the square of the VEV of $t$. Consequently 
the mass of this field can be made parametrically small tuning the value of $t$. 
Thus, we can summarize our approach to identify the light fields as follows:
(1) drop an infinite set of deformations $s$ which are massive via 
\eqref{eq:volvar} at each point in the complex structure moduli space, 
(2) include any brane deformation that has vanishing \eqref{eq:volvar} 
at some point in the closed string moduli space. 
These remaining deformations are not necessarily massless at higher orders in the 
complex structure deformations, or at higher $\epsilon$ order when expanding 
$\text{Vol}(\Sigma_{\epsilon s})$. This induces a five-brane superpotential $W$ which 
can be computed using the blow-up proposal as we will show for a number of examples in 
section \ref{ch:CalcsBlowUp}.\footnote{The critical locus of $W$ will either set the VEV 
$t$ back to zero promoting $s$ to a unobstructed deformation or will leave a discrete 
set of holomorphic curves.}

\subsubsection{Brane Deformations II: Analytic Families of Holomorphic Curves}
\label{branedeformationsII}

Let us now present the standard account on deformations of holomorphic curves 
\cite{kodaira1962theorem}. The basic question in this context is, as in the case 
of complex structure deformations, whether a given infinitesimal deformation can 
be integrated. Mathematically, finite deformations are described by the existence 
of an analytic family of compact submanifolds, in our context of curves. An analytic 
family of curves is a fiber bundle over a complex base or parameter manifold $M$ 
with fibers of holomorphic curves $\Sigma_u$ in $Z_3$ over each point $u\in M$.

Given a single curve $\Sigma$ in $Z_3$ one can ask the reverse question, namely 
under which conditions does an analytic family of curves exist? The answer to this 
question was formulated by Kodaira \cite{kodaira1962theorem}. In general an analytic 
family of holomorphic curves\footnote{Kodaira considered the general case of a 
compact complex submanifold in an arbitrary complex manifold.} exists if the obstructions 
$\psi$, that are elements in $H^1(\Sigma,N_{Z_3}\Sigma)$, vanish at every order $m$, 
which is of course trivially the case if $H^1(\Sigma,N_{Z_3}\Sigma)=0$. Then 
$\underline{u}$ are coordinates of points $u$ in $M$ and a basis of holomorphic 
sections in $H^0(\Sigma_u,N_{Z_3}\Sigma_u)$ is given by the tangent space of $TM_u$ 
via the isomorphism\footnote{This map is called the infinitesimal displacement of 
$\Sigma_u$ along $\frac{\partial}{\partial u^a}$ \cite{kodaira1962theorem}.}
\begin{equation} \label{eq:Kodairaopen}
  	\varphi^z_*:\, \frac{\partial}{\partial u^a}\ \longmapsto \ \frac{\partial \varphi^i(z^i;u)}{\partial u^a}
\end{equation}
at every point $u$ in $M$. Here, $\varphi^i$, $i=1,2$, are local normal coordinates 
to $\Sigma_u$, cf.~eqn. \eqref{eq:displacedcurve}. In other words, in this case every 
deformation $H^0(\Sigma,N_{Z_3}\Sigma)$ corresponds to a finite direction $u^a$ in the 
complex parameter manifold $M$ of the analytic family of curves. 

This theorem can be understood locally \cite{kodaira1962theorem} but is somewhat technical. 
Starting with the single holomorphic curve $\Sigma$ we introduce patches $U_i$ on $Z_3$ 
covering $\Sigma$ with coordinates $y^{i}_1$, $y^{i}_2$, $z^{i}$. Then $\Sigma$ is described 
as $y^{i}_1=y^{i}_2=0$ and $z^{i}$ is tangential to $\Sigma$. A deformation 
$\Sigma_{\underline{u}}$ of $\Sigma=\Sigma_0$ is described by finding functions 
$\varphi_l^{i}(z^{i};\underline{u})$, $l=1,2$, with the boundary condition 
$\varphi_l^{i}(z^{i};0)=0$ such that $\Sigma_{\underline{u}}$ reads
\begin{equation} \label{eq:displacedcurve}
 	y^{i}_1=\varphi^{i}_1(z^{i};\underline{u})\,,\quad y^{i}_2=\varphi^{i}_2(z^{i};\underline{u})\,
\end{equation}
upon introducing parameters $\underline{u}$ for convenience chosen in polycylinders 
$||\underline{u}||<\epsilon$. Furthermore, the first derivatives 
$\frac{\partial}{\partial u_a}\varphi^{i}_k\vert_{\underline{u}=0}$ should form a 
basis $s^a$ of $H^0(\Sigma,N_{Z_3}\Sigma)$. In addition, these functions have to obey 
specific consistency conditions, that we now discuss. As in the complex structure case, 
these functions are explicitly constructed as a power series 
\begin{equation} \label{eq:powerseriesopen}
 \varphi^{i}(z^i;\underline{u})=\varphi^{i}(0)+\varphi^{i}_{1}(\underline{u})
 +\varphi^{i}_{ 2}(\underline{u})+\ldots,\quad \norm{\underline{u}}<\epsilon\,,
\end{equation}
where we suppress the dependence on $z^{i}$ and further denote a homogeneous polynomial in 
$u$ of degree $n$ by $\varphi^{i}_n(\underline{u})$. The first order deformation is defined as
\begin{equation}
 	\varphi^{i}_{1}(\underline{u})=\sum_a u^a s^{(i)}_a(z^{i})\,,
\end{equation}
where $a=1,\ldots, h^0(N\Sigma)$ in the basis $s_a$ of $H^0(\Sigma,N_{Z_3}\Sigma)$ so that 
\eqref{eq:Kodairaopen} is obviously an isomorphism. 

Then the $m^{\text{th}}$ obstructions $\psi^{ik}(z^{k};\underline{u})$ are homogeneous polynomials 
of order $m+1$ taking values in $\check{\text{C}}$ech 1-cocycles on the intersection 
$U_i\cap U_j\cap U_k$ of the open covering of $\Sigma$ with coefficients in $N_{Z_3}\Sigma$. This 
means that the collection of local section $\psi^{ik}(z^{k};\underline{u})$ defines an element in 
the $\check{\text{C}}$ech-cohomology $H^1(\Sigma,N_{Z_3}\Sigma)$. It expresses the possible 
mismatch in gluing together the $\varphi^{i}(z^{i};\underline{u})$ defined on open patches 
$U_i\cap \Sigma_{\underline{u}}$ consistently to a global section on $\Sigma_{\underline{u}}$ 
at order $m+1$ in $\underline{u}$. In other words if the obstruction at $m^{\text{th}}$ order is 
trivial and we consider \eqref{eq:displacedcurve} on $U_i$ and $U_k$,
\begin{equation}
 	U_i:\, y^{i}_l=\varphi^{i}_l(z^{i};\underline{u})\,, \qquad 
 	U_k:\,y^{k}_l=\varphi_l^{k}(z^k;\underline{u})\,,\qquad (l=1,2)\,,
\end{equation}
then there exist functions $f^{ik}$ and $g^{ik}$ with $y^{i}_l=f_l^{ik}(\underline{y}^{k},z^{k})$, 
$z^{i}=g^{ik}(\underline{y}^{k},z^{k})$ so that 
\begin{eqnarray} 
y^{i}&=&\varphi^{i}(z^{i};\underline{u})=\varphi^{i}(g^{ik}(\underline{y}^{k},z^{k});\underline{u})
=\varphi^{i}(g^{ik}(\varphi^{k}(z^{k};\underline{u}),z^{k});\underline{u})\nonumber\\
y^{i}&=&f^{ik}(\underline{y}^{k},z^{k})=f^{ik}(\varphi^{k}(z^{k};\underline{u}),z^{k})\nonumber\\
&\Rightarrow&\varphi^{i}(g^{ik}(\varphi^{k}(z^{k};\underline{u}),z^{k});\underline{u})
=f^{ik}(\varphi^{k}(z^{k};\underline{u}),z^{k})
\end{eqnarray}
holds at order $m+1$ in $\underline{u}$. Here we suppressed the index $l$ labeling the coordinates 
$y^i_1$, $y^i_2$. Then $\psi^{ik}(z;\underline{u})$ is the homogeneous polynomial of degree $m+1$
\begin{equation}
 	\psi^{ik}(z^k;\underline{u}):=\big[\varphi^i(g^{ik}(\varphi^k(z^k;\underline{u}),z^k);\underline{u})
 	-f^{ik}(\varphi^k(z^k;\underline{u}),z^k)\big]_{m+1}
\end{equation}
where we expand $\varphi^i$, $\varphi^k$ to order $m$ in $\underline{u}$. It can be shown to have 
the transformation 
\begin{equation}
 	\psi^{ik}(z^k;\underline{u})=\psi^{ij}(z^j;\underline{u})+F^{ij}(z^j)\cdot\psi^{jk}(z^k;\underline{u})\,,
\end{equation}
where $F^{ij}(z^j)$ is the complex $2\times 2$ transition matrices on $N_{Z_3}\Sigma$ at a point 
$z^j$ in $\Sigma_u$ that acts on the two-component vector $\psi^{jk}\equiv(\psi_1^{jk},\psi_2^{jk})$.
This equation identifies the $\psi^{ik}$ as elements in $H^1(N_{Z_3}\Sigma)$ which can be identified 
by the Dolbeault theorem with $\bar{\partial}$-closed $(0,1)$-forms taking values in $N_{Z_3}\Sigma$, 
$H^1(N_{Z_3}\Sigma)=H^{(0,1)}_{\bar\partial}(NZ_3)$. Assuming that all $\psi^{ik}$ are trivial in 
cohomology and further proving the convergence of the power series \eqref{eq:powerseriesopen}, the 
analytic family of holomorphic curves is constructed. 

In principle one can calculate the obstructions $\psi^{ik}$ according to this construction at any 
order $m$. However, the obstructions are precisely encoded in the superpotential of the effective 
theory of a five-brane on $\Sigma$. This superpotential is in general a complicated function of both 
the brane and bulk deformations. Thus, determining the superpotential is equivalent to solving the 
deformation theory of a pair given by the curve $\Sigma$ and the Calabi-Yau threefold $Z_3$ containing 
it. It is this physical ansatz that we will take in the following. 

\subsection{Formalizing Five-Brane Superpotentials}
\label{sec:fivebranesuperpotential}

In this section we discuss, in the light of currents and brane deformations, the perturbative 
superpotentials both in the type IIB as well as in the heterotic theory. 
Using the expansion \eqref{eq:globalForms} the superpotential can conveniently be written as
\begin{equation} \label{eq:unifiedSuperpots}
	W_{\rm IIB}=\int_{Z_3}\Omega\wedge F_3\,\qquad W_{\rm het}=\int_{Z_3}\Omega\wedge H_3\,,
\end{equation}
where both $F_3$ and $H_3$ have to be considered appropriately as currents obeying \eqref{eq:dF_3}. 
According to the expansion \eqref{eq:globalForms} this contains now both the conventional flux 
superpotential, the heterotic Chern-Simons functional and the five-brane superpotential.
Let us put special emphasis on the part contributed by the brane $W_{\rm brane}$. 

The brane superpotential has the following properties. It depends holomorphically on the complex 
structure moduli of $Z_3$, as well as on the first order deformations corresponding to holomorphic 
sections of $N_{Z_3}\Sigma$. More precisely, we expect a superpotential $W_{\rm brane}=(u^a)^{n+1}$ 
if the deformation along the direction $s_a$ is obstructed at order $n$ \cite{Kachru:2000ih,Kachru:2000an}. 
Furthermore, the F-term supersymmetry conditions or critical values of $W_{\rm brane}$ in mathematical 
terms precisely correspond to holomorphic curves. Demanding these basic properties of $W_{\rm brane}$ 
was sufficient to determine the brane superpotential\footnote{Alternatively, $W_{\rm brane}$ can be 
directly deduced by dimensional reduction of the D5-brane action as presented in section \ref{sec:N=1dataD5} 
\cite{Grimm:2008dq}.} in \cite{Witten:1997ep} in the context of M-theory 
on a Calabi-Yau threefold $Z_3$ with a spacetime-filling M5-brane supported on a curve $\Sigma$,
\begin{equation}
\label{eq:chain}
 	W_{\rm brane}=\int_{\Gamma({\underline{u}})}\Omega(z)\,.
\end{equation}
Here $\Gamma({\underline{u}})$ denotes a three-chain bounded by the deformed curve $\Sigma_{\underline{u}}$ 
and the reference curve $\Sigma_{0}$ that is in the same homology class. Indeed, $W_{\rm brane}$ depends on 
both the deformations ${\underline{u}}$ of the five-brane on $\Sigma$ as well as the complex structure moduli 
$\underline{z}$ of $Z_3$ due to the holomorphic three-form $\Omega$. In the language of currents
the chain integral can be rewritten in the form \eqref{eq:unifiedSuperpots} using 
the expansion \eqref{eq:globalForms} as
\begin{equation}
	W_{\rm brane}=\int_{Z_3-\mathcal{U}_{\epsilon}^{(4)}(\Sigma)}\Omega\wedge
        \rho e_3.
\label{eq:chainonopenset}  
\end{equation}
We note that this is gauge invariant under SO$(4)$ gauge transformations on $N_{Z_3}\Sigma$ and 
has only support on a small neighborhood of $\Sigma$ so that it localizes on $\Sigma$ as expected. 

In the next section \ref{sec:5braneblowupsanddefs} we present the lift of the superpotential \eqref{eq:unifiedSuperpots}
to the blow-up threefold $\hat{Z}_3$, with a focus on the flux and brane superpotential. We employ
the geometrization of bulk-brane deformations on $\hat{Z}_3$ and the extension of the geometrical tools 
familiar from the study of the complex structure moduli space of Calabi-Yau threefolds, cf.~section 
\ref{sec:CSModuliSpace+PFO}, to the non-Calabi-Yau threefold $\hat{Z}_3$ to determine the superpotential 
explicitly. Finally in chapter \ref{ch:su3structur}, we even understand both the closed superpotential 
$W_{\rm flux}$ as well as the brane superpotential $W_{\rm brane}$ as part of the generalized flux 
superpotential on the $SU(3)$-structure manifold $\hat{Z}_3$.

\section{Five-Brane Blow-Ups and Unification of Deformations}
\label{sec:5braneblowupsanddefs}

In section \ref{sec:N=1branes} we started from the Bianchi identities for $F_3$ 
and $H_3$ and explained how five-brane sources on curves $\Sigma$ are properly 
described by delta-currents. Using the language of currents it was further 
explained, how to relate period and chain integrals on $Z_3$ to regularized integrals
on the open manifold $Z_3-\Sigma$. The crucial point of the blow-up proposal \cite{Grimm:2008dq} 
presented in this section is to replace the open manifold $Z_3-\Sigma$ by a physically 
equivalent geometry $\hat{Z}_3$ with a distinguished divisor $E$. In addition the blow-up
proposal naturally yields a flux $F_2=[\Sigma]$ on $E$ which can be understood as the 
partially dissolved five-brane charge. Furthermore, this implies an embedding of the open
and closed deformations of the geometry $Z_3$ and the brane on $\Sigma$ into pure 
complex structure deformations on $\hat{Z}_3$.

In the first part of this section, section \ref{sec:geometricblowups}, we construct the 
manifold $\hat Z_3$ by blowing up a $\mathbb{P}^1$-bundle along $\Sigma$. This introduces a 
new divisor $E$ in $\hat{Z}_3$, the exceptional divisor, that is by construction a ruled 
surface over $\Sigma$. Away from the exceptional divisor the open manifolds $Z_3-\Sigma$ 
and $\hat Z_3-E$ are biholomorphic. Using this fact and the formalism of section \ref{sec:N=1branes} 
we can evaluate the open integrals on $\hat{Z}_3-E$. Furthermore it becomes possible to extend 
all open integrals, in particular the brane superpotential $W_{\rm brane}$, the forms 
$H_3$ and $F_3$ as well as the closed periods\footnote{See section \ref{sec:CSModuliSpace+PFO}
for a review of the definition of periods on a Calabi-Yau threefold $Z_3$.} $\Pi^k(\underline{z})$ 
from $Z_3-\Sigma=\hat{Z}_3-E$ to $\hat Z_3$ by constructing local completions of these quantities 
in the vicinity of the divisor $E$. As explained in section \ref{sec:unificationofdeformations} 
for the case that $\Sigma$ is given as a complete intersection, the blow-up proposal unifies 
the description of the closed and open deformations of $(Z_3,\Sigma)$, which both become 
complex structure deformations on $\hat Z_3$. Of particular importance for the superpotential 
\eqref{eq:unifiedSuperpots} and deformation theory on $\hat{Z}_3$ is the pullback $\hat{\Omega}$
of the Calabi-Yau three-form $\Omega$ of $Z_3$ to $\hat Z_3$, which we construct in section 
\ref{sec:hatomega}. In section \ref{sec:potentialhatZ3minusD} we describe how the 
superpotentials concretely map to the blow-up $\hat{Z}_3$. 

Ultimately as explained in chapter \ref{ch:su3structur}, the flux and the brane superpotentials 
on $\hat Z_3$ of section \ref{sec:fivebranesuperpotential} are unified to a flux superpotential 
on $\hat Z_3$. The latter structure requires in addition to the extension of $\Omega$ also the 
extension of the K\"ahler form $J$ and the flux $H_3$ from $Z_3$ to $\hat Z_3$. Our formalism 
as presented in this section can be understood as the first step in the full geometrization of 
the five-brane and prepares the approach of chapter \ref{ch:su3structur} to consider the 
flux-geometry  $\hat Z_3$, $F_2$ as a string background with $SU(3)$-structure.

\subsection{Geometric Properties of the Blow-Up along $\Sigma$}
\label{sec:geometricblowups} 

Given a $k$-dimensional complex submanifold $\Sigma_k$ 
in an $n$-dimensional complex manifold $Z_n$, it is a standard technology 
in algebraic geometry~\cite{Griffiths:1978yf} to blow-up along $\Sigma_k$ to 
obtain a new $n$-dimensional complex manifold $\hat Z_n$. This directly applies 
to a supersymmetric five-brane on a holomorphic curve\footnote{For the use of the blow-up proposal 
to analyze non-holomorphic deformations of $\Sigma$ cf.~section \ref{sec:unificationofdeformations}.} 
$\Sigma=\Sigma_1$ inside a Calabi-Yau threefold $Z_3$.  

 \begin{figure}[htb]
 \begin{center}
 \includegraphics[width=.5\textwidth]{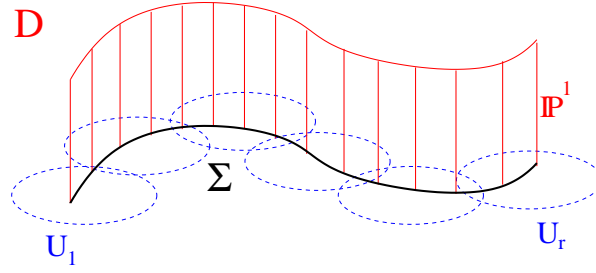}
 \begin{quote}
 \caption{Blow up of the curve $\Sigma$ to the ruled surface $E$} \label{blow-up}
 \end{quote}
 \end{center}
 \end{figure}

Since the blow-up is a local operation, it can be described near the complex submanifold within 
neighborhoods $U_\alpha$ with the topology of a disk, which cover $\Sigma$. Let $y_{\alpha,\,i}$, 
$i=1,\ldots,3$ be local coordinates\footnote{Given the submanifold locally by $h_i(\underline{x})=0$, 
$i=1,2$ in generic $x_1,\ldots,x_3$ coordinates of $U_\alpha$ this choice is fixed by the inverse 
function theorem stating that for every point $x_0\in \C^3$ with 
$(\partial_k h_1\partial_l h_2-\partial_l h_1\partial_k h_2)\vert_{x_0}\neq0$ for $k,l\neq j$, 
there exists a local parameterization of $\cC$ near $x_0$ as a graph over $x_j$. 
In particular, the blow-up is independent of the coordinates used, cf.~p. 603 \cite{Griffiths:1978yf}.} 
on $U_\alpha$ in which $\Sigma$ is specified by  the intersection of two divisors 
$D_{\alpha,\, 1}\cap D_{\alpha,\,2}$ i.e. by $y_{\alpha,\, i}=0, i=1,2$. 
  
The local blow-up can be described as a hypersurface constraint~\cite{Griffiths:1978yf} 
\begin{equation}  \label{eq:def-hatU}
\hat U_\alpha=\{(y_{\alpha,1},y_{\alpha,2}, y_{\alpha,3},
(l_{\alpha,1}:l_{\alpha,2}))\subset U_\alpha\times \mathbb{P}^1:
y_{\alpha,2} l_{\alpha,1}-y_{\alpha,1} l_{\alpha,2} =0\}\ .
\end{equation}  
Here $(l_{\alpha,1}:l_{\alpha,2})$ are projective coordinates of the $\mathbb{P}^1$ over the local patch 
$U_\alpha$. We define a projection map $\pi_\alpha:\hat{U}_\alpha\rightarrow U_\alpha$ by discarding the 
direction of the $\P^1$, 
\begin{equation}
	\pi_\alpha(y_{\alpha,1},y_{\alpha,2}, y_{\alpha,3},
(l_{\alpha,1}:l_{\alpha,2}))=(y_{\alpha,1},y_{\alpha,2}, y_{\alpha,3})\,.
\end{equation} 
Obviously, $\hat U_\alpha-\pi^{-1}_\alpha(\Sigma)$ is biholomorphic to $U_\alpha-\Sigma$, 
as we can eliminate the $l_{\alpha,i}$ in the hypersurface \eqref{eq:def-hatU} outside of the locus 
$y_{\alpha,1}=y_{\alpha,2}=0$ that defines $\Sigma$. Conversely, the set $E_\alpha:=
\pi^{-1}_\alpha(0,0,y_{\alpha,3})=\pi^{-1}_{\alpha}(\Sigma)$ is described as follows. Over a point 
$(0,0,y_{\alpha,3})\in\Sigma$ in $U_\alpha$ the fibres of the projection $\pi_\alpha$ are canonically, 
i.e. independently of the coordinate system, identified with lines in the projectivized normal bundle 
$\mathbb{P}(N_{U_\alpha}\Sigma)=\mathbb{P}(\mathcal{O}(D_{\alpha,1})\oplus \mathcal{O}(D_{\alpha,2}))$, 
\begin{equation} 
(0,0,y_{\alpha,3}, (l_{\alpha,1}:l_{\alpha,2}))\,\mapsto\, l_{\alpha,1}
\frac{\partial}{\partial y_{\alpha,1}} + l_{\alpha,2} \frac{\partial}{\partial y_{\alpha,2}} \ . 
\end{equation}
This allows to glue the open sets $\hat{U}_\alpha$ to obtain $\hat{Z}_3$ and the $E_\alpha$ to obtain 
a divisor $E=\pi^{-1}(\Sigma)$ which is identified with
\begin{equation} \label{eq:Eglobal}
	E=\mathbb{P}(N_{Z_3}\Sigma)\,.
\end{equation} 
Similarly we obtain a unique global projection map $\pi$ that is trivially extended to all open sets 
$\hat U_{\alpha}$ and thus to $\hat{Z}_3$ as the identity map on $\hat{Z}_3-E=Z_3-\Sigma$. The divisor $E$ 
is the exceptional divisor and it is a $\mathbb{P}^1$-ruled surface over $\Sigma$ by \eqref{eq:Eglobal}.

Essential facts that are intensively used in this paper are the biholomorphism 
\begin{equation}
\label{eq:biholomorphism}
\pi:(\hat Z_3 - E)\rightarrow (Z_3- \Sigma)
\end{equation}
and the statement, that all relevant aspects of the blow-up can be analyzed locally in patches 
near $\Sigma$, except for the non-triviality of $N_{Z_3} \Sigma$, which is captured by the Thom 
class $\frac{e_3}{2}$. 

The relation between the local and global construction is particularly easy if $Z_3$ is given by a 
family hypersurface $P(\underline{x},\underline{z}) =0$ and $\Sigma$ is constructed  as a complete 
intersection of $P(\underline{x},\underline{z})=0$ and divisors $D_i$ given by 
$h_{i}(\underline{x},\underline{u})=0$, $i=1,2$ in $\mathbb{P}_\Delta$. Here the ambient space is in 
general a toric variety $\mathbb{P}_\Delta$ with homogeneous coordinates $\underline{x}$ and the 
variables $\underline{z}$, $\underline{u}$ parameterize the complex structure and brane moduli 
respectively. Then we can choose in any patch $U_\alpha$ coordinates so that 
$y_{\alpha,1}=h_1(\underline{x})|_{U_\alpha}$, $y_{\alpha,2}=h_2(\underline{x})_{U_\alpha}$ and 
$y_{\alpha,3}$ is a coordinate along $\Sigma$. Now $N_{U^{\alpha}}\Sigma$ is globally given as the 
sum of two line bundles, $N_{Z_3}\Sigma =\mathcal{O}(D_1)\oplus\mathcal{O}(D_2)$, and $\hat Z_3$ is 
given globally as the complete intersection in the total space of the projective bundle
\begin{equation} \label{eq:Wambient}
\mathcal{W}=\mathds{P}(\mathcal{O}(D_1)\oplus\mathcal{O}(D_2)).
\end{equation} 
Indeed, using the projective coordinates $(l_1,l_2)\sim
\lambda(l_1,l_2)$ on the $\mathds{P}^1$-fiber of the blow-up, $\hat{Z}_3$ can be
written as
\begin{equation}
        P(\underline{x},\underline{z})= 0\ ,\qquad 	
        Q\equiv l_1 h_2(\underline{x},\underline{u}) - l_2 h_1(\underline{x},\underline{u}) = 0 \, ,
\label{eq:blowup}
\end{equation}
in the projective bundle $\mathcal{W}$. 

We conclude  by mentioning selected geometrical properties of the blow-up threefold $\hat{Z}_3$ and 
refer to appendix \ref{App:topoHatZ_3} for details. First of all we note that $\hat{Z}_3$ does not meet 
the Calabi-Yau condition. This is a consequence of the general formula for the Chern classes of the 
blow-up of a complex manifold along a curve \cite{Griffiths:1978yf}, $c_1(\hat Z_3)=-K_{\hat Z_3}=\pi^*(c_1(Z_3))-[E]=-[E]\neq 0$
where we used that $Z_3$ is a Calabi-Yau manifold in the last equality. In addition, $\hat Z_3$ is not 
a Fano variety. One important consequence from this is that complex structure deformations in 
$H^{1}(\hat{Z}_3,T\hat{Z}_3)$ are not necessarily isomorphic to $H^{(2,1)}(\hat{Z}_3)$, since the 
holomorphic three-form $\hat{\Omega}$ vanishes along $E$. Secondly, it is crucial to emphasize that the 
blow-up is the minimal description of the deformations of the pair $(Z_3,\Sigma)$ in the sense that in the 
blow-up procedure no \textit{new} degrees of freedom associated to deformations of $E$ are introduced. 
Indeed by construction the normal bundle to $E$ in $\hat Z_3$ is the tautological bundle $T$ 
\cite{Griffiths:1978yf}, see appendix \ref{App:topoHatZ_3} for details, which is a negative bundle on
$E$. Thus $E$ has no deformations, $H^0(E,N_{\hat Z_3}E)=\emptyset$, i.e.~$E$ is isolated. Furthermore, 
it can be shown mathematically rigorously that all deformations of $Z_3$ and the curve $\Sigma$, that 
are deformations of complex structures of $Z_3$ and deformations of $\Sigma$ in $Z_3$, map to complex 
structure deformations of $\hat{Z}_3$.\footnote{We thank Daniel Huybrechts for a detailed explanation 
of the equivalence of the two deformation theories.} The K\"ahler sector of $Z_3$ maps to that of 
$\hat{Z}_3$ that contains one additional class of the exceptional divisor $E$.

\subsection{Unification of Open and Closed Deformations on Blow-Up Threefolds}
\label{sec:unificationofdeformations}  

In the first part of this section we present the key points of the blow-up proposal suggested in 
\cite{Grimm:2008dq}, namely the unification of open and closed deformations, and its use to analyze 
the geometrical dynamics of five-branes. Then, in a second part we study the deformation space of 
branes wrapping rational curves via the complete intersection curves \eqref{eq:blowup} and the 
blow-up of the latter. 

\subsubsection{Matching Deformations and Obstructions: a Mathematical Proposal}
\label{unificationofdeformationsI}

We have described  in section \ref{branedeformationsII} that the 
infinitesimal elements $\varphi_*^z(\frac{\partial}{\partial u_a})=\partial_{u_a} \varphi(z;\underline{u})$ 
span the tangent space to the open deformations space  and 
live in $H^0(\Sigma,N_{Z_3} \Sigma)$, while it was reviewed 
in section \ref{sec:CSModuliSpace+PFO} that deformation of 
the closed complex structure deformations of a 
manifold $M$ live\footnote{Most statements about the complex structure
deformations apply to $Z_3$ and $\hat Z_3$. We denote both complex manifolds
by $M$ in the following.} in $H^1(M,T_{M})$. 
Up to global automorphisms of the toric ambient space $\mathbb{P}_\Delta$, 
which are compatible with the torus action, this cohomology  can be represented by 
the infinitesimal deformations $\delta_z$ of the parameters $\underline{z}$  
multiplying monomials in $P(\underline{x},\underline{z})= 0$ 
of the hypersurface. Likewise for the 
complete intersection \eqref{eq:blowup} elements in $H^1(M,T_{M})$ 
can be represented by infinitesimal deformations $\delta_{(z,u)}
=:\delta_{\hat z}$ of the parameters in \eqref{eq:blowup}, 
modulo global automorphisms of $\cal W$.  
Using these facts, it is easy to check 
for the complete intersections  description 
\eqref{eq:blowup} that  the moduli $\underline{u}$ of $\Sigma$ 
described by the coefficients of the monomials in 
$h_i(\underline{x},\underline{u})$, $i=1,2$ turn into 
complex structure moduli $\hat{\underline{z}}$ of $\hat{Z}_3$ since
$h_i(\underline{x},\underline{u})$, $i=1,2$  
enter the defining equations of $\hat{Z}_3$ 
via $Q$ in \eqref{eq:blowup}.

\begin{figure}[htb]
 \begin{center}
 \includegraphics[width=0.9\textwidth]{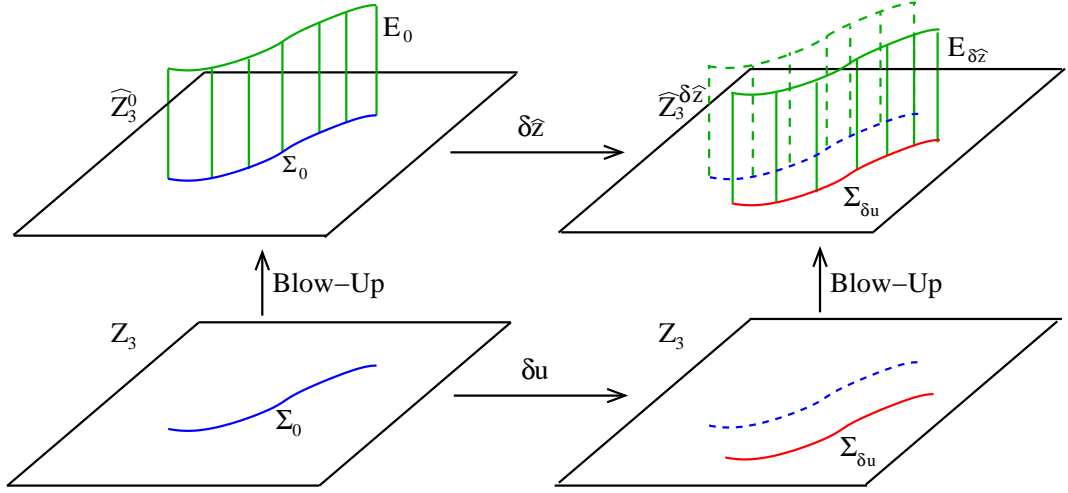}
 \begin{quote}
 \caption{Map of brane to complex structure deformations. A displacement $\Sigma_{\delta u}$ 
 of a holomorphic curve $\Sigma_0$ yields a new blow-up $\hat{Z}^{\delta \hat{z}}_3$ with 
 induced complex structure deformation $\delta \hat{z}$ and divisor $E_{\delta \hat{z}}$. 
 The divisor $E_0$, which is holomorphic in $\hat{Z}_3^{0}$, is not holomorphic in 
 $\hat{Z}_3^{\delta \hat z}$.} \label{fig:defsblow-up}
 \end{quote}
 \end{center}
 \end{figure}

As noted below \eqref{eq:Eisolated}, the divisor $E$ is isolated in $\hat Z_3$, i.e. 
on $\hat Z_3$ there are no deformations associated to $E$. From this follows more 
illustratively, that blowing up along $\Sigma$ for different values of $\underline{u}$ yields 
diffeomorphic blow-ups $\hat{Z}_3$ which just differ by a choice of complex structure. 
The situation is visualized in figure \ref{fig:defsblow-up}. 
Mathematically, the equivalence even of the full deformation theory 
of $(Z_3,\Sigma)$ and $\hat Z_3$ are expected in general. This means that
not only the deformations of $(Z_3,\Sigma)$, counted by elements in $H^{1}(Z_3,TZ_3)$
and $H^{0}(\Sigma,N_{Z_3}\Sigma)$, agree with the complex structure deformations of 
$\hat{Z}_3$, that are in $H^{1}(\hat{Z}_3,T\hat{Z}_3)$, but also the obstruction problems 
of both geometries. In particular, this implies an order-by-order match of the obstructions 
on both sides of the correspondence. 

As detailed in the section \ref{sec:CSModuliSpace+PFO} and
\ref{branedeformationsII} possible obstructions to the closed 
deformation space live in $H^2(M,T_M)$ while possible obstructions 
to the open deformations space live in $H^1(\Sigma,N_{Z_3} \Sigma)$. 
While one can conclude from the vanishing of these homology groups 
that the corresponding deformations are unobstructed, it is not 
necessarily true that the deformation problems are obstructed, if 
these homology groups do not vanish. In particular the complex 
structure deformations of Calabi-Yau spaces, such as $Z_3$, are unobstructed 
despite the fact that $H^2(Z_3,T_{Z_3})\neq 0$.
However the deformations of the curve $\Sigma$ can in general be obstructed by
elements $H^{1}(\Sigma,N_{Z_3}\Sigma)$ at some order. Given the equivalence of the
obstruction problems we expect that these are precisely matched by the obstructions to 
complex structure deformations on $\hat{Z}_3$ in $H^2(\hat{Z}_3,T\hat{Z}_3)$. 
In physical terms, the obstruction problem is in general expressed by a superpotential, 
which in the case of the obstruction problem of $(Z_3,\Sigma)$ is given by the superpotential 
\eqref{eq:unifiedSuperpots}. Our strategy to investigate the possible obstructions of 
$(Z_3,\Sigma)$ will be to match the calculation of the superpotential \eqref{eq:unifiedSuperpots} 
before and after the blow-up. Here it will be crucial to understand the lift of the brane 
superpotential $W_{\rm brane}$ of \eqref{eq:chain} under the blow-up, that will be replaced 
by a specific flux superpotential on $\hat{Z}_3$. This flux superpotential induces obstructions 
to complex structure deformations on $\hat{Z}_3$ that are equivalent to the original obstructions 
on moving the brane on $\Sigma$ expressed by $W_{\rm brane}$. We will discuss the match of the 
superpotentials in a two step procedure in section \ref{sec:potentialhatZ3minusD} and in chapter 
\ref{ch:su3structur}. In particular the blow-up $\hat{Z}_3$ will yield an easy calculational 
scheme of the superpotential as explained and applied to specific examples in section 
\ref{ch:CalcsBlowUp}.

\subsubsection{Matching Deformations and Obstructions: Concrete Examples}
\label{unificationofdeformationsII}

Before we proceed we have to explain how we use the blow-up $\hat{Z}_3$ constructed as the complete 
intersection \eqref{eq:blowup} to calculate the superpotentials $W_{\rm brane}$ for five-branes on 
rational curves. This is crucial since the families of holomorphic curves themselves defined by the 
complete intersection of complex equations $h_1=h_2=0$ are unobstructed. Similarly on $\hat Z_3$ 
the corresponding complex structure deformations are unobstructed and the deformation problem and 
the corresponding superpotentials are trivial.

The general statement for the moduli 
space of holomorphic curves\footnote{In the following we will use the term of a 'moduli space' of 
holomorphic curves in $Z_3$ to denote an analytic family of holomorphic curves as introduced in 
section \ref{branedeformationsII}.} on Calabi-Yau threefolds is that its virtual deformation space 
is zero-dimensional~\cite{Kachru:2000ih, Kachru:2000an, Katz:1992vx, Hori:2003ic}.
Naively this could be interpreted as the statement, that generically holomorphic curves in a Calabi-Yau 
threefold never occur in families. However, this conclusion is not true as one can learn already 
from the case of rational curves\footnote{A rational curve is birationally equivalent to a line i.e.~a 
curve of genus zero which is a $\P^1$.} in the quintic as explained in~\cite{Katz:1992vx}.
Rational curves in a generic Calabi-Yau manifold $Z_3$, like the quintic with a constraint $P=0$ 
including $101$ complex structure parameters $\underline{z}$ at generic values, are isolated and 
have a moduli space consisting of points, which we denote by ${\cal M}^{\underline z}(\mathbb{P}^1)=pts.$  
However, at special loci $\underline{z}_0$ in the complex structure moduli space, which correspond 
to specially symmetric Calabi-Yau constraints $P=0$ like the Fermat point
\begin{equation} \label{eq:Fermatlocus}
	P=x_1^5+x_2^5+x_3^5+x_4^5+x_5^5\,
\end{equation}
of the quintic, a family of
curves parametrized by a finite dimensional moduli space 
${\cal M}^{{\underline  z}_0}(\mathbb{P}^1)$ can appear. 
Physically this means that the open superpotential 
becomes a constant of the brane moduli and  the scalar potential has a flat direction 
along ${\cal M}^{{\underline z}_0}(\mathbb{P}^1)$.
However it can be generally argued~\cite{Kachru:2000ih} 
that in the vicinity of the special loci in the closed string 
deformation space $\underline{z}_0$ a superpotential develops 
for the rational curves. In agreement with \eqref{eq:csinducedmass} the superpotential starts 
linear in the closed string deformation $t\propto \delta_{z}$ away from $\underline{z}_0$ and is 
of arbitrary order in the open string moduli so that it has
$(-1)^{{\rm dim}({\cal M}^{{\underline  z}_0}(\mathbb{P}^1))} 
\chi({\cal M}^{\underline{z}_0}(\mathbb{P}^1))$ minima\footnote{This formula 
follows from complex deformation invariance of the BPS numbers associated 
to holomorphic curves~\cite{Katz:1999xq}.}. 
We note that this is precisely the most interesting physical situation, 
as $t$ can be made parametrically small against the compactification scale, as explained below 
equation \eqref{eq:csinducedmass}.

There is one important caveat in order when working with concrete algebraical curves. A given 
family of holomorphic curves in a specific algebraic representation $P=0$ of $Z_3$ can become 
obstructed due to the presence of non-algebraic complex structure deformations, i.e.~those $\underline{z}$ 
that are not contained in $P=0$. For example, this situation occurs in the Calabi-Yau hypersurfaces 
$Z_3$ of degree $2+2(n_1+n_2+n_3)$ in weighted projective spaces of the type 
$\mathbb{P}^4(1,1,2 n_1, 2 n_2, 2 n_3)$ with $n_i\in \mathds{Z}$ as discussed in \cite{Kachru:2000ih}.
This realization of the Calabi-Yau manifold $Z_3$ contains always a ruled surface, i.e. an 
$\mathbb{P}^1$ fibered over a (higher genus) Riemann surface\footnote{The same Riemann surface is 
identified with the moduli space ${\cal M}^{{\underline z}_0}(\mathbb{P}^1)$ in this case.}. The 
embedding of $Z_3$ in this particular ambient space is such that the generic obstructed situation, 
which corresponds to a non-vanishing superpotential, is not accessible using the algebraic deformations, 
i.e.~upon tuning the parameters in the Calabi-Yau constraint $P=0$. The absence of these deformations 
as algebraic deformations in $P=0$ happens since the corresponding monomials are not compatible with 
the symmetries of the ambient space.

Let us next describe the obstructed deformation problem of rational curves and the relation to the 
complete intersection curves and the blow-up \eqref{eq:blowup}. The basic idea is to map the obstructed 
deformations of the rational curves to the algebraic moduli space of the complete intersection in the 
following way. As mentioned in \eqref{eq:blowup} the algebraic deformations parametrized by the closed 
moduli $\underline{z}$ and the open moduli ${\underline u}$ are unobstructed. Let us denote the corresponding 
open and closed moduli space of the complete intersection $\Sigma$, defined by $P(\underline{z})=0$ and  
$h_{i}(\underline{x},\underline{u})=0$, $i=1,2$, by ${\cal M}(\Sigma)$ and the open moduli space of 
$\Sigma$ for fixed closed moduli $\underline{z}$ by ${\cal M}^z(\Sigma)$. The generic dimension of this 
open moduli space $h^0(N \Sigma)$ is positive. The idea is to consider a representation of $Z_3$ which is 
compatible with a discrete symmetry group $G$. This symmetry group allows us to identify lower degree 
and genus curves with the complete intersection $\Sigma$ at a special sublocus of the moduli space 
${\cal M}(\Sigma)$. In our main examples in section \ref{sec:ToricBraneBlowup} and \ref{sec:ToricBraneBlowupII} 
these are rational curves, i.e.~curves of degree one. Let us denote this sublocus by ${\cal M}_{\mathbb{P}^1}(\Sigma)$. 
This sublocus is determined by the requirement that the algebraic constraints $P$, $h_i$ degenerate so that 
they can be trivially factorized as powers of linear constraints,
\begin{equation} \label{eq:sublocus}
 	{\cal M}_{\mathbb{P}^1}(\Sigma):\quad P(\underline z)=h_i(\underline{u})=0\,\quad \Leftrightarrow\quad 
 	\prod_k\sum_l a^{(s)}_{lk} x_l=0\,,\,\, s=1,2,3\,,
\end{equation}
where the different linear factors $L^{(s)}_k=\sum_l a^{(s)}_{lk} x_l$ are identified by the discrete 
group $G$, $L^{(s)}_{k_1}\leftrightarrow L^{(s)}_{k_2}$. Then the right hand side of this identification 
describes rational curves
\begin{equation} \label{eq:rationalcurves}
  L^{(1)}_{k_1}=L^{(2)}_{k_1}=L^{(3)}_{k_1}=0
\end{equation}
modulo $G$ in the ambient space and in $Z_3$ since $P=0$ is trivially fulfilled. For a concrete situation 
we refer to section \ref{sec:ToricBraneBlowup}. In particular this identification embeds the moduli space 
${\cal M}^{{\underline z}_0}(\mathbb{P}^1)$ into ${\cal M}^{{\underline z}_0}(\Sigma)$ and more trivially 
the (discrete) ${\cal M}^{{\underline z}}(\mathbb{P}^1)$ into ${\cal M}^{{\underline z}}(\Sigma)$. 

\begin{figure}[htb]
 \begin{center}
 \includegraphics[width=.7\textwidth]{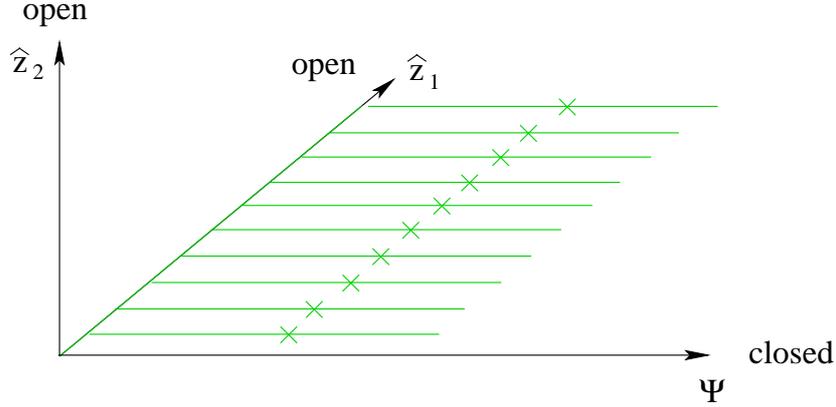}
 \begin{quote}
 \caption{Picture of the moduli space of ${\cal M}(\Sigma)$ given by the quintic 
 modulo $\mathds{Z}_5^3$ and the loci ${\cal M}_{\mathbb{P}^1}(\Sigma)$ where $\Sigma$ degenerates to (holomorphic) 
 rational curves. All lines in the $\hat{z}_2=0$ plane correspond to the embedding of the moduli space of rational 
 curves ${\cal M}_{\mathbb{P}^1}(\Sigma)\subset {\cal M}(\Sigma)$, cf.~\eqref{eq:RSquintic}. At $\Psi=0$ the $\hat{z}_1$ 
 direction opens up as a modulus of a family of $\mathbb{P}^1$'s over 
 the genus $6$ curve ${\cal M}^{\Psi=0}(\mathbb{P}^1)$. For generic values of $\Psi$ 
 only $(-1)^{{\rm dim}({\cal M}^{\Psi=0}(\mathbb{P}^1))} \chi({\cal M}^{\Psi=0}(\mathbb{P}^1))=10$ 
 points belong to ${\cal M}^{\Psi}(\mathbb{P}^1)$. Away from ${\cal M}_{\mathbb{P}^1}(\Sigma)$ the holomorphic 
 configuration in $Z_3$ is an irreducible higher genus curve, which corresponds to non-holomorphic $S^2$'s in 
 $Z_3$.} \label{fig:moduli}
 \end{quote}
 \vspace{-1cm}
 \end{center}
 \end{figure}

More generally, i.e.~away from the sublocus ${\cal M}_{\mathbb{P}^1}(\Sigma)$ defined by \eqref{eq:sublocus}, 
this embedding implies that the \textit{obstructed} deformation space of the rational curves \eqref{eq:rationalcurves} 
is identified with the \textit{unobstructed} moduli space ${\cal M}(\Sigma)$. This can be compared to the method 
presented in \cite{Jockers:2008pe,Jockers:2009mn,Alim:2009rf,Alim:2009bx} where the obstructed deformations of 
a curve are identified with the unobstructed moduli of an appropriate divisor. For the curves we consider we 
depict the embedding of the deformation spaces of the rational curves into the moduli space ${\cal M}(\Sigma)$ 
of complete intersection $\Sigma$ in figure \ref{fig:moduli}, where we introduce new open moduli $\hat{z}^1$, $\hat{z}^2$ 
that are functions of the $u^i$.
The point is that away from ${\cal M}_{\mathbb{P}^1}(\Sigma)$ 
the identification \eqref{eq:sublocus} of the complete intersection curve $\Sigma$ with the holomorphic rational 
curves in the locus $P=0$ fails. We can understand this failure in two different ways emphasizing different aspects 
of the identification of the deformation space of rational curves with the true moduli space ${\cal{M}}(\Sigma)$. 
If we analyze the identification of $\Sigma$ with holomorphic rational 
curves infinitesimally close to ${\cal M}_{\mathbb{P}^1}(\Sigma)$  one can either keep the linear constraints 
\eqref{eq:rationalcurves} and relax the condition that these rational curves lie identically on the $P=0$ locus 
or we linearize the equations $P=h_i=0$ such that the rational curves keep lying in the $P=0$ locus. 
However, since the trivial factorization \eqref{eq:sublocus} and the identification of the linear factors 
$L^{(s)}_k$ modulo $G$ fails, the latter possibility introduces non-holomorphic equations 
with nontrivial branching. This implies that the rational curve away from ${\cal M}_{\P^1}(\Sigma)$ inside 
${\cal M}(\Sigma)$ is not holomorphic in $Z_3$. In particular, the process of turning on 
$t=\delta_{z}$ around $\underline{z}_0$ for values of the open moduli, that are only close to 
${\cal M}^{ \underline{z}_0+\delta_{z} }(\mathbb{P}^1)$, can be understood as deforming the analytic family 
${\cal{M}}^{\underline{z}_0}(\P^1)$ of rational curves inside the $P=0$ locus to rational curves, which are 
non-holomorphic. For concreteness, for the later example \eqref{eq:RSquintic} of the quintic, we identify 
$\delta z=\Psi$ with the complex structure of the mirror quintic deforming the Fermat locus \eqref{eq:Fermatlocus} 
as
\begin{equation}
 	P=x_1^5+x_2^5+x_3^5+x_4^5+x_5^5-5\Psi x_1x_2x_3x_4x_5\,.
\end{equation}
Then, $\delta z\neq 0$ deforms away from the one-dimensional moduli space of rational curves 
${\cal M}^{ \underline{z}_0}(\mathbb{P}^1)$ in $P=0$, that exists at the Fermat locus $\underline{z}_0=0$.

As noted before, a brane on a non-holomorphic curve is not supersymmetric and thus violates the 
F-term supersymmetry condition that is expressed by the superpotential $W_{\rm brane}$. In the following, 
in particular in the examples of sections \ref{sec:ToricBraneBlowup} and \ref{sec:ToricBraneBlowupII}, 
we will consider five-branes on rational curves in a given Calabi-Yau threefold $Z_3$ and their excitation 
about the supersymmetric minimum that correspond, in geometric terms, to possibly obstructed deformations 
about holomorphic curves. The brane excitations we consider correspond on the one hand to light fields 
that are generically obstructed and that become, as discussed in section \ref{branedeformationsI}, massless 
at some point in the complex structure moduli space. On the other hand we include fields that parameterize 
non-holomorphic deformations for all values of the closed moduli. In figure \ref{fig:moduli} the first type 
of fields corresponds to $\hat{z}_1$, which becomes massless at $\Psi=0$, and the second type of fields 
corresponds to $\hat{z}_2$. However, the crucial point for the consideration of this deformation space, as 
noted above, is the identification with the moduli space ${\cal M}(\Sigma)$ of complete intersection curves 
$\Sigma$. This identification and the unification of the open-closed moduli space of $(Z_3,\Sigma)$ in the 
blow-up \eqref{eq:blowup} of the complete intersection curve $\Sigma$ will enable us to calculate the 
superpotential $W_{\rm brane}$ for branes wrapping rational curves. We determine the periods on the complex 
structure moduli space of the blow-up $\hat{Z}_3$ that physically describe the closed and open superpotential 
$W_{\rm brane}$ upon turning on an appropriate flux on $\hat{Z}_3$. For this purpose we will describe 
$W_{\rm brane}$ explicitly by chain and flux integrals on the blow-up $\hat{Z}_3$ respectively in section 
\ref{sec:potentialhatZ3minusD} and chapter \ref{ch:su3structur}. Finally, we note that the periods on 
$\hat{Z}_3$ can equivalently be understood as a definition of the concept of periods on the 
\textit{brane moduli space} ${\cal M}(\Sigma)$, extending the familiar notion of periods on the complex 
structure moduli space of a Calabi-Yau manifold.

\section{Probing the Open-Closed Deformation Space}
\label{sec:hatomega}

The key in  describing  the deformations of complex structures on $\hat
Z_3$ are the construction and the properties of the pull-back $\hat \Omega=\pi^*(\Omega)$ of 
the  holomorphic three-form $\Omega$ from the Calabi-Yau threefold $Z_3$ to $\hat
Z_3$. Since the blow-up is a local procedure $\hat \Omega:= \pi^*(\Omega)$ will first be constructed 
in the local patches $\hat{U}_\alpha$ and then be globalized as a residue integral for the complete 
intersection \eqref{eq:blowup}. From this we obtain differential equations, the Picard-Fuchs equations, 
which determine the full complex structure dependence of $\hat{\Omega}$ and its periods.

Let us summarize the results of the actual calculation, which is done 
in appendix \ref{App:Local}. As in section \ref{sec:geometricblowups} we assume that $\Sigma$ is 
represented as a complete intersection of divisors $D_i$, $i=1,2$, in $Z_3$ given by constraints 
$h_i(\underline{x},\underline{u})=0$ for coordinates $\underline{x}$. If we consider a patch $U_\alpha$ 
on $Z_3$ near $\Sigma$, then the holomorphic three-form $\Omega$ is locally given by 
\begin{equation} \label{eq:omegaLocal}
 \Omega=\dd x_1\wedge \dd x_2 \wedge \dd x_3 =\det J^{-1} \dd y_{\alpha,1} \wedge \dd y_{\alpha,2} \wedge \dd
y_{\alpha,3}\,,
\end{equation}
where $J$ is the Jacobian matrix for choosing coordinates $y_{\alpha,i}=h_i(\underline{x})$, $i=1,2$, 
and $y_{\alpha,3}=x_3$, starting with generic coordinates $\underline{x}$. This expression is pulled 
back via the projection map $\pi:\hat{Z}_3\rightarrow Z_3$ to the patch $\hat U_\alpha$ defined in  
\eqref{eq:def-hatU}. We introduce coordinates
\begin{equation}
l_1\neq 0\,:\quad z_{\alpha,1}^{(1)}=y_{\alpha,1}\,,\quad z_{\alpha,2}^{(1)}=\frac{l_2}{l_1}
=\frac{y_{\alpha,2}}{y_{\alpha,1}}\,,\quad  
z_{\alpha,3}^{(1)}=y_{\alpha,3} 
\end{equation}
on $\hat{U}_\alpha$ for $l_1\neq 0$ to obtain  
\begin{equation} \label{eq:pullbacklocal}
\hat \Omega=\pi^*(\Omega)=z_{\alpha,1}^{(1)} \det
J^{-1} \dd z_{\alpha,1}^{(1)} \wedge 
\dd z_{\alpha,2}^{(1)} \wedge \dd z_{\alpha,3}^{(1)}\,.
\end{equation}
Here the subscript $*_\alpha$ and the superscript $*^{(1)}$ label the patches $U_\alpha$ on $Z_3$ 
as well as the patch $l_1\neq 0$ on the exceptional $\P^1$ with projective coordinates 
$(l_1:l_2)$\footnote{We drop the label $\alpha$ on the coordinates $l_i$ in order to shorten our formulas.}.
We obtain a similar expression on the second patch $l_2\neq 0$ of $\P^1$ using local coordinates 
$z_{\alpha,1}^{(2)}=\frac{l_1}{l_2}$, $z_{\alpha,2}^{(2)}=y_{\alpha,2}$ and  $z_{\alpha,3}^{(2)}=y_{\alpha,3}$. 
    
Now one can show that the pull-back map $\pi^*$ on $\Omega$ can be written as the residue
\begin{equation} \label{eq:HatOmegaLocal}
\hat{\Omega}=\int_{S^1_Q}\frac{h_i}{l_i}\frac{\Delta_{\P^1}}{Q}\wedge \Omega\,,\quad i=1,2\,.
\end{equation}
On can easily check that this is globally well-defined on both patches $l_i\neq 0$, $i=1,2$, covering 
the $\mathbb{P}^1$ using to the blow-up constraint $Q$ in \eqref{eq:def-hatU} respectively \eqref{eq:blowup}. 
Here we insert the local expression \eqref{eq:omegaLocal} for $\Omega$ and 
\begin{equation} 
\Delta_{\mathbb{P}^1}=l_1 \dd l_2- l_2 \dd l_1\,,
\end{equation}
which is the invariant top-form on $\P^1$.  
In fact, the residuum \eqref{eq:HatOmegaLocal} specializes correctly to the local expressions 
\eqref{eq:pullbacklocal} of $\hat{\Omega}$ in every chart. This ensures that the residuum expression on 
$\hat{U}_\alpha$ can be globalized to $\hat{Z}_3$. We use the standard residuum expression for the holomorphic 
three-form $\Omega$ given in \eqref{eq:residueZ3} to replace the local expression \eqref{eq:omegaLocal} by
\begin{equation}
\label{eq:ResZhat}
\hat \Omega =
\int_{S^1_P}\int_{S^1_Q} \frac{h_1}{l_1}\frac{\Delta}{P Q}=\int_{S^1_P}\int_{S^1_Q} \frac{h_2}{l_2}\frac{\Delta}{P Q}\ ,
\end{equation}
where $P,Q$ are the two constraints of \eqref{eq:blowup}. The five-form $\Delta$ denotes an invariant holomorphic 
top-form on the five-dimensional ambient space $\mathcal{W}$ defined in \eqref{eq:Wambient} and $S^1_P$, $S^1_Q$ 
are small loops around $\{P=0\}$, $\{Q=0\}$ encircling only the corresponding poles. 
The measure $\Delta$ is given explicitly in section \ref{sec:mirror_toric_branes}.
For the example of a trivial fibration it takes the schematic form 
\begin{equation}
 	\Delta=\Delta_{\mathbb{P}_{\Delta}}\wedge \Delta_{\mathbb{P}^1}\, .
\end{equation}
where $\Delta_{\mathbb{P}_{\Delta}}$ denotes the invariant top-form on the toric basis $\mathbb{P}_{\Delta}$. 

Let us now discuss the essential properties of $\hat{\Omega}$ and of the residue integral expression 
\eqref{eq:ResZhat}. By construction of $\hat{Z}_3$, in particular by the isomorphism 
$H^{(3,0)}(\hat{Z}_3)\cong H^{(3,0)}(Z_3)$, $\hat{\Omega}$ is the unique generator of $H^{(3,0)}(\hat{Z}_3)$ 
\cite{Grimm:2008dq}. In general $\hat{\Omega}$ varies under a deformation of the complex structure on 
$\hat{Z}_3$. This is due to the fact that the notion of holomorphic and anti-holomorphic coordinates 
changes when changing the complex structure. More rigorously, this is described by the variation of Hodge 
structures, where the split 
\begin{equation}
  H^3(\hat{Z}_3)=\bigoplus_{i=0}^3 H^{(3-i,i)}(\hat{Z}_3)
\end{equation}
by the Hodge type $(p,q)$ is analyzed over the complex structure moduli space ${\cal M}(\hat{Z}_3)$ of 
$\hat{Z}_3$. Then $\hat{\Omega}$ is a holomorphic section of the locally constant vector bundle 
$H^3(\hat{Z}_3)$ over ${\cal M}(\hat{Z}_3)$ and is of type $(3,0)$ at a fixed point $\underline{\hat{z}}_0$ 
in ${\cal M}(\hat{Z}_3)$. When moving away from $\underline{\hat{z}}_0$ by an infinitesimal displacement 
$\delta_{z}$ the Hodge type of $\hat{\Omega}$ changes according to the diagram
\begin{equation} \label{eq:HodgeClosed}
         H^{(3,0)}(\hat{Z}_3)={\mathcal F}^3\, \overset{\delta_z}{\longrightarrow} \,
         \mathcal F^2 \, \overset{\delta_z}{\longrightarrow} \, 
        \mathcal F^1 \, \overset{\delta_z}{\longrightarrow} \,
                {\mathcal F}^0=H^3(\hat{Z}_3)
\end{equation}
where we define the holomorphic vector bundles $\mathcal{F}^p$ over $\underline{\hat{z}}$ in ${\cal M}(\hat{Z}_3)$ by
\begin{equation} \label{eq:Hodgefilt}
 	\mathcal{F}^p\vert_{\underline{\hat{z}}}=\bigoplus_{i\geq p}H^{(i,3-i)}(\hat{Z}_3)\vert_{\underline{\hat{z}}}\,.
\end{equation}
Most importantly, this implies the existence of differential equations, the Picard-Fuchs equations, on 
${\cal M}(\hat{Z}_3)$ since the diagram \eqref{eq:HodgeClosed} terminates at fourth order in $\delta_z$. 
These can be explicitly obtained from the residue integral representation \eqref{eq:ResZhat} of $\hat{\Omega}$ 
by applying the Griffiths-Dwork reduction method. The Picard-Fuchs system in turn determines the full moduli 
dependence of $\hat{\Omega}$ and its periods.

Our general strategy will be to calculate all integrals 
relevant for the evaluation of the superpotential, discussed in section \ref{sec:fivebranesuperpotential}, 
on $\hat Z_3$ using this Picard-Fuchs system. In fact, in the toric examples of sections \ref{sec:ToricBraneBlowup}, 
\ref{sec:ToricBraneBlowupII} we obtain a GKZ-system whose solutions are the periods of $\Omega$ and the brane 
superpotential $W_{\rm brane}$. This, as explained in \ref{sec:unificationofdeformations}, unifies the closed 
and open deformations of $(Z_3,\Sigma)$, but, as we will see later in more detail, also the expression for 
individual pieces of the superpotential into a flux superpotential on $\hat{Z}_3$.  
Indeed, the unification of open-closed deformations, as mentioned before, as complex structure deformations 
on $\hat{Z}_3$ guarantees, that the study of variations of pure Hodge structures \eqref{eq:HodgeClosed} is 
sufficient \cite{Grimm:2008dq}. This is true despite the fact that $\hat{\Omega}$ vanishes\footnote{This can 
be directly seen, using the fact that $E$ is given in local coordinates by $z^{(1)}_{\alpha,1}=0$ respectively 
$z^{(2)}_{\alpha,2}=0$, from the local expression \eqref{eq:HatOmegaLocal} and its global counterpart \eqref{eq:ResZhat}.} 
as a section of $K\hat{Z}_3=E$ along the exceptional divisor $E$. Consequently $\hat{\Omega}$ naturally defines an 
element in open cohomology $H^3(\hat{Z}_3-E)\cong H^3(\hat{Z}_3,E)$, which in general carries a mixed Hodge 
structure \cite{Voisin2002}. However, the analysis of variations of this mixed Hodge structure reduces to the 
variation of the pure Hodge structure \eqref{eq:HodgeClosed} on the graded weight $Gr_3^WH^3(\hat{Z}_3-E)=H^3(\hat{Z}_3)$ 
since $E$ has no deformation\footnote{In the diagram of variations of mixed Hodge structures, the downward arrows 
corresponding to deforming $E$ do not exist, cf.~equation (4.40) of \cite{Grimm:2008dq}.}. To derive the 
superpotential as a solution of the Picard-Fuchs equations we also have to use an appropriate chain integral or 
current as explained below.

We conclude by discussing the expected structure of the Picard-Fuchs equations on ${\cal M}(\hat{Z}_3)$ from 
the residue \eqref{eq:ResZhat}. For a detailed discussion along the lines of concrete examples we refer to 
sections \ref{sec:ToricBraneBlowup} and \ref{sec:ToricBraneBlowupII}. In general all periods of $\hat{\Omega}$ 
over closed three-cycles in $\hat{Z}_3$ are solutions to this Picard-Fuchs system. First, we note that the 
Calabi-Yau constraint $P(\underline{x},\underline{a})$ appears both in the residues \eqref{eq:ResZhat} on 
$\hat{Z}_3$ as in well as in \eqref{eq:residueZ3}. The parameters $\underline{a}$ multiplying monomials in 
$P$ are identified, modulo the symmetries of the toric ambient space $\mathbb{P}_{\Delta}$, with complex 
structure moduli $\underline{z}$. Since monomials in $Q(\underline{x},\underline{b})$ are multiplied by 
independent parameters $\underline{b}$, the Picard-Fuchs operators $\mathcal{L}_k(\underline{a})$ annihilating 
$\Omega$, expressed by derivatives w.r.t.~$\underline{a}$, annihilate $\hat{\Omega}$ as well. Second, since 
$\mathbb{P}_{\Delta}$ is the basis of the $\P^1$-fibration ${\cal W}$ the toric symmetries of $\mathbb{P}_{\Delta}$ 
are contained in those of ${\cal W}$. They act on $P(\underline{x},\underline{a})$ in the same way in ${\cal W}$ 
as in $\mathbb{P}_{\Delta}$, but also act on the parameters $\underline{b}$ in the constraint $Q$. Therefore, the 
operators $\mathcal{Z}_i(\underline{a})$, expressing the toric symmetries of $\mathbb{P}_{\Delta}$ on $Z_3$, are 
lifted to $\hat{Z}_3$ as
\begin{equation}	\hat{\mathcal{Z}}_i(\underline{a},\underline{b})=\mathcal{Z}_i(\underline{a})+\mathcal{Z}'_i(\underline{b})\,,
\label{eq:Zslift}
\end{equation}
where the first operator is as before on $Z_3$ and the second operator just contains differentials of the 
$\underline{b}$. Then, \eqref{eq:Zslift} is easily solved by choosing the coordinates $\underline{z}$ as on 
$Z_3$. The new torus symmetries of $\mathcal{W}$, that correspond to its $\P^1$-fiber, do not involve the 
$\underline{a}$, but only the $\underline{b}$. They merely fix the variables $\underline{u}$ in terms of 
the parameters $\underline{a}$ and $\underline{b}$. Consequently, dividing out all toric symmetries of 
$\mathcal{W}$, the form $\hat\Omega$ depends on $\hat{\underline{z}}\equiv(\underline{z},\underline{u})$ only,
\begin{equation}
	\hat{\Omega}(\underline{a},\underline{b})\equiv \hat\Omega(\underline{z},\underline{u})=\hat{\Omega}(\hat{\underline{z}})
\end{equation}
and the differential operators $\mathcal{L}_k(\underline{a})$ take, possibly after a factorization to operators 
$\mathcal{D}_k(\underline{\hat{z}})$ of lower degree, the schematic form
\begin{equation}
	\mathcal{D}_k({\underline{\hat{z}}})=\mathcal{D}_k^{Z_3}(\underline{z})+\mathcal{D}'_k(\underline{z},\underline{u})
	\label{eq:LInZHat}
\end{equation}
in these coordinates. Here $\mathcal{D}_k^{Z_3}(\underline{z})$ are the Picard-Fuchs operators of $Z_3$ in 
the coordinates $\underline{z}$ and $\mathcal{D}'_k(\underline{z},\underline{u})$ are at least linear in 
derivatives w.r.t.~$\underline{u}$. It follows immediately that the Picard-Fuchs system for $\hat \Omega(\hat{\underline{z}})$ 
contains the Picard-Fuchs system for $\Omega(\underline{z})$, as determined by the $\mathcal{D}_k(\underline{\hat{z}})$, 
as a closed subsystem. Consequently the periods $\Pi^l(\underline{z})$ of $Z_3$ over closed three-cycles, that fulfill 
the differential equations $\mathcal{L}_k^{Z_3}(\underline{z})$, are also solutions to \eqref{eq:LInZHat}. We note that 
there are new operators ${\mathcal{L}}_m(\hat{\underline{z}})$ due to the constraint $Q$ that do not have a counterpart 
on $Z_3$. They form, together with \eqref{eq:LInZHat}, a complete differential system on ${\cal M}(\hat{Z}_3)$. However, 
the ${\mathcal{L}}_m(\hat{\underline{z}})$ are again at least linear in differentials of $\underline{u}$, thus act 
trivially on functions independent of $\underline{u}$, such as $\Pi^l(\underline{z})$.

Geometrically the lift of the periods of $\Omega$ to $\hat{Z}_3$ is a consequence of the 
isomorphism of $\pi:Z_3- \Sigma\rightarrow \hat Z_3- E$ and the fact that $\hat \Omega$ vanishes on $E$. 
The analog lift of the more interesting open periods on $Z_3$, in particular $W_{\rm brane}$ is discussed 
in the next section, section \ref{sec:potentialhatZ3minusD}, where we provide the corresponding 
expressions for the lifted superpotential on $\hat{Z}_3$. 

We conclude by mentioning that the structure \eqref{eq:LInZHat} allows to directly determine the inhomogeneous 
Picard-Fuchs equations obeyed by the domain wall tension $\mathcal{T}(\underline{z})$ between two five-branes,
\begin{equation}
 	\mathcal{D}_k^{Z_3}(\underline{z})\mathcal{T}(\underline{z})=f_k(\underline{z})\,.
\end{equation}
The tension $\mathcal{T}$ is obtained
as $\mathcal{T}(\underline{z})=W_{\rm brane}(\underline{z},\underline{u}^{rm c})-W_{\rm brane}(\underline{z},\underline{u}^{rm c}_0)$ 
and the inhomogeneity $f(\underline{z})$ upon evaluating $f_k(\underline{z})=\mathcal{D}'_k(\underline{z},\underline{u})W_{\rm brane}(\underline{z},\underline{u})\vert_{\underline{u}^{\rm c}}$, where $\underline{u}^{\rm c}$, $\underline{u}^{\rm c}_0$ 
are critical points of $W_{\rm brane}$. This inhomogeneous system was obtained from residues in 
\cite{Walcher:2009uj,Krefl:2008sj,Morrison:2007bm,Walcher:2006rs}.

\section{Lift of the  Superpotentials} 
\label{sec:potentialhatZ3minusD}

We have seen at the end of the previous section from the discussion of the Picard-Fuchs equations 
on $\hat{Z}_3$ that the periods $\Pi^l(\underline{z})$ over closed three-cycles lift from $Z_3$ 
to $\hat Z_3$. Illustratively this is clear because these integrals depend only on the geometry 
of $Z_3$ with the five-brane removed, cf.~section \ref{sec:N=1branes}. Thus, in order to lift the 
flux superpotential in \eqref{eq:fluxpot}, \eqref{eq:unifiedSuperpots}, which is just a linear 
combination of the periods $\Pi^l(\underline{z})$, we just need to lift the flux data.  
This is straightforward as the third cohomology of $\hat{Z}_3$ is given by \cite{Grimm:2008dq},
cf.~appendix \ref{App:topoHatZ_3}, 
\begin{equation}
	H^3(\hat{Z}_3)=\pi^* H^3(Z_3)\oplus H^3(E)
\label{eq:thirdcohom}
\end{equation}
so that any flux $G_3$ on $Z_3$ has a counterpart $\hat{G}_3=\pi^* G_3$ on $\hat{Z}_3$. Thus we 
readily obtain the lift of the flux superpotential to $\hat{Z}_3$ as
\begin{equation}
	W_{\rm flux}=\int_{\hat{Z}_3}\hat{\Omega}\wedge \hat{G}_3.
\label{eq:hatWflux}
\end{equation} 
Again the precise integral basis of cycles on $\hat{Z}_3$ for which this integral can be expanded 
in terms of periods of $\hat{\Omega}$ with integral coefficients has to be obtained by matching 
the classical terms at large radius and by assuring that the monodromy acts by integral transformations. 
Obviously, on $\hat Z_3$ flux configurations associated to the new three-cycles in $E$, which are 
not promoted from fluxes on $Z_3$ can be considered.

In order to lift the five-brane superpotential $W_{\rm brane}$ let us first make a local 
heuristic analysis, which casts already much of the general structure. Locally on a patch 
$U_\alpha$ we can write $\Omega=\dd \omega$ and evaluate the integral over the chain 
$\Gamma(u)$ leading to the five brane superpotential localized to the boundary 
$\partial \Gamma(u)=\Sigma-\Sigma_0$, 
\begin{equation}
	W_{\rm brane}=\int_{\Sigma}\omega\ .
\end{equation} 
Here we suppress the integral over the fixed reference curve $\Sigma_0$ in the same homology 
class as $\Sigma$, as it gives only rise to an irrelevant constant in $W_{\rm brane}$. 
Then we use the fact that the original curve $\Sigma$ is contained in the second cohomology 
$H^2(E)$ by its Poincare dual class \cite{Grimm:2008dq}. Thus we can write
\begin{equation}
	W_{\rm brane}=\int_E\pi^*(\omega)\wedge F_2
\label{eq:intoverD}
\end{equation}
where $[F_2]$  is the class of $\Sigma$ in the exceptional divisor $E$, i.e.
\begin{equation} 
[F_2]=\Sigma \ \ \ \text{in} \ \  E\ 
\end{equation} 
We note that at the supersymmetric minimum $F_2$ is equal to the K\"ahler form of the Fubini 
study metric on the $\mathbb{P}^1$, i.e.~$F_2=\omega_{FS}$. Locally \eqref{eq:intoverD} can be 
written as an integral on $\hat Z_3$
\begin{equation}
	W_{\rm brane}=\int_{\Gamma_5}\hat{\Omega}\wedge F_2
	\label{eq:hatWbrane}
\end{equation}
over a five-chain $\Gamma_5$ with $\partial\Gamma_5=E-E_0$ on which we extend $F_2$. $E_0$ 
denotes a reference divisor in the same homology as $E$, e.g. the blow-up of $\Sigma_0$, to match 
the constant contributions.

To prove this more rigorously it is instructive to consider the lift of the Bianchi identity to 
$\hat Z_3$. The formalism is equal for the three-form R-R field strength $F_3$ in type IIB and 
the three-form NS--NS field strength $H_3$ in the heterotic string. For the following 
analysis of the Bianchi identities, which is local along the curve $\Sigma$, one can focus on 
the source term $\delta_\Sigma$ of one five-brane neglecting the other terms in \eqref{eq:dF_3} 
respectively \eqref{eq:dH3}. The only aspect of the geometry which cannot be seen locally in a 
patch near a point in $\Sigma$ is the non triviality of the bundle $N_{\hat Z_3} E$, which is 
captured by its Thom class $\frac{e_1}{2}$. As can be calculated explicitly by evaluating the 
pull-back $\pi^*$ to the blow-up $\hat{Z}_3$, the form $H_3$ in \eqref{eq:dH3} is replaced 
by\footnote{Note that mathematically $\delta_E \wedge F_2 \equiv \cT(F_2)$ is given by the Thom 
isomorphism $\cT: H^\bullet(E) \rightarrow H^{\bullet +2}(Y)_{cpt}$ of the normal bundle 
$N_{\hat Z_3}E$ in $\hat Z_3$. $\cT$ maps cohomology classes on $E$ to compactly supported classes 
in $\hat Z_3$.} 
\begin{equation} \label{eq:dhatH_3}
  d\hat H_3 = \delta_E \wedge F_2=
\dd \rho \wedge \frac{e_1}{2} \wedge F_2 \ ,
\end{equation}
where the limit $\epsilon\rightarrow 0$ is implicit and we used 
\begin{equation}
\label{regdeltaD}
\lim_{\epsilon\rightarrow 0}\dd \rho \wedge \frac{e_1}{2}=\delta_E\ .
\end{equation}
Formally \eqref{eq:dhatH_3} can be integrated in the language of currents to 
\begin{equation}
\label{eq:hatH3}
\hat H_3=
\dd \rho \wedge \frac{e^{(0)}_{0}}{2} F_2+\dd B_2\ ,
\end{equation}
where $e_1=\dd e^{(0)}_0$ and a possible term $\rho \frac{e_1}{2}\wedge F_2$ is neglected by 
the requirement of regularity of $\hat{H}_3$. Thus, by pulling back both $H_3$ and $\Omega$ to 
$\hat{Z}_3$ we lift the superpotential \eqref{eq:unifiedSuperpots} as
\begin{equation} 
\label{eq:liftedbranesuperpotetial}
W_{\rm brane}=\lim_{\epsilon\rightarrow 0}\int_{\hat Z_3} \hat \Omega \wedge  \hat H_3
=\lim_{\epsilon\rightarrow 0}\int_{\hat{Z}_3}\hat{\Omega}\wedge F_2\wedge \rho e_1 
\end{equation}
where we restrict to the singular part \eqref{eq:hatH3} of $\hat{H}_3$ only. By \eqref{regdeltaD} 
and the identity \eqref{eq:hatH3} for $\hat{H}_3$ we see that this is equivalent to \eqref{eq:intoverD} 
and hence to \eqref{eq:hatWbrane}. We note that we can easily switch between the open manifold 
$\hat{Z}_3-E$ and $\hat{Z}_3$ in \eqref{eq:liftedbranesuperpotetial} since $\hat{\Omega}\vert_E=0$.
Mathematically, this match of \eqref{eq:liftedbranesuperpotetial} and the original superpotential 
$W_{\rm brane}$ in \eqref{eq:chainonopenset} follows more geometrically by the canonical identification 
$\hat \Omega|_{\hat Z_3- E}=\Omega|_{Z_3- \Sigma}$ under the biholomorphism $Z_3-\Sigma\cong \hat{Z}_3-E$, 
by $\hat{\Omega}\vert_E=0$ and the lift of the Thom classes, $\rho e_3\cong \rho e_1\wedge F_2$. 

Finally, we conclude by arguing that $W_{\rm brane}$ is a solution to the Picard-Fuchs system on 
$\hat{Z}_3$. In fact, this is confirmed for the examples of sections \ref{sec:ToricBraneBlowup}, 
\ref{sec:ToricBraneBlowupII} using the corresponding open-closed GKZ-system on $\hat{Z}_3$. All we 
have to ensure is that integration of $\hat{\Omega}$ over $\hat{H}_3$ given in \eqref{eq:hatH3}, which 
is not a closed form, commutes with the application of the Picard-Fuchs operators, that annihilate 
$\hat{\Omega}$. In addition, the whole integral has to be annihilated as well. Since $\hat{H}_3$ does 
not depend on the complex structure on $\hat{Z}_3$, all differential operators indeed commute with 
integration. Furthermore, for the GKZ-system of the form discussed below \eqref{eq:Zslift}, \eqref{eq:LInZHat}, 
the operators $\mathcal{L}_k(\underline{\hat{z}})$, $\hat{\mathcal{L}}_m(\underline{\hat{z}})$ annihilate 
$\hat{\Omega}$ identically and consequently also $W_{\rm brane}$ in \eqref{eq:liftedbranesuperpotetial}. 
However, the operators $\hat{\mathcal{Z}}_i(\underline{a},\underline{b})$ expressing the toric symmetries 
of $\mathcal{W}$ obey in general \cite{Batyrev:1994hm,cox1999mirror}
\begin{equation}
 	\hat{\mathcal{Z}}_i(\underline{a},\underline{b})\hat{\Omega}(\underline{a},\underline{b})=d\hat{\alpha}\,
\end{equation}
for a two-form $\hat{\alpha}$. This can potentially lead to a non-zero result in $W_{\rm brane}$ by partial 
integration since $d\hat{H}_3\neq 0$. In fact we exploit \eqref{eq:dhatH_3} to rewrite this as in integral over $E$,
\begin{equation}
 	\hat{\mathcal{Z}}_i(\underline{a},\underline{b})W_{\rm brane}=-\int_E \iota^*(\hat{\alpha})\wedge F_2\,,
\end{equation}
where $\iota:E\hookrightarrow \hat{Z}_3$ denotes the embedding of $E$. Fortunately, the pull-back $\iota^*(\hat{\alpha})$ 
vanishes on $E$ by the fact that we are dealing with variations of pure Hodge structures \eqref{eq:HodgeClosed} 
and thus can not reach the cohomology of $E$\ \footnote{As noted below \eqref{eq:Hodgefilt}, the cohomology of $E$ 
is included in the variations of mixed Hodge structures on $H^3(\hat{Z}_3-E)$ and can only be reached by deformations 
of $E$. These do not exist in our case since $E$ is rigid.}. Thus we have argued that $W_{\rm brane}$ is a solution of 
the GKZ-system on $\hat{Z}_3$. This will be further confirmed for the examples in section \ref{ch:CalcsBlowUp}.

\chapter{Superpotentials from Brane Blow-Ups}
\label{ch:CalcsBlowUp}

In this chapter we apply the blow-up proposal to a selection of examples of five-branes in compact 
Calabi-Yau threefolds. We consider two different Calabi-Yau threefolds, the one parameter example 
of the quintic and the two parameter Calabi-Yau in $\P^4(1,1,1,6,9)$, where the latter was the main 
example in chapter \ref{ch:Calcs+Constructions} \cite{Grimm:2009ef}. The type of branes we consider 
have the geometric interpretation of five-branes on rational curves, i.e.~holomorphic curves 
$\Sigma=\P^1$, at special loci in the open-closed deformation space. The number of open deformations 
is two. It is the common feature of both geometries that there is for generic values of the closed 
string moduli only a discrete number of such lines \cite{Katz:1992vx}, however, for special values 
of the moduli, at the Fermat point, an one parameter family of holomorphic curves. 

We start, mainly following \cite{Grimm:2010gk}, our discussion in section \ref{sec:ToricBraneBlowup} 
for the case of the quintic Calabi-Yau threefold. We show that five-branes on rational curves in the 
quintic can be described as toric branes $\Sigma$ at a special sublocus ${\cal M}(\P^1)$ of their 
moduli space $\mathcal{M}(\Sigma)$. This establishes the mapping of the moduli space 
$\mathcal{M}(\Sigma)$ with the obstructed deformations space of rational curves $\tilde{\P}^1$ as 
discussed in section \ref{sec:unificationofdeformations}. This is crucial since it enables us to 
work with a well-defined complex moduli space of $(Z_3,\Sigma)$ to describe the space of 
\textit{anholomorphic}, obstructed deformations of $\tilde{\P}^1$. 
From this description of the rational curve as a toric brane $\Sigma$ we readily construct the blow-up 
$\hat{Z}_3$. We study the complex structure moduli space of $\hat{Z}_3$ in section 
\ref{sec:toricbranesP11111-GKZ} by exploiting that $\hat{Z}_3$ is also governed by new toric data that 
is canonically related to the toric data of $(Z_3,\Sigma)$. First we construct the pull-back form 
$\hat{\Omega}$ on $\hat{Z}_3$ along the lines of section \ref{sec:hatomega}. Then we read of a toric 
GKZ-system that is associated to the underlying toric data of the blow-up $\hat{Z}_3$. From the 
GKZ-system we derive a complete Picard-Fuchs-system that we solve in section \ref{sec:toricbranesP11111-W} 
at various loci in the moduli space. These are the large radius point of $Z_3$ and the five-brane and selected 
discriminant loci of the Picard-Fuchs system for $\hat{Z}_3$. 
Since the periods of $\hat{\Omega}$ are also understood as periods on $\mathcal{M}(\Sigma)$, cf.~section 
\ref{sec:unificationofdeformations}, we finally obtain the brane superpotential $W_{\rm brane}$ for the 
line $\tilde{\P}^1$ as a linear combination of solutions fixed by an appropriate flux on $\hat{Z}_3$ that 
encodes the information about the five-brane. In particular we use open mirror symmetry to obtain 
predictions for the disk invariants at large volume, that match and extend independent results in the 
literature. Similarly, the disk instantons are obtained for a different brane phase in section 
\ref{sec:toricbranesP11111-W-LV}. These are the first available results for branes with \textit{two} open 
string deformations.

In section \ref{sec:generaltoricstructure} we discuss the toric structure of $\hat{Z}_3$ obtained by 
blowing up toric curves $\Sigma$. We give a general recipe in section \ref{sec:generaltoricstructure} 
to obtain a toric polyhedron $\Delta_7^{\hat Z}$ from the charge vectors of the geometry $Z_3$ and 
the curve $\Sigma$, that can be efficiently used to obtain the open-closed GKZ-system for the brane 
deformation problem associated to arbitrary toric curves $\Sigma$. We comment on a connection to 
Calabi-Yau fourfolds which applies to special choices of toric branes, but \textit{not} in general.
Next in section \ref{sec:ToricBraneBlowupII} we consider another two parameter example of the 
elliptically fibered Calabi-Yau threefold in $\P^6(1,1,1,6,9)$ with a five-brane supported on a 
rational curve. There we make similar use of the toric GKZ-system on $\hat{Z}_3$ that we solve completely
and from which we construct the five-brane superpotential. Under toric 
mirror symmetry we obtain the disk instanton invariants of the dual A-model geometry, cf.~appendix \ref{sec:instantonsLV11169}. 
In addition, we comment on the connection to heterotic/F-theory duality.

Finally in section \ref{sec:heteroticF+blowup} we connect, following \cite{Grimm:2009sy}, the 
construction of the blow-up threefold with heterotic/F-theory duality as reviewed in section 
\ref{sec:HetFDuality} and applied in section \ref{sec:SuperpotsHetF}. We demonstrate that the 
blow-up $\hat{Z}_3$ for a horizontal five-brane on a curve $\Sigma$ in the heterotic theory 
can be understood as the heterotic dual to the blow-up in the F-theory fourfold $X_4$ due to 
the presence of the horizontal heterotic five-brane. 
More concretely we start in section \ref{sec:heterotic_blowup} by blowing up the heterotic 
threefold $Z_3$ along the five-brane curve $\Sigma$ into $\hat Z_3$ \cite{Grimm:2008dq} which 
we realize as a complete intersection. Then, according to the discussion of section \ref{sec:5braneblowupsanddefs},
the non-Calabi-Yau threefold $\hat{Z}_3$ contains the five-brane moduli as a subsector of its complex
structure moduli. We propose, and explicitly demonstrate for examples, that the F-theory geometry 
$X_4$ can in turn be entirely constructed from $\hat{Z}_3$. In particular, this identification becomes 
apparent when also realizing $X_4$ as a complete intersection. Moreover, in this way the complex structure 
moduli of $\hat{Z}_3$ naturally form a subsector of the complex structure moduli of $X_4$.
Finally in section \ref{sec:non-CYblowup} we exploit the matching of the geometrical data of $\hat{Z}_3$
and $X_4$, together with the lift of the heterotic superpotential to the blow-up $\hat{Z}_3$ as in section 
\ref{sec:potentialhatZ3minusD}, to match the heterotic superpotential with the F-theory flux superpotential.
We conclude with a formal matching of the fluxes.
In summary, the general idea of this discussion is to reformulate and slightly
extend the heterotic/F-theory duality map schematically as:
\begin{equation*}
	\xymatrix @C=-1in {
	\parbox{7cm}{\centering Heterotic string on CY threefold $Z_3$,\\ vector bundle $E$, five-brane on $\Sigma$} \ar[rd]\ar@{<->}[rr] & & \parbox{6cm}{\centering F-theory on CY fourfold $ X_4$\\ blown up along $\Sigma$, $G_4$-flux}\\
	& \parbox{7cm}{\centering non-Calabi-Yau $\hat Z_3$ blown up along $\Sigma$,\\ vector bundle $\hat E$, flux $F_2=[\Sigma]_E$} \ar[ru]&
	}
\end{equation*}
where the horizontal arrow indicates the action of heterotic/F-theory duality.

Before we start, let us summarize some other approaches to find open-closed Picard-Fuchs equations
for the brane superpotential. The most thoroughly studied cases are non-compact toric Calabi-Yau 
manifolds with Harvey-Lawson type branes \cite{Aganagic:2000gs,Aganagic:2001nx}. It was first proposed  
and demonstrated for non-compact toric Calabi-Yau threefolds in \cite{Mayr:2001xk,Lerche:2001cw,Lerche:2002ck,Lerche:2002yw} 
to use open-closed Picard-Fuchs systems to obtain the brane superpotential.
For D5-branes in compact geometries the superpotential was calculated in 
\cite{Walcher:2009uj,Krefl:2008sj,Morrison:2007bm,Walcher:2006rs,Knapp:2008uw,Knapp:2008tv} by deriving and solving 
inhomogeneous Picard-Fuchs equations. Finally the methods of \cite{Lerche:2002ck,Lerche:2002yw} have
been extended to compact geometries in \cite{Jockers:2008pe} and further applied in 
\cite{Alim:2009rf,Alim:2009bx,Aganagic:2009jq,Li:2009dz,Jockers:2009mn,Baumgartl:2010ad,Alim:2010za,Fuji:2010uq,Shimizu:2010us}.

\section{Open-Closed Picard-Fuchs Systems on the Quintic}
\label{sec:ToricBraneBlowup}

We now apply the blow-up proposal of section \ref{sec:5braneblowupsanddefs} to the case of 
toric curves\footnote{We emphasize that the five-brane does \textit{not} wrap $\Sigma$ for 
generic values of the closed and open moduli.} $\Sigma$ on the one parameter quintic. As 
explained in section \ref{sec:unificationofdeformations} the moduli space ${\cal M}(\Sigma)$ 
of $\Sigma$ is identified with the obstructed deformation space of rational curves $\tilde{\P}^1$, 
on which we wrap a five-brane. The holomorphic $\P^1$ is directly visible from the complete 
intersection description of $\Sigma$ at a special sublocus ${\cal M}(\P^1)$ of  ${\cal M}(\Sigma)$ 
where $\Sigma$ degenerates appropriately modulo the action of the quintic orbifold 
$G=(\mathds{Z}_5)^3$. This way, we understand the five-brane on a rational curve $\P^1$ as a 
special case of a toric brane and consequently apply open mirror symmetry along the lines of 
section \ref{sec:mirror_toric_branes}. Thus, we start in section \ref{sec:toricbranesP11111} 
with the toric curve $\Sigma$, then determine the sublocus ${\cal M}(\P^1)$ and identify the 
wrapped rational curve $\P^1$ that we represent via the standard Veronese embedding in $\P^4$. 
After this definition of the deformation problem, we determine the Picard-Fuchs system on the 
deformation space of the rational curves $\tilde{\P}^1$ from the GKZ-system on the complex 
structure moduli space of the blow-up $\hat{Z}_3$ along $\Sigma$ in section 
\ref{sec:toricbranesP11111-GKZ}. The solutions of this system including the large volume 
expressions for the flux and brane superpotential, the open-closed mirror map and the disk 
instantons are summarized in section \ref{sec:toricbranesP11111-W}. Furthermore, we find 
the solutions of the Picard-Fuchs system at its discriminant loci, namely in the vicinity 
of the sublocus ${\cal M}(\P^1)$ and at the involution brane, where we also identify the 
superpotential $W_{\rm brane}$. Finally in section \ref{sec:toricbranesP11111-W-LV} we obtain 
the disk invariants for a different brane geometry at large volume.

\subsection{Branes on Lines in the Quintic and the Blow-Up}
\label{sec:toricbranesP11111}

The quintic Calabi-Yau $\tilde{Z}_3$ is given as the general quintic hypersurface $\tilde{P}$ in 
$\P^4$. It has 101 complex structure moduli corresponding to the independent coefficients of the
monomials entering $\tilde{P}$. Its K\"ahler moduli space is one-dimensional generated by the 
unique K\"ahler class of the ambient $\P^4$. Toric Lagrangian submanifolds of this geometry were 
discussed in \cite{Aganagic:2000gs} along the lines of section \ref{sec:mirror_toric_branes}.

The mirror quintic threefold $Z_3$ is given as the hypersurface
\begin{equation}
 Z_3\,:\quad P=x_1^5+x_2^5+x_3^5+x_4^5+x_5^5-5\Psi x_1x_2x_3x_4x_5\,,
\label{eq:mirrorquintic}
\end{equation}
where $\Psi$ denotes its complex structure modulus. It is obtained via \eqref{eq:HVmirror} from 
the toric data
\begin{equation} \label{eq:quinticTD}
 	\begin{array}{rlllll}
 	 	 \ell^{(1)}=(-5, & 1, & 1,& 1,& 1, &1) \\ \hline \hline
		y_0 & y_1 & y_2 & y_3 & y_4 & y_5\\ 
		x_1x_2x_3x_4x_5 &x_1^5 & x_2^5 & x_3^5  & x_4^5 & x_5^5 
 	\end{array}
\end{equation}
where the $y_i$ corresponding to the entries $\ell^{(1)}_i$ of the charge vector are given as monomials 
in the projective coordinates $x_i$ constructed using the formula \eqref{eq:Z3typeIIB}. In addition we 
divide by an orbifold group $G=(\mathds{Z}_5)^3$ that acts on the coordinates so that $x_1x_2x_3x_4x_5$ 
is invariant. A convenient basis of generators $g^{(i)}$ is given by $v^{(i)}=(1,-1,0,0,0) \mod 5$ and 
all permutations of its entries where we use
\begin{equation}
	g^{(i)}:\,\,x_k\mapsto e^{2\pi i v^{(i)}_k/5} x_k\,.
\label{eq:quinticorbifold}
\end{equation}
We note that the Fermat point $\Psi=0$ is a point of enhanced symmetry where $G$ enhances to $(\mathds{Z}_5)^4$. 
As required by mirror symmetry we have $h^{(2,1)}(Z_3)=1$, $h^{(1,1)}(Z_3)=101$. 
 
Next we introduce an open string sector by putting a five-brane on a line $\P^1$ in the quintic $Z_3$. 
Following the above logic we first construct a toric curve $\Sigma$ to define the deformation space 
$\tilde{\P}^1$ of the rational curve. The holomorphic $\P^1$ is then obtained at the sublocus ${\cal M}(\P^1)$ 
where $\Sigma$ degenerates accordingly. Up to a relabeling of the projective coordinates $x_i$ of $\P^4$ 
we consider the toric curves $\Sigma$ given by
\begin{eqnarray} \label{eq:RSquintic}
	\Sigma&:& P=0\,,\quad h_1\equiv \beta^5x_3^5-\alpha^5x_4^5=0\,,\quad h_2\equiv\gamma^5x_3^5-\alpha^5x_5^5=0\,, \\
	&&\hat{\ell}^{(1)}=(0,0,0,1,-1,0)\,,\quad\hat{\ell}^{(2)}=(0,0,0,1,0,-1)\,,\nonumber
\end{eqnarray} 
where the brane charge vectors $\hat{\ell}^{(i)}$ correspond to the constraints $h_i$ using \eqref{eq:BBrane} 
and the toric data \eqref{eq:quinticTD}. The complete intersection \eqref{eq:RSquintic} describes for all 
values of the parameters $\alpha$, $\beta$ and $\gamma$, that take values in $\P^2$, an analytic family of 
holomorphic curves in the quintic. Consequently, $\alpha$, $\beta$ and $\gamma$ parameterize the unobstructed 
moduli space of $\Sigma$ on which we introduce coordinates $u^1=\frac{\beta^5}{\alpha^5}$ and 
$u^2=\frac{\gamma^5}{\alpha^5}$. 

The obstructed deformation problem is defined by the definition of a non-holomorphic family $\tilde{\P}^1$ and 
the identification of the locus ${\cal M}(\P^1)$. As discussed before in section 
\ref{sec:unificationofdeformations} the obstructed deformation space of lines is identified with the moduli space 
${\cal M}(\Sigma)$. For generic values of the moduli in \eqref{eq:RSquintic} the curve $\Sigma$ is an irreducible 
higher genus Riemann surface. However, we can always linearize \eqref{eq:RSquintic} for generic values of the moduli,
\begin{equation} \label{eq:anholoC}
	\tilde{\P}^1:\quad \eta_1 x_1+\sqrt[5]{x_2^5+x_3^3 m(x_1,x_2,x_3)}=0\,,\quad  \eta_2\beta x_3-\alpha x_4=0\,,\quad \eta_3\gamma x_3-\alpha x_5=0\,,
\end{equation} 
Here we inserted $h_1$ and $h_2$ into $P$, introduced fifths roots of unity $\eta_i^5=1$ and the polynomial
\begin{equation}
	m(x_1,x_2,x_3)=\frac{\alpha^5+\beta^5+\gamma^5}{\alpha^{5}}x_3^2-5\Psi\frac{\beta\gamma}{\alpha^2} x_1x_2\,.  
\label{eq:quinticDiv}
\end{equation} 
This equation \eqref{eq:anholoC} is evidently non-holomorphic due to the non-trivial branching of the fifth root, 
in other word $\tilde{\P}^1$ is a non-holomorphic family of rational curves in the quintic. 
However at special loci $\Sigma$ degenerates as follows. We rewrite \eqref{eq:RSquintic} as 
\begin{equation}
 \Sigma\,:\quad x_1^5+x_2^5+x_3^3 m(x_1,x_2,x_3)=0 \,,\quad \alpha^5 x_4^5-\beta^5 x_3^5=0\,,\quad \alpha^5 x_5^5- \gamma^5 x_3^5=0\,.
\end{equation}
Whereas $h_1$, $h_2$ can be linearized for generic values of the moduli, $m(x_1,x_2,x_3)$ forbids a holomorphic 
linearization of \eqref{eq:RSquintic} and accordingly to take the fifths root in \eqref{eq:anholoC}. However, at 
the sublocus
\begin{equation}
	\mathcal{M}_{\P^1}(\Sigma)\,:\quad \alpha^5+\beta^5+\gamma^5=0\,,\quad \Psi\alpha\beta\gamma=0\,
\label{eq:modulicond}
\end{equation}
the polynomial $m$ vanishes identically and the Riemann surface $\Sigma$ in \eqref{eq:RSquintic} degenerates to
\begin{equation}
	\Sigma:\quad h_0\equiv x_1^5+x_2^5\,,\quad  h_2=\beta^5x_3^5-\alpha^5x_4^5=0\,,\quad h_2=\gamma^5x_3^5-\alpha^5x_5^5=0\,.
\label{eq:tbquintic_h0}
\end{equation}
This can be trivially factorized as in the general discussion \eqref{eq:sublocus} in linear factors that differ 
only by fifths roots of unity $\eta_i$, that are the 125 solutions to \eqref{eq:anholoC}. In other words, at 
the locus ${\cal M}(\P^1)$ the curve $\Sigma$ degenerates to $125$ lines corresponding to each choice of $\eta_i$ 
in the three constraints $h_i$. However, in contrast to \eqref{eq:RSquintic} which is invariant under the 
orbifold $G$, the linearized equations do transform under $G$. In fact, all the $125$ different lines are 
identified modulo the action of $G=(\mathds{Z}_5)^3$ so that \eqref{eq:tbquintic_h0} describes a single line 
on the quotient by $G$,
\begin{equation}
	 {\cal M}(\P^1):\quad\eta x_1+x_2=0\,,\quad \alpha x_4-\beta x_3=0\,,\quad \alpha x_5-\gamma x_3=0\,.
\label{eq:constlines}	
\end{equation}
Equivalently, these lines are given parametrically in $\P^4$ in terms of homogeneous coordinates $U$, $V$ on $\P^1$ 
as the Veronese mapping
\begin{equation}
 	\quad(U,V)\,\,\mapsto\,\, (U,-\eta U,\alpha V,\beta V,\gamma V)\,,\quad \eta^5=1\,.
\label{eq:paramlines}
\end{equation}
This way, the family $\Sigma$ contains the holomorphic lines \eqref{eq:paramlines} at the sublocus ${\cal M}(\P^1)$ 
of \eqref{eq:modulicond} defined by the vanishing of $m(x_1,x_2,x_3)$. In summary this shows that a five-brane 
wrapping the line \eqref{eq:paramlines} falls in the class of toric branes at the sublocus ${\cal M}(\P^1)$ of 
their moduli space. We emphasize again that \eqref{eq:paramlines} is not invariant under the orbifold group $G$ 
and that the identification of the 125 distinct solutions to \eqref{eq:anholoC} under $G$ is essential to match 
an in general higher genus Riemann surface $\Sigma$ with a rational curve of genus $g=0$.

This picture is further confirmed from the perspective of the rational curve \eqref{eq:paramlines} since the 
constraint \eqref{eq:modulicond} defining ${\cal M}(\P^1)$ is precisely the condition for the line 
\eqref{eq:paramlines} to lie holomorphically in the quintic constraint $P$. Thus, the sublocus ${\cal M}(\P^1)$ 
defined in \eqref{eq:modulicond} is precisely the moduli space of the five-brane wrapping the holomorphic lines 
in the quintic. For generic $\Psi\neq 0$ this moduli space is only a number of discrete points whereas at the 
Fermat point $\Psi=0$ there is a one-dimensional moduli space of lines in the quintic parametrized by a Riemann 
surface\footnote{The first constraint in \eqref{eq:modulicond} is a quintic constraint in $\P^2$ describing a 
Riemann surface of genus $g=6$.} of genus $g=6$, cf.~figure \ref{fig:moduli}. In the language of superpotentials, 
we understand ${\cal M}(\P^1)$ as the critical locus of $W_{\rm brane}$ at which the five-brane on the rational 
curve is supersymmetric. Conversely, deforming away from the critical locus ${\cal M}(\P^1)$ in ${\cal M}(\Sigma)$ 
is obstructed, inducing a non-trivial superpotential. Thus, we consider in the following the deformation space 
defined by $\tilde{\P}^1$, more precisely by the coefficients of $m$,
\begin{equation}
 	\hat{z}^1=\frac{\beta\gamma}{\alpha^2}=(u^1u^2)^{\frac{1}{5}}\,,\quad \hat{z}^2=\frac{\alpha^5+\beta^5+\gamma^5}{\alpha^{5}}=1+u^1+u^2\,, 
\end{equation}
which agrees with the choice of variables used in figure \ref{fig:moduli}. As noted before, we can canonically 
identify this deformation space with the moduli space ${\cal M}(\Sigma)$ of $\Sigma$ in $Z_3$ by dividing out 
the orbifold group $G$ and working with the holomorphic constraint \eqref{eq:RSquintic} instead of \eqref{eq:anholoC}.

Most importantly for the blow-up procedure, the description of the toric curve $\Sigma$ of \eqref{eq:RSquintic} 
is precisely in the form used in section \ref{sec:geometricblowups} to construct the blow-up geometry $\hat{Z}_3$. 
In particular, we can easily read off the normal bundle $N_{Z_3}\Sigma$ of $\Sigma$ in the quintic which is  
$N_{Z_3}\Sigma=\mathcal{O}(5)\oplus\mathcal{O}(5)$ by simply noting the degree of the divisors $h_1=0$, $h_2=0$. 
Then the blow-up $\hat{Z}_3$ is given by the complete intersection \eqref{eq:blowup}, which in the case at hand reads
\begin{equation}
	\hat{Z}_3\,:\quad P=0\,,\quad Q=l_1(u^2 x_3^5-x_5^5)-l_2(u^1 x_3^5-x_4^5)=0\,.
\label{eq:BUquintic}
\end{equation}   
Since both the closed modulus $\Psi$ as well as the open moduli $u^1$, $u^2$ enter \eqref{eq:BUquintic}, we 
formally obtain the embedding of the open-closed moduli space of $(Z_3,\Sigma)$, and equivalently the obstructed 
deformation space of $(Z_3,\tilde{\P}^1)$, into the complex structure moduli space of $\hat{Z}_3$. In particular 
this trivially embeds the moduli space of the rational curves \eqref{eq:paramlines} by restricting to the critical 
locus \eqref{eq:modulicond}. 

On the blow-up $\hat{Z}_3$ this embedding as well as the obstructions can be understood purely geometrically. 
First of all we note that the action of the quintic orbifold $G$ directly lifts to $\hat{Z}_3$. Then by deforming 
away from the critical values \eqref{eq:modulicond} we change the topology of the blow-up divisor $E$ from a 
ruled surface over $\P^1$ to a ruled surface $E$ over a Riemann-surface $\Sigma$ of higher genus. The one-cycles 
of the Riemann-surface in the base lift to new three-cycles on the blow-up $\hat{Z}_3$ that correspond to new 
non-algebraic complex structure deformations\footnote{Although related these new complex structure deformations 
should not be confused with the parameters entering $Q$ since these are algebraic by definition.}, compare to the 
similar discussion of \cite{Kachru:2000ih,Kachru:2000an}. Upon switching on flux on these three-cycles turns on 
higher order obstructions for the complex structure of $\hat{Z}_3$ destroying the ruled surface $E$ and thus 
driving us back to the critical locus where $\Sigma$ degenerates to $\P^1$. This way the flux obstructs the 
complex structure in \eqref{eq:BUquintic} which is expressed by a flux superpotential on $\hat{Z}_3$ that is the 
sought for superpotential $W_{\rm brane}$. 

In the following we will use the complete intersection \eqref{eq:BUquintic} to analyze the open-closed deformation 
space $(Z_3,\tilde{\P}^1)$. The crucial point is that we are working with a well-defined complex moduli space of 
$(Z_3,\Sigma)$ respectively of complex structures on $\hat{Z}_3$ to describe the space of \textit{anholomorphic} 
deformations $\tilde{\P}^1$. In this context this is another reason for the effectiveness of the blow-up $\hat{Z}_3$ 
for the description of the obstructed brane deformations $\tilde{\P}^1$. 

\subsection{Toric Branes on the Quintic: the Open-Closed GKZ-System}
\label{sec:toricbranesP11111-GKZ}

In the following we analyze the open-closed deformation space of $(\tilde{\P}^1,Z_3)$ embedded in 
the complex structure moduli of $\hat{Z}_3$ augmented by appropriate flux data. We perform this analysis 
by toric means, i.e.~the GKZ-system. Thus we supplement the polyhedron $\Delta^{\tilde Z}_4$ and the 
charge vectors $\ell^{(1)}$, $\hat{\ell}^{(1)}$, $\hat{\ell}^{(2)}$ of the quintic Calabi-Yau 
\eqref{eq:mirrorquintic} and the toric brane \eqref{eq:RSquintic},
\begin{equation}\label{quinticpoly}
	\begin{pmatrix}[c|cccc|c|l||cc]
	    	&   &  \Delta_4^{\tilde Z} &   &    &  \ell^{(1)} &   &\hat{\ell}^{(1)}&\hat{\ell}^{(2)}\\ \hline
		\tilde{v}_0 & 0 & 0 & 0 & 0 	&  -5    & y_0 = x_1x_2x_3x_4x_5 & 0&0\\
		\tilde{v}_1 &-1 &-1 &-1 & -1 	&  1     & y_1 = x_1^5 & 0&0\\
		\tilde{v}_2 & 1 & 0 & 0 & 0 	&  1     & y_2 = x_2^5 & 0&0\\
		\tilde{v}_3 & 0 & 1 & 0 & 0 	&  1     & y_3 = x_3^5 & 1&1\\
		\tilde{v}_4 & 0 & 0 & 1 & 0 	&  1     & y_4 = x_4^5 & -1&0\\
	  \tilde{v}_5 & 0 & 0 & 0 & 1 	&  1     & y_5 = x_5^5 & 0&-1
	\end{pmatrix}.
\end{equation}
The points of the dual polyhedron $\Delta^{ Z}_4$ are given by $v_1=(-1,-1,-1,-1)$, $v_2=(4,-1,-1,-1)$, 
$v_3=(-1,4,-1,-1)$, $v_4=(-1,-1,4,-1)$ and $v_5=(-1,-1,-1,4)$. These monomials both enter the constraints 
$P$ and $h_i$ according to \eqref{eq:Z3typeIIB} and \eqref{eq:BBrane} yielding
\begin{equation}
	Z_3\,:\quad P=\sum_{i=1}^5a_ix_i^5+a_0 x_1x_2x_3x_4x_5\,,\quad \Sigma\,:\quad h_1=a_6 x_3^5+a_7x_4^5\,,\quad h_2=a_8 x_3^5+a_9x_5^5\,,
\label{eq:toricquintic}
\end{equation}
where we introduced free complex-valued coefficients $\underline{a}$.\footnote{Conversely to the 
conventions in \eqref{eq:Zslift}, we denote the parameters $b_i$ by $a_{5+i}$ for convenience.} 
From the polyhedron \eqref{quinticpoly} we readily obtain the standard toric GKZ-system for $Z_3$ 
along the lines of eqs. \eqref{eq:pfo} and \eqref{eq:Zs},
\begin{eqnarray}\label{eq:GKZquinticclosed}
 	&\mathcal{Z}_0=\sum_{i=0}^5\vartheta_i+1\,,\quad \mathcal{Z}_i=\vartheta_{i+1}-\vartheta_1\,\,(i=1,\ldots,4)\,,\nonumber&\\
	&\displaystyle\mathcal{L}_{1}=\prod_{i=1}^5\frac{\partial}{\partial a_i}-\frac{\partial^5}{\partial a_0^5}\,,\displaystyle&
\end{eqnarray}
where we use the logarithmic derivative $\vartheta_i=a_i\frac{\partial}{\partial a_i}$. The 
$\mathcal{Z}_i$ express the coordinate rescalings leaving the measure $\Delta$ and the monomial 
$x_1x_2x_3x_4x_5$ in \eqref{eq:residueZ3} invariant. They express infinitesimal rescalings of the 
parameters $\underline{a}$ and the coordinates $\underline{x}$ entering $P$. For example the 
rescaling $(x_1,x_2)\mapsto (\lambda^{1/5} x_1,\lambda^{-1/5}x_2)$ combined with 
$(a_1,a_2)\mapsto (\lambda^{-1} a_1,\lambda a_2)$ leaves $P$ invariant and consequently the 
periods have the symmetry $\Pi^k(a_0,\lambda^{-1}a_1,\lambda a_2,a_3,a_4,a_5)=\Pi^k(\underline{a})$. 
The corresponding generator of this symmetry is $\mathcal{Z}_1$. These homogeneity properties of 
the $\Pi^k(a_i)$ imply that they are functions of only a specific combination of the $\underline{a}$, 
which in the case of the quintic takes the form
\begin{equation}
	z^1=-\frac{a_1a_2a_3a_4a_5}{a_0^5}\,.
\end{equation}
This is perfectly consistent with \eqref{eq:algCoords} and the charge vector $\ell^{(1)}$.

The analysis of the combined system of the quintic and the curve \eqref{eq:toricquintic} is 
performed by replacing $(Z_3,\Sigma)$ by the blow-up $(\hat{Z}_3,E)$ given by the family of 
complete intersections in $\mathcal{W}=\P(\mathcal{O}(5)\oplus \mathcal{O}(5))\cong \P(\mathcal{O}\oplus \mathcal{O})$,
\begin{equation}
	\hat{Z}_3\,:\quad P=0\,,\quad Q=\ell_1(a_8 x_3^5+a_9x_5^5)-\ell_2(a_6x_3^5+a_7x_4^5)\,.
\end{equation}
Then the holomorphic three-form $\hat{\Omega}$ is constructed using the residue \eqref{eq:ResZhat}. 
Using this explicit residue integral expression, it is straightforward to find the Picard-Fuchs 
system on the blow-up $\hat{Z}_3$ that encodes the complex structure dependence of $\hat{\Omega}$. 
As it can be directly checked the GKZ-system is given by $\mathcal{L}_{1}$ as in \eqref{eq:GKZquinticclosed} 
complemented to the system 
\begin{eqnarray}\label{eq:GKZquinticopen}
 	&\mathcal{Z}_0=\sum_{i=0}^5\vartheta_i+1\,,\quad \mathcal{Z}_1=\sum_{i=6}^9\vartheta_i\,,\quad \mathcal{Z}_2=\vartheta_2-\vartheta_1\,,\quad \mathcal{Z}_3=\vartheta_3-\vartheta_1+\vartheta_6+\vartheta_8\,,&\nonumber\\[0.7Em]
 	& \mathcal{Z}_4=\vartheta_4-\vartheta_1+\vartheta_7\,,\quad \mathcal{Z}_5=\vartheta_5-\vartheta_1+\vartheta_9\,,\quad \mathcal{Z}_6=\vartheta_8+\vartheta_9-\vartheta_6-\vartheta_7\,,\displaystyle&
\end{eqnarray}
encoding the torus symmetries of $\mathcal{W}$ and the actual Picard-Fuchs operators 
\begin{eqnarray}
	&\displaystyle\mathcal{L}_{1}=\prod_{i=1}^5\frac{\partial}{\partial a_i}-\frac{\partial^5}{\partial a_0^5}\,,\quad\mathcal{L}_{2}=\frac{\partial^2}{\partial a_3\partial a_7}-\frac{\partial^2}{\partial a_4\partial a_6}\,,\quad \mathcal{L}_{3}=\frac{\partial^2}{\partial a_3\partial a_9}-\frac{\partial^2}{\partial a_5\partial a_8}\,.\displaystyle
\end{eqnarray} 
We emphasize that there are two new second order differential operators $\mathcal{L}_{2}$, 
$\mathcal{L}_{3}$ that annihilate $\hat{\Omega}$ identically and that incorporate the deformations 
$a_i$, $i=6,7,8,9$ associated to the curve $\Sigma$. It is clear from the appearance of the 
$\P^1$-coordinates $(l_1,l_2)$ in the constraint $Q$ that there are no further operators 
$\mathcal{L}_{a}$ on $\hat{Z}_3$ of minimal degree. Let us briefly explain the origin of the 
operators $\mathcal{Z}_k$. The first two are simply associated to an overall rescaling of the 
two constraints $P\mapsto \lambda P$, $Q\mapsto \lambda' Q$ which acts on $\hat{\Omega}(\underline{a})$ 
as $\hat{\Omega}(\lambda a_0,\ldots, \lambda a_6,a_7,\ldots, a_{10})=\lambda\hat{\Omega}(\underline{a})$ 
and $\hat{\Omega}(a_0,\ldots, a_6,\lambda'a_7,\ldots, \lambda'a_{10})=\hat{\Omega}(\underline{a})$. 
For the rescaling of $Q$ the factor $\lambda'$ is compensated by the non-trivial prefactor $h_i/\ell_i$ 
in \eqref{eq:ResZhat}. The third to sixth operators are associated to the torus symmetries of the 
$\P^4$ as before, $(x_1,x_j)\mapsto(\lambda_j x_1,\lambda_j^{-1}x_j)$, $j=2,\ldots,5$, and the last 
operator $\mathcal{Z}_6$ is related to the torus symmetry $(l_1,l_2)\mapsto (\lambda l_1,\lambda^{-1}l_2)$ 
of the exceptional $\P^1$. It is important to note that the operators $\mathcal{Z}_i$ of the $\P^4$ 
are altered due to the blow-up $\hat{Z}_3$, i.~e.~due to the presence of the five-brane, as compared 
to the closed string case of \eqref{eq:GKZquinticclosed}. 

Before delving into the determination of the solutions to this differential system let us reconsider 
the operators we just found from a slightly different perspective. This will in particular allow for 
a straightforward systematization of the constructions of GKZ-system.
Comparing \eqref{eq:GKZquinticopen} to the closed GKZ-system \eqref{eq:GKZquinticclosed} associated 
to $Z_3$ we recover a very similar structure. Indeed the above differential system governing 
the complex structure on $\hat{Z}_3$ defines a new GKZ-system with exponent $\beta$. To obtain the 
set of integral points $\hat{v}_i$ associated to this GKZ-system we apply the general definition of 
the $\mathcal{Z}_i$ in \eqref{eq:Zs} backwards to obtain
\begin{equation}\label{blowupPolyquintic}
	\begin{pmatrix}[c|ccccccc|ccc|l]
	    		     &  &   &   & \Delta_7^{\hat Z}&&&   		  &\hat{\ell}^{(1)} &\hat{\ell}^{(2)} &\hat{\ell}^{(3)}  &          \\ \hline
		\hat{v}_0    &1 & 0 & 0 & 0 & 0 & 0 & 0   &-5   & 0  		  &	0	&	 \hat{y}_0 = x_1x_2x_3x_4x_5 \\
		\hat{v}_1    &1 & 0 &-1 &-1 &-1 &-1 & 0	  & 1   & 0  		  & 0	&	 \hat{y}_1 = x_1^5           \\
		\hat{v}_2    &1 & 0 & 1 & 0 & 0 & 0 & 0	  & 1   & 0  		  & 0	&	 \hat{y}_2 = x_2^5           \\
		\hat{v}_3    &1 & 0 & 0 & 1 & 0 & 0 & 0	  & 3   & -1  		  & -1	&	 \hat{y}_3 = x_3^5           \\
		\hat{v}_4    &1 & 0 & 0 & 0 & 1 & 0 & 0   & 0   &1  		  & 0	&  \hat{y}_4 = x_4^5            \\
	  \hat{v}_5    &1 & 0 & 0 & 0 & 0 & 1 & 0	  & 0   & 0  		  & 1	&	 \hat{y}_5 = x_5^5            \\
		\hat{v}_6    &0 & 1 & 0 & 1 & 0 & 0 &-1	  & -1   &1  		  & 0	&	 \hat{y}_6 =l_1 \hat{y}_3        \\
		\hat{v}_7    &0 & 1 & 0 & 0 & 1 & 0 &-1	  & 1   & -1  		  & 0	&	 \hat{y}_7 =l_1 \hat{y}_4        \\
		\hat{v}_8    &0 & 1 & 0 & 1 & 0 & 0 & 1	  & -1   & 0  		  & 1	&	 \hat{y}_8 =l_2 \hat{y}_3         \\
		\hat{v}_9    &0 & 1 & 0 & 0 & 0 & 1 & 1	  & 1   & 0  		  & -1	&  \hat{y}_9 =l_2 \hat{y}_5                   
	\end{pmatrix}.                                                                        
\end{equation}
Here we have displayed the points $\hat{v}_i$, the corresponding monomials $\hat{y}_i$ and a basis 
of relations $\hat{\ell}^{(i)}$, that we obtain as a Mori cone of a triangulation of the polyhedron 
$\Delta_7^{\hat Z}$. We emphasize that besides the closed string charge vectors of $Z_3$ embedded 
as $\ell^{(1)}=\hat{\ell}^{(1)}+\hat\ell^{(2)}+\hat\ell^{(3)}$, the brane charge vectors $\hat{\ell}^{(a)}$ 
are among the charge vectors $\hat{\ell}^{(j)}$ of $\Delta_7^{\hat{Z}}$ as well. Furthermore, for 
the above triangulation of $\Delta_7^{\hat{Z}}$ we immediately obtain the full GKZ differential 
system $\mathcal{L}_{a}$, $\mathcal{Z}_i$ of \eqref{eq:GKZquinticopen} by the standard formulas for 
the \textit{standard} GKZ-system in \eqref{eq:pfo}, \eqref{eq:Zs} using the points $\hat{v}_i$ and 
relations $\hat{\ell}^{(j)}$ from $\Delta^{\hat{Z}}_7$ with exponent $\beta=(-1,0,0,0,0,0,0)$.

This GKZ-system defines coordinates $\hat{z}^a$ on the complex structure moduli space of $\hat{Z}_3$ 
as before. We apply the closed string formula \eqref{eq:algCoords} for the charge vectors 
$\hat{\ell}^{(a)}$ of $\Delta^{\hat{Z}}_7$ to obtain the three coordinates
\begin{equation} \label{eq:zOpenquintic}
 	\hat{z}^1=-\frac{a_1a_2a_3^3a_7a_9}{a_0^5a_6a_8}\,,\quad \hat{z}^2=\frac{a_4a_6}{a_3a_7}\,,\quad \hat{z}^3=\frac{a_5a_8}{a_3a_9}\,.
\end{equation}
We obtain a complete system of differential operators $\mathcal{D}_a$, the Picard-Fuchs operators, 
by adding to \eqref{eq:GKZquinticopen} further operators $\mathcal{L}_a$ associated to scaling 
symmetries specified by integer positive linear combinations of the charge vectors $\hat{\ell}^{(a)}$ 
in \eqref{blowupPolyquintic}. By factorizing these operators $\mathcal{L}_a$ expressed in the 
coordinates \eqref{eq:zOpenquintic} we obtain the differential system generated by 
\begin{eqnarray}	
\mathcal{D}_1&=&\theta _1 \theta_2 \theta_3 \left(3 \theta _1\!-\!\theta _2\!-\!\theta _3\!\right)-
5 \prod_{i=1}^4 \left(5 \theta _1\!-\!i\right)\hat{z}^1 \hat{z}^2 \hat{z}^3\,,\nonumber\\
\mathcal{D}_i&=&\left(\theta _1-\theta _i\right) \theta _i+ \left(1+\theta _1-\theta _2\right) \left(1+3 \theta _1-\theta _2-\theta _3\right)\hat{z}^i\,,\quad i=2,3\,,
\label{eq:GKZquinticz}
\end{eqnarray}
where we introduced $\theta_i=\hat{z}^i\frac{\partial}{\partial \hat{z}^i}$ and further rescaled 
the holomorphic three-form $\hat{\Omega}$ by $a_0$. Each of these three operator $\mathcal{D}_a$ 
corresponds to a linear combination of the charge vectors $\hat{\ell}^{(i)}$, whose integer 
coefficients can be read off from the powers of $\hat{z}^i$ in the last term of $\mathcal{D}_a$.
Obviously the deformation problem is symmetric under exchange of $\hat{z}^2$ and $\hat{z}^3$. 
While $\mathcal{D}_1$ is symmetric under that symmetry, $\mathcal{D}_2$ and $\mathcal{D}_3$ map 
onto each other under $\hat{z}^2\leftrightarrow \hat{z}^3$. 

This Picard-Fuchs system is perfectly consistent with the expected structure from section 
\eqref{eq:LInZHat}, that in particular implies that the periods of $\Omega$ directly lift to 
the blow-up $\hat{Z}_3$. Upon the identification of the coordinate  $z^1=\hat{z}^1\hat{z}^2\hat{z}^3$ 
on the complex structure moduli space of the quintic\footnote{This is perfectly consistent 
with the embedding of the quintic charge vector as 
$\ell^{(1)}=\hat{\ell}^{(1)}+\hat\ell^{(2)}+\hat\ell^{(3)}$.}, and keeping $\hat{z}^2$, $\hat{z}^3$ 
unchanged, we rewrite the operators \eqref{eq:GKZquinticz} as
\begin{eqnarray}
 	\mathcal{D}_1&=&\mathcal{D}_1^{Z_3}+[\theta_1(\theta_1+\theta_3)(\theta_1-\theta_2-\theta_3)\theta_2+(\theta_2\leftrightarrow \theta_3)]\,,\nonumber\\
	\mathcal{D}_i&=&-[\theta_1+\theta_i-z_i(\theta_1-\theta_2-\theta_3)]\cdot\theta_i\,,
\end{eqnarray}
where we write $\theta_1=z^1\frac{\partial}{\partial z^1}$ by abuse of notation.
The first operator $\mathcal{D}_{1}$ splits into the well-known fourth order quintic operator 
$\mathcal{D}_1^{Z_3}=\theta _1^4-5 \prod_{i=1}^4 \left(5 \theta _1\!-\!i\right)z^1$ and a term 
linear in the derivatives $\theta_2$, $\theta_3$. The other operators $\mathcal{D}_{2}$, 
$\mathcal{D}_{3}$ are proportional to $\theta_2$, $\theta_3$. Consequently, it is ensured that 
the solutions to \eqref{eq:GKZquinticz} contain the closed string periods $\Pi^k(z^1)$ of the 
quintic as the unique solutions independent of the open string parameters $\hat{z}^2$, $\hat{z}^3$.

Thus, we summarize by emphasizing that the complete information for the study of complex structure 
variations in the family $\hat{Z}_3$ of complete intersection threefolds $P=Q=0$ just reduces to 
the determination of the toric data $\Delta_7^{\hat{Z}}$ and the associated GKZ-system.

\subsection{Superpotentials from Blow-Up Threefolds}
\label{sec:toricbranesP11111-W}

The complex structure moduli space of the blown up quintic orbifold $\hat{Z}_3$ described above is 
the model for our open/closed deformation space and \eqref{eq:GKZquinticz} is 
the Picard-Fuchs system annihilating its periods.  We will analyze and interpret
the global properties of the deformations space and the solutions at special 
points  in the deformation space. First we analyze the solutions at the locus $\hat z_i=0$.
Different than for systems that can be embedded in a Calabi-Yau fourfold, as the 
one in sections \ref{sec:ToricBraneBlowupII}, we find at $\hat z_i=0$ no maximal unipotent monodromy. 
Rather the indicial equations of the system \eqref{eq:GKZquinticz} have  the solutions 
$(0,0,0)^{12},(\frac{1}{3},0,0),(\frac{2}{3},0,0,0)$, $(\frac{1}{2},\frac{1}{2},0), (\frac{1}{2},0,\frac{1}{2})$. 
So in total we find  $16$ solutions. The twelve-times degenerate solution $((0,0,0)^{12}$ gives rise 
to one power series  solution 
\begin{equation} 
X^{(0)}_1=1+120 z+113400 z^2+ 168168000 z^3 + {\cal O}(z^4)\ , 
\end{equation}
where $z=\hat{z}_1\hat{z}_2\hat{z}_3$ is the quintic complex structure 
parameter near the point of maximal unipotent monodromy in its moduli space. This solution 
is identified with the fundamental period $X_0$ of the quintic. Denoting 
$\hat l_i:=\log(\hat z_i)$ we get additional eleven logarithmic solutions 
\begin{eqnarray} \label{eq:leadinglogsquintic_b1}
		X^{(1)}_i\,:\,& \hat{l}_1\,,\,\,\hat{l}_2\,,\,\,\hat{l}_3\,,&\\[0.3Em]
		X^{(2)}_{\alpha}\,:\,& \frac{1}{2}\hat{l}_1^2\,,\,\, \hat{l}_2(\frac12\hat{l}_2+ \hat{l}_1)\,,\,\,\hat{l}_3(\frac12\hat{l}_3 +\hat{l}_2 )\,,\,\,\hat{l}_2\hat{l}_3&\nonumber\\[0.3Em]
		X^{(3)}_{\beta}\,:\,& \frac16 \hat{l}_1^3\,,\,\,\frac16 \hat{l}_2^3+\frac12 \hat{l}_1^2 \hat{l}_2  +\frac12 \hat{l}_2^2 \hat{l}_1\,,\,\,\frac16 \hat{l}_3^3+\frac12 \hat{l}_1^2 \hat{l}_3  +\frac12 \hat{l}_3^2 \hat{l}_1\,,\,\,\frac12 \hat{l}_2^2 \hat{l}_3 +\frac12 \hat{l}_3 \hat{l}_3^2+ \hat{l}_1 \hat{l}_2\hat{l}_3\, . &\nonumber
\end{eqnarray}
The  single logarithmic solutions are 
\begin{eqnarray}
		X^{(1)}_1&=& X^{(0)}\log(\hat{z}_1)-60 \hat{z}_1(\hat{z}_2 +\hat{z}_3)+770 z+ 9450 {\hat z}_1^2({\hat z}_2^2+{\hat z_3}^2)
                    +60 {\hat z}_1 ({\hat z}_2^2{\hat z}_3 +{\hat z}_2 {\hat z}_3^2)+\mathcal{O}(\underline{\hat{z}}^5)\,,\nonumber\\
		X^{(1)}_2&=& X^{(0)}\log(\hat{z}_2)+60 \hat{z}_1 \hat{z}_3-9450 {\hat z}_1^2 {\hat z_3}^2
                    -60 {\hat z}_1 {\hat z}_2^2{\hat z}_3 +  \mathcal{O}(\underline{\hat{z}}^5)   \,,\\
		X^{(1)}_3&=& X^{(0)}\log(\hat{z}_3) +60 \hat{z}_1 \hat{z}_2-9450 {\hat z}_1^2 {\hat z_2}^2
                    -60 {\hat z}_1 {\hat z}_3^2{\hat z}_2 +\mathcal{O}(\underline{\hat{z}}^5)\,.\nonumber
\end{eqnarray}
It is easy to check that the single logarithmic period of the mirror quintic is obtained as 
$\sum_i X^{(1)}_i$. Similarly we have chosen the normalization of \eqref{eq:leadinglogsquintic_b1} 
so that $\sum_\alpha X^{(2)}_{\alpha}$  and $\sum_\beta X^{(3)}_{\beta}$ are double and triple 
logarithmic solutions of the Picard-Fuchs equation of the mirror quintic $Z_3$. Using the 
information about the classical terms of the mirror quintic~\cite{Candelas:1990rm,Hosono:1994ax} 
one can identify the precise combination of periods corresponding to a basis of $H_3(Z_3,\mathds{Z})$.

Notable are the four fractional power series solutions to the remaining indices, 
\begin{eqnarray}
X^{(0)}_2&=&\hat{z}_1^\frac{1}{3}+\hat{z}_1^\frac{1}{3}(\frac{1}{2}\hat{z}_2+\frac{1}{2}\hat{z}_3+\frac{6545}{2592}\hat{z}_1)+{\cal O}(\underline{\hat{z}}^\frac{7}{3})\,,  \\
X^{(0)}_3&=&\hat{z}_1^\frac{2}{3}+\hat{z}_1^\frac{2}{3}(4\hat{z}_2+4 \hat{z}_3+\frac{86944}{10125}\hat{z}_1)+{\cal O}(\underline{\hat{z}}^\frac{7}{3})\,,  \nonumber\\
X^{(0)}_4&=&\sqrt{\hat{z}_1 \hat{z}_2} +\sqrt{\hat{z}_1 \hat{z}_2} \hat{z}_3-\frac{5005}{72} (\hat{z}_1\hat{z}_2)^\frac{3}{2}+ {\cal O}(\underline{\hat{z}}^4)\,, \nonumber\\ 
X^{(0)}_5&=&\sqrt{\hat{z}_1 \hat{z}_3} +\sqrt{\hat{z}_1 \hat{z}_3} \hat{z}_2-\frac{5005}{72} (\hat{z}_1\hat{z}_3)^\frac{3}{2}+ {\cal O}(\underline{\hat{z}}^4)\,.\nonumber 
\end{eqnarray}

Let us discuss now the global properties of the moduli space of the branes 
on the quintic orbifold defined by \eqref{eq:RSquintic}. As discussed in section \ref{sec:toricbranesP11111}
there are critical points, where the unobstructed deformation problem of the complete 
intersection \eqref{eq:RSquintic} gives rise to superpotentials associated to obstructed deformation 
problems such as the lines in the quintic orbifold. Clearly these loci must occur at the 
discriminant of the Picard-Fuchs equation determined by $\mathcal{D}_1,\mathcal{D}_2,\mathcal{D}_3$ 
described  in the last section. We find
\begin{eqnarray}
\Delta&=&(1 + \hat{z}_2) (1 - \hat{z}_2) (1+\hat{z}_3) (1-\hat{z}_3) (1-\hat{z}_2-\hat{z}_3) (1+2 \hat{z}_2-\hat{z}_3)(1+2 \hat{z}_3-\hat{z}_2) 
 \,  \\
& &  \times(4+ 5^5\hat{z}_1\hat{z}_2 (1-\hat{z}_3)^2) (4+5^5 \hat{z}_1 \hat{z}_3 (1-\hat{z}_2)^2) (1 - 5^5 \hat{z}_1 \hat{z}_2\hat{z}_3) (27 + 5^5 \hat{z}_1 (1 - \hat{z}_2 - \hat{z}_3)^3)\ .\nonumber
\end{eqnarray}
We expect to get a degeneration of the holomorphic curve $\Sigma$ of \eqref{eq:RSquintic} at the 
discriminate locus and thus obstructed deformation problems that can be characterized by appropriate 
flux quantum numbers. Let us consider two discriminant loci of particular interest.

At the locus $\hat{z}_2=-1$ and $\hat{z}_3=-1$\footnote{We note that $\hat{z}_2=\hat{z}_3=-1$ 
agrees with $u^1=u^2=1$ in the notation of \eqref{eq:BBrane} since $\hat{z}_a=-u^a$.} the complete 
intersection becomes holomorphic in the quintic and in fact the toric A-brane, which is mirror 
to the holomorphic constraint, becomes compatible with the involution brane, i.e.  the fixpoint 
locus of the involution 
\begin{equation}
 (x_1,x_2,x_3,x_4,x_5)\rightarrow (\bar x_1,\bar x_2 ,\bar x_3, \bar x_4, \bar x_5)\ . 
\end{equation}
More precisely the toric $A$-branes is given by the constraints \eqref{eq:ABranes} 
defined by the charge vectors $\hat{\ell}^{(a)}$ in \eqref{eq:toricquintic} with vanishing 
relative K\"ahler/Wilson line parameters $c^a=0$. Comparing the solutions at that locus we 
obtain a two open parameter deformation of the brane discussed in~\cite{Walcher:2006rs}. 
The relevant periods at the involution brane point are trivially obtained 
from the solutions at the large complex structure point by analytic continuation. 
In particular the solutions at large complex structure, which are at most linear in the 
logarithms of the $\hat{z}_a$ converge in the variables $(v_1=z_1,v_2=(1+\hat{z}_2),v_3=(1+\hat{z}_3))$.
The solutions with the square root cuts $X^{(0)}_{4}$ and $X^{(0)}_{5}$ are expected to 
specialize to the superpotential for the involution brane, if the open moduli $v_2$ 
and $v_3$  are set to zero. Indeed, if we symmetrize in the two square root solutions,
we find up to a normalization worked out in~\cite{Walcher:2006rs} the series
\begin{eqnarray}  \label{eq:involutionW}
W^{quant}&=&\displaystyle{\frac{30}{4 \pi^2}\biggl(v_1^{1/2}+\frac{5005}{9} v_1^{3/2}+\frac{52055003}{75}v_1^{5/2}+
v_1^{1/2}( \frac{1}{2}(v_2+ v_3)- \frac{1}{16}(v_2^2-4v_1 v_2 + v_3^2))}\nn\\ 
&+&\displaystyle{\frac{5005}{6} v_1^{3/2} (v_2+v_3)+{\cal O}(\underline{v}^{7/2})\biggl)}\ .
\end{eqnarray} 
In particular we note that for $v_2=v_3=0$ this superpotential is exactly the 
one for the involution brane obtained in~\cite{Walcher:2006rs}. Using the mirror map of the 
quintic it is possible to obtain from \eqref{eq:involutionW} at $v_2=v_3=0$ the disk instantons 
for the involution brane. We expect that the scalar potential induced by \eqref{eq:involutionW}
has a minimum along the $v_2=v_3=0$ direction. However to see this minimalization explicitly 
requires the construction of the K\"ahler potential, a choice of flat coordinates and a choice of 
the gauging of the superpotential as a section in the K\"ahler line bundle. We note that the 
above discussion of the involution brane is similar to the one of~\cite{Alim:2009rf} in the 
context of a one open parameter family of a toric brane on the quintic.  

A similarly interesting locus is the $(1-\hat{z}_2-\hat{z}_3)=0$ and $\frac{1}{\hat{z}_1}=0$. 
According to the discussion in section \ref{sec:toricbranesP11111} this is 
the locus $\mathcal{M}_{\P^1}(\Sigma)$ of \eqref{eq:modulicond}, 
$\frac{(\alpha^5+\beta^5+\gamma^5)}{\alpha^5}=0$ and $\frac{\psi \beta \gamma}{\alpha^2}=0$, where 
the constraints \eqref{eq:RSquintic} factorize and the holomorphic lines occur. 
We expect the superpotential to vanish at this locus. Indeed if we expand in 
$(w_1=\frac{1}{\hat{z}_1}, w_2=(1-\hat{z}_2-\hat{z}_3),w_3=\hat{z}_2-\hat{z}_3)$ we find 16 solutions 
having the indicials $(\frac{k}{5},i,j)$, where $k=1,\ldots,4$ and 
$(i,j)=(0,0),(1,0),(0,1),(1,1)$. Thus, the solutions vanish with 
$\hat{z}_1^{-\frac15},\,\hat{z}_1^{-\frac25},\,\hat{z}_1^{-\frac35},\,\hat{z}_1^{-\frac45}$ for 
$\hat{z}_1^{-\frac15}=\frac{\psi\beta\gamma}{\alpha^2}$. This is compatible with the vanishing of the 
superpotential at the locus of the holomorphic lines. 
Again one would  need  the flat coordinates and the gauge choice in order to 
to perform a detailed local analysis of the orbifold 
superpotential. 

In summary, from the two examples above it is clear that the discriminant of 
the Picard-Fuchs equation contains the expected information about the 
degeneration of the two open parameter brane system are special loci, 
where the problem can be related to obstructed deformation problems. 
We expect this also to be true at the other loci of the discriminant, where a brane interpretation
is not yet available.

\subsection{Brane Superpotential at Large Volume: Disk Instantons}
\label{sec:toricbranesP11111-W-LV}

In this section we apply the blow-up $\hat{Z}_3$ to a different five-brane on the quintic. The following 
analysis is focused on the determination of the disk instanton invariants at large radius of the A-model 
and thus brief at several points for the sake of brevity.

The Calabi-Yau geometry of the B-model is given by the one parameter mirror quintic with the constraint 
$P$ as in \eqref{eq:toricquintic}. We add an open string sector of a five-brane, that we describe as 
before under the identification of the moduli space $\mathcal{M}(\Sigma)$ with the deformation space 
$\tilde{P}^1$, by the toric curve $\Sigma$ specified by brane charge vectors $\hat{\ell}^{(a)}$ as
\begin{eqnarray} \label{eq:toricquintic_b2}
 \Sigma&:& P=0\,,\quad h_1\equiv a_6x_1x_2x_3x_4x_5+a_7x_1^5=0\,,\quad h_2\equiv a_8x_1^5+a_9x_2^5=0\,, \nonumber\\
& & \hat{\ell}^{(1)}=(-1,1,0,0,0,0)\,,\quad\hat{\ell}^{(2)}=(0,-1,1,0,0,0)\,.
\end{eqnarray}

For this geometry we readily construct the blow-up $\hat{Z}_3$ as the complete intersection in the toric 
variety $\mathcal{W}=\mathds{P}(\mathcal{O}(5)\oplus\mathcal{O}(5))\cong \P(\mathcal{O}\oplus\mathcal{O})$,
\begin{equation}
	\hat{Z}_3\,:\, \quad P=0\,,\quad Q=l_1(a_8x_1^5+a_9x_2^5)-l_2(a_6x_1x_2x_3x_4x_5+a_7x_1^5)\,.	
\end{equation}
From these constraints we construct the holomorphic three-form $\hat{\Omega}$ as the residue \eqref{eq:ResZhat} 
from which we read off the GKZ-system for $\hat{Z}_3$ as
\begin{eqnarray}\label{eq:GKZquinticopen_b2}
 	&\mathcal{Z}_0=\sum_{i=0}^5\vartheta_i+1\,,\qquad \mathcal{Z}_1=\sum_{i=6}^9\vartheta_i\,,\qquad \mathcal{Z}_2=\vartheta_2-\vartheta_1-\vartheta_7-\vartheta_8+\vartheta_9\,,\nonumber\\ &\mathcal{Z}_i=\vartheta_i-\vartheta_1-\vartheta_7-\vartheta_8\,,\,\, i=3,4,5\,,\qquad \mathcal{Z}_6=\vartheta_8+\vartheta_9-\vartheta_6-\vartheta_7\,,\displaystyle&\nonumber\\
	&\displaystyle\mathcal{L}_{1}=\prod_{i=1}^5\frac{\partial}{\partial a_i}-\frac{\partial^5}{\partial a_0^5}\,,\qquad\mathcal{L}_{2}=\frac{\partial^2}{\partial a_1\partial a_6}-\frac{\partial^2}{\partial a_0\partial a_7}\,,\qquad \mathcal{L}_{3}=\frac{\partial^2}{\partial a_2\partial a_8}-\frac{\partial^2}{\partial a_1\partial a_9}\,,\displaystyle
\end{eqnarray} 
for the logarithmic derivative $\vartheta_i=a_i\frac{\partial}{\partial a_i}$.
Again there are two second order differential operators $\mathcal{L}_2$, $\mathcal{L}_3$ that include 
the curve moduli $a_i$, $i=6,7,8,9$, and one fifth order operator $\mathcal{L}_1$ which is lifted from 
the quintic Calabi-Yau to the blow-up. There are no further operators of minimal degree.
We obtain the GKZ-system \eqref{eq:GKZquinticopen_b2} from the following toric data of $\hat{Z}_3$
\begin{equation}\label{blowupPolyquintic_b2}
	\begin{pmatrix}[c|ccccccc|ccc|l]
	    		     &  &   &   & \Delta_7^{\hat Z}&&&   		  &\hat{\ell}^{(1)} &\hat{\ell}^{(2)} &\hat{\ell}^{(3)}  &          \\ \hline
		\hat{v}_0    &1 & 0 & 0 & 0 & 0 & 0 & 0   &-3   & -1  		  &	0	&	 \hat{y}_0 = x_1x_2x_3x_4x_5 \\
		\hat{v}_1    &1 & 0 &-1 &-1 &-1 &-1 & 0	  & 0   & 1  		  & -1	&	 \hat{y}_1 = x_1^5           \\
		\hat{v}_2    &1 & 0 & 1 & 0 & 0 & 0 & 0	  & 0   & 0  		  & 1	&	 \hat{y}_2 = x_2^5           \\
		\hat{v}_3    &1 & 0 & 0 & 1 & 0 & 0 & 0	  & 1   & 0  		  & 0	&	 \hat{y}_3 = x_3^5           \\
		\hat{v}_4    &1 & 0 & 0 & 0 & 1 & 0 & 0   & 1   &0  		  & 0	&  \hat{y}_4 = x_4^5            \\
	  \hat{v}_5    &1 & 0 & 0 & 0 & 0 & 1 & 0	  & 1   & 0  		  & 0	&	 \hat{y}_5 = x_5^5            \\
		\hat{v}_6    &0 & 1 & 0 & 1 & 0 & 0 &-1	  & -2   &1  		  & 0	&	 \hat{y}_6 =l_1 \hat{y}_0        \\
		\hat{v}_7    &0 & 1 & 0 & 0 & 1 & 0 &-1	  & 2   & -1  		  & 0	&	 \hat{y}_7 =l_1 \hat{y}_1        \\
		\hat{v}_8    &0 & 1 & 0 & 1 & 0 & 0 & 1	  & -1   & 0  		  & 1	&	 \hat{y}_8 =l_2 \hat{y}_1         \\
		\hat{v}_9    &0 & 1 & 0 & 0 & 0 & 1 & 1	  & 1   & 0  		  & -1	&  \hat{y}_9 =l_2 \hat{y}_2                   
	\end{pmatrix}.                                                                        
\end{equation}
We note that the second and third charge vector realize the brane charge vectors \eqref{eq:toricquintic_b2} 
and the closed string charge vector of the quintic is embedded as 
$\ell^{(1)}=\hat{\ell}^{(1)}+2\hat{\ell}^{(2)}+\hat{\ell}^{(3)}$. Here the generators of the Mori cone 
$\hat{\ell}^{(a)}$ are obtained as a triangulation of the polyhedron $\Delta_7^{\hat{Z}}$.

The GKZ-system \eqref{eq:GKZquinticopen_b2} defines three local coordinates $\hat{z}^a$ on the complex 
structure moduli space of $\hat{Z}_3$, that are chosen according to the basis of charge vectors in 
\eqref{blowupPolyquintic_b2},
\begin{equation} \label{eq:zOpenquintic_b2}
 	\hat{z}^1=-\frac{a_3a_4a_5a_7^2a_9}{a_0^3a_6^2a_8}\,,\quad \hat{z}^2=\frac{a_1a_6}{a_0a_7}\,,\quad \hat{z}^3=\frac{a_2a_8}{a_1a_9}\,.
\end{equation}
The complete system of differential operators $\mathcal{D}_a$ constituting the Picard-Fuchs system are 
found by linear combinations of the charge vectors $\hat{\ell}^{(a)}$ in \eqref{blowupPolyquintic_b2}. 
They are obtained by factorizing the corresponding operators $\mathcal{L}_a$, that are directly 
associated to the scaling symmetries of the charge vectors. We obtain the system
\begin{eqnarray}
\mathcal{D}_1&=&\theta _1^3 \left(2 \theta _1-\theta _2\right) \left(\theta _1-\theta _3\right)+\left(2 \theta _1-\theta _2-2\right) \left(\theta _1-\theta_3-1\right)\prod_{i=0}^{i=2} \left(3 \theta _1+\theta _2-i\right) \hat{z}^1\,,\\
\mathcal{D}_2&=&\left(2 \theta _1-\theta _2\right) \left(\theta _2-\theta _3\right)+\left(2 \theta _1-\theta _2+1\right) \left(3 \theta _1+\theta _2\right) \hat{z}^2\,,\nonumber\\
\mathcal{D}_3&=&\left(\theta _1-\theta _3\right) \theta _3-\left(\theta _1-\theta _3+1\right) \left(-\theta _2+\theta _3-1\right) \hat{z}^3\,,\nonumber\\
\mathcal{D}_4&=&-\theta _1^3 \left(2 \theta _1-\theta _2\right) \theta _3-\left(2 \theta _1-\theta _2-1\right) \prod_{i=0}^3\left(3 \theta _1+\theta _2-i\right) \hat{z}^1\hat{z}^2\hat{z}^3\,,\nonumber\\
\mathcal{D}_5&=&\theta _1^3 \left(\theta _1-\theta _3\right) \left(-\theta _2+\theta _3\right) \left(1-\theta _2+\theta _3\right)+\left(-1+\theta _1-\theta _3\right)\prod_{i=0}^4\left(3 \theta _1+\theta _2-i\right) \hat{z}^1(\hat{z}^2)^2\,,\nonumber\\
\mathcal{D}_6&=&\theta _1^3 \left(-\theta _2+\theta _3\right) \theta _3+\prod_{i=0}^4\left(3 \theta _1+\theta _2-i\right) \hat{z}^1(\hat{z}^2)^2\hat{z}^3\,,\nn
\end{eqnarray}
where the corresponding linear combination of the $\hat{\ell}^{(a)}$ can be read off from the powers of 
the $\hat{z}^a$. We note that this system has the structure advertised in eq. \eqref{eq:LInZHat} and 
thus the periods $\Pi^k(z^1)$ of the quintic $Z_3$ with $z^1=\hat{z}^1(\hat{z}^2)^2\hat{z}^3$ are solutions to it.

Indeed, we identify 12 solutions of the following form at $\hat{z}^i\rightarrow 0$. There is one solution 
$X^{(0)}$ with a power series expansion, three single logarithmic solutions $X^{(1)}_i$, four double 
logarithmic solutions $X^{(2)}_{\alpha}$ and four triple logarithmic solutions $X^{(3)}_{\beta}$. The 
unique power series solution starts with a constant, that we normalize to $1$,
\begin{equation}
	X^{(0)}=1+120 z^1+113400 (z^1)^2+168168000 (z^1)^3+305540235000 (z^1)^4+\mathcal{O}((z^1)^5)\,,
\end{equation}
where we set $z^1=\hat{z}^1 (\hat{z}^2)^2 \hat{z}^3$. Thus, we identify this as the fundamental period 
$\Pi^0(z^1)$ of the quintic. We recover the three other quintic periods by first noting that the leading 
logarithms of the solutions are given by
\begin{eqnarray} \label{eq:leadinglogsquintic_b2}
		X^{(1)}_i\,:\,& \hat{l}_1\,,\,\,\hat{l}_2\,,\,\,\hat{l}_3\,,&\\[0.3Em]
		X^{(2)}_{\alpha}\,:\,& \frac{1}{2}\hat{l}_1^2\,,\,\, \hat{l}_2(\hat{l}_1-2 \hat{l}_3)\,,\,\,\hat{l}_3(\hat{l}_1 +2\hat{l}_2 +\frac{1}{2}\hat{l}_3)\,,\,\,\hat{l}_2(\frac{1}{2}\hat{l}_2+\hat{l}_3)&\nonumber\\[0.3Em]
		X^{(3)}_{\beta}\,:\,& \frac16 \hat{l}_1^3\,,\,\,\frac12 \hat{l}_1^2 \hat{l}_2-\frac13 \hat{l}_2^3-2 \hat{l}_1 \hat{l}_2 \hat{l}_3-3 \hat{l}_2^2 \hat{l}_3- \hat{l}_2 \hat{l}_3^2\,,& \nonumber\\
		&\,\,\frac12 \hat{l}_1^2 \hat{l}_3+2 \hat{l}_1 \hat{l}_2 \hat{l}_3+2 \hat{l}_2^2 \hat{l}_3+\frac12 \hat{l}_1 \hat{l}_3^2+ \hat{l}_2 \hat{l}_3^2+\frac16 \hat{l}_3^3\,,\,\,\frac12 \hat{l}_1 \hat{l}_2^2+\frac12 \hat{l}_2^3+\hat{l}_1 \hat{l}_2 \hat{l}_3+\frac32 \hat{l}_2^2 \hat{l}_3+\frac12 \hat{l}_2 \hat{l}_3^2\,,&\nonumber
\end{eqnarray}
where we used the abbreviation $\log(\hat{z}^i)=\hat{l}_i$. We immediately observe that all quintic 
periods $\Pi^k(z^1)$ with leading logarithms $l_1$, $\frac12 l_1^2$ and 
$\frac16l_1^3$ for $l_1=\hat{l}_1+2\hat{l}_2+\hat{l}_3$ are indeed contained in the leading logarithms 
\eqref{eq:leadinglogsquintic_b2} of the solutions on $\hat{Z}_3$. We readily check that the complete 
$z^1$-series of the quintic periods $\Pi^k(z^1)$ are reproduced as well on the blow-up.

The remaining six logarithmic solutions are related to the open string sector. In particular, we can 
cross-check this statement by finding the brane superpotential $W_{\rm brane}$ by its A-model 
interpretation at large volume as a generating functional for disk instantons. First we interpret the 
single logarithms in \eqref{eq:leadinglogsquintic_b2} as the mirror map of the open-closed system at 
$z\rightarrow 0$ defining the flat coordinates via $\hat{t}_i=X^{(1)}_i/X^{(0)}$,
\begin{eqnarray}
		X^{(1)}_1&=& X^{(0)}\log(\hat{z}_1)+2 \hat{z}_2-\hat{z}_2^2-60 \hat{z}_1 \hat{z}_2^2+\frac{2\hat{z}_2^3 }{3}-\frac{\hat{z}_2^4}{2}+\frac{2\hat{z}_2^5}{5}-48 \hat{z}_1 \hat{z}_2 \hat{z}_3+462 z_1+\mathcal{O}(\underline{\hat{z}}^6)\,,
		\nonumber\\
		X^{(1)}_2&=& X^{(0)}\log(\hat{z}_2)-\hat{z}_2+\frac{\hat{z}_2^2}{2}-\frac{\hat{z}_2^3}{3}+\frac{\hat{z}_2^4}{4}-\frac{\hat{z}_2^5}{5}+24 \hat{z}_1 \hat{z}_2 \hat{z}_3+154 z_1-360 \hat{z}_1 \hat{z}_2^3 \hat{z}_3+\mathcal{O}(\underline{\hat{z}}^6)\,,\nonumber\\
		X^{(1)}_3&=& X^{(0)}\log(\hat{z}_3)+60 \hat{z}_1 \hat{z}_2^2-9450 \hat{z}_1^2 \hat{z}_2^4+75600 \hat{z}_1^2 \hat{z}_2^4 \hat{z}_3-60 \hat{z}_1 \hat{z}_2^2 \hat{z}_3^2+\mathcal{O}(\underline{\hat{z}}^8)\,.
\end{eqnarray}
Here we omit a factor of $\frac{1}{2\pi i}$ in front of the logarithms for brevity\footnote{We also 
label the variables $\hat{z}_i$ by a subscript instead of a superscript in order to shorten the expressions.}.
This is perfectly consistent with the mirror map of the quintic that is obtained as 
$t=\hat{t}_1+2\hat{t}_2+\hat{t}_3$ or as $\Pi^{1}(z_1)=X^{(1)}_1+2X^{(1)}_2+X^{(1)}_3=X^{(0)}\log(z_1)+770 z_1+\ldots$
as required by the charge vectors $\hat{\ell}^{(a)}$ in \eqref{blowupPolyquintic_b2}. Upon inversion 
of the mirror map, we obtain the $\hat{z}^i$ as a series of $q_a=e^{2\pi i \hat{t}_a}$, that we readily 
insert into the double logarithmic solutions in \eqref{eq:leadinglogsquintic_b2}. 
Then we construct a linear combination of the double logarithmic solutions in \eqref{eq:leadinglogsquintic_b2} as 
\begin{equation}
 	W_{\rm brane}=(2X^{(2)}_1+4X^{(2)}_2+a X^{(2)}_3+4X^{(2)}_4)/X^{(0)}
\end{equation}
in which we insert the inverse mirror map to obtain 
\begin{equation}
	W_{\rm brane}=2 t^2+2 \hat{t}_2^2+\frac{1}{2} (4-a) \hat{t}_3^2-t\hat{t}_2-(4-a) t\hat{t}_3-\frac{1}{4\pi}\sum_{n_i}n_{d_1,d_2,d_3}\text{Li}_2(q_1^{d_1} q_2^{d_2}q_2^{d_3})\,,
\end{equation}
where $a$ denotes a free complex parameter.
This has the expected integrality properties of the Ooguri-Vafa Li$_2$-double cover formula, such that 
we obtain the disk instantons $n_{d_1,d_2,d_3}$. Selected invariants $n_{j,i+j,j}$ are summarized in 
table \ref{tab:instantonsLVquintic_b2}, where the rows and columns are labeled by $i$ and $j$, respectively. 
\begin{table}[!ht]
\centering
$
\scriptstyle
 \begin{array}{|c|rrrrrr|}
\hline
\rule[-0.2cm]{0cm}{0.6cm}  i&j=0&j=1&j=2&j=3&j=4&j=5\\
\hline
 0 & 0 & -320 & 13280 & -1088960 & 119783040 & -15440622400 \\
 1 & 20 & 1600 & -116560 & 12805120 & -1766329640 & 274446919680 \\
 2 & 0 & 2040 & 679600 & -85115360 & 13829775520 & -2525156504560 \\
 3 & 0 & -1460 & 1064180 & 530848000 & -83363259240 & 16655092486480 \\
 4 & 0 & 520 & -1497840 & 887761280 & 541074408000 & -95968626498800 \\
 5 & 0 & -80 & 1561100 & -1582620980 & 931836819440 & 639660032468000 \\
 6 & 0 & 0 & -1152600 & 2396807000 & -1864913831600 & 1118938442641400 \\
 7 & 0 & 0 & 580500 & -2923203580 & 3412016521660 & -2393966418927980 \\
 8 & 0 & 0 & -190760 & 2799233200 & -5381605498560 & 4899971282565360 \\
 9 & 0 & 0 & 37180 & -2078012020 & 7127102031000 & -9026682030832180 \\
 10& 0 & 0 & -3280 & 1179935280 & -7837064629760 & 14557931269209000 \\
 11& 0 & 0 & 0 & -502743680 & 7104809591780 & -20307910970428360 \\
 12& 0 & 0 & 0 & 155860160 & -5277064316000 & 24340277955510560 \\
 13& 0 & 0 & 0 & -33298600 & 3187587322380 & -24957649473175420 \\
 14& 0 & 0 & 0 & 4400680 & -1549998228000 & 21814546476229120 \\
 15& 0 & 0 & 0 & -272240 & 597782974040 & -16191876966658500 \\
 16& 0 & 0 & 0 & 0 & -178806134240 & 10157784412551120 \\
 17& 0 & 0 & 0 & 0 & 40049955420 & -5351974901676280 \\
 18& 0 & 0 & 0 & 0 & -6332490480 & 2348019778753280\\
\hline
\end{array}
$
\caption{Disk instanton invariants $n_{j, i+j, j}$ on the quintic at large volume. These results agree 
with \cite{Alim:2009bx}.}
\label{tab:instantonsLVquintic_b2}
\end{table}
We note that the parameter $a$ does not affect these instantons, however, it does affect the classical 
terms\footnote{The classical term $4t^2-2t_2^2$ of \cite{Alim:2009bx} can not be reproduced by tuning 
the parameter $a$. The ``closest'' match is $2 (t-\hat{t}_2)^2$ for $a=4$, for which the only 
non-vanishing disk instantons are those in table \ref{tab:instantonsLVquintic_b2}.}. It should be fixed 
by the determination of the symplectic basis on the blow-up $\hat{Z}_3$. 

\section{Open-Closed GKZ-Systems from Blow-Up Threefolds}
\label{sec:generaltoricstructure}

In the following section we present a general recipe to easily obtain the toric GKZ-system of an 
arbitrary toric brane $\Sigma$ in an arbitrary toric Calabi-Yau hypersurface $Z_3$. 

Motivated by the above example it is possible in a simple manner to construct the toric data 
$\Delta_7^{\hat{Z}}$ right from the original polyhedron $\Delta^{\tilde{Z}}_3$ and the toric curve 
$\Sigma$ as specified by the charge vectors $\hat{\ell}^{(1)}$, $\hat{\ell}^{(2)}$. We denote the 
vertices of $\Delta_3^{\tilde{Z}}$ by $\tilde{v}_i$, $i=1,\ldots, n$, with $\tilde{v}_0$ the origin, 
its charge vectors by $\ell^{(i)}$, $i=1,\ldots,n-4$ and the two brane vectors by $\hat{\ell}^{(1)}$, 
$\hat{\ell}^{(2)}$ as before. We define $n+5$ points $\hat{v}_i$ spanning a seven-dimensional 
polyhedron $\Delta_7^{\hat{Z}}$ as
\begin{eqnarray}\label{eq:genConstruction}
 	Z_3:& \quad\,\ \,\,\hat{v}_i=&(1,0,\tilde{v}_i,0)\,,\qquad\quad i=0,\ldots,n\,,\nonumber\\ \hat{\ell}^{(1)}:&\quad \hat{v}_{n+1}=&(0,1,v^{(-)}_1,-1)\,,\,\,\,\hat{v}_{n+2}=(0,1,v^{(+)}_1,-1)\,,\nonumber\\
	\hat{\ell}^{(2)}:&\quad\hat{v}_{n+3}=&(0,1,v^{(-)}_2,1)\,,\,\,\,\hat{v}_{n+4}=(0,1,v^{(+)}_2,1)\,,
\end{eqnarray}
where we use the abbreviations
\begin{equation}
 	v^{(+)}_1=\sum_{\hat{\ell}^{(1)}_i>0}\hat{\ell}^{(1)}_i\tilde{v}_i\,,\,\, v^{(-)}_1=-\sum_{\hat{\ell}^{(1)}_i<0}\hat{\ell}^{(1)}_i\tilde{v}_i\,,\,\,
	v^{(+)}_2=\sum_{\hat{\ell}^{(2)}_i>0}\hat{\ell}^{(2)}_i\tilde{v}_i\,,\,\,\quad v^{(-)}_2=-\sum_{\hat{\ell}^{(2)}_i<0}\hat{\ell}^{(2)}_i\tilde{v}_i\,.
\end{equation}
The first line of \eqref{eq:genConstruction} simply embeds the original toric data associated to $Z_3$ 
into $\hat{Z}_3$, whereas the second and third line translate the brane data into geometric data of 
$\hat{Z}_3$. The structure of the points $\hat{v}_i$ is quite generic for the description of a toric 
complete intersection, cf.~\cite{Hosono:1994ax} for the Calabi-Yau case.

In our context this structure in addition reflects the distinction between the closed and open string 
sector. It is encoded by the two canonical hyperplanes in the first and second row of the $\hat{v}_i$. 
Points in the hyperplane $H_1=\{(1,0,w_1,w_2,w_3,w_4,w_5)\}$ correspond to the closed string sector, 
i.e.~the geometry of the Calabi-Yau encoded in the constraint $P$, whereas points in the hyperplane 
$H_2=\{(0,1,w_1,w_2,w_3,w_4,w_5)\}$ contribute to the open string sector as encoded in the constraints 
$h_1$, $h_2$ of $\Sigma$. On the blow-up $\hat{Z}_3$, that we construct as a complete intersection 
\eqref{eq:blowup}, this translates to the rule, which monomial $y_i(\underline{x})$ contributes to which 
of the constraints $P$, $Q$. Points $\hat{v}_i$ in $H_1$ contribute to $P$, whereas those in $H_2$ 
contribute to $Q$. We summarize this in the following table
\begin{equation}\label{blowupDelta7}
	\begin{pmatrix}[c|cccc||ccc|cc|cc]
	    		        &        &       &   \Delta_7^{\hat Z}     &   & \hat{\ell}^{(1)}&\ldots&\hat{\ell}^{(n-4)}&\hat{\ell}^{(n-3)}&\hat{\ell}^{(n-2)} & &\text{monomials}\\ \hline
		\hat{v}_0       & 1      & 0     & \tilde{v}_0    & 0      & \vert    & \ldots & \vert        & \vert           &\vert           & &\hat{y}_0=y_0 \\
	        \ldots          & \vdots & \vdots& \vdots & \vdots & \ell^{(1)}& \ldots & \ell^{(n-4)}& \hat{\ell}^{(1)}&\hat{\ell}^{(2)}& P:&\ldots \\
		\hat{v}_n       & 1      & 0     & \tilde{v}_n    &0       & \vert     & \ldots & \vert       & \vert           &\vert           & &\hat{y}_{n-4}=y_{n-4}\\ \hline
		\hat{v}_{n+1}   &0 & 1   & v^{(-)}_1  & -1         & 0         & \ldots & 0           & 1               & 0              & &\hat{y}_{n-3}=l_1\prod_{\hat{\ell}^{(1)}_i<0}y_i^{-\hat{\ell}^{(1)}_i}\\
		\hat{v}_{n+2}   &0 & 1   & v^{(+)}_1  & -1	   & 0         & \ldots & 0           &-1               & 0              & Q:&\hat{y}_{n-2}=l_1\prod_{\hat{\ell}^{(1)}_i>0}y_i^{\hat{\ell}^{(1)}_i}    \\
		\hat{v}_{n+3}   &0 & 1   & v^{(-)}_2  & 1          & 0         & \ldots & 0           & 0               & 1              & &\hat{y}_{n-1}=l_2\prod_{\hat{\ell}^{(1)}_i<0}y_i^{-\hat{\ell}^{(2)}_i}      \\
		\hat{v}_{n+4}   &0 & 1   & v^{(+)}_2  & 1          & 0         & \ldots & 0           & 0               & -1             & &\hat{y}_{n}=l_2\prod_{\hat{\ell}^{(1)}_i>0}y_i^{\hat{\ell}^{(2)}_i}     \\
	\end{pmatrix}.                                                                        
\end{equation}
Here we displayed besides the points $\hat{v}_i$ of \eqref{eq:genConstruction} also a natural choice 
of basis $\hat{\ell}^{(a)}$ of the lattice of relations of $\Delta_7^{\hat{Z}}$. In this basis the 
first $n-4$ charge vectors are identical to those of $\Delta_4^{\tilde{Z}}$ up to the extension by 
four further entries $0$. More importantly the last two charge vectors naturally contain the two 
original brane vectors $\hat{\ell}^{(a)}$ extended by four further entries with $\pm1$, $0$. 
As before their entries sum up to zero. In the last row we associated monomials $\hat{y}_i$ to the 
points $\hat{v}_i$ where the $y_i(x_j)$ merely denote the polynomials on the original geometry of 
$\Delta_4^{\tilde{Z}}$ computed by the Batyrev formula \eqref{eq:Z3typeIIB}. The coordinates $l_1$, $l_2$ 
denote the homogeneous coordinates on $\P^1$. We note that the form of the polynomials $\hat{y}_i$ 
associated to the four new points $\hat{v}_{n},\ldots,\hat{v}_{n+4}$ reflects the definition of the 
brane constraints $h_1$, $h_2$ defined via \eqref{eq:BBrane}. As mentioned before and indicated in 
\eqref{blowupDelta7} the constraints $P$ and $Q$ are given by
\begin{equation} \label{eq:blowupConstrains}
	\hat{Z}_3\,:\,\quad P=\sum_{i=0}^{n-4}a_i\hat{y_i}\equiv \sum_{i=0}^{n-4}a_iy_i\,,\qquad Q=\sum_{i=n-3}^n a_i\hat{y}_i\,,
\end{equation}
where the $\underline{a}$ denote free complex parameters.
As can be easily checked the general toric data in \eqref{blowupDelta7} immediately reproduce the 
toric data of the blow-up \eqref{blowupPolyquintic} of the curve $\Sigma$ in the quintic $Z_3$. 
Similarly we obtain the toric data \eqref{blowupPolyCand} of our second example in section 
\ref{sec:ToricBraneBlowupII}

To the general form of the toric data \eqref{blowupDelta7} of $\hat{Z}_3$ we associate a GKZ-system 
on the complex structure moduli space of $\hat{Z}_3$ by the standard formulas
\begin{eqnarray} \label{eq:GKZoopen}
 	\mathcal{L}_{a}&=&\prod_{\hat{\ell}^{(a)}_i>0}\left(\frac{\partial}{\partial a_i}\right)^{\hat{\ell}^{(a)}_i}-\prod_{\hat{\ell}^{(a)}_i<0}\left(\frac{\partial}{\partial a_i}\right)^{-\hat{\ell}^{(a)}_i}\,,\qquad a=1,\ldots n-2\,,\\
	\mathcal{Z}_{j}&=&\sum_i(\hat{v}_i)^j\vartheta_i-\beta_j\,,\qquad j=1,\ldots,7\,.\nonumber
\end{eqnarray}
This immediately yields a natural choice of complex coordinates given by
\begin{equation}
	\hat{z}^a=(-)^{\hat{\ell}^{(a)}_0}\prod_{i=0}^{n+4} a_i^{\hat{\ell}^{(a)}_i}\,,\quad a=1,\ldots n-2.
\end{equation} 
We readily apply these formulas to reproduce the GKZ-system \eqref{eq:GKZquinticopen} and the 
coordinates \eqref{eq:zOpenquintic} for the choice of charge vectors $\hat{\ell}^{(a)}$ as in 
\eqref{blowupPolyquintic}. The same applies for the GKZ-system in the second example discussed 
in section \ref{sec:ToricBraneBlowupII}.

Let us conclude by mentioning some remarkable properties of the geometry of $\hat{Z}_3$ as 
encoded by $\Delta^{\hat Z}_7$. First, the last row in \eqref{blowupDelta7} is associated 
to the toric symmetries of the exceptional $\P^1$ in the blow-up divisor $E$. In fact, this 
$\P^1$ can be made directly visible in $\Delta_7^{\hat{Z}}$ by projection on the ray 
$(0,0,0,0,0,0,w_1)$. Second, one might be tempted to map the toric data $\Delta_7^{\hat{Z}}$ 
of the complete intersection \eqref{eq:blowupConstrains} to toric data of a hypersurface defined 
by six-dimensional vectors obtained by adding the first and second row, 
$((\hat{v}_i)^1,(\hat{v}_i)^2,\ldots)\,\mapsto\,((\hat{v}_i)^1+(\hat{v}_i)^2,\ldots)\equiv (1,v_i')$, 
where we note that $(\hat{v}_i)^1+(\hat{v}_i)^2=1$. This defines a five-dimensional polyhedron 
$\Delta_5$ with points $v_i'$. Clearly, this polyhedron has one further charge vector 
$\hat{\ell}^{(n-1)}$ so that the dimension of a corresponding toric variety is five-dimensional. 
Furthermore, for special choices of the brane charge vectors $\hat{\ell}^{(a)}$, but by far not 
for all choices\footnote{The polyhedron $\Delta_5$ defined by the $v_l'$ is not generically 
reflexive. This is the case in section \ref{sec:toricbranesP11111-W-LV}.}, this toric data 
defines a mirror pair of \textit{compact} Calabi-Yau fourfolds $\tilde{X}_4$, $X_4$ with 
the hypersurface constraint given by the standard Batyrev formalism \eqref{eq:Z3typeIIB}. In 
combination with the first observation about the universal presence of the $\P^1$ in 
$\Delta_7^{\hat{Z}}$, the geometry of $\tilde{X}_4$ will contain this very $\P^1$ as the basis 
of a Calabi-Yau threefold fibration with generic fiber $\tilde{Z}_3$. This is precisely the 
geometric structure we encountered in the Calabi-Yau geometries used in \cite{Grimm:2009ef} 
and in the context of heterotic/F-theory duality in \cite{Grimm:2009sy}. In fact, the 
Calabi-Yau fourfold $X_4$ we obtain from $\Delta_7^{\hat{Z}}$ for the example of the next 
section \ref{sec:ToricBraneBlowupII} precisely agrees with the F-theory dual fourfold of a 
heterotic setup with horizontal five-branes as predicted by heterotic/F-theory. We suspect 
that heterotic/F-theory duality for horizontal five-branes is in general the reason for the 
occurrence of a Calabi-Yau fourfold geometry associated to some blow-up threefolds $\hat{Z}_3$. 
This nicely completes the discussion of the application of the blow-up proposal for heterotic 
five-branes and heterotic/F-theory duality in \cite{Grimm:2009sy}. Let us conclude by 
emphasizing that only parts of the fourfold geometry $X_4$, if present, are intrinsically related 
to the original five-brane as e.g.~signaled by the additional charge vector 
$\hat{\ell}^{(n-1)}$.\footnote{On the heterotic side, the additional data of $X_4$ is related 
to the heterotic bundle.} However, the toric data $\Delta_7^{\hat{Z}}$ of the blow-up $\hat{Z}_3$ 
should by construction always carry the minimal amount
of information in order to study the open-closed system of the five-brane in $Z_3$.

A technical similar but differently motivated method to obtain toric data and a GKZ-system 
governing the deformations of toric D5-branes was presented in 
\cite{Mayr:2001xk,Lerche:2001cw,Jockers:2008pe,Jockers:2009mn,Alim:2009rf,Alim:2009bx} and 
formulated mathematically rigorously in \cite{Li:2009dz}.

\section{Open-Closed Picard-Fuchs Systems on $\P^4(1,1,1,6,9)[18]$}
\label{sec:ToricBraneBlowupII}

As a second demonstration of the application of the blow-up proposal we consider a two-parameter 
Calabi-Yau threefold $Z_3$. Since the discussion are similar to the case of the quintic 
we keep it as brief as possible. First in section \ref{sec:toricbranesP11169} we consider branes on 
lines in $Z_3=\P^4(1,1,1,6,9)[18]$ and identify the toric curve $\Sigma$, that degenerates appropriately at
the critical locus ${\cal M}(\P^1)$. We construct the blow-up $\hat{Z}_3$ along $\Sigma$ as a 
complete intersection and construct the Picard-Fuchs system in section \ref{sec:GKZP11169}. Finally 
in section \ref{sec:toricbranesP11169-W-LV} we determine the solutions at large volume
and identify the brane superpotentials from which we extract compact disk instanton invariants,
cf.~appendix \ref{sec:instantonsLV11169}, that match the local results of \cite{Aganagic:2001nx}.

\subsection{Branes on Lines in $\P^4(1,1,1,6,9)[18]$ and the Blow-Up}
\label{sec:toricbranesP11169}

The Calabi-Yau threefold $Z_3$\footnote{We will abbreviate this simply by $Z_3$.} is 
defined as the mirror of the Calabi-Yau hypersurface $\tilde Z_3$ in $\mathds{P}^4(1,1,1,6,9)$ 
which admits $h^{(2,1)}(Z_3)=2$ complex structure moduli and is an elliptic fibration over $\P^2$. 
In the conventions of \cite{Candelas:1994hw} the two complex structures denoted $\Psi_1$, $\Psi_2$ 
enter the constraint $P$ as
\begin{equation}
	P=y^2+x^3+u_1^{18}+u_2^{18}+u_3^{18}-3\Psi_1 (u_1u_2u_3)^6-18\Psi_2 xyu_1u_2u_3\,,
\label{eq:Cand2param}
\end{equation}
where we introduce the homogeneous coordinates $(u_i,x,y,z)$, $i=1,2,3$, for the $\P^2$-base and 
the elliptic fiber $\P^{2}(1,2,3)$, respectively. Note however that we are working in an affine 
patch $z=1$ of the elliptic fiber\footnote{Strictly speaking one has to include the divisor $z$ 
to resolve a curve of $\mathds{Z}_3$-singularity in $\P^4(1,1,1,6,9)$ \cite{Candelas:1994hw}.}. 
This is reflected in the toric data used to obtain $P$,
\begin{equation} \label{eq:P11169TD}
 	\begin{array}{cclllll}
 	 	 \ell^{(1)}=(\phantom{-}0, & -3,\phantom{i} & 1,& 1,& 1, &0 &0) \\ 
		\ell^{(2)}=(-6, & 1, & 0,& 0,& 0, &2 &3) \\\hline \hline
		y_0 & y_1 & y_2 & y_3 & y_4 & y_5& y_6\\ 
		zxyu_1u_2u_3 &(zu_1u_2u_3)^6 & z^6u_1^{18} & z^6u_2^{18}  & z^6u_3^{18} & x^3 & y^2 
 	\end{array}
\end{equation}
where the $y_i$ corresponding to the entries $\ell^{(j)}_i$ of the charge vectors are monomials in 
the homogeneous coordinates on $\mathds{P}^4(1,1,1,6,9)$.
This hypersurface data is augmented by the action of a discrete orbifold group $G$ which is 
$G=\mathds{Z}_{6}\times\mathds{Z}_{18}$ generated by $v^{(i)}=(0,1,3,2,0) \text{ mod } 6$, 
$v'^{(j)}=(1,-1,0,0,0)\text{ mod } 18$ acting on the coordinates as 
\begin{equation}
	g^{(i)}:\,\,x_k\mapsto e^{2\pi i v^{(i)}_k/6} x_k\,,\quad g'^{(j)}:\,\,x_k\mapsto e^{2\pi i v'^{(j)}_k/18} x_k\,.
\label{eq:GCand}
\end{equation}

To this setup we add five-brane wrapping a rational curve $\P^1$ in $Z_3$ that will be of similar 
type as the lines \eqref{eq:paramlines} considered in the quintic. As before we use the moduli 
space $\mathcal{M}(\Sigma)$ as a model for the deformation space $\tilde{P}^1$ of this line, 
where we specify the analytic family of curves $\Sigma$ in the form of toric branes 
\begin{eqnarray} \label{eq:constlinesCand}
	\Sigma&:& P=0\,,\quad h_1\equiv \beta^{12}(u_1u_2u_3)^6-\alpha^6\gamma^6 u_2^{18}=0\,,\quad 
	h_2\equiv\gamma^{12}(u_1u_2u_3)^6-\alpha^6\beta^6u_3^{18}=0\,,\nonumber\\
	&&\hat{\ell}^{(1)}=(0,-1,0,1,0,0,0)\,,\quad\quad\hat{\ell}^{(2)}=(0,-1,0,0,1,0,0)\,.
\end{eqnarray}
The brane charge vectors $\hat{\ell}^{(i)}$ are used to construct the constraint $h_i$ via 
\eqref{eq:BBrane}. This basis of constraints and parameters might look inconvenient, is however 
justified by noting the convenient, equivalent form 
\begin{equation}
	\Sigma\,:\quad P=0\,,\quad\gamma^{18}u_2^{18}-\beta^{18}u_3^{18}=0\,,\quad 
	\gamma^{18}u_1^{18}-\alpha^{18}u_3^{18}=0\,
\label{eq:constlinesCandNF}
\end{equation}
upon a trivial algebraic manipulation.  
We introduce affine coordinates parameterizing this analytic family of curves in $Z_3$ that we 
choose to be $u^1=\frac{\beta^{18}}{\gamma^{18}}$ and $u^2=\frac{\alpha^{18}}{\gamma^{18}}$.
 
Next we proceed by linearizing \eqref{eq:constlinesCandNF} to describe a non-holomorphic 
deformation $\tilde{P}^1$ of rational curves given by
\begin{equation} \label{eq:anholoCCand}
	\tilde{\P}^1:\quad \eta_1 x+\sqrt[3]{y^2+ m(u_3,x,y)}=0\,,\quad  \eta_2\gamma u_2-\beta u_3=0\,,\quad 
	\eta_3\gamma u_1-\alpha u_3=0\,.
\end{equation} 
Here we inserted $h_1$ and $h_2$ into $P$ and introduced the third and eighteenth roots of 
unity $\eta_1$, respectively $\eta_2$, $\eta_3$ as well as the polynomial
\begin{equation}
 	m(u_3,x,y)=\frac{\alpha^{18}+\beta^{18}+\gamma^{18}}{\gamma^{18}}-3\Psi_1\left(\frac{\alpha\beta}{\gamma^2}\right)^6u_3^{18}-18\Psi_2\frac{\beta\alpha}{\gamma^2}xyu_3^{18}\,.
\label{eq:CandDiv}
\end{equation}
At the critical locus of the parameter space $\alpha$, $\beta$, $\gamma$ where the polynomial 
$m$ vanishes identically, the generically higher genus Riemann surface $\Sigma$ degenerates. 
This locus reads
\begin{equation}
	\mathcal{M}_{\P^1}(\Sigma)\,:\quad\alpha^{18}+\beta^{18}+\gamma^{18}-3\Psi_1\alpha^6\beta^6\gamma^6=0\,,
	\quad \Psi_2\alpha\beta\gamma=0\,.
\label{eq:modulicondCand}
\end{equation}
At this locus the Riemann surface $\Sigma$ in \eqref{eq:constlinesCandNF} degenerates to
\begin{equation}
	\Sigma\,:\quad h_0\equiv y^2+x^3\,,\quad\gamma^{18}u_2^{18}-\beta^{18}u_3^{18}=0\,,
	\quad \gamma^{18}u_1^{18}-\alpha^{18}u_3^{18}=0\,.
\end{equation}
Modulo the action of $G$ identifying the different solutions in \eqref{eq:anholoCCand} we can 
solve \eqref{eq:anholoCCand} holomorphically and consistent with the weights of $\P^2(1,2,3)$ 
at the locus $\mathcal{M}_{\P^1}(\Sigma)$ by the Veronese embedding of a line in $\P^{4}(1,1,1,6,9)$,
\begin{equation}
	(U,V)\,\mapsto\,(\alpha U,\beta U,\gamma U,-{\eta_1}^2 V^6,V^9)\,,\quad \eta_1^3=1.
\label{eq:paramlinesCand}
\end{equation}
From the perspective of this line, the constraint \eqref{eq:modulicondCand} on the parameters 
$(\alpha,\beta,\gamma)$ is precisely the condition for it to lie holomorphically in the 
Calabi-Yau constraint $P$ of $\hat{Z}_3$. This implies that at the point $\Psi_2=0$ there is 
an analytic family of lines in $Z_3$ and otherwise, for $\Psi_2\neq 0$, only a discrete number 
of lines.

To study the open-closed system defined by the five-brane in $Z_3$ we construct the blow-up 
threefold $\hat{Z}_3$. As explained above, cf.~section \ref{sec:unificationofdeformations}, 
we use the holomorphic description by the toric curve $\Sigma$ of the anholomorphic brane 
deformations $\P^1$ in order to construct the blow-up. Before we construct $\hat{Z}_3$, we 
switch to a full toric description.

\subsection{Toric Branes on $\P^4(1,1,1,6,9)[18]$: the Open-Closed GKZ-System}
\label{sec:GKZP11169}

We begin the analysis of the open-closed moduli space using the toric means. First of all 
let us recall the toric construction of the Calabi-Yau $Z_3$ by giving its constraint as 
well as the curve $\Sigma$,
\begin{eqnarray} \label{eq:constraintsCand}
	Z_3:& P=a_6y^2+a_5x^3+a_1u_1^6u_2^6u_3^6+a_2u_1^{18}+a_3u_2^{18}+a_4u_3^{18}+a_0xyu_1u_2u_3\,,\\
	\Sigma:& h_1=a_7u_1^6u_2^6u_3^6+a_8u_2^{18}\,,\quad h_2=a_9u_1^6u_2^6u_3^6+a_{10}u_3^{18}\,. \nonumber
\end{eqnarray}
The coefficients $\underline{a}$ redundantly parameterize the complex structure of 
$Z_3$ respectively the moduli of the curve $\Sigma$ in $Z_3$. The information in 
\eqref{eq:constraintsCand} is directly encoded in the toric data specifying 
$(Z_3,\Sigma)$ via the polyhedron $\Delta^{\tilde{Z}}_4$ and the two brane-vectors 
$\hat{\ell^{(1)}}$, $\hat\ell^{(2)}$ reading
\begin{equation}\label{eq:3foldellp2blowup}
	\begin{pmatrix}[c|cccc|cc|l||cc]
	    	&   &  \Delta_4^{\tilde Z} &   &    &  \ell^{(1)} & \ell^{(2)} &   &\hat{\ell}^{(1)}&\hat{\ell}^{(2)}\\ \hline
		\tilde{v}_0 & 0 & 0 & 0 & 0 	  &  0  &  -6 & y_0 = zxyu_1u_2u_3 & 0&0\\
		\tilde{v}_1 & 0 & 0 & 2 & 3 	  &  -3   &1  & y_1 = z^6 u_1^6 u_2^6 u_3^6 & -1&-1\\
		\tilde{v}_2 & 1 & 1 & 2 & 3 	&  1   & 0  & y_2 = z^6 u_1^{18} & 0&0\\
		\tilde{v}_3 &-1 & 0 & 2 & 3 	&  1   & 0  & y_3 = z^6 u_2^{18} & 1&0\\
		\tilde{v}_4 & 0 &-1 & 2 & 3 	&  1   & 0  & y_4 = z^6 u_3^{18} & 0&1\\
	        \tilde{v}_5 & 0 & 0 &-1 & 0 	&  0   & 2  & y_5 = x^3 & 0&0\\
		\tilde{v}_6 & 0 & 0 & 0 &-1 	&  0   & 3  & y_6 = y^2 & 0&0 
	\end{pmatrix}.
\end{equation}
The points of the dual polyhedron $\Delta_4^Z$ are given by $v_1=(-12,6,1,1)$, 
$v_2=(6,-12,1,1)$, $v_3=(6,6,1,1)$, $v_4=(0,0,-2,1)$ and $v_5=(0,0,1,-1)$, where 
the point $(0,0,1,1)$ corresponds to the $z$-coordinate on the elliptic fiber that 
we set to $1$. The Calabi-Yau as well as the two brane constraints of 
\eqref{eq:constraintsCand} are then associated to this data via \eqref{eq:HVmirror} 
and \eqref{eq:BBrane}, respectively.

Accordingly, the variational problem of complex structures on $Z_3$ can be studied 
by exploiting the existence of the GKZ-system \eqref{eq:pfo}, \eqref{eq:Zs} 
associated to $\Delta_4^{\tilde Z}$. For the example at hand it reads
\begin{eqnarray}\label{eq:GKZclosed}
 	&\mathcal{Z}_0=\sum_{i=0}^6\vartheta_i+1\,,\quad \mathcal{Z}_i=\vartheta_2-\vartheta_{i+2}\,\, (i=1,2)\quad \mathcal{Z}_3=2\displaystyle\sum_{i=1}^4\vartheta_i-\vartheta_5\,,\displaystyle\quad
	\displaystyle\mathcal{Z}_4=3\sum_{i=1}^4\vartheta_i-\vartheta_6\,,\displaystyle\nonumber\\
	\displaystyle&\displaystyle\mathcal{L}_{1}=\prod_{i=2}^4\frac{\partial}{\partial a_i}-\frac{\partial^3}{\partial a_1^3}\,,\qquad \mathcal{L}_{2}=\frac{\partial^6}{\partial a_1\partial a_5^2\partial a_6^3}-\frac{\partial^6}{\partial a_0^6}\,,\displaystyle&\displaystyle
\end{eqnarray}
for $\vartheta_i=a_i\frac{\partial}{\partial a_i}$ as before. The differential system 
$\mathcal{Z}_i$ then determines two algebraic coordinates $z^1$, $z^2$ on the 
complex structure moduli space that are given in terms of the Mori generators 
$\ell^{(i)}$ according to \eqref{eq:algCoords} as 
\begin{equation} 
		z^1=\frac{a_2a_3a_4}{a_1^3}\,,\quad z^2=\frac{a_1a_5^2a_6^3}{a_0^6}\,.
\end{equation}

To analyze the open-closed system $(Z_3,\Sigma)$ described by \eqref{eq:constraintsCand} 
we now apply the blow-up proposal, i.e.~construct the geometry $(\hat{Z}_3,E)$ given as 
the family of complete intersections \eqref{eq:blowup} in 
$\mathcal{W}=\P(\mathcal{O}(18H)\oplus \mathcal{O}(18H))\cong \P(\mathcal{O}\oplus\mathcal{O})$ 
which now reads
\begin{eqnarray}
 	\hat{Z}_3:& P=0\,,\quad Q=l_1(a_7u_1^6u_2^6u_3^6+a_8u_2^{18})-l_2(a_9u_1^6u_2^6u_3^6+a_{10}u_3^{18})\,.
\end{eqnarray}
We define the three-form $\hat{\Omega}$ by the residue expression \eqref{eq:ResZhat} and 
determine a system of differential operators, the Picard-Fuchs operators, for the family 
$\hat{Z}_3$. First we determine the GKZ-system on the complex structure moduli space of 
the blow-up $Z_3$. We check that $\hat{\Omega}$ is identically annihilated by the two 
differential operators $\mathcal{L}_{1}$ and $\mathcal{L}_{2}$ of \eqref{eq:GKZclosed} 
that are complemented to the system 
\begin{eqnarray} \label{eq:GKZopenCand}
	&\displaystyle\mathcal{Z}_0=\sum_{i=0}^6\vartheta_i+1\,,\quad \mathcal{Z}_1=\sum_{i=7}^{10}\vartheta_i\,,\quad \mathcal{Z}_2=\vartheta_2-\vartheta_3-\vartheta_8\,,\quad \mathcal{Z}_3=\vartheta_2-\vartheta_4-\vartheta_{10}\,,&\nonumber\\
	&\displaystyle\mathcal{Z}_4=2\Big(\sum_{i=1}^4+\sum_{i=7}^{10}\Big)\vartheta_i-\vartheta_5\,,\quad  \mathcal{Z}_5=3\Big(\sum_{i=1}^4+\sum_{i=7}^{10}\Big)\vartheta_i-\vartheta_6\,,\quad \mathcal{Z}_6=\vartheta_9+\vartheta_{10}-\vartheta_7-\vartheta_8\,.\nonumber \displaystyle&\\
\displaystyle&\displaystyle\mathcal{L}_{1}\,,\qquad\mathcal{L}_2\,,\qquad\mathcal{L}_{3}=\frac{\partial^2}{\partial a_3\partial a_7}-\frac{\partial^2}{\partial a_1\partial a_8}\,,\qquad \mathcal{L}_{4}=\frac{\partial^2}{\partial a_4\partial a_9}-\frac{\partial^2}{\partial a_1\partial a_{10}}\displaystyle&\displaystyle
\end{eqnarray}
There are two additional second order differential operators $\mathcal{L}_{3}$ and 
$\mathcal{L}_{4}$ that annihilate $\hat{\Omega}$ and incorporate the parameters 
$a_7,\ldots a_{10}$ that are associated to the moduli of $\Sigma$. Clearly, there 
are no further operators of low minimal degree. The operators $\mathcal{Z}_k$ are 
related to the symmetries of $\mathcal{W}$. The first two operators are associated to 
an overall rescaling of the two constraints $P\mapsto \lambda P$, $Q\mapsto \lambda' Q$. 
The third and fourth operator describe the torus symmetries of the $\P^2$-base, 
$(u_1,u_j)\mapsto(\lambda_j u_1,\lambda_j^{-1}u_j)$, $j=2,3$, the fifth and sixth 
operator are due to the $\P^2(1,2,3)$-fiber symmetries, 
$(x,y,z)\mapsto(\lambda'_1x,y,{\lambda'_1}^{-1}z)$, $(x,y,z)\mapsto(x,\lambda'_2y,{\lambda'_2}^{-1}z)$ 
and the last operator $\mathcal{Z}_6$ is related to the torus symmetry 
$(l_1,l_2)\mapsto (\lambda l_1,\lambda^{-1}l_2)$ of the exceptional $\P^1$. All 
operators $\mathcal{Z}_i$ of the original system \eqref{eq:GKZclosed} are altered 
due to the lift to the blow-up $\hat{Z}_3$. 

As discussed in detail in section \ref{sec:generaltoricstructure} the constructed 
GKZ-system can be obtained as a GKZ-system associated to toric data of the blow-up $\hat{Z}_3$.
The set of integral points $\hat{v}_i$ reads
\begin{equation}\label{blowupPolyCand}
	\begin{pmatrix}[c|ccccccc|cccc|l]
	    		     &  &   &   & \Delta_7^{\hat Z}&&&   		  &\hat{\ell}^{(1)} &\hat{\ell}^{(2)} &\hat{\ell}^{(3)}  &\hat{\ell}^{(4)}   &          \\ \hline
		\hat{v}_0    &1 & 0 & 0 & 0 & 0 & 0 & 0   & 0   &-6  		  &	0	&	0	& \hat{y}_0 = zxyu_1u_2u_3          \\
		\hat{v}_1    &1 & 0 & 0 & 0 & 2 & 3 & 0	  &-2   & 0  		  &    -1	&	1	& \hat{y}_1 = z^6 u_1^6 u_2^6 u_3^6 \\
		\hat{v}_2    &1 & 0 & 1 & 1 & 2 & 3 & 0	  & 1   & 0  		  &	0	&	0	& \hat{y}_2 = z^6 u_1^{18}          \\
		\hat{v}_3    &1 & 0 &-1 & 0 & 2 & 3 & 0	  & 0   & 0  		  &	1	&	0	& \hat{y}_3 = z^6 u_2^{18}          \\
		\hat{v}_4    &1 & 0 & 0 &-1 & 2 & 3 & 0   & 1   & 1  		  &	0	&      -1	& \hat{y}_4 = z^6 u_3^{18}          \\
	        \hat{v}_5    &1 & 0 & 0 & 0 &-1 & 0 & 0	  & 0   & 2  		  &	0	&	0	& \hat{y}_5 = x^3            	      \\
		\hat{v}_6    &1 & 0 & 0 & 0 & 0 &-1 & 0	  & 0   & 3  		  &	0	&	0	& \hat{y}_6 = y^2                   \\
		\hat{v}_7    &0 & 1 & 0 & 0 & 2 & 3 &-1	  &-1   & 0  		  &	1	&	0	& \hat{y}_7 =l_1 \hat{y}_1                   \\
		\hat{v}_8    &0 & 1 &-1 & 0 & 2 & 3 &-1	  & 1   & 0  		  &    -1	&	0	& \hat{y}_8 =l_1 \hat{y}_3                   \\
		\hat{v}_9    &0 & 1 & 0 & 0 & 2 & 3 & 1	  & 0   & 1  		  &	0	&      -1	& \hat{y}_9 = l_2\hat{y}_1                   \\
		\hat{v}_{10} &0 & 1 & 0 &-1 & 2 & 3 & 1	  & 0   &-1  		  &	0	&	1	& \hat{y}_{10} = l_2\hat{y}_4                \\
	\end{pmatrix}.                                                                        
\end{equation}
Here we have displayed the points $\hat{v}_i$, the basis of relations $\hat{\ell}^{(i)}$ 
and the corresponding monomials $\hat{y}_i$. We emphasize that besides the closed 
string charge vectors of $Z_3$ embedded as $\ell^{(1)}=\hat{\ell}^{(1)}+\hat{\ell}^{(3)}$, 
$\ell^{(2)}=\hat{\ell}^{(2)}+\hat{\ell}^{(4)}$ the brane charge vectors $\hat{\ell}^{(a)}$ 
are among the $\hat{\ell}^{(i)}$ of $\Delta_7^{\hat{Z}}$ as well. We note that this toric 
data is completely consistent with the general formula \eqref{eq:genConstruction} to obtain 
$\Delta_7^{\hat{Z}}$. Similarly the associated GKZ-system precisely reproduces 
\eqref{eq:GKZopenCand} upon using the general formula of the GKZ-system \eqref{eq:GKZoopen}. 
We confirm as mentioned in section \ref{sec:generaltoricstructure} that this polyhedron 
can be mapped to the five-dimensional polyhedron with an associate compact Calabi-Yau 
fourfold by adding the first and second row. This agrees with the heterotic/F-theory dual 
fourfold when considering the elliptic $Z_3$ as a heterotic compactification \cite{Grimm:2009sy}. 
Furthermore, the GKZ-system \eqref{eq:GKZopenCand} is a closed and more restrictive subsystem 
of GKZ-system for the Calabi-Yau fourfold.

The GKZ-system \eqref{eq:GKZopenCand} defines four coordinates $\hat{z}^a$ on the complex 
structure moduli space of $\hat{Z}_3$ that we calculate, according to the triangulation 
\eqref{blowupPolyCand} of $\Delta_7^{\hat{Z}}$, as
\begin{equation} \label{eq:zOpenCand}
 	\hat{z}^1=\frac{a_2a_4a_8}{a_1^2a_7}\,,\quad \hat{z}^2=\frac{a_4a_5^2a_6^3a_9}{a_0^6a_{10}}\,,\quad \hat{z}^3=\frac{a_3a_7}{a_1a_8}\,,\quad\hat{z}^4=\frac{a_1a_{10}}{a_4a_9}\,.
\end{equation}
We obtain a complete system of linear differential operators $\mathcal{D}_a$, the Picard-Fuchs system, 
by considering operators $\mathcal{L}_a$ associated to linear combinations of the charge 
vectors $\hat{\ell}^{(a)}$ forming the lattice of relations of $\Delta_7^{\hat{Z}}$. By factorizing these operators as 
expressed in the algebraic coordinates \eqref{eq:zOpenCand} we obtain the differential system 
\begin{eqnarray}
\mathcal{D}_{1}&=&\theta _1 \left(\theta _1-\theta _3\right) \left(\theta _1+\theta _2-\theta _4\right)+\left(\theta _1-\theta _3-1\right) \prod_{i=1}^2\left(2 \theta _1+\theta _3-\theta _4-i\right) \hat{z}^1\,,\\
\mathcal{D}_{2}&=&\theta _2 \left(\theta _2-\theta _4\right) \left(\theta _1+\theta _2-\theta _4\right)+12 \left(6 \theta _2-5\right) \left(6 \theta _2-1\right) \left(\theta _2-\theta _4-1\right) \hat{z}^2\,,\nonumber\\
\mathcal{D}_{3}&=&\left(\theta _1-\theta _3\right) \theta _3-\left(1+\theta _1-\theta _3\right) \left(2 \theta _1+\theta _3-\theta _4-1\right) \hat{z}^3\,,\nonumber\\
\mathcal{D}_{4}&=&\left(\theta _2-\theta _4\right) \left(-2 \theta _1-\theta _3+\theta _4\right)+\left(1+\theta _2-\theta _4\right) \left(1+\theta _1+\theta _2-\theta _4\right) \hat{z}^4\,,\nonumber\\
\mathcal{D}_{5}&=&\theta _1 \theta _3 \left(\theta _1+\theta _2-\theta _4\right)+\prod_{i=1}^3\left(2 \theta _1+\theta _3-\theta _4-i\right) \hat{z}^1\hat{z}^3\,,\nonumber\\
\mathcal{D}_{6}&=&\theta _1 \left(\theta _1-\theta _3\right) \left(\theta _2-\theta _4\right)+\left(\theta _1-\theta _3-1\right) \left(1+\theta _2-\theta _4\right) \left(2 \theta _1+\theta _3-\theta _4-1\right) \hat{z}^1\hat{z}^4\,,\nonumber\\
\mathcal{D}_{7}&=&\theta _2 \left(2 \theta _1+\theta _3-\theta _4\right)+12 \left(6 \theta _2-5\right) \left(6 \theta _2-1\right) \hat{z}^2\hat{z}^4\,,\nonumber\\
\mathcal{D}_{8}&=&\theta _1 \theta _3 \left(\theta _2-\theta _4\right)- \left(-\theta _2+\theta _4-1\right)\prod_{i=1}^2\left(2 \theta _1+\theta _3-\theta _4-i\right)  \hat{z}^1\hat{z}^3\hat{z}^4\,.\nonumber
\label{eq:GKZCandz}
\end{eqnarray}
The linear combination of charge vectors corresponding to each of these operators can be read 
off from the powers of the $\hat{z}^a$ in the last term of the $\mathcal{D}_a$.
We note that this system has the expected structure advertised in eq. \eqref{eq:LInZHat} of 
section \ref{sec:hatomega}. Consequently the periods $\Pi^k(z^1,z^2)$ lift to the blow-up 
$\hat{Z}_3$ upon the identification $z^1=\hat{z}^1\hat{z}^3$, $z^2=\hat{z}^2\hat{z}^4$.

\subsection{Brane Superpotential at Large Volume: Disk Instantons}
\label{sec:toricbranesP11169-W-LV}

In this section we solve the Picard-Fuchs system \eqref{eq:GKZCandz} at the point of maximal 
unipotent monodromy $\hat{z}^{a}\rightarrow 0$ in the complex structure moduli space of 
$\hat{Z}_3$. In addition we exploit the local limit $K_{\P^2}=\mathcal{O}_{\P^2}(-3)$ of $Z_3$, 
which is given by a decompactification of the elliptic fiber $t_2\rightarrow i\infty$, to 
determine the compact brane superpotential $W_{\rm brane}$ and the compact disk instanton 
invariants.

We find 16 solutions at large volume, that split into one power series solution $X^{(0)}$, 
four single logarithmic solutions $X^{(1)}_i$, seven double logarithmic solutions $X^{(2)}_\alpha$ 
and four triple logarithmic solutions $X^{(3)}_\beta$. As already expected from the observation 
made below \eqref{blowupPolyCand}, that $\hat{Z}_3$ connects to a compact Calabi-Yau fourfold, 
these are the only solutions. In particular, there are now square root and third root at large 
volume. The unique power series solution reads in the chosen normalization as
\begin{equation}
 	X^{(0)}=1+60 z_2+13860 z_2^2+4084080 z_2^3+24504480 z_1 z_2^3 +1338557220 z_2^4+\mathcal{O}(\underline{\hat{z}}^{10})\,,
\end{equation}
where we identify $z_1=\hat{z}_1\hat{z}_3$ and $z_2=\hat{z}_2\hat{z}_4$ as the complex structure 
moduli of $Z_3$ corresponding to $\ell^{(1)}$, $\ell^{(2)}$ in \eqref{eq:3foldellp2blowup}. 
Thus, $X^{(0)}$ is precisely the fundamental period $\Pi^0(z_1,z_2)$ of $Z_3$, 
cf.~\cite{Candelas:1994hw,Hosono:1993qy}. In addition we recover the other five periods of 
$\P^4(1,1,1,6,9)[18]$ as linear combinations of the solutions of the GKZ-system on $\hat{Z}_3$ 
with leading logarithms
\begin{eqnarray} \label{eq:leadinglogs11169}
		X^{(1)}_i\,:\,& \hat{l}_1\,,\,\,\hat{l}_2\,,\,\,\hat{l}_3\,,\,\,\hat{l}_4&\\[0.3Em]
		X^{(2)}_{\alpha}\,:\,& \frac{\hat{l}_1^2}{2}\,,\,\,  \hat{l}_2 (\hat{l}_1 +2 \hat{l}_4) \,,\,\,
\hat{l}_3(\hat{l}_1 +\frac{\hat{l}_3}{2})\,,\,\,\hat{l}_4(\hat{l}_1 +\hat{l}_4)\,,\,\,\frac{\hat{l}_2^2}{2}\,,\,\,\hat{l}_2 (\hat{l}_3+ \hat{l}_4)\,,\,\, \hat{l}_4(\hat{l}_3 +\frac{\hat{l}_4}{2} )&\nonumber\\[0.3Em]
		X^{(3)}_{\beta}\,:\,& \frac{1}{2} \hat{l}_1^2 \hat{l}_2+\frac{1}{2} \hat{l}_2^2 \hat{l}_3+\frac{1}{2} \hat{l}_1^2 \hat{l}_4+2 \hat{l}_1 \hat{l}_2 \hat{l}_4+\frac{1}{2} \hat{l}_2^2 \hat{l}_4+\hat{l}_1 \hat{l}_4^2+2 \hat{l}_2 \hat{l}_4^2+\frac{2 \hat{l}_4^3}{3} \,,\,\, \hat{l}_1 \hat{l}_2^2-\frac{3}{2} \hat{l}_2^2 \hat{l}_3+\frac{1}{2} \hat{l}_2^2 \hat{l}_4\,,& \nonumber\\
		&\!\!\!\!\hat{l}_1 \hat{l}_2 \hat{l}_3+\hat{l}_2^2 \hat{l}_3+\frac{1}{2} \hat{l}_2 \hat{l}_3^2+\hat{l}_1 \hat{l}_2 \hat{l}_4+\hat{l}_2^2 \hat{l}_4+\hat{l}_1 \hat{l}_3 \hat{l}_4+3 \hat{l}_2 \hat{l}_3 \hat{l}_4+\frac{1}{2} \hat{l}_3^2 \hat{l}_4+\frac{1}{2} \hat{l}_1 \hat{l}_4^2+\frac{5}{2} \hat{l}_2 \hat{l}_4^2+\frac{3}{2} \hat{l}_3 \hat{l}_4^2+\frac{5 \hat{l}_4^3}{6}\,,\!\!\!\nonumber\\ &  \frac{\hat{l}_2^3}{3}+\frac{1}{2} \hat{l}_2^2 \hat{l}_3+\frac{1}{2} \hat{l}_2^2 \hat{l}_4 \,,&\nonumber
\end{eqnarray}
where we used the abbreviation $\hat{l}_i=\log(\hat{z}_i)$. Indeed the threefold periods 
$l_1=\hat{l}_1+\hat{l}_3$, $l_2=\hat{l}_2+\hat{l}_4$, $l_1^2$, $l_1l_2+\frac{3}{2}l_2^2$ 
and $l_2(\frac12 l_1^2+\frac32 l_1 l_2+\frac32 l_2^2)$ are linear combinations of the 
solutions in \eqref{eq:leadinglogs11169}. Furthermore, we readily check that the complete 
$(z_1,z_2)$-series of the periods agree with the solutions on the blow-up.

Next we obtain the disk instanton invariants of the A-model dual to the five-brane on $Z_3$ 
from the local limit geometry $K_{\P^2}$. First we use the single logarithms of 
\eqref{eq:leadinglogs11169} as the mirror map $\hat{t}_i=\frac{X_i^{(1)}}{X^{(0)}}$ at 
large volume, where we have the series expansions
\begin{eqnarray}
		X^{(1)}_1&\!\!\!=&\!\!\! X^{(0)}\log(\hat{z}_1)-4 \hat{z}_1 \hat{z}_3+120 \hat{z}_2 \hat{z}_4+60 \hat{z}_2 \hat{z}_3 \hat{z}_4+30 \left(\hat{z}_1^2 \hat{z}_3^2+4 \hat{z}_1 \hat{z}_2 \hat{z}_3 \hat{z}_4+1386 \hat{z}_2^2 \hat{z}_4^2\right) +\mathcal{O}(\underline{\hat{z}}^5),\!\!
		\nonumber\\
		X^{(1)}_2&\!\!\!=&\!\!\! X^{(0)}\log(\hat{z}_2)
-60 \hat{z}_2-3080 \hat{z}_2^2 \left(442 \hat{z}_2+9 \hat{z}_4\right)+6 \hat{z}_2 \left(1155 \hat{z}_2+62 \hat{z}_4\right)+\mathcal{O}(\underline{\hat{z}}^4)\,,\nonumber\\
		X^{(1)}_3&\!\!\!=&\!\!\! X^{(0)}\log(\hat{z}_3)-2 \hat{z}_1 \hat{z}_3+60 \hat{z}_2 \hat{z}_4-60 \hat{z}_2 \hat{z}_3 \hat{z}_4+15 \left(\hat{z}_1^2 \hat{z}_3^2+4 \hat{z}_1 \hat{z}_2 \hat{z}_3 \hat{z}_4+1386 \hat{z}_2^2 \hat{z}_4^2\right)+\mathcal{O}(\underline{\hat{z}}^5),\!\nonumber\\
		X^{(1)}_4&\!\!\!=&\!\!\! X^{(0)}\log(\hat{z}_4)+60 \hat{z}_2-6930 \hat{z}_2^2+2 \hat{z}_1 \hat{z}_3-60 \hat{z}_2 \hat{z}_4+3080 \hat{z}_2^2 \left(442 \hat{z}_2+9 \hat{z}_4\right)+\mathcal{O}(\underline{\hat{z}}^4)\,.
\end{eqnarray}
Here we omit a factor $2\pi i$ for brevity. This confirms the consistency with the mirror 
of the threefold $Z_3$, $t_1=\hat{t}_1+\hat{t}_3$ and $t_2=\hat{t}_2+\hat{t}_4$, since 
the periods agree as 
$\Pi^{(1)}(\underline{z})=X^{(1)}_1+X^{(1)}_3=X^{(0)}\log(z_1)-6 z_1 + 45 z_1^2+\ldots$ 
and $\Pi^{(2)}(\underline{z})=X^{(1)}_1+X^{(1)}_3=X^{(0)}\log(z_2)+2 z_1+312 z_2+\ldots$. 
Upon inversion of the mirror map, we obtain the $\hat{z}^i$ as a series of 
$\hat{q}_a=e^{2\pi i \hat{t}_a}$, that we readily insert into the double logarithmic solutions 
in \eqref{eq:leadinglogs11169}. Then we construct a linear combination of these solutions 
to match the disk instantons in \cite{Aganagic:2000gs,Aganagic:2001nx} of the local geometries. 
Since $q_2=e^{2\pi it_2}\rightarrow 0$ in the local limit this means that we match, as a 
first step, only the part of the $q$-series, that is independent of $q_2$. Morally speaking, 
this procedure fixes part of the flux on $\hat{Z}_3$ specifying the five-brane. Indeed we 
obtain a perfect match of the disk instantons for both brane phases $I/II$, $III$ considered in 
\cite{Aganagic:2001nx} for the choices of superpotential
\begin{eqnarray} \label{eq:W11169}
 	W^{I/II}_{\rm brane}&=&\Big((\tfrac12(a_4-a_7)+a_3-\tfrac12)X^{(2)}_1+\sum_{i=2}^7 a_iX^{(2)}_i\Big)/X^{(0)}\,,\\
	W^{III}_{\rm brane}&=&\Big(a_1X^{(2)}_1+\tfrac{1}{3}(1+a_4+a_5-2 a_6+a_7)X^{(2)}_1+\sum_{i=3}^7 a_iX^{(2)}_i\Big)/X^{(0)}\,,
\end{eqnarray}
where we in addition fix the parameters $a_3=a_7=0$ to switch off the contribution of the 
closed instantons of $Z_3$ through its double logarithmic periods. We note that the two 
choices corresponding to \eqref{eq:W11169} are compatible with each other so that we can 
also find a single superpotential matching both phases of \cite{Aganagic:2001nx} simultaneously, 
corresponding to the parameters 
$a_1=-1+\tfrac{3}{2}a_2+a_3-\tfrac{1}{2}a_5+a_6-a_7,a_4=-1+3 a_2-a_5+2 a_6-a_7$.
Most notably, this match and the match in \eqref{eq:W11169} of the disk instantons of the 
local geometry already predicts the parts of the disk instantons in the \textit{compact} 
threefold $Z_3$. The compact disk instantons according to the Ooguri-Vafa multi-covering 
formula are listed in the tables in appendix \ref{sec:instantonsLV11169}.

\section{Heterotic/F-Theory Duality and Blow-Up Threefolds}
\label{sec:heteroticF+blowup}

In this section we apply the blow-up procedure of section \ref{sec:5braneblowupsanddefs}  
in the context of heterotic/F-theory duality.
This allows us to relate the geometry of the blow-up $\hat{Z}_3$ of an elliptic Calabi-Yau 
threefold $Z_3$ as well as the heterotic superpotential, in particular the superpotential induced 
by fluxes and horizontal five-branes, to the dual F-theory fourfold $X_4$ and flux superpotential. 
Following \cite{Grimm:2009sy} we demonstrate explicitly in section \ref{sec:heterotic_blowup} that the 
geometry of the blow-up $\hat{Z}_3$, constructed as a complete intersection, can be used to obtain 
the dual Calabi-Yau fourfold $X_4$ on the F-theory side, which is also given as a complete intersection. 
In particular we recover the toric data defining the F-theory fourfold. Then in section 
\ref{sec:non-CYblowup} we use the the blow-up description $\hat{Z}_3$ of the heterotic string 
with horizontal five-branes for a direct geometric map of the heterotic flux and five-brane superpotential 
to the F-theory dual flux superpotential. In particular we show how to formally identify the 
four-form flux $G_4$ dual to the five-brane from the perspective of the blow-up $\hat{Z}_3$. This 
all agrees with the observations of section \ref{sec:ToricBraneBlowupII}, where it was noted that the GKZ-system 
of $\hat{Z}_3$ is related to the toric data of the dual F-theory fourfold $X_4$.

\subsection{F-theory Fourfolds from Heterotic Blow-Up Threefolds}
\label{sec:heterotic_blowup}

In this section we discuss the example of section \ref{sec:Example1} 
employing the blow-up procedure of chapter \ref{ch:blowup}.
We find that the geometry of the F-theory Calabi-Yau fourfold $X_4$ is 
naturally obtained from $\hat Z_3$ by an additional $\P^1$-fibration and in particular
identical to the fourfold $\hat{X}_4$ considered in section \ref{sec:Example1}, despite the fact 
that it is now realized as a complete intersection. 

As in section \ref{sec:Example1} the starting point is the elliptic 
fibration $\tilde{Z}_3$ over $B_2=\P^2$  with a five-brane wrapping the 
hyperplane class of the base. Let us describe the explicit construction of $\hat Z_3$.
Recall that $\tilde{Z}_3$ is a hypersurface $\{p_0=0\}$ in a toric variety $\P_{\tilde{\Delta}}$ and 
assume that the curve $\Sigma$ is given as a complete intersection of two hypersurfaces in 
$\tilde{Z}_3$ that takes for a horizontal curve the form
\begin{equation}
	h_1 \equiv\tilde{z}=0\ ,\quad h_2\equiv g_5=0\ .
\end{equation}	
Here $\{\tilde{z}=0\}$ restricts to the base $B_2$ and $g_5$ specifies $\Sigma$ within
$B_2$, cf.~section \ref{sec:F_blowup}. 
Assume further that the charge vectors of $\P_{\tilde{\Delta}}$ are given by $\{\ell^{(i)}\}$ with $i=1,\dots,k$.
We are aiming to torically construct a five-dimensional toric variety which is given by 
$\hat{\P}_{\tilde{\Delta}}=\P(N_{\P_{\tilde{\Delta}}}\Sigma)$ and in which the blow-up $\hat{Z}_3$ 
is given by the complete intersection \eqref{eq:blowup}. Let us denote the divisor classes defined by 
$h_i$ as $H_i$ and the charges of $h_i$ by $\mu_i=(\mu_{i}^{(1)},\dots\mu_{i}^{(k)})$.
Then, the coordinates $l_i$ of $N_{\P_{\tilde{\Delta}}}H_i$ transform with charge $\mu_i^{(m)}$
under the $k$ scaling relations.
Since we have to projectivize $N_{\P_{\tilde{\Delta}}} \Sigma$, we have to include another 
$\C^*$-action with charge vector $\ell^{(k+1)}_{\hat{\P}_{\tilde{\Delta}}}$ acting non-trivially only 
on the new coordinates $l_i$.
The new charge vectors of $\hat{\P}_{\tilde{\Delta}}$ are thus given by the following table:
\begin{center}
	\begin{tabular}{|c||c|c|c|}
		\hline\T  & coordinates of $\P_{\tilde{\Delta}}$ & $l_1$ & $l_2$ \B\\
		 \hline  $\ell^{(1)}_{\hat{\P}_{\tilde{\Delta}}}$ & $\ell^{(1)}$ & $\mu_{1}^{(1)}$ & $\mu_{2}^{(1)}$ \B\\
		 \hline\T \vdots & \vdots & \vdots & \vdots \B\\
		 \hline\T $\ell^{(k)}_{\hat{\P}_{\tilde{\Delta}}}$ & $\ell^{(k)}$ & $\mu_{1}^{(k)}$ & $\mu_{2}^{(k)}$\B \\
		 \hline\T $\ell^{(k+1)}_{\hat{\P}_{\tilde{\Delta}}}$ & 0 & 1 & 1 \B\\ \hline
	\end{tabular}
\end{center}
Then the blown-up threefold $\hat{Z}_3$ is given by the complete intersection \eqref{eq:blowup} with $P\equiv p_0$ 
in this toric space.

To apply this to the elliptic fibration over $\P^2$ with the polyhedron \eqref{eq:3foldellp2}, one 
picks the curve $\Sigma$ given by $h_1:=$ and $g_5:=x_1$.
$\Sigma$ is a genus zero curve and we readily infer that the exceptional divisor $E$ is the first 
del Pezzo surface $dP_1$ in accord with the general formulas for the topology of $E$ in appendix 
\ref{App:topoHatZ_3}. We construct the five-dimensional ambient manifold $\hat{\P}_{\tilde\Delta}$
as explained above,
\begin{equation}
	\centering
	\Delta_5^{\hat Z}=\left(
	\begin{tabular}{ccccc|c}
		-1 & 0 & 0 & 0& 0 & $3B+3D+9H$\\
		0 & -1 & 0 & 0 & 0 &$2B+2D+6H$\\
		3 & 2& 0 & 0 & 0 & $B$\\
		3 & 2 & 1 & 1 & 1 & $H$\\
		3 & 2 & -1 & 0 & 0 & $H$\\
		3 & 2 & 0 & -1 & 0 & $H$\\
		3 & 2 & 0 & 0 & -1 & $E$\\
		0 & 0 & 0 & 0 & -1 & $H-E$
	\end{tabular}
	\right).
	\label{eqn:blowup-poly}
\end{equation}
Note that one has to include the inner point $(3,2,0,0,0)$ 
which corresponds to the base of the elliptic fibration $\hat Z_3$.
Furthermore, one readily shows that the point $(0,0,0,0,1)$, required for the above 
scalings, can be omitted since the associated divisor does not intersect the complete intersection 
$\hat Z_3$. Explicitly the complete intersection $\hat Z_3$ is given by a generic constraint in 
the class 
\begin{equation}
	\hat Z_3: \quad (6B+6E+18H) \cap H\ ,
	\label{eqn:ci-blow-up}
\end{equation}
where $H,\,B,\,E$  are the divisor classes of the ambient space \eqref{eqn:blowup-poly}.
The first divisor in \eqref{eqn:ci-blow-up} is the sum of the first seven divisors in \eqref{eqn:blowup-poly} 
and corresponds to the original Calabi-Yau constraint $p_0=0$ of $Z_3$.  The second divisor in 
\eqref{eqn:ci-blow-up} is the sum of the last two divisors and is the class of the second constraint $Q$ 
in \eqref{eq:blowup}. This complete intersection threefold has 
$\chi(\hat Z_3)=-538=\chi(\tilde{Z}_3)-\chi(\P^1)+\chi(dP_1)$, and one checks that the exceptional 
divisor $E$ has the characteristic data of a del Pezzo surface $dP_1$. 
This means that we have replaced the hyperplane which is isomorphic to $\P^1$ in the base by the 
exceptional divisor which is $dP_1$. It can be readily checked that the first Chern class of $\hat Z_3$ is 
non-vanishing and equals $-E$.

Having described the blow-up geometry $\tilde{Z}_3$, we now turn to the construction
of the fourfold $\hat{X}_4$ for F-theory.
This fourfold will also be constructed as complete intersection, but it will
be the same manifold as the fourfold described in section \ref{sec:Example1} by equation \eqref{eq:fourfold1}.
Geometrically we fiber an additional $\P^1$ over $\hat{\P}_{\tilde{\Delta}}$ which is only non-trivially fibered 
along the exceptional divisor.
This is analogous to the construction of the dual fourfold in heterotic/F-theory duality 
where one also fibers a $\P^1$ over the base $B_2$ of the Calabi-Yau threefold to obtain the base $B_3$ of the F-theory 
fourfold. Here we proceed in a similar fashion but construct a $\P^1$-fibration over the base of the 
non-Calabi-Yau manifold $\hat Z_3$. This base is a complete intersection and thus leads to a realization of  
$\hat{X}_4$ as a complete intersection. Concretely, we have the following polyhedron
\begin{equation}
	\Delta^{\hat{X}}_6=\left(
	\begin{array}{cccccc|c|c}
		-1 &  0 &  0  & 0  & 0  & 0  & 3D+3B+9H+6K & D_1 \\
		0  & -1 &  0  & 0  & 0  & 0  & 2D+2B+6H+4K & D_2  \\
		3  &  2 &  0  & 0  & 0  & 0  & B & D_3 \\
		3  &  2 &  1  & 1  & 1  & 0  & H& D_4 \\
		3  &  2 &  -1 & 0  & 0  & 0  & H& D_5 \\
		3  &  2 &  0  & -1 & 0  & 0  & H& D_6 \\
		3  &  2 &  0  & 0  & 0  & 1  & K& D_7 \\
		3  &  2 &  0  & 0  & 0  & -1 & K+E& D_8 \\ \hline
		0  &  0 &  0  & 0  & -1 & 1  & E& D_9 \\
		0  &  0 &  0  & 0  & -1  & 0 & H-E& D_{10} 
	\end{array}
	\right).
	\label{eqn:ci-fourfoldpoly}
\end{equation}
The fourfold $\hat X_4$ is then given as the complete intersection in the class
\begin{equation}
	\hat{X}_4:\quad (6B+6E+18H+12K)\cap H\ .
	\label{eqn:ci-fourfold}
\end{equation}

Note that this fourfold indeed obeys Calabi-Yau condition as can be checked explicitly by analyzing the toric 
data \eqref{eqn:ci-fourfoldpoly}. For complete intersections the Calabi-Yau constraint is realized 
via the two partitions, so-called nef partitions, in \eqref{eqn:ci-fourfoldpoly} as in 
\cite{Batyrev:1994pg,Batyrev:2002fd}. The first nef partition yields the sum of the first eight 
divisors $\sum_{i=1}^{8}D_i$ in \eqref{eqn:ci-fourfoldpoly} and gives the 
first constraint in \eqref{eqn:ci-fourfold}. The second nef partition yields the sum 
of the last two divisors $D_9+D_{10}$ in \eqref{eqn:ci-fourfoldpoly}
and determines the second constraint in  \eqref{eqn:ci-fourfold}.
The divisors $D_7$ and $D_8$ correspond to the $\P^1$-fiber in the 
base of $\hat X_4$ obtained by dropping the first two columns in 
\eqref{eqn:ci-fourfoldpoly}. This fibration is only non-trivial over the 
exceptional divisors $D_9=E$ in the second nef partition of \eqref{eqn:ci-fourfoldpoly}. 
Note that if one simply drops $K$ from the expression \eqref{eqn:ci-fourfold}
one formally recovers the constraint \eqref{eqn:ci-blow-up} of $\hat Z_3$.
To check that the complete intersection $\hat{X}_4$ is precisely the 
fourfold constructed in section \ref{sec:Example1}, one has to compute the 
intersection ring and Chern classes. In particular, it is not hard to show that 
also \eqref{eqn:ci-fourfoldpoly} has three triangulations matching the 
result of section \ref{sec:Example1}.

In summary, we have found that there is a natural construction of $\hat{X}_4$ as
complete intersection with the base obtained from the heterotic non-Calabi-Yau threefold 
$\hat Z_3$. Let us stress that this construction will straightforwardly generalize to dual 
heterotic/F-theory setups with other toric base spaces $B_2$ and different types 
of bundles. For example, to study the bundle configurations on $Z_3$ of section \ref{sec:Example1} 
with $\eta_{1,2} = 6 c_1(B_2) \pm k H,\ k=0,1,2$ one has 
to replace 
\begin{equation}
  D_4 \rightarrow (3,2,1,1,k)\ ,\qquad D_4  \rightarrow  (3,2,1,1,1,k)\ ,
\end{equation}
in the polyhedra \eqref{eq:fourfold1} and \eqref{eqn:ci-fourfoldpoly}, respectively.
Moreover, also bundles
which are not of the type $E_8 \times E_8$ can be included by generalizing the form 
of the $\P^1$-fibration just as in the standard construction of dual F-theory fourfolds.

\subsection{Superpotentials and $G_4$-fluxes from Heterotic Blow-Ups}
\label{sec:non-CYblowup}

In this concluding section we show how the geometric matching of section \ref{sec:heterotic_blowup} 
can be used to map the heterotic flux and brane superpotential to the F-theory side. In particular we 
use the geometry of $\hat{Z}_3$ to specify the corresponding $G_4$-flux on $X_4$.

For concreteness, let us again consider an elliptic Calabi-Yau threefold $Z_3$ described as the
hypersurface $\{p_0 = 0\}$ in a projective or toric ambient space $\P_{\Delta}$. 
Then, the blown-up threefold $\hat Z_3$ is given by the complete
intersection \eqref{eq:blowup} in the projective bundle $\mathcal{W}$. 
Next we use the lift of both the flux as well as the brane superpotential discussed in section 
\ref{sec:potentialhatZ3minusD}, where its complete moduli dependence is entirely encoded
in the complex structure dependence of the lifted three-form $\hat{\Omega}$.
On $\hat{Z}_3$ the combined heterotic superpotential can be written as
\begin{equation}
 	W_{\rm flux}+W_{\rm brane}=\int_{\hat{Z}_3} \hat{H}_3\wedge
\hat{\Omega}
\end{equation}
where $\hat{H}_3$ contains the current \eqref{eq:dhatH_3} and also a smooth flux part encoding 
$W_{\rm flux}$. 

Now we can in principle proceed as in section \ref{sec:ToricBraneBlowup} and derive  
Picard-Fuchs differential equations for $\hat{\Omega}$ by studying its complex structure 
dependence explicitly. Then the closed periods of $Z_3$ as well as the superpotential $W_{\rm brane}$
are expected to be among the solutions to this system as argued in sections \ref{sec:hatomega} and 
\ref{sec:potentialhatZ3minusD}. However, instead of performing a direct analysis on the heterotic side 
we follow the route of heterotic/F-theory duality to calculate the superpotential.

In accord with the discussion of section \ref{sec:heterotic_blowup} the
dual fourfold $X_4$ can in general be realized as a complete intersection 
blown up along the five-brane curves, cf.~also section \ref{sec:F_blowup} for the 
general construction.
We want to match this description with the heterotic theory on $\hat Z_3$.
Indeed, one can now identify the blow-up constraints on the heterotic and on the F-theory side as
\begin{equation}
	\xymatrix{Q_{\rm het} = l_1 g_5(\underline{u}) - l_2 \tilde{z} \ar@{|->}[r]& 
	Q_{\rm F}= l_1 g_5(\underline{u}) - l_2 k} \ , \qquad \qquad \xymatrix{\tilde z \ar@{|->}[r]& k}\ ,
	\label{eq:Q-map}
\end{equation}
where $\underline{u}$ denote coordinates on the base $B_2$, 
$(\tilde z=0)$ defines the base $B_2$ in $Z_3$, which is realized in $X_4$ as 
$(z=0)\cap (k=0)$.\footnote{Note that the $\P^1$-fibration 
$B_3 \rightarrow B_2$ has actually two sections, one of which is an isolated section and the other
comes in a family of sections. As in section \ref{sec:F_blowup}, $k=0$ 
is one of the two sections, say, the zero section.}  
The map \eqref{eq:Q-map} is possible since both $Z_3$ and $X_4$ share the twofold base 
$B_2$ with the curve $\Sigma$. The identification of $\tilde z$ with $k$
corresponds to the fact that in heterotic/F-theory duality the elliptic fibration 
of $Z_3$ is mapped to the $\P^1$-fibration of $B_3$, see section \ref{sec:het-Fdual}.
Clearly, the map \eqref{eq:Q-map} identifies the deformations of $\Sigma$ realized 
as coefficients in the constraint $Q_{\rm het}$ of $\hat Z_3$ with the complex structure deformations of 
$X_4$ realized as coefficient in $Q_{\rm F}$. 

We also have 
to match the remaining constraints $p_0$ and $P$ of $\hat Z_3$ and 
$X_4$, respectively. Clearly, there will not be a general match. 
However, as was argued in \cite{Berglund:1998ej} for Calabi-Yau fourfold hypersurfaces, 
one can split $P=p_0+ V_E$ yielding a map 
\begin{equation}
	\xymatrix{p_0 + V_E\ar@{|->}[r]& P}\ ,
	\label{eq:P-map}
\end{equation}
where $V_E$ describes the spectral cover of the dual heterotic bundles $E=E_1 \oplus E_2$.
Again, this requires an identification of $\tilde z$ and $k$. For $SU(1)$-bundles this map 
was given in \eqref{eq:Q-map}, but can be generalized for non-trivial bundles.
Note that the maps \eqref{eq:Q-map} and \eqref{eq:P-map} can also be formulated in terms of the GKZ-systems 
of the complete intersections $\hat Z_3$ and $X_4$. It implies that the charge vectors $\ell^{(a)}_i$ of $X_4$
contain charge vectors of $Z_3$ and the five-brane charge vectors $\hat{\ell}^{(a)}$, which is precisely 
the situation encountered for the blow-up GKZ-system constructed in section \ref{sec:generaltoricstructure} and in 
\cite{Alim:2009rf,Alim:2009bx,Grimm:2009ef,Aganagic:2009jq,Li:2009dz}.

To match the superpotentials $W_{\rm het}$ and $W_{G_4}$ as in \eqref{eq:supermatch} one finally has to identify 
the flux quanta on the blow-up $\hat{Z}_3$ including the five-brane superpotential\footnote{Formally these integrals
are defined by the cohomology of currents \cite{Griffiths:1978yf} or its dual the relative homology group 
$H_3(\hat{Z}_3,E)=H_3(\hat{Z}_3-E)$ \cite{Lerche:2001cw,Lerche:2002ck,Lerche:2002yw}.} with elements of $H^{4}( X_4,\Z)$
and show that the periods of $\hat \Omega$ as well as $W_{\rm brane}$ can 
be identified with a subset of the fourfold periods of $\Omega_4$. In order 
to do that, one compares the residue integrals \eqref{eq:ResZhat}  and \eqref{eq:residuumform}
for both $\hat \Omega$ and $\Omega_4$ in the case of complete intersections. 
Using the maps \eqref{eq:Q-map} and \eqref{eq:P-map} one then shows that 
each Picard-Fuchs operator annihilating $\hat \Omega$ also annihilates 
$\Omega_4$. Hence, also a subset of the solutions to the Picard-Fuchs equations can 
be matched accordingly. As a minimal check, one finds employing the arguments of section \ref{sec:hatomega} 
that the periods of $\Omega$ on $Z_3$ before the blow-up arise as a subset of the periods of $\Omega_4$ \cite{Grimm:2009ef}. The map between the cohomologies 
$H^{3}(\hat Z_3-E,\Z) \hookrightarrow H^{4}(X_4,\Z)$ is 
then best formulated in terms of operators $\cR_p^{(i)}$ applied to 
the forms $\hat \Omega$ and $\Omega_4$, 
\begin{equation} 
	\xymatrix{
\left. {\cal R}_{p}^{(i)}\tilde \Omega_3(\underline{t}^c,\underline{t}^o)\right|_{\underline{t}^{c}=\underline{t}^{o}=0}\quad
\ar@{|->}[r]& \quad \left. {\cal R}_{p}^{(i)}\Omega_4(\underline{z})\right|_{\underline z=0}
	}\ .\label{eq:fluxmap}
\end{equation}
Here we used a notation from the context of mirror symmetry on fourfolds, where horizontal fluxes are described by derivatives of $\Omega_4$, see section \ref{sec:FFMirrors}, also for the blow-up $\hat{Z}_3$. 
The map \eqref{eq:fluxmap} implies the assumption that the heterotic flux $\hat{H}_3$ can be represented
by ${\cal R}_{p}^{(i)}\tilde \Omega_3$.
We conclude by noting that the preimage of this map will in general contain derivatives with
respect to the closed and  open string variables $\underline{t}^c$, $\underline{t}^o$.
One then finds that by identifying the heterotic and F-theory moduli at the large complex
structure point $\underline{z}=0$, one obtains an embedding map of the integral basis on $\hat{Z}_3$.

\chapter{SU(3)-Structure on Brane Blow-Ups}
\label{ch:su3structur}

In the previous chapters we have argued that five-brane deformations
can be equivalently described on $\hat Z_3$ with a two-form flux 
localized on the blow-up divisor $E$. The aim of this section 
is to delocalize the flux further to three-form flux and propose an 
$SU(3)$-structure on the open manifold $\hat Z_3-E$ and on $\hat Z_3$. 
Let us note that this chapter is independent of any explicit 
toric construction and the reader only interested in the enumerative 
aspects of the superpotential can safely skip this section. 
We hope to provide concrete proposals which should only be viewed 
as first steps to identify a complete back-reacted vacuum. 

We begin our discussion, following \cite{Grimm:2010gk}, 
in section \ref{sec:SU(3)rev} by recalling some basic facts about 
$SU(3)$-structure manifolds and the generalized flux superpotential. 
As an intermediate step, we show next how the blow-up threefold
$\hat Z_3$ can be endowed with a K\"ahler structure 
in section \ref{sec:blow-up_as_Kaehler}. However, it is 
well-known that a supersymmetric vacuum with background 
three-form fluxes requires that the internal space is non-K\"ahler (section \ref{sec:SU(3)rev}).
Furthermore, there exists no globally defined, nowhere vanishing holomorphic three-form on $\hat Z_3$. 
We propose a resolution to these issues in two steps in section \ref{sec:non-Kaehlertwist}. 
In a first step we argue that there is a natural non-K\"ahler structure $\hat J$ on the 
open manifold $\hat Z_3 - E$ which, in a supersymmetric vacuum, 
matches the flux via $i(\bar \partial - \partial)\hat H_3 = d\hat J$.
While $(\hat J, \hat \Omega, \hat H_3)$ are well-defined forms 
on the open manifold, they have poles, as it is the case for $\hat J$ and $\hat H_3$, 
and zeros, as we have seen for $\hat \Omega$, when extended to all of $\hat Z_3$. 
Hence, in a second step, we argue that the poles and zeros can be removed by an appropriate
local logarithmic transformation yielding new differential forms $(J',\Omega',H_3')$ on $\hat Z_3$ . 
In fact, the new global forms are defined 
such that the zeros and poles precisely cancel in the superpotential which 
can then be evaluated on $\hat Z_3$.

\section{Brief Review on SU(3)-Structures and the Superpotential} 
\label{sec:SU(3)rev} 

To begin with, let us recall some basic facts about compactifications 
on non-Calabi-Yau manifolds $\hat Z_3$. In order that the four-dimensional effective theory 
obtained in such compactifications has $\cN=1$ supersymmetry one 
demands that $\hat Z_3$ has $SU(3)$-structure \cite{Koerber:2010bx}.
$SU(3)$-structure manifolds can be characterized by the
existence of two no-where vanishing forms, a real two-form $J'$ and a real three-form $\rho'$. 
Following \cite{Hitchin:2000jd} one demands that $J'$ and $\rho'$ are stable forms, 
i.e.~are elements of open orbits under the action of general linear transformations $GL(6,\bbR)$ at 
every point of the tangent bundle $T\hat Z_3$. Then one can set $\hat\rho'=*\rho'$ and show that 
$\hat \rho'$ is a function of $\rho'$ only \cite{Hitchin:2000jd}.
These forms define a reduction of the structure group from $GL(6,\bbR)$ to
$SU(3)$ if they satisfy $J \wedge \Omega' = 0$,
with a nowhere vanishing three-form $\Omega'=\rho'+i\hat\rho'$ and $J'$.

By setting $I_m^{\ n} = J'_{mp}g^{pn}$ one defines 
an almost complex structure with respect to which the metric $g_{mn}$ is hermitian.
The almost complex structure allows to introduce a $(p,q)$ grading of forms. Within this 
decomposition the form  $J'$ is of type $(1,1)$ while $\Omega'$ is of type $(3,0)$. In general, 
neither $J'$ nor $\Omega'$ are closed. The non-closedness is parameterized 
by five torsion classes $\cW_i$ which transform as 
$SU(3)$ irreducible representations \cite{Chiossi:2002tw,Cardoso:2002hd}. One has 
\begin{eqnarray}\label{dJ}
    dJ'&=&\tfrac{3}{2}\I (\bar \WV_1\Omega')+\WV_4\wedge J'+\WV_3\nn\\
     d\Omega'
   &=&\WV_1 J' \wedge J'+\WV_2\wedge J'+\overline\WV_5\wedge\Omega' \ ,
\end{eqnarray}
with constraints $J'\wedge J'\wedge\WV_2=J'\wedge\WV_3=\Omega' \wedge\WV_3=0$.
The pattern of vanishing torsion classes defines the properties of 
the manifold $\hat Z_3$. In a supersymmetric vacuum the 
pattern of torsion classes is constraint by the superpotential.  

Let us first discuss the pure flux superpotential for heterotic and Type IIB orientifolds with $O5$-planes. 
Recall from section \ref{sec:fivebranesuperpotential} that the pure flux superpotential of these theories 
is of the form $W_{\rm flux} = \int \Omega \wedge H_3$ and $W_{\rm flux} = \int \Omega \wedge F_3$.
It is easy to check that there are no supersymmetric flux vacua for Calabi-Yau compactifications.
In fact, in the absence of branes $W_{\rm flux}$ is the only perturbative superpotential for a 
Calabi-Yau background. The supersymmetry conditions are 
\begin{equation} \label{DW=0}
  D_{z^k} W_{\rm flux} = 0 \ , \qquad W_{\rm flux}= 0 \ ,
\end{equation}
where the latter condition arises from the fact that $D_S W_{\rm flux} = K_S W_{\rm flux} = 0$,
for other moduli $S$ which do not appear in $W_{\rm flux}$. One easily checks that the 
first condition in \eqref{DW=0} implies that $H_3$ cannot be of type $(2,1)+(1,2)$, while 
the second condition implies that it cannot be of type $(3,0)+(0,3)$. This implies that 
$H_3$ has to vanish and there are no flux vacua in a Calabi-Yau compactification.

The situation changes for non-Calabi-Yau compactifications since the superpotential in 
this case is of more general form. 
More precisely, denoting by $\hat Z_3$ a generic $SU(3)$-structure manifold it takes
the form \cite{Behrndt:2000zh,LopesCardoso:2003af,Benmachiche:2008ma}
\begin{equation}
   W = \int_{\hat Z_3} \Omega' \wedge (H'_3+idJ')\ .
\end{equation}
It is straightforward to evaluate the supersymmetry conditions for 
this superpotential. Firstly, we note that in a supersymmetric background 
the compact manifold $\hat{Z}_3$ is complex, thus the torsion classes vanish, $\cW_1 = \cW_2=0$.
Second the superpotential $W$
is independent of the dilaton superfield and hence one 
evaluates in the vacuum that $W=0$, which implies that the $(0,3)$
part of $H'_3 + idJ'$ has to vanish.
However, since the $(3,0)+(0,3)$ component of $dJ'$ vanishes 
for a vanishing $\cW_1$, one concludes that also $H_3'+i dJ'$ has no 
$(3,0)$ component, and hence can be non-zero in the $(2,1)$ and $(1,2)$ directions.
The K\"ahler covariant derivative of $\Omega'$ yields 
the condition that $H_3'+idJ'$ has only components along the
$(2,1)$ direction. Hence, using the fact that $J'$ is of type $(1,1)$ 
one finally concludes  
\begin{equation} \label{dJ=H}
   H'_3 = i (\bar \partial - \partial) J'\ .
\end{equation}
This matches the long-known relation found in \cite{Strominger:1986uh} for
a supersymmetric vacuum of the heterotic string with background 
three-form flux. It should be stressed that 
there will be additional conditions which have to be respected 
by the heterotic vacuum. These involve a non-constant dilaton 
and cannot be captured by a superpotential. 

\section{The Blow-Up Space as a K\"ahler Manifold} 
\label{sec:blow-up_as_Kaehler}

Before introducing an $SU(3)$-structure on $\hat Z_3$, it is necessary to 
recall that every variety $\hat Z_3$ obtained by blowing up a holomorphic curve 
naturally admits a K\"ahler structure \cite{Griffiths:1978yf,Voisin2002}. We `twist' 
this K\"ahler structure to obtain a non-K\"ahler $SU(3)$-structure in 
section \ref{sec:non-Kaehlertwist}. It is crucial, however, to look first at the geometry of 
$\hat Z_3$ near $E$ more closely, and introduce the K\"ahler structure very explicitly. 
As a preparation we consider in section \ref{sec:Kahleronpointblowup} the blow-up of
the origin in $\C^2$ and construct its canonical K\"ahler form. This can be viewed as the 
normal bundle $N_{Z_3}\Sigma$ at a point in $\Sigma$ and thus as local version 
of the blow-up of $Z_3$ along $\Sigma$. Indeed the K\"ahler structure on $\hat{Z}_3$ is 
defined in section \ref{sec:KaehleronBlowupThreefold} in a formally very similar way.

\subsection{K\"ahler Geometry on the Blow-Up: Warm-Up in Two Dimensions} 
\label{sec:Kahleronpointblowup}

To warm up for the more general discussion, let us first consider a simpler 
example and blow up a point in a complex surface. In a small patch $U_\epsilon$ 
around this point this looks like blowing up the 
the origin in $\bbC^2$ into an exceptional divisor $E = \P^1$. 
Let us denote by $B_{\bbC^2}$
the space obtained after blowing up as in section \ref{sec:geometricblowups}. 
Our aim is to explicitly define a K\"ahler form $\tilde J$ on $B_{\bbC^2}$ 
following ref.~\cite{Griffiths:1978yf}.

To define $\tilde J$ the key object we will study is the line bundle $\cL \equiv \cO_{B_{\bbC^2}}(E)$, 
or rather $\cL^{-1} \equiv \cO_{B_{\bbC^2}}(-E)$. To get a clearer picture of this bundle, we 
give a representation of $\cL$ near $E$. As in subsection \ref{sec:geometricblowups} we first introduce the patch
$\hat U_\epsilon = \pi^{-1}(U_\epsilon)$.  
One can embed 
the fibers of $\cL$ into $\hat U_{2\epsilon} = \{U_{2\epsilon} \times \P^1: y_2 l_1 - y_1 l_2=0\}$ as
\begin{equation} \label{expl_cL}
   \cL_{(y,l)} = \{\lambda\cdot (l_1 ,l_2), \, \lambda \in \bbC \}\ ,
\end{equation}
where $(l_1,l_2)$ are the projective coordinates of $\P^1$ and $y$ collectively 
denote the coordinates on $U_\epsilon$. This implies that 
holomorphic sections $\sigma$ of $\cL$ are locally specified by $\sigma \equiv \lambda(\underline{y},\underline{l})$.
To explicitly display the expressions we introduce local coordinates on patches 
$\hat U_{2 \epsilon}^{(1)}$ and $\hat U_{2 \epsilon}^{(2)}$, which cover the 
$\P^1$ such that $l_i \neq 0 $ on $\hat U_{2 \epsilon}^{(i)}$.
We set 
\begin{equation} \label{local_coords_uell}
   \hat U_{2 \epsilon}^{(1)}: \quad u_1 = y_1,\quad \ell_1 = \frac{l_2}{l_1} \ ,\qquad \quad  
    \hat U_{2 \epsilon}^{(2)}: \quad u_2 = y_2,\quad \ell_2 = \frac{l_1}{l_2} \ ,
\end{equation}
Using the blow-up relation one finds the following 
coordinate transformation on the overlap
\begin{equation} \label{coord_trans}
  \hat U_{2 \epsilon}^{(1)} \cap \hat U_{2 \epsilon}^{(2)}:\qquad \quad (\ell_2,u_2) = (1/\ell_1 , \ell_1 u_1)\ ,
\end{equation}
which shows that $B_{\bbC^2}$ is identified with $\cO_{\P^1}(-1)$. Let us point out that 
this matches the local description presented in appendix \ref{App:Local} if we interpret $B_{\bbC^2}$ as 
a local model of $(N_{\hat Z_3} \Sigma)_p$ at a point $p$ on $\Sigma$.\footnote{To 
avoid cluttering of indices we introduce the new notation for $z_1^{(1)}\equiv u_1,$ $z_2^{(1)} \equiv \ell_1$ and
$z_2^{(2)} \equiv u_2$, $z_2^{(2)}\equiv \ell_2$.}

One now can introduce a metric $|| \simga ||$ for sections $\sigma$ of 
the line bundle $\cL$ as follows. Since $\cL$ is non-trivial
one cannot simply specify $|| \simga ||$ by using a single global holomorphic 
section $\sigma$. Each such global holomorphic section will have either poles of zeros.
However, we can specify $||\cdot||$ on local holomorphic sections patchwise and 
glue these local expressions together.
Let us define the local expression on $B_{\bbC^2}-E$ by evaluation on a global holomorphic section $\sigma_{(0)}$ with 
zeros along $E$. One defines
\begin{equation} \label{patch1}
  B_{\bbC^2} - E:\qquad ||\sigma_{(0)}||_1 := 1 \ .
\end{equation}
On the patches $\hat U_{2\epsilon}$ covering $E$ one needs to 
use other sections which are non-vanishing also along $E$.  
Using the explicit realization of $\cL$ as in \eqref{expl_cL} 
with sections $\sigma = \lambda$ one can 
specify $||\cdot||_2$ in $\hat U_{2\epsilon}$ 
setting
\begin{equation}\label{patch2}
  \hat U_{2\epsilon}:\qquad ||\sigma||_2:= |\lambda|\, (|l_1|^2 + |l_2|^2)^{1/2} \ .  
\end{equation}
Note that the section $\sigma_{(0)}$ is also defined in $\hat U_{\epsilon}-E$ and can 
be given in 
the representation \eqref{expl_cL} of $\cL$. In the same representation we 
can also define local sections $\sigma_{(i)}$ near $E$, such that 
\begin{equation} \label{loc_sections_near_D}
  \hat U_{2\epsilon}: \quad  \sigma_{(0)} = \frac{y_1}{l_1} = \frac{y_2}{l_2}\ , 
     \qquad  \qquad \hat U_{2\epsilon}^{(i)}: \quad \sigma_{(i)} = \frac{1}{l_i}\ ,   
\end{equation}
Recall that the $y_i$, as also introduced in subsection \ref{sec:geometricblowups}, 
specify the point which is blown up as $y_1 =y_2 =0$. The metric 
\eqref{patch2} for these sections is simply given by 
\begin{equation}  \label{expl_metric_2}
   \hat U_{\epsilon}^{(i)}:\quad || \sigma_{(i)}||_2 = (1 + |\ell_i|^2)^{1/2}\ ,\qquad \quad
    || \sigma_{(0)}||_2 = |u_i|\cdot (1 + |\ell_i|^2)^{1/2}  \ ,\\
\end{equation} 
where we have used the local coordinates \eqref{local_coords_uell}.
To give the global metric one next splits 
$B_{\bbC^2}$ into patches $(B_{\bbC^2} - \hat U_\epsilon, \hat U_{2\epsilon})$,
and introduces a partition of unity $(\rho_1,\rho_2)$. The local expressions 
\eqref{patch1} and \eqref{patch2} are glued together as 
\begin{equation} \label{patched_metric}
   ||\cdot ||:=\rho_1 ||\cdot||_1 + \rho_2 ||\cdot||_2\ .
\end{equation}

Using this metric one can now determine the Chern curvature form $\frac{i}{2}\Theta $ of the 
line bundle $\cL$.
Locally one has to evaluate
\begin{equation} 
  \frac{i}{2}\Theta = - \frac{1}{4\pi}\partial \bar \partial \log || \sigma||^2\ ,
\end{equation} 
for holomorphic sections $\sigma$ which have no poles or zeros in the considered patch. 
Using \eqref{patch1} and \eqref{expl_metric_2} one finds that $\cL$ is trivial on $B_{\bbC^2}-\hat U_{2\epsilon}$, 
but non-trivial in the patches $\hat U_\epsilon^{(i)}$:
\begin{eqnarray} \label{Theta_patches}
  B_{\bbC^2} - \hat U_{2\epsilon}: \quad \Theta &=& 0 \ , \\
 \hat U_\epsilon^{(i)}: \quad \Theta &=& - \frac{i}{2\pi} \partial \bar \partial  \log \big(1+|\ell_i|^2 \big) = - \omega_{FS} \ , \nn
\end{eqnarray}
where $\omega_{FS}$ is the Fubini-Study metric. 
One notes that the form $\Theta$ is strictly negative on $E$, since
it is given by minus the pull-back of the Fubini-Study metric under the restriction map to $\P^1$. 
Hence, one finds that $-\Theta > 0$ on $E$, and $-\Theta \geq 0$ on $\hat U_\epsilon$.
In the region $\hat U_{2\epsilon}-\hat U_\epsilon$ the form $\Theta$ interpolates
in a continuous way. Finally, we can give the K\"ahler form $\tilde J$  on $B_{\bbC^2}$.
Since $-\Theta \geq 0$ on $\hat U_\epsilon$ and $\Theta=0$ on 
$B_{\bbC^2} - \hat U_{2\epsilon}$ continuity implies that $-\Theta$ is bounded from below. 
This fact can be used to define a K\"ahler form on $B_{\bbC^2}$ by setting
\begin{equation} \label{Kahleronblowup}
    \tilde J =\pi^* J - v_{\rm bu} \Theta\ ,
\end{equation}
where $\pi^* J$ is the pull-back of the K\"ahler form on $\bbC^2$.
$\tilde J$ is a closed $(1,1)$-form, and positive for a sufficiently small blow-up volume $v_{\rm bu}$. 
In other words, one finds that the manifold $B_{\bbC^2}$ is naturally endowed with a 
K\"ahler structure. Since the blow-up is a local operation, one uses
this construction to blow up a point in any K\"ahler surface identifying $J$ in 
\eqref{Kahleronblowup} with the K\"ahler form before the blow-up. 

\subsection{Blow-Up Threefolds as K\"ahler Manifolds}
\label{sec:KaehleronBlowupThreefold}

Having discussed the K\"ahler structure on the blow-up of a point in a K\"ahler surface, we 
can now generalize this construction to blowing up curves in $Z_3$. 
This is again textbook knowledge \cite{Voisin2002} 
and we can be brief. 

As in subsection \ref{sec:Kahleronpointblowup} we 
study the line bundles $\cL \equiv \cO_{\hat Z_3}(E)$ and 
$\cL^{-1} \equiv \cO_{B_{\hat Z_3}}(-E)$. One proceeds as in 
the two-dimensional example and first examines the bundle 
near $E$. One shows that restricted to $E$ one obtains  
\begin{equation}
   \cL^{-1}|_{E} \ \cong\ N^*_{\hat Z_3} E\ \cong\ \cO_E(1)\ ,
\end{equation}
where $ N^*_{\hat Z_3} E$ is the co-normal bundle to $E$ in $\hat Z_3$.
The bundle $E=\P^1 (N_{Z_3} \Sigma) \rightarrow \Sigma$ 
admits a natural closed $(1,1)$-form which is positive on the fibers. 
This form can be obtained from a hermitian metric induced from 
$N_{Z_3} \Sigma$ and is obtained from the Chern curvature of $\cO_E(1)$.
As in the previous section this curvature can be extended to $\cL^{-1}$ by 
a partition of unity as in \eqref{patched_metric}.
Let us denote the Chern curvature of $\cL$ again by $\frac{i}{2}\Theta$.
The K\"ahler form on $\hat Z_3$ is then given by 
$\tilde J= \pi^* J   - v_{\rm bu} \Theta $, for sufficiently 
small blow-up volume $v_{\rm bu}$.
While the construction of $\Theta$ depends on the 
explicit metric $||\cdot||$ on $\cL$ one can also 
evaluate the corresponding cohomology classes which are topological in nature. Clearly, one has
\begin{equation}
  c_1(\cL) = \tfrac{1}{\pi} [\Theta]\ , \qquad \quad c_1 (\cL)|_E = c_1(N_{\hat Z_3} E) = [E]_{E}\ ,
\end{equation}
where we have also displayed the restriction to the exceptional divisor.

The above discussion identifies $\hat Z_3$ as a K\"ahler manifold.
If one directly uses this K\"ahler structure, however, 
one finds that $\hat Z_3$ cannot arise as an actual 
supersymmetric flux background. Recall that, for example, in heterotic compactifications 
with background fluxes $H_3$ the internal manifold has to be non-K\"ahler to satisfy \eqref{dJ=H} as
shown in \cite{Strominger:1986uh}.
In the following we will show that there actually exists a natural 
$SU(3)$-structure on $\hat Z_3$ which renders it to be non-K\"ahler and allows to 
identify a supersymmetric flux vacuum on $\hat Z_3$.

\section{Defining the SU(3)-Structure: the Non-K\"ahler Twist}
\label{sec:non-Kaehlertwist}

In the following we propose an $SU(3)$-structure on the open manifold $\hat Z_3 - E$ and the 
blow-up space $\hat Z_3$ in sections \ref{sec:SU3_on_open} and \ref{sec:SU(3)onblowup}, respectively. 
Before turning to the three-dimensional case we first discuss 
the complex two-dimensional case $B_{\bbC^2}$ in section \ref{sec:Hopf}. 

\subsection{The Non-K\"ahler Twist: Warm-Up in two complex Dimensions}
\label{sec:Hopf}

To warm up for the more general discussion, let us first introduce a 
non-K\"ahler structure on the simpler
two-dimensional example $B_{\bbC^2}$. Recall from \eqref{coord_trans} that $B_{\bbC^2}$
is the total space of the universal line bundle $\cO_{\P^1}(-1)$ over $\P^1$.
There are two geometries related to $B_{\bbC^2}$ which admit a non-trivial 
non-K\"ahler structure. Firstly, to render 
$B_{\bbC^2}$ into a compact space $\hat B_{\bbC^2}$ one can replace each fiber $\bbC$ 
of the line bundle $B_{\bbC^2}$ by a two-torus $\bbC^*/\bbZ$.\footnote{More precisely,
parameterizing a fiber by $\lambda = r e^{i\theta}$ one removes the origin $r=0$ 
and identifies $r \cong r +1$.} Secondly, one can consider the open manifold $B_{\bbC^2} - E$,
where one simply removes the origin and the attached exceptional blow-up divisor $E = \P^1$. 
On the geometries $\hat B_{\bbC^2}$ and $B_{\bbC^2}$ 
on can introduce coordinates $(\ell_i, u_i)$ as in \eqref{local_coords_uell}. 
The new geometries have been modified from $B_{\bbC^2}$ such that, in particular, one 
has $u_i \neq 0 $. For the compact space $\hat B_{\bbC^2}$ one 
further has a periodicity since $u_i \in \bbC^*/\bbZ$.
Note that one inherits from the blow-up the coordinate transformation \eqref{coord_trans}
on the overlaps $\hat U^{(1)} \cap \hat U^{(2)}$ covering the north and 
south pole of the $\P^1$.

One observes that with the transformations \eqref{coord_trans} the surface $\hat B_{\bbC^2}$ can 
be identified as the \textit{Hopf surface} $S^1 \times S^3$. This surface does not admit a K\"ahler 
structure since $h^{(1,1)}=0$, while $h^{(0,0)}=h^{(1,0)}=h^{(2,1)}=h^{(2,2)}=1$. 
Note however, that $\hat B_{\bbC^2}$ admits a natural globally defined no-where vanishing 
$(1,1)$-form $\hat J$. The construction of $\hat J$ was given in ref.~\cite{Gualtieri:2010fd}. 
In fact, one can also introduce 
a three-form flux $\hat H_3$ such that $i(\bar \partial - \partial) \hat J = \hat H_3$, just as in 
the supersymmetry conditions \eqref{dJ=H}. 

Keeping this connection with the Hopf surface in mind, we aim 
to introduce $\hat J$ and $\hat H_3$ on $B_{\bbC^2} - E$. 
Firstly, we introduce on $U^{(i)}_{\epsilon}$ the B-field 
\begin{eqnarray} \label{hatBi}
  U^{(i)}_{\epsilon}-E:\quad  \hat B_i &=& \tfrac{1}{4 \pi} \big(\partial \log (1+ |\ell_i|^2) \wedge \partial \log u_i + \bar \partial \log (1+ |\ell_i|^2) \wedge \bar \partial \log \bar u_i  \big)\ ,\\
   &=& \tfrac{1}{2 \pi} \text{Re}\big(\partial \log || \sigma_{(i)}||^2 \wedge \partial \log ||\sigma_{(0)}||^2 \big) \ . \nn 
\end{eqnarray}
Here we have inserted the definitions \eqref{loc_sections_near_D} of the sections $\sigma_{(0)},\sigma_{(i)}$ and 
the norm $||\cdot||$ given in \eqref{expl_metric_2}.
One realizes that these $\hat B_i$ are of type $(2,0)+(0,2)$. They extend continuously to
$B_{\bbC^2} - E$ and one checks that $\hat B_i=0$ on $B_{\bbC^2} - \hat U_{2 \epsilon}$, since 
$|| \sigma_{(0)}|| =1 $ outside $\hat U_{2 \epsilon}$. 
However, note that the $ \hat B_i$ do not patch together on the overlap 
$\hat U^{(1)}_\epsilon \cap \hat U^{(2)}_\epsilon$ as a form, but rather satisfy
\begin{equation}
  \hat U^{(1)}_\epsilon \cap \hat U^{(2)}_\epsilon: \qquad  \hat B_2- \hat B_1 = F_{21}\ ,  
\end{equation}
where one identifies 
\begin{equation} 
   F_{21} = - \tfrac{1}{2\pi} \text{Re} \big( d \log \ell_1 \wedge d \log u_1\big)\ ,
\end{equation}
The $(2,0)+(0,2)$ form $F_{21}$ on $\hat  U^{(1)}_\epsilon \cap \hat U^{(2)}_\epsilon$ can be used to 
define a Hermitian line bundle on this overlap. In mathematical terms $F_{01}$ and $\hat B_i$ define a 
gerbe with curvature 
\begin{equation} \label{def-H3_surf}
  \hat H_3 = d \hat B_i = - \tfrac{i}{2} \omega_{FS} \wedge (\bar \partial - \partial) \log |u_i|^2\ , 
\end{equation}
where we have introduced the Fubini-Study metric $\omega_{FS} = \tfrac{i}{2\pi} \partial \bar \partial   \log || \sigma_{(i)}||^2$ on $\P^1$.
One can now check that indeed on the Hopf surface $\hat B_{\bbC^2} \cong S^1 \times S^3$ one 
has a non-vanishing integral of $\hat H_3$ over the $S^3$. Similarly, 
one can evaluate the integral on the open manifold $B_{\bbC^2} - E$, where the integral of 
$\hat H_3$ is performed on $\P^1$ and the $S^1$ encircling the zero section which has been removed.

It is now straightforward to read off the non-K\"ahler $(1,1)$-form $\hat J$, satisfying the 
supersymmetry condition $i(\bar \partial - \partial) \hat J = \hat H_3$.
There are two immediate choices. One could set
\begin{equation} \label{def-J_ex1}
 \hat U_\epsilon^{(1)}-E: \quad   
   \hat J = \tfrac{1}{2 \pi} \text{Im} \big(\partial \log ||\sigma_{(1)}||^2 \wedge \bar \partial \log \bar u_1 \big) \ ,
\end{equation}
which is the choice used on the Hopf surface in \cite{Gualtieri:2010fd}.
To evaluate $\hat J$ in the second patch $U_1$ one uses the coordinate transformation \eqref{coord_trans} and 
transforms $\hat J$ like a standard differential form. However, the second choice 
\begin{equation} \label{def-J_ex2}
   \hat U_\epsilon^{(1)}-E: \quad   
   \hat J =  - \tfrac{1}{2} \log |u_1|^2 \, \omega_{FS}\ ,  
\end{equation}
is more appropriate to the blow-up case. The reason is that 
\eqref{def-J_ex2} can be extended to $B_{\bbC^2}-E$ by replacing 
$\omega_{FS} \rightarrow - \Theta$, where $\Theta$ is proportional to 
the Chern curvature of the line bundle $\cL = \cO(E)$ introduced in subsection \ref{sec:Kahleronpointblowup}. 
As noted in \eqref{Theta_patches} the Chern curvature of $\cL$ is localized near 
$E$ and vanishes outside $\hat U_{2 \epsilon}$. It will be the 
expression \eqref{def-J_ex2} which we will extend to the three-dimensional 
case in the next subsection.

Note that for the expressions in \eqref{hatBi}-\eqref{def-J_ex2}
to define well-behaved differential forms, one has to exclude the origin $u_i =0$ in $\hat U^{(i)}_\epsilon$.
This is precisely the reason why we considered the restriction of the blow-up space $B_{\bbC^2}$
to the open manifold $B_{\bbC^2}-E$. The extension of this structure 
to $B_{\bbC^2}$ will be discussed in subsection \ref{sec:SU(3)onblowup} below. However, 
one can extend the above constructions to $\bbC^2$ and $B_{\bbC^2}$ 
by including currents as in sections \ref{sec:N=1branes} and \ref{sec:potentialhatZ3minusD}.
Note that on $\bbC^2$ the delta-current $\delta_{\{0\}}=\bar \partial T_\beta$ 
localizing on the origin is defined using the Cauchy kernel 
\begin{equation} \label{Cauchykernel}
   \beta = -\tfrac{1}{4\pi}\partial \log (|x_1|^2 + |x_2|^2) \wedge \partial \bar \partial   \log (|x_1|^2 + |x_2|^2) \ ,
\end{equation}
where $x_i$ are the coordinates on $\bbC^2$. Here one has to use the same formalism as in section 
\ref{sec:N=1branes}. The Cauchy kernel $\beta$ can be lifted to the blow-up space $B_{\bbC^2}$ using 
the blow-down map $\pi: B_{\bbC^2} \rightarrow \bbC^2$
as 
\begin{equation}
  \pi^* \beta =  i \partial \log u_i \wedge \omega_{FS} \ ,
\end{equation}
where $u_i$ is identified with the coordinate on 
the fiber of $B_{\bbC^2}$ viewed as a line bundle over $\P^1$, and 
$\omega_{FS}$ is 
the Fubini-Study metric on $\P^1$ defined below \eqref{def-H3_surf}. Hence, we see that 
$\hat H_3$ as defined 
in \eqref{def-H3_surf} is not a form on the full space $B_{\bbC^2}$.
Rather the $(2,1)$-part of $\hat H_3$ is seen to be the pull-back of the Cauchy 
kernel \eqref{Cauchykernel}, $\hat H_3^{(2,1)} = - i \partial \hat J = \pi^* \beta$.
One readily evaluates 
\begin{equation}
  d \hat H_3 = \pi^* d \beta = \pi^* \delta_{\{ 0\}} = \omega_{FS} \wedge \delta_{E}\ ,
\end{equation}
where $\delta_E$ is the delta-current localizing on the exceptional divisor $E$ as
in section \ref{sec:potentialhatZ3minusD}.
Therefore, we can reduce the integrals involving the so-defined $\hat H_3$ to chain integrals over 
a one-chain ending on $\{ 0 \}$ in $\bbC^2$ since 
\begin{equation} \label{chain_rewrite}
   \int_{\Gamma} \gamma = \int_{\bbC^2} \gamma \wedge \bar \beta = \int_{B_{\bbC^2}} \pi^* \gamma \wedge \hat H_3^{(1,2)} 
   = \tfrac12 \int_{B_{\bbC^2}} \pi^* \gamma \wedge (\hat H_3 + i d \hat J) \ , 
\end{equation}
for a compactly supported $(1,0)$-form $\gamma$.
Note that this is the analog of the computation performed in section \ref{sec:potentialhatZ3minusD}, 
where we have shown how the five-brane superpotential is translated to a superpotential on the 
blow-up space $\hat Z_3$. We are now in the position to introduce an $SU(3)$-structure on the open 
manifold $\hat Z_3 - E$.

\subsection{The SU(3)-Structure on the Open Manifold $\hat Z_3 - E$}
\label{sec:SU3_on_open}

We are now in the position to discuss the $SU(3)$-structure on the open manifold $\hat Z_3 - E$
by introducing forms $\hat \Omega, \hat J$ of type $(3,0)$ and $(1,1)$, respectively.
We show that these forms satisfy 
\begin{equation} \label{dhatJ}
   d \hat \Omega = 0 \ ,\qquad d \hat J = \cW_4 \wedge \hat J + \cW_3 \ ,
\end{equation}
for non-trivial $\cW_4$ and $\cW_3$.
Note that $\hat \Omega = \pi^* \Omega$ has been already discussed in detail in section \ref{sec:hatomega}.
It was noted that $\hat \Omega$ has a zero along $E$, but is a well-defined, nowhere vanishing form on the 
open manifold $\hat Z_3 - E$. The basic idea to introduce 
$\hat J$ is to extend the definition \eqref{def-J_ex2} to $\hat Z_3 - E$ by 
using the sections and metric on $\hat Z_3$ defined in section \ref{sec:blow-up_as_Kaehler}. It is further 
illuminating to view the construction of subsection \ref{sec:Hopf} as a local model for $N_p \Sigma$. 

Let us recall that on patches $\hat U_{2 \epsilon}$ one can introduce 
the holomorphic sections $\sigma_{(0)}$ and $\sigma_{(1)},\sigma_{(2)}$ of $\cL= \cO_{\hat Z_3}(E)$ 
as in \eqref{loc_sections_near_D}. 
One recalls that $\sigma_{(0)}$ is a global section which has zeros along 
$E$ just as the form $\hat \Omega$. We therefore work on $\hat Z_3 - E$. Moreover, 
we first consider a situation where a supersymmetric five-brane curve 
has been blown up, and hence the flux $\hat H_3$ on $\hat Z_3$ is supersymmetric and 
satisfies $i (\bar \partial - \partial) \hat J = \hat H_3$.
Following the steps of subsection \ref{sec:Hopf} we first introduce the 
B-fields
\begin{equation}
  \hat U^{(i)}_{2\epsilon} - E: \qquad  \hat B_i =  \tfrac{1}{2\pi} \text{Re} \big( \partial  \log || \sigma_{(i)}||^2 \wedge \partial \log || \sigma_{(0)}||^2 \big) \ ,
\end{equation}
where $\hat U^{(i)}_{2\epsilon}$
are the patches which cover $\Sigma$  and the blow-up $\P^1$'s such that $\sigma_{(i)}$ is well-defined. 
Note that this implies that we have to refine the cover $\hat U^{(i,\alpha)}_{2\epsilon}$ as in section 
\ref{sec:geometricblowups}. For simplicity we will drop these additional indices in the following.
Evaluating $\hat H_3 = d \hat B_i$ this implies 
\begin{eqnarray}
  \hat U^{(i)}_{2\epsilon} - E:\qquad   
   \hat H_3 & = & - \tfrac{i}{2} \Theta \wedge ( \bar \partial - \partial)  \log \big(|| \sigma_{(0)}||^2 / || \sigma_{(i)}||^2 \big) \qquad  \\
    &=&  - \tfrac{i}{2} \Theta \wedge ( \bar \partial - \partial)  \log | z^{(i)}_i|^2 \ , \nn
\end{eqnarray}
where $\frac{i}{2}\Theta$ is the Chern curvature of $\cL$ as introduced 
in section \ref{sec:blow-up_as_Kaehler}, and  we inserted the sections as given in \eqref{loc_sections_near_D}
using local coordinates $z^{(i)}_i = y_i$ in $\hat U^{(i)}_{2\epsilon}$. Note that $E$ is 
given in $\hat U^{(i)}_{2\epsilon}$ as $z^{(i)}_i=0$ and has been excluded from $\hat U^{(i)}_{2\epsilon}$.
One can further evaluate $\hat B_2 - \hat B_1 = F_{21}$ on the overlap $ \hat U^{(i)}_{\epsilon}\cap \hat U^{(i)}_{\epsilon}$
and show in local coordinates that $dF_{21}=0$.

With this preparation one next defines $\hat J$ using the 
above supersymmetric flux.
One finds the natural extension of \eqref{def-J_ex2} given by 
\begin{equation} \label{def-hatJ}
  \hat U^{(1)}_{2\epsilon} - E :\qquad \hat J = \pi^* J - (1- \log |y_1|^2 ) \Theta \ ,   
\end{equation}
where $J$ is the K\"ahler form on $Z_3$ and $\pi$ is the blow-down map.
In order to transform this $\hat J$ into the other patches  $\hat U^{(2)}_{2\epsilon}$ covering $\P^1$ 
one transforms $\hat J$ as a differential form. Note that $\Theta$ vanishes outside 
$\cup \hat  U_{2 \epsilon}$ and one finds
\begin{equation} \label{oldJ}
  \hat Z_3 -\cup \hat  U_{2 \epsilon}:\qquad \hat J = \pi^* J\ .
\end{equation}
In other words, the departure from the original K\"ahler structure only 
arises in a small neighborhood around $E$. 
Using the explicit form of $\hat J$ one checks 
that $d \hat J$ has non-trivial torsion classes $\cW_3,\cW_4$ in \eqref{dhatJ}.
In conclusion one finds that near $E$ we introduced a non-K\"ahler geometry on the open manifold. 

Let us close this section by noting that the constructed structure $(\hat J,\hat \Omega)$ 
and $\hat H_3$ are only well-defined differential forms on the open manifold $\hat Z_3 - E$.
Furthermore, one finds on $\hat Z_3 -E$ that $d \hat H_3= 2i  \partial \bar \partial \hat J =0$, 
which implies that there is no source term for the blown-up five-brane. To include such a source 
one has to work on the whole manifold $\hat Z_3$ using currents as in section \ref{sec:potentialhatZ3minusD}. 
One then finds 
\begin{equation}
  d \hat H_3 =2 i \partial \bar \partial \hat J = \delta_E \wedge \Theta\ ,   
\end{equation}
in accord with \eqref{eq:dhatH_3}. Analog to \eqref{chain_rewrite} one also rewrites the 
chain integral on $Z_3$ to $\hat Z_3$ as 
\begin{equation} 
   \int_{\Gamma} \chi = \int_{\hat Z_3} \pi^* \chi \wedge (\hat H_3 + i d\hat J)  \ ,
\end{equation}
where $\Gamma$ is a three-chain ending on $\Sigma$. 
Here we have used $\hat H_3 = i (\bar \partial - \partial) \hat J $ for a supersymmetric 
configuration. Clearly, upon varying the complex structure of $\hat Z_3$ this relation 
will no longer hold, since $(\hat \Omega,\partial\hat J)$ are still forms of type $(3,0)$ and $(2,1)$
in the new complex structure, respectively. The flux $\hat H_3$, however, is fixed as a form but
changes its type under a complex structure variation. Thus in the supersymmetric configuration 
the complex structure on $\hat{Z}_3$ is adjusted so that $\hat{H}_3$ is of type $(2,1)+(1,2)$. 

Note that the described structure is not yet satisfying, since $(\hat \Omega,\hat J,\hat H_3)$
are no well-defined differential forms on all of $\hat Z_3$.
In the next subsection we resolve this issue by proposing a redefinition 
which allows us to work with differential forms on the full space $\hat Z_3$.

\subsection{The SU(3)-Structure on the Manifold $\hat Z_3$}
\label{sec:SU(3)onblowup}

In this subsection we will finally completely resolve the 
five-brane into a non-K\"ahler geometry. Recall that 
so far we had to work on the open manifold if we wanted to 
use forms, while on $\hat Z_3$ we had to use currents due 
to the singularities of $\hat H_3$ and $\hat J$ along the 
exceptional divisor $E$.
In the following we will introduce an $SU(3)$-structure on $\hat Z_3$
by specifying a new $(1,1)$-form $J'$, a non-singular three-form flux 
$H_3'$ and a non-where vanishing $(3,0)$-form $\Omega'$.\footnote{Recent 
constructions of non-K\"ahler geometries can be found in ref.~\cite{Carlevaro:2009jx,Larfors:2010wb,Chen:2010bn,Chen:2010ssa}.
The constructions of \cite{Carlevaro:2009jx,Larfors:2010wb,Chen:2010bn,Chen:2010ssa} share sensible similarities with our approach. 
We hope to come back to this issue in future works.}

To begin with we note that there exists no holomorphic no-where vanishing 
$(3,0)$ form on $\hat Z$ since $K\hat Z_3$ is non-trivial. 
Hence, $\Omega'$ has to be non-holomorphic and 
we will find non-vanishing torsion classes $\cW_3,\cW_4,\cW_5$ such that  
\begin{equation}
  d \Omega' = \bar \cW_5 \wedge \Omega'\ , \qquad d J' = \cW_4 \wedge J' + \cW_3\ . 
\end{equation} 
The basic idea to define $\Omega'$ and $J'$ is rather simple. We first note 
that $\hat H_3$ and $d\hat J$ have a pole of order one along $E$, while $\hat \Omega$ has a first 
order zero along $E$. Then we want to cancel the zero against the 
pole and introduce $(J',\Omega')$ such that
\begin{equation} \label{match_super}
   \int_{\hat Z_3 - E} (\hat H_3 + i d \hat J) \wedge \hat \Omega = \int_{\hat Z_3} (H'_3 +i d J') \wedge \Omega' \ , 
\end{equation}
which is essential in the matching of the superpotentials.

In order to implement this cancellation we note that $d\hat J$ is proportional to $d|y_1|/|y_1|$
in $\hat U^{(1)}_\epsilon$. We thus define 
\begin{equation}  
 \hat U^{(1)}_\epsilon:\quad \Omega' = \frac{1}{|y_1|} \hat \Omega\ ,
\end{equation}
and transform $\Omega'$ to the other patches like a differential form.
Note that $\Omega'$ no longer admits a zero along $E$, but also is no longer 
holomorphic. Evaluating $d\Omega'$ one finds a non-vanishing real $\cW_5 = - d \log |y_1|$ in 
$\hat U^{(1)}_\epsilon$. Let us stress, however, that by construction of $\hat{Z}_3$,
\begin{equation} 
  \hat Z_3 - \cup \hat U_{2 \epsilon}:\qquad \Omega' = \pi^* \Omega, \quad \cW_5 = 0 \ . 
\end{equation} 
This implies that we need to 
`glue in' the non-holomorphic dependence of $\Omega'$ using a partition 
of unity for $(\hat Z_3 -\cup \hat U_\epsilon,\hat U_{2 \epsilon})$.

Let us comment on a more global approach to define $\Omega'$ on 
the blow-up $\hat Z_3$ realized as in \eqref{eq:blowup}.
Instead of starting with $\hat \Omega$ we 
begin with a globally defined and no-where vanishing $\tilde \Omega$.
However, this $\tilde \Omega$ is not a differential from, but rather  
a section of $\cL^{-1} \otimes K\hat Z_3$ given by the residue expression 
\begin{equation}
   \tilde \Omega = \int_{\epsilon_1}\int_{\epsilon_2} \frac{\Delta}{P Q}\ .
\end{equation} 
Comparing this expression with $\hat \Omega$ given in \eqref{eq:ResZhat} 
one immediately sees that $\tilde \Omega$ does not vanish along $E$. 
This $\tilde \Omega$ captures the complex structure dependence of the 
$\hat Z_3$. It also satisfies the same Picard-Fuchs equations as 
$\hat \Omega$, and hence can be given in terms of the same periods.
By the blow-up construction it depends on the complex structure 
deformations of $Z_3$ and the five-brane deformations.
However, $\tilde \Omega$ transforms under $\cL^{-1}$ since the scaling-weight 
of $\Delta$ is not entirely canceled.  In other words, if one insists on 
holomorphicity in the coordinates of $\hat Z_3$ one can either work with $\hat \Omega$
which has zeros along $E$, or with $\tilde \Omega$ which transforms also
under $\cL^{-1}$ and thus is not a three-form. A natural global definition
of the non-holomorphic $\Omega'$ is then given by the residue integral
\begin{equation} \label{globalOmega'}
  \Omega' = \int_{\epsilon_1}\int_{\epsilon_2} \frac{\Delta}{P Q} \frac{h_i}{l_i} \frac{|l_i|}{|h_i|}\ .
\end{equation}
Note that this expression is invariant under phase transformations of the 
global projective coordinates $(l_1,l_2)$ and $x_k$ entering the constraints $h_i$ as in \eqref{eq:blowup}.
In contrast $\Omega'$ would scale under real scalings of the global coordinates. However, 
these are fixed by the conditions defining the K\"ahler volumes. For example, the blow-up volume $v_{\rm bu}$
fixes the real scalings of the $(l_1,l_2)$ 
\begin{equation}
   |l_1|^2 + |l_2|^2 = v_{\rm bu}\ .
\end{equation}
Similarly, the real scalings of the $h_i(x)$ are fixed by the definitions
of the K\"ahler moduli of $Z_3$. It would be very interesting to check 
if the conjecture \eqref{globalOmega'} for a global $\Omega'$ 
can be used to derive Picard-Fuchs equations for $\hat Z_3$, probably including 
anti-holomorphic derivatives.

Finally, let us turn to the definition of the non-K\"ahler form 
$J'$. In a patch $\hat U^{(1)}_{\epsilon}$ it is natural to identify  
\begin{equation}
  \hat U^{(1)}_{\epsilon}:\quad J' = \pi^* J  - v_{\rm bu} (1-|y_1|) \Theta\ ,
\end{equation}
which agrees with $\hat J$ given in \eqref{def-hatJ} up to the logarithmic singularity 
along $E$. The non-trivial torsion classes from $dJ'$ are again $\cW_3,\cW_4$.
Since we expect the non-K\"ahlerness to be localized near $E$ one has to have 
\begin{equation}
  \hat Z_3 - \cup \hat U_{2 \epsilon}:\qquad J' = \pi^* J\ .
\end{equation}
This condition is indeed satisfied for any ansatz for $J'$ which involves the curvature 
$\Theta$ of $\cL$, since $\Theta$ vanishes outside a patch covering $E$. 
For a blow-up of a curve wrapped by a holomorphic five-brane one infers 
the flux $H_3'$ using $H'_3 = i (\bar \partial - \partial) J'$. $H'_3$ is 
a differential form on all of $\hat Z_3$, and appears in the superpotential 
\begin{equation} \label{SU(3)super}
   W = \int_{\hat Z_3} (H_3' + idJ') \wedge \Omega'\ ,
\end{equation}
which is now valid also for complex structure variation yielding a setup departing from a 
supersymmetric configuration. 
Clearly, by construction one can use \eqref{match_super} to equate the superpotential on the 
open manifold $\hat Z_3 - E$ with the expression \eqref{SU(3)super}.

Let us close this section by noting that the above construction should be considered 
as a first step in finding a fully back-reacted solution of the theory which 
dissolves the five-brane into flux. It will be interesting to extend these considerations 
to include the remaining supersymmetry conditions of \cite{Strominger:1986uh} which are 
not encoded by a superpotential. In particular, this requires a careful treatment of the 
dilaton $\phi$ and the warp factor already in $Z_3$, 
where $e^{\phi}$ becomes infinite near the five-brane.

\part{Conclusion and Appendix}

\chapter{Conclusions and Outlook}
\label{ch:conclusion}

In this work we have studied non-perturbative four-dimensional $\mathcal{N}=1$ 
string compactifications with spacetime-filling branes. As a starting point of 
this analysis we have derived the effective action of a D5-brane in a generic 
Type IIB Calabi-Yau orientifold compactification.
Then we have continued with a more involved study of non-perturbative string 
compactifications, namely heterotic string or heterotic M-theory Calabi-Yau 
threefold compactifications with five-branes and F-theory Calabi-Yau fourfold 
compactifications. We used techniques from mirror symmetry and heterotic/F-theory
duality to calculate the exact flux and brane superpotentials for these compactifications. 
Furthermore we have presented a novel geometric description of five-brane dynamics 
by a dual blow-up threefold $\hat{Z}_3$. We provided, by constructing an $SU(3)$-structure 
on $\hat{Z}_3$, first evidence for an interpretation of $\hat{Z}_3$ as a valid flux 
compactification of the Type IIB and heterotic string with flux due to the dissolved five-brane. 
We have verified this duality on the level of the effective flux and five-brane 
superpotentials, that we calculated from open-closed Picard-Fuchs equations on $\hat{Z}_3$.

In chapter \ref{ch:EffActCYOrieCompact} we started with a systematic 
review of Calabi-Yau orientifold compactifications of Type II string 
theory. We readily focused on Type IIB orientifolds, that are $O3/O7$- 
and $O5/O9$-orientifolds, for which we reviewed the full $\mathcal{N}=1$ 
effective action. This served as a preparation for our actual derivation 
of the full effective action of a spacetime-filling D5-brane wrapping 
an internal curve $\Sigma$ in chapter \ref{ch:EffActD5}, which contains 
the bulk $O5/O9$-orientifold action in the case that all brane degrees 
of freedom are frozen out. After a brief discussion of D-branes in Calabi-Yau 
manifolds in section \ref{sec:DbranesinCY3Orie}, we performed in section \ref{sec:D5branes}
the Kaluza-Klein reduction of the six-dimensional Dirac-Born-Infeld and Chern-Simons 
actions of the D5-brane coupled to the ten-dimensional bulk supergravity action in the 
most generic situations allowing background and brane fluxes. Then we read 
off the entire $\mathcal{N}=1$ characteristic data in section \ref{sec:N=1dataD5}. 
Most notably we derived 
the explicit corrections of the D5-brane fields to the bulk $O5/O9$-fields to form new 
$\mathcal{N}=1$ chiral coordinates, the D5-brane superpotential, 
a gauging of a bulk chiral multiplet by the D5-brane gauge field inducing 
a new D-term and the kinetic mixing of the bulk and D5-brane gauge fields, cf.~appendix 
\ref{app:PartEffActions} for the combined bulk-brane gauge kinetic function. 
Of particular conceptual interest was the 
derivation of the $\mathcal{N}=1$ scalar potential which was shown to consist 
of both F- and D-term contributions. One D-term needed to be canceled 
by the tension of the $O5$-planes in order for the setup to be stable 
which reflects, on the level of the effective action, tadpole 
cancellation. The two further terms were induced by gaugings of chiral 
fields, one of which encodes the BPS-calibration condition requiring 
equality of the NS--NS B-field with the D5-brane gauge flux and the other 
one forbidding non-trivial NS--NS three-form flux. In order to calculate the complete 
F-term potential in our purely bosonic reduction, we had to consider the couplings of four-dimensional 
non-dynamical three-forms. After performing 
a formal dualization of these fields into constants\footnote{In fact, the correct interpretation 
of these constants being the flux quantum numbers of $F_3$ was given in \cite{Beasley:2002db} as labeling 
quantum mechanical states of the system.} we were able to derive 
the complete scalar potential in the presence of the D5-brane and of background 
R--R three-form flux $F_3$.  
Furthermore we extended in section \ref{sec:extensionInfinite} the considerations 
of \cite{Grimm:2008dq} to the full geometric deformation modes of the D5-brane, 
including an infinite set of massive modes. This led to a more complete study of 
the geometric $\mathcal{N}=1$ open-closed deformation modes and to the derivation 
of some special relations between its light modes. In addition we showed that the 
massive modes are generically obstructed by an at least quadratic potential.

In the second part of this work on string dualities we have studied and calculated the holomorphic 
flux superpotential in F-theory compactifications on Calabi-Yau fourfolds using techniques from 
mirror symmetry and string dualities. This was used to derive both the flux and seven-brane 
superpotential of the associated IIB theory as well as the superpotential of a dual heterotic 
compactification. More concretely we started with a systematic discussion of heterotic/F-theory 
duality in chapter \ref{ch:HetFThyFiveBranes}. In our presentation of heterotic string and heterotic 
M-theory compactifications in section \ref{sec:HetString+Fivebranes} we readily included 
non-perturbative five-branes on curves $\Sigma$. We put special emphasis on the small instanton 
transition of a smooth heterotic gauge bundles $E$ to a singular, maximally localized limit, which 
is appropriately described by a five-brane. This led in particular to the deduction of the 
heterotic five-brane superpotential from the holomorphic Chern-Simons functional, which localizes in
the singular small instanton limit of $E$ to a chain-integral over the Calabi-Yau form $\Omega$.
Then, after the introduction of the basic concepts and constructions of F-theory in section 
\ref{sec:FTheoryCompactifications}, we discussed in detail in section \ref{sec:HetFDuality}, that in 
the duality between the heterotic string on an elliptic Calabi-Yau threefold $Z_3$ and F-theory on a 
$K3$-fibered elliptic fourfold $X_4$ horizontal heterotic five-branes are geometrized in 
F-theory via a blow-up along $\Sigma$. As argued further this implied a map of heterotic five-brane 
moduli and superpotentials to complex structure moduli respectively the flux superpotential in F-theory. 
Counting F-theory complex structure moduli and computing the F-theory flux superpotential in comparison 
with all heterotic moduli and superpotentials, being a sum of the flux, Chern-Simons and five-brane superpotential,  
thus served, on the one hand, as an ideal testing ground for heterotic/F-theory duality. On the other hand, 
the controllable geometric setup of the F-theory fourfold allowed to compute the 
heterotic superpotentials explicitly, as demonstrated in chapter \ref{ch:Calcs+Constructions} for 
the flux and five-brane superpotential. 

The essential geometrical tools for the described analysis were toric
geometry, briefly reviewed in section \ref{sec:mirror_toric_branes}, 
and mirror symmetry in combination with enumerative geometry as presented in chapter 
\ref{ch:MirrorSymm+FiveBranes}. Mirror symmetry on fourfolds (section \ref{sec:FFMirrors}) 
is in general less studied and understood as in the Calabi-Yau threefold case (section 
\ref{sec:CSModuliSpace+PFO}), which goes mainly back to the less restrictive $\mathcal{N}=1$ 
structure of the fourfold complex structure moduli space, like e.g.~the absence of $\mathcal{N}=2$ 
special geometry, and an incomplete understanding of its global structure and monodromies. Essential 
progress in this direction was presented in section \ref{sec:FFMirrors} where the analytic continuation 
to the universal conifold was performed for the example of the sextic fourfold. A novel 
$\mathds{Z}_2$-monodromy around the conifold divisor was observed that in turn allowed, by the 
requirement of integral monodromy, to determine the classical terms in the periods at the 
large complex structure/large volume point. This was essential for fixing an integral basis
that is for example needed to specify a quantized four-flux which in turn induces 
the effective flux superpotential. Furthermore, we made use of the unified description of open
and closed moduli in F-theory to explicitly compute in chapter \ref{ch:Calcs+Constructions}, 
section \ref{sec:Superpots+MirrorSymmetry}, the flux and seven-brane superpotentials of 
the underlying Type IIB compactification from the F-theory flux superpotential for a selection of Calabi-Yau 
fourfold geometries, cf.~also appendix \ref{app:PartDetailsExamplesTables}. For this matching, the enumerative meaning
both of the flux superpotentials on Calabi-Yau three- and fourfolds and of the brane superpotential for the
A-model was essential, as discussed in section \ref{sec:EnumGeo}. In particular we calculated both closed 
and open Gromov-Witten invariants of compact Calabi-Yau threefolds and compact branes that extend the 
well established results from local
branes consistently. In addition we used in section \ref{sec:SuperpotsHetF} physically independent 
arguments from heterotic/F-theory duality to explain the fact that the periods of the threefold $Z_3$ as well 
as the five-brane superpotential are contained in the fourfold periods on $X_4$.

In the third part of this work we studied directly the dynamics  
of five-brane wrapped on curves $\Sigma$ in a compact Calabi-Yau threefold $Z_3$. 
Here we focused on NS5-branes and D5-branes in $\cN=1$ heterotic and 
orientifold compactifications. We analyzed the geometric deformations of 
the five-brane curve $\Sigma$ in $Z_3$ by replacing the geometry 
by a new threefold $\hat Z_3$ which is no longer Calabi-Yau and
canonically obtained by blowing up along $\Sigma$ in $Z_3$. It was shown that
this replacement naturally unifies the open-closed deformations of $(\Sigma,Z_3)$
as pure complex structure deformations of $\hat{Z}_3$. We used this
description explicitly to derive open-closed Picard-Fuchs differential
equations that are precisely solved by the effective flux and five-brane 
superpotential of the original compactification. In particular we showed that
the blow-up threefold $\hat{Z}_3$ is in agreement with the familiar blow-up 
in F-theory, which is the dual of a heterotic five-brane in 
heterotic/F-theory duality. We concluded by constructing
an $SU(3)$-structure on $\hat{Z}_3$, which provides first evidence to use $\hat{Z}_3$ 
as a consistent flux background, that naturally takes into account the five-brane backreaction.

In more detail we started in chapter \ref{ch:blowup} with a systematic and 
formal analysis of five-brane dynamics. We emphasized in section \ref{sec:N=1gensection} 
that a rigorous treatment of the five-brane backreaction in the Bianchi identity for $C_2$ respectively $B_2$
naturally requires to work on the open manifold $Z_3-\Sigma$ and also that the 
five-brane superpotential has to be treated more appropriately in the language 
of currents. Equipped with this mathematical observations we were naturally
led to the replacement of the original setup by the blow-up 
threefold $\hat{Z}_3$ in chapter \ref{sec:5braneblowupsanddefs}. There we also presented 
an explicit construction of $\hat{Z}_3$ for Calabi-Yau hypersurfaces
and the geometrization of closed and open deformations of $(Z_3,\Sigma)$ on $\hat{Z}_3$
was explained, first in general and then via a concrete example of the quintic
threefold with a curve $\Sigma$ realized as a complete intersection. Finally in sections
\ref{sec:hatomega} and \ref{sec:potentialhatZ3minusD} we used the pullback $\hat{\Omega}$ of 
the Calabi-Yau form $\Omega$ on $Z_3$ to probe the complex structure of $\hat{Z}_3$. This 
readily yielded open-closed Picard-Fuchs equations on $\hat{Z}_3$ that are precisely solved by both the flux 
superpotential as encoded by the closed periods of $\Omega$, and the five-brane superpotential $W_{\rm brane}$.
In this context it was essential to note the general structure of the expected Picard-Fuchs differential equations
and to rigorously formulate the lift of the five-brane superpotential from $Z_3$ to $\hat{Z}_3$ 
using currents.

While the proposal to study deformations of a holomorphic curve $\Sigma$
in a complex variety $Z_3$ via the blown up manifold ${\hat Z}_3$ is very general,
we concretely obtained in chapter \ref{ch:CalcsBlowUp} the corresponding 
open-closed Picard-Fuchs equations for the deformation problem in some 
generality for hypersurfaces in toric varieties in the presence 
of toric branes (section \ref{sec:generaltoricstructure}). We focused on two examples, namely branes
in the mirror quintic in section \ref{sec:ToricBraneBlowup} and in the mirror of the degree-18 
hypersurface in $\bbP^4(1,1,1,6,9)$ in section \ref{sec:ToricBraneBlowupII}.
We noted that there exist maps from
the unobstructed deformation problem of a complex higher genus curve $\Sigma$ 
in $Z_3$, which is realized as a complete intersection
with the Calabi-Yau threefold $Z_3$, to generically obstructed
configurations of branes on rational curves or the involution
brane studied in~\cite{Walcher:2006rs}. This map was defined around
a critical locus in  the moduli space of $\Sigma$ where it degenerates 
holomorphically to e.g.~the rational curves. Away from this locus one has to identify the rational curves
using a non-holomorphic map which has branch cuts and
the corresponding obstructions are encoded, upon wrapping these cycles by a five-brane, 
in a superpotential that we calculate from $\hat{Z}_3$. This picture was further confirmed by noting
that the discriminant components of the Picard-Fuchs system for $\hat{Z}_3$ 
factorizes into several components, each of which corresponding to the critical locus of
an underlying deformation problem of a five-brane on an in general obstructed curve.
We checked our results for the superpotential by
extraction of integral disk instanton invariants at large volume
that agree with the available results in the literature or partially extend them. 
We concluded our calculations by a different cross-check in section \ref{sec:heteroticF+blowup}, 
namely by the application of the blow-up proposal 
to heterotic Calabi-Yau threefolds $Z_3$ that in addition admit an F-theory dual description.
Indeed we were able to construct the dual F-theory fourfold $X_4$ as a toric complete
intersection directly from the blow-up threefold $\hat{Z}_3$ of the heterotic
string with five-branes, that was also realized as a toric complete intersection.
Furthermore, we established a map of the heterotic flux and five-brane superpotential 
to the flux superpotential on $X_4$ by formally constructing the dual four-flux $G_4$
directly from $\hat{Z}_3$.      

In the final chapter \ref{ch:su3structur} of this work we 
presented a proposal for formulating an $SU(3)$-structure 
on the blow-up threefold $\hat Z_3$, in order to view $\hat{Z}_3$
as a dual physical description of the backreacted five-brane.
The basic idea was to first construct a no-where 
vanishing $(3,0)$-form $\Omega'$, which was roughly obtained by canceling 
the first order zero of $\hat{\Omega}$ along the exceptional 
divisor $E$ and the first order pole of $d\hat H_3$ along $E$ with
each other.  This could only be achieved by introducing a 
non-holomorphic form $\Omega'$, which is no longer closed 
near $E$. In addition to a local construction in sections \ref{sec:SU(3)rev} and \ref{sec:blow-up_as_Kaehler}
we proposed a global residuum representation for $\Omega'$
in the case that $\hat Z_3$ is given as a toric hypersurface in section \ref{sec:non-Kaehlertwist}.
Simultaneously, this procedure yielded a smooth three-form flux $H_3'$.
In addition we defined a new $(1,1)$-form, a non-K\"ahler-form $J'$, by 
smoothing out a logarithmic singularity in the pullback K\"ahler form
$\hat{J}$, which is dictated by the supersymmetry condition $id^c \hat{J}=\hat{H}_3$.
Finally, employing these definitions the complete effective superpotential takes
the form $\int_{\hat Z_3}  (H_3' + i dJ') \wedge \Omega' $ as familiar from flux compactifications
on $SU(3)$-structure manifolds.

We end our conclusion by mentioning some directions for future research. First, we note that the
derived generic D5-brane effective action allows for various phenomenological applications. 
Examples include the study of D5-branes for mechanism of inflation, e.g.~via D5-branes on 
the vanishing $\P^1$ of the conifold \cite{Becker:2007ui} or via the D5-brane Wilson line moduli 
\cite{Avgoustidis:2006zp}, or for dynamical supersymmetry breaking, e.g.~by extending the non-compact
setups of \cite{Aganagic:2007py} to D5-branes on vanishing two-cycles in compact Calabi-Yau orientifolds.
A different application is provided by explicit GUT model building in Type IIB compactifications, where it
would be of particular importance to include matter from intersecting D5-branes using similar techniques as developed 
for intersecting D7-branes \cite{BBGW,Grimm:2008ed}\footnote{See \cite{Oikonomou:2011ba} for an analysis of the fermionic modes on the intersection curve of D7-branes.}. Furthermore, the effective action allows also some general
conclusions about moduli stabilization. From the independence of the Type IIB flux\footnote{The usual dilaton-depended 
flux combination $G_3=F_3-\tau H_3$ reduces to $F_3$ in $O5/O9$-orientifolds.} and brane superpotential
on the dilaton multiplet $S$ as well as from the positive definite induced F-term scalar potential due to the no-scale 
structure of the $\mathcal{N}=1$ data we deduce that the dilaton can not be stabilized perturbatively. 
It might be interesting to break the no-scale structure due to non-perturbative 
corrections our by treating the backreaction of the fluxes and the D5-brane on the geometry more thoroughly,
which naturally leads to more general non-Calabi-Yau backgrounds or strong warp effects.  

Possible applications of the knowledge of the full F-theory flux superpotential are 
provided by the use to systematically calculate $g_S$-corrections
to the underlying type IIB compactification or to directly stabilize complex structure moduli in F-theory. 
In the light of GUT model building in F-theory this might allow to analyze whether phenomenologically 
preferred settings, e.g.~like in the study of local geometries initiated in \cite{Donagi:2008ca,Beasley:2008dc,Beasley:2008kw}, 
can indeed be stabilized by fluxes in a global setup. Furthermore, one can use the 
explicit knowledge of the F-theory flux superpotential to analyze the question of moduli stabilization
in the dual heterotic string. This is of particular interest since in F-theory all complex structure 
moduli can generically be stabilized by fluxes which in the heterotic dual would correspond to a stabilization
of complex structure, bundle and five-brane moduli. In addition one might wonder about extracting, like in \cite{Jockers:2009ti}, 
the heterotic Chern-Simons functional for non-trivial bundles $E$, e.g.~to analyze its integral structure, or 
to deduce K\"ahler corrections from the elliptic fiber of $Z_3$ to the heterotic superpotential. Furthermore it would be very 
interesting to extend the calculations to the heterotic K\"ahler potential, which might be a tractable task in heterotic/F-theory 
duality since the F-theory K\"ahler potential is computable as a function of the fourfold periods. More conceptually on might 
analyze the structure of the fourfold periods at other points of the moduli space including a general analysis of monodromies. 
Physical interpretations of their different structure compared to the threefold case might 
shed some light on additional massless states that are tightly related to singularities in the moduli space.
Furthermore it would be challenging to identify and calculate the geometric quantity in F-theory that 
encodes the second holomorphic $\mathcal{N}=1$ coupling, namely the seven-brane gauge kinetic function.  
This would be of particular interest in the context of open mirror symmetry by extending the local results of
\cite{Bouchard:2007ys} and probably the Bergmann kernel to compact Calabi-Yau compactifications.

Finally it would be very interesting to improve the understanding of the blow-up threefold $\hat{Z}_3$ as
defining a string background of a backreacted five-brane. This includes a more thorough treatment of the dilaton and 
the warp factor, that both might no longer be constant, and a better understanding of the size modulus of the blow-up $\P^1$. 
This is essential in order to establish a physical duality, where in particular all new fields need to be identified 
with new quantum degrees of freedom. In the heterotic compactifications, for example, it is natural to identify the 
K\"ahler modulus of the $\P^1$ with the positions of the heterotic five-branes in the interval of heterotic M-theory. 
It would be interesting to make this map explicit and clarify the interpretation of the blow-up mode in the Type II setups.
Moreover, the computation of disk instanton numbers via the superpotential is only the simplest check of this duality. 
One can attempt to compute amplitudes of higher genus and with more boundaries like the annulus. It would be of interest to investigate 
how these can be derived on the blow-up space $\hat Z_3$ as well. From a more technical point of view
it would further be essential to gain a more complete picture of the open-closed field space as encoded 
by the complex structure moduli space of $\hat Z_3$. This might require to find a general method 
to fix the integral symplectic basis of the third cohomology on $\hat Z_3$ which might allow an identification of the open-closed 
mirror map similar to the Calabi-Yau threefold case. For example, the knowledge of the flat open coordinates  
is particularly relevant to determine critical values of the brane superpotential in the open-closed moduli space.

\appendix

\chapter{Effective Actions}
\label{app:PartEffActions}

\section{The $\cN=2$ Gauge-Kinetic Coupling Function} 
\label{app:complexMatrixM}

In this appendix we collect some useful formulas applied in the derivation of
the $\cN=1$ scalar potential and the $\cN=1$ gauge-kinetic function for the
bulk vectors.  Both quantities depend on the complex structure deformations of 
the internal Calabi-Yau manifold $Z_3$. In the underlying $\cN=2$ theory the complex
structure deformations are in vector multiplets together with vectors
$\underline{V}$ in the expansion $C_4 = V^K \alpha_K + \ldots$ where
$K=0,\ldots,h^{(2,1)}$ labels the symplectic basis of $H^{3}(Z_3,\mathds{Z})$ and 
$V^0$ is identified with the graviphoton in the $\mathcal{N}=2$ gravity
multiplet, cf.~table \ref{tab:N=1SpecOrie}. The four-dimensional $\cN=2$ action 
for the vectors $\underline{V}$ is of the form
\begin{equation}
   S_{\underline{V}} = \int \big[ \tfrac14 \I \cM_{KL} dV^K \wedge * dV^L + 
   \tfrac14 \text{Re} \cM_{KL} dV^K \wedge dV^L\big]\ .
\end{equation}
The complex matrix $\cM_{KL}$ can be expressed in terms of the periods
$(X^K,\cF_K)$ in the expansion $\Omega = X^K \alpha_K - \cF_K \beta^K$ as 
\begin{equation}
         \mathcal{M}_{KL}=\bar{\mathcal{F}}_{KL}
         +2i\frac{(\text{Im}\mathcal{F})_{KL}X^M(\text{Im}\mathcal{F})_{LN}X^N}{X^N(\text{Im}\mathcal{F})_{NM}X^M}\ ,
        \label{eq:matrix-a-b-m}
\end{equation}
where $\cF_{KL}= \partial_{X^K} \cF_L$.
To derive this expression one uses the natural scalar product on the 
cohomology $H^3(Z_3)$. This can be encoded in the following matrix \cite{Craps:1997gp}
\begin{equation}
        E= \begin{pmatrix}
                \int \alpha_K\wedge *\alpha_L & \int \alpha_K\wedge *\beta^L\\
                \int\beta^K\wedge * \alpha_L & \int \beta^K\wedge * \beta^L
        \end{pmatrix}
       =\begin{pmatrix}
                -(A+BA^{-1}B) & -BA^{-1}\\
                -A^{-1}B & -A^{-1}
        \end{pmatrix}\ ,
\end{equation}
where $A=\text{Im}\mathcal{M}$ and $B=\text{Re}\mathcal{M}$. 
A matrix of this form can be easily inverted where the inverse matrix reads
\begin{equation}
        E^{-1}=\begin{pmatrix}
                -A^{-1} & A^{-1}B\\
                BA^{-1} & -(A+BA^{-1}B)
        \end{pmatrix}=\begin{pmatrix}
                \int \beta^K\wedge * \beta^L &  -\int\beta^K\wedge * \alpha_L\\
                -\int \alpha_K\wedge *\beta^L & \int \alpha_K\wedge *\alpha_L
        \end{pmatrix}\ .
        \label{eq:matrix-d-2}
\end{equation}
These matrices will be used in the derivation of the $\cN=1$
scalar potential in section \ref{sec:scalarpotderivation}, where the indices $K=0,\ldots,
h^{(2,1)}_+$ are in the positive eigenspace $H^3_+(Z_3)$. 
The complex matrix $\mathcal{M}$ will also appear in the
$\mathcal{N}=1$ gauge-kinetic coupling function in section \ref{sec:gaugeKin+rest} 
where now the indices $k=1,\ldots,h^{(2,1)}_-$ are in the negative eigenspace 
$H^{3}_-(Z_3)$.

\section{Kinetic Mixing of Bulk and Brane Gauge Fields}
\label{app:kinmix}

The reduction of the Chern-Simons action to the effective Lagrangian
\eqref{eq:CS} contains mixing terms between the bulk vector 
fields $\underline{V}$, $\underline{U}$ and the D5-brane U(1)-field $F$. 
Since the vectors $\underline{U}$ are the magnetic duals to the vector 
$\underline{V}$, a dualization procedure has to be performed in order 
to reveal the effective action for the propagating fields. Here, 
we will present this dualization in detail and demonstrate how it affects the 
kinetic term of the D5-brane vector $F$ such that a further 
intertwining between open and closed moduli appears.

First, we have to collect all terms of the effective action that are relevant 
for the dualization procedure. These are the kinetic terms from the bulk vectors 
$\underline{V}$, $\underline{U}$ of the bulk supergravity action, the 
kinetic as well as instanton term of the D5-brane vector $F$ given in the 
Dirac-Born-Infeld action \eqref{eq:DBI} and the Chern-Simons action \eqref{eq:CS},
respectively, and mixing terms between bulk and brane vectors of
\eqref{eq:CS}. Thus, the starting point of the dualization is the action
\begin{equation}
 S_{\rm vec}= - \int\big[ \tfrac18 d\vec{V}^T\wedge\ast E\, d\vec{V}+ 
\tfrac{1}{2}\mu_5\ell^2 \big(v^{\Sigma} e^{-\phi}F\wedge\ast F-c^\Sigma
F\wedge F\big) +\tfrac12\mu_5\ell\vec{\hat{\mathcal{N}}}^T\, d\vec{V}\wedge F
        \big] \ , \label{eq:Lbeforedualization}
\end{equation}
where we again used the matrix $E$ introduced in \eqref{eq:e-matrix} and the convenient 
shorthand notation
\begin{equation}
 \vec{V}=\begin{pmatrix}
                \underline{V}\\
                \underline{U}
         \end{pmatrix},\qquad \qquad
\vec{\hat{\mathcal{N}}}=\hat\zeta^{\mathcal{A}}\begin{pmatrix}
                \mathcal{N}_{\mathcal{A}k}\\
                \mathcal{N}_{\mathcal{A}}^{l}
                \end{pmatrix}
                =
                \begin{pmatrix}
                \mathcal{N}_{k}\\
                \mathcal{N}^{l}
         \end{pmatrix}. 
\end{equation}
Next we have to add the Lagrange multiplier term $\tfrac14 dV^{k}\wedge F_{k}$ to 
the Lagrangian \eqref{eq:Lbeforedualization} in order to integrate out the magnetic
field strength $F_{k}=dU_{k}$. However, the equations of motion for the vectors 
$\underline{V}$ and their duals $\underline{U}$ are not compatible with each 
other after the naive addition of this term. In order to restore consistency of 
the equations of motion, we have to shift the field strengths $dV^{k}$, $dU_{k}$ 
in the kinetic terms appropriately by
\begin{equation}
    dV^{k}\;\rightarrow\; \tilde F^{k}:=dV^{k}-2\mu_5\ell \cN^{k}F\ ,\qquad \qquad 
    dU_l\;\rightarrow\; \tilde F_{l}:=dU_{l}-2\mu_5\ell \cN_{l}F.
\end{equation}
Now, we can integrate out the magnetic dual $\tilde F_{l}$ consistently and obtain 
\begin{eqnarray}
S_{\rm vec}&=& \int \big[\tfrac14\text{Im}\mathcal M_{kl}  F^{k}\wedge\ast  F^{l} 
 +\tfrac14\text{Re}\mathcal M_{kl}  F^{k} \wedge F^{l}\nonumber \\ 
 &-& \tfrac12 \mu_5\ell^2\big((v^{\Sigma} e^{-\phi}+2\mu_5\text{Im}\mathcal{M}_{kl}(N^{k}
 +\bar{N}^{k})(N^{l}+\bar{N}^{l}))F\wedge\ast F\nonumber \\
 &+&(c^\Sigma+i\mu_5\text{Im}\mathcal{M}_{kl}(N^k N^l-\bar{N}^k\bar{N}^l)) F\wedge F\big)\nn \\
&+& \mu_5\ell\big(\text{Im}\mathcal M_{kl} \ast F+\text{Re}\mathcal M_{kl}F \big)\wedge F^k\big(N^l+\bar{N}^l\big)\big]\, . 
\label{eq:dualizedL}
\end{eqnarray}
Here we introduced the complex fields 
\begin{equation}
        N^{k}=\int_{\Sigma_-}\zeta\lrcorner \beta^{k},\qquad \bar{N}^{k}
        =\int_{\Sigma_-}\bar{\zeta}\lrcorner\beta^{k}.
\end{equation}
The crucial point of this dualization is the change of the gauge-kinetic term in 
\eqref{eq:dualizedL} compared to the form in \eqref{eq:Lbeforedualization} before 
dualization.

\section{Derivation of the F-term Potential: Massless Modes}
\label{app:F-termscalarpot}

The calculation of the F-term contribution of the scalar potential \eqref{eq:N=1pot}
using the superpotential \eqref{eq:effsuperpot} and K\"ahler potential
\eqref{eq:kaehler-pot} is straightforward but tedious. To simplify this 
computation it is convenient to exploit one of the
shift symmetries of the K\"ahler potential $S \rightarrow S + i \Lambda$ and
dualize the chiral multiplet with bosonic scalar $S$ into a 
linear multiplet with bosonic components $(L,C_2)$. Here $L$ is a real scalar associated to $\text{Re} S$ while
$C_2$ is a two-form dual to $\text{Im} S$. In the context of $O5$-orientifolds without
D5-brane moduli this dualization has been carried out in refs.~\cite{Grimm:2004uq,Grimm:2005fa}, and
we refer the reader to these references for more details on the linear
multiplet formalism and references. Here we will be mainly interested in the scalar potential 
in the new scalar variables $L$ and $M^I=(\underline{P},\underline{a},\underline{t},\underline{\zeta})$. 
First we express the K\"ahler potential \eqref{eq:kaehler-pot} in terms of the 
new variables $L=-K_S = \frac12 e^\phi\mathcal{V}^{-1}$ and $M^I$ such that 
\begin{equation} \label{eq:KL_pot}
  K = -\ln\big[ -i\int\Omega\wedge\bar\Omega \Big] 
  		-\ln \big[ \tfrac{1}{48} \cK_{\alpha \beta \gamma} \Xi^\alpha\,
      \Xi^\beta\, \Xi^\gamma \big] + \ln [L]\ ,
\end{equation}
where $\Xi^\alpha$ is given in \eqref{eq:xi-def}. The kinetic terms in the 
effective action with a linear multiplet are then obtained as derivatives 
of the kinetic potential 
\begin{equation} \label{eq:tildeK_pot}
  \tilde K(L,M^I,\bar M^I) = K + (S + \bar S)\cdot L \ , 
\end{equation}
where $S+\bar S\equiv (S+\bar{S})(L,M^I)$ is now a function of $(L,M^I)$. In fact, this is just a 
Legendre transformation of the function $K$ with respect to $S+\bar S$ to 
obtain the Legendre transformed $\tilde K$ as a function of $L$. 
In terms of this data the scalar potential takes the general form 
\begin{equation} \label{eq:V_lin}
  V = e^{K}(\tilde K^{IJ} D_I W D_{\bar J} \bar W - (3-L K_L) |W|^2)\ ,
\end{equation}
where $D_I W = \partial_I W + K_I W$ and $K_L = \partial_L K$. Note that in
front of $|W|^2$ as well as in $D_I W$ only the derivatives of the 
K\"ahler potential \eqref{eq:KL_pot} appear.

With this formalism at hand we evaluate the scalar potential. We first take
derivatives of \eqref{eq:KL_pot} and \eqref{eq:tildeK_pot} such that
\begin{eqnarray}        
        K_{t_\alpha}= -\frac{e^{\phi}}{4\mathcal V}\mathcal K _\alpha\; , 
        \qquad K_{P_a}=0\; ,\qquad
        K_{a_I}= 0\; ,\qquad
        K_{\zeta^A}= \tfrac12\mu_5 e^{\phi} \mathcal{G}_{A\bar B}\bar \zeta^{\bar B}\; ,
\label{eq:firstderivatives}
\end{eqnarray}
when $K\equiv K(L,M^I)$ is viewed as a function of $L$ instead of $S+\bar{S}$.\footnote{It is a basic fact
that a Legendre transformation leaves the derivatives with respect to the untransformed variables invariant, 
i.e.~$\partial_{M^I}K=\partial_{M^I}\tilde{K}$ where both sided have to be evaluated in the same variables 
\textit{after} differentiation.} Similarly we obtain 
\begin{eqnarray}        
        \tilde{K}_{t_\alpha}= \frac{e^{\phi}}{4\mathcal V}(\mathcal K _{\alpha ab}\mathcal{B}^a\mathcal{B}^b
        -\mathcal K _\alpha)\; , 
        \quad \tilde{K}_{P_a}=-\frac{e^{\phi}}{2\mathcal V} \mathcal B ^a\; ,\quad
        \tilde{K}_{a_I}= \frac{\mu_5\ell^2 e^{\phi}}{\mathcal V} \mathcal C ^{I\bar J}\bar a _{\bar J}\; ,\quad
        \tilde{K}_{\zeta^A}= K_{\zeta^A}\; .
\label{eq:firstderivatives_2}
\end{eqnarray}
where we made use of the solution
\begin{equation}
	S+\bar{S}=\frac{1}{L}-\frac14(\text{Re}\Theta)^{ab}(P+\bar{P})_a(P+\bar{P})_b+2\mu_5\ell^2\mathcal{C}^{I\bar J}a_I\bar{a}_{\bar J}\,.
\end{equation}
From this we can easily determine the metric $\tilde{K}_{I\bar J}$ for the remaining fields which is 
block-diagonal with one block $\tilde{K}_{a_I\bar a_{\bar J}}=\mu_5\ell^2e^\phi\mathcal{V}^{-1} C^{I\bar J}$ 
for the Wilson lines and another block of the following type
\begin{eqnarray}
 \tilde{K}_{I\bar J}=
\begin{pmatrix}
        A+B^\dagger GB &-B^\dagger G & 0\\
-GB& G+D^\dagger C D&D^\dagger C\\
0&C D& C
\end{pmatrix}
\label{eq:tildemetric}
\end{eqnarray}
for the moduli $(\underline{\zeta},\underline{t},\underline{P})$.
Its inverse $\tilde{K}^{I\bar J}$ is then given by
\begin{eqnarray}
 \tilde{K}^{I\bar J}=
\begin{pmatrix}
        A^{-1} &A^{-1}B^\dagger  & -A^{-1}B^\dagger D^\dagger\\
BA^{-1}& G^{-1}+BA^{-1}B^\dagger&-(G^{-1}+BA^{-1}B^\dagger) D^\dagger\\
-DBA^{-1}&-D(G^{-1}+BA^{-1}B^\dagger)& C^{-1}+D(G^{-1}+BA^{-1}B^\dagger) D^\dagger\\
\end{pmatrix}.
\label{eq:tildemetricinverse}
\end{eqnarray}
Here, we abbreviated the various matrices as follows,
\begin{eqnarray}
        A=\tfrac12e^\phi\mu_5\mathcal{G}_{A\bar B}\,,\quad G=e^{2\phi}G_{\alpha\beta}\,,
        \quad B=\mu_5 \mathcal{L}^\alpha_{A\bar B}\bar{\zeta}^{\bar B}\,,\quad C=-\frac{e^\phi}{2\mathcal{V}}(\text{Re} \Theta)_{ab}\,,
        \quad D=\tfrac12\mathcal{K}_{ab\alpha}\mathcal{B}^b,\nn\\
\end{eqnarray}
where the matrix $\mathcal G$ is defined in \eqref{eq:metrics} and we introduced the K\"ahler metric 
on the K\"ahler moduli space in \eqref{eq:KaehlerMetricOrie},
\begin{equation}
        G_{\alpha\beta}=\frac{1}{4\mathcal{V}}\left(\frac{\mathcal{K}_{\alpha}
        \mathcal K_{\beta}}{4\mathcal V}-\mathcal K_{\alpha\beta}\right).
\end{equation}
Now we use this to compute the F-term scalar potential.
First we note the no-scale structure of $K$ and $W$. 
The superpotential does not depend on the moduli $(S,\underline{a},\underline{P})$ 
as well as on $\underline{t}$ such that the covariant derivative 
$D_I=\partial_I+K_I$ reduces just to $K_I$.
Moreover, for the dual linear multiplet to $S$  we find a contribution $1\cdot |W|^2$ to the 
scalar potential $V$ which is an immediate consequence of $K_L L=1$ in \eqref{eq:V_lin}.
The block matrix for the Wilson lines $\underline{a}$ does not contribute to $V$ since $K_{a_I}=0$. 
However, the block for the moduli $(\underline{\zeta},\underline{t},\underline{P})$ yields a 
contribution of the form
\begin{equation}
        D_{(\zeta,t,P)}WD_{(\bar \zeta,\bar t,\bar P)}\bar W \tilde K^{(\zeta, t,P)(\bar \zeta,\bar t,\bar P)}
        =\frac{\mathcal K_\alpha (G_{\text{ks}})^{\alpha\beta}\mathcal K_\beta}{(4\mathcal V)^2} |W|^2+2\mu_5 \left(\int_{\Sigma_+}s_A\lrcorner \Omega\int_{\Sigma_+}\bar s_{\bar B}\lrcorner \bar \Omega\right) e^{-\phi}\mathcal G^{A\bar B}.
        \label{eq:str-derivative-kaehler-metric-2}
\end{equation}
Using the various intersection matrices $v^\alpha=\int J\wedge \tilde \omega^\alpha$, 
$\mathcal{K}_{\alpha}$, $\mathcal{K}_{\alpha\beta}$ and its formal inverse $\mathcal{K}^{\alpha\beta}$ 
as well as the inverse metric
\begin{equation}
        G_{\text{ks}}^{\alpha\beta}=2v^\alpha v^\beta-4\mathcal V\mathcal{K}^{\alpha\beta}\ ,
\end{equation}
we deduce the useful relation 
\begin{equation}
        \mathcal K_\alpha (G_{\text{ks}})^{\alpha\beta}\mathcal K_\beta 
        = (8\mathcal V)^2v^\alpha(G_{\text{ks}})_{\alpha\beta}v^\beta=3(4\mathcal V)^2. 
        \label{eqn:kaehler-moduli-identity}
\end{equation}
Finally, we obtain the F-term contribution to the scalar potential $V$ of the form
\begin{eqnarray}
        V 
&=&\frac{ie^{4\phi}}{2\mathcal V^2 \int\Omega\wedge\bar\Omega}\left[ \left| W\right|^2
+D_{z^\kappa}WD_{\bar z^{\bar\kappa}}\bar W G^{\kappa\bar\kappa} 
+2\mu_5 e^{-\phi}\mathcal G^{A\bar B}\int_{\Sigma_+}s_A\lrcorner \Omega\int_{\Sigma_+}\bar s_{\bar B}\lrcorner \bar \Omega \right]\ .
        \label{eq:other-form-of-v}
\end{eqnarray}

\section{Derivation of the F-term Potential: Massive Modes}
\label{app:potcalc}

This section provides the necessary background to perform the calculation of the F-term potential 
\eqref{eq:Ftermpot}. The following calculation extends the analysis made in chapter \ref{ch:EffActD5} \cite{Grimm:2008dq} 
for deformations associated to holomorphic sections $H^0(\Sigma,N_{Z_3}\Sigma)$ to the case of the 
infinite dimensional space of deformations $\mathcal{C}^{\infty}(\Sigma,N_{Z_3}\Sigma)$. 
In particular, the K\"ahler metric for the open string deformations is accordingly generalized.

First we need the general form of the K\"ahler potential of the D5-brane action that is given by \cite{Grimm:2008dq}
\begin{equation} \label{eqn:kaehler-pot}
    K =-\ln\big[ -i\int\Omega\wedge\bar\Omega \Big]+K_q\ ,\qquad  K_q=-2\ln\big[
      \sqrt{2}e^{-2\phi}\mathcal V \Big]\ ,
\end{equation}
that immediately implies that 
\begin{equation} \label{eq:e^K}
e^K=\frac{ie^{4\phi}}{2\mathcal{V}^2\int\Omega\wedge\bar\Omega}.
\end{equation}
In order to evaluate the K\"ahler metric the potential $K$ has to to be expressed as a function 
of the $\mathcal{N}=1$ complex coordinates. For the purpose of our discussion in section 
\ref{sec:extensionInfinite} we only need the part of the K\"ahler metric for the open string
deformations $u^a$. 

We straight forwardly extend the K\"ahler metric deduced in section \ref{sec:N=1coords+KpotD5} 
for the fields associated to $H^0(\Sigma,N_{Z_3}\Sigma)$ to the infinite dimensional space 
$\mathcal{C}^{\infty}(\Sigma,N_{Z_3}\Sigma)$.  This is possible since the condition of holomorphicity 
of sections does not enter the calculations of section \ref{sec:D5branes}. Thus, we obtain the inverse 
K\"ahler metric
\begin{equation} \label{eqn:Kaehlerpart}
       K^{a\bar b}=2\mu_5^{-1}e^{-\phi}\mathcal{G}^{a\bar b}\,
\end{equation}
cf.~appendix \ref{app:F-termscalarpot} for the process of inversion of the full K\"ahler metric, 
where the matrix $\mathcal{G}^{a\bar b}$ is the inverse of 
\begin{equation}
	\mathcal{G}_{a\bar b}=\frac{-i}{2\mathcal{V}}\int_{\Sigma}s_a\lrcorner \bar{s}_{\bar b}\lrcorner (J\wedge J).
\end{equation} 
Here we introduce a basis $s_a$ of $\mathcal{C}^{\infty}(\Sigma,N_{Z_3}\Sigma)$ so that a generic 
section $s$ enjoys the expansion $s=u^as_a$. Next we Taylor expand the superpotential \eqref{eq:chainIIB} 
in the open string deformations $u^a$ around the supersymmetric vacuum of the holomorphic curve 
$\Sigma=\Sigma_h$ as
\begin{equation}
\displaystyle
W_{\rm brane}=\displaystyle\int_{\Gamma_u}\Omega\equiv\int_{\Sigma_{0}}^{\Sigma+\delta \Sigma_u}\Omega=\int_{\Gamma_h}\Omega+\int_{\Sigma}u\lrcorner \Omega+\frac12 u^au^b\int_{\Sigma}s_a\lrcorner d s_b\lrcorner\Omega)+\mathcal{O}(u^3)\,.
\end{equation}
To evaluate the derivatives $\frac{\partial^k}{\partial^k u^a} W_{\rm brane}\vert_{u=0}$ we use that 
for every $\frac{\partial}{\partial u^a}$ the Lie-derivative $\mathcal{L}_{s_a}=d s_a\lrcorner +s_a\lrcorner d$ 
acts on the integrand $\Omega$, where we further denote the interior product with a vector $s_a$ by 
$s_a\lrcorner$. In addition we use that on the holomorphic curve $\Sigma$ there are no $(2,0)$-forms 
such that the linear term in $u^a$ in the Taylor expansion vanishes identically. Then the derivative 
with respect to $u^a$ is obtained as
\begin{equation}
	\partial_{u_a}W_{\rm brane}=-\mu_5\int_{\Sigma}\bar\partial s\lrcorner s_a\lrcorner \Omega
\end{equation}
where we perform a partial integration on $\Sigma$. In addition, we rescale the superpotential by the 
D5-brane charge, $W_{\rm brane}\mapsto \mu_5 W_{\rm brane}$, as in \cite{Grimm:2008dq}. Now we can 
calculate the F-term potential 
\begin{equation}
	V=\frac{2\mu_5}{e^{\phi}}e^K\int_{\Sigma}\bar\partial s\lrcorner s_a\lrcorner\Omega\ \mathcal{G}^{a\bar b}\int_{\Sigma}\partial \bar{s}\lrcorner \bar{s}_{\bar b}\lrcorner\bar{\Omega}\,.
\end{equation}
To further evaluate this we have to rewrite the matrix $\mathcal{G}_{a\bar b}$ as follows.
Consider the integral 
\begin{equation}
 	I_{a\bar{b}}:=\int_\Sigma \left(s_a\lrcorner\Omega\right)_{ij}\left(\bar{s}_{\bar{b}}\lrcorner\bar{\Omega}\right)^{ij}\iota^\ast\left(J\right).
\end{equation}
The contracted indices $i$, $j$ denote the coordinates on $Z_3$, one of which is tangential and two 
are normal to $\Sigma$. Then using $\bar{s}_{\bar a}\lrcorner\Omega=0$, 
$\Omega\wedge\bar\Omega=\tfrac{\int\Omega\wedge\bar\Omega}{6\mathcal V}J^3$ where $\mathcal{V}$ 
denotes the compactification volume and the rule 
$s_a\lrcorner (\alpha\wedge\beta)=(s_a\lrcorner\alpha)\wedge\beta+(-1)^p\alpha\wedge (s_a\lrcorner \beta)$ 
for a $p$-form $\alpha$ we can rewrite this as
\begin{eqnarray}
 	I_{a\bar{b}}&=&-\int_\Sigma \left(s_a\lrcorner\bar{s}_{\bar{b}}\lrcorner\left(\Omega\wedge\bar{\Omega}\right)\right)_{ij}^{ij}\iota^\ast\left(J\right)=-\tfrac{\int\Omega\wedge\bar\Omega}{6\mathcal V}\int_\Sigma\left(s_a\lrcorner s_{\bar b}\lrcorner\left(J^3\right)\right)_{ij}^{ij}\iota^\ast\left(J\right)\nonumber\\
&=&-\tfrac{\int\Omega\wedge\bar\Omega}{6\mathcal V}\int_\Sigma\left[3\left( s_a\lrcorner s_{\bar b}\lrcorner J\right)J^2-6\left(\bar{s}_{\bar b}\lrcorner J\right)\wedge \left(s_a\lrcorner J\right)\wedge J\right]_{ij}^{ij}\iota^\ast\left(J\right)\nonumber\\
&=&\tfrac{\int\Omega\wedge\bar\Omega}{4\mathcal V}\int_\Sigma s_a\lrcorner s_{\bar b}\lrcorner J^2=\tfrac{i\int\Omega\wedge\bar\Omega}{2}\mathcal{G}_{a\bar b}.
\end{eqnarray}
Consequently, introducing the abbreviation $\Omega_a=s_a\lrcorner \Omega$ we can write the matrix 
$\mathcal{G}_{a\bar b}$ as
\begin{equation}
	\mathcal{G}_{a\bar b}=\frac{-i}{2\mathcal{V}}\int_{\Sigma}s_a\lrcorner \bar{s}_{\bar b}\lrcorner (J\wedge J)=\frac{-2i}{\int\Omega\wedge \bar\Omega}\int_{\Sigma}(\Omega_a)_{ij}(\bar{\Omega}_{\bar b})^{ij} \iota^\ast(J)\,.
\end{equation}
Next we use the basis $\Omega_a$ of $\Omega^{(1,0)}(\Sigma,N_{Z_3}\Sigma)$ to expand the section
\begin{equation} \label{eq:basisexpansion}
	\bar\partial s\lrcorner J=c^{\bar b}\bar\Omega_{\bar b}\,.
\end{equation}
The coefficients $c^{\bar{b}}$ are determined by contraction of \eqref{eq:basisexpansion} with 
$\Omega_{a}^{\bar\imath\bar\jmath}$, where we have raised the form indices using the hermitian 
metric on $Z_3$. This way we obtain a function on $\Sigma$ that we can integrate over $\Sigma$ 
using the volume form $\vol_{\Sigma}=\iota^*(J)$ to determine the $c^{\bar b}$ so that
\begin{equation} 
	\bar\partial s\lrcorner J=\frac{-2i}{\int\Omega\wedge \bar\Omega}\,\mathcal{G}^{a\bar b}\,\bar{\Omega}_{\bar b}\int_{\Sigma}\bar{\partial} s\lrcorner \Omega_{a}\,.
\end{equation} 
With this expansion we immediately obtain the desired form of the F-term superpotential
\begin{equation}
 	V=\frac{2e^K\mu_5}{e^{\phi}}\int_{\Sigma}\bar\partial s\lrcorner \big(\Omega_a\mathcal{G}^{a\bar b}\int_{\Sigma}\partial \bar{s}\lrcorner \bar{\Omega}_b\big)=\frac{-\mu_5e^{3\phi}}{2\mathcal{V}^2}\int_{\Sigma}\bar\partial s\lrcorner \partial \bar s\lrcorner J
	= \frac{\mu_5e^{3\phi}}{2\mathcal{V}^2}\int_{\Sigma}||\bar\partial s||^2 \iota^\ast(J)\,.
\end{equation}
This perfectly matches the potential \eqref{eq:VDBI} obtained from the reduction of the DBI-action 
of the D5-brane on $\Sigma$ 

\chapter{Geometrical Background}
\label{app:PartGeoBackground}

\section{Topology of Elliptic Calabi-Yau's and Ruled Threefolds} 
\label{app:chernEllipticCY}

In this section we summarize useful relation when working with elliptically fibered Calabi-Yau 
manifolds $X_n$ given by a Weierstrass constraint
\begin{equation}
 	y^2 = x ^3 + g_2(\underline{u}) x z^4 + g_3(\underline{u}) z^6\ 
\end{equation}
in homogeneous coordinates $(z,x,y)$ of a projective bundle 
$\mathds{P}(\mathcal{O}_{B_{n-1}}\oplus \mathcal{L}^2\oplus \mathcal{L}^3)$ as in sections 
\ref{sec:spectralcover} and \ref{sec:ellFourfolds}. Here $\mathcal{L}$ denotes an a priori 
undetermined line bundle on $B_{n-1}$ and $g_2(\underline{u})$, $g_3(\underline{u})$ are 
sections of  $\mathcal{L}^4$ and $\mathcal{L}^6$ for \eqref{eq:Weierstrass} to be a well-defined constraint equation.
Locally on the base $B_{n-1}$ they are functions in local coordinates $\underline{u}$ on $B_{n-1}$.

Next, we determine the line bundle $\mathcal{L}$ by the requirement of $X_n$ being Calabi-Yau.
First, we note that the total Chern classes of the projective bundle $\mathcal{W}$
and of $X_n$ read \cite{Friedman:1997ih,Friedman:1997yq,Andreas:1998zf}
\begin{eqnarray}
	c(\mathcal{W})&=&c(B_{n-1})(1+r)(1+2r+2l)(1+3r+3l)\,,\\[0.5Em]  
	c(X_n)&=&c(B_{n-1})\frac{(1+r)(1+2r+2l)(1+3r+3l)}{1+6r+6l}\,,
\label{eq:ChernclassW}
\end{eqnarray}
where we used  that the class of $X_n$ is $[X_n]=6r+6l$ as well as the adjunction formula 
\cite{Griffiths:1978yf} for $X_n$ in $\mathcal{W}$. 
Here we used $c_1(\mathcal{L})=l$ and $r=c_1(\mathcal{O}_{\mathds{P}^2(1,2,3)}(1))$ to denote 
the first Chern class of the hyperplane in the fiber $\mathds{P}^2(1,2,3)$. Then the Calabi-Yau 
condition on $X_n$ implies $l=c_1(B)$ or equivalently $\mathcal{L}=K^{-1}_{B_3}$. Alternatively
we draw the same conclusion by noting that the normal bundle to $\sigma$ in $Z_3$ is given by
$\mathcal{L}^{-1}=N_{Z_3}\sigma$ which implies by adjunction $K_{Z_3}\vert_\sigma=K_\sigma\otimes \mathcal{L}$
the condition $c_1(B_2)=l$ as well.

In order to calculate the second Chern-class of $X_n$ we first note that $r(2r+2l)(3r+3l)=0$
and since the class of $X_n$ is given as $6r+6c_1(L)=0$ we obtain in the intersection ring of 
$X_n$ the relation
\begin{equation}
	\sigma^2=-c_1(\mathcal{L})\sigma\,,
\label{eq:sigma^2}
\end{equation}
which is generally valid for projective bundles \cite{Griffiths:1978yf}.
Here we identified $r=c_1(\mathcal{O}(\sigma))=(z=0)$ where $\sigma$ denotes the section of 
$\pi:\,X_n\rightarrow B_{n-1}$. The second Chern class of $X_n$ is determined by expanding
\eqref{eq:ChernclassW} to second order using \eqref{eq:sigma^2} to obtain 
\begin{eqnarray}
	c_2(X_2)&=&12 c_1(B_2)\sigma\,,\\
	\nn c_2(X_3)&=&12 c_1(B_2)\sigma+11 c_1(B_2)^2+c_2(B_2)\\
	\nn c_2(X_4)&=&12 c_1(B_2)\sigma+11 c_1(B_2)^2+c_2(B_2)\,.
\label{eq:c2EllipticThreefold}
\end{eqnarray}

We conclude by a discussion of the Chern class of the projective bundle $B_3=\P(\mathcal{O}_{B_2}\oplus L)$
for some arbitrary line bundle $L$ over $B_2$ with $c_1(L)=t$ that reads
\begin{equation}
	c(\P(\mathcal{O}_{B_2}\oplus L))=c(B_2)(1+r)(1+r+t).
\label{eq:ChernclassB_3}
\end{equation}
This implies $c_1(B_3)=c_1(B_2)+2r+t$ and $c_2(B_3)=c_2(B_2)+c_1(B_2)(t+2r)$.

\section{A Local Study of the Blow-Up Threefold $\hat{Z}_3$}
\label{App:Local}

In this appendix we study the geometry of the blow-up $\hat{Z}_3$ in more detail in a local analysis. 
The obtained results provide insights in the blow-up process that immediately apply for the global 
discussion of section \ref{sec:5braneblowupsanddefs} and allow for a derivation of the expressions 
used for $\hat{Z}_3$ as a complete intersection and in particular $\hat{\Omega}$ as a residue integral.

The following discussion bases on the general lore in algebraic geometry that the process of blowing up 
is local in nature, i.e.~just affects the geometry near the subvariety $\Sigma$ which is blown-up leaving 
the rest of the ambient space invariant. Thus the geometrical properties of the blow up geometry can be 
studied in a completely local analysis in an open neighborhood around $\Sigma$. In particular, this 
applies for the case at hand, the blow-up of the curve $\Sigma$ in the compact Calabi-Yau threefold $Z_3$ 
into a divisor $E$ in the threefold $\hat{Z}_3$.

Starting from a given open covering of $Z_3$ by local patches $U_k\cong \C^3$ we choose a neighborhood 
$U$ centered around the curve $\Sigma$. Thus, this local patch can be modeled by considering just $\C^3$ 
on which we introduce local coordinates $x_1$,$x_2$, $x_3$. The holomorphic three-form $\Omega$ on $Z_3$ 
takes then simply the local form $\Omega=dx_1\wedge dx_2\wedge dx_3$. The curve $\Sigma$ is accordingly 
described as the complete intersection
\begin{equation}
 	D_1\cap D_2=\{h_1(x_i)=0\}\cap\{h_2(x_i)=0\}\,,
\end{equation}
for two given polynomials $h_i$ with corresponding divisor classes $D_i$ in $\C^3$. 

To construct the blow-up along $\Sigma$, denoted by $\hat{\C}^3$, we have to consider the new ambient 
space of the projective bundle $\mathcal{W}=\P(\mathcal{O}(D_1)\oplus\mathcal{O}(D_2))$. This is locally 
of the form ``$\C^3\times \P^1$'' as necessary for the blow-up procedure described in standard textbooks, 
see e.g.~p. 182 and 602 of \cite{Griffiths:1978yf}. Next we introduce homogeneous coordinates $(l_1,l_2)$ 
on the $\P^1$ to obtain the blow-up $\hat{\C}^3$ as the hypersurface 
\begin{equation} \label{eq:blowuplocal}
 	Q\equiv l_1h_2-l_2h_1=0 
\end{equation}
in $\cW$ as before in \eqref{eq:blowup}. 

In the following we construct and study the pullback form $\hat{\Omega}=\pi^*(\Omega)$ that is a section 
of the canonical bundle $K\hat{\C}^3$, cf.~p. 187 of \cite{Griffiths:1978yf}. To simplify the 
calculations we first perform a coordinate transformation to coordinates $y_i$ such that $y_1=h_1(x_i)$, 
$y_2=h_2(x_i)$ and $y_3=x_j$ for appropriate\footnote{This choice is fixed by the inverse function theorem 
stating that for every point $x_0\in \C^3$ with 
$(\partial_k h_1\partial_l h_2-\partial_l h_1\partial_k h_2)\vert_{x_0}\neq0$ for $k,l\neq j$, there 
exists a local parameterization of $\cC$ near $x_0$ as a graph over $x_j$. In particular, the blow-up is 
not independent of the coordinates used, cf.~p.~603 of \cite{Griffiths:1978yf}.} $j$. For notational 
convenience we relabel the coordinates such that $y_3=x_3$. Thus, we obtain
\begin{equation}
 	Q=l_1y_2-l_2y_1\,,\qquad\Omega=\det{J}^{-1}dy_1\wedge dy_2\wedge dy_3
\end{equation}
in the coordinates $y_i$, where $J=\frac{\partial y_i}{\partial x_j}$ denotes the Jacobian of the coordinate 
transformation that is generically non-zero by assumption of a complete intersection $\Sigma$.

Now we perform the blow-up on the two local patches on the $\P^1$-fiber of $\mathcal{W}$ that are defined 
as usual by $U_i=\{l_i\neq0\}$ for $i=1,2$. In the patch $U_1$, for example, we introduce coordinates 
$z^{(1)}_1$, $z^{(1)}_2$ on $\hat{\C}^3$ as
\begin{equation} \label{eq:coordsU1}
 	z^{(1)}_1=y_1\,,\qquad z^{(1)}_2=-\frac{l_2}{l_1}=\frac{y_2}{y_1}\,,\qquad z^{(1)}_3=y_3
\end{equation}
which allows us to evaluate the pullback map as
\begin{eqnarray}\label{eq:OmegaU1}
 	\pi^*(\Omega)&=&\det J^{-1}\pi^*(dy_1\wedge dy_2\wedge dy_3)=\det J^{-1}dz^{(1)}_1\wedge d(z^{(1)}_2z^{(1)}_1)\wedge dz_3^{(1)}\nonumber\\&=& z_1^{(1)}\det J^{-1}dz^{(1)}_1\wedge dz^{(1)}_2\wedge dz_3^{(1)}\,.
\end{eqnarray}
From this expression we can read off the canonical bundle $K\hat{\C}^3$, cf.~p. 608 of \cite{Griffiths:1978yf}, 
by determining the zero-locus of $\pi^*(\Omega)$. Indeed we obtain $K\hat{\C}^3=E$ as mentioned in section 
\ref{sec:geometricblowups} and in \cite{Grimm:2008dq} since $z^{(1)}_1=0$ describes the exceptional divisor 
$E$ in $U_1$ and $\det J$ is non-zero by assumption. 
Analogously, we obtain a similar expression in the patch $U_2$ for local coordinates
\begin{equation} \label{eq:coordsU2}
 	z^{(2)}_1=\frac{l_1}{l_2}=\frac{y_1}{y_2}\,,\qquad z^{(2)}_2=y_2\,,\qquad z^{(2)}_3=y_3
\end{equation}
that reads
\begin{equation} \label{eq:OmegaU2}
 	\pi^*(\Omega)=z_2^{(2)}\det J^{-1}dz^{(2)}_1\wedge dz^{(2)}_2\wedge dz_3^{(2)}\,,
\end{equation}
which as well vanishes on $E$ since $E=\{z^{(2)}_2=0\}$ in $U_2$. Thus we extract the transition functions 
$g_{ij}$ of $\mathcal{O}(E)$ on $U_1\cap U_2$ to be given as
\begin{equation} 
 	g_{ij}=z^{(j)}_i=\frac{y_i}{y_j}=\frac{l_i}{l_j}
\end{equation}
which reflects the fact that $E=\P(O(D_1)\oplus O(D_2))$ over $\Sigma$ with fiber $\P^1$. Additionally, 
the transition functions $g_{ij}$, when restricted to $E$, are just those of $\mathcal{O}_{\P^1}(-1)$ on 
the $\P^1$ fiber in $E$ and thus we obtain $\mathcal{O}(E)\vert_E=\mathcal{O}(-1)$ as used in chapter 
\ref{ch:su3structur} and in \cite{Grimm:2008dq,Grimm:2009ef,Grimm:2009sy}.

Let us now take a different perspective on the pullback-form $\pi^*(\Omega)$ of the blow-up $\hat{\C}^3$ 
that is more adapted for the global geometry of $\hat{Z}_3$ as a complete intersection $P=Q=0$ in \eqref{eq:blowup}. 
The key point will be the description of the blow-up $\hat{\C}^3$  as the hypersurface \eqref{eq:blowuplocal} 
in $\mathcal{W}$ that will allow for a residue integral representation of $\pi^*(\Omega)$. In particular, 
the advantage of this residue expression in contrast to the local expressions \eqref{eq:OmegaU1}, 
\eqref{eq:OmegaU2} is the fact that it can straight forwardly be extended to a global expression on $\hat{Z}_3$ 
as used in \eqref{eq:ResZhat}. 

Let us start with an ansatz $\hat{\Omega}$ for $\pi^*(\Omega)$,
\begin{equation} \label{eq:OmegaAnsatz}
	\hat{\Omega}=\int_{S^1}A(x_i,l_j)\frac{dx_1\wedge dx_2\wedge dx_3\wedge\Delta_{\P^1}}{Q}\,,
\end{equation}
where $\Delta_{\P^1}=l_1dl_2-l_2dl_1$ is the measure on $\P^1$ obtained from \eqref{eq:MeasurePn} and 
$S^1_Q$ denotes a loop in $\mathcal{W}$ centered around $Q=0$. The function $A(x_i,l_j)$ of the coordinates 
$x_i$, $l_j$ is fixed by its scaling behavior w.r.t.~the $\C^*$-action $(l_1,l_2)\mapsto \lambda(l_1,l_2)$. 
Since $Q\mapsto \lambda Q$ and $\Delta_{\P^1}\mapsto\lambda^2\Delta_{\P^1}$ we demand 
$A(x_i,\lambda l_j)=\lambda^{-1}A(x_i,l_j)$, i.e.~it is a section of $\mathcal{O}(E)$. Thus we make the ansatz 
\begin{equation}
 	A(x_i,l_j)=a_1\frac{h_1}{l_1}+a_2\frac{h_2}{l_2}=-a_1\frac{Q}{l_1l_2}+(a_2+a_1)\frac{h_2}{l_2}=a_2\frac{Q}{l_1l_2}+(a_1+a_2)\frac{h_1}{l_1}\,,
\end{equation}
which we insert in $\hat{\Omega}$ of \eqref{eq:OmegaAnsatz} to obtain
\begin{equation} \label{eq:HatOmegaLocalapp}
 	\hat{\Omega}=(a_1+a_2)\int_{S^1_Q}\frac{h_i}{l_i}\frac{dx_1\wedge dx_2\wedge dx_3\wedge\Delta_{\P^1}}{Q}\,,\quad i=1,2\,.
\end{equation}
Now we show that this precisely reproduces the local expressions \eqref{eq:OmegaU1}, \eqref{eq:OmegaU2}, 
this way fixing the free parameters $a_i$.

Let us perform the calculations in the local patch $U_1$. Then the measure on $\P^1$ reduces to 
$\Delta_{\P^1}=(l_1)^2dz_2^{(1)}$ with $z^{(1)}_2=l_2/l_1$ and we obtain, after a change of coordinates to $y_i$, 
\begin{eqnarray}
 	\hat{\Omega}&=&(a_1+a_2)\int_{S^1_Q}\det{J}^{-1} \frac{y_il_1}{l_i}\frac{dy_1\wedge dy_2\wedge dy_3\wedge dz^{(1)}_2}{-z^{(1)}_2 y_1+y_2}\\
		    &=&-(a_1+a_2) (\det{J}^{-1}\frac{y_il_1}{l_i}dy_1\wedge dy_3\wedge dz^{(1)}_2)\vert_{y_2=z^{(1)}_2y_1}\,.
\end{eqnarray}
In the last line we indicated that the residue localizes on the locus $Q=0$. This implies that 
$z^{(1)}_2=\frac{y_2}{y_1}=\frac{l_2}{l_1}$ as before in \eqref{eq:coordsU1} and we put $z^{(1)}_1=y_1$, 
$z^{(1)}_3=y_3$ as well. Next, we evaluate $\hat{\Omega}$ for $i=1$ for which the prefactor reduces to 
$\frac{y_1l_1}{l_1}=z^{(1)}_1$, and for $i=2$, for which we get $\frac{y_2l_1}{l_2}=z^{(1)}_1$. Thus, 
the two expressions in \eqref{eq:HatOmegaLocalapp} for $i=1,2$ yield one unique form $\hat{\Omega}$ 
after evaluating the residue integral,
\begin{equation}
 	\hat{\Omega}=(a_1+a_2)\det J^{-1}z^{(1)}_1dz_1^{(1)}\wedge dz_2^{(1)}\wedge dz_3^{(1)}\,,
\end{equation}
which agrees with $\pi^*(\Omega)$ on $U_1$ for $a_1+a_2=1$. Thus, we propose the global residue expression
\begin{equation} \label{eq:RedidueHatC3}
 	\hat{\Omega}\equiv\pi^*(\Omega)=\int_{S^1_Q}\frac{h_1}{l_1}\frac{\Omega\wedge\Delta_{\P^1}}{Q}=\int_{S^1_Q}\frac{h_2}{l_2}\frac{\Omega\wedge\Delta_{\P^1}}{Q}
\end{equation}
The transition from the local chart $U\cong\C^3$ on $Z_3$ to the global threefold $Z_3$ is then 
effectively performed by replacing the  three-form $\Omega=dx_1\wedge dx_2\wedge dx_3$ on $\C^3$ by 
the residue integral $\Omega=\text{Res}_P(\frac{\Delta_{\P_{\Delta}}}{P})$ of \eqref{eq:residueZ3} 
on $Z_3$ in \eqref{eq:RedidueHatC3}. This way the local analysis motivates and proves the global 
expression of \eqref{eq:ResZhat} for the pullback form $\hat{\Omega}\equiv\pi^*(\Omega)$ of the 
holomorphic three-form $\Omega$ to $\hat{Z}_3$ and its properties mentioned there.

We conclude this analysis by briefly checking that this residue also reproduces $\pi^*(\Omega)$ on 
$U_2$ as evaluated in \eqref{eq:OmegaU2}. We set $z^{(2)}_1=\frac{l_1}{l_2}$ for which the measure 
reduces as $\Delta_{\P^1}=-l_2^2dz_1^{(2)}$ which we readily insert into \eqref{eq:RedidueHatC3} to 
obtain
\begin{eqnarray}
 	\hat{\Omega}&=&\int_{S^1_Q}\det{J}^{-1}\frac{y_il_2}{l_i}\frac{dy_1\wedge dy_2\wedge dy_3\wedge dz_1^{(2)}}{y_1-z_1^{(2)}y_2}\\
	&=& (\det{J}^{-1}\frac{y_il_2}{l_i}dy_2\wedge dy_3\wedge dz_1^{(2)})\vert_{y_1=z^{(2)}_1y_2}=\det{J}^{-1}z^{(2)}_2  dz_1^{(2)}\wedge dz^{(2)}_2\wedge dz^{(2)}_3\,. 
\end{eqnarray}
Here we again use $z^{(2)}_1=\frac{y_1}{y_2}=\frac{l_1}{l_2}$ on $Q=0$ and introduce the local 
$z^{(2)}_2=y_2$, $z^{(2)}_3=y_3$ as in \eqref{eq:coordsU2}. The result is in perfect agreement 
with the local expression \eqref{eq:OmegaU2} on $U_2$.

\section{Topology of the Blow-Up Threefold $\hat{Z}_3$}
\label{App:topoHatZ_3}

In this appendix we summarize the basic topological data of the blow-up threefold $\hat{Z}_3$ 
of the Calabi-Yau threefold $Z_3$ along the curve $\Sigma$. For more details we refer to \cite{Grimm:2008dq}
as well as a standard reference \cite{Griffiths:1978yf}. See also \cite{Rares}.

We start with the cohomology of $E$. The cohomology ring of $E$ is generated by
$\eta=c_1(T)$, the first Chern-class of the tautological bundle\footnote{The tautological 
bundle can be defined on any projectivization of a vector bundle, like $E=\mathbb{P}(N_{Z_3}\Sigma)$, 
by the defining property that $T$ restricted to each fiber agrees with the universal 
bundle $\mathcal{O}(1)$ on projective space \cite{Griffiths:1978yf}.} on $E$, as 
an $H^\bullet(\Sigma)$-algebra, i.e.~$H^\bullet(E)=H^\bullet(\Sigma)\langle \eta\rangle$.
Thus, for a single smooth curve $\Sigma$ of genus $g$ the non-vanishing Hodge numbers 
of $E$ are determined to be
\begin{equation}
       h^{(0,0)}= h^{(2,2)}=1\ , \qquad h^{(1,0)}=g\ ,\qquad h^{(1,1)}=2
       \label{eqn:hodge-exceptional-divisor}
\end{equation}
as usual for a ruled surface $E$ over a genus $g$ curve. 
This is due to the fact the $\mathbb{P}^1$-fibration of $E$ does not degenerate and thus 
the Hodge numbers of $E$ are equal to those of $\mathbb{P}^1\times \Sigma$. Thus the elements in
$H^{(1,0)}(E)$ are precisely the holomorphic Wilson lines $a_i$ of $\Sigma$ and the three-forms in 
$H^{(2,1)}(E)$ take the form $a_i\wedge \eta$. 
Since, by construction, the normal bundle to $E$ in $\hat Z_3$ is the tautological bundle $T$
the class $\eta$, being one of the two classes in $H^{(1,1)}(E)$, is induced from the
ambient space $\hat Z_3$ and given by $\eta=c_1(N_{\hat Z_3}E)$. The second
element spanning $H^{(1,1)}(E)$ is given by the Poincar\'e dual $[\Sigma]_E$ of the
curve $\Sigma$ in $E$, $[\Sigma]_E =c_1(N_E \Sigma)$. It is related to the first Chern
class $c_1(\Sigma)$ and thus to the genus as
\begin{equation}
	c_1(N_E\Sigma)=-c_1(\Sigma)-2\eta\,,
\end{equation}
by using the adjunction formula in $\hat{Z}_3$. Note that as an exceptional divisor, 
$E$ is rigid in $\hat Z_3$. This is a crucial feature of the blow-up procedure, that 
guarantees that no \textit{new} degrees of freedom associated to deformations of $E$ are introduced.
By construction the normal bundle to $E$ in $\hat Z_3$ 
is $T$, which is a negative bundle on $E$. Thus we have
\begin{equation} \label{eq:Eisolated}
	H^0(E,N_{\hat{Z}_3}E)=\emptyset\,.
\end{equation}

Next we note that the cohomology of the blow-up $\hat{Z}_3$ takes the form
\begin{equation}
	H^\bullet(\hat{Z}_3)=\pi^*H^\bullet(Z_3)\oplus H^\bullet(E)/\pi^* H^\bullet(\Sigma)\,,
\label{eq:cohomZhat}
\end{equation}
where $\pi:\,Z_3\rightarrow Z_3$ denotes the blow-down map.
This in particular implies $H^{(3,0)}(\hat{Z}_3)=\pi^*H^{(3,0)}(Z_3)$ since $H^{(3,0)}(\Sigma)=
H^{(3,0)}(E)$ for dimensional reasons. Thus, $h^{(3,0)}(\hat{Z}_3)=1$ as in the Calabi-Yau case.
Then, the intersection ring on $\hat{Z}_3$ has the following 
relations on the level of intersection curves 
\begin{equation}
   E^2 = - \pi^* \Sigma  - \chi(\Sigma) F\ , \qquad E\cdot \pi^* D = (\Sigma \cdot D) F\ ,
\end{equation}
where $D$ is any divisor in $Z_3$, $F$ is the class of the $\P^1$-fiber of $E$, 
and $\chi(\Sigma)$ is the Euler number of the blow-up curve $\Sigma$.
The intersection numbers read
\begin{equation}
  E^3 = \chi(\Sigma)\ , \quad  F  \cdot E= -1\ , \quad F \cdot \pi^* D = 0 \ , \quad 
   E \cdot \pi^* \tilde \Sigma  = F \cdot \pi^* \tilde \Sigma = 0\ ,
\end{equation}
where $\tilde \Sigma$ is any curve in $Z_3$.
We conclude with the first and second Chern class of $\hat{Z}_3$, that are in general affected by the blow-up as
\begin{eqnarray}
	c_1(\hat Z_3) &=&\pi^*(c_1(Z_3))-c_1(N_{\hat Z_3}E)\ ,\\
         c_2 (\hat Z_3) &=& \pi^*(c_2(Z_3) + [\Sigma]) - \pi^*(c_1(Z_3)) E\ .
\label{eqn:c1-ztilde}
\end{eqnarray}
Clearly, if $Z_3$ is a Calabi-Yau manifold one can use $c_1(Z_3)=0$ to find
\begin{equation}
   c_1 (\hat Z_3) = - \eta \ , \qquad   c_2 (\hat Z_3) = \pi^*(c_2(Z_3) + [\Sigma]) \ ,
\end{equation}
in particular that $\hat{Z}_3$ is no more Calabi-Yau.
This implies that the lift of the holomorphic three-form $\Omega$ vanishes as
\begin{equation}
	\hat{\Omega}=\pi^*(\Omega)\,,\quad \hat{\Omega}|_E=0\,.
\label{eq:proptr}
\end{equation}

\chapter{Details, Examples and Tables}
\label{app:PartDetailsExamplesTables}

\section{Topological Data of the Main Example} 
\label{app:FurtherP2}

Here we supply the topological data of the fourfold $\tilde{X}_4$ that was omitted 
in the main text for convenience. Besides the intersection rings we will also present 
the full Picard-Fuchs system at the large radius/large complex structure point. These 
determine as explained in section \ref{sec:FFMirrors} the primary vertical subspace 
$H^{(p,p)}_{V}(\tilde X_4)$ of the A-model. 

As was mentioned before there are four triangulations whereas only three yield 
non-singular varieties. Again we restrict our exposition to the two triangulations 
mentioned in section \ref{sec:ffConstruction}. For the following we label the points in the 
polyhedron $\Delta_5^{\tilde X}$ given in \eqref{eq:vl-for-p2} consecutively by 
$\nu_i$, $i=0,\ldots,9$ and associated coordinates $x_i$ to each $\nu_i$. Then the 
toric divisors are given by $D_i:=\{x_i=0\}$. 

\textbf{Phase I:} In phase I of the toric variety defined by the polyhedron 
$\Delta_5^{\tilde X}$ in \eqref{eq:vl-for-p2} one has the following Stanley-Reisner 
ideal 
\begin{equation}
	SR=\{D_3 D_8, D_7 D_9, D_8 D_9, D_1 D_5 D_6,
        D_2 D_3 D_4, D_2 D_4 D_7\}\ .
\end{equation}
From this we compute by standard methods of toric geometry the intersection numbers 
\begin{align} \label{IntX_TextEx}
	\nn\mcal C_0&= 
                 J_4(J_1^2 J_2 + J_1 J_3 J_2 + J_3^2 J_2 + 3 J_1 J_2^2 + 3 J_3 J_2^2+9 J_2^3) + J_1^2 J_3 J_2  
                 + J_1 J_3^2 J_2 +  J_3^3 J_2 \\
 \nn&\quad +2 J_1^2 J_2^2  + 4 J_1 J_3 J_2^2  + 4 J_3^2 J_2^2 + 11 J_1 J_2^3 + 15 J_3 J_2^3 + 46 J_2^4,\\
 \mcal C_2&= 24J_1^2+36 J_1 J_4 + 48 J_1 J_3 + 36 J_4 J_3 + 48J_3^2+ 12 8 J_1 J_2 +102 J_2 J_4 \\
    &\phantom{=}+172 J_2 J_3 + 530 J_2^2 \ , \nn \\
 \nn\mcal C_3&= -660 J_1-540J_4-900 J_3-2776 J_2\ .
\end{align}
Here we denoted generators of the K\"ahler cone of \eqref{eq:KCPhaseI} dual to the 
Mori cone by $J_i$ as before. The notation for the $\mathcal{C}_k$ is as follows. 
Denoting the dual two-forms to $J_i$  by $\omega_i$ the coefficients of the top 
intersection ring $\mcal C_0$ are the quartic intersection numbers 
$J_i\cap J_j\cap J_k\cap J_l=\int_{\tilde X_4}\omega_i\wedge\omega_j\wedge\omega_k\wedge \omega_l$, 
while the coefficients of $\mcal C_2$ and $\mcal C_3$ are $[c_2({\tilde X_4})]\cap J_i \cap J_j
= \int_{\tilde X_4} c_2\wedge \omega_i\wedge \omega_j$ and  $[c_3({\tilde X_4})]\cap J_i
= \int_{\tilde X_4} c_3\wedge\omega_i$ respectively. 

As reviewed in section \ref{sec:PFO+3foldMirrorSymmetry} in the threefold case, that readily applies
to the fourfold case as well, the Picard-Fuchs operators of the mirror 
fourfold $X_4$ at the large complex structure point are calculated by the methods 
described in \cite{Hosono:1993qy}. In the appropriate coordinates $z_i$ defined 
by \eqref{eq:algCoords} and evaluated in \eqref{eq:zI} we obtain the full Picard-Fuchs 
system on $\tilde X_4$ given by
\begin{eqnarray} \label{PFOFFP2T1}
	\nn\mcal D_1^I&=& -\theta_1^2 (\theta_1+\theta_4-\theta_3)\\
	\nn&\phantom{=}&-(-1+\theta_1-\theta_3) (-2+2 \theta_1+\theta_4+\theta_3-\theta_2) (-1+2 \theta_1+\theta_4+\theta_3-\theta_2) z_1\,,\\
	\mcal D_2^I&=& \theta_2 (-2 \theta_1-\theta_4-\theta_3+\theta_2)-12 (-5+6 \theta_2) (-1+6 \theta_2) z_2\,,\\
	\nn\mcal D_3^I&=& (\theta_1-\theta_3) (-\theta_4+\theta_3)-(1+\theta_1+\theta_4-\theta_3) (-1+2 \theta_1+\theta_4+\theta_3-\theta_2) z_3\,,\\
	\nn\mcal D_4^I&=& \theta_4 (\theta_1+\theta_4-\theta_3)-(-1+\theta_4-\theta_3) (-1+2 \theta_1+\theta_4+\theta_3-\theta_2) z_4\,.	
\end{eqnarray}
Now we calculate the ring $\mathcal{R}$ given by the orthogonal complement of the 
ideal of Picard-Fuchs operators defined in \eqref{eq:ring}. Using the isomorphism 
$\theta_i\mapsto J_i$ discussed in section \ref{sec:matching} we obtain the topological 
basis of $H^{(p,p)}_{V}(\tilde X_4)$ by identification with the graded ring $\mcal{R}^{(p)}$. 
Since $J_i$ form the trivial basis of $H^{(1,1)}(\tilde X_4)$ and $H^{(3,3)}(\tilde X_4)$ 
is fixed by duality to $H^{(1,1)}(\tilde X_4)$, the non-trivial part is the cohomology 
group $H^{(2,2)}_{V}(X_4)$. We calculate the ring $\mathcal{R}^{(2)}$ by choosing the basis
\begin{eqnarray}
\nn	& \cR_1^{(2)}=\theta_1^2,\quad \cR_2^{(2)}=\theta_4(\theta_1 +\theta_3),\quad \cR_3^{(2)}
=\theta_3( \theta_1+\theta_3),\quad \cR_4^{(2)}=\theta_2(\theta_1+2 \theta_2),\quad \\
	& \cR_5^{(2)}=\theta_2 (\theta_4+\theta_2),\quad \cR_6^{(2)}=\theta_2(\theta_3+\theta_2)\,.&
\end{eqnarray}
Then we can use the intersection ring $\mathcal{C}_0$ to determine the topological metric 
$\eta^{(2)}$ of \eqref{eq:amodeltopologicametric} given by
\begin{equation}
	\eta^{(2)}_I=
\left(
\begin{array}{cccccc}
 0 & 0 & 0 & 4 & 3 & 3 \\
 0 & 0 & 0 & 14 & 6 & 8 \\
 0 & 0 & 0 & 18 & 10 & 10 \\
 4 & 14 & 18 & 230 & 124 & 137 \\
 3 & 6 & 10 & 124 & 64 & 73 \\
 3 & 8 & 10 & 137 & 73 & 80
\end{array}
\right)\,.
\end{equation}
The entries are just the values of the integrals
\begin{equation}
 	\cR^{(2)}_\alpha \cR^{(2)}_\beta
 	=\int_{\tilde{X}_4}(\cR^{(2)}_\alpha \cR^{(2)}_\beta)|_{\theta_i\mapsto J_i}\,,
\end{equation}
where we think of it in terms of the Poincar\'e duals and the quartic intersections are 
given as the coefficients of monomials in $\mathcal{C}_0$. The basis $\cR^{(3)}_i$ at 
grade $p=3$ is determined by requiring $\eta^{(3)}_{ab}=\delta_{a,h^{(1,1)}-b+1}$ where 
$h^{(1,1)}=4$ for the case at hand. Then the basis reads 
\begin{eqnarray}
& \cR_1^{(3)}=\theta_1(-\theta_1 \theta_4-\theta_2 \theta_4+\theta_2 \theta_3),\quad \cR_2^{(3)}
=\theta_1(-\theta_1 \theta_4+\theta_1 \theta_2+\theta_2 \theta_4- \theta_2 \theta_3),&\nn \\ 
& \cR_3^{(3)}=\theta_1^2 \theta_4,\quad \cR_4^{(3)}=\theta_1(-2 \theta_1 \theta_4-\theta_1 \theta_2+\theta_2 \theta_3)\,.&
\end{eqnarray}
Finally, we choose a basis of $\cR^{(4)}$ by 
$\cR^{(4)}=\tfrac{1}{103}\mathcal C_0|_{J_i\mapsto \theta_i}$ such that $\eta^{(4)}_{a_0,b_0}=1$ 
for $\cR^{(0)}=1$.

\textbf{Phase II:} Turning to the phase II of \eqref{eq:vl-for-p2} the Stanley-Reisner 
ideal and the intersection numbers read
\begin{eqnarray}
	\nn SR&=& \{D_1 D_7,D_7 D_9,D_8 D_9, D_1 D_5 D_6, D_2 D_3 D_4, D_2  D_4 D_7, D_3 D_5 D_6 D_8\},\\
  \nn\mathcal C_0&=& J_1^2 J_4 J_3+2 J_1^2 J_3^2+3 J_1 J_4 J_3^2+12 J_1 J_3^3+9 J_4 J_3^3+54 J_3^4+J_1^2 J_2 J_4+2 J_1^2 J_3 J_2\\
\nn&\phantom=&+3 J_1 J_2 J_3 J_4+12 J_1 J_3^2 J_2 +9 J_2 J_3^2 J_4+54 J_3^3 J_2+2 J_1^2 J_2^2+3 J_1 J_4 J_2^2+12 J_1 J_3 J_2^2\\
 &\phantom=&+9 J_4 J_3 J_2^2+54 J_3^2 J_2^2+11 J_1 J_2^3+9 J_4 J_2^3+51 J_3 J_2^3+46 J_2^4\,, \nn \\
\nn\mathcal C_2 &=&24 J_1^2+36 J_1 J_4+138 J_1 J_3+102 J_4 J_3+618 J_3^2+128 J_1 J_2 +102 J_2 J_4\\
 &&+588 J_3 J_4+530 J_4^2\,,\\
\nn\mathcal C_3&=& 660J_1- 540 J_4-3078 J_3-2776 J_2\,,
\end{eqnarray}
where the K\"ahler cone generators were given in \eqref{eq:KCPhaseII}.

The complete Picard-Fuchs system consists of four operators given by
\begin{eqnarray} \label{PFOFFP2T2}
 	\nn\mcal D_1^{II}&=&-\theta _1^2 \left(\theta _1+\theta _2-\theta _3\right)\\
	&\nn\phantom{=}&- \left(-3+3 \theta _1-\theta _3+2 \theta _4\right) \left(-2+3 \theta _1-\theta _3+2 \theta _4\right) \left(-1+3 \theta _1-\theta _3+2 \theta _4\right)z_1\,,\\
	\nn\mcal D_2^{II}&=&-\theta _2 \left(\theta _1+\theta _2-\theta _3\right) \left(\theta _2-\theta _3+\theta _4\right)-12 \left(-5+6 \theta _2\right) \left(-1+6 \theta _2\right) \left(-1+\theta _2-\theta _3\right)z_2\,, \\
	\mcal D_3^{II}&=&-\left(\theta _2-\theta _3\right) \left(-3 \theta _1+\theta _3-2 \theta _4\right)-\left(1+\theta _1+\theta _2-\theta _3\right) \left(1+\theta _2-\theta _3+\theta _4\right)z_3\,,\\
	\nn\mcal D_4^{II}&=&
\theta _4 \left(\theta _2-\theta _3+\theta _4\right)-\left(-2+3 \theta _1-\theta _3+2 \theta _4\right) \left(-1+3 \theta _1-\theta _3+2 \theta _4\right)z_4 \,.
\end{eqnarray}
This enables us to calculate $H^{(p,p)}_{V}(\tilde X_4)$ as before. 
The basis at grade $p=2$ reads
\begin{eqnarray}
\nn	& \cR_1^{(2)}=\theta_1^2,\quad \cR_2^{(2)}=\theta_2(2 \theta_1+6 \theta_3),\quad \cR_3^{(2)}=\theta_3(\theta_1+3 \theta_3),\quad \cR_4^{(2)}=\theta_1 \theta_4,\quad \\
	& \cR_5^{(2)}=\theta_2^2,\quad \cR_6^{(2)}=\theta_3(2 \theta_2 +2 \theta_3+\theta_4)+\theta_2 \theta_4\,,&
\end{eqnarray}
for which the topological metric $\eta^{(2)}$ is given by
\begin{equation}
 	\eta^{(2)}=\left(
\begin{array}{cccccc}
 0 & 12 & 6 & 0 & 2 & 10 \\
 12 & 2240 & 1120 & 20 & 328 & 1512 \\
 6 & 1120 & 560 & 10 & 174 & 756 \\
 0 & 20 & 10 & 0 & 3 & 12 \\
 2 & 328 & 174 & 3 & 46 & 228 \\
 10 & 1512 & 756 & 12 & 228 & 1008
\end{array}
\right)\,.
\end{equation}
Again the basis of $H^{(3,3)}(\tilde X_4)$ is fixed by  
$\eta^{(3)}_{ab}=\delta_{a,h^{(1,1)}-b+1}$ to be
\begin{eqnarray}
& \cR_1^{(3)}=-\tfrac{1}{91} \left(182 \theta_1^2+25 \theta_2^2+\theta_1 (-225 \theta_2+85 \theta_3)\right) (\theta_1+\theta_2+\theta_3+\theta_4)\,,&\nn \\ 
& \cR_2^{(3)}=\tfrac{1}{91} \left(91 \theta_1^2+10 \theta_2^2+\theta_1 (\theta_2-57 \theta_3)\right) (\theta_1+\theta_2+\theta_3+\theta_4)\,, &\\\
&\cR_3^{(3)}=-\theta_1 (\theta_2-\theta_3) (\theta_1+\theta_2+\theta_3+\theta_4)\,,\nn&\\
& \nn \cR_4^{(3)}=-\tfrac{1}{91} \left(273 \theta_1^2+23 \theta_2^2+\theta_1 (-207 \theta_2+60 \theta_3)\right) (\theta_1+\theta_2+\theta_3+\theta_4)\,.&
\end{eqnarray}
We conclude with the basis of $H^{(4,4)}(\tilde X_4)$ fixed by 
$\cR^{(0)}=1$ as $\cR^{(4)}=\tfrac{1}{359}\mcal{C}_0|_{J_i\mapsto \theta_i}$.

\section{Further Examples of Fourfolds}
\label{app:furtherexamples}

Here we consider a broader class of Calabi-Yau fourfolds $(\tilde{X}_4,X_4)$ 
that are constructed as described in section \ref{sec:ffConstruction} by fibering 
Calabi-Yau threefolds $\tilde{Z}_3$ over $\mathds{P}^1$. The threefolds we 
consider here are itself elliptically fibered over the two-dimensional base 
of the Hirzebruch surfaces $\mathds{F}_n$ for $n=0,1$,
\begin{equation}
 \begin{array}{cccc} \mathds{F}_n  & \rightarrow & \tilde{Z}_3 & \\
         && \downarrow &\\ 
         && \mathds{P}^1 & 
   \end{array}\,
\end{equation} 
Therefore, we will distinguish the constructed mirror pairs $(\tilde{X}_4,X_4)$ 
by the two-dimensional base $\mathds{F}_n$ we used to construct the threefold $\tilde{Z}_3$. 

In the following we will present the toric data of the threefolds $\tilde{Z}_3$ 
and fourfolds $\tilde{X}_4$ including some of their topological quantities. 
Then we will determine the complete system of Picard-Fuchs differential 
operators at the large complex structure point of the mirror Calabi-Yau fourfold 
and calculate the holomorphic prepotential $F^0$. From this we extract the 
invariants $n^g_\beta$ which are integer in all considered cases. Furthermore 
we show that there exists a subsector for these invariants that reproduces the 
closed and open Gromov-Witten invariants of the local Calabi-Yau threefolds
obtained by a suitably decompactifying the elliptic fiber of the original compact 
threefold. This matching allows us to determine the four-form flux $G_4$ for the 
F-theory compactification on these fourfolds such that the F-theory flux superpotential 
\eqref{eq:fluxpotFourfoldExpanded} admits the split \eqref{eq:SuperpotLimit} into flux and brane 
superpotential of the Type IIB theory. 

\subsection{Fourfold with $\mathds{F}_0$} \label{FF0}

We start with an elliptically fibered Calabi-Yau threefold $\tilde{Z}_3$ with base 
given by the toric Fano basis of the zeroth Hirzebruch surface 
$\mathds{F}_0=\mathds{P}^1\times \mathds{P}^1$. Its polyhedron and charge vectors read
\begin{equation}\label{3foldellf0}
	\begin{pmatrix}[c|cccc|ccc]
	    	&   &  \Delta_4^{\tilde Z} &   &   	&  \ell^{(1)} & \ell^{(2)}& \ell^{(3)}\\ \hline
		v_0 & 0 & 0 & 0 & 0 	& -6   & 0 & 0  \\
		v^b_1 & 0 & 0 & 2 & 3 	&  1   &-2 &-2  \\
		v^b_2 & 1 & 0 & 2 & 3 	&  0   & 1 & 0  \\
		v^b_3 &-1 & 0 & 2 & 3 	&  0   & 1 & 0  \\
		v^b_4 & 0 & 1 & 2 &  3 	&  0   & 0 & 1  \\
		v^b_5 & 0 &-1 & 2 & 3 	&  0   & 0 & 1  \\
	        v^1   & 0 & 0 &-1 & 0 	&  2   & 0 & 0  \\
		v_2   & 0 & 0 & 0 &-1 	&  3   & 0 & 0
	\end{pmatrix}\,,
\end{equation}
where points in the base are again labeled by a superscript $^b$. There is one 
triangulation for which the Stanley-Reisner ideal in terms of the toric divisors 
$D_i=\{x_i=0\}$ takes the form
\begin{equation} 
 	SR=\{D_2D_3,D_4 D_5,D_1 D_6D_7\}.
\end{equation}
This threefold $\tilde{Z}_3$ has Euler number $\chi=-480$, $h_{1,1}=3$ and $h_{2,1}=243$, 
where the three K\"ahler classes correspond to the elliptic fiber and the two 
$\mathds{P}^1$'s of the base $\mathds{F}_0$. The intersection ring for this Calabi-phase in 
terms of the K\"ahler cone generators
\begin{equation}
 	J_1=D_1+2D_2+2D_4,\quad J_2=D_2,\quad J_3=D_4
\end{equation}
reads $\mathcal{C}_0=8J_1^3+2J_1^2J_3+2J_1^2J_2+J_1J_2J_3$ and $\mathcal{C}_2=92J_1+24J_2+24J_3$.

In the local limit $K_{\mathds{F}_0}\rightarrow \mathds{F}_0$ Harvey-Lawson type branes described 
by the brane charge vectors $\hat\ell^{(1)}=(-1,0,1,0,0)$ and $\hat\ell^{(1)}=(-1,0,0,1,0)$ 
were studied in \cite{Aganagic:2001nx}. To construct the Calabi-Yau fourfold $\tilde{X}_4$ 
we use the construction of section \ref{sec:ffConstruction} with the brane vector $\hat\ell^{(1)}$ 
and expand $\Delta_4^{\tilde Z}$ to the polyhedron $\Delta_5^{\tilde X}$ and determine the Mori 
cone generators $\ell^{(i)}$ with $i=1,\ldots5$ for the four different triangulations of the 
corresponding Calabi-Yau phases. Here we display one of the four triangulations on which we 
focus our following analysis:

\begin{equation}
	\begin{pmatrix}[c|ccccc|ccccc]
	    	&   &   &\Delta_5^{\tilde X}  &   &   & \ell^{(1)} & \ell^{(2)} & \ell^{(3)} & \ell^{(4)} & \ell^{(5)} \\ \hline
		v_0 & 0 & 0 & 0 & 0 & 0          &-6 & 0   & 0 & 0 & 0  \\
		v_1 & 0 & 0 & 2 & 3 & 0          & 1 &-1   &-2 &-1 &-1  \\
		v_2 & 1 & 0 & 2 & 3 & 0          & 0 & 1   & 0 & 0 & 0  \\
		v_3 &-1 & 0 & 2 & 3 & 0          & 0 & 0   & 0 & 1 &-1  \\
		v_4 & 0 & 1 & 2 & 3 & 0          & 0 & 0   & 1 & 0 & 0  \\
		v_5 & 0 &-1 & 2 & 3 & 0          & 0 & 0   & 1 & 0 & 0  \\
		v_6 & 0 & 0 &-1 & 0 & 0          & 2 & 0   & 0 & 0 & 0  \\
		v_7 & 0 & 0 & 0 &-1 & 0          & 3 & 0   & 0 & 0 & 0  \\
		v_8 &-1 & 0 & 2 & 3 &-1          & 0 & 1   & 0 &-1 & 1  \\
		v_9 & 0 & 0 & 2 & 3 &-1          & 0 &-1   & 0 & 1 & 0  \\
		v_{10} & 0 & 0 & 2 & 3&1         & 0 & 0   & 0 &0 & 1  
	\end{pmatrix}\,.
	\label{eqn:vl-for-f0}
\end{equation}
In this triangulation the Stanley-Reisner ideal takes the form
\begin{equation}
	SR=\{\Div_2 \Div_3,~ \Div_2 \Div_8,~ \Div_3 \Div_9,~ \Div_4 \Div_5,~ \Div_8 \Div_{10},~ \Div_9 \Div_{10},~ \Div_1 \Div_6 \Div_7\}.
\end{equation}
The generators of the K\"ahler cone of the fourfold $\tilde{X}_4$ in the given triangulation are 
\begin{equation}
	\Jiv_1 = \Div_1+2\Div_{10}+\Div_2+\Div_3+2\Div_4,\quad \Jiv_2 = \Div_{10},\quad \Jiv_3 = \Div_4,\quad \Jiv_4 = \Div_{10}+\Div_3,\quad \Jiv_5 = \Div_2\,,
\end{equation}
for which the intersections are determined to be
\begin{align} \label{intsFFF0}
	\nonumber \mcal C_0&=42 J_1^4+8 J_1^3 J_2+7 J_1^3 J_3+2 J_1^2 J_2 J_3+12 J_1^3 J_4+2 J_1^2
J_2 J_4+3 J_1^2 J_3 J_4+J_1 J_2 J_3 J_4\nn\\
	&\quad +2 J_1^2 J_4^2+J_1 J_3 J_4^2+8
J_1^3 J_5+2 J_1^2 J_2 J_5+2 J_1^2 J_3 J_5+J_1 J_2 J_3 J_5+2 J_1^2 J_4
J_5+J_1 J_3 J_4 J_5,\\
 	\nonumber \mcal C_2&=92\Jiv_1\Jiv_2+486\Jiv_1^2+24\Jiv_2\Jiv_3+82\Jiv_1\Jiv_3+24\Jiv_3\Jiv_5+92\Jiv_1\Jiv_5+24\Jiv_2\Jiv_5\\
	&\quad +24\Jiv_2\Jiv_4+138\Jiv_1\Jiv_4+36\Jiv_3\Jiv_4+24\Jiv_4\Jiv_5+24\Jiv_4^2,\\
	\nonumber \mcal C_3&=-2534\Jiv_1-480\Jiv_2-420\Jiv_3-720\Jiv_4-480\Jiv_5.
\end{align}
We calculate the core topological quantities to be
\begin{equation}
 	\chi=15408\ ,\quad h_{3,1}=2555\ , \quad h_{2,1}=0\ ,\quad h_{1,1}=5\ .
\end{equation}
 
Furthermore, we note that these intersections reveal the fibration structure of $\tilde{X}_4$. 
We recognize the Euler number of the threefold $\tilde{Z}_3$ as the coefficient of $J_2$ and 
$J_5$ in $\mathcal{C}_3$ and the fact that both $J_2$ and $J_5$ appear at most linear in 
$\mathcal{C}_0$, $\mathcal{C}_2$. This is consistent with the fact that the fiber $F$ of a 
fibration has intersection number $0$ with itself which implies $c_3(F)=c_3(\tilde{X}_4)$ 
using the adjunction formula as well as $c_1(F)+c_1(N_{\tilde{X}_4} F)=c_1(N_{\tilde{X}_4}F)=0$ 
for $\tilde{X}_4$ Calabi-Yau. Thus we observe a fibration of $\tilde{Z}_3$ represented by the 
classes $J_2$ and $J_5$ over the base curves corresponding to $\ell^{(2)}$, $\ell^{(5)}$, 
respectively.   

The Picard-Fuchs operators are determined as before and read
\begin{eqnarray}
	\nn\mcal D_1 &=& \theta _1 \left(\theta _1-\theta _2-2 \theta _3-\theta _4-\theta _5\right)-12  \left(-5+6 \theta _1\right) \left(-1+6 \theta _1\right)z_1,\\
	\nn\mcal D_2 &=& \theta _2 \left(\theta _2-\theta _4+\theta _5\right)-\left(-1+\theta _2-\theta _4\right) \left(-1-\theta _1+\theta _2+2 \theta _3+\theta _4+\theta _5\right)z_2 ,\\
	\mcal D_3 &=& \theta _3^2-\left(1+\theta _1-\theta _2-2 \theta _3-\theta _4-\theta _5\right) \left(2+\theta _1-\theta _2-2 \theta _3-\theta _4-\theta _5\right)z_3 ,\\
	\nn\mcal D_4 &=& \left(\theta _2-\theta _4\right) \left(\theta _4-\theta _5\right)- \left(1+\theta _2-\theta _4+\theta _5\right) \left(-1-\theta _1+\theta _2+2 \theta _3+\theta _4+\theta _5\right)z_4,\\
	\nn\mcal D_5 &=& \theta _5 \left(\theta _2-\theta _4+\theta _5\right)- \left(1+\theta _1-\theta _2-2 \theta _3-\theta _4-\theta _5\right) \left(1+\theta _4-\theta _5\right)z_5\ .
\end{eqnarray}
Then we can proceed with fixing the basis of $H^{(p,p)}_{V}(\tilde X_4)$ at each grade $p$ 
by determining the ring $\mathcal{R}$ of \eqref{eq:ring}.
We choose a basis at grade $p=2$ as
\begin{eqnarray}
\nn &\cR_1^{(2)}=\theta_1 \left(\theta_1+\theta_5\right),\quad \cR_2^{(2)}=\theta_1 \left(\theta_1+\theta_2\right),\quad \cR_3^{(2)}=\theta_1 \left(2 \theta_1+\theta_3\right),\quad \cR_4^{(2)}=\theta_1 \left(\theta_1+\theta_4\right), &\\
& \cR_5^{(2)}=\theta_2 \theta_3,\quad \cR_6^{(2)}= \left(\theta_2+\theta_4\right) \left(\theta_4+\theta_5\right), \quad \cR_7^{(2)}=\theta_3 \theta_4, \quad \cR_8^{(2)}=\theta_3 \theta_5\,.&
\end{eqnarray}
The basis of solution dual to this basis choice is given by
\begin{eqnarray} \label{FFF0dual}
 	\nn&\mathbb{L}^{(2)}_1=\tfrac{1}{8} l_1 \left(l_1-l_2-2 l_3-l_4+7 l_5\right)\ ,\quad \mathbb{L}^{(2)}_2=\tfrac{1}{8} l_1 \left(l_1+7 l_2-2 l_3-l_4-l_5\right)\ , &\\ \nn
&\mathbb{L}^{(2)}_3=\tfrac{1}{4} l_1 \left(l_1-l_2+2 l_3-l_4-l_5\right)\ ,\quad \mathbb{L}^{(2)}_4=\tfrac{1}{8} l_1 \left(l_1-l_2-2 l_3+7 l_4-l_5\right)\ ,\quad \mathbb{L}^{(2)}_5=l_2 l_3\ , &\\ &\mathbb{L}^{(2)}_6=\tfrac{1}{4} \left(l_2+l_4\right) \left(l_4+l_5\right)\ ,\quad
\mathbb{L}^{(2)}_7=l_3 l_4\ ,\quad \mathbb{L}^{(2)}_8=l_3 l_5\ .&
\end{eqnarray}
The topological two-point coupling between the $\cR^{(2)}_\alpha$ in the chosen basis reads
\begin{equation}
\eta^{(2)}=\left(
\begin{array}{cccccccc}
 58 & 60 & 109 & 64 & 3 & 8 & 4 & 2 \\
 60 & 58 & 109 & 64 & 2 & 8 & 4 & 3 \\
 109 & 109 & 196 & 118 & 4 & 20 & 6 & 4 \\
 64 & 64 & 118 & 68 & 3 & 8 & 4 & 3 \\
 3 & 2 & 4 & 3 & 0 & 0 & 0 & 0 \\
 8 & 8 & 20 & 8 & 0 & 0 & 0 & 0 \\
 4 & 4 & 6 & 4 & 0 & 0 & 0 & 0 \\
 2 & 3 & 4 & 3 & 0 & 0 & 0 & 0
\end{array}
\right).
\end{equation}
The basis of $\mathcal{R}^{(3)}$ determining $H^{(3,3)}(\tilde{X}_4)$ that is fixed by 
Poincar\'e duality to the K\"ahler cone generators satisfying 
$\eta^{(3)}_{ab}=\delta_{a,h^{(1,1)}-b+1}$ is chosen to be
\begin{eqnarray}
\nn	&\cR_1^{(3)}=\tfrac{1}{4}(9 \theta_1 \theta_5- 2\theta_1 \theta_3-\theta_3^2)\theta_3+\theta_2 \theta_3^2 -\theta_1 \theta_2 \theta_5,&\\
\nn&\quad \cR_2^{(3)}=\tfrac{1}{8}( \theta_1 \theta_3+2\theta_3^2-10 \theta_1  \theta_5)\theta_3-\theta_2\theta_3^2 -\theta_1 \theta_2 \theta_5,\quad \cR_3^{(3)}=\theta_1 (\tfrac{1}{2} \theta_3^2- \theta_3 \theta_5-2 \theta_2 \theta_5),&\\
& \cR_4^{(3)}=\theta_1 \theta_2 \theta_5,\quad \cR_5^{(3)}=\tfrac{1}{8} \theta_3(2\theta_3^2-3\theta_1 \theta_3-10 \theta_1  \theta_5-4 \theta_2\theta_3 )-\theta_1 \theta_2 \theta_5\ .&
\end{eqnarray}
We choose the basis of $H^{(4,4)}(\tilde{X}_4)$ such that the volume is normalized as 
$\eta^{(4)}_{a_0,b_0}=1$ for $\cR^{(0)}=1$, i.e.~$\cR^{(4)}=\tfrac{1}{96}\mcal C_0|_{J\mapsto \theta}$.

In order to fix the integral basis of $H^{(2,2)}_V(\tilde X_4)$ we again match the threefold 
periods from the fourfold periods via \eqref{eq:matchPrepot}. The first step is to identify the 
K\"ahler classes of $\tilde Z_3$. As discussed above $J_5$ represents the class of the 
Calabi-Yau fiber $\tilde Z_3$. The intersections of $\tilde Z_3$ are obtained from 
\eqref{intsFFF0} upon the identification 
\begin{equation} \label{FFF0match}
 	J_1\ \leftrightarrow \ J_1(\tilde Z_3)\quad J_2+J_4\ , \leftrightarrow \ J_2(\tilde Z_3)\ ,\quad J_3\ \leftrightarrow \ J_3(\tilde Z_3)\,.
\end{equation}
With this in mind we calculate the leading logarithms $\mathbb{L}_\alpha( Z_3)$ on the 
threefold given by
\begin{equation}
 	\mathbb{L}_1( Z_3)=\tfrac12X_0(2 \tilde{l}_1+\tilde{l}_2) (2 \tilde{l}_1+\tilde{l}_3)\ ,\quad \mathbb{L}_2( Z_3)=\tfrac12X_0\tilde{l}_1 (\tilde{l}_1+\tilde{l}_3)\ , \quad \mathbb{L}_3( Z_3)=\tfrac12X_0\tilde{l}_1 (\tilde{l}_1+\tilde{l}_2)\  .
\end{equation}
This together with the requirement of matching the instanton numbers\footnote{We note here 
that by just matching the threefold instantons the solution on the fourfold could not be 
fixed. The two free parameters could only be determined by matching the classical terms, too.} 
$n_{d_1,d_2,d_3}$ of $\tilde Z_3$ via $n_{d_1,d_2,d_3,d_2,0}$ on $\tilde X_4$ fixes unique 
solutions of the Picard-Fuchs system
\begin{equation}
 	\mathbb{L}^{(2)}_1=\tfrac12X_0(2 l_1+l_3) (2 l_1+l_2+l_4)\ ,\quad \mathbb{L}^{(2)}_6= \tfrac12X_0l_1 (l_1+l_3)\ ,\quad \mathbb{L}^{(2)}_8=\tfrac12X_0l_1 (l_1+l_2+l_4)\, ,
\end{equation}
that upon \eqref{FFF0match} coincide with the threefold solutions.
This fixes three ring elements $\tilde{\cR}^{(2)}_{\alpha}$, $\alpha=1,6,8$, by the map 
induced from \eqref{eq:intersecsLL} that we complete to a new basis 
\begin{eqnarray}
 	\nn&\tilde{\cR}^{(2)}_1=\tfrac{1}{8} \theta_1 \left(\theta_1-\theta_2-2 \theta_3-\theta_4+7 \theta_5\right)\ ,\quad \tilde{\cR}^{(2)}_2=\tfrac{1}{8} \theta_1 \left(\theta_1+7 \theta_2-2 \theta_3-\theta_4-\theta_5\right)\ , &\\ \nn
&\tilde{\cR}^{(2)}_3=\tfrac{1}{4} \theta_1 \left(\theta_1-\theta_2+2 \theta_3-\theta_4-\theta_5\right)\ ,\quad \tilde{\cR}^{(2)}_4=\tfrac{1}{8} \theta_1 \left(\theta_1-\theta_2-2 \theta_3+7 \theta_4-\theta_5\right)\ ,  &\\ &\tilde{\cR}^{(2)}_5=\theta_2 \theta_3\ ,\quad\tilde{\cR}^{(2)}_6=\tfrac{1}{4} \left(\theta_2+\theta_4\right) \left(\theta_4+\theta_5\right)\ ,\quad
\tilde{\cR}^{(2)}_7=\theta_3 \theta_4\ ,\quad \tilde{\cR}^{(2)}_8=\theta_3 \theta_5\ .&
\end{eqnarray}
Then the integral basis elements are given by
\begin{equation}
 	\hat{\gamma}^{(2)}_1=\tilde{\cR}^{(2)}_1\Omega_4|_{z=0}\ ,\quad \hat{\gamma}^{(2)}_6=\tilde{\cR}^{(2)}_6\Omega_4|_{z=0}\ ,\quad \hat{\gamma}^{(2)}_8=\tilde{\cR}^{(2)}_8\Omega_4|_{z=0}\, ,
\end{equation}
where again the new grade $p=2$ basis is obtained by replacing $l_i\leftrightarrow \theta_i$ 
in the dual solutions of \eqref{FFF0dual}. We conclude by presenting the leading logarithms 
of the periods $\Pi^{(2)\, \alpha}$ when integrating $\Omega_4$ over the duals $\gamma^{(2)\, \alpha}$ 
for $\alpha=1,6,8$. They are then as well given by $\mathbb{L}^{(2)\, 1}=X_0l_1 \left(l_1+l_5\right)$, 
$\mathbb{L}^{(2)\, 6}=X_0\left(l_2+l_4\right)\left(l_4+l_5\right)$ and $\mathbb{L}^{(2)\, 8}=X_0l_3l_5$.

Finally we determine a $\hat \gamma$ flux in $H^{(2,2)}_{H}(X_4)$ such that we match the disk 
invariants of \cite{Aganagic:2001nx} for both classes of the local geometry 
$K_{\mathds{F}_0}\rightarrow \mathds{F}_0$ with the brane class. Furthermore we reproduce the closed 
invariants of \cite{Haghighat:2008gw} for the two $\mathds{P}^1$-classes for zero brane 
winding $m=0$. First we identify in the polyhedron \eqref{eqn:vl-for-f0} the vector 
$\ell^{(4)}$ as corresponding to the brane vector. Then we expect to recover the disk invariants 
from the fourfold invariants $n_{0,d_1,d_2,d_1+m,0}$. Then the flux $\hat{\gamma}$ deduced this 
way still contains a freedom of three parameters and takes the form
\begin{eqnarray}
 	\hat{\gamma}=(-\cR^{(2)}_5+\tfrac14\cR^{(2)}_6+\cR^{(2)}_7+\tfrac12\cR^{(2)}_8)\Omega_4|_{z=0}
\end{eqnarray}
where we choose the free parameters $\underline{a}$ in front of $\cR^{(2)}_1$, $\cR^{(2)}_2$, $\cR^{(2)}_3$ 
and $\cR^{(2)}_4$ to be zero. Note that $a_7=1$ is fixed by the requirement of matching the disk 
invariants. For this parameter choice the leading logarithmic structures of the corresponding 
period $\int_\gamma\Omega_4$ and of the solution matching the invariants are respectively given by
\begin{equation}
 	\mathbb{L}^{(2)\, \gamma}=X_0\left(l_2+l_4\right) \left(l_4+l_5\right)\ ,\quad \mathbb{L}^{(2)}_{\gamma}=\tfrac{1}{2} X_0l_1 \left(4 l_1+3 l_2+2 l_3+l_4\right)\ .
\end{equation}

\subsection{Fourfold with $\mathds{F}_1$} \label{FF1}

Here we consider an elliptically fibered Calabi-Yau threefold $\tilde{Z}_3$ with base twofold 
given by $\mathds{F}_1=\mathds{P}(\mathcal{O}\oplus \mathcal{O}(1))$ which is the blow-up of $\mathds{P}^2$ 
at one point. The polyhedron and charge vectors read
\begin{equation}\label{3foldellf1}
	\begin{pmatrix}[c|cccc|ccc]
	    	&   &  \Delta_4^{\tilde Z} &   &   	&  \ell^{(1)} & \ell^{(2)}& \ell^{(3)}\\ \hline
		v_0 & 0 & 0 & 0 & 0 	&  0   &-6 & 0  \\
		v^b_1 & 0 & 0 & 2 & 3 	& -1   & 0 &-2  \\
		v^b_2 & 1 & 1 & 2 & 3 	&  1   & 0 & 0  \\
		v^b_3 &-1 & 0 & 2 & 3 	&  1   & 0 & 0  \\
		v^b_4 & 0 & 1 & 2 &  3 	& -1   & 0 & 1  \\
		v^b_5 & 0 &-1 & 2 & 3 	&  0   & 0 & 1  \\
	        v^1   & 0 & 0 &-1 & 0 	&  0   & 2 & 0  \\
		v_2   & 0 & 0 & 0 &-1 	&  0   & 3 & 0
	\end{pmatrix}.
\end{equation}
where the labels by a superscript $^b$ again denote points in the base. There are two Calabi-Yau 
phases and for the triangulation given above the Stanley-Reisner ideal reads
\begin{equation} 
 	SR=\{D_2 D_3,D_4 D_5,D_1 D_6 D_7\}.
\end{equation}
This threefold has Euler number $\chi=480$, $h_{1,1}=3$ and $h_{2,1}=243$, where the three K\"ahler 
classes correspond to the elliptic fiber and the two $\mathds{P}^1$'s of the base $\mathds{F}_1$. The 
intersection ring for this Calabi-phase in terms of the K\"ahler cone generators
\begin{equation}
 	J_1=D_2,\quad J_2=D_1+3D_2+2D_4,\quad J_3=D_2+D_4
\end{equation}
reads $\mathcal{C}_0=2J_1J_2^2+8J_2^3+J_1J_2J_3+3J_2^2J_3+J_2J_3^2$ and $\mathcal{C}_2=24J_1+92J_2+36J_3$.

For the second Calabi-Yau phase we have the following data:
\begin{eqnarray}
\nn &
	\left(
\begin{array}{l|rrrrrrrrrrr}
 \ell^{(1)} & -6 & 0 & 1 & 1 & -1 & 0 & 2 & 3 \\
 \ell^{(2)} & 0 & -3 & 1 & 1 & 0 & 1 & 0 & 0 \\
 \ell^{(3)} & 0 & 1 & -1 & -1 & 1 & 0 & 0 & 0
\end{array}
\right), &\\
&SR=\{D_1\cdot D_4,D_4\cdot D_5,D_1\cdot D_6\cdot D_7,D_2\cdot D_3\cdot D_5,D_2\cdot  D_3\cdot D_6\cdot D_7\},&\\
\nn & J_1=D_1+3D_2+2D+4,\quad J_2=D_2+D_4,\quad J_3=D_1+3D_2+3D_4, &\\
\nn & \mathcal C_0=8J_1^3+3J_1^2J_2+J_1J_2^2 + 9J_1^2J_3+3J_1J_2J_3+J_2^2J_3+9J_1J_3^2+3J_2J_3^2+9J_3^3, &\\
\nn & \mathcal C_2=92J_1+36J_2+102J_3. &
\end{eqnarray}

Harvey-Lawson type branes were considered in \cite{Aganagic:2001nx} for the brane charge vectors 
$\hat\ell^{(1)}=(-1,1,0,0,0)$ and $\hat\ell^{(1)}=(-1,0,0,1,0)$ for the non-compact model 
$K_{\mathds{F}_1}\rightarrow \mathds{F}_1$. The Calabi-Yau fourfold $\tilde{X}_4$ is constructed from the 
brane vector $\hat\ell^{(1)}$ for which there are eleven triangulations.  Again we restrict our 
attention to one triangulation with the following data
\begin{eqnarray} \label{ellFFF1}
 	&	\left(
\begin{array}{l|rrrrrrrrrrr}
 \ell^{(1)} & 0 & -1 & 0 & -1 & 0 & 0 & 0 & 0 & 1 & 0 & 1 \\
 \ell^{(2)} & 0 & -1 & 0 & 1 & 0 & 0 & 0 & 0 & -1 & 1 & 0 \\
 \ell^{(3)} & 0 & 0 & 1 & 0 & -1 & 0 & 0 & 0 & 1 & -1 & 0 \\
 \ell^{(4)} & 0 & -2 & 0 & 0 & 1 & 1 & 0 & 0 & 0 & 0 & 0 \\
 \ell^{(5)} & -6 & 1 & 0 & 0 & 0 & 0 & 2 & 3 & 0 & 0 & 0
\end{array}
\right), &\nn\\
&SR=\{D_2\cdot D_3,~ D_2\cdot D_8,~ D_3\cdot D_9,~ D_4\cdot D_5,~ D_8\cdot D_{10},~ D_9\cdot D_{10},~ D_1\cdot D_6\cdot D_7\}\ ,&\nn\\
&\Jiv_1 = \Div_2,~ \Jiv_2 = \Div_1+2\Div_{10}+\Div_2+\Div_3+2\Div_4,~ \Jiv_3 = \Div_4,~ \Jiv_4 = \Div_{10},~ \Jiv_5 = \Div_{10}+\Div_3&\nn
\end{eqnarray}
with intersections
\begin{align} \label{intsFFF1}
 	\mcal C_0&=J_1 J_2 J_4 J_5 + J_2^2 J_4 J_5 + J_1 J_3 J_4 J_5 + J_2 J_3 J_4 J_5 + J_1 J_4^2 J_5 + J_2 J_4^2 J_5 \\
\nn&\quad +2 J_1 J_2 J_5^2 + 2 J_2^2 J_5^2 + 2 J_1 J_3 J_5^2 +2 J_2 J_3 J_5^2 + 3 J_1 J_4 J_5^2 + 4 J_2 J_4 J_5^2  \\
\nn &\quad +2 J_3 J_4 J_5^2 + 2 J_4^2 J_5^2 + 8 J_1 J_5^3  +12 J_2 J_5^3 + 8 J_3 J_5^3 + 11 J_4 J_5^3 + 42 J_5^4,	\\
\nn \mcal C_2 &= 24J_1J_2+24J_2^2+24J_1J_3 +24J_2J_3+36J_1J_4+48J_2J_4+24J_3J_4+24J_4^2\\
\nn &\quad +92J_1J_5+138J_2J_5+92J_3J_5+128J_4J_5+486J_5^2, \\
\nn \mcal C_3 &= -480J_1-270J_2-480J_3-660J_4-2534J_5\ .
\end{align}
Furthermore, we determine 
$\chi=15408$, $h^{(3,1)}=2555$, $h^{(2,1)}=0$ and $h^{(1,1)}=5$.

Again the Euler number of  the threefold $\tilde{Z}_3$ appears in $\mcal C_3$ in front of $J_1$ 
and $J_3$ confirming the fibration structure. By comparing the coefficient polynomial of $J_1$, 
$J_3$ with the threefold intersection rings presented in appendix \ref{FF0}, \ref{FF1} we infer 
that $J_1$ is precisely $\tilde Z_3=\cE\rightarrow \mathds{F}_1$, whereas $J_3$ is 
$\tilde Z_3'=\cE\rightarrow \mathds{F}_0$. Since we discussed $\mathds{F}_0$ in detail before we will just 
concentrate on the fibration structure involving $\mathds{F}_1$. 

The Picard-Fuchs operators of $X_4$ read as
\begin{eqnarray}
	\nn\mcal D_1&=& \theta_1 (\theta_1-\theta_2+\theta_3)-(-1+\theta_1-\theta_2) (-1+\theta_1+\theta_2+2 \theta_4-\theta_5) z_1,\\
	\nn\mcal D_2&=& (\theta_1-\theta_2) (\theta_2-\theta_3)-(1+\theta_1-\theta_2+\theta_3) (-1+\theta_1+\theta_2+2 \theta_4-\theta_5) z_2,\\
	\mcal D_3&=& -\theta_3 (\theta_1-\theta_2+\theta_3)-(1+\theta_2-\theta_3) (-1+\theta_3-\theta_4) z_3,\\
	\nn\mcal D_4&=& \theta_4 (-\theta_3+\theta_4)-(-2+\theta_1+\theta_2+2 \theta_4-\theta_5) (-1+\theta_1+\theta_2+2 \theta_4-\theta_5) z_4,\\
	\nn\mcal D_5&=& \theta_5 (-\theta_1-\theta_2-2 \theta_4+\theta_5)-12 (-5+6 \theta_5) (-1+6 \theta_5) z_5\,,
\end{eqnarray}
from which we determine the basis of $\cR^{(2)}$ as
\begin{eqnarray}
	 \nn&\cR_1^{(2)}=\left(\theta_1+\theta_2\right)\left(\theta_2+\theta_3\right),\quad \cR_2^{(2)}
	 =\theta_1 \theta_4,\quad \cR_3^{(2)}=\theta _5 \left(\theta _1+\theta _5\right),\quad \cR_4^{(2)}
	 =\theta_2 \theta_4,&\\
	&\quad \cR_5^{(2)}=\theta _5 \left(\theta _2+\theta _5\right),\quad \cR_6^{(2)}=\theta _4 \left(\theta _3+\theta _4\right),
	\quad \cR_7^{(2)}=\theta_3 \theta_5,\quad \cR_8^{(2)}=\theta _5 \left(\theta _4+2 \theta _5\right)\ \ \ \ \ \ &
\end{eqnarray}
with the two-point coupling
\begin{equation}
	\eta^{(2)}=	\left(
\begin{array}{cccccccc}
 0 & 0 & 8 & 0 & 8 & 0 & 0 & 20 \\
 0 & 0 & 3 & 0 & 4 & 0 & 1 & 7 \\
 8 & 3 & 58 & 5 & 64 & 6 & 10 & 114 \\
 0 & 0 & 5 & 0 & 5 & 0 & 1 & 9 \\
 8 & 4 & 64 & 5 & 68 & 6 & 10 & 123 \\
 0 & 0 & 6 & 0 & 6 & 0 & 0 & 8 \\
 0 & 1 & 10 & 1 & 10 & 0 & 0 & 18 \\
 20 & 7 & 114 & 9 & 123 & 8 & 18 & 214
\end{array}
\right)\ .
\end{equation}
The dual basis of solutions reads
\begin{eqnarray}
\nn&\mathbb{L}^{(2)\,1}=\tfrac{1}{4} (l_1+l_2) (l_2+l_3)\ ,\quad\mathbb{L}^{(2)\,2}=l_1 l_4\ ,
\quad\mathbb{L}^{(2)\,3}=\tfrac{1}{7} l_5 (6 l_1-l_2-2 l_4+l_5)\ ,&\\\nn& \mathbb{L}^{(4)\,1}=l_2 l_4\ ,
\quad \mathbb{L}^{(2)\,5}=\tfrac{1}{7} l_5 (-l_1+6 l_2-2 l_4+l_5)\ ,\quad\mathbb{L}^{(2)\,6}=\tfrac{1}{2} l_4 (l_3+l_4)\ ,
\quad\mathbb{L}^{(2)\,7}=l_3 l_5\ ,&\\ &\mathbb{L}^{(2)\,8}=\tfrac{1}{7} l_5 (-2 l_1-2 l_2+3 l_4+2 l_5)\ .&
\end{eqnarray}
We determine $H^{(3,3)}(\tilde{X}_4)$ by duality to the canonical basis of $H^{(1,1)}(\tilde{X}_4)$ by the 
basis choice of $\cR^{(3)}$ given as
\begin{eqnarray}
 	&\cR_1^{(3)}=\theta_1 \theta_2 \theta_4,\quad \cR_2^{(3)}=-2 \theta_1 \theta_2 \theta_4 + \theta_1 \theta_2 \theta_5,\quad \cR_3^{(3)}
 	=-\theta_1 \theta_2 \theta_5 + 
 \theta_2 \theta_4 \theta_5 - \theta_3 \theta_4 \theta_5,&\nn\\&\cR_4^{(3)}
 =-\theta_1 \theta_2 \theta_4 + \theta_1 \theta_4 \theta_5 - 
 \theta_2 \theta_4 \theta_5 + \theta_3 \theta_4 \theta_5,\quad \cR_5^{(3)}
 =-\theta_1 \theta_2 \theta_4 - \theta_1 \theta_4 \theta_5 + 
 \theta_2 \theta_4 \theta_5 \ .&\nn
\end{eqnarray}
Our choice for a basis of $H^{(4,4)}(\tilde{X}_4)$ is given by 
$\cR^{(4)}=\tfrac{1}{106}\mathcal C_0|_{J_i\mapsto \theta_i}$.

Again we fix the integral basis of $H^{(2,2)}(\tilde X_4)$ by the requirement of recovering 
the threefold periods from the fourfold ones. We readily identify the K\"ahler classes of the 
threefold $\tilde Z_3$ among the fourfold classes as
\begin{equation} \label{matchFF1}
 	J_2+J_3\ \leftrightarrow\ J_1(\tilde Z_3)\,,\quad J_5\ \leftrightarrow\ J_2(\tilde Z_3)\,,\quad J_4\ \leftrightarrow\ J_3(\tilde Z_3)\,,
\end{equation}
which matches the threefold intersections by identifying $J_1\equiv \tilde Z_3$ in the 
fourfold intersections \eqref{intsFFF1}. Then we calculate the classical terms of the 
threefold periods to be
\begin{equation}
 	\mathbb{L}_1(\tilde Z_3)=\tilde{l}_2 (\tilde{l}_2+\tilde{l}_3 )\ , \quad \mathbb{L}_2(\tilde Z_3)=\tfrac{1}{2} (2\tilde{l}_2+\tilde{l}_3) (2 \tilde{l}_1+4 \tilde{l}_2+\tilde{l}_3)\ ,\quad \mathbb{L}_3(\tilde Z_3)=\tfrac{1}{2} l_2 (2 l_1+3 l_2+2 l_3)\,.
\end{equation}
On the fourfold $X_4$ we determine the periods that match this leading logarithmic structure. 
They are given by 
\begin{eqnarray}
 	\nn&\mathbb{L}^{(2)}_1=X_0l_5(l_4+l_5)\ ,\quad \mathbb{L}^{(2)}_2=\tfrac{1}{2}X_0 (l_4+2 l_5) (2 (l_2+ l_3)+l_4+4 l_5)\ ,&\\
&\quad \mathbb{L}^{(2)}_3=\tfrac{1}{2} X_0l_5 (2 (l_2+ l_3)+2 l_4+3 l_5)\,& 
\end{eqnarray}
and immediately coincide with the threefold result using \eqref{matchFF1}. It can be shown 
explicitly that the instanton series contained in the corresponding full solution matches 
the series on the threefold as well. The threefold invariants $n_{d_1,d_2,d_3}$ are obtained 
as $n_{0,d_1,d_1,d_3,d_2}$ from the fourfold invariants. To these solutions we associate 
using \eqref{eq:intersecsLL} ring elements $\cR^{(2)}_\alpha$, $\alpha=1,3,2$, that we complete 
to a new basis as
\begin{eqnarray}
 	\nn&\tilde{\cR}^{(2)}_1=\tfrac{1}{4} (\theta _1+\theta _2) (\theta _2+\theta _3)\ ,\quad\tilde{\cR}^{(2)}_2=\theta _1 \theta _4\ ,\quad\tilde{\cR}^{(2)}_3=\tfrac{1}{7} \theta _5 (6 \theta _1-\theta _2-2 \theta _4+\theta _5)\ ,&\\ \nn
	&\tilde{\cR}^{(2)}_4=\theta _2 \theta _4\ ,\quad\tilde{\cR}^{(2)}_5=\tfrac{1}{7} \theta _5 (-\theta _1+6 \theta _2-2 \theta _4+\theta _5)\ ,\quad\tilde{\cR}^{(2)}_6=\tfrac{1}{2} \theta _4 (\theta _3+\theta _4)\ ,&\\
&\tilde{\cR}^{(2)}_7=\theta _3 \theta _5\ ,\quad\tilde{\cR}^{(2)}_8=\tfrac{1}{7} \theta _5 (-2 \theta _1-2 \theta _2+3 \theta _4+2 \theta _5)\ ,&
\end{eqnarray}
where we again note that the basis of dual solutions and the new ring basis coincide by 
$l_i\leftrightarrow \theta_i$. Then the integral basis elements read
\begin{equation}
 	\hat{\gamma}^{(2)}_1=\tilde{\cR}^{(2)}_1\Omega_4|_{z=0}\ ,\quad \hat{\gamma}^{(2)}_2=\tilde{\cR}^{(2)}_2\Omega_4|_{z=0}\ ,\quad \hat{\gamma}^{(2)}_3=\tilde{\cR}^{(2)}_3\Omega_4|_{z=0}\ ,
\end{equation}
such that we obtain the full solution with the above leading parts $\mathbb{L}^{(2)}_\alpha$ 
as $\Pi^{(2)}_\alpha=\int\Omega_4\wedge\hat{\gamma}_\alpha$. The leading behavior of the 
periods $\Pi^{(2)\,\alpha}$ is then given as $\mathbb{L}^{(2)\, 1}=X_0(l_1+l_2)(l_2+l_3)$, 
$\mathbb{L}^{(2)\, 2}=X_0l_1 l_4$, $\mathbb{L}^{(2)\, 3}=X_0l _5 (l_1+l_5)$, respectively

We conclude by determining the flux element $\hat \gamma$ in $H^{(2,2)}_H(X_4)$
that reproduces the disk invariants in the phase II of \cite{Aganagic:2001nx}, where the 
local geometry $K_{\mathds{F}_1}\rightarrow \mathds{F}_1$ is considered. First we identify $\ell^{(2)}$ 
of the toric data in \eqref{ellFFF1} as the vector encoding the brane physics. Therefore, 
we expect the fourfold invariants $n_{0,m+d_1,d_1,d_2,0}$ to coincide with the disk 
invariants what can be checked in a direct calculation. The ring element yielding this 
result reads $\hat{\gamma}= \cR_4^{(2)}$ where the free coefficients in front of the other 
ring elements were chosen to vanish. The leading logarithmic parts of the period 
$\int_\gamma \Omega_4$ and of the solution $\Pi^{(2)}_{\gamma}=\int\Omega_4\wedge\hat{\gamma}\equiv W_\text{brane}$ 
respectively read
\begin{equation}
 	\mathbb{L}^{(2)}_{\gamma}=X_0l_5 (l_1+l_2+l_3+l_4+2 l_5)\ ,\quad \mathbb{L}^{(2)\,\gamma}=X_0l_2l_4\, .
\end{equation}

\clearpage
\section{Compact Disk Instantons on $\P^4(1,1,1,6,9)[18]$}
\label{sec:instantonsLV11169}

\begin{table}[!ht]
\centering
$
\scriptscriptstyle
 \begin{array}{|c|rrrrrrrr|}
\hline
\rule[-0.2cm]{0cm}{0.6cm}  i&\!j=0&\!\!j=1&\!\!j=2&\!\!j=3&\!\!j=4&\!\!j=5&\!\!j=6&\!\!j=7\\
\hline
 0&\!0 &\!\! 1 &\!\! 0 &\!\! 0 &\!\! 0 &\!\! 0 &\!\! 0 &\!\! 0 \\
 1&\!2 &\!\! -1 &\!\! -1 &\!\! -1 &\!\! -1 &\!\! -1 &\!\! -1 &\!\! -1 \\
 2&\!-8 &\!\! 5 &\!\! 7 &\!\! 9 &\!\! 12 &\!\! 15 &\!\! 19 &\!\! 23 \\
 3&\!54 &\!\! -40 &\!\! -61 &\!\! -93 &\!\! -140 &\!\! -206 &\!\! -296 &\!\! -416 \\
 4&\!-512 &\!\! 399 &\!\! 648 &\!\! 1070 &\!\! 1750 &\!\! 2821 &\!\! 4448 &\!\! 6868 \\
 5&\!5650 &\!\! -4524 &\!\! -7661 &\!\! -13257 &\!\! -22955 &\!\! -39315 &\!\! -66213 &\!\! -109367 \\
 6&\!-68256 &\!\! 55771 &\!\! 97024 &\!\! 173601 &\!\! 312704 &\!\! 559787 &\!\! 989215 &\!\! 1719248 \\
 7&\!879452 &\!\! -729256 &\!\! -1293185 &\!\! -2371088 &\!\! -4396779 &\!\! -8136830 &\!\! * &\! * \\
 8&\!-11883520 &\!\! 9961800 &\!\! 17921632 &\!\! 33470172 &\!\! * &\!\! * &\!\! * &\!\! * \\
 9&\!166493394 &\!\! -140747529 &\!\! * &\!\! * &\!\! * &\!\! * &\!\! * &\!\! *\\ 
\hline
\end{array}
$
\caption{$k=0$: Disk instanton invariants $n_{i,k, i+j, k}$ on $\P^4(1,1,1,6,9)[18]$ at large volume. $i$, $k$ label the classes $t_1$, $t_2$ of $Z_3$, where $j$ labels the brane winding. These results for $k=0$ agree with phase $I$, $II$ of \cite{Aganagic:2001nx}. Entries $*$ exceed the order of our calculation.}
\end{table}
\begin{table}[!ht]
\centering
$
\scriptstyle
 \begin{array}{|c|rrrrrr|}
\hline
\rule[-0.2cm]{0cm}{0.6cm}  i&j=1&j=2&j=3&j=4&j=5&j=6\\
\hline
 0& ** & 0 & 0 & 0 & 0 & 0 \\
 1& 300 & 300 & 300 & 300 & 300 & 300 \\
 2& -2280 & -3180 & -4380 & -5880 & -7680 & -9780 \\
 3& 24900 & 39120 & 61620 & 95400 & 144060 & 211800 \\
 4& -315480 & -526740 & -892560 & -1500900 & -2477580 & -3996780 \\
 5& 4340400 & 7516560 & 13329060 & 23641980 & 41421000 & 71240400 \\
 6& -62932680 & -111651720 & -204177600 & -375803820 & -686849280 & * \\
 7& 946242960 & 1707713040 & 3192621180 & * & * & *      \\
\hline
\end{array}
$
\caption{$k=1$: Compact disk instanton invariants $n_{i,k, i+j, k}$ for brane phase $I$, $II$ on $\P^4(1,1,1,6,9)[18]$ at large volume. $i$, $k$ label the classes $t_1$, $t_2$ of $Z_3$, where $j$ labels the brane winding. The entry $**$ as well as the $j=0$ column could not be fixed by our calculation.}
\end{table}
\begin{table}[!ht]
\centering
$
 \begin{array}{|c|rrrrrr|}
\hline
\rule[-0.2cm]{0cm}{0.6cm}  i&\!j=2&\!\!j=3&\!\!j=4&\!\!j=5&\!\! j=6&\!\! j=7\\
\hline
 0&\! ** &\!\! 0 &\!\! 0 &\!\! 0 &\!\! 0 &\!\! 0 \\
 1&\! -62910 &\!\! -62910 &\!\! -62910 &\!\! -62910 &\!\! -62910 &\!\! -62910 \\
 2&\!778560 &\!\! 1146690 &\!\! 1622580 &\!\! 2206530 &\!\! 2898240 &\!\! 3698010 \\
 3&\! -12388860 &\!\! -20596140 &\!\! -33454530 &\!\! -52626780 &\!\! -80081460 &\!\! -118092960 \\
 4&\! 208471080 &\!\! 368615070 &\!\! 645132360 &\!\! 1103916150 &\!\! 1838367780 &\!\! 2976756210 \\
 5&\! -3588226470 &\!\! -6587809920 &\!\! -12083913000 &\!\! -21840712470 &\!\! * &\!\! * \\
 6&\! 62538887280 &\!\! 117754228980 &\!\! * &\!\! * &\!\! * &\!\! *\\
\hline
\end{array}
$
\caption{$k=2$: Compact disk instanton invariants $n_{i,k, i+j, k}$ for brane phase $I$, $II$ on $\P^4(1,1,1,6,9)[18]$ at large volume. $i$, $k$ label the classes $t_1$, $t_2$ of $Z_3$, where $j$ labels the brane winding. The entry $**$ as well as the $j=0,\,1$ columns could not be fixed by our calculation.}
\end{table}
\begin{table}[!ht]
\centering
$
 \begin{array}{|c|rrrrrrrr|}
\hline
\rule[-0.2cm]{0cm}{0.6cm}  i&j=0&j=1&j=2&j=3&j=4&j=5& j=6&j=7\\
\hline
  0&0 & -1 & 0 & 0 & 0 & 0 & 0 & 0 \\
  1&** & 2 & 1 & 1 & 1 & 1 & 1 & 1 \\
  2&**& -5 & -4 & -3 & -4 & -5 & -7 & -9 \\
  3&**& 32 & 21 & 18 & 20 & 26 & 36 & 52 \\
  4&**& -286 & -180 & -153 & -160 & -196 & -260 & -365 \\
  5&** & 3038 & 1885 & 1560 & 1595 & 1875 & 2403 & 3254 \\
  6&**& -35870 & -21952 & -17910 & -17976 & -20644 & -25812 & -34089 \\
  7&** & 454880 & 275481 & 222588 & 220371 & 249120 & * & * \\
  8&**& -6073311 & -3650196 & -2926959 & * & * & * & * \\
  9&**& 84302270 & * & * & * & * & * & *\\
\hline
\end{array}
$
\caption{$k=0$: Disk instanton invariants $n_{i,k, i, k+j}$ on $\P^4(1,1,1,6,9)[18]$ at large volume. $i$, $k$ label the classes $t_1$, $t_2$ of $Z_3$, where $j$ labels the brane winding. These results for $k=0$ agree with phase $III$ of \cite{Aganagic:2001nx}.}
\end{table}
\begin{table}[!ht]
\centering
$
 \begin{array}{|c|rrrrrrr|}
\hline
\rule[-0.2cm]{0cm}{0.6cm}  i&j=1&j=2&j=3&j=4&j=5& j=6&j=7\\
\hline
 0&\! 0 &\! 0 &\! 0 &\! 0 &\! 0 &\! 0 &\! 0 \\
 1&\! -540 &\! -300 &\! -300 &\! -300 &\! -300 &\! -300 &\! -300 \\
 2&\! 2160 &\! 1620 &\! 1680 &\! 2280 &\! 3180 &\! 4380 &\! 5880 \\
 3&\! -18900 &\! -12960 &\! -12300 &\! -15000 &\! -21060 &\! -31200 &\! -47220 \\
 4&\! 216000 &\! 140940 &\! 126240 &\! 142380 &\! 185280 &\! 261300 &\! 386160 \\
 5&\! -2800980 &\! -1775520 &\! -1535160 &\! -1653900 &\! -2046060 &\! -2750280 &\! -3896760 \\
 6&\! 39087360 &\! 24316200 &\! 20544720 &\! 21489780 &\! 25725600 &\! * &\! * \\
 7&\! -572210460 &\! -351319680 &\! -292072920 &\! * &\! * &\! * &\! * \\
 8&\! 8663561280 &\! * &\! * &\! * &\! * &\! * &\! *\\
\hline
\end{array}
$
\caption{$k=1$: Compact disk instanton invariants $n_{i,k, i, k+j}$ for brane phase $III$ on $\P^4(1,1,1,6,9)[18]$ at large volume. $i$, $k$ label the classes $t_1$, $t_2$ of $Z_3$, where $j$ labels the brane winding. The $j=0$ column could not be fixed by our calculation.}
\end{table}
\begin{table}[!ht]
\centering
$
 \begin{array}{|c|rrrrrr|}
\hline
\rule[-0.2cm]{0cm}{0.6cm}  i&\!j=1&\!j=2&\!j=3&\!j=4&\!j=5&\! j=6\\
\hline
 0&\! 0 &\! 0 &\! 0 &\! 0 &\! 0 &\! 0 \\
 1&\! ** &\! 62910 &\! 62910 &\! 62910 &\! 62910 &\! 62910 \\
 2&\! -430110 &\! -413640 &\! -557010 &\! -836340 &\! -1223730 &\! -1718880 \\
 3&\! 5190480 &\! 3923100 &\! 4415580 &\! 6237810 &\! 9720180 &\! 15561180 \\
 4&\! -76785570 &\! -52941600 &\! -52475850 &\! -65786040 &\! -93752550 &\! -143003760 \\
 5&\! 1227227760 &\! 806981670 &\! 747944550 &\! 869842800 &\! 1154721060 &\! *\\
 6&\! -20387141100 &\! -13027278600 &\! -11592978930 &\! * &\! * &\! * \\
 7&\! 346430247840 &\! * &\! * &\! * &\! * &\! *\\
\hline
\end{array}
$
\caption{$k=2$: Compact disk instanton invariants $n_{i,k, i, k+j}$ for brane phase $III$ on $\P^4(1,1,1,6,9)[18]$ at large volume. $i$, $k$ label the classes $t_1$, $t_2$ of $Z_3$, where $j$ labels the brane winding. The entry $**$ and the $j=0$ column could not be fixed by our calculation.}
\end{table}
\clearpage

{
\bibliography{mybib}{}
\bibliographystyle{utphys}
\addcontentsline{toc}{chapter}{\numberline{}{Bibliography}}
}

\end{document}